\documentclass[twocolumn,preprint,twocolappendix]{aastex63}
\usepackage{graphicx}

\begin{document}

\newcommand{\hst}{\textit{HST}}
\newcommand{\mbarj}{$\overline m_{110}$}
\newcommand{\Mbarj}{$\overline M_{110}$}
\newcommand\gminz{\ensuremath{(g{-}z)}}
\newcommand\jminh{\ensuremath{(J{-}H)}}
\newcommand\gzacs{\ensuremath{(g_{475}{-}z_{850})}}
\newcommand\JHwfc{\ensuremath{(J_{110}{-}H_{160})}}
\newcommand\JminH{\ensuremath{(J{-}H)}}
\defcitealias{Jensen2015}{J15}

\title{Infrared Surface Brightness Fluctuation Distances for MASSIVE \\ and Type Ia Supernova Host Galaxies \footnote{Based on observations with the NASA/ESA \textit{Hubble Space Telescope}, obtained at the Space Telescope Science Institute, which is operated by AURA, Inc., under NASA contract NAS 5-26555. These observations are associated with GO Programs \#11711, \#11712, \#12450, \#14219, \#14654, \#14771, \#14804, \#15265, and \#15329}}

\author[0000-0001-8762-8906]{Joseph B. Jensen}
\affil{Department of Physics, Utah Valley University, 800 W. University Parkway, MS 179, Orem, UT 84058, USA}  

\author[0000-0002-5213-3548]{John P. Blakeslee}
\affiliation{Gemini Observatory and NSF’s NOIRLab, 950 N. Cherry Ave., Tucson, AZ 85719, USA}

\author[0000-0002-4430-102X]{Chung-Pei Ma}
\affiliation{Department of Astronomy and Department of Physics, University of California, Berkeley, CA 94720, USA}

\author[0000-0002-0370-157X]{Peter A. Milne}
\affiliation{University of Arizona, Steward Observatory, 933 N. Cherry Avenue, Tucson, AZ 85721, USA}

\author[0000-0001-6272-5507]{Peter J. Brown}
\affiliation{George P. and Cynthia Woods Mitchell Institute for Fundamental Physics and Astronomy, Texas A\&M University, College Station, TX 77843, USA}

\author[0000-0003-2072-384X]{Michele Cantiello}
\affiliation{INAF Osservatorio Astronomico d’Abruzzo, via Maggini, snc, 64100, Italy}

\author[0000-0003-4069-2817]{Peter M. Garnavich}
\affiliation{Department of Physics, University of Notre Dame, Notre Dame, IN 46556, USA}

\author[0000-0002-5612-3427]{Jenny E. Greene}
\affiliation{Department of Astrophysical Sciences, Princeton University, Princeton, NJ 08544, USA}

\author[0000-0002-9748-961X]{John R. Lucey}
\affiliation{Centre for Extragalactic Astronomy, University of Durham, Durham DH1 3LE, UK}

\author{Anh Phan}
\affiliation{Department of Physics, Utah Valley University, 800 W. University Parkway, MS 179, Orem, UT 84058, USA}

\author[0000-0002-9291-1981]{R. Brent Tully}
\affiliation{Institute for Astronomy, University of Hawaii, 2680 Woodlawn Dr., Honolulu, Hawaii, USA}

\author[0000-0003-4773-4602]{Charlotte M. Wood}
\affil{Department of Physics, University of Notre Dame, Notre Dame, IN 46556, USA}

\correspondingauthor{Joseph B. Jensen}
\email{jjensen@uvu.edu}


\begin{abstract}
We measured high-quality surface brightness fluctuation (SBF) distances for a sample of 63 massive early-type galaxies using the WFC3/IR camera on the Hubble Space Telescope. The median uncertainty on the SBF distance measurements is 0.085~mag, or 3.9\% in distance. Achieving this precision at distances of 50 to 100 Mpc required significant improvements to the SBF calibration and data analysis procedures for WFC3/IR data. Forty-two of the galaxies are from the MASSIVE Galaxy Survey, a complete sample of massive galaxies within ${\sim}100$~ Mpc; the SBF distances for these will be used to improve the estimates of the stellar and central supermassive black hole masses in these galaxies. Twenty-four of the galaxies are Type Ia supernova hosts, useful for calibrating SN Ia distances for early-type galaxies and exploring possible systematic trends in the peak luminosities. Our results demonstrate that the SBF method is a powerful and versatile technique for measuring distances to galaxies with evolved stellar populations out to 100~Mpc and constraining the local value of the Hubble constant.
\\
\end{abstract}

\section{Introduction}

To understand the expansion history and contents of the universe, we must be able to measure accurate extragalactic distances with high precision and low systematic uncertainty well out into the Hubble flow. We have measured high-precision surface brightness fluctuation \citep[SBF,][]{Tonry1988} distances to 63 galaxies out to 100~Mpc to answer specific questions related to the most important issues in the extragalactic distance scale \citep[e.g.,][]{Blakeslee2021,Verde2019,Cantiello2018}. The SBF technique uses the Poisson statistics of discrete stars to determine the mean brightness of red giant branch (RGB) stars in a distant galaxy, even when such stars cannot be individually resolved \citep{Tonry1988,Jensen1998,Blakeslee2009}.

The SBF technique is particularly powerful for determining the distances to massive elliptical galaxies compared to other methods \citep{Jensen2003,Blakeslee2009}. It does not require the serendipitous discovery of a supernova explosion or assumptions about the relative velocities and distances of elliptical and spiral galaxies in a given group or galaxy cluster \citep[][for example]{Riess2021}. SBF reaches distances far greater than Cepheid variable stars or other techniques that depend on resolving individual stars such as the tip of the red giant branch \citep[][and references therein]{Freedman2020}. It is also independent of the dynamics or mass of the target galaxy. 

By measuring the power in the spatial Fourier power spectrum of an early-type galaxy (ETG) with globular clusters and background galaxies removed, we can determine the mean apparent luminosity of RGB stars in the galaxy statistically without having to measure them individually \citep{Tonry1988,Jensen2003}. The distance to the galaxy can then be computed using an empirical SBF calibration from which the absolute magnitude of the fluctuations has been determined as a function of the age and metallicity of the galaxy's stellar population \citep{Jensen2003,Jensen2015}. Because RGB stars are very bright in the near-IR, the SBF signal at these wavelengths is very strong and measurable out to ${\sim}100$~Mpc with \hst\ \citep{Jensen2001} in modest exposure times.
 IR SBF with WFC3/IR has now been developed as a technique capable of achieving ${\sim}5$\% distances out to 70~Mpc in a single \hst\ orbit, as outlined by \citet[][hereafter \citetalias{Jensen2015}]{Jensen2015} and demonstrated in a few individual cases \citep{Cantiello2018,Nguyen2020,Liepoldetal2020}. 

The IR SBF technique for \emph{HST} WFC3/IR was calibrated by \citetalias{Jensen2015} using 16 galaxies in the Fornax and Virgo clusters with well-measured optical SBF distances. The optical SBF distances were taken from the large Advanced Camera for Surveys (ACS) Virgo  and Fornax Cluster Surveys (ACSVCS and ACSFCS) \citep{Mei2007,Blakeslee2009} based on Cepheid variable star distances to the Large Magellanic Cloud (LMC), as described below in Section~\ref{calibrationsection}. 
In this paper we describe additional improvements to the \citetalias{Jensen2015} procedures to further reduce the uncertainties of the SBF distance measurements, including (i) better measurements of the point-spread function (PSF); (ii) improved measurements of the instrumental background via reference to galaxy profiles from the 2MASS atlas images; (iii) more accurate measurements of the globular cluster and background galaxy luminosity functions; (iv) additional optical and IR color measurements from PanSTARRS, SDSS, and 2MASS for calibration of the absolute fluctuation magnitude; and (v) an updated distance to the LMC and Cepheid-derived SBF zero point.  These improvements are described in detail below.
We have now demonstrated the efficacy of the IR SBF methodology out to distances of 100~Mpc, taking advantage of the intrinsic IR brightness of the red giant branch stars that give rise to the fluctuations to reach much greater distances than achievable with optical SBF. 

\section{The IR SBF Survey Sample}

There are two main subsamples in the current study. First, we observed 42 massive early-type galaxies in the MASSIVE sample \citep{Ma2014} for the purpose of better constraining the relationship between black hole mass and the bulk dynamics of the galaxy. Galaxy evolution is linked to the growth of the central black hole in most galaxies, but the scatter in those relationships is often dominated by distance uncertainties. The second large subsample of this project was observed to explore how type Ia supernova luminosity correlates with galaxy mass or environment by measuring accurate distances to early-type SN host galaxies \citep{Milne2015}. The SN host distance measurements will help address the disagreement between values of the Hubble constant $H_0$ measured using SN Ia and Cepheids in spiral galaxies, and values inferred from the era of recombination \citep{Riess2019,Planck2020}. Both of these topics are discussed in detail below.
The target galaxies were all observed using WFC3/IR in the F110W filter for one or two orbits, sufficient for a precise distance measurement to 70 or 100~Mpc, respectively. The specific targets and programs are:
\begin{itemize}
    \item GO-11711: NGC~4874 in the Coma cluster (PI J. Blakeslee).
    \item GO-12450: NGC~3504 (PI C. Kochanek), previously published by \citet{Nguyen2020}.
    \item GO-14219: 35 early-type galaxies in the MASSIVE survey sample (PI J. Blakeslee).
    \item GO-14645: 19 type Ia SN early-type host galaxies (PI P. Milne).
    \item GO-15265: 6 additional distant MASSIVE sample galaxies (PI J. Blakeslee).
    \item GO-14771, GO-14804, and GO-15329: NGC~4993, host of the merging neutron stars detected through gravitational waves (PIs N. Tanvir, A. Levan, and E. Berger), previously published by \citet{Cantiello2018}. 
\end{itemize}
In addition to the galaxy distances from these programs, we also reprocessed the 16 Virgo and Fornax cluster galaxies used for the SBF calibration (GO-11712, PI J. Blakeslee), previously published by \citetalias{Jensen2015}.

\subsection{The MASSIVE Survey}
The largest component (42/63) of our SBF distance database are the MASSIVE sample galaxies. MASSIVE is a volume-limited survey \citep{Ma2014} of the ${\sim}100$ most massive ETGs within a distance of 108 Mpc in the northern sky ($\delta > -6$ degrees).  Within this volume, the sample is complete to an absolute $K$-band magnitude of $M_K\,{=}\,-25.3$ mag, or a stellar mass of $M_{*}\,{\sim}\,10^{11.5}\ {\rm M}_\odot$, with no selection cuts on galaxy velocity dispersion, size, or environment.  The MASSIVE survey goal is to obtain a comprehensive census of all the major constituents---luminous and dark---of local galaxies in the highest-mass regime.   To achieve these goals, the MASSIVE team has acquired extensive data using optical integral-field spectrographs (IFS) on both sub-arcsecond and arc-minute scales, and obtained uniform measurements of spatially resolved stellar kinematics \citep{Vealeetal2017b,Vealeetal2017a,Vealeetal2018,Eneetal2018,Eneetal2019,Eneetal2020}, stellar populations \citep{Greeneetal2015,Greeneetal2019} and ionized gas kinematics \citep{Pandyaetal2017}. They also obtained radio and X-ray data to investigate the properties of cold molecular gas \citep{Davisetal2016,Davisetal2019} and hot X-ray gas in these massive ETGs \citep{Gouldingetal2016,Voitetal2018}.

Using the homogeneous set of sub-arcsecond and wide-field IFS and photometric data as constraints, the MASSIVE collaboration is performing dynamical mass modeling of the supermassive black holes (SMBH), stars, and dark matter components for a sample of MASSIVE galaxies \citep{Liepoldetal2020, Quenneville2021}. Accurate and precise distances are essential to measuring the most fundamental properties of massive ETGs and their central black holes, including their masses. The largest ETGs contain the most massive black holes (BHs), the most extreme stellar initial mass functions \citep[IMFs,][]{Cappellari2012, vanDokkum2010},  
and the most dramatic size evolution over cosmic time \citep[e.g.,][]{Collins2009,DeLucia2007}. The primary goals of MASSIVE include precisely constraining the SMBH-galaxy scaling relations, the stellar IMF, and the late-time assembly histories of elliptical galaxies. 

The purpose of the current study with respect to MASSIVE is to use high-resolution F110W (1.1 $\mu$m) infrared images to determine high-precision SBF distances \citep{Tonry1988} for all galaxies in the MASSIVE sample within 70~Mpc for which SBF distances were not previously available. 
Accurate SBF distances are needed to remove potentially large errors from peculiar velocities, especially in the high-density galaxy cluster regions typically occupied by massive ETGs, or systematic errors from heterogeneous distance measurement methods that would otherwise be used to determine the SMBH-galaxy mass scaling relations. Distance errors affect BH masses and galaxy properties in dissimilar ways, and thus can bias both the \emph{scatter} and \emph{slopes} of the scaling relations. The measured distances and central profiles will be used to constrain the stellar, dark matter, and BH masses to high precision, giving unprecedented insight into the formation and evolution of the most massive galaxies in the local universe.

The IR SBF distance technique was recently used to measure the stellar dynamical mass of the SMBH in NGC~1453, one of the MASSIVE targets in the current SBF sample \citep{Liepoldetal2020}; the 5\% SBF distance from this study reported therein was instrumental to the interpretation of the SMBH mass being consistent with the mass and galaxy scaling relations in the potential of that galaxy. Another recent MASSIVE study by \citet{Goullaud2018} used the high-resolution WFC3/IR F110W images from this study to determine the central luminosity profiles of 35 MASSIVE galaxies, greatly reducing the degeneracy between $M/L$ and BH masses in dynamical orbit modeling. Another example of how SBF distances can help constrain BH masses even in some later-type galaxies is the recent molecular gas dynamical measurement by \citet{Nguyen2020} using the SBF technique to determine the distance and BH mass in NGC~3504. Prior to the SBF measurement, the best Tully Fisher distance to NGC~3504 was 13.6~Mpc \citep{Russell2002}, which is more than a factor of two smaller than the SBF distance. Using the shorter distance would have led to a significant difference in the derived BH mass.

\subsection{Type Ia Supernovae}
The SBF distance database also includes 24 Type Ia supernova (SN) host galaxies (five of which are also MASSIVE galaxies). These galaxies were observed to better understand the consistency of SN Ia luminosity as a function of galaxy type and environment.
SN Ia are one of the highest-precision distance indicators for cosmology and have been used to measure the Hubble constant \citep{Riess2021, Freedman2020, Khetan2021, Beaton2018}. Precise and accurate distance measurements are critical to resolving the discrepancy between local values of $H_0$ and those inferred from high-redshift cosmic microwave background radiation measurements \citep{Riess2019,Planck2020}. Resolving this discrepancy could reveal systematic errors in the distance measurements, provide new insight into the physics of the early universe, or both.

\startlongtable
\begin{deluxetable*}{lccccCcl}
\tabletypesize{\footnotesize}
\tablecaption{WFC3/IR F110W SBF Observation and Morphology Data \label{obstable}}
\tablehead{
\colhead{Galaxy} &
\multicolumn{2}{c}{\emph{HST} Program} &
\colhead{$t_{\rm exp}$} &
\colhead{Extinction} &
\colhead{Background} &
\colhead{Type} &
\colhead{Notes} \\
& \colhead{ID} & 
\colhead{sample} &
\colhead{(s)} & 
\colhead{(mag)} &
\colhead{(e$^-$s$^{-1}$pix$^{-1}$)} &
\colhead{$T$} & 
}
\colnumbers
\startdata
IC~2597  & 14654 & SN & 2496 & 0.062 & 0.88 \pm 0.12 & $-3.9$ & \\
NGC~0057 & 14219 & M  & 2496 & 0.068 & 1.28 \pm 0.08 & $-4.9$ & \\
NGC~0315 & 14219 & M  & 2496 & 0.057 & 1.20 \pm 0.23 & $-4.1$ & ND \\
NGC~0383 & 14219 & M  & 2496 & 0.062 & 0.88 \pm 0.11 & $-2.9$ & C \\
NGC~0410 & 14219 & M  & 2496 & 0.052 & 1.80 \pm 0.07 & $-4.3$ & Sh \\
NGC~0495 & 14654 & SN & 2496 & 0.063 & 1.00 \pm 0.40 & $+0.2$ & SB \\
NGC~0507 & 14219 & M  & 2496 & 0.055 & 1.80 \pm 0.16 & $-3.3$ & Sh, C \\
NGC~0524 & 14654 & SN & 2496 & 0.073 & 1.60 \pm 0.40 & $-1.2$ & Sh, ND \\
NGC~0533 & 14219 & M  & 2496 & 0.027 & 1.76 \pm 0.08 & $-4.9$ & \\
NGC~0545 & 14219 & M  & 2496 & 0.036 & 1.44 \pm 0.06 & $-2.9$ & C \\
NGC~0547 & 14219 & M  & 2496 & 0.036 & 1.59 \pm 0.06 & $-4.8$ & C \\
NGC~0665 & 14219 & M  & 2496 & 0.078 & 1.28 \pm 0.08 & $-2.0$ & SB, D \\
NGC~0708 & 14219 & M  & 2496 & 0.079 & 1.36 \pm 0.08 & $-4.9$ & ND, C \\
NGC~0741 & 14219 & M  & 2496 & 0.046 & 1.28 \pm 0.12 & $-4.8$ & C \\
NGC~0777 & 14219 & M  & 2496 & 0.041 & 1.44 \pm 0.08 & $-4.8$ & \\
NGC~0809 & 14654 & SN & 2496 & 0.021 & 1.40 \pm 0.04 & $-1.8$ & R \\
NGC~0890 & 14219 & M,SN & 2496 & 0.068 & 1.88 \pm 0.20 & $-2.6$ & \\
NGC~0910 & 14654 & SN & 5293 & 0.051 & 1.42 \pm 0.06 & $-4.1$ & \\
NGC~1016 & 14219 & M  & 2496 & 0.027 & 1.32 \pm 0.14 & $-4.9$ &\\
NGC~1060 & 14219 & M  & 2496 & 0.171 & 1.56 \pm 0.15 & $-3.0$ & \\
NGC~1129 & 14219 & M  & 2496 & 0.099 & 1.56 \pm 0.09 & $-4.9$ & C \\
NGC~1167 & 14219 & M  & 2496 & 0.160 & 1.60 \pm 0.08 & $-2.4$ & Sp, D \\
NGC~1200 & 14654 & SN & 2496 & 0.064 & 1.24 \pm 0.08 & $-2.9$ &  \\
NGC~1201 & 14654 & SN & 1997 & 0.014 & 1.30 \pm 0.15 & $-2.6$ & \\
NGC~1259 & 14654 & SN & 5293 & 0.140 & 1.34 \pm 0.04 & $-3.0$ & \\
NGC~1272 & 14219 & M,SN & 2496 & 0.142 & 1.16 \pm 0.04 & $-4.0$ & C \\
NGC~1278 & 14654 & SN & 2496 & 0.145 & 1.40 \pm 0.12 & $-4.8$ & C \\
NGC~1453 & 14219 & M  & 2496 & 0.093 & 1.74 \pm 0.16 & $-4.7$ & \\
NGC~1573 & 14219 & M  & 2496 & 0.121 & 3.00 \pm 0.14 & $-4.8$ & \\
NGC~1600 & 14219 & M  & 2496 & 0.038 & 1.60 \pm 0.19 & $-4.6$ & \\
NGC~1684 & 14219 & M  & 2496 & 0.051 & 1.56 \pm 0.08 & $-3.9$ & ND \\
NGC~1700 & 14219 & M  & 2496 & 0.038 & 1.44 \pm 0.18 & $-4.7$ & \\
NGC~2258 & 14219 & M,SN & 2496 & 0.113 & 2.08 \pm 0.12 & $-2.0$ & C \\ 
NGC~2274 & 14219 & M  & 2496 & 0.092 & 1.56 \pm 0.19 & $-4.8$ & \\
NGC~2340 & 15265 & M  & 2812 & 0.065 & 1.32 \pm 0.36 & $-4.8$ & Sh \\
NGC~2513 & 14219 & M  & 2496 & 0.020 & 2.00 \pm 0.06 & $-4.9$ & \\
NGC~2672 & 14219 & M  & 2496 & 0.019 & 1.64 \pm 0.19 & $-4.8$ & C \\
NGC~2693 & 14219 & M  & 2496 & 0.017 & 1.40 \pm 0.26 & $-4.8$ & C \\
NGC~2765 & 14654 & SN & 2496 & 0.028 & 1.92 \pm 0.20 & $-1.9$ & R \\
NGC~2962 & 14654 & SN & 2496 & 0.051 & 1.76 \pm 0.08 & $-1.1$ & R \\
NGC~3158 & 15265 & M,SN & 2612 & 0.012 & 1.26 \pm 0.19 & $-4.8$ & \\
NGC~3392 & 14654 & SN & 2695 & 0.012 & 1.16 \pm 0.04 & $-3.7$ & ND \\
NGC~3504 & 12450 &    & 1398 & 0.023 & 1.64 \pm 0.14 & $+2.1$ & Sp, D \\
NGC~3842 & 15265 & M  & 2612 & 0.019 & 1.34 \pm 0.19 & $-4.9$ & C \\
NGC~4036 & 14654 & SN & 2695 & 0.021 & 1.63 \pm 0.19 & $-2.5$ & R \\
NGC~4073 & 15265 & M  & 2612 & 0.024 & 2.53 \pm 0.19 & $-4.1$ & \\
NGC~4386 & 14654 & SN & 2695 & 0.034 & 1.22 \pm 0.19 & $-2.0$ & R \\
NGC~4839 & 15265 & M  & 2612 & 0.009 & 1.34 \pm 0.19 & $-3.9$ & C \\
NGC~4874 & 11711 & M  & 4794 & 0.008 & 1.56 \pm 0.10 & $-3.6$ & C \\
NGC~4914 & 14219 & M  & 2496 & 0.012 & 0.92 \pm 0.08 & $-4.0$ & \\
NGC~4993 & 14771+14804 & BNS & 893 & 0.109 & 4.38\tablenotemark{a} & $-3.0$ & Sh, D \\
         & 15329 &    & 1012 &       & 4.00\tablenotemark{a} & & \\
NGC~5322 & 14219 & M  & 2496 & 0.012 & 1.60 \pm 0.20 & $-4.8$ & \\
NGC~5353 & 14219 & M,SN & 2496 & 0.011 & 1.32 \pm 0.21 & $-2.0$ & R, C\\
NGC~5490 & 14654 & SN & 2496 & 0.024 & 0.92 \pm 0.16 & $-4.9$ & \\
NGC~5557 & 14219 & M  & 2496 & 0.011 & 1.16 \pm 0.08 & $-4.8$ & \\
NGC~5839 & 14654 & SN & 2496 & 0.046 & 1.64 \pm 0.08 & $-2.0$ & Sp, R \\
NGC~6482 & 14219 & M  & 2496 & 0.089 & 0.88 \pm 0.04 & $-4.9$ & \\
NGC~6702 & 14654 & SN & 2496 & 0.094 & 0.84 \pm 0.08 & $-4.9$ & ND \\
NGC~6964 & 14654 & SN & 2496 & 0.087 & 1.76 \pm 0.08 & $-4.5$ & \\
NGC~7052 & 14219 & M  & 2496 & 0.108 & 0.88 \pm 0.28 & $-4.9$ & \\
NGC~7242 & 15265 & M  & 2612 & 0.133 & 1.38 \pm 0.19 & $-4.0$ & C \\
NGC~7619 & 14219 & M  & 2496 & 0.072 & 1.56 \pm 0.06 & $-4.8$ & \\
ESO125-G006 & 14654 & SN & 5293 & 0.200 & 1.06 \pm 0.06 & $-3.5$ & Sh \\
\enddata
\vspace{0.5cm}
\tablenotemark{a}{The elevated background for the NGC~4993 observations resulted from the unusually small solar angle required to observe this galaxy during the period when the kilonova optical afterglow was still visible \citep{Cantiello2018}.}
\tablenotetext{}{Column notes: 
(1) Galaxy name; 
(2) \emph{HST} Program ID; 
(3) SBF sub-sample: M=MASSIVE, SN=supernova host, BNS=binary neutron star merger;
(4) Exposure time; 
(5) F110W extinction from \citet{SF2011} retrieved from NED;
(6) F110W instrumental background; 
(7) Galaxy T-type from the HyperLEDA database \citep{leda2014}, where $T\,{<}\,-3$ are elliptical galaxies, $-3\,{<}\,T\,{<}0$ are S0 galaxies, and spirals have $T\,{>}\,0$; 
(8) Morphology notes: C=bright companions, D=dust, ND=nuclear dust, Sh=dynamical shells or arcs, Sp=spiral structure, SB=Barred spiral, R=rings and other residual disk galaxy subtraction artifacts, such as from edge-on disks or boxy structures.}
\end{deluxetable*}

There are several motivations for measuring distances to SN~Ia in early-type hosts.
The width of SN~Ia light curves is correlated with host galaxy type, with slowly-declining, bright SN~Ia typically found in late-type hosts while fast-declining, subluminous SN~Ia are often found in early-type hosts \citep{Hamuy1995}. While light curve fitting techniques that correct for decline rate do not show significant residuals with host type, there is clear evidence that host galaxy properties have a systematic effect on SN~Ia $H_0$ residuals \citep[e.g.,][]{Lampeitl2010}. SN~Ia luminosities are typically calibrated using distances derived from Cepheids, which are found in young stellar populations in star-forming spiral galaxies. However, SN~Ia in the Hubble flow occur in a wide range of galaxy types and local environments. \citet{Rigault2020,Rigault2015} found that $H_0$ can be biased up to ${\sim}5$\% if the SN in the Hubble flow are preferentially located in older stellar populations than those used for Cepheid calibration; the effect is smaller with different sample selection cuts \citep{Jones2018}. There are also differences in SN~Ia properties between passively and actively star-forming hosts \citep[e.g.,][]{Sullivan2010, Kang2016}. In addition, \citet{Milne2013} discovered two previously unrecognized sub-populations of normal SN Ia, one of which preferentially occurs in early-type hosts. Therefore an independent determination of SN Ia distances in early-type hosts is important to understanding the luminosity variations arising from host galaxy type.

It is currently unclear if all faint, fast-declining SN~Ia, which are most common in early-type hosts, can be described along a continuum from normal SN~Ia to an extreme of faint 1991bg-like SN~Ia or if the faint events represent an entirely separate class \citep{Kattner2012, Burns2014, Dhawan2017}. This sample of 24 distance measurements to early-type SN host galaxies is an important step towards understanding the role of galaxy type in the SN distance scale.

One published example of how IR SBF has been used to better understand the properties of a stellar explosion is the 2017 binary neutron star merger in NGC~4993 that was first observed as a gravitational wave event. \citet{Cantiello2018} used WFC3/IR SBF measurements to provide the highest precision distance measurement of the galaxy to-date, which provided a better constraint on the viewing angle with respect to the rotation axis of the merging binary pair and the orientation of its relativistic jets.

The following section describes in detail the procedures used to make the SBF distance measurements to the galaxies reported in this study. Section~\ref{calibrationsection} discusses the SBF calibration zero point. Section~\ref{colortransformations} contains transformations for galaxy colors, which are needed to correct fluctuation magnitudes for population effects and compute distances. Uncertainties are discussed in Section~\ref{uncertaintysection} and the SBF distances are presented in Section~\ref{distancesection}. Alternative calibrations and methods for determining SBF distances using other photometric systems are included in the Appendix.

\section{SBF Analysis Procedures for WFC3/IR}

The SBF method for measuring extragalactic distances was first described by \citet{Tonry1988} and the basic procedures as applied in the infrared were outlined by \citet{Jensen1998,Jensen2001,Jensen2003}. \citetalias{Jensen2015} provided the calibration and procedures for WFC3/IR, which are updated below.

\subsection{Initial Pipeline Processing and Analysis}

The basic image processing steps followed those described in detail by \citetalias{Jensen2015} with a few modifications to improve reliability and  reduce uncertainties. The first step in the SBF data reduction process was to produce a combined, cleaned, background-subtracted image. To avoid correlated noise between pixels, which leads to a variable noise component in the spatial power spectrum, we chose to perform the SBF analysis using the native WFC3/IR pixels and integer pixel shifts when combining, even though doing so results in a $\sim$10\% difference between the $x$ and $y$ plate scales (due to the focal plane distortion from the WFC3/IR optics). We used the raw images from the \hst\ archive without the geometric correction usually applied using \emph{astrodrizzle}. Processed images are shown in Figure~\ref{galaxyfig}. The uncorrected images were  used in all subsequent SBF analysis steps except when comparing galaxy surface brightness profiles with other survey images.

Approximately 40\% of the F110W images in our study were affected by the diffuse upper-atmosphere helium emission line at 1.083 $\mu$m \citep{Brammer2014}. During each MULTIACCUM exposure,\footnote{See \url{https://www.stsci.edu/hst/instrumentation/wfc3} for a description of the WFC3/IR readout modes.} multiple reads of the detector allowed us to determine the signal in each pixel at multiple points during each exposure, and therefore identify pixels that were affected by saturation or cosmic rays, which can be identified as a break in the count rate of electrons being detected in a pixel. The He emission, which appears as a flickering variable background, also causes the signal rate to change during a MULTIACCUM exposure. Gabriel Brammer at STScI has written a specialized Python program to correct for variable backgrounds and events like satellite passages in MULTIACCUM sequences that affect the flux slope in a large number of pixels at once.\footnote{\url{https://github.com/gbrammer/wfc3}}
The correction algorithm compares the count rate over a large region of the detector in the first and last reads of a MULTIACCUM sequence, and then flattens the sequence of reads so that each individual subframe has the same average background count rate.
We used this algorithm to identify individual subframes that had varying background and re-processed all images to have flattened (linear) background count rates. The new \emph{flt} files were then combined using integer pixel offsets to produce the final images used for SBF analysis.

\subsection{Background Determination and Subtraction}

The next step in the SBF procedure was to determine the correct residual background level. The fluctuation power must be normalized by the mean galaxy brightness at a particular location, so any uncertainty in the background level affects the SBF amplitude measurement and its uncertainty. Measuring the background level was an iterative process that included several different complementary methods, most of which are described in detail by \citet{Goullaud2018} and summarized here.

We started with an estimate of the maximum possible sky value by measuring the mean level in the corners of the image, and then used $r^{1/4}$ galaxy fits to make a more realistic estimate of the residual galaxy flux in the corners. We made an elliptical model of each galaxy using the ELLIPROF routine \citep{Tonry1997, Jordan2004} by determining the surface brightness in annuli that were allowed to vary in ellipticity, position, and orientation.
The galaxy model is also a key ingredient in the SBF measurement, since the fluctuations are normalized by the galaxy surface brightness model. The surface brightness profiles from the elliptical fits were compared to $r^{1/4}$ profiles to determine the likely residual background level. Several galaxies have bright neighbor galaxies in the WFC3/IR field of view that also had to be removed by iteratively fitting and subtracting elliptical models for each galaxy one at a time, and masking all other compact galaxies and point sources.

Many of the galaxies do not follow a $r^{1/4}$ profile, so an improved sky estimate was determined by matching the F110W surface brightness profiles (in their geometrically-corrected forms from the \emph{HST} pipeline for consistent comparison) to the 2MASS $J$-band radial profiles \citep{Jarrett2000}. 
The 2MASS profiles were measured in elliptical annuli after subtracting the sky background level measured well away from the galaxies, much farther than possible in the limited field of view of the WFC3/IR frames (the MASSIVE sample galaxies extend well beyond the edges of the WFC3/IR field of view, making direct sky estimation difficult).
The F110W profiles generally agree very well with the 2MASS profiles; in some cases they disagreed due to the presence of companions or other morphological oddities.
Adjustments to the background estimates were made to improve agreement with 2MASS profiles, and the results were published by \citet{Goullaud2018}.

To ensure the best possible SBF measurements, the background levels reported by \citet{Goullaud2018} were revised to take into account the radial behavior of the SBF signal. The SBF amplitude depends on distance and stellar population.
Making the justifiable assumption that all the stars in the galaxy are at the same distance and that any trend in stellar population age or metallicity with radius is revealed by a \gminz\ color gradient (if any), we refined the background levels from \citet{Goullaud2018} to minimize the scatter between the radial regions in which the SBF amplitude was measured. The updated background levels are listed in Table~\ref{obstable}.
The background subtraction uncertainties for the MASSIVE galaxies were taken from \citet{Goullaud2018} and computed for the rest of the galaxies using the technique therein.
The average F110W background level for these data, revised as described above and excluding NGC~4993, is 1.50 e$^-$s$^{-1}$, with a minimum of 0.88 and maximum of 3.0 e$^-$s$^{-1}$ (the NGC~4993 observations were scheduled with an unusually small solar angle to observe the fading optical afterglow of the merging neutron star hypernova explosion). This corresponds to an average surface brightness of 21.95 mag arcsec$^{-2}$ and a range of 21.16 to 22.50 mag arcsec$^{-2}$. The mean background level reported by \citet{Goullaud2018} was 1.78 e$^-$s$^{-1}$ pix$^{-1}$, or 21.77 AB mag arcsec$^{-2}$.
These values are also consistent with the \emph{HST} published values for the F110W background \citep{Pirzkal2014} and the mean background level of 1.3 e$^-$ s$^{-1}$ pix$^{-1}$ measured by \citetalias{Jensen2015}, given the standard deviation of 0.4 e$^-$s$^{-1}$pix$^{-1}$ between individual observations.

\subsection{Creating a Smooth Galaxy Model}

After background subtraction, each galaxy in our sample was fitted using the ELLIPROF isophotal fitting routine
a second time. To get the best possible fits for the galaxies with bright companions, we computed the fit to each of the galaxies iteratively and independently, starting at smaller radii and progressing to larger distances from the center, removing companions by degrees and carefully masking other compact objects. The iterations yielded clean fits to most of the galaxies. The smooth galaxy models were used to plot $r^{1/4}$ profiles, check for residual non-zero background levels, and saved for later use to normalize the SBF power spectrum.
In a few cases, neighboring galaxies, central dust lanes, shell structure, and spiral structure or bars left significant residual patterns after fitting and removing the elliptical profile, which we removed using a spline fitting routine on a grid with 8 to 16 nodes in each pixel dimension. This residual fit procedure usually cleaned up the residuals, but we avoided fitting the SBF spatial power spectrum in regions with persistent residuals and for wavenumbers less than 25 to 30 pix$^{-1}$ because the residual fit artificially suppresses power at these scales. For such galaxies, a fit to the lower wavenumbers in the power spectrum would not have been possible anyway.

\begin{figure*}
\begin{center}
\includegraphics[scale=0.2]{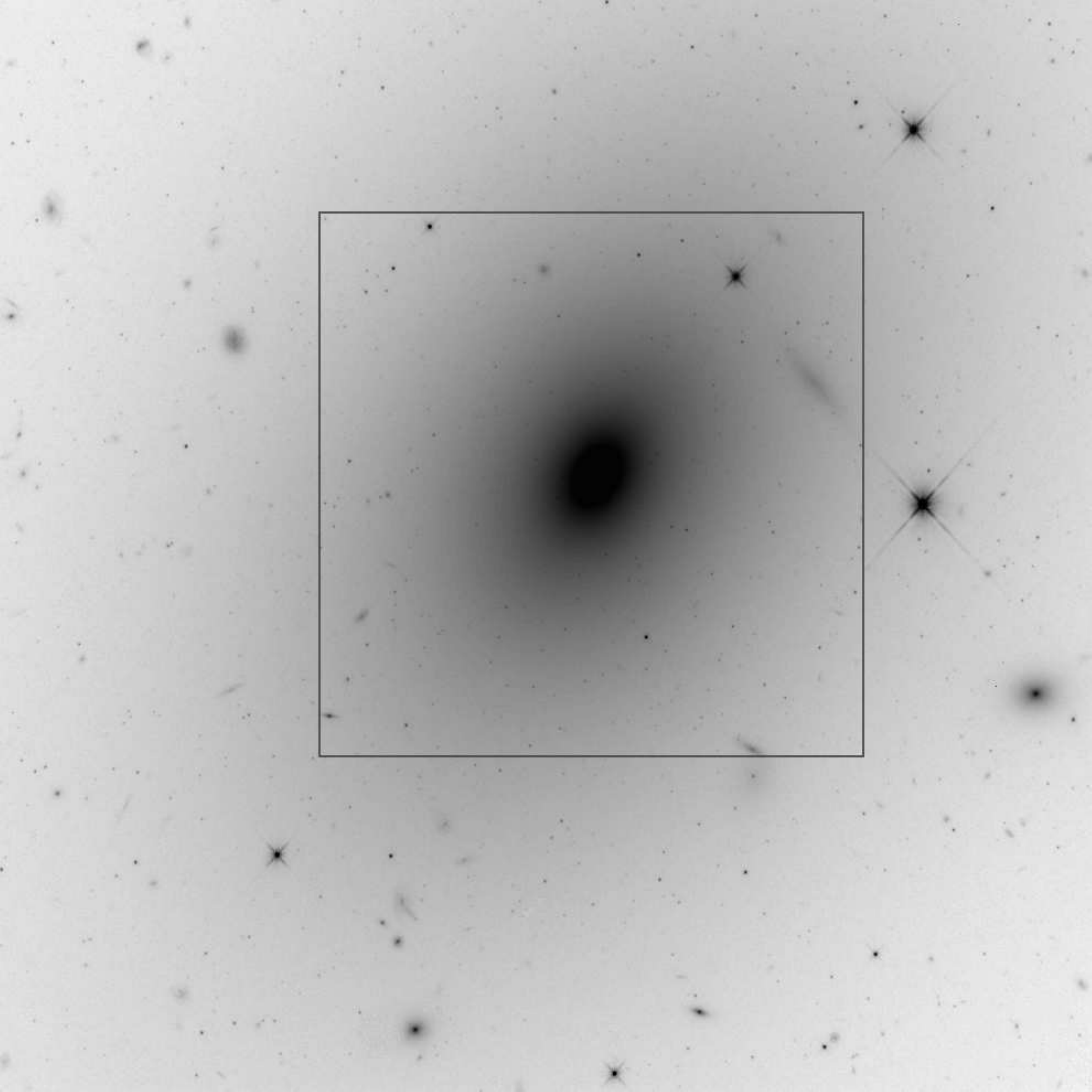}
\includegraphics[scale=0.4]{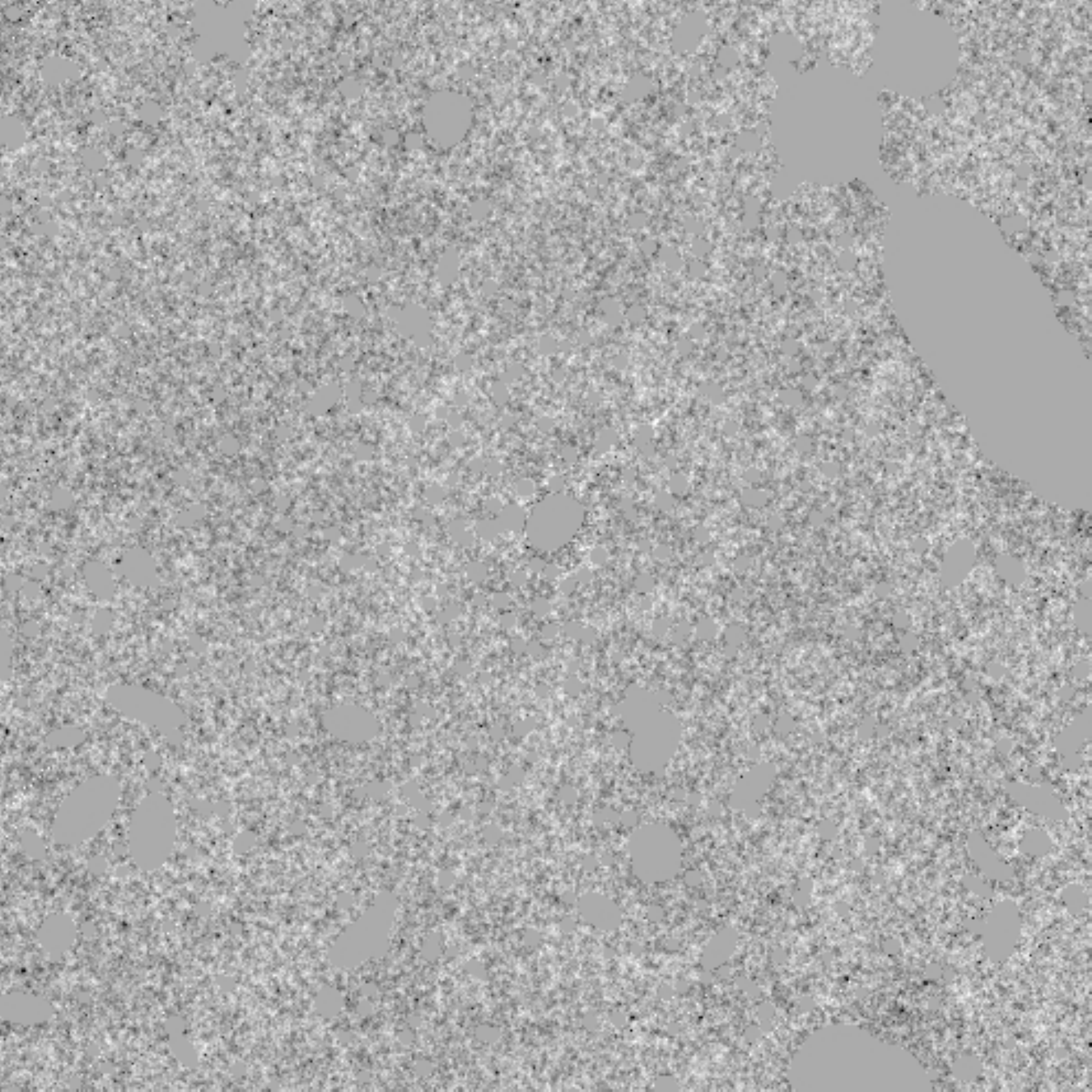} \\
\vspace{10pt}
\includegraphics[scale=0.4]{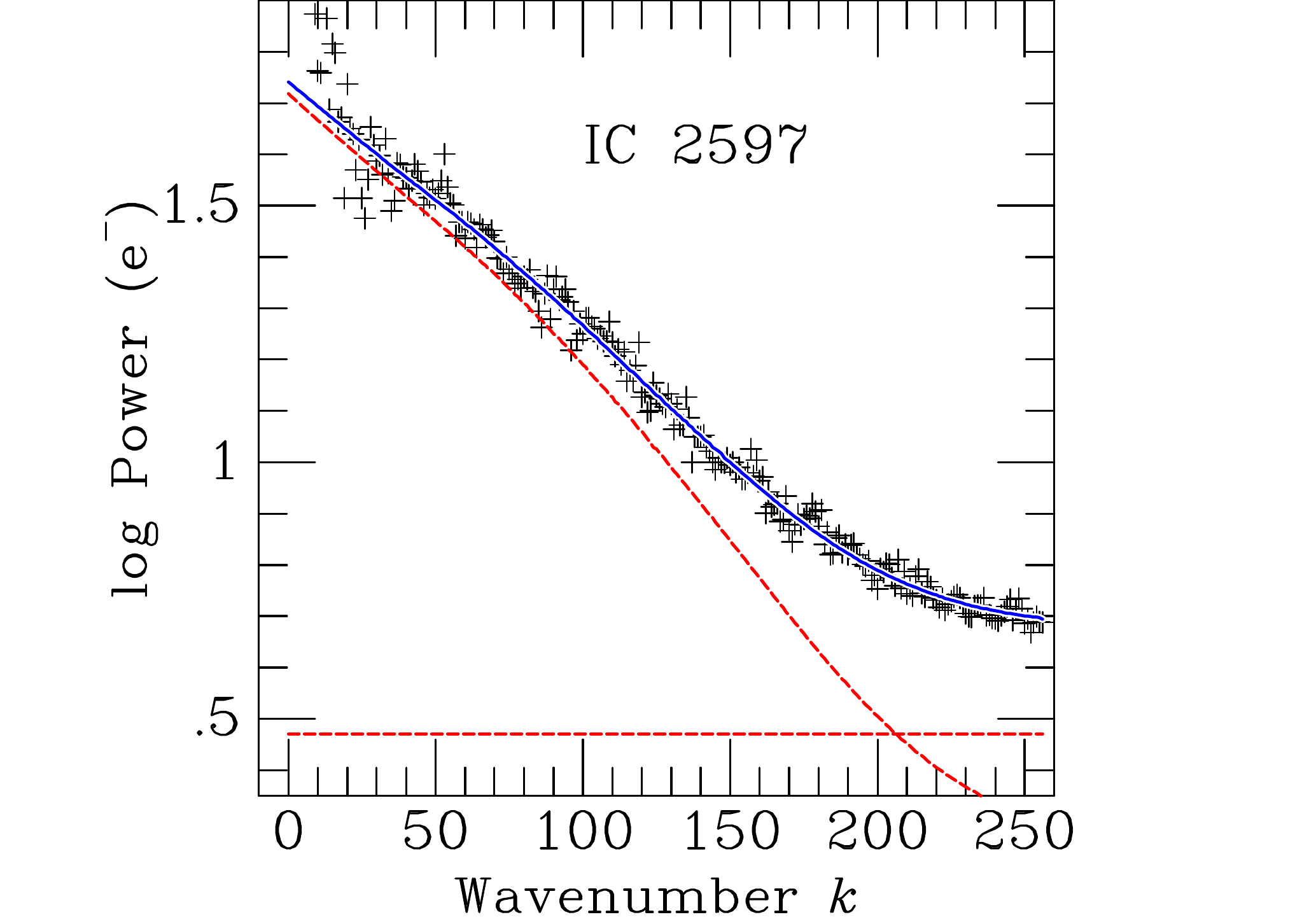}
\hspace{-25pt}
\includegraphics[scale=0.4]{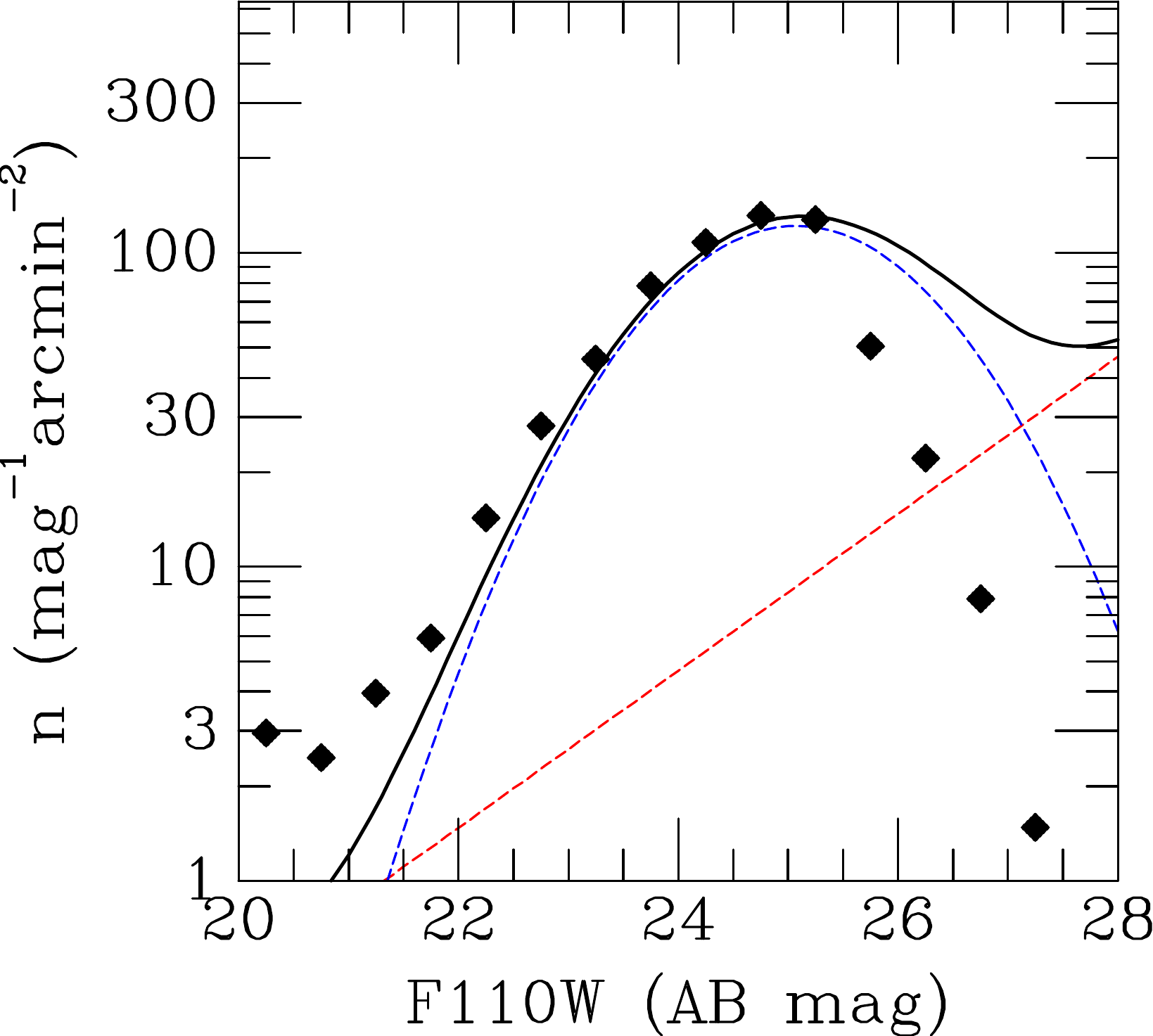}
\caption{Combined figure for IC~2597. The upper left image is the full field of view (127~arcsec on a side). The upper right image shows the central $63{\times}63$ arcsec region (the region in the box in the left image) with the smooth galaxy model subtracted, scaled by the square root of the galaxy model, and with globular clusters and background galaxies masked. The lower left plot is the spatial power spectrum (crosses) in units of total electrons detected in the full exposure fitted to the scaled expectation power spectrum $P_0{\times}E(k)$ and white noise floor $P_1$ (red dashed lines). The solid blue line is the sum of the two, fitted to the residual power spectrum. The lower right panel shows the fit to the GCLF (Gaussian, blue dashed line) and background galaxy luminosity function (power law, red dashed line) used to estimate the $P_r$ contribution from undetected sources. The black solid line is the sum of the two luminosity functions as fitted to the points. The diamond points in the lower right panel include both globular clusters and background galaxies. This is the first of the set of 61 figures available in the online version of the paper. All the galaxy images use the same logarithmic stretch and limiting values; the residual images are plotted with linear stretch and the same upper and lower limits. The closer galaxies show larger amplitude fluctuations. Comparable images and power spectra for NGC~3504 and NGC~4993 have already been published by \citet{Nguyen2020} and \citet{Cantiello2018}.
\label{galaxyfig}}
\end{center}
\end{figure*}

\subsection{Measuring the Point-Spread Function}

Another ingredient needed for SBF measurement is an accurate representation of the point-spread function (PSF) for the geometrically-uncorrected WFC3/IR F110W images. The spatial Fourier power spectrum is a convolution of the pixel-to-pixel stellar Poisson variations with the PSF plus a white noise component resulting from the background flux and detector readout noise. The SBF signal, in the absence of instrumental blurring, would simply be the Poisson variance between pixels due to the varying number of stars per pixel. The real images are blurred by the instrumental PSF, however, so the SBF signal is the sum of the white noise component and the SBF signal convolved with the PSF. Accurately fitting the SBF spatial power spectrum therefore requires a very high $S/N$ measurement of the PSF.

A library of reference PSF star images was compiled by extracting the best clean, bright point sources (Galactic foreground stars) from the residual images of a dozen galaxies (with background and galaxy removed). We also extracted many more intermediate brightness stars and combined the best-centered ones to construct additional composite PSF references. The extracted PSFs were used to compute the fluctuation magnitudes for many different galaxies, compared to each other, and compared to the composite PSFs selected for their individual centering. This provided a library of 12 individual PSF stars and several composite PSFs from four different \hst\ programs. During the creation of the PSF library, we were careful to choose stars within approximately 100 pixels of the vertical location of the center of the galaxy, where the geometrical distortion is ${\lesssim}1$\% and the resulting PSF normalization is less subject to the pixel scale variation in WFC3/IR. The central region is also the area on the detector where the highest-$S/N$ SBF measurements are made. In the final error analysis we used all the PSF stars to measure each galaxy, determined the best-fitting PSFs, and used the standard deviation of the full set of more than a dozen individual and combined PSFs to determine the PSF-fitting uncertainty.

This new library of PSF stars yielded much more reliable and consistent SBF measurements than the limited set of fainter stars used for the original WFC3 calibration \citepalias{Jensen2015}, which allowed us to reduce the uncertainty due to PSF normalization from ${\sim}0.05$ mag in the \citetalias{Jensen2015} study to 0.015 mag for the current study, with a systematic offset of ${\sim}0.05$ mag compared to \citetalias{Jensen2015}, as described by \citet{Cantiello2018}. We also compared the empirical PSFs to Tiny Tim models \citep{TinyTim};\footnote{\url{http://tinytim.stsci.edu/cgi-bin/tinytimweb.cgi}} the latter were symmetrical, unlike the PSFs extracted from the uncorrected images, but with modest Gaussian softening they still matched the SBF measurements reasonably well. A quantitative analysis of the PSF fits is presented below in the analysis of the uncertainties.

\subsection{Residual Power from Undetected Globular Clusters and Background Galaxies}

In addition to the stellar SBF component, the spatial Fourier power spectrum has a contribution from globular clusters (GC) and background galaxies superimposed on it, which can be the dominant source of uncertainty at large distances. 
The established procedure for removing their contribution to the spatial power spectrum requires fitting a Gaussian luminosity function to the GC population and a power law luminosity function to the background galaxies that \emph{can} be detected, and then integrating those functions beyond the photometric completeness limit to estimate the residual contribution from fainter undetected objects (see Figure~\ref{galaxyfig}). 

At distances beyond 50 Mpc, like the majority of the galaxies in this study, even when the SBF signal is still easy to detect, the point source sensitivity and resulting GC and galaxy luminosity function fits are usually the limiting factor in measuring accurate SBF distances.
For giant elliptical galaxies that have significant populations of GCs, it is very important to measure as much of the globular cluster luminosity function (GCLF) as possible, determine the photometric completeness limit, and correct the SBF signal for the undetected objects beyond the completeness limit. 
In most cases, the GCLF can be measured well enough by reaching within 0.5~mag of the peak of the GCLF. We found that when this sensitivity limit was reached, the contribution to the SBF signal from unmasked, undetected objects is less than ${\sim}10$\%, and the uncertainty in the correction was low enough (see Section~\ref{uncertaintysection}) that the resulting SBF distance is good to 5\% out to 70~Mpc in a single orbit, and to 10\% out to ${\sim}100$~Mpc in one orbit.

To test the robustness of the GCLF and background galaxy luminosity function measurements, we compared two different photometry programs. We used SExtractor \citep{sextractor}, which uses a combination of aperture photometry and elliptical model fits for point sources and extended sources, respectively, and an updated version of DoPHOT \citep{Schechter1993} modified by J.\ Alonso-Garcia \citep{Alonso2012}, to determine magnitudes using PSF-fitting of identified objects. To ensure that the photometric performance of the two programs was consistent for point sources of the appropriate brightness (i.e., near the peak of the GCLF for galaxies from 50 to 100~Mpc), we made detailed comparisons using the F110W and F160W observations of the GC population around NGC~4874 (GO-11711), which were deeper than most of the other observations in this study (two orbits at F110W). We compared our SExtractor aperture magnitudes to those from \citet{Cho2016} to determine the aperture correction for the PSF-fit magnitudes from DoPHOT: the aperture correction Cho et al.\ used for F160W was $-0.259$ mag for 0.6 arcsec diameter apertures, and we found that the same correction worked well for F110W. As expected, the agreement between Cho's and our independent analyses of the F110W data using the same data and same photometry package was excellent. We then used the SExtractor results to determine the aperture corrections for the PSF-fit magnitudes from DoPHOT. Because DoPHOT determines the magnitude in a smaller region near the core of the PSF, the aperture corrections were larger ($-0.539$ mag at F110W), with a small additional correction for objects fainter than 24 mag AB.

For the purpose of finding objects superimposed on a bright galaxy, we started by removing a smooth fit to all the large galaxies in the field of view and masking bright foreground stars. 
The resulting residual images do not have uniform noise because of the galaxy subtraction, and so a variance map was created and used with both SExtractor and DoPHOT to prevent them from identifying spurious objects near the galaxy center, or, alternatively, from missing objects farther away where galaxy subtraction noise was not significant. 
As in past SBF studies, we added an additional component (the galaxy model scaled by a constant) to the variance map to prevent SExtractor and DoPHOT from detecting the surface brightness fluctuations themselves as point sources. The empirically-determined scale factor we used with SExtractor is ${\sim}7$ at 20~Mpc and decreases to 1.5 times the mean galaxy surface brightness at distances of 70 Mpc and beyond \citep[c.f.][]{Jordan2004}; the factor used with DoPHOT was typically twice as large since DoPHOT is capable of detecting fainter compact objects near the center of the galaxy than SExtractor. 

While both SExtractor and DoPHOT were successful at detecting and measuring faint point sources, SExtractor did a better job at measuring the faint \emph{extended} objects (primarily background galaxies) that abound in deep \hst\ images. DoPHOT, on the other hand, did a better job measuring faint point sources (mostly GCs), particularly near the centers of the galaxies. The PSF fitting procedure used by DoPHOT would often interpret faint extended objects as multiple faint point sources, however. DoPHOT therefore finds significantly more faint objects than SExtractor and fewer brighter ones, as expected if the flux from one extended object is divided among many fainter ones. 

To get the best possible measurements of both the GC and background galaxy populations, we adopted a hybrid approach for this study that utilized the strengths of both systems. The extended objects (those with SExtractor {\tt CLASS\_STAR} $< 0.7$)  were extracted from the SExtractor output catalog and used to make a mask of those objects, and the extended object mask was applied to the DoPHOT input image to remove faint galaxies and isolate the point sources. DoPHOT was then used to photometer the point sources using PSF-fitting, and the output catalog was then merged with the SExtractor extended source catalog. The result was generally an excellent fit to both the GC and background galaxy luminosity functions, with a fainter limiting cutoff magnitude that avoided the contamination caused by extended objects being broken up into many smaller ones. Individual and combined GCLF and galaxy luminosity function fits are shown in Fig.~\ref{galaxyfig}. The separate independent catalogs were also retained to help identify problematic regions or fits, as described below.

Using the merged SExtractor+DoPHOT photometric catalog, the contribution to the total SBF signal from undetected globular clusters and galaxies was determined by fitting the GCLF assuming a Gaussian width of 1.2 to 1.4~mag, appropriate for giant ellipticals and early-type galaxies with rich GC populations. The GCLF was shifted to the distance of each galaxy by adopting an absolute peak magnitude of $M_J=-9.18$ \citep{Nantais2006}.
An initial first guess at the distance was used to get an approximate SBF distance, which was then used to make a more precise GCLF fit, followed by the final SBF distance measurement. (The SBF distance measurement process was done blindly until this second iteration where the distance was needed to constrain the GCLF peak brightness.)
The background galaxy population was assumed to follow a power-law distribution with a slope of 0.25 and normalized at the faint end to match the GOODs and HUDF surveys \citep{Retzlaff2010, Windhorst2011}.
The two luminosity functions (galaxies and GC) were then combined and integrated beyond the photometric completeness limit, typically ${\sim}25$ AB mag for single-orbit F110W measurements, to determine how much the undetected sources contribute to the stellar SBF signal ($P_r$). The correction to the SBF signal was typically largest nearer the centers of the galaxies, where the median was 9\% of the total SBF signal ($P_0$), and smaller in the outer regions (${\sim}5$\%) where GC were less numerous and the completeness limit was fainter. For the most distant galaxies in our sample (80 to 100~Mpc), values were 25\% of the total SBF power $P_0$ near the center and 10\% in the outer regions.
The uncertainty in this correction, as measured by repeating the SBF analysis using a variety of plausible fit parameters and cutoff limits, is shown in Table~\ref{errorbudget}, and was determined by taking half the difference between the correction determined using SExtractor and DoPHOT and combining in quadrature with 25\% of the fractional correction ($P_r/P_0$). The separate GCLF and galaxy luminosity function fits are shown in the lower right panels of the combined figure set (see Fig.~\ref{galaxyfig}).

The calibration galaxies from \citetalias{Jensen2015} were near enough that the corrections for undetected GCs and background galaxies were negligible (${<}1$\%). For galaxies beyond 50~Mpc, the power in undetected sources is a strong function of both distance and position relative to the center of the galaxy. Near the center, the galaxy brightness prevents detection of the faint sources near the peak of the GCLF. Because fewer sources are detected, the GCLF fit is less reliable. The regions farther from the center are better for detecting GCs and galaxies, but the lower galaxy surface brightness makes detection of the SBF signal less reliable. To balance the two sources of uncertainty (correction and background subtraction), we optimized both by measuring the radial behavior of the stellar SBF signal and the residual source correction in a series of radial annuli as described below. 

\subsection{Fluctuation Magnitudes \label{sbfmags}}

The SBF fluctuation power $P_0$ was determined by fitting the measured two-dimensional Fourier spatial power spectrum of the residual image $P(k)$, normalized by the mean galaxy brightness, with expectation power spectrum $E(k)$, the normalized power spectrum of the PSF reference star convolved with the mask window function:
\begin{equation}
P(k) = P_0{\times}E(k) + P_1
\end{equation}
where the scale factor $P_0$ is the SBF signal and is the white noise component. As long as the noise between pixels is uncorrelated, which is true for images combined using integer pixel shifts to avoid pixel interpolation, the term is constant with wavenumber $k$.
$P_0$ has units of flux (total electrons detected), which is converted to an apparent ``fluctuation magnitude''
\begin{equation}
\overline{m}_{110} = -2.5\log(P_0-P_r) + 26.8223 - A_{110}
\end{equation}
where is the correction for the contribution from undetected GC and background galaxies and $A_{110}$ is the foreground extinction in the F110W bandpass from \citet{SF2011} and adapted for F110W using the NASA Extragalactic Database extinction lists (Table~\ref{obstable}). 

We measured the fluctuations independently in four radial annuli spanning 32--64, 64--128, 128--256, and 256--512 pixels. Fitting the power spectrum to get a reliable value for \mbarj\ is a trade-off between stronger $S/N$ near the center where the galaxy is brightest, but the number of pixels is smaller and the contribution from is larger, and the regions farther out, where the galaxy surface brightness is lower relative to the background, and is therefore subject to larger background variations. The four annular regions were treated independently. We measured the fluctuation magnitudes in each annulus individually, and then combined the final results into a single distance measurement for each galaxy by taking the uncertainty-weighted average of the good regions. 

Annuli were excluded from the average for a variety of reasons. Some were affected by nearby bright companion galaxies, dust lanes, shells, or other galaxy fitting defects that rendered the fluctuation measurements unreliable. We also excluded from the weighted average any region of the galaxy where the total \mbarj\ uncertainty was greater than 0.15~mag (roughly twice the average uncertainty). A region was also excluded when the difference between the independent SExtractor and DoPHOT fits exceeded 0.15~mag. 
The final distance measurement made use of the hybrid SExtractor and DoPHOT photometry only when the two were independently consistent within 0.15 mag. We also required that the photometric completeness limit for point sources be no more than 0.5~mag brighter than the GCLF peak. When the photometry does not go deep enough to catch most of the GCs close to the peak of the GCLF, the correction becomes unreliable and \mbarj\ is potentially biased. The central annulus was the most likely to be excluded because point sources become increasingly difficult to detect when superimposed on the bright galaxy profile. 
Other exclusions included annuli where the fraction of masked pixels in the annulus was greater than 50\%, the background level exceeded the galaxy surface brightness, or the SBF $S/N = (P_0-P_r)/P_1 < 5$. These latter conservative criteria have been established through experience with WFC3/IR SBF measurements \citepalias{Jensen2015}.

Some of the galaxies in this sample are lenticular and their projected shapes are very elongated or disky. If these galaxies have population gradients, then using elliptical annuli instead of circular ones could produce more consistent distances. To check this possibility, we repeated the \mbarj\ measurements for a subsample of 14 galaxies using elliptical annuli that were completely contained within the original (default) circular annuli to prevent cross-talk 
\citep{Phan2020}. Seven galaxies were highly elongated and seven were nearly round. Eight of the galaxies showed less than 0.01~mag differences between elliptical and circular annuli, and included some of the most elongated galaxies in our entire sample (including NGC~2765 and NGC~4036, for example). All but one galaxy differed by less than 0.03~mag, and the only exception was NGC~495, a barred S0 galaxy where neither elliptical nor circular apertures match the light distribution well. Given that the differences were significantly less than the individual annular uncertainties in \mbarj, we did not repeat the \mbarj\ measurements using elliptical annuli for the rest of the sample. This result is consistent with similar tests done using optical $I$-band SBF measurements \citep{Cantiello2011}.

\section{Calibration of SBF Distances \label{calibrationsection}}

Determining the distance to a galaxy using the apparent fluctuation magnitude \mbarj\ requires knowledge of the absolute magnitude \Mbarj, which was determined empirically by \citetalias{Jensen2015} using Cepheid variable stars to determine the distance zero point and fitting variations in \Mbarj\ as a function of galaxy optical or IR color to correct for variations in stellar population age and metallicity. Since SBF are typically measured in giant elliptical galaxies and Cepheids in spiral galaxies, the SBF calibration must take into account the spatial relationship between the elliptical and spiral galaxies in the Fornax and Virgo clusters. The \citetalias{Jensen2015} F110W and F160W SBF calibration was performed using optical F850LP SBF distances (both individual and mean cluster distances) for a subset of 16 galaxies in the extensive ACS SBF surveys of the Virgo and Fornax clusters \citep{Blakeslee2009}. 
The SBF distances for those large surveys were derived from the Cepheid measurements and tied to the LMC Cepheid distance modulus of 18.50 mag \citep{Freedman2001,Freedman2010}. For this study we have adopted the updated LMC distance of 18.477~mag \citep{Piet2019}. The systematic zero point uncertainty arising from this calibration chain was estimated to be 0.10~mag by \citetalias{Jensen2015}, \citet{Cantiello2018}, and \citet{Blakeslee2010}. Improvements in the distance to the LMC reduce this uncertainty to 0.09~mag \citep[for details, see Section 2.4 of][]{Blakeslee2021}.

This study includes several other procedural improvements and updates from the detailed calibration presented by \citetalias{Jensen2015}, including changes to the initial \emph{HST} pipeline reduction, correction for sky background and undetected sources, choice of PSF reference stars, etc. These changes have little or no effect on the measurement of \mbarj\ for the nearby galaxies in the Virgo and Fornax clusters where the SBF signal is large and is small, but are important for the more distant sample here. Some of the changes to the calibration zero point have been described by \citet{Cantiello2018}. A small change in the ACS calibration led to a 0.004~mag shift in the ACS \gminz\ colors, resulting in a 0.009~mag shift in \Mbarj. A more significant change of 0.05~mag to the zero point was the result of the expanded WFC3/IR PSF library that we used for NGC~4993 and all the galaxies in this survey. The increased $S/N$ of the brighter PSF stars also reduced the individual statistical uncertainty considerably. 

\section{Color Measurements and Transformations \label{colortransformations}}

Measuring accurate SBF distances also requires knowledge of the galaxy's color to correct the absolute fluctuation magnitude \Mbarj\ for variations in the age and metallicity of the underlying stellar population \citep{Jensen2015,Jensen2003,Tonry1997}. Optical \gminz\ colors are effective for constraining population variations, particularly for SBF measurements at wavelengths near 1~$\mu$m where the effects of age and metallicity are largely degenerate. In this section we establish the transform for determining the colors needed for WFC3/IR SBF measurements from the ground-based Panoramic Survey Telescope and Rapid Response System (PanSTARRS) as an alternative to ACS F475W and F850LP photometry \citep{PanSTARRSimaging}. Similar transformations for the Sloan Digital Sky Survey (SDSS) and 2-Micron All-Sky Survey (2MASS) can be found in the Appendix. 

The IR SBF calibration galaxies from \citetalias{Jensen2015} and the observations of NGC~4993 \citep{Cantiello2018} had ample ACS F475W and F850LP imaging for determination of the \gzacs\ colors and deep F110W and F160W observations useful for determining \JHwfc. This is not generally the case for most \hst\ observations, however, and the subsequent IR SBF projects described in this paper rely on ground-based color measurements instead of using valuable \hst\ orbits for color measurements. For the purpose of establishing a uniform WFC3/IR SBF database, we have carefully cross-calibrated the WFC3/IR F110W SBF distance calibration using PanSTARRS survey DR1 images \citep{Chambers2019,Waters2020} and 2MASS images \citep[2MASS,][]{2MASS2006} for \JminH\ as an alternative when \gminz\ is unavailable or unreliable due to high optical foreground extinction. In all cases we used the survey images to consistently measure colors in the specific annuli used for SBF, and did not rely on catalog photometry for the galaxies.

The PanSTARRS images are preferred for measuring \gminz\ colors in support of WFC3/IR SBF because they have higher spatial resolution and greater depth than SDSS or 2MASS, making color measurements more reliable in the outer annuli. PanSTARRS \gminz\ colors were used for all the distances in this paper with the exception of ESO~125-G006, for which no \gminz\ data was available. To measure colors using PanSTARRS, we retrieved large $g$ and $z$-band images (12.5~arcmin square) from the public archive.\footnote{\url{http://panstarrs.stsci.edu}} We determined the background level for each image from regions well away from the target galaxies. We then constructed masks of stars and other galaxies by first subtracting a smooth model of the galaxy and then using SExtractor \citep{sextractor} on the galaxy-subtracted images to identify and mask objects. We manually masked any undetected objects and regions where the background subtraction left sharp edges or other processing artifacts. The next step was to apply the correct photometric calibration for each image (in AB mag) and scale by the relative exposure times in the headers, correct for extinction \citep{SF2011}, and create the color map image. 

Once we had a color map for each galaxy, we applied a series of annular masks to each map and measured the median color in the same regions of the galaxies that we used to measure the SBF magnitudes in the WFC3/IR images (the color maps were not remapped to the WFC3/IR focal plane geometry, so there is a distortion of a few percent in the $y$-axis in the outer annuli). The majority of the galaxies in this survey are giant elliptical or massive S0 galaxies that have little or no color gradient in the regions used for SBF analysis, but this step was still important to ensure that we could measure SBF distances completely independently in several radial regions of each galaxy and use the radial variations in the SBF magnitudes to further constrain background subtraction uncertainties and distinguish such from population variations in the galaxies.

Translating the PanSTARRS colors into the relevant ACS photometric system used for the \citetalias{Jensen2015} calibration required a statistically significant overlap sample with which to make the comparison.  The \citetalias{Jensen2015} sample contains 16 galaxies total, eight in the Fornax cluster and eight in the Virgo cluster. Four of the Virgo galaxies were low-luminosity blue dwarf galaxies; not only were they outside the calibration color range (their inclusion in the \citetalias{Jensen2015} sample was primarily for studying stellar population variations), but they were also poorly measured in the ground-based optical survey images. The Fornax cluster is too far  to be included in the PanSTARRS survey. The remaining four galaxies from \citetalias{Jensen2015} are too few to get a robust color calibration link, so we expanded the sample by including 68 Virgo cluster galaxies from the ACS Virgo Cluster Survey \citep{Mei2007,Blakeslee2009}. These galaxies do not have IR SBF measurements, but they do allow us to link the \gminz\ colors from PanSTARRS to the ACS \gzacs\ color system for 68 galaxies.

Multiple color maps of the same galaxies from four different sources (ACS, PanSTARRS, SDSS, and 2MASS) allowed us to compare radial luminosity and color profiles and check consistency (see the Appendix). We found that the ACS, PanSTARRS, and SDSS \gminz\ color profiles are consistent near the centers of the galaxies, in the innermost SBF measurement annulus; the PanSTARRS profiles remain consistent with ACS farther out, but the SDSS and 2MASS are not as deep and the color measurements become inconsistent and noisy in the outer annuli. We therefore computed the calibration offsets for the central region and use the PanSTARRS \gminz\ measurements for the SBF measurements in all the annuli. 

\begin{figure}
\begin{center}
\plotone{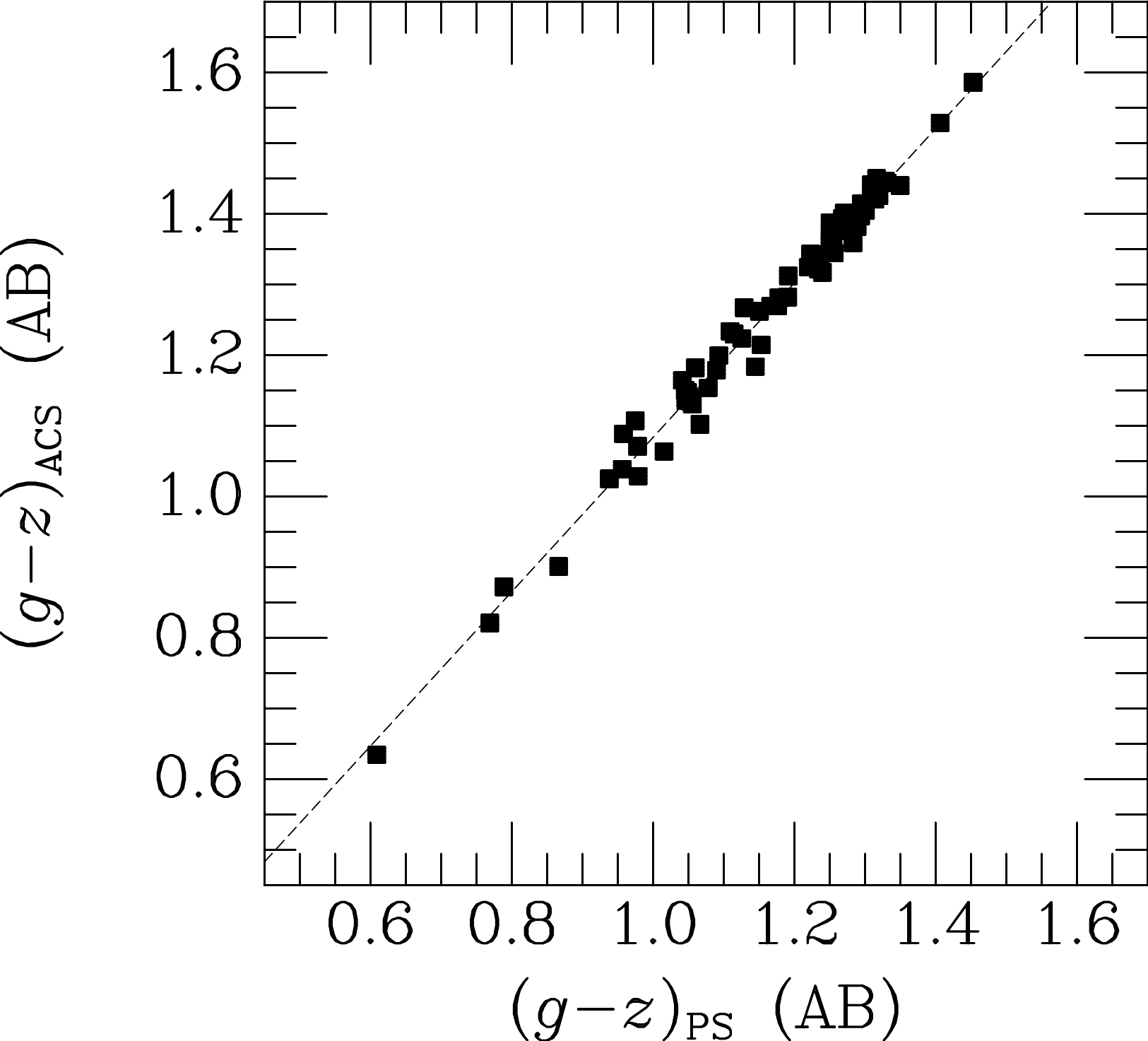}
\caption{Relationship between PanSTARRS and ACS \gminz\ color photometry for 68 galaxies in the ACSVCS sample.}
\label{colorcalibfig}
\end{center}
\end{figure}

The PanSTARRS \gminz\ colors differ from the ACS system by about 10\% with a modest slope with galaxy color. The relation we measured for 68 calibrators in the inner, high surface brightness region is:
\begin{equation}
\label{psacsequation}
(g_{475}{-}z_{850})_{ACS} = 1.092(g{-}z)_{PS} - 0.009
\end{equation}
in AB mags (Fig.~\ref{colorcalibfig}), with a standard deviation of 0.016~mag measured among the subset of 33 color calibrators with \gzacs\ between 1.3 and 1.5~mag, the same color range as the SBF sample galaxies.  The color calibration is nearly identical in the next two larger annuli, where the scatter is 0.016 and 0.018~mag, respectively. The fit is shown in Figure~ \ref{colorcalibfig}; since all of the target SBF galaxies have \gminz\ colors in a narrow range, the distance measurements are not very sensitive to the slope in Equation~\ref{psacsequation}. We adopted the standard deviation of the color calibrator galaxies with $\gzacs >1.3$ as the uncertainty on the PanSTARRS color measurements, added in quadrature with the extinction uncertainty (see Section~\ref{uncertaintysection}). 

For IR SBF, when foreground extinction is high or \gminz\ colors are unavailable, \JminH\ can be used instead (e.g., ESO~125-G006 in this study). In the near IR, the range of \JminH\ color spanned by different stellar populations is smaller, however, and the leverage on population variations is therefore reduced and the corresponding uncertainty is larger. The \JminH\ distances are discussed in the Appendix.

\section{SBF Uncertainties \label{uncertaintysection}}

The apparent fluctuation magnitudes \mbarj\ are subject to several sources of uncertainty that
were measured empirically by repeating the SBF measurements with different 
input parameters in different independent annular regions around the galaxy center. 
These uncertainties, which were treated as independent and added in quadrature, include: 
\begin{enumerate}
    \item Power spectrum fit uncertainty. The process of fitting the PSF power spectrum to the residual data results in a statistical fit uncertainty of ${\sim}2$\%. The quality of the power spectrum fit is affected by the number of good pixels in the annulus, the spatial structure of the mask, and patterns in the residual image due to shells, bars, spiral arms, or tidal interaction features. Residual features often have low-wavenumber power that can be filtered in the power spectrum fit.
    \item Background subtraction. The background subtraction exhibits scatter due to uncertain correction for the He emission, instrumental background, and the fit of the galaxy model. Once the background was determined as described above, we offset the background subtraction by ${\pm}1\,\sigma$ and repeated the entire SBF measurement in each annulus, and used the offsets in the uncertainty calculation. Background subtraction uncertainty increases substantially in the outer annuli where the galaxy surface brightness is faintest.
    \item PSF fit. As described above, we developed a library of PSF reference stars. For any given galaxy, some PSF stars fit better than others (depending on the details of the fine guidance and dithering accuracy). After choosing the best PSF for a particular observation, we then repeated the SBF measurements in all four annuli with 12 individual PSF stars, and used the standard deviation of the measurements as the PSF fitting uncertainty for each annulus.
    \item Globular cluster and background galaxy correction. To determine the uncertainty on the correction for background galaxies and undetected globular clusters, we used the difference between the results achieved using SExtractor and DoPHOT independently. We adopted a correction uncertainty of one half of the difference, added in quadrature with a minimum estimated uncertainty of 25\% of the fractional contribution $P_r/P_0$. 
    \item Extinction correction. The extinction in the F110W filter is quite low for the majority of the galaxies (see Table~\ref{obstable}); 10\% of the extinction was added to the uncertainty in \mbarj. 
\end{enumerate}

The presence of dust in a galaxy also adds to the uncertainty in \mbarj, but in ways that are not easily quantified. Clumpy dust can add to the fluctuation power quite dramatically, while a uniform screen of dust would reduce the apparent brightness of the galaxy. The effects of dust are greatly reduced when we use the optical images and color maps to identify and mask dusty regions of the galaxy. We have included the uncertainty in the foreground Galactic extinction as part of the error budget but do not add an additional contribution from dust in the target galaxy. Instead, we mask the dusty regions and flag galaxies with apparent dust when computing distances. Computations of the Hubble constant are not significantly affected by including or excluding galaxies with dust \citep{Blakeslee2021}.

Our final \mbarj\ was computed by taking the uncertainty-weighted mean of the SBF measurements in the good annuli (usually two or three of the four) as discussed in Section~\ref{sbfmags}. The combined uncertainty in \mbarj\ used in calculating the distances was the weighted average of the uncertainties in the good annuli (with the extinction uncertainty treated as a systematic uncertainty common to all annuli), which has a median value of 0.057~mag. Median values and ranges for the individual sources of uncertainty are listed in Table~\ref{errorbudget}.

\begin{deluxetable}{lcc}
\tablecaption{SBF Uncertainties \label{errorbudget}}
\tablehead{
\colhead{Uncertainty Source} &
\colhead{Median} &
\colhead{Range} \\
\colhead{$\sigma$} & \colhead{(mag)} & \colhead{(mag)}
}
\startdata
Power spectrum fit ($P_0$) & 0.020 & 0.01--0.08 \\
Background subtraction & 0.020 & 0.01--0.11 \\
PSF fit                & 0.015 & 0.012--0.062 \\
GCLF + galaxy LF fits ($P_r$) & 0.027 & 0.001--0.119 \\
F110W extinction correction & 0.004 & 0.0004--0.014 \\
Combined \mbarj\ uncertainty\tablenotemark{a} & 0.057 & 0.021--0.131 \\
\hline
\gminz\ extinction correction  & 0.011 & 0.001--0.036 \\
Total \gminz\ uncertainty    & 0.021 & 0.016--0.040 \\
Stellar population scatter\tablenotemark{b} & 0.06 & 0.05--0.06 \\
Combined \Mbarj\ uncertainty\tablenotemark{c} & 0.075 & 0.060--0.105 \\
\hline
Total $(m{-}M)$ uncertainty  & 0.085 & 0.072--0.195 \\
\enddata
\vspace{0.1cm}
\tablenotemark{a}{Individual sources of uncertainty added in quadrature for each annulus.}
\tablenotetext{b}{The scatter due to stellar population variations was measured by \citet{Jensen2015} and \citet{Cantiello2018}.}
\tablenotetext{c}{The combined uncertainty on \Mbarj\ includes the total \gminz\ uncertainty times 2.16 added in quadrature with the stellar population scatter.}
\end{deluxetable}

The uncertainty associated with the absolute fluctuation magnitude \Mbarj\ includes the color measurement uncertainty and intrinsic scatter due to population variations. Although the old massive galaxies in this sample typically show very little color gradient, we still computed \Mbarj\ for each annulus using the \gminz\ or \JminH\ color measured in each region. The weighted average distance for each galaxy made use of the individual \Mbarj\ values. The combined uncertainty in the weighted average included the rms scatter for the color conversions added in quadrature with 10\% of the difference between the $g$ and $z$ extinction values. The median \gminz\ uncertainty calculated in this way was 0.021~mag (Table~\ref{errorbudget}). The combined color uncertainty was added in quadrature (multiplied by the calibration slope of 2.16) with the \Mbarj\ calibration uncertainty of 0.06~mag arising from stellar population variations \citep{Blakeslee2009,Jensen2015,Cantiello2018}. Combining the \mbarj\ and \Mbarj\ uncertainties in quadrature gives a median uncertainty on our distance moduli of 0.085~mag, or ${\sim}3.9$\% in distance (see Table~\ref{errorbudget}).




\startlongtable
\begin{deluxetable*}{lCCCCCCCCCC}
\tabletypesize{\footnotesize}
\tablecaption{F110W SBF Distances and Velocities\label{distances}}
\tablehead{
\colhead{Galaxy} &
\colhead{$(g{-}z)$} &
\colhead{\mbarj} &
\colhead{$(\overline{m}{-}\overline{M})$} &
\colhead{$d_{\rm SBF}$} &
\colhead{$v_{\rm gal}$} &
\colhead{$v_{\rm grp}$} &
\colhead{$v_{\rm flow}$} & \colhead{$v_{2M++}$} &
\colhead{$v_{\rm rms}$} &
\colhead{$N_{\rm grp}$} \\
& 
\colhead{(mag)} & 
\colhead{(mag)} & 
\colhead{(mag)} & 
\colhead{(Mpc)} & 
\colhead{(km/s)} & 
\colhead{(km/s)} &
\colhead{(km/s)} & \colhead{(km/s)} &
\colhead{(km/s)} & 
}
\colnumbers
\decimals
\startdata
IC~2597  &	1.509	\pm	0.021	&	31.044	\pm	0.031	&	33.673	\pm	0.082	&	54.3	\pm	2.1	&	3334	&	4099	&	4003	& 4227 &	648	&	85	\\
NGC~0057 &	1.613	\pm	0.022	&	31.723	\pm	0.053	&	34.126	\pm	0.094	&	66.9	\pm	2.9	&	5088	&	5278	&	5875	& 5485 &	339	&	\phn6	\\
NGC~0315 &	1.561	\pm	0.022	&	31.650	\pm	0.027	&	34.166	\pm	0.081	&	68.1	\pm	2.5	&	4635	&	4819	&	4707	& 4938 &	327	&	14	\\
NGC~0383 &	1.513	\pm	0.021	&	31.481	\pm	0.037	&	34.101	\pm	0.084	&	66.1	\pm	2.6	&	4802	&	4900	&	4555	& 5087 & 432	&	48	\\
NGC~0410 &	1.538	\pm	0.020	&	31.371	\pm	0.034	&	33.937	\pm	0.082	&	61.3	\pm	2.3	&	5002	&	4900	&	4579	& 5086 &	432	&	48	\\
NGC~0495 &	1.456	\pm	0.021	&	31.306	\pm	0.030	&	34.049	\pm	0.081	&	64.5	\pm	2.4	&	3831	&	4626	&	4181	& 4547 &	424	&	47	\\
NGC~0507 &	1.497	\pm	0.021	&	31.298	\pm	0.032	&	33.953	\pm	0.081	&	61.7	\pm	2.3	&	4651	&	4626	&	4196	& 4548 &	424	&	47	\\
NGC~0524 &	1.565	\pm	0.024	&	29.703	\pm	0.043	&	32.212	\pm	0.090	&	27.7	\pm	1.1	&	2068	&	2166	&	2429	& 2229 &	171	&	\phn9	\\
NGC~0533 &	1.514	\pm	0.018	&	31.693	\pm	0.029	&	34.310	\pm	0.077	&	72.8	\pm	2.6	&	5240	&	5073	&	5013	& 5177 &	351	&	\phn8	\\
NGC~0545 &	1.426	\pm	0.019	&	31.682	\pm	0.083	&	34.489	\pm	0.110	&	79.0	\pm	4.0	&	5162	&	5083	&	5214	& 5184 &	460	&	43  \\
NGC~0547 &	1.459	\pm	0.019	&	31.653	\pm	0.063	&	34.391	\pm	0.096	&	75.5	\pm	3.3	&	5162	&	5083	&	5214	& 5184 &	460	&	43	\\
NGC~0665 &	1.525	\pm	0.024	&	31.380	\pm	0.027	&	33.973	\pm	0.084	&	62.3	\pm	2.4	&	5127	&	5074	&	5283	& 5164 &	168	&	\phn9	\\
NGC~0708 &	1.581	\pm	0.024	&	31.476	\pm	0.029	&	33.945	\pm	0.085	&	61.5	\pm	2.4	&	4601	&	4687	&	4182	& 4521 &	520	&	63	\\
NGC~0741 &	1.560	\pm	0.020	&	31.696	\pm	0.028	&	34.214	\pm	0.080	&	69.6	\pm	2.6	&	5280	&	5144	&	5308	& 5172 &	189	&	\phn8   \\
NGC~0777 &	1.564	\pm	0.019	&	31.652	\pm	0.036	&	34.161	\pm	0.082	&	68.0	\pm	2.6	&	4758	&	4988	&	4851	& 5265 &	148	&	10	\\
NGC~0809 &	1.311	\pm	0.017	&	31.320	\pm	0.084	&	34.376	\pm	0.109	&	75.0	\pm	3.8	&	5080	&	5148	&	5469	& 5302 &	\phn\phn0	&	\phn1	\\
NGC~0890 &	1.399	\pm	0.023	&	30.430	\pm	0.022	&	33.296	\pm	0.081	&	45.6	\pm	1.7	&	3745	&	3782	&	3361	& 3248 &	\phn\phn0	&	\phn1	\\
NGC~0910 &	1.532	\pm	0.021	&	31.880	\pm	0.055	&	34.459	\pm	0.093	&	77.9	\pm	3.3	&	4995	&	5597	&	6051	& 5993 &	656	&	54	\\
NGC~1016 &	1.531	\pm	0.018	&	32.089	\pm	0.047	&	34.670	\pm	0.085	&	85.9	\pm	3.4	&	6428	&	6449	&	6533	& 6350 &	512	&	16	\\
NGC~1060 &	1.572	\pm	0.040	&	31.160	\pm	0.016	&	33.653	\pm	0.108	&	53.8	\pm	2.7	&	4977	&	4788	&	4270	& 4668 &	474	&	21	\\
NGC~1129 &	1.557	\pm	0.027	&	31.580	\pm	0.043	&	34.105	\pm	0.095	&	66.2	\pm	2.9	&	5009	&	5247	&	5244	& 5218 &	683	&	46	\\
NGC~1167 &	1.463	\pm	0.037	&	30.922	\pm	0.050	&	33.650	\pm	0.113	&	53.7	\pm	2.8	&	4757	&	4808	&	4234	& 4514 &	105	&	\phn4	\\
NGC~1200 &	1.467	\pm	0.022	&	30.940	\pm	0.026	&	33.660	\pm	0.080	&	54.0	\pm	2.0	&	3805	&	3793	&	3822	& 3721 &	104	&	\phn4	\\
NGC~1201 &	1.425	\pm	0.017	&	28.538	\pm	0.021	&	31.347	\pm	0.074	&	18.6	\pm	0.6	&	1494	&	1500	&	1856	& 1754 &	\phn\phn0	&	\phn1	\\
NGC~1259 &	1.477	\pm	0.033	&	31.668	\pm	0.046	&	34.365	\pm	0.105	&	74.6	\pm	3.6	&	5653	&	5281	&	5031	& 5199 &	962	&	180	\\
NGC~1272 &	1.529	\pm	0.034	&	31.671	\pm	0.022	&	34.256	\pm	0.099	&	71.0	\pm	3.2	&	3655	&	5281	&	4984	& 5194 &	962	&	180	\\
NGC~1278 &	1.512	\pm	0.035	&	31.580	\pm	0.044	&	34.202	\pm	0.106	&	69.2	\pm	3.4	&	5931	&	5281	&	4980	& 5193 &	962	&	180	\\
NGC~1453 &	1.532	\pm	0.026	&	30.967	\pm	0.022	&	33.546	\pm	0.085	&	51.2	\pm	2.0	&	3778	&	3946	&	3867	& 3879 &	193	&	13	\\
NGC~1573 &	1.504	\pm	0.030	&	31.374	\pm	0.027	&	34.014	\pm	0.094	&	63.5	\pm	2.7	&	4161	&	4408	&	4411	& 4594 &	320	&	17	\\
NGC~1600 &	1.486	\pm	0.020	&	31.600	\pm	0.038	&	34.277	\pm	0.083	&	71.7	\pm	2.7	&	4620	&	4502	&	4450	& 4570 &	401	&	30	\\
NGC~1684 &	1.412	\pm	0.020	&	31.151	\pm	0.025	&	33.989	\pm	0.079	&	62.8	\pm	2.3	&	4378	&	4511	&	4471	& 4610 &	226	&	15	\\
NGC~1700 &	1.368	\pm	0.019	&	30.657	\pm	0.022	&	33.590	\pm	0.076	&	52.2	\pm	1.8	&	3870	&	3953	&	3783	& 3808 &	143	&	\phn7	\\
NGC~2258 &	1.541	\pm	0.029	&	31.221	\pm	0.034	&	33.781	\pm	0.094	&	57.0	\pm	2.5	&	4055	&	3979	&	4102	& 4053 &	105	&	\phn3	\\
NGC~2274 &	1.499	\pm	0.026	&	31.545	\pm	0.025	&	34.196	\pm	0.086	&	69.1	\pm	2.7	&	5171	&	5207	&	4861	& 5233 &	136	&	\phn6	\\
NGC~2340 &	1.537	\pm	0.022	&	31.947	\pm	0.067	&	34.514	\pm	0.102	&	79.9	\pm	3.8	&	6007	&	6061	&	6292	& 6071 &	512	&	29	\\
NGC~2513 &	1.484	\pm	0.018	&	31.576	\pm	0.033	&	34.258	\pm	0.078	&	71.1	\pm	2.6	&	4903	&	4724	&	4623	& 4802 &	318	&	\phn7	\\
NGC~2672 &	1.497	\pm	0.017	&	31.450	\pm	0.040	&	34.103	\pm	0.081	&	66.2	\pm	2.5	&	4611	&	4502	&	4067	& 4503 &	129	&   \phn5 	\\
NGC~2693 &	1.501	\pm	0.017	&	31.609	\pm	0.065	&	34.256	\pm	0.096	&	71.0	\pm	3.1	&	5098	&	5248	&	5483	& 5271 & \phn97 & \phn6	\\
NGC~2765 &	1.346	\pm	0.018	&	30.744	\pm	0.041	&	33.725	\pm	0.083	&	55.6	\pm	2.1	&	4086	&	4130	&	4239	& 4184 &	\phn\phn0	&	\phn1	\\
NGC~2962 &	1.465	\pm	0.020	&	29.789	\pm	0.040	&	32.532	\pm	0.084	&	32.1	\pm	1.2	&	2300	&	2354	&	2485	& 2490 &	\phn57	&	\phn2	\\
NGC~3158 &	1.548	\pm	0.017	&	32.261	\pm	0.069	&	34.806	\pm	0.099	&	91.5	\pm	4.2	&	7170	&	7185	&	7490	& 7120 &	388	&	13	\\
NGC~3392 &	1.351	\pm	0.016	&	30.890	\pm	0.054	&	33.861	\pm	0.088	&	59.2	\pm	2.4	&	3340	&	3456	&	3953	& 3751 &	121	&	\phn2	\\
NGC~3504\tablenotemark{a} &	1.312 \pm	0.023 & 29.630 \pm	0.039 & 32.683 \pm	0.095 & 34.4 \pm 3.8 & 1829 & 1760 & 1981   & 2009 & 110	&	\phn2 \\
NGC~3842 &	1.532	\pm	0.017	&	32.132	\pm	0.073	&	34.711	\pm	0.102	&	87.5	\pm	4.1	&	6561	&	6987	&	6665	& 6656 &	707	&	61	\\
NGC~4036 &	1.447	\pm	0.018	&	29.013	\pm	0.031	&	31.776	\pm	0.078	&	22.7	\pm	0.8	&	1574	&	1442	&	1580	& 1693 &	124	&	\phn3	\\
NGC~4073 &	1.529	\pm	0.019	&	32.060	\pm	0.121	&	34.646	\pm	0.141	&	85.0	\pm	5.5	&	6268	&	6380	&	6152	& 6154 &	358	&	13	\\
NGC~4386 &	1.442	\pm	0.019	&	29.653	\pm	0.034	&	32.427	\pm	0.080	&	30.6	\pm	1.1	&	1721	&	1661	&	1807	& 1940 &	188	&	\phn7	\\
NGC~4839 &	1.553	\pm	0.017	&	32.267	\pm	0.065	&	34.800	\pm	0.096	&	91.2	\pm	4.0	&	7610	&	7331	&	7202	& 7324 &	866	&	136	\\
NGC~4874 &	1.545	\pm	0.012	&	32.430	\pm	0.107	&	34.981	\pm	0.128	&	99.1	\pm	5.8	&	7436	&	7331	&	7230	& 7336 &	866	&	136	\\
NGC~4914 &	1.317	\pm	0.018	&	31.090	\pm	0.023	&	34.133	\pm	0.075	&	67.1	\pm	2.3	&	4896	&	4962	&	5228	& 4813 &	\phn\phn2	&	\phn2	\\
NGC~4993\tablenotemark{b} &	1.329	\pm	0.027	&	30.024	\pm	0.043	&	33.023	\pm	0.076	&	40.2	\pm	1.4	&	3215	&	2924 &	2989 & 2888	& 143	&	\phn8 \\
NGC~5322 &	1.432	\pm	0.017	&	29.695	\pm	0.013	&	32.490	\pm	0.072	&	31.5	\pm	1.0	&	1857	&	2071	&	2365	& 2366 &	213	&	\phn8  \\
NGC~5353 &	1.525	\pm	0.017	&	30.116	\pm	0.028	&	32.711	\pm	0.076	&	34.8	\pm	1.2	&	2510	&	2645	&	2712	& 2827 &	160	&	18 \\
NGC~5490 &	1.490	\pm	0.018	&	31.597	\pm	0.036	&	34.267	\pm	0.080	&	71.4	\pm	2.6	&	5092	&	5357	&	5307	& 5043 &	280	&	\phn8	\\
NGC~5557 &	1.378	\pm	0.017	&	30.546	\pm	0.015	&	33.458	\pm	0.072	&	49.2	\pm	1.6	&	3390	&	3295	&	3667	& 3363 &	158	&	\phn4	\\
NGC~5839 &	1.386	\pm	0.019	&	29.475	\pm	0.027	&	32.369	\pm	0.078	&	29.8	\pm	1.1	&	1417	&	1825	&	1534	& 1821 &	280	&	13  \\
NGC~6482 &	1.534	\pm	0.026	&	30.994	\pm	0.040	&	33.570	\pm	0.091	&	51.8	\pm	2.2	&	3859	&	3871	&	3824	& 3908 &	121	&	\phn3 	\\
NGC~6702 &	1.397	\pm	0.026	&	31.121	\pm	0.029	&	33.992	\pm	0.087	&	62.9	\pm	2.5	&	4592	&	4648	&	4751	& 4781 &	\phn\phn0	&	\phn1	\\
NGC~6964 &	1.421	\pm	0.025	&	30.867	\pm	0.052	&	33.685	\pm	0.096	&	54.6	\pm	2.4	&	3494	&	3698	&	3622	& 3811 &	253	&	\phn8	\\
NGC~7052 &	1.511	\pm	0.028	&	31.333	\pm	0.031	&	33.957	\pm	0.092	&	61.9	\pm	2.6	&	4362	&	4412	&	4353	& 4561 &	\phn\phn0	&	\phn1	\\
NGC~7242 &	1.529	\pm	0.033	&	31.918	\pm	0.044	&	34.504	\pm	0.104	&	79.6	\pm	3.8	&	5439	&	5610	&	6284	& 5703 &	389	&	17	\\
NGC~7619 &	1.516	\pm	0.023	&	30.727	\pm	0.023	&	33.341	\pm	0.081	&	46.6	\pm	1.7	&	3391	&	3043	&	3237	& 3102 & 306 &	16	\\
\medskip
ESO125-G006     &       0.22 \pm 0.024\tablenotemark{c} &       31.670  \pm     0.056   &       34.896  \pm     0.195   &       95.3    \pm     8.6     &       6762    &       6883    &       6766 & 6752     & \phn\phn0 &   \phn1   \\ 
\enddata
\vspace{0cm}
\tablenotetext{a}{F110W measurements from \citet{Nguyen2020} adjusted by --0.023 mag.}
\tablenotetext{b}{F110W measurements from \citet{Cantiello2018} adjusted by --0.023 mag.}
\tablenotetext{c}{The color used for ESO~125-G006 is \JminH, not \gminz.}
\tablenotetext{}{Column notes: 
(1) galaxy name; 
(2) \gminz\ colors from PanSTARRS transformed to the ACS \gzacs\ system using Equation~\ref{psacsequation}, with the exception of ESO~125-G006, which is \JminH\ from 2MASS transformed to $(J_{110}-H_{160})$ using Equation~\ref{2masswfc3equation};
(3) Apparent fluctuation magnitude;
(4) SBF distance modulus;
(5) SBF distance;
(6) individual and (7) group velocities in the CMB frame are taken from the Cosmic Flows 3 Extragalactic Database \citep[][see also \url{http://edd.ifa.hawaii.edu}]{Tully2015,edd} except NGC~4993 \citep{Cantiello2018}; 
(8) flow-corrected velocities derived from the linear velocity model used in Cosmic Flows 3 \citep{Graziani2019} using the calculator described by \citet{Kourkchi2020a}; 
(9) redshift velocities from the 2M++ compilation of \citet{Carrick2015};
(10) rms velocity dispersion of the cluster or group to which the galaxy belongs (0 for solitary galaxies); and 
(11) the number of galaxies in the cluster or group, both from the Cosmic Flows 3 Extragalactic Database.}
\end{deluxetable*}

\section{Distance Calculations \label{distancesection}}

Distance moduli for each annulus were computed using the absolute magnitude
\begin{equation}
  \overline M_{110} = -2.864 + 2.16[(g{-}z)_{\rm ACS} - 1.4],
\end{equation}
which is Equation~1 from \citet{Cantiello2018} using $(g_{475}-z_{850})$ values from ACS and including a correction of $-0.023$~mag to adjust to the LMC distance of 18.477 mag \citep{Piet2019}. Similarly,
\begin{equation}
  \overline M_{110} = -2.841 + 2.36[(g{-}z)_{\rm PS} - 1.3],
\end{equation}
can be used with \gminz\ from PanSTARRS as described above in Section~\ref{colortransformations}.
The distances listed in Table~\ref{distances} are uncertainty-weighted averages of all good annuli with reliable SBF measurements, free of dust or other galaxy subtraction residuals, and with the background level corrected as described above. The individual galaxies are shown in Figure~\ref{galaxyfig}, which includes the residual fluctuation images, the spatial power spectra fits, and the GCLF and background galaxy fits.

The nearby \citetalias{Jensen2015} calibration galaxies in the Virgo and Fornax clusters required no $k$-corrections to account for redshift; the more distant galaxies in this sample could require a redshift-dependent filter correction, which would have to be calculated using theoretical SBF models as a function of age and metallicity.  The most recent SBF models with published $k$-corrections are those of \citet{Liu2000}. For the standard $J$ filter (similar to F110W), the $k$-corrections are quite small: 0.003~mag at 50~Mpc and 0.006~mag at 100~Mpc. These model-dependent values are small enough to be insignificant for our distance measurements. The SBF calibration from \citetalias{Jensen2015} was compared with several newer stellar population models to explore the sensitivity of the calibration to the spread in age and metallicity as a function of the optical \gzacs\ and infrared \JHwfc\ colors. The SBF signals measured by \citetalias{Jensen2015} were consistent for old, metal-rich stellar populations but showed significant scatter for bluer, younger, and metal-poor galaxies. The \citetalias{Jensen2015} comparisons also showed significant scatter among the different models compared, but $k$-corrections were not calculated for those models. Improved IR SBF models will be required before reliable $k$-corrections can be calculated for distances greater than 100~Mpc.

\section{Applications}
\subsection{MASSIVE Galaxies and Black Hole Masses}
The new SBF distances presented here for the MASSIVE galaxies are on average 8\% shorter than the redshift distances adopted by \citet{Ma2014} at the beginning of the MASSIVE project. 
The original distance estimates were derived assuming a Hubble constant of $H_0\,{=}\,70$ kms$^{-1}$Mpc$^{-1}$ and using group velocities in the CMB frame from \citet{Crook2007}. The new SBF distance measurements presented here imply a value of $H_0\,{\sim}\,75$ kms$^{-1}$Mpc$^{-1}$, as reported by \citet{Verde2019} using the same velocities; the \citet{Ma2014} velocities included the velocity field corrections of \citet{Mould2000}, which lead to a somewhat larger value of $H_0$ than current best estimates \citep{Blakeslee2021}. 
The new SBF distances obtained here will help reduce one of the major systematic uncertainties in ongoing efforts to determine dynamical masses of black holes in MASSIVE galaxies \citep[e.g.][]{Liepoldetal2020}.

\subsection{Supernova Luminosity Calibration}
This study includes 24 Type Ia supernova host galaxies that can be used to extend the calibration of SN absolute magnitude to earlier galaxy types and nearer the centers of rich clusters. These SBF distances, along with other published SBF distances, is the subject of two companion papers by Garnavich et al.\ (2021) and Milne et al.\ (2021) addressing these and other issues related to the SN distance scale. The new SBF distances are also part of a concurrent study by Milne et al.\ (2021) that explores the dependence of supernova luminosity on galaxy type for the ``narrow-normal'' SN~Ia population. These SBF distances are crucial to extending the SN distance calibration to early-type galaxies for which Cepheid distances are not available, and for determining the relationship between supernova luminosity and UV color \citep{Milne2013,Milne2015,brown2017,brown2019}.

\subsection{The Hubble Constant}
The new SBF distances to a substantial sample of galaxies reaching 100~Mpc provides an opportunity to measure the Hubble constant independently of the SN distance scale, which is reported in detail in the paper by \citet{Blakeslee2021}. The combined sample of SBF distances gives a robust determination of the Hubble Constant $H_0\,{=}\,73.3$ kms$^{-1}$Mpc$^{-1}$, where the difference between the current best estimate and that from \citet{Verde2019} is the improvement to the correction for bulk flows \citep[for details, see][]{Blakeslee2021}. The new SBF distances provide important insights into the current disagreement between the values of $H_0$ derived from SN and cosmic microwave background radiation measurements \citep{Riess2019, Planck2020}.

\section{Summary}
This paper presents an updated description of the SBF methodology of \citet{Jensen2015} that we used to measure extragalactic distances out to 100~Mpc with WFC3/IR F110W. The revised calibration of \Mbarj\ includes the updated distance to the LMC \citep{Piet2019} and corrections to the PSF normalization \citep{Cantiello2018}. We also provide transformations between PanSTARRS, SDSS, and ACS \gminz\ colors, and between 2MASS and WFC3/IR \jminh, which are required to determine the SBF distances. The revised procedures and calibration are applied to WFC3/IR F110W observations of 63 galaxies from a variety of \hst\ programs to provide a uniform database of IR SBF distances that can now be used for a number of new studies related to the extragalactic distance scale, including a measurement of the Hubble constant \citep{Blakeslee2021}.

\acknowledgments
This project was supported by NASA grants HST-GO-14219, HST-GO-14654, and HST-GO-15265 from the Space Telescope Science Institute, which is operated by AURA, Inc., under NASA contract NAS 5-26555. 
J.~Blakeslee was supported in part by the International Gemini Observatory, a program of NSF’s NOIRLab, which is managed by the Association of Universities for Research in Astronomy (AURA) under a cooperative agreement with the National Science Foundation, on behalf of the Gemini partnership of Argentina, Brazil, Canada, Chile, the Republic of Korea, and the United States of America.
The MASSIVE Survey is supported in part by NSF grants AST-1815417 and AST-1817100. C.-P.~Ma acknowledges support from the Heising-Simons Foundation and the Miller Institute for Basic Research in Science.
M. Cantiello acknowledges support from MIUR, PRIN 2017 (grant 20179ZF5KS).
J. Lucey was supported by the Science and Technology Facilities Council through the Durham Astronomy Consolidated Grants ST/P000541/1 and ST/T000244/1.

The Pan-STARRS1 Surveys (PS1) and the PS1 public science archive have been made possible through contributions by the Institute for Astronomy, the University of Hawaii, the Pan-STARRS Project Office, the Max-Planck Society and its participating institutes.

This project used data from the Sloan Digital Sky Survey.
Funding for the SDSS and SDSS-II has been provided by the Alfred P. Sloan Foundation, the Participating Institutions, the National Science Foundation, the U.S. Department of Energy, the National Aeronautics and Space Administration, the Japanese Monbukagakusho, the Max Planck Society, and the Higher Education Funding Council for England. 

This publication makes use of data products from the Two Micron All Sky Survey, which is a joint project of the University of Massachusetts and the Infrared Processing and Analysis Center/California Institute of Technology, funded by the National Aeronautics and Space Administration and the National Science Foundation.

This research has made use of the NASA/IPAC Extragalactic Database (NED), which is funded by the National Aeronautics and Space Administration and operated by the California Institute of Technology.

\facilities{HST(WFC3/IR, ACS), PanSTARRS, 2MASS, SDSS}

\software{SExtractor, DoPHOT, ELLIPROF, TinyTim}


\vfill \eject
\appendix
\section{Alternate Color Transformations and Distances \label{colorappendix}}

The WFC3/IR SBF calibrations from \citet{Cantiello2018} and \citetalias{Jensen2015}, shifted by $-0.023$~mag to take into account the updated distance to the LMC \citep{Piet2019} are:
\begin{equation}
    \overline M_{110} = -2.864 + 2.16[(g_{475}{-}z_{850})_{\rm ACS} - 1.4]
    \label{equationA1}
\end{equation}
\begin{equation}
    \overline M_{160} = -3.617 + 2.13[(g_{475}{-}z_{850})_{\rm ACS} - 1.4]
\end{equation}
\begin{equation}
    \overline M_{110} = -2.891 + 6.7[(J_{110}{-}H_{160})_{\rm WFC3} - 0.27]
    \label{equationA3}
\end{equation}
\begin{equation}
    \overline M_{160} = -3.645 + 7.1[(J_{110}{-}H_{160})_{\rm WFC3} - 0.27]
\end{equation}

The preferred calibration of \Mbarj\ presented by \citetalias{Jensen2015} depends on the \gzacs\ colors measured in the F475W and F850LP filters using ACS on \hst. 
In this Appendix we provide the color transformations required to make WFC3/IR SBF distance measurements using \gminz\ colors from the Sloan Digital Sky Survey (SDSS) and \JminH\ colors from 2-Micron All-Sky Survey (2MASS) images as we did in Section~\ref{distancesection} for the PanSTARRS case.

The SDSS images we used for this study (including both the SBF targets and the ACSVCS galaxies) were taken from the SDSS DR9 data release\footnote{\url{http://sdss.org}}  and processed using the ``ubercal'' pipeline \citep{Padmanabhan2008}. The ubercal algorithm achieves ${\sim}1$\% photometric calibration across the full survey area using relative stellar fluxes in overlapping observations. 
The native SDSS flux units were offset by 0.02 mag in the $z$ filter (AB mags) as recommended on the SDSS website\footnote{\url{http://www.sdss.org/dr16/algorithms/fluxcal/}} and extinction corrected \citep{SF2011}.

To create color maps from SDSS images, a similar procedure was followed as was used for the PanSTARRS images (see Section~\ref{colortransformations}). A smooth galaxy model was subtracted, objects were identified and masked using SExtractor, and the background level checked. 
The galaxy luminosity and color profiles were compared with ACS and PanSTARRS to confirm that the process yielded consistent results.

\begin{figure}
\begin{center}
\plotone{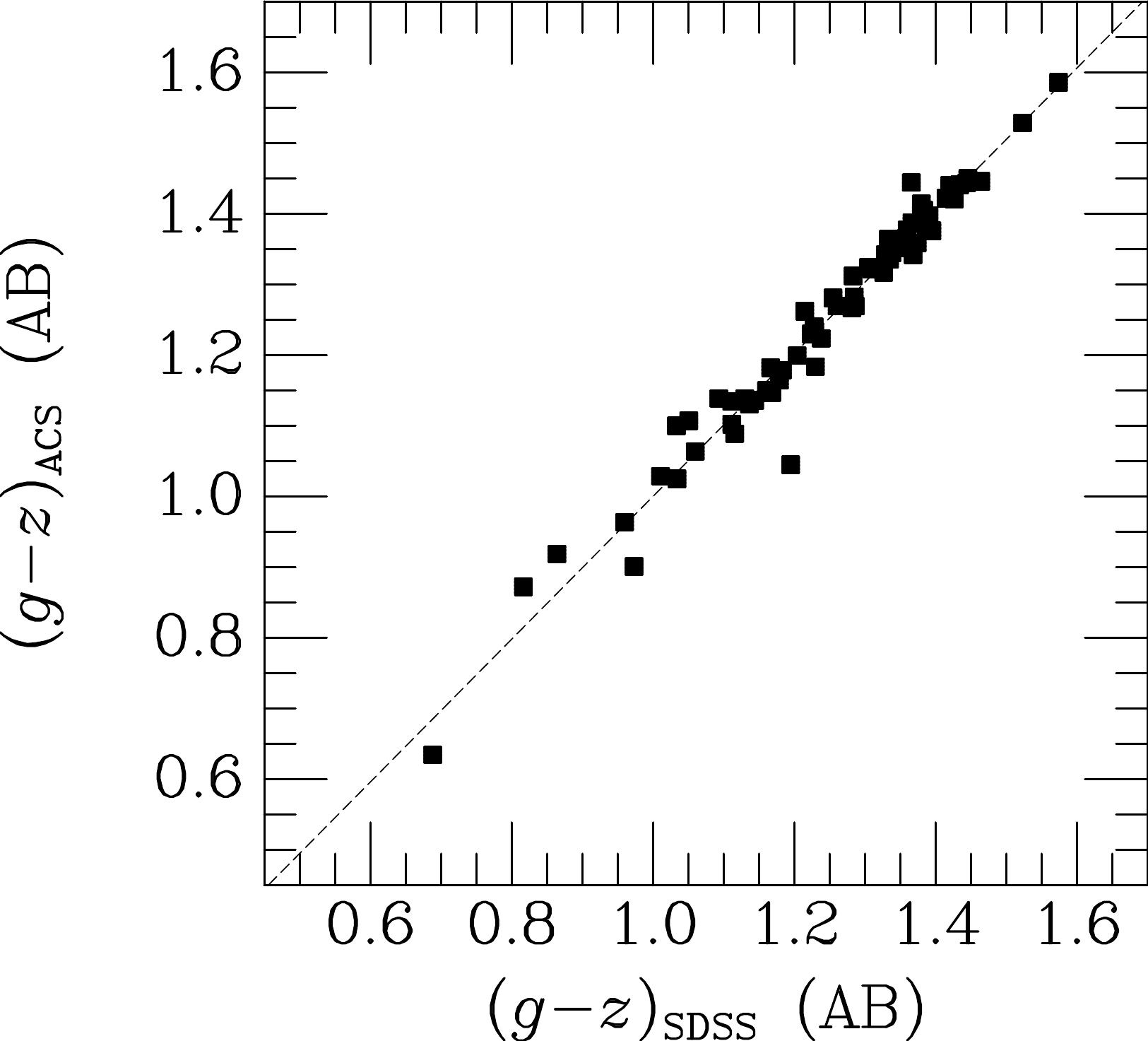}
\caption{Comparison of SDSS and ACS \gminz\ color photometry for 72 galaxies in the ACSVCS sample. The dashed line is the fit from Equation~\ref{sdssacsequation}. \label{appcolorcalibfig1}} 
\end{center}
\end{figure}

The SDSS \gminz\ color measurements for 72 ACSVCS galaxies are very consistent with the ACS \gminz\ colors, with almost no offset or slope, but with somewhat higher scatter than the PanSTARRS measurements (0.019 mag in the inner SBF annulus, increasing to 0.040 and 0.098 mag moving to the outer annuli). The relationship shown in Fig.~\ref{appcolorcalibfig1} is 
\begin{equation}
\label{sdssacsequation}
(g{-}z)_{ACS} = 1.010(g{-}z)_{SDSS} - 0.010.
\end{equation}
For the red early-type galaxies in this sample, the relationship between SDSS and PanSTARRS can be written as
\begin{equation}
    (g{-}z)_{SDSS} = 1.081(g{-}z)_{PS}+0.001.
\end{equation}

To use SDSS \gminz\ for computing distances, the SBF calibration is therefore
\begin{equation}
    \overline M_{110} = -2.855 + 2.18[(g{-}z)_{\rm SDSS} - 1.4].
\end{equation}

\begin{figure}
\begin{center}
\plotone{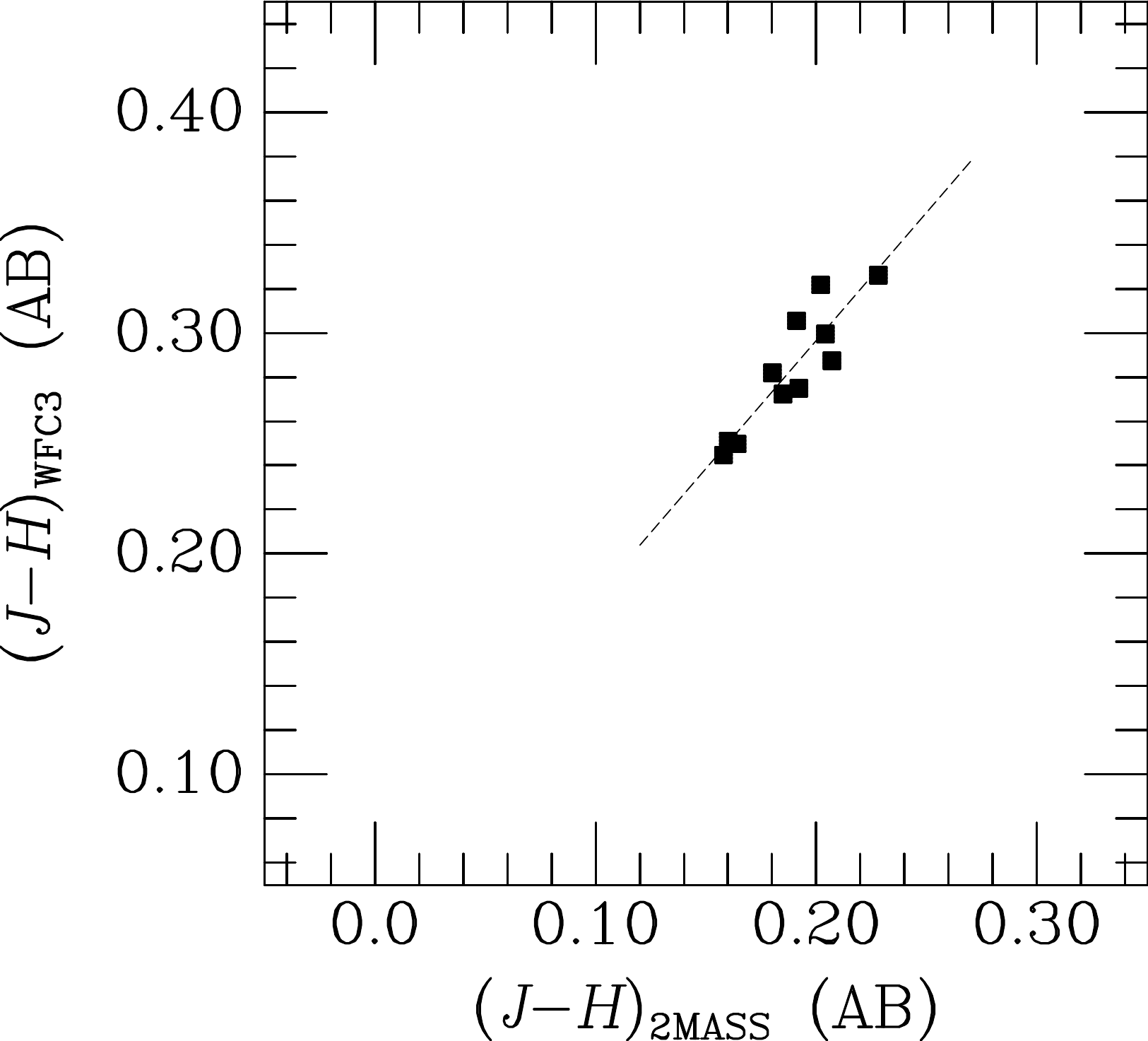}
\caption{Comparison of WFC3/IR and 2MASS \JminH\ colors for 11 calibrator galaxies from \citetalias{Jensen2015}. The dashed line is the relation from Equation~\ref{2masswfc3equation}. The figure axis limits are set so that the color ranges reflect the same range in derived distance as in Figures~\ref{colorcalibfig} and \ref{appcolorcalibfig1}, thus the relative scatter in these figures as they affect the distance measurements may be compared directly. The distance calibration using \JminH\ has larger scatter because of the limited leverage provided by the relatively small range in \JminH\ spanned by the calibrator galaxies (and early-type galaxies in general).
\label{appcolorcalibfig2}} 
\end{center}
\end{figure}

For SBF targets for which ACS, PanSTARRS, or SDSS \gminz\ data are not available, 2MASS \JminH\ is a viable (though less precise) alternative.\footnote{\url{http://irsa.ipac.caltech.edu/Missions/2mass.html}} 
The SBF distances calculated using \gminz\ are superior to the near-IR \JminH\ colors for two reasons: first, the PanSTARRS optical images are deeper and the \gminz\ colors are more accurately measured than 2MASS $J$ and $H$; and second, the dependence of \Mbarj\ on color is steeper in the near-IR than it is for \gminz.
The slope of \Mbarj\ with \gminz\ is 2.16 (Equation~\ref{equationA1}); this means an uncertainty in color of 0.01~mag will result in 0.022~mag uncertainty in distance modulus. Alternatively, when using \JHwfc, the calibration has a slope of 6.7 (Equation~\ref{equationA3}) because of the greater age and metallicity sensitivity at longer wavelengths, and so generally IR colors result in less precise distance measurements. For 2MASS, the SBF distance uncertainties are typically twice as large as those for PanSTARRS \gminz. 

2MASS \JminH\ colors were similarly translated into the WFC3/IR (F110W--F160W) system for SBF measurements using the eleven red galaxies in the Virgo/Fornax calibration sample \citepalias{Jensen2015}. For the 2MASS images, the magnitude zero points were retrieved from the image headers and the background level checked \citep[][Sec. 2.2]{Goullaud2018}. 
The 2MASS \JminH\ colors differ from WFC3/IR \JHwfc\ by about 10\%, with a color translation relationship of
\begin{equation}
\label{2masswfc3equation}
(J_{110}{-}H_{160})_{WFC3} = 1.158(J{-}H)_{2MASS} + 0.065
\end{equation}
as shown in Figure~\ref{appcolorcalibfig2}.
The corresponding equation for calculating distance moduli is
\begin{equation}
   \overline M_{110} = -2.891 + 7.76[(J{-}H)_{\rm 2MASS} - 0.177]
\end{equation}
where the \JminH\ color is in AB mag, which can be computed from the equivalent Vega \JminH\ by subtracting 0.492 mag.
While the distances measured using \JminH\ are less precise, it is still useful for southern hemisphere galaxies for which PanSTARRS data is unavailable, and in regions where foreground extinction makes optical color measurements less reliable. 

\begin{figure}
    \centering
    \plotone{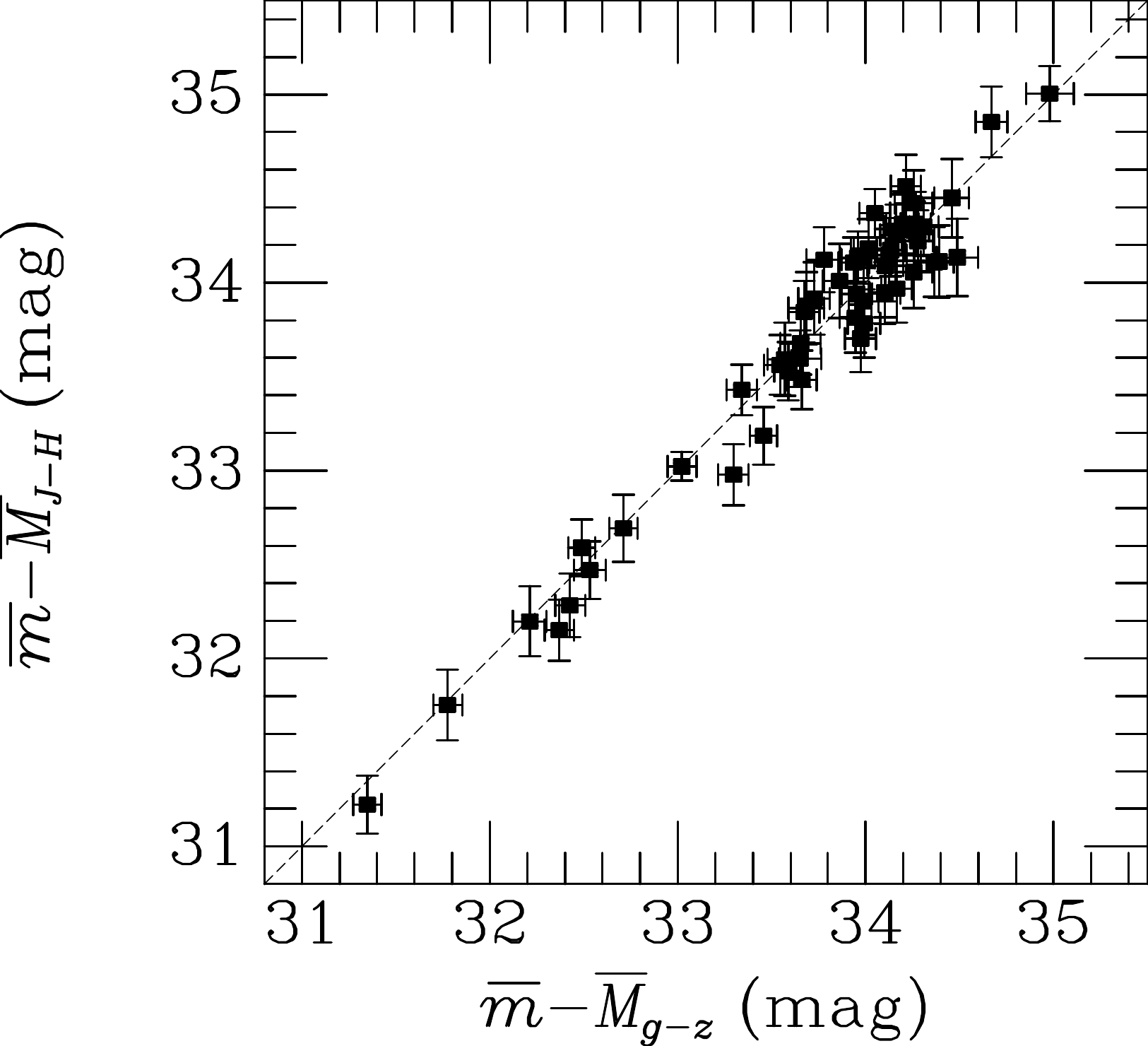}
    \caption{SBF distance moduli computed using \JminH\ plotted against the \gminz\ SBF distance moduli for the 54 galaxies for which we have reliable 2MASS colors. The dotted line is the 1:1 line, and is not a fit to the data. The \jminh-calibrated distances are less precise but show no significant offset from the distances derived from the PanSTARRS \gminz\ color calibration. With one exception (ESO~125-G006), the distances presented in this paper are all derived using the PanSTARRS \gminz\ calibration.}
    \label{colorcomparison}
\end{figure}

\begin{figure}
    \centering
    \plotone{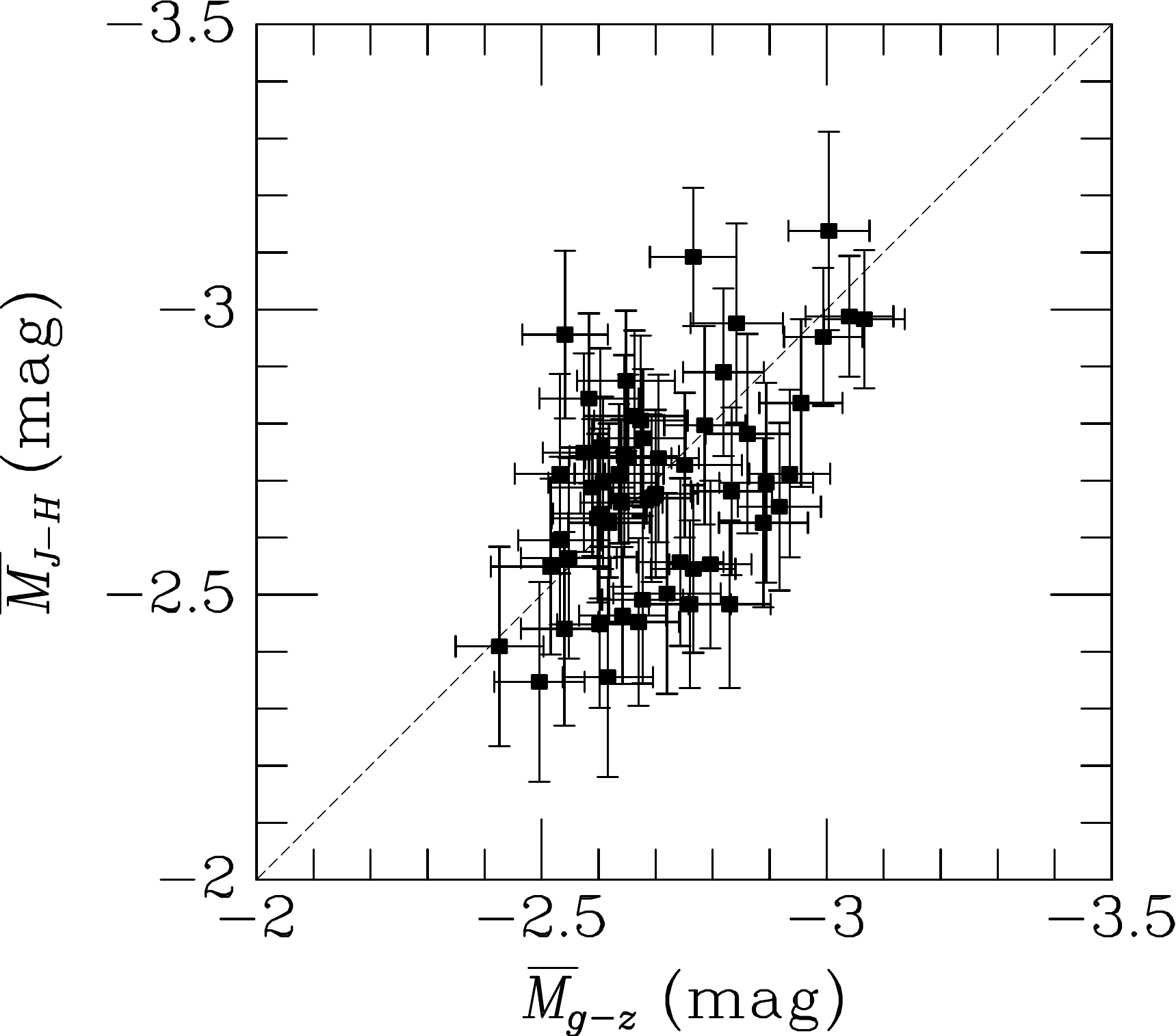}
    \caption{Absolute \Mbarj\ derived from \JminH\ and \gminz\ colors compared to each other. This plot is similar to Figure~\ref{colorcomparison}, but is distance independent. The dotted line is the 1:1 line, and is not a fit to the data.}
    \label{MbarvsMbar}
\end{figure}

We can use the current sample to establish the reliability of the \JminH\ measurements by looking for systematic offsets between the optical and IR color systems and as a function of foreground extinction. The 2MASS colors are completely independent of the PanSTARRS and SDSS \gminz\ colors, and are less sensitive to dust absorption and Galactic foreground extinction. 
We plotted the \gminz\ and \JminH\ SBF distance moduli and \Mbarj\ values against each other in Figures~\ref{colorcomparison} and \ref{MbarvsMbar}. We found that the two systems agree with an insignificant median offset between the \JminH\ and \gminz\ calibrations of 0.011~mag in distance modulus (measured in individual annuli) and 0.004~mag in \Mbarj\ (weighted averages) for 54 galaxies, which is less than the typical uncertainty on an individual \jminh\ distance measurement (0.17~mag) divided by the square root of 54. We computed the reduced ${\chi}^2/N=0.97$ relative to the one-to-one line (with no degrees of freedom), indicating that our total uncertainties on the distance moduli are statistically robust. 

\begin{figure}
    \centering
    \plotone{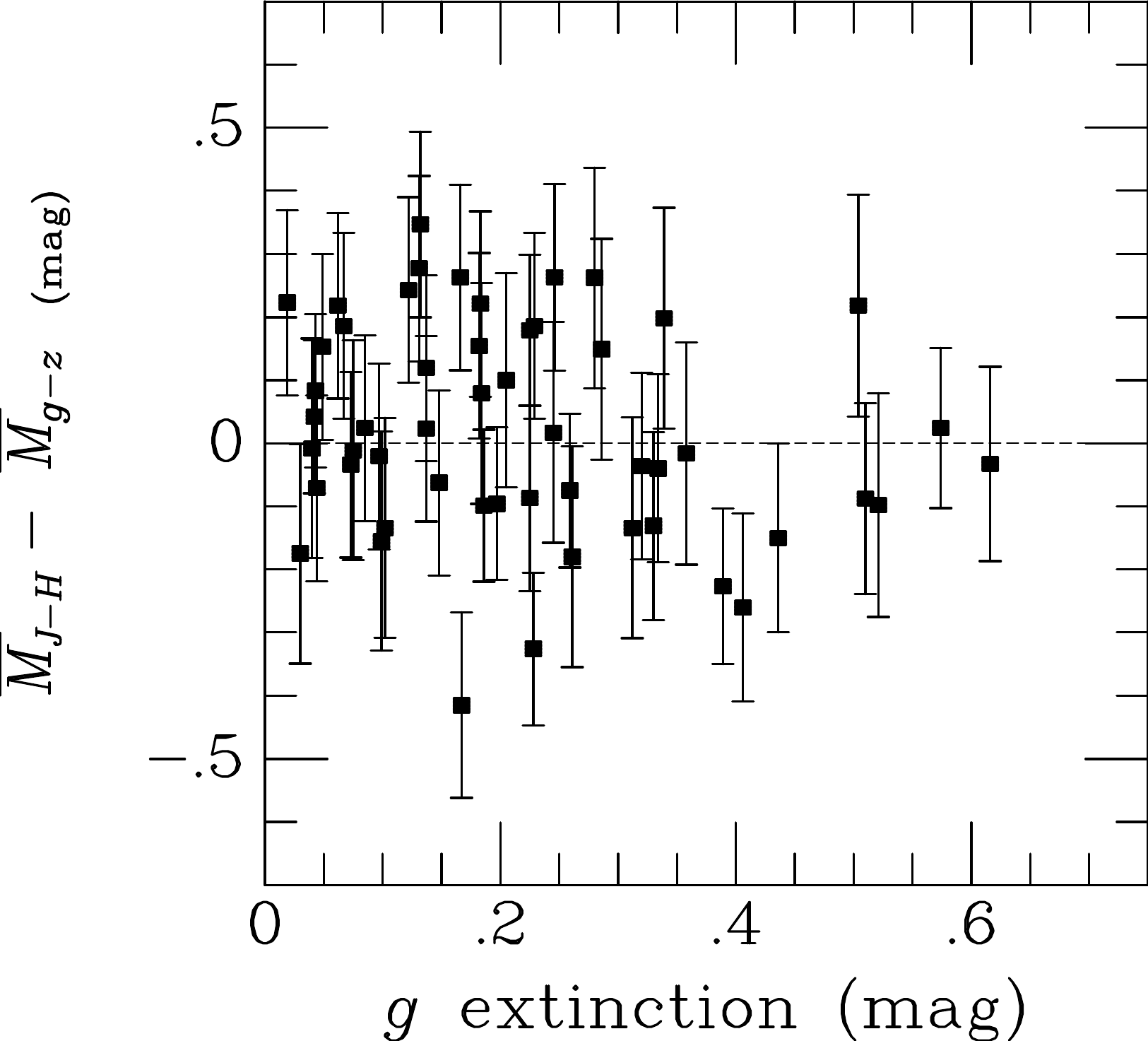}
    \caption{The difference between the \JminH\ and \gminz\ distances are plotted as a function of optical $g$-band extinction. Given the relatively large scatter in \Mbarj\ determined from the \gminz\ and \JminH\ colors, we looked for systematic trends in the absolute magnitude differences as a function of extinction and found none.}
    \label{MbarvsExt}
\end{figure}

We also checked for systematic offsets between the distances derived using optical and IR colors in Figure~\ref{MbarvsExt}. There is no evidence of any systematic offset in the \gminz-derived distances compared to the \JminH\ values for $g$-band extinctions up to 0.6 mag, nor any evidence that the scatter in the calibration has any dependence on foreground extinction.

\clearpage
\bibliographystyle{aasjournal}
\bibliography{jensen2021}
\bigskip


These figures are included as an online figure set in the published version. They are reproduced here for the convenience of the reader. \\

\begin{figure*}
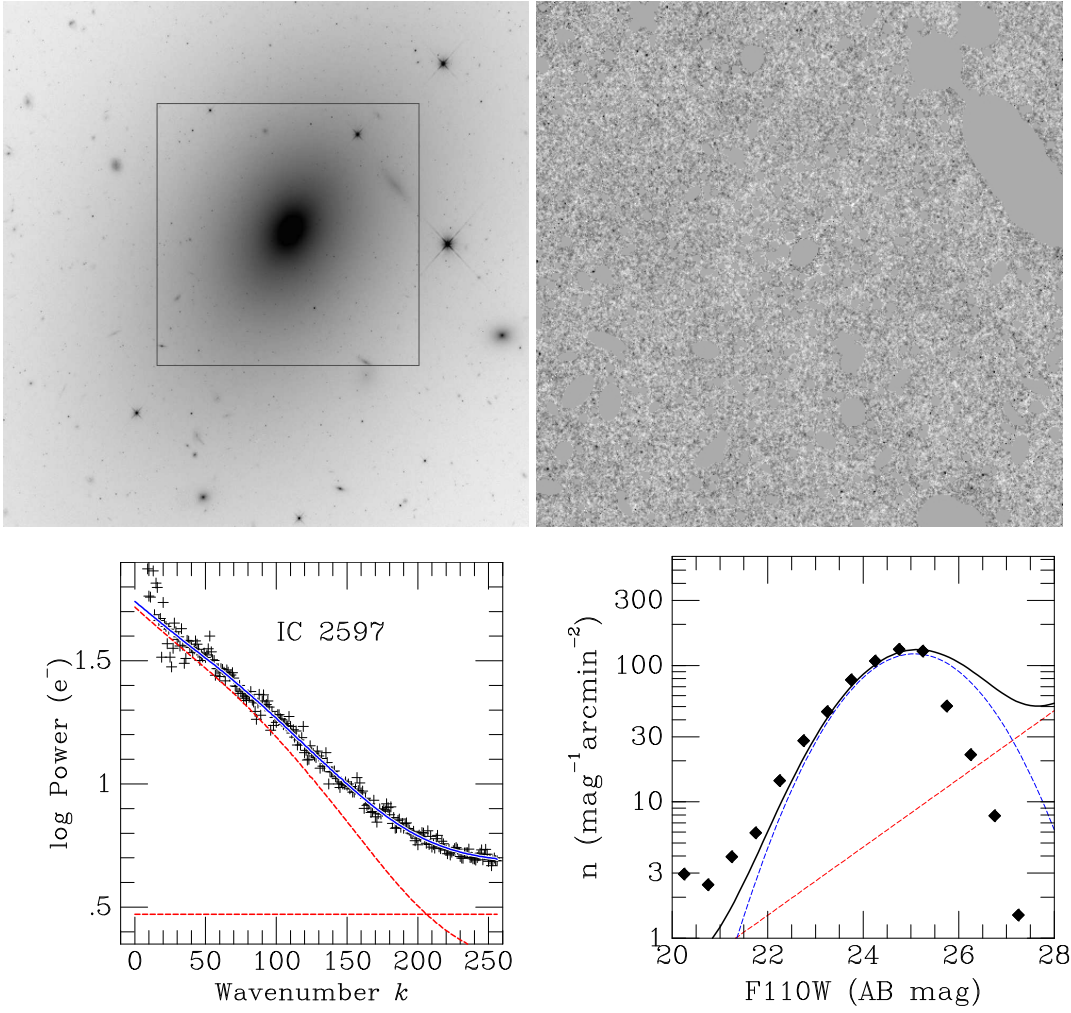

\begin{center}
\includegraphics[scale=0.2]{ic2597j}
\includegraphics[scale=0.4]{ic2597jresid} \\
\vspace{10pt}
\includegraphics[scale=0.4]{ic2597jc2}
\hspace{-25pt}
\includegraphics[scale=0.4]{ic2597jlkn6}
\caption{Combined figure for IC~2597.}
\end{center}
\end{figure*}
\clearpage

\begin{figure*}
\begin{center}
\includegraphics[scale=0.2]{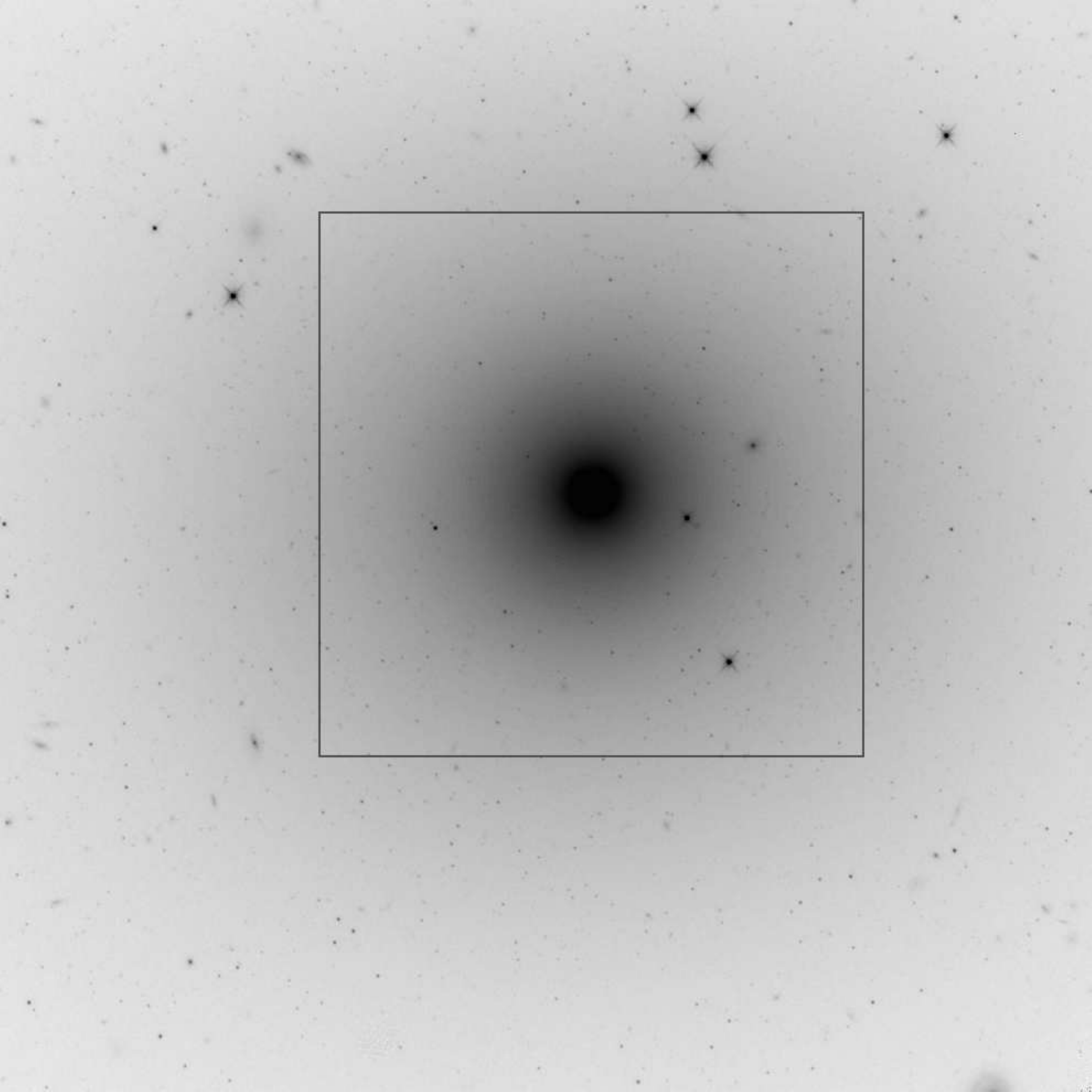}
\includegraphics[scale=0.4]{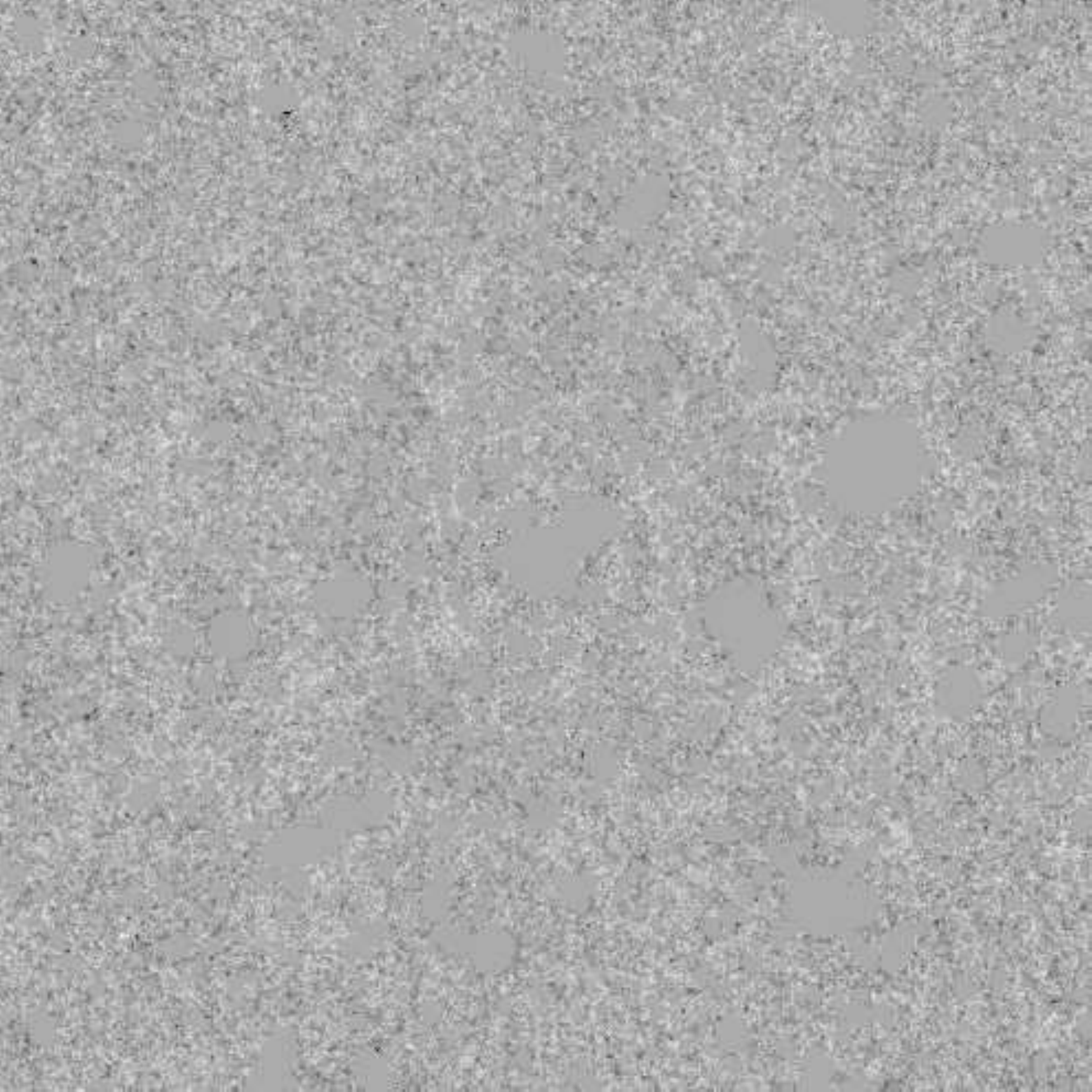} \\
\vspace{10pt}
\includegraphics[scale=0.4]{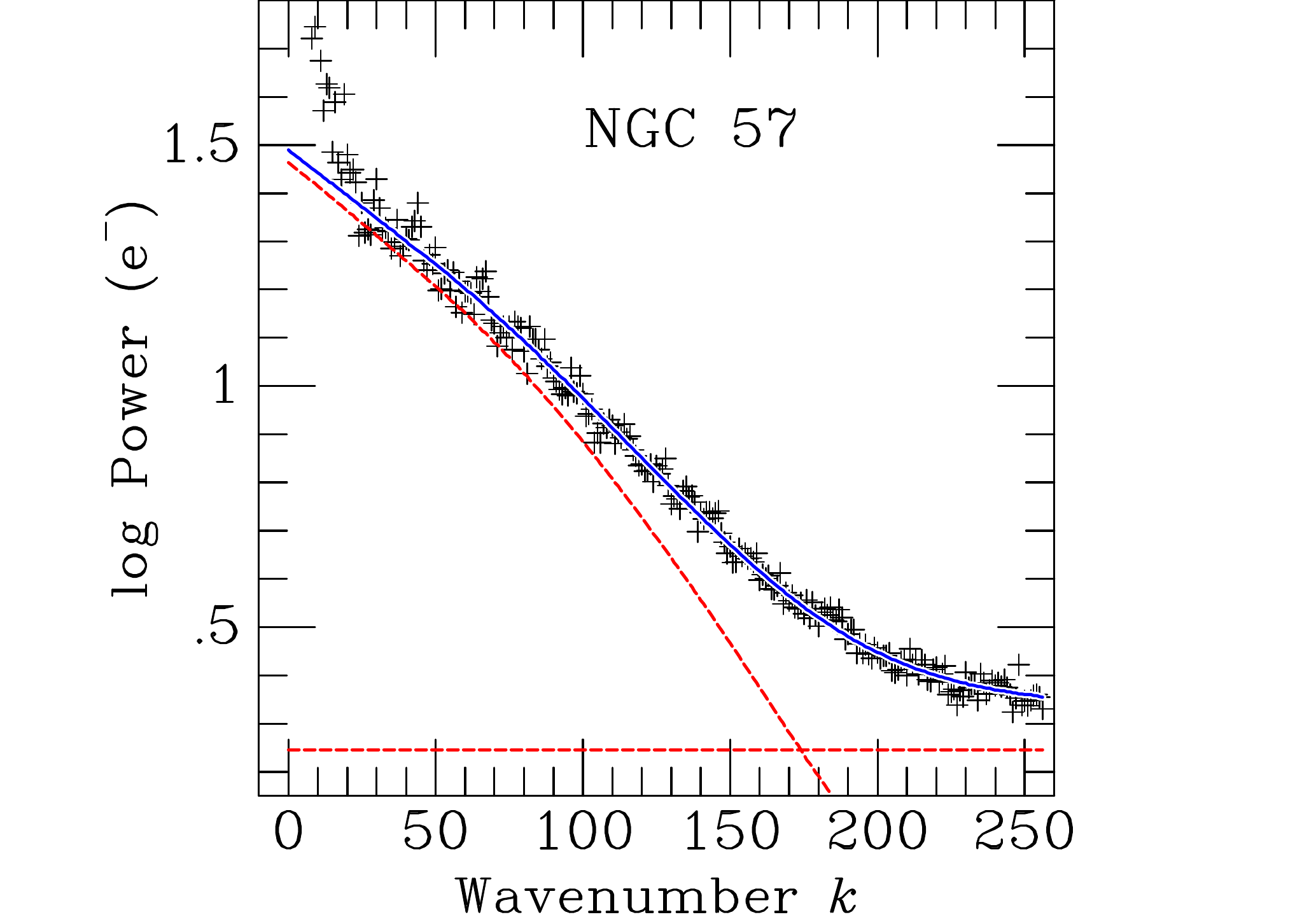}
\hspace{-25pt}
\includegraphics[scale=0.4]{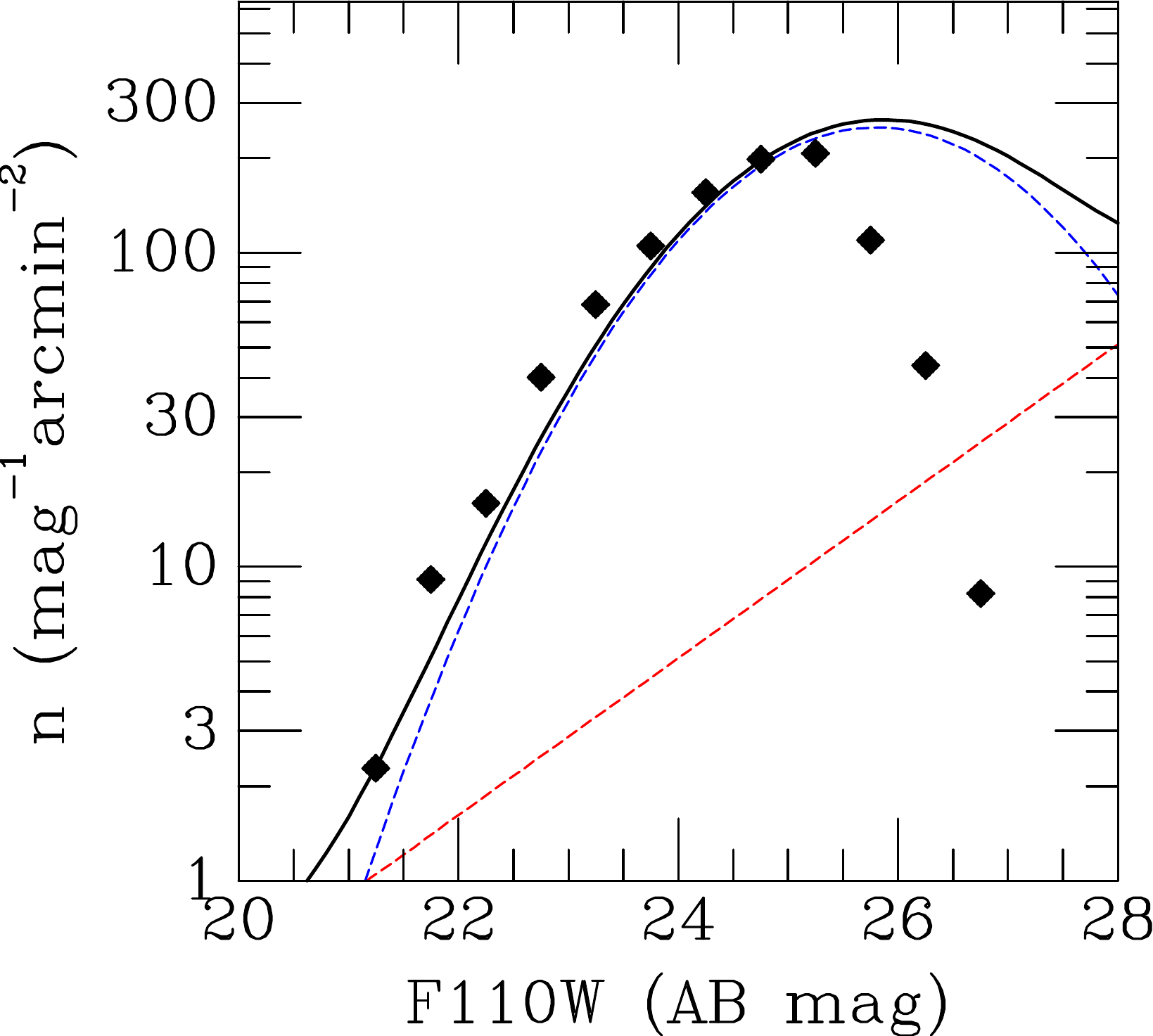}
\caption{Combined figure for NGC~57.}
\end{center}
\end{figure*}
\clearpage

\begin{figure*}
\begin{center}
\includegraphics[scale=0.2]{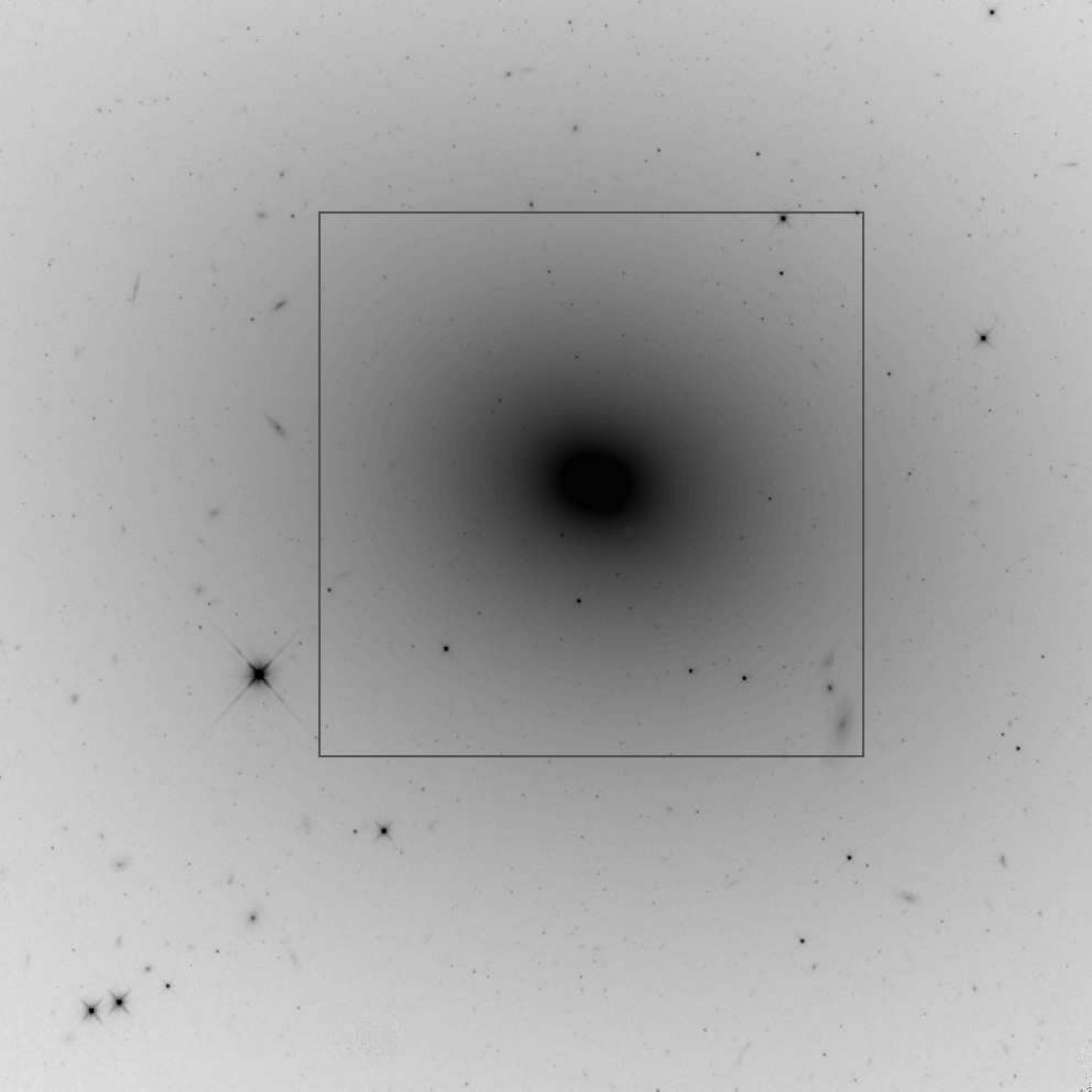}
\includegraphics[scale=0.4]{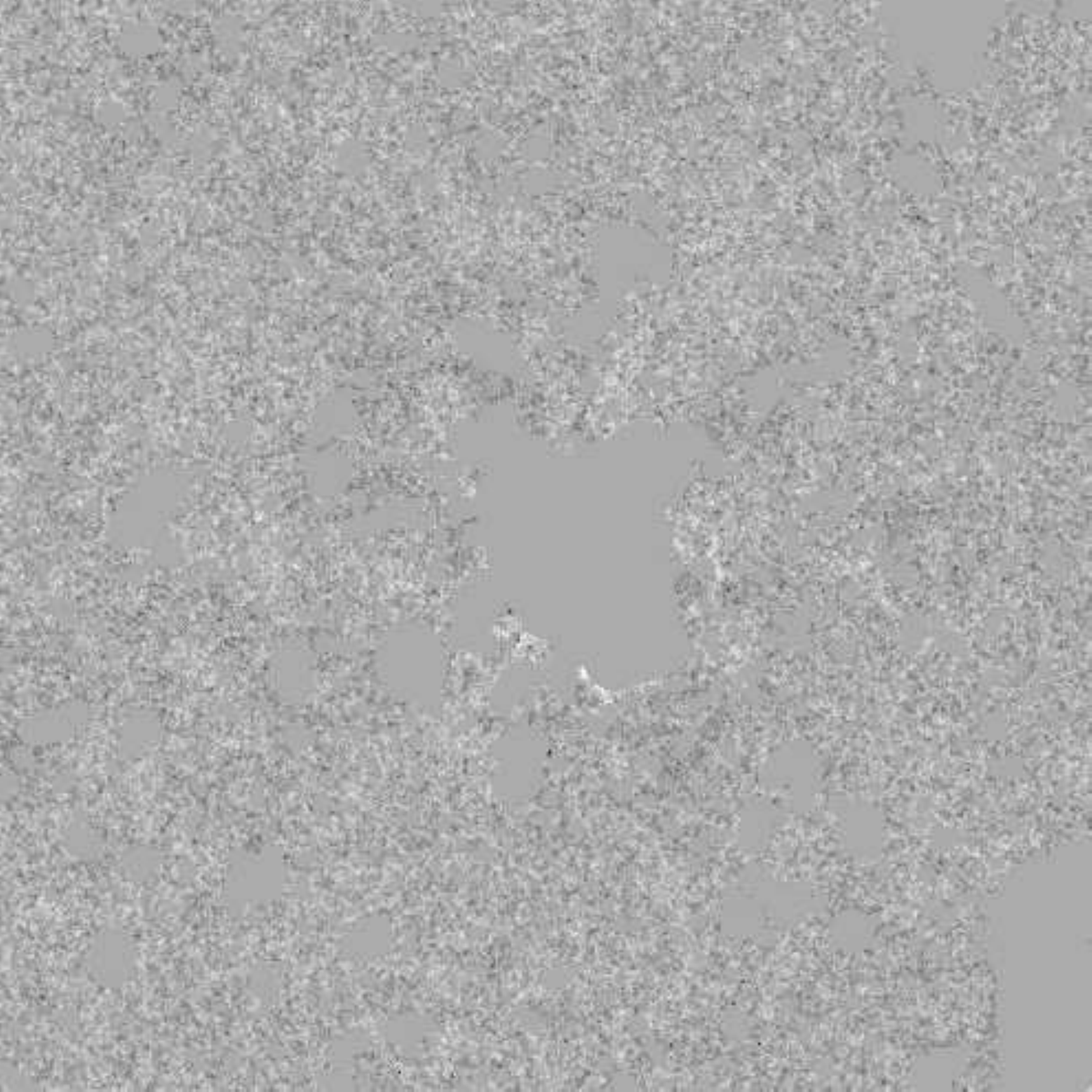} \\
\vspace{10pt}
\includegraphics[scale=0.4]{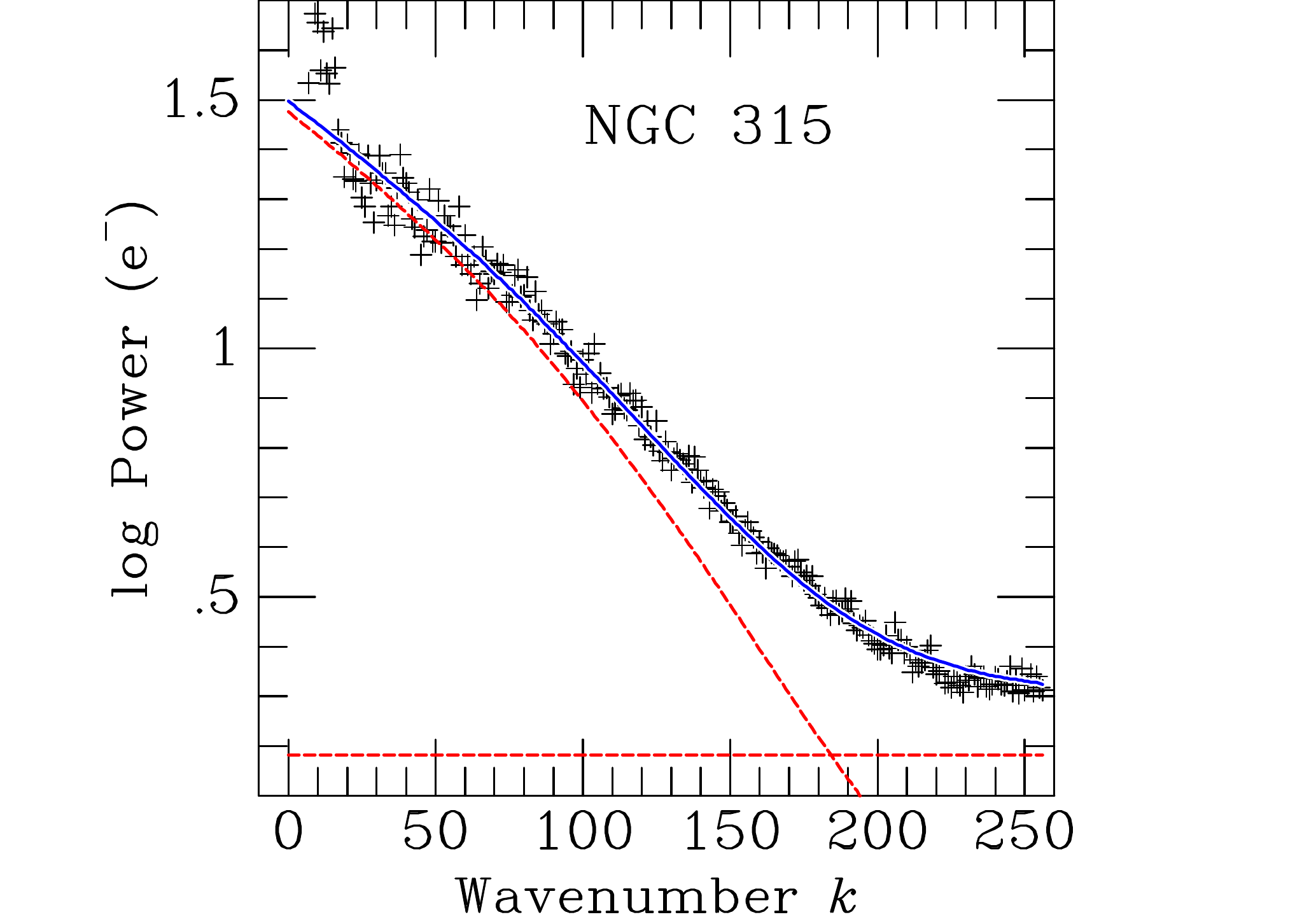}
\hspace{-25pt}
\includegraphics[scale=0.4]{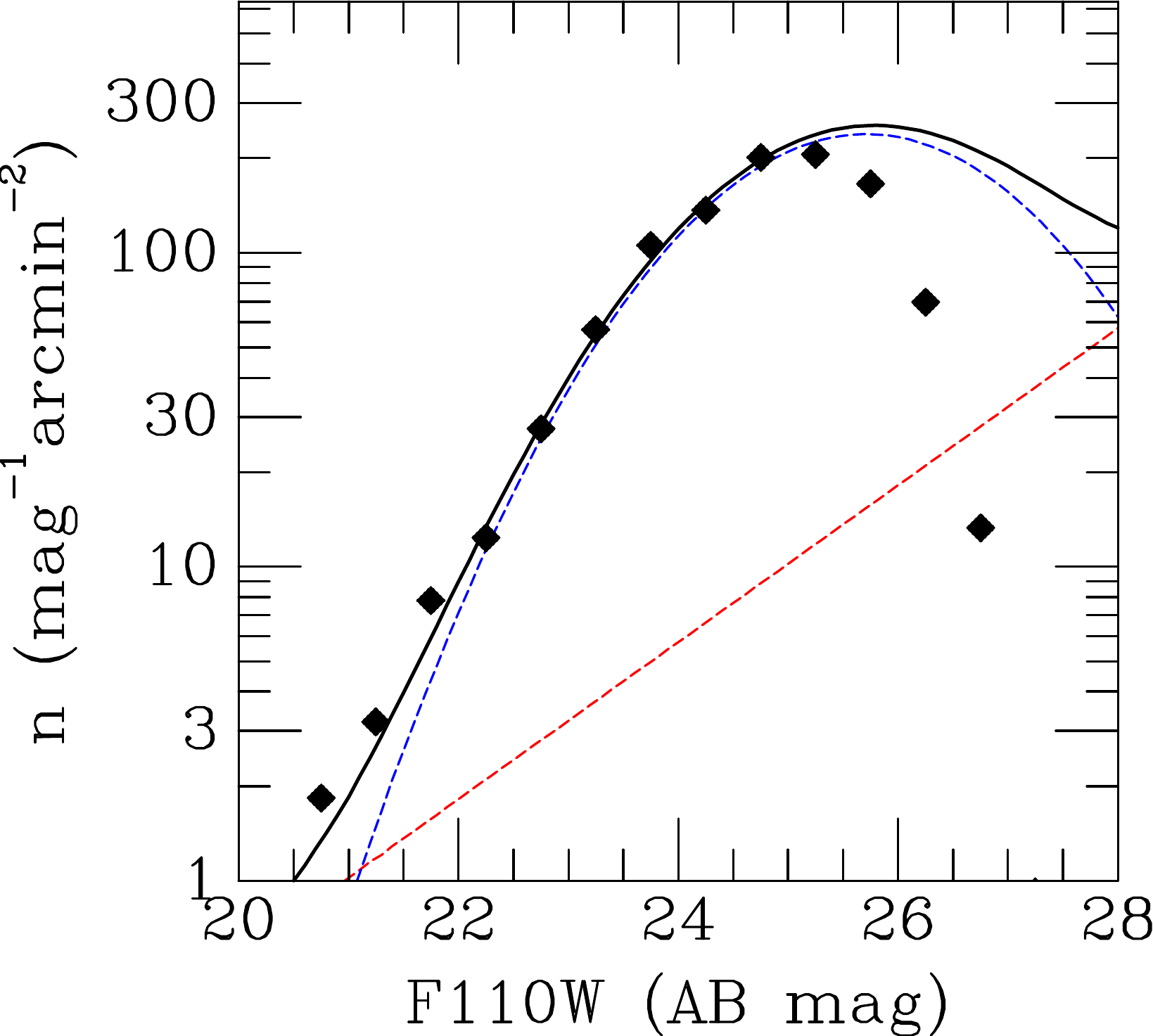}
\caption{Combined figure for NGC~315.}
\end{center}
\end{figure*}
\clearpage

\begin{figure*}
\begin{center}
\includegraphics[scale=0.2]{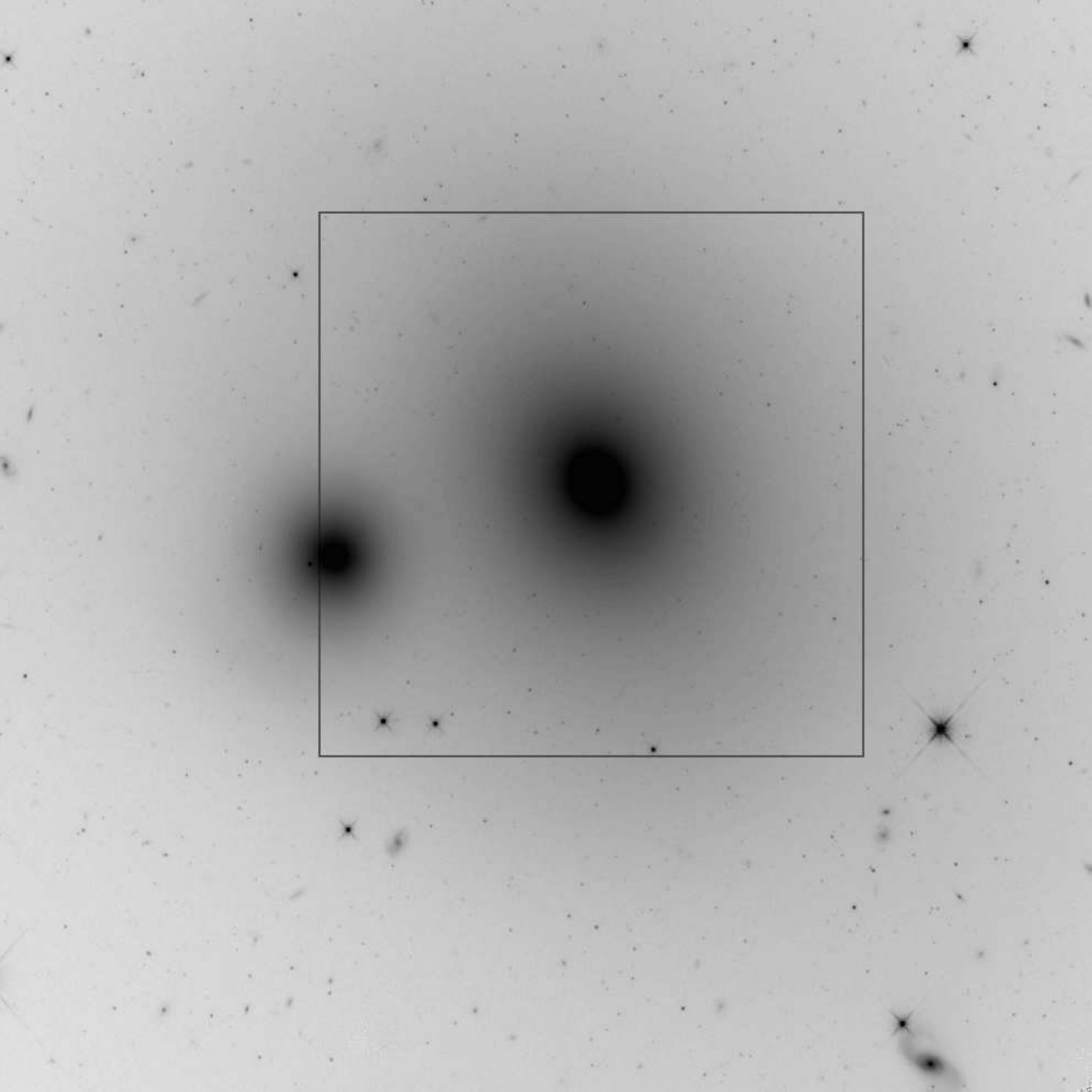}
\includegraphics[scale=0.4]{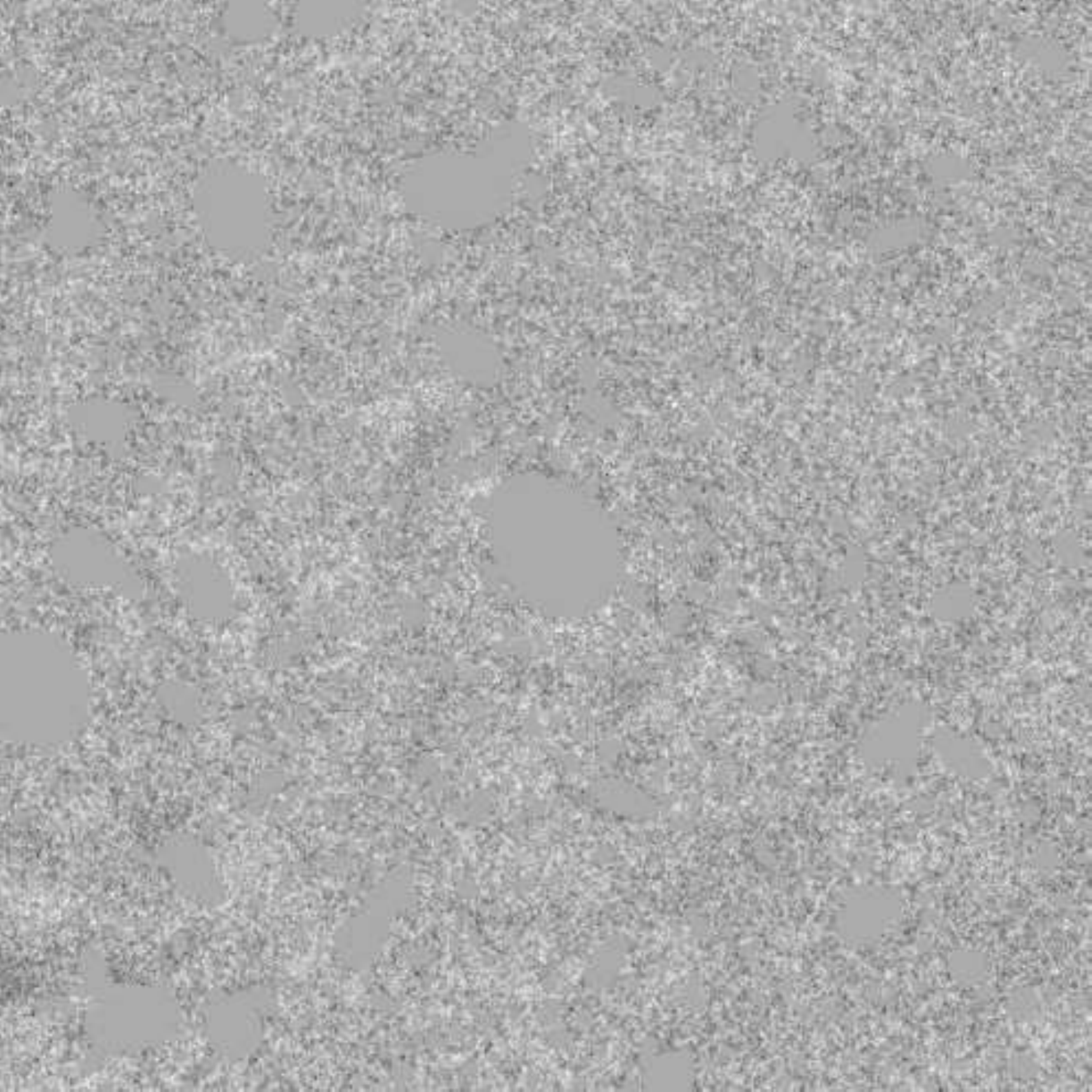} \\
\vspace{10pt}
\includegraphics[scale=0.4]{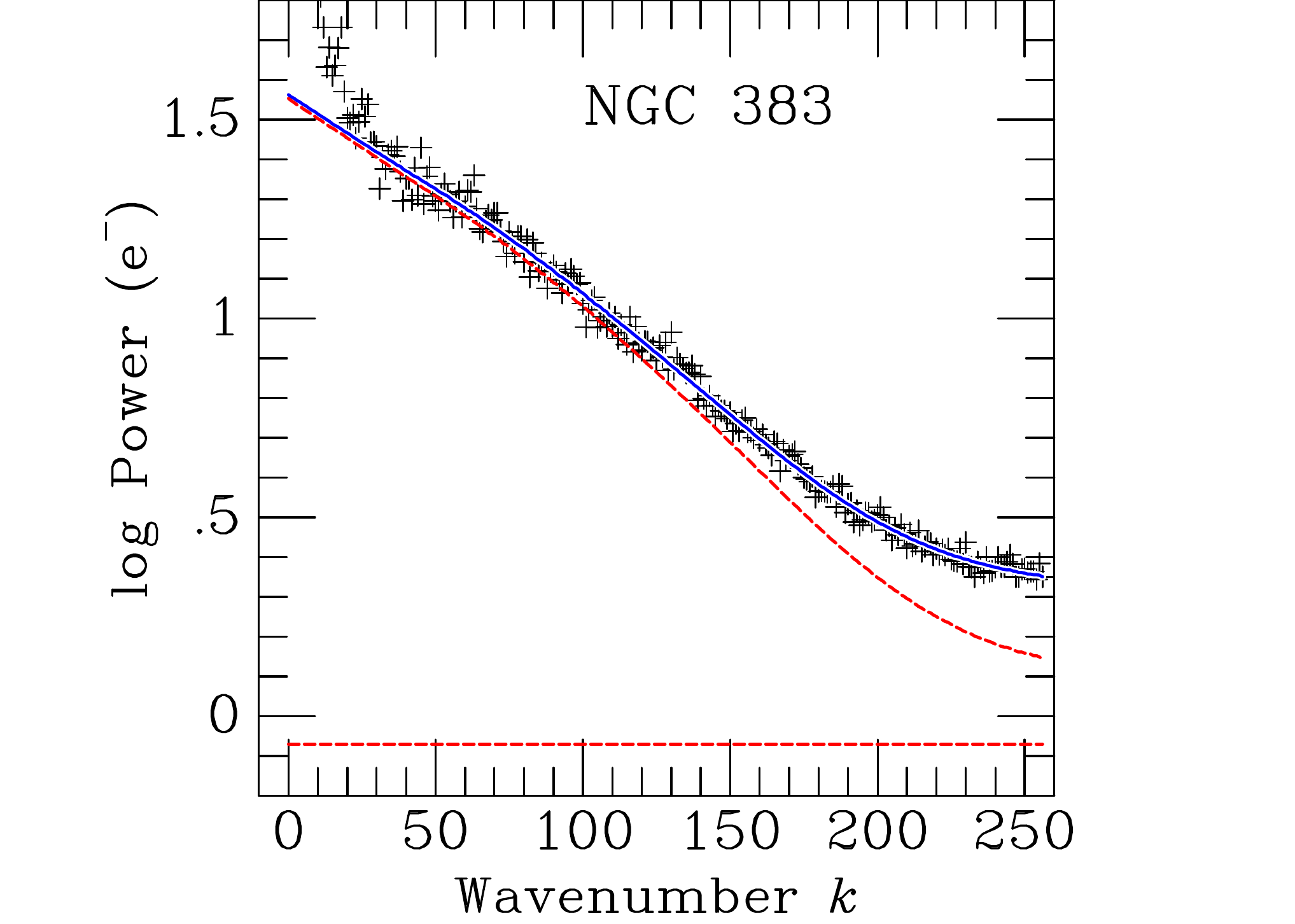}
\hspace{-25pt}
\includegraphics[scale=0.4]{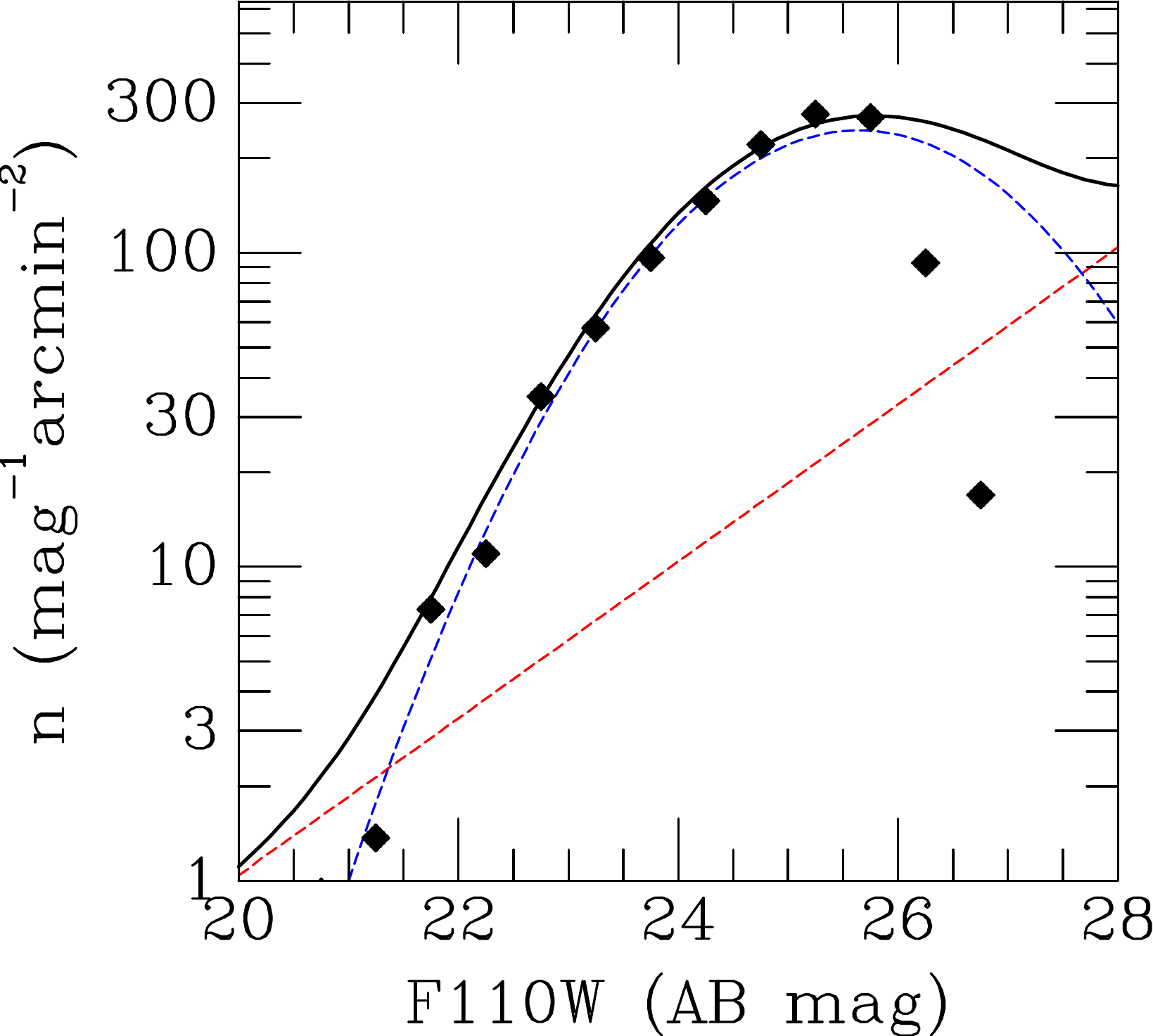}
\caption{Combined figure for NGC~383.}
\end{center}
\end{figure*}
\clearpage

\begin{figure*}
\begin{center}
\includegraphics[scale=0.2]{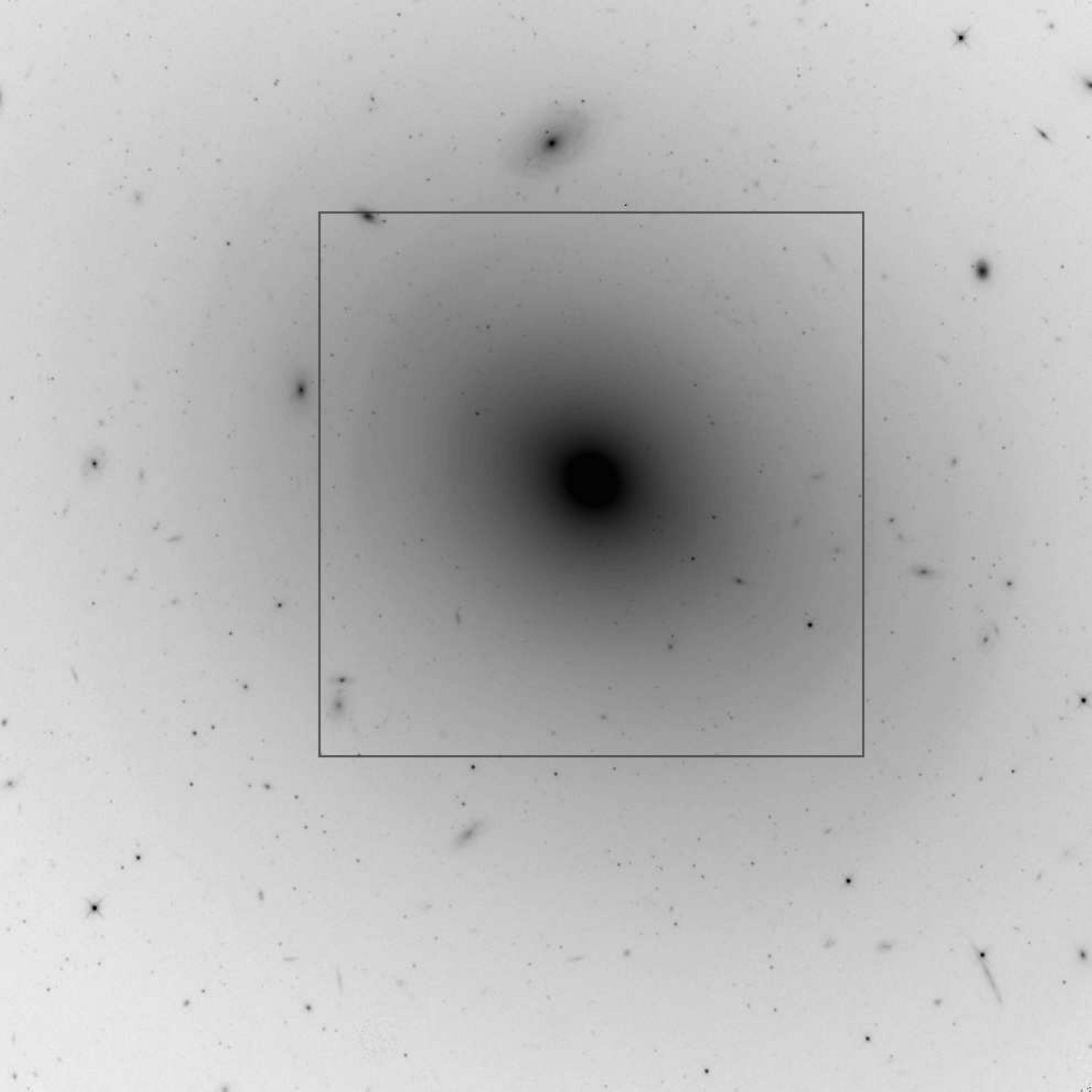}
\includegraphics[scale=0.4]{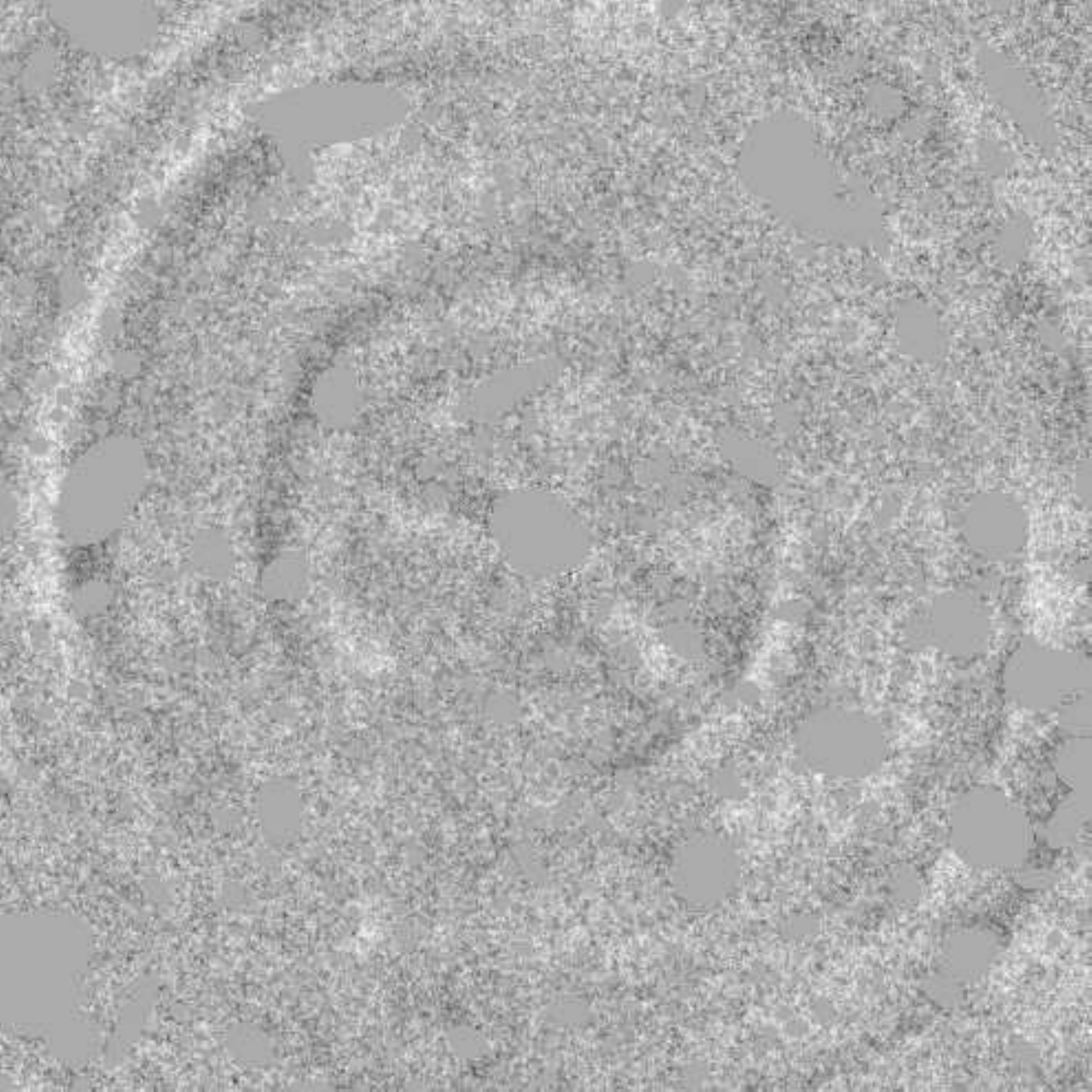} \\
\vspace{10pt}
\includegraphics[scale=0.4]{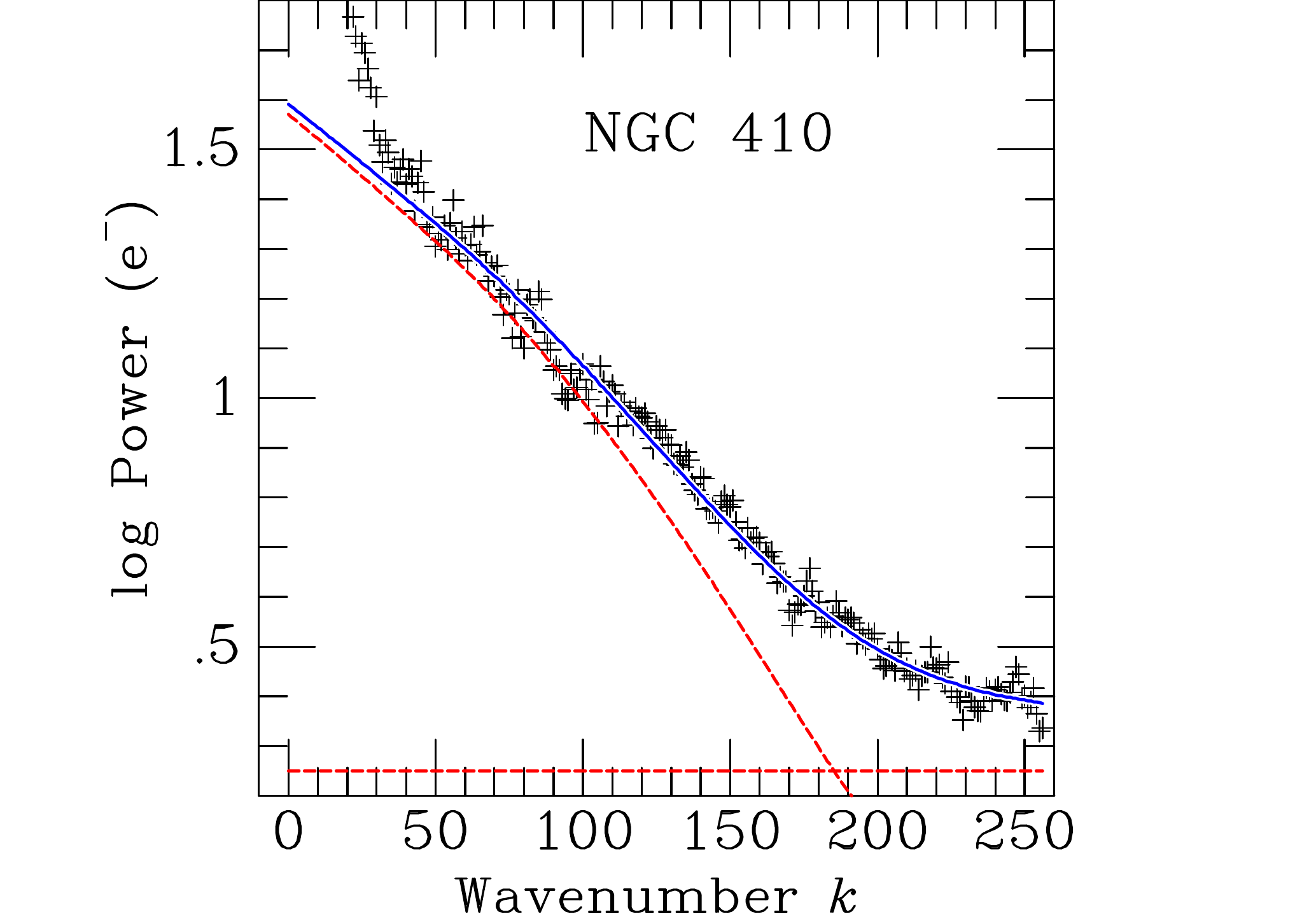}
\hspace{-25pt}
\includegraphics[scale=0.4]{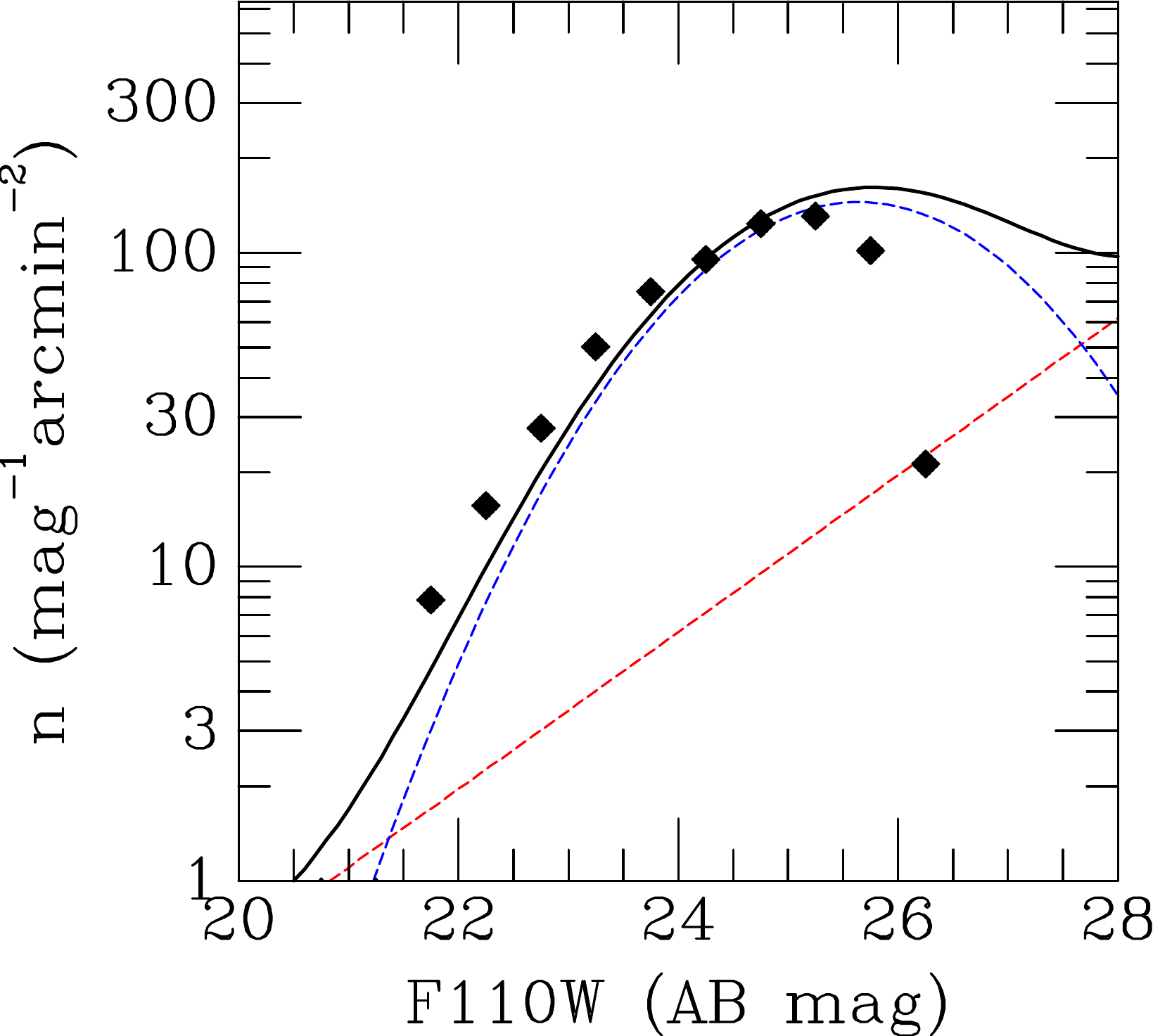}
\caption{Combined figure for NGC~410.}
\end{center}
\end{figure*}
\clearpage

\begin{figure*}
\begin{center}
\includegraphics[scale=0.2]{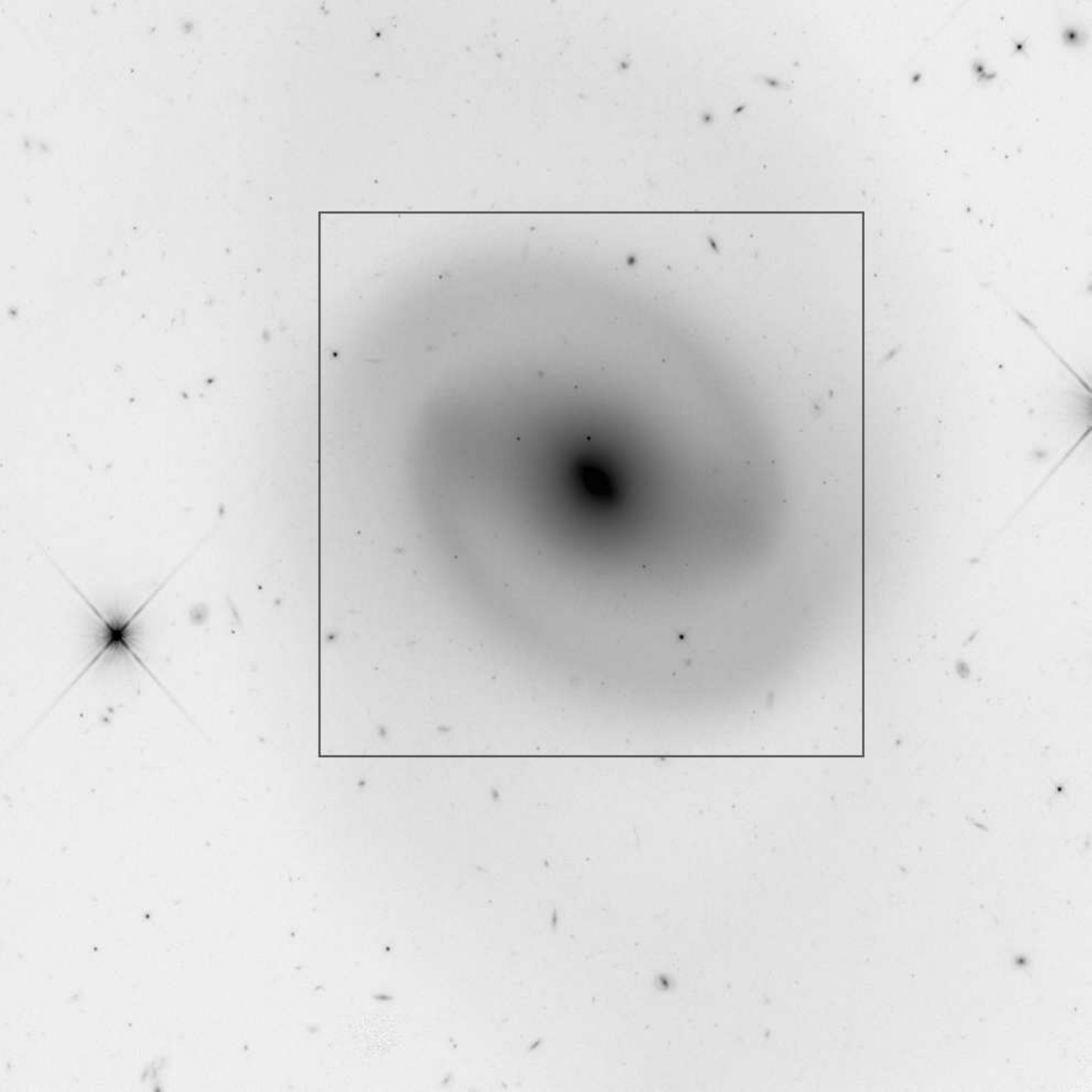}
\includegraphics[scale=0.4]{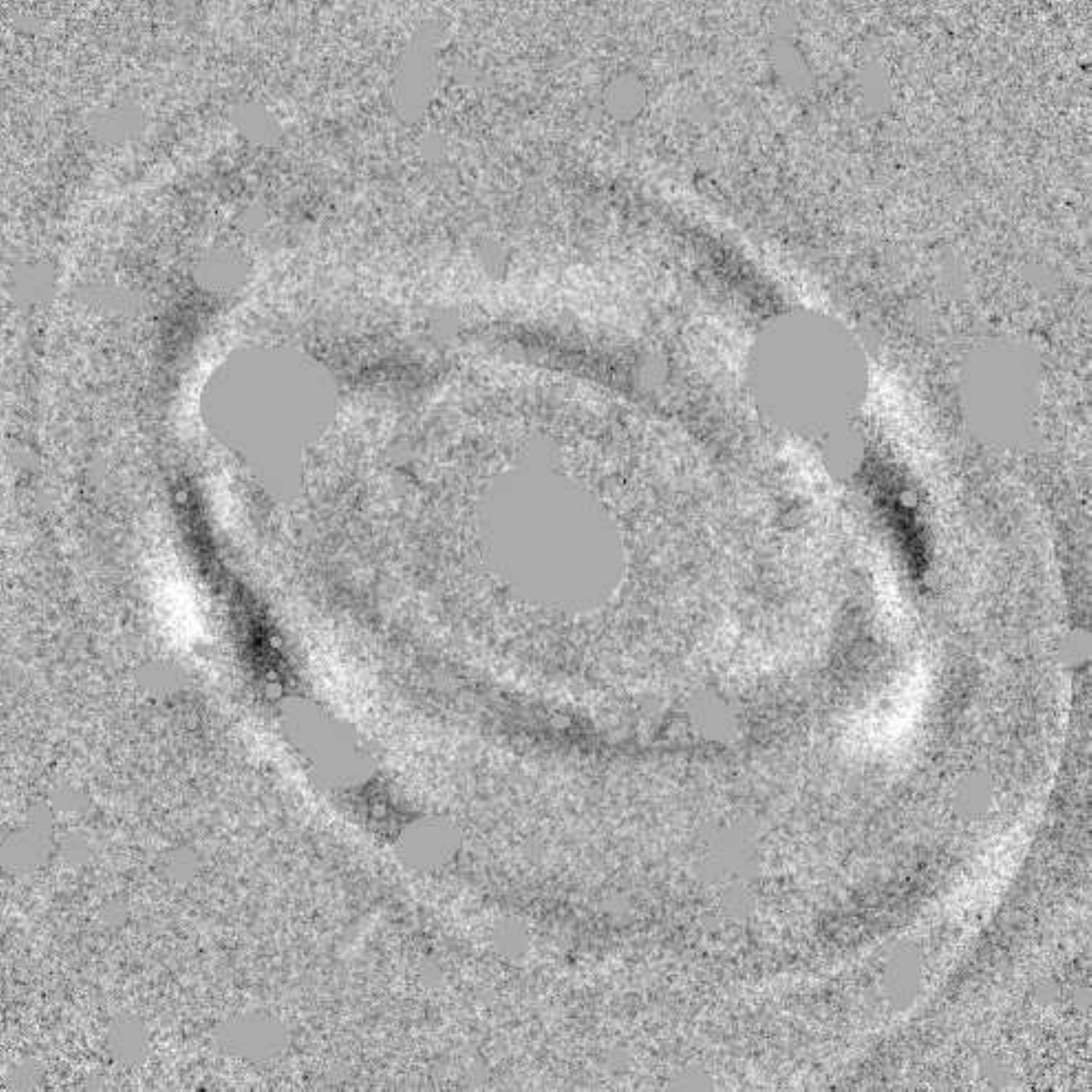} \\
\vspace{10pt}
\includegraphics[scale=0.4]{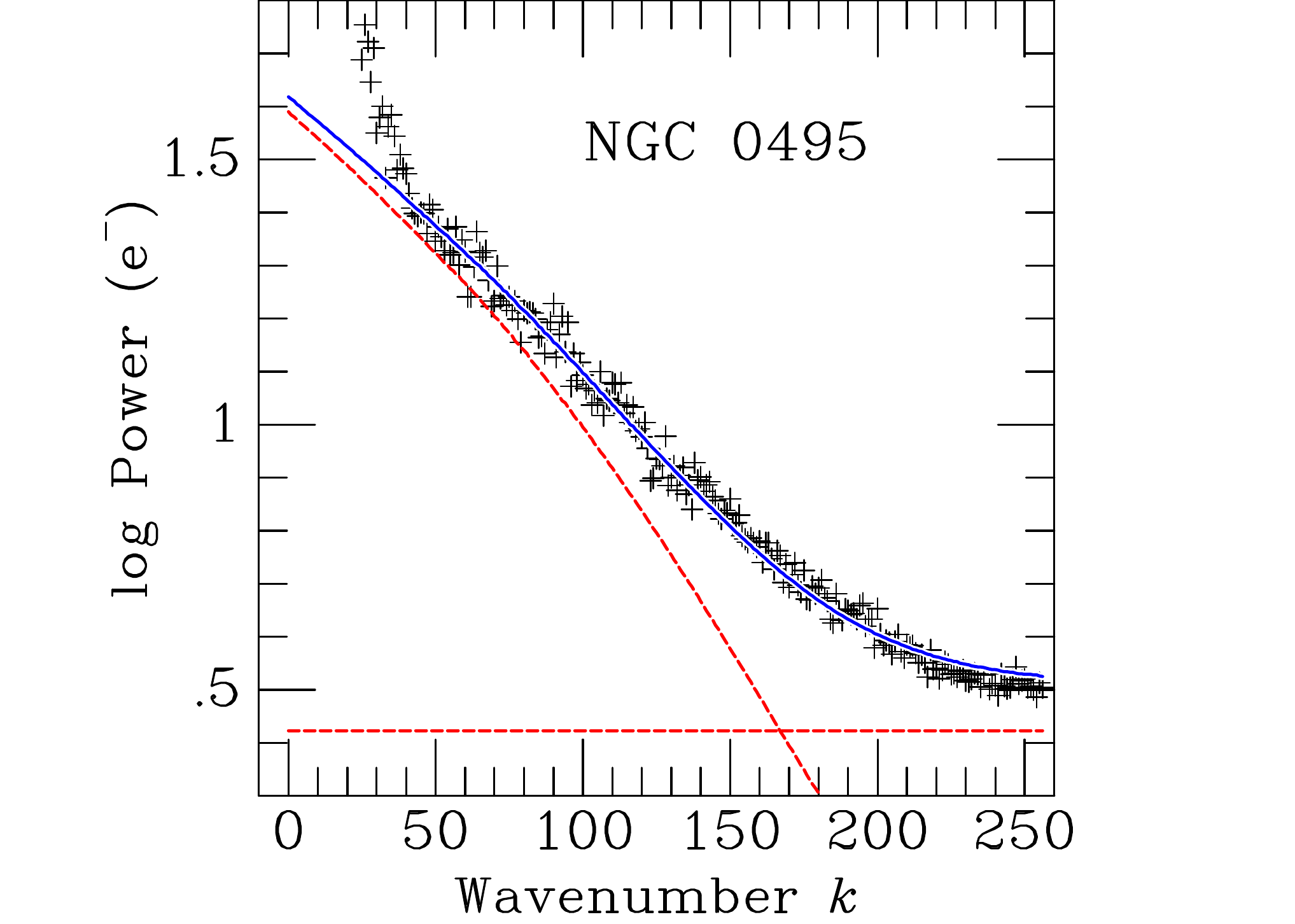}
\hspace{-25pt}
\includegraphics[scale=0.4]{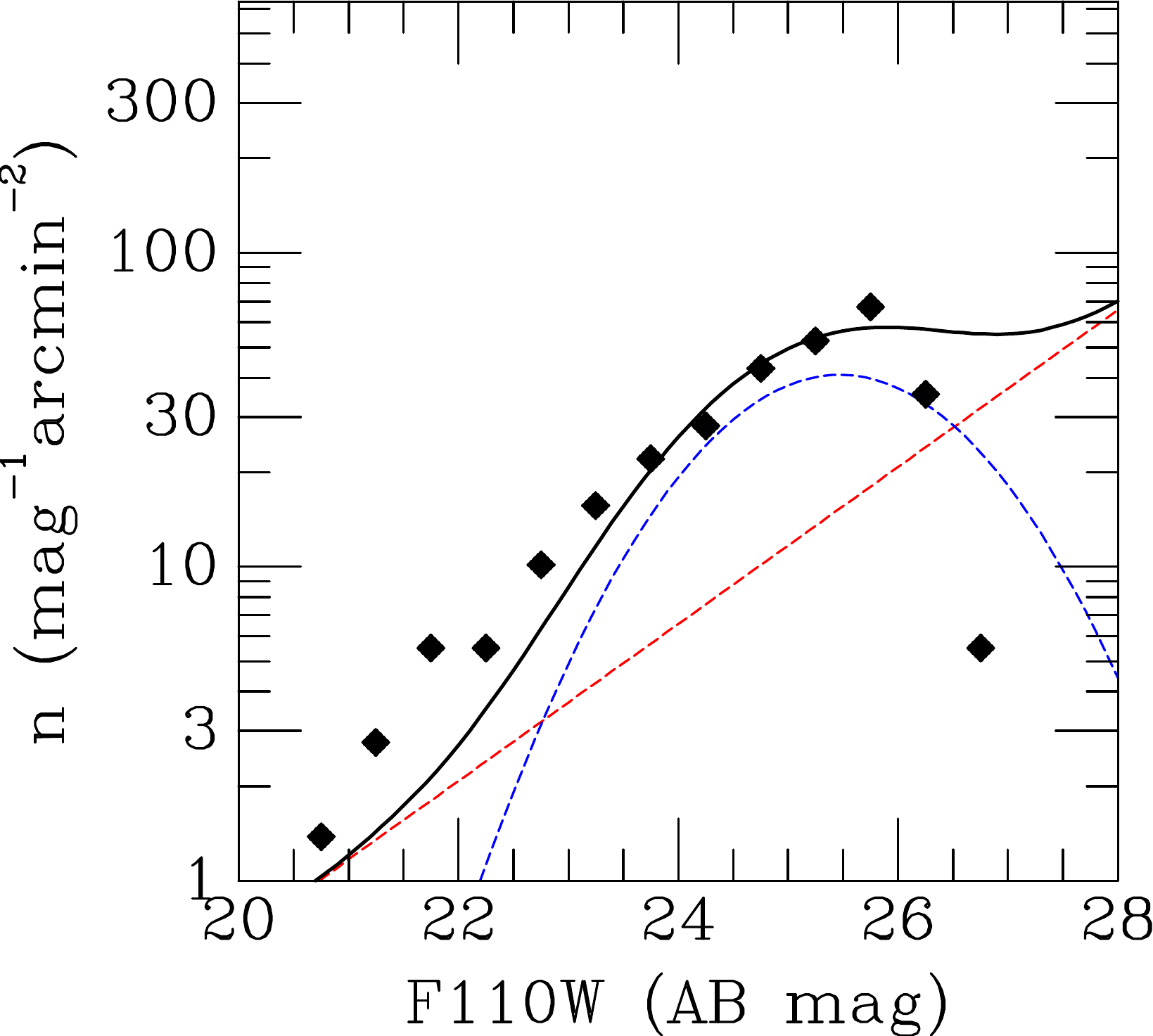}
\caption{Combined figure for NGC~495.}
\end{center}
\end{figure*}
\clearpage

\begin{figure*}
\begin{center}
\includegraphics[scale=0.2]{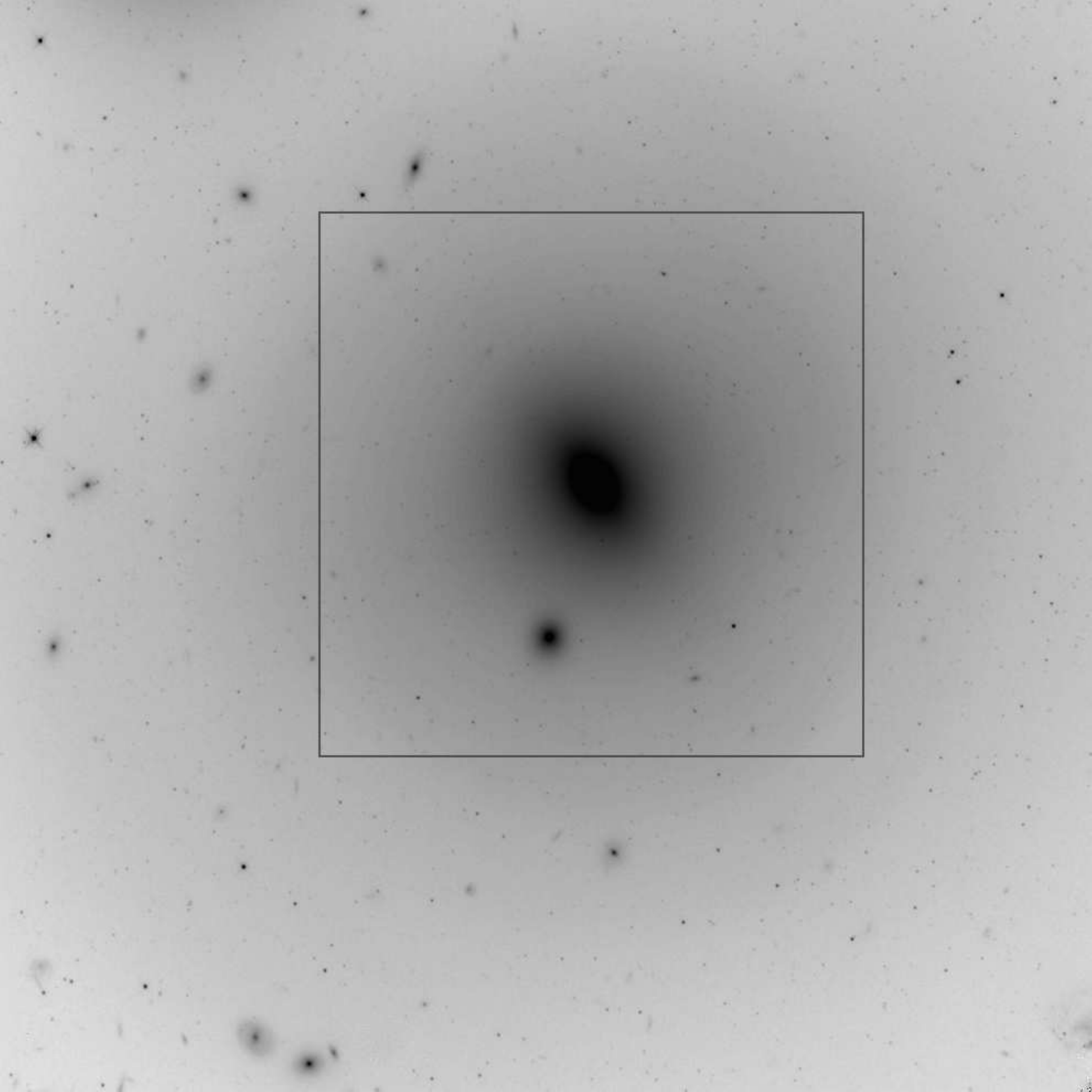}
\includegraphics[scale=0.4]{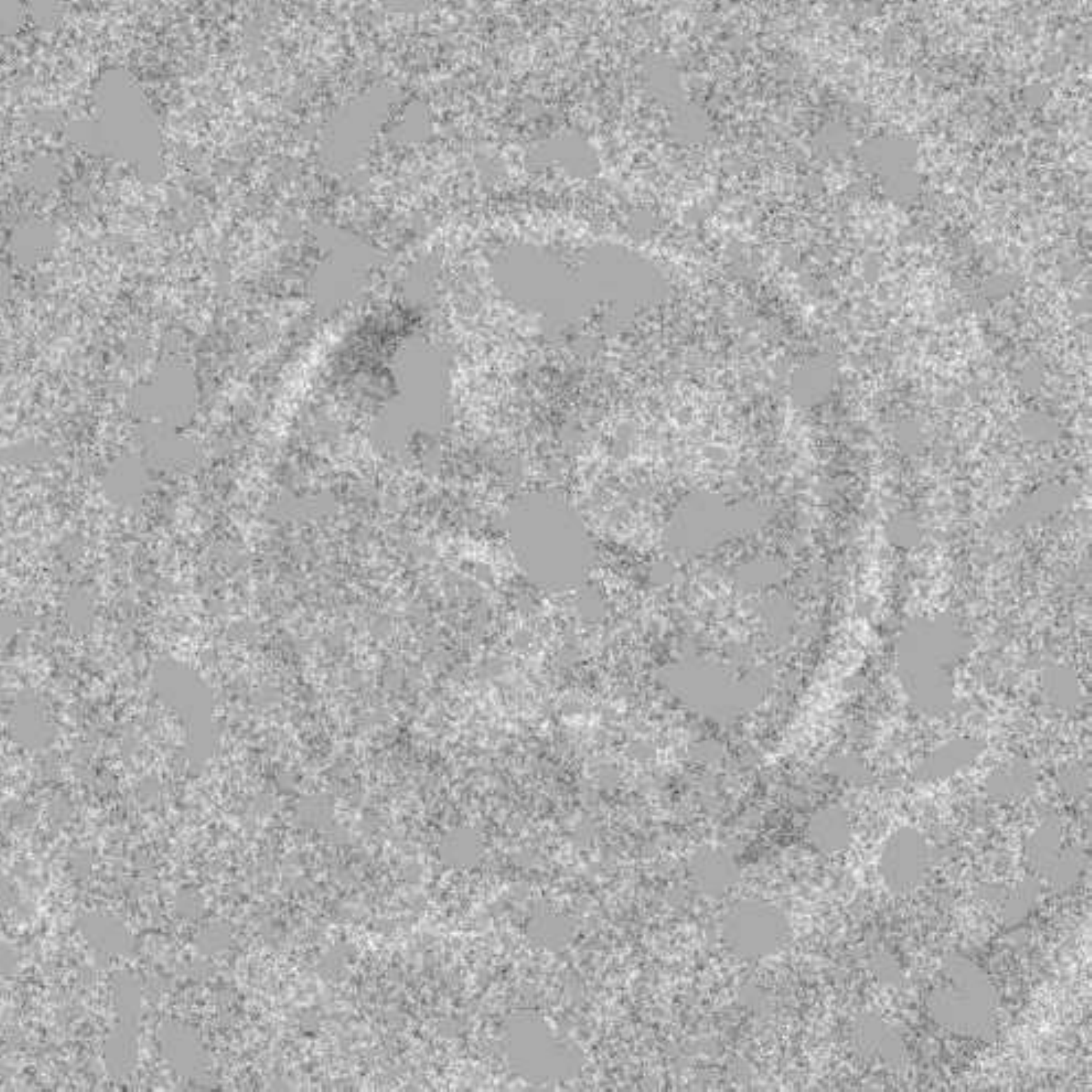} \\
\vspace{10pt}
\includegraphics[scale=0.4]{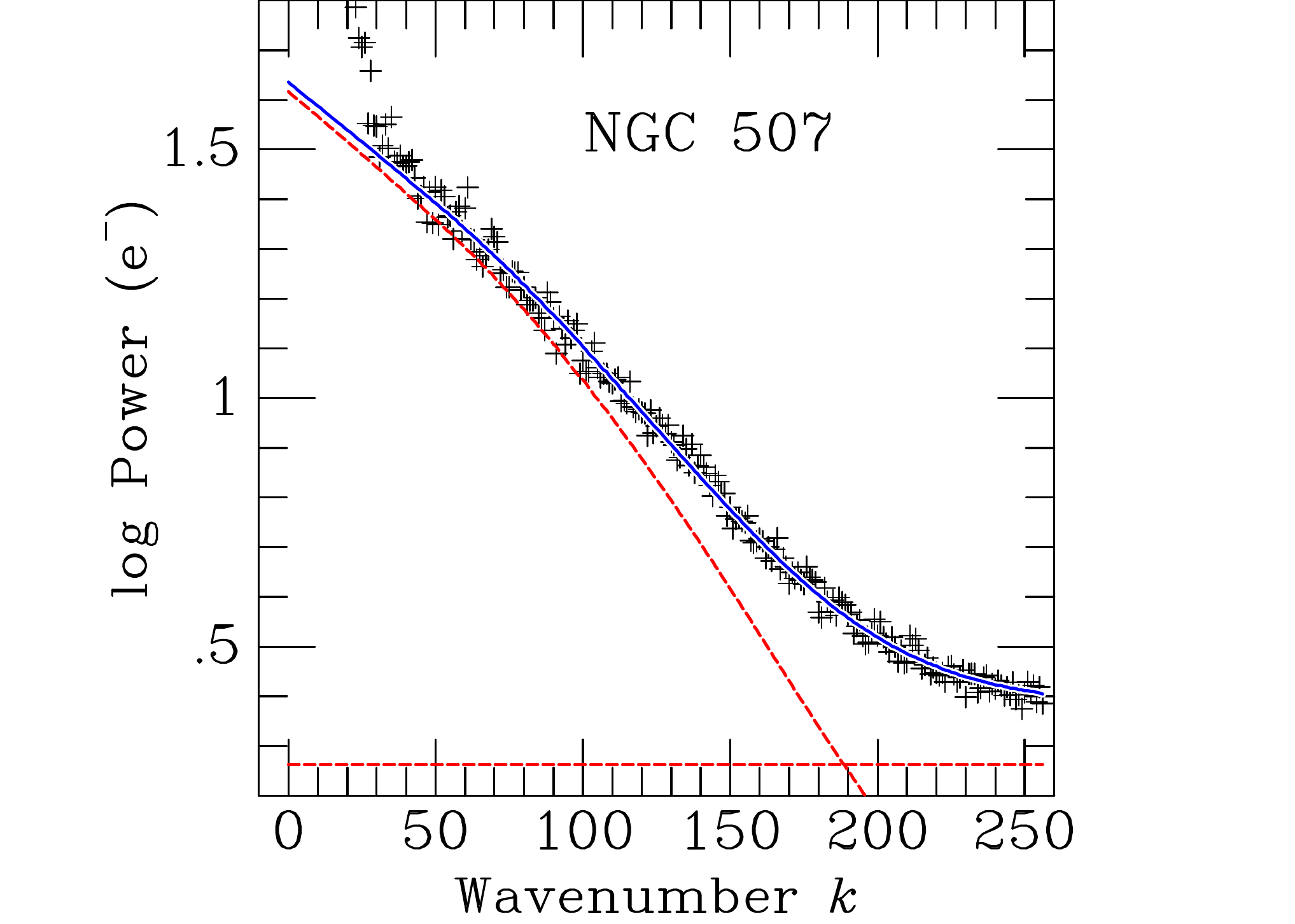}
\hspace{-25pt}
\includegraphics[scale=0.4]{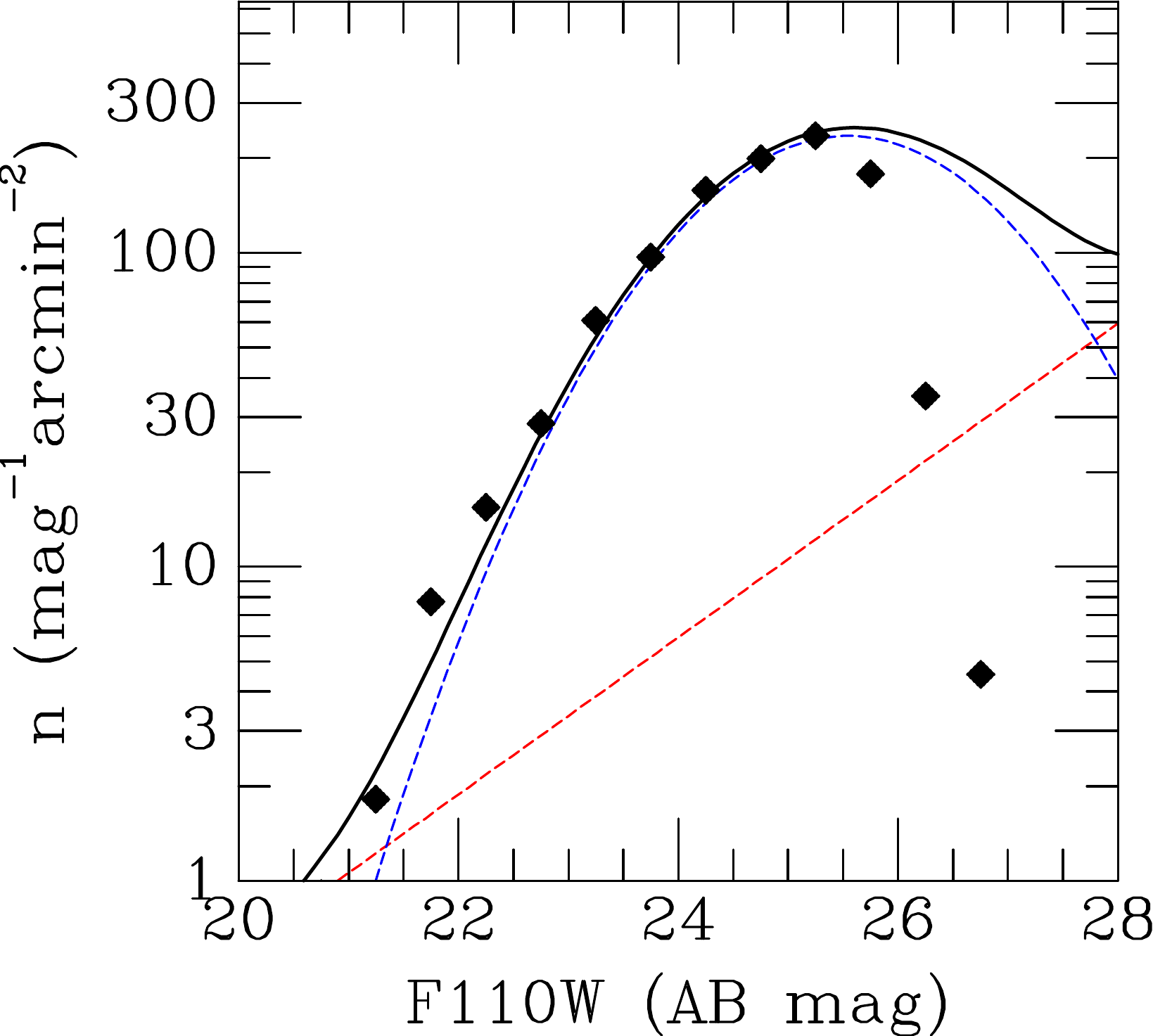}
\caption{Combined figure for NGC~507.}
\end{center}
\end{figure*}
\clearpage

\begin{figure*}
\begin{center}
\includegraphics[scale=0.2]{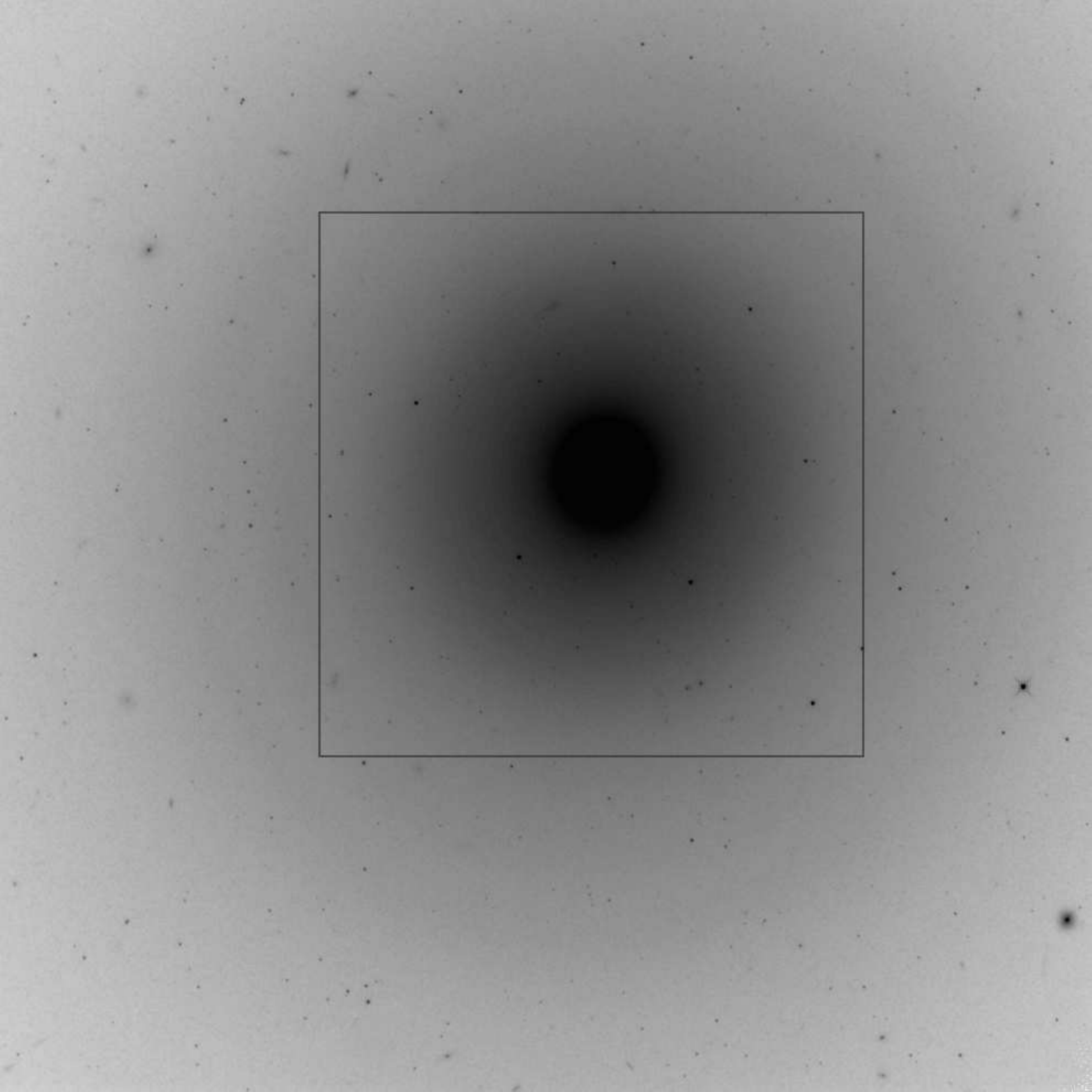}
\includegraphics[scale=0.4]{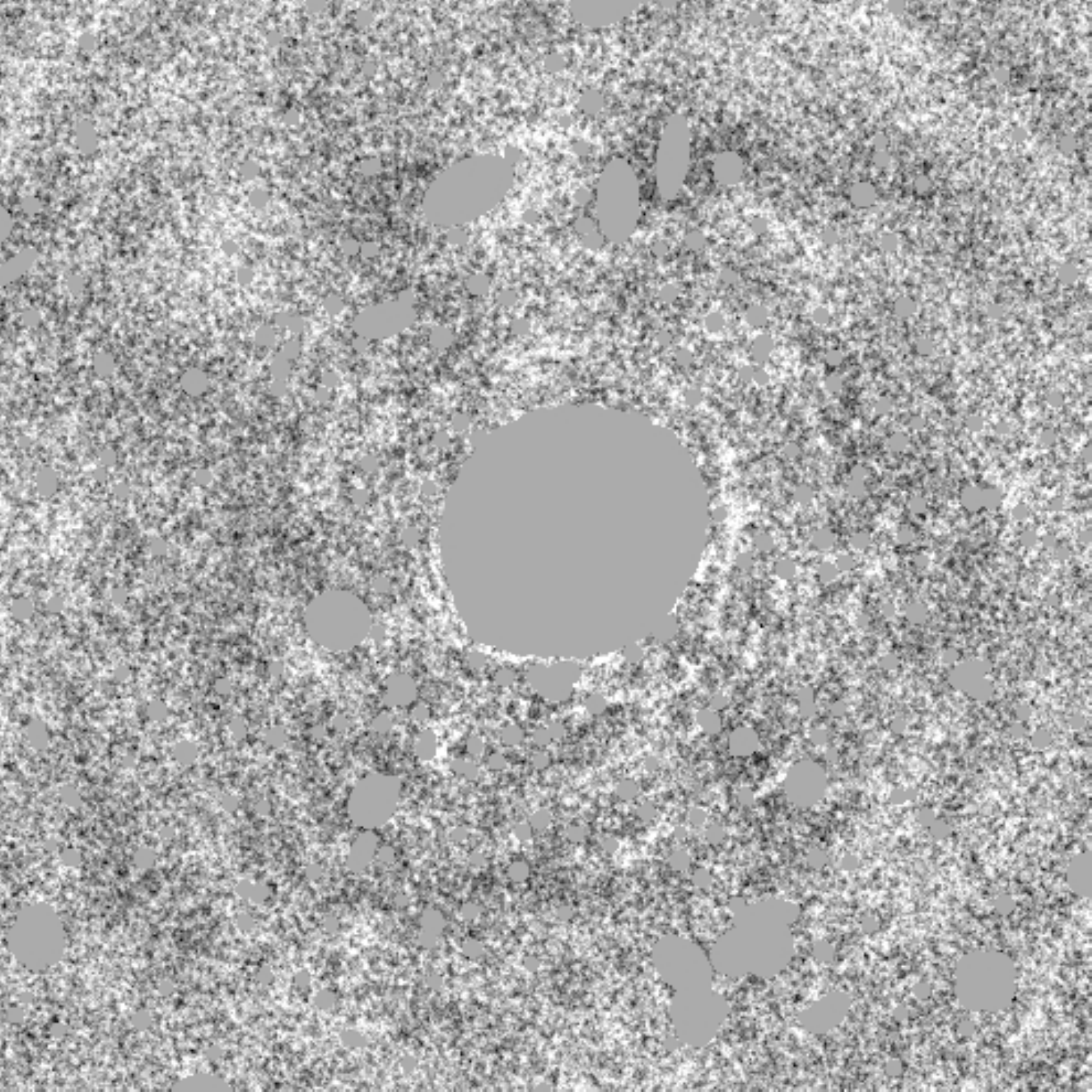} \\
\vspace{10pt}
\includegraphics[scale=0.4]{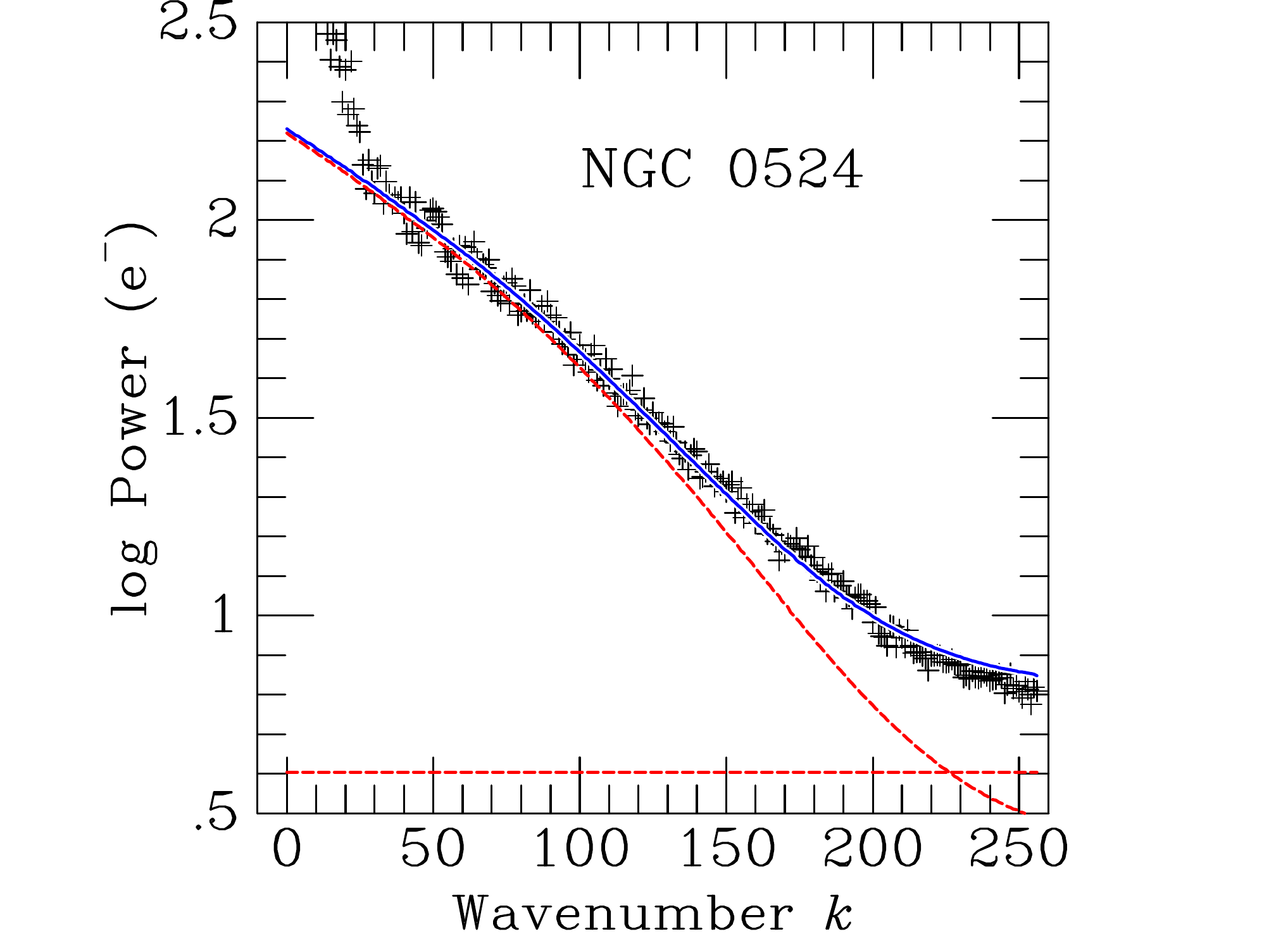}
\hspace{-25pt}
\includegraphics[scale=0.4]{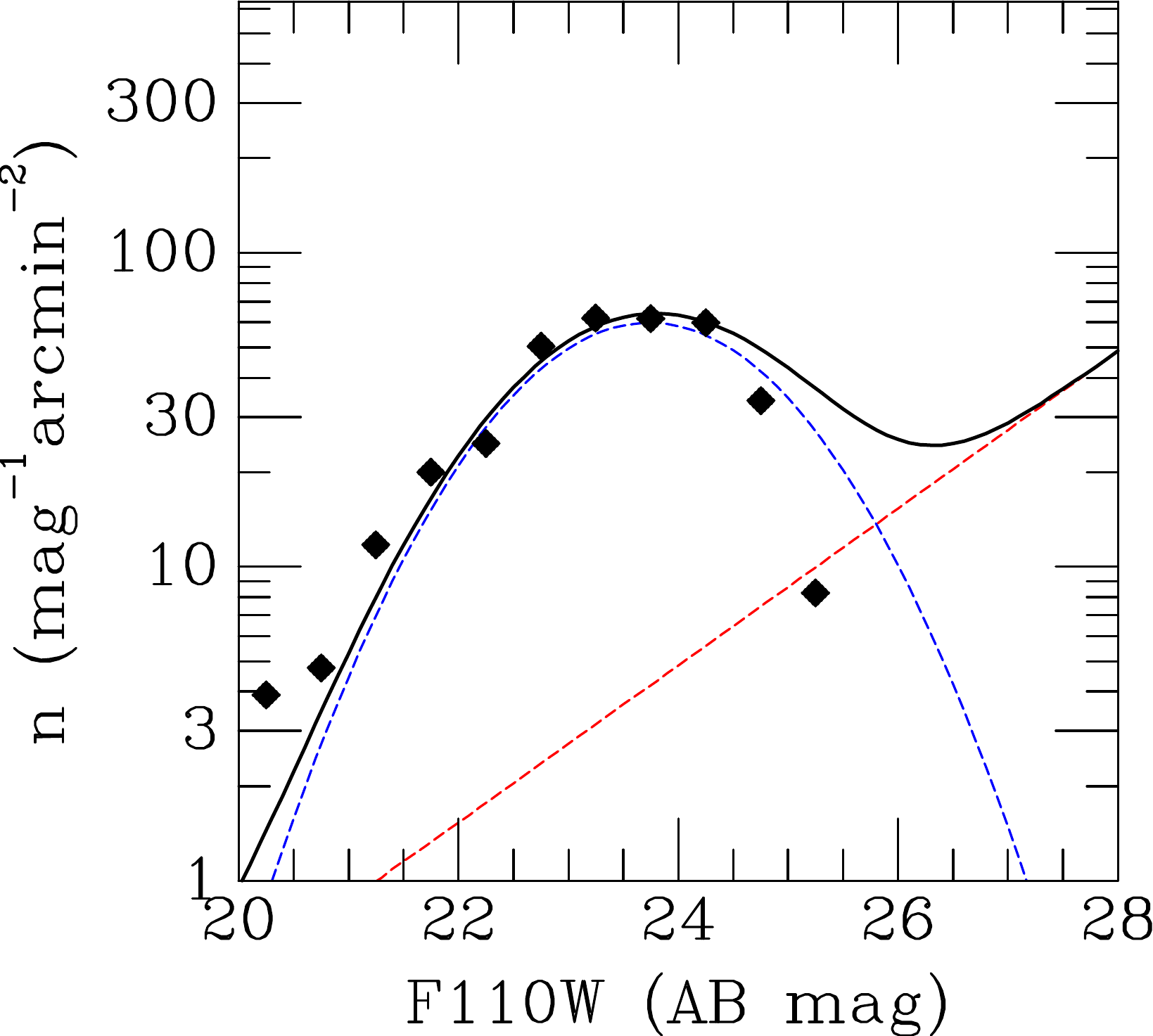}
\caption{Combined figure for NGC~524.}
\end{center}
\end{figure*}
\clearpage

\begin{figure*}
\begin{center}
\includegraphics[scale=0.2]{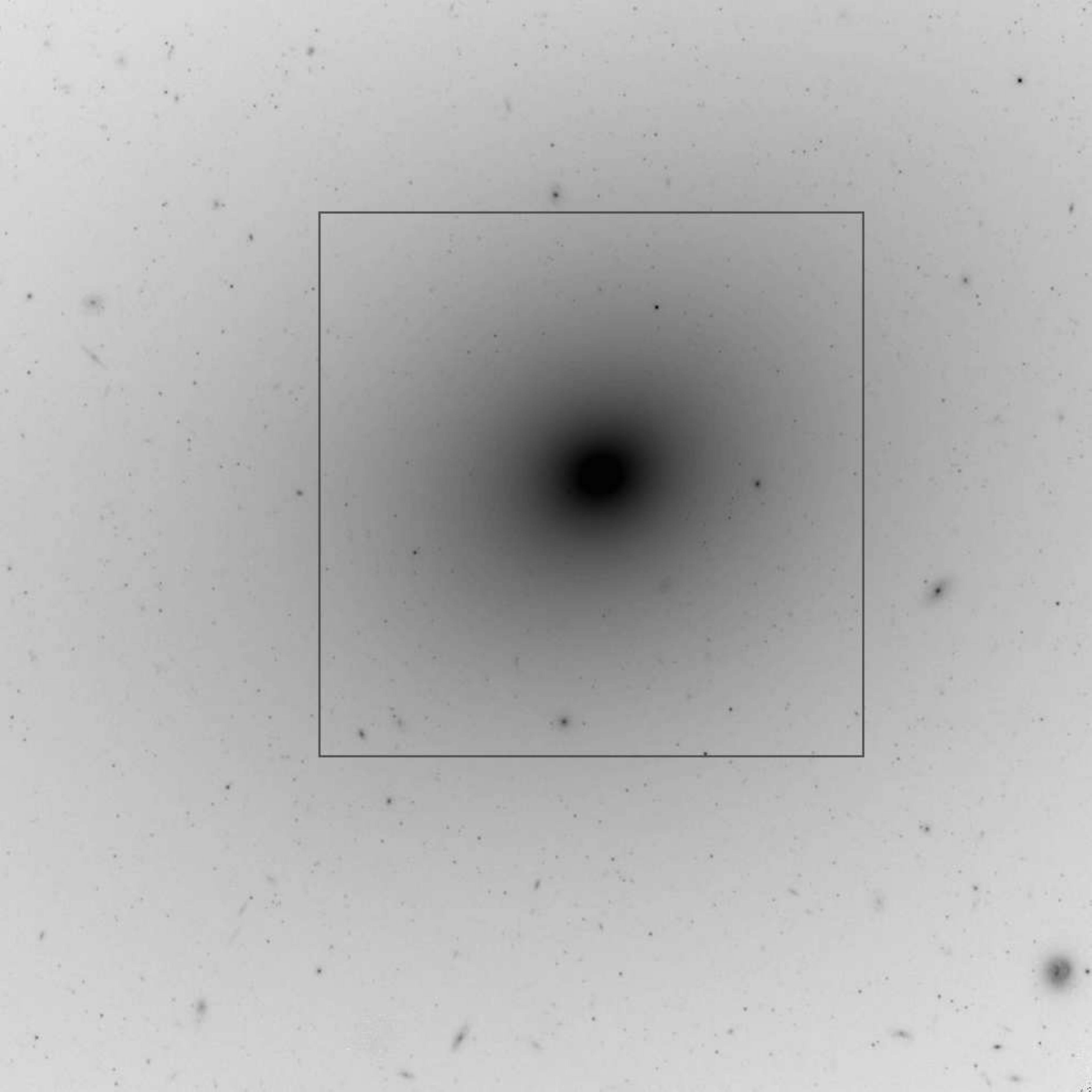}
\includegraphics[scale=0.4]{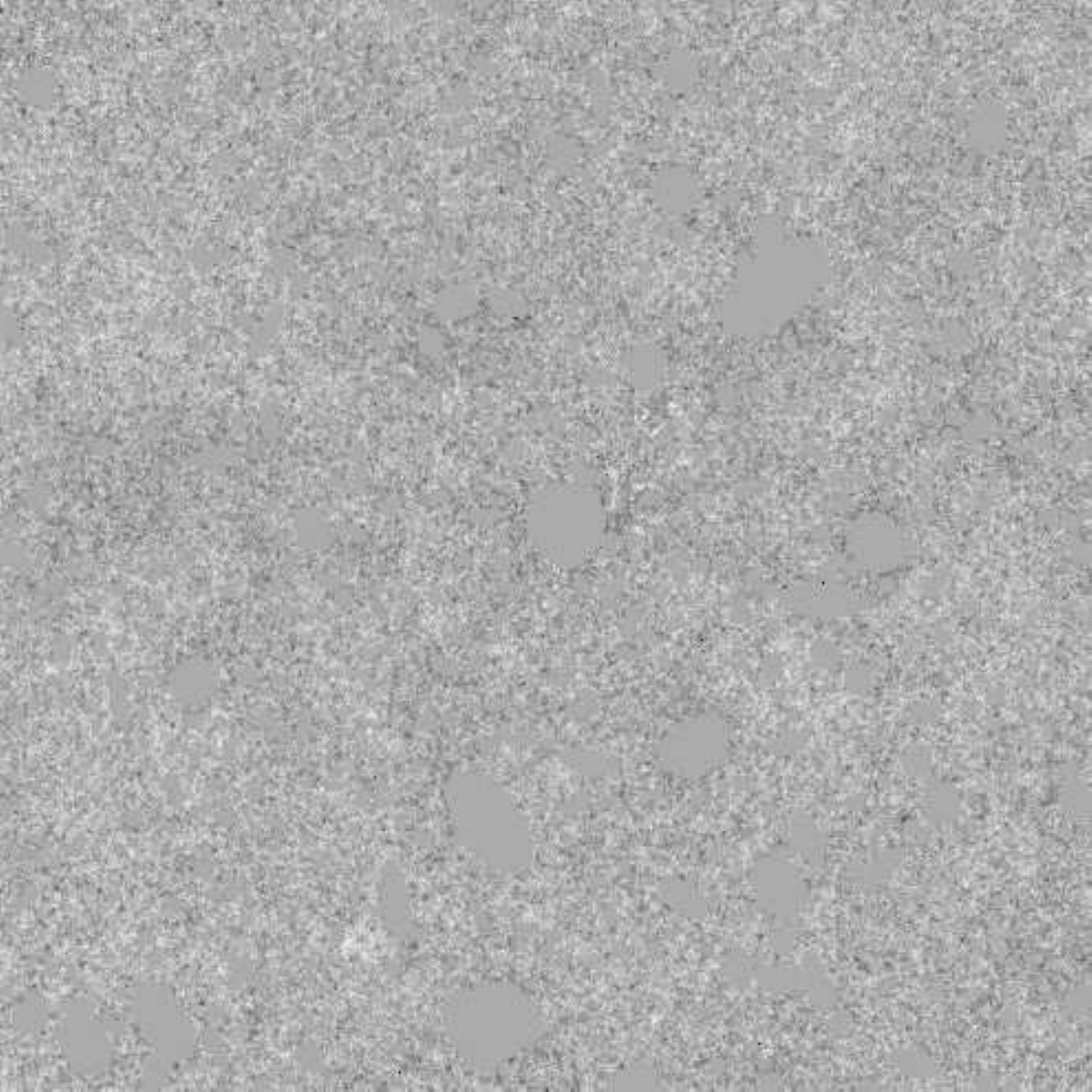} \\
\vspace{10pt}
\includegraphics[scale=0.4]{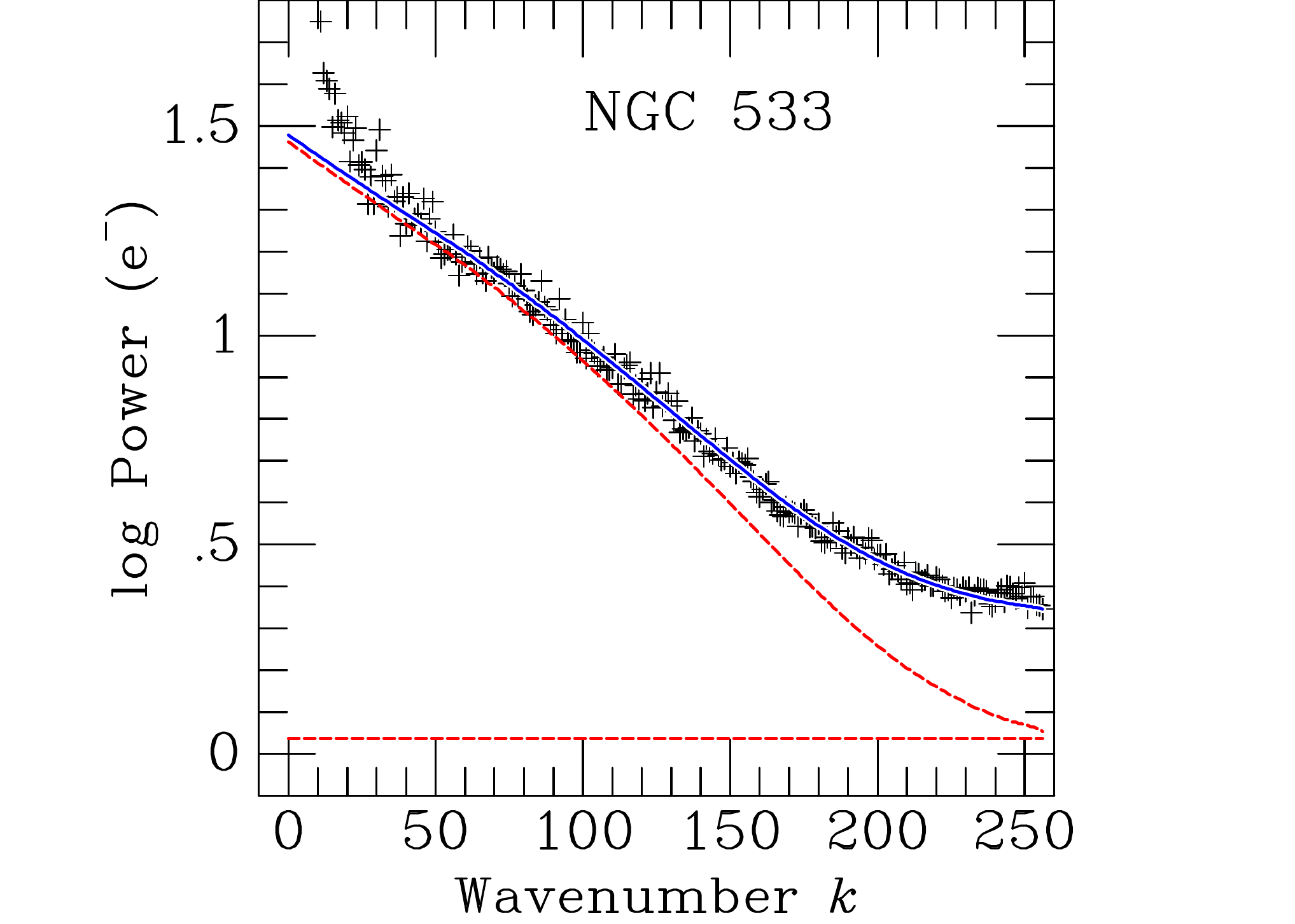}
\hspace{-25pt}
\includegraphics[scale=0.4]{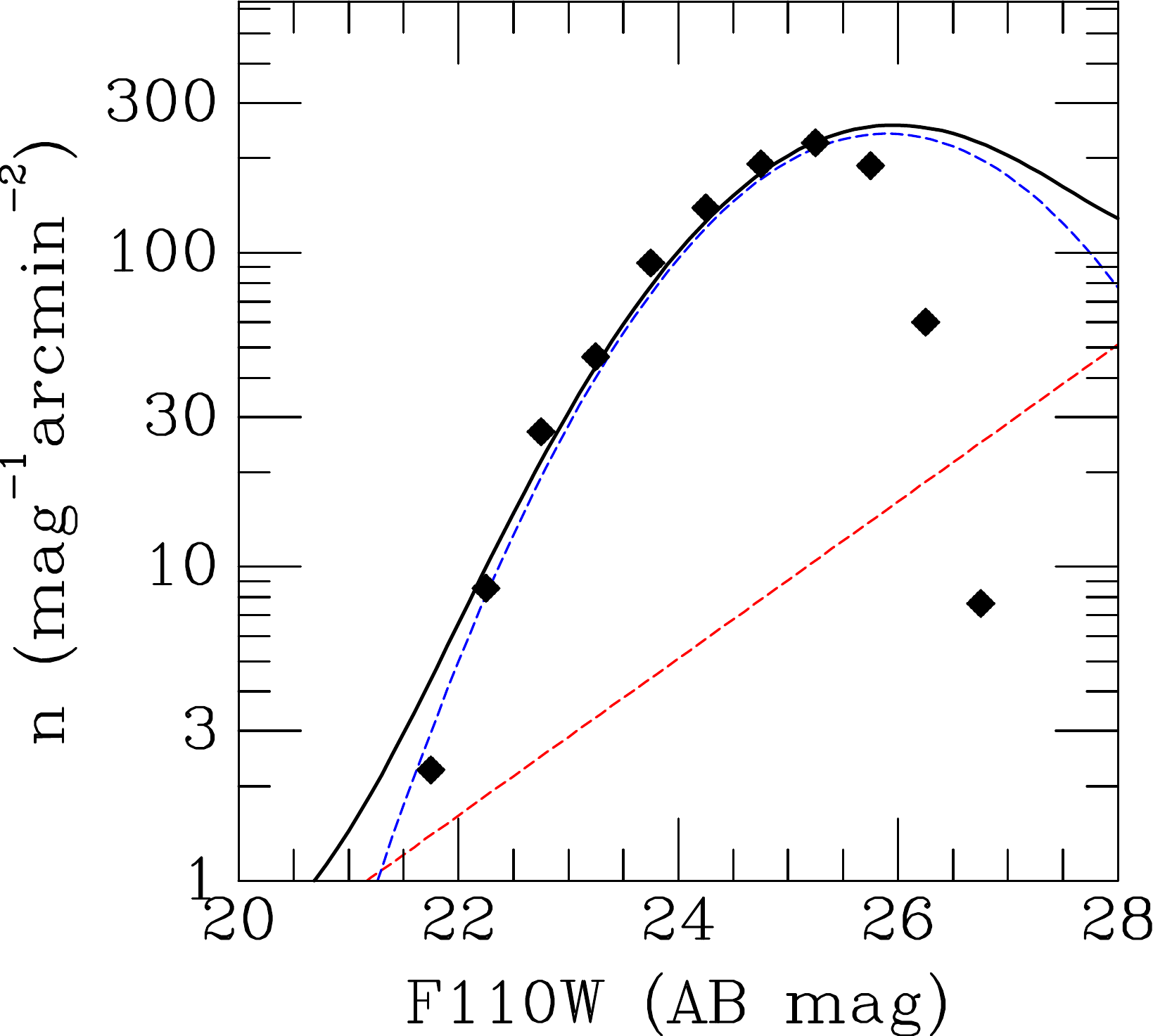}
\caption{Combined figure for NGC~533.}
\end{center}
\end{figure*}
\clearpage

\begin{figure*}
\begin{center}
\includegraphics[scale=0.2]{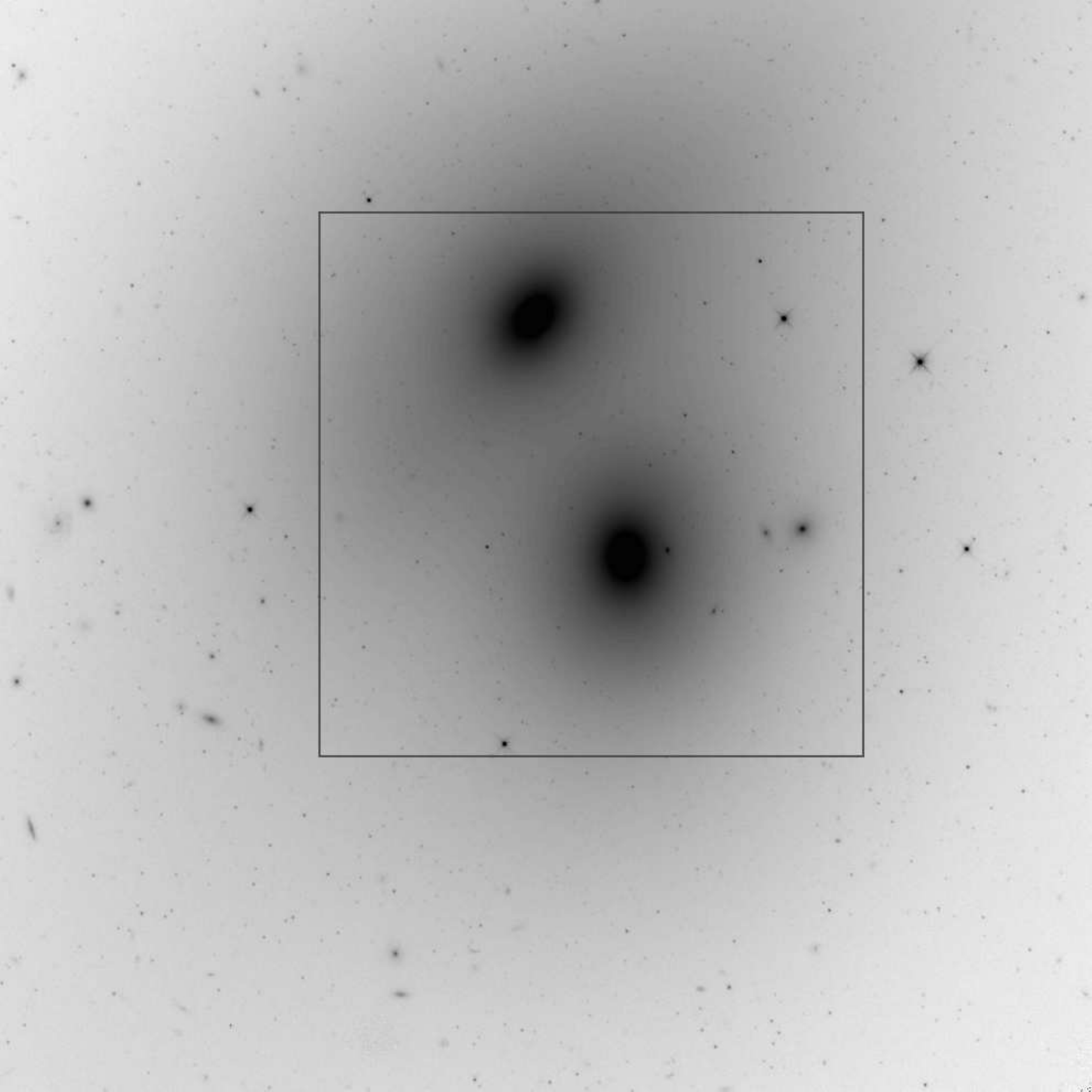}
\includegraphics[scale=0.4]{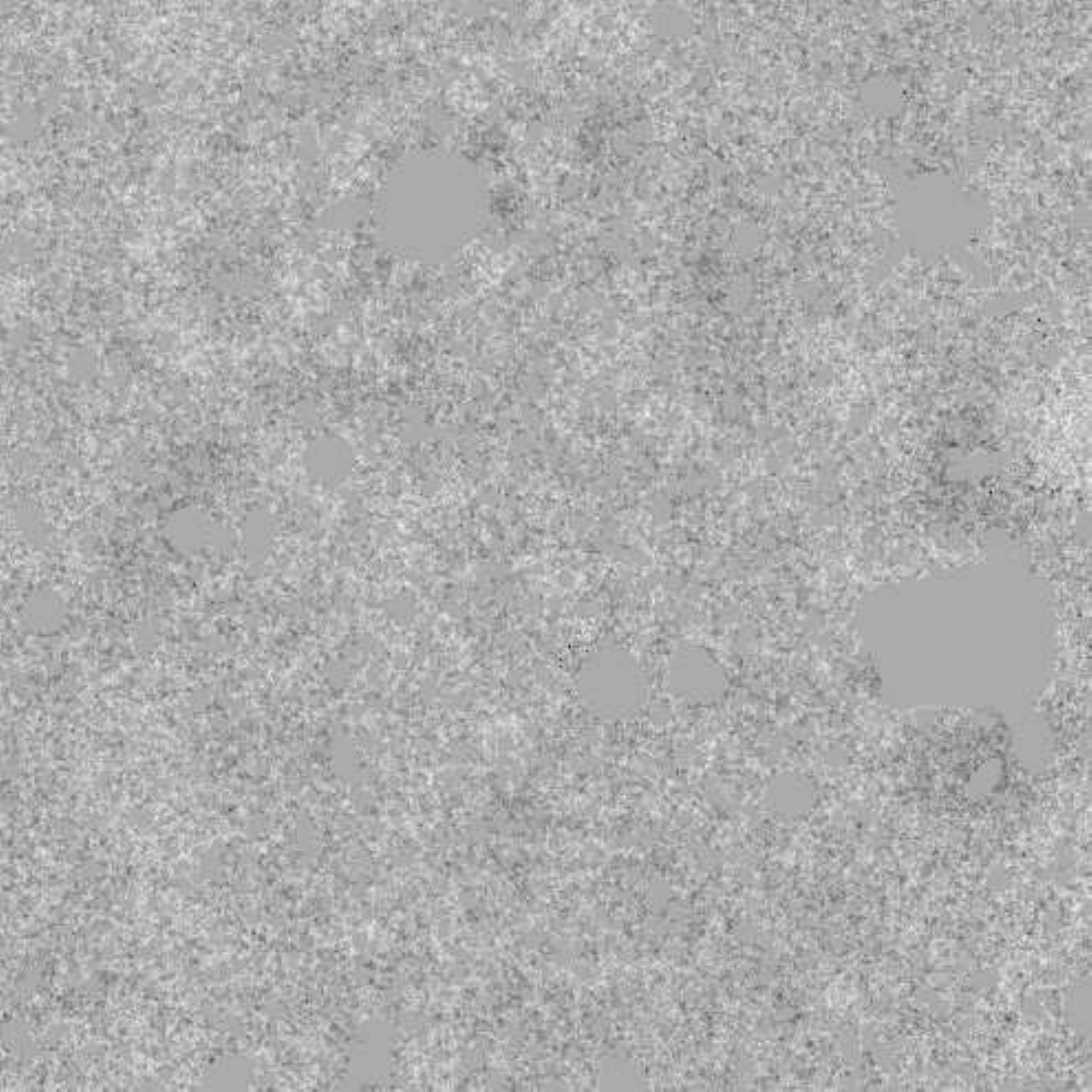} \\
\vspace{10pt}
\includegraphics[scale=0.4]{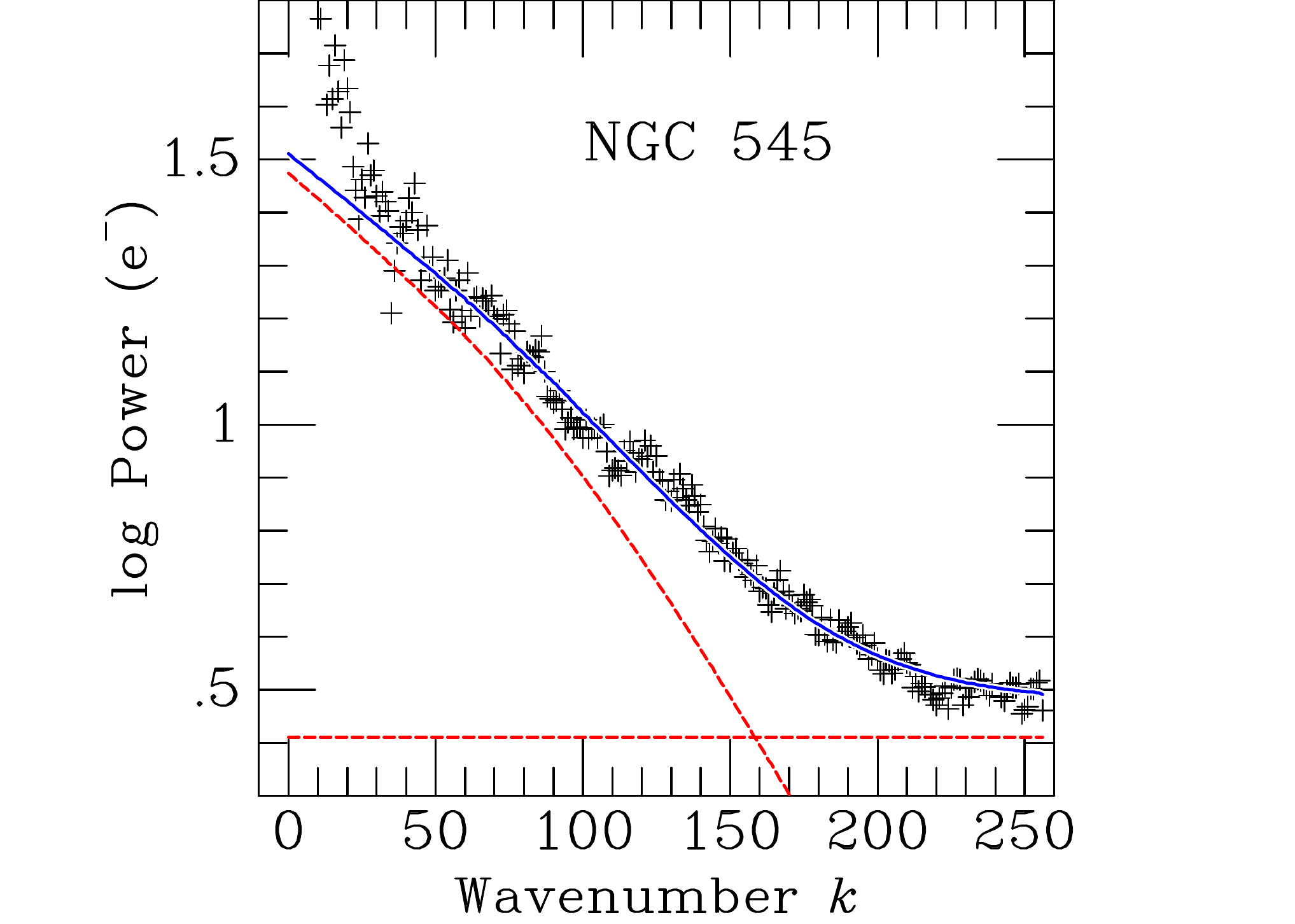}
\hspace{-25pt}
\includegraphics[scale=0.4]{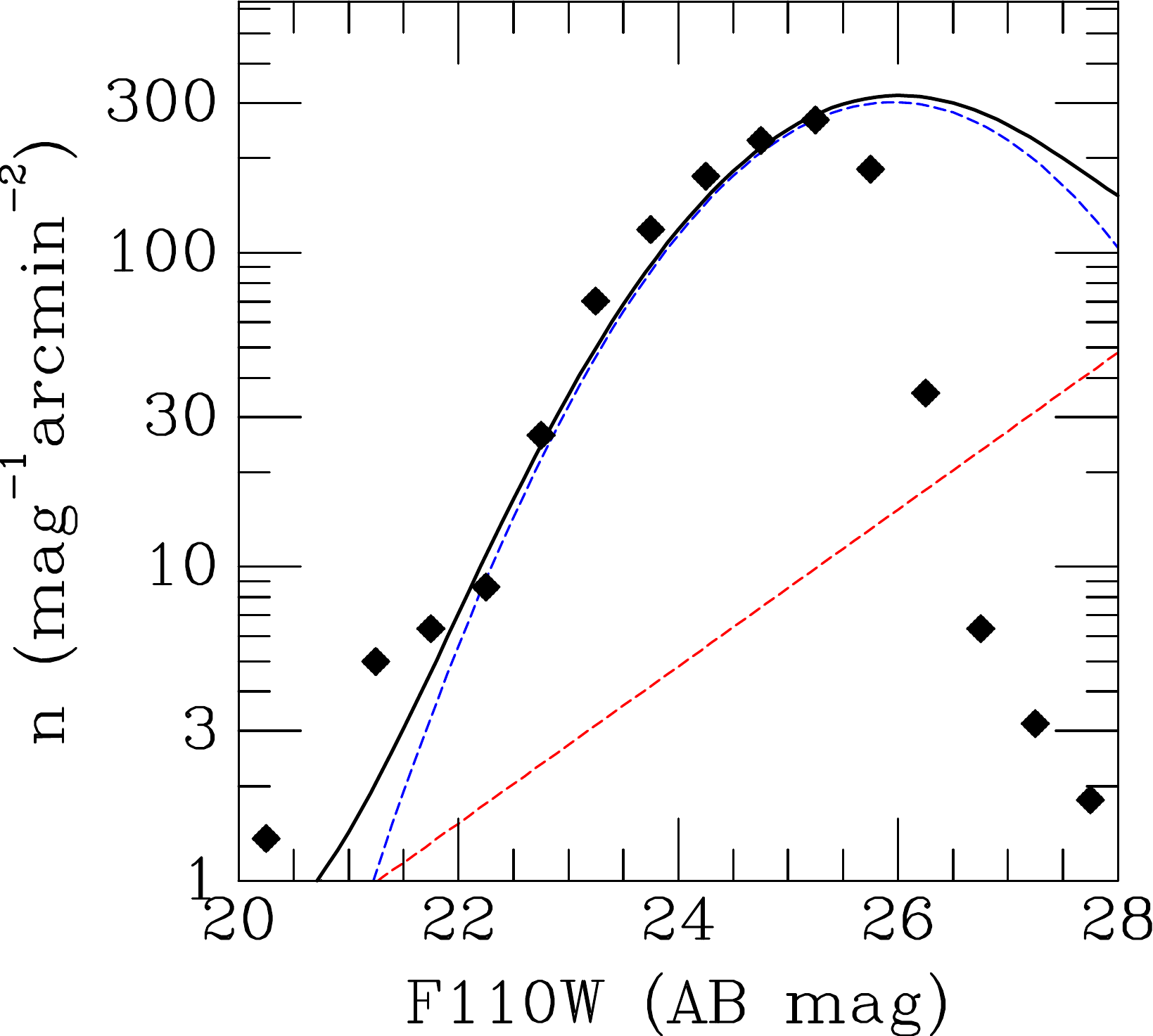}
\caption{Combined figure for NGC~545 (top; NGC~547 is also visible.)}
\end{center}
\end{figure*}
\clearpage

\begin{figure*}
\begin{center}
\includegraphics[scale=0.2]{n0547j}
\includegraphics[scale=0.4]{n0547jresid} \\
\vspace{10pt}
\includegraphics[scale=0.4]{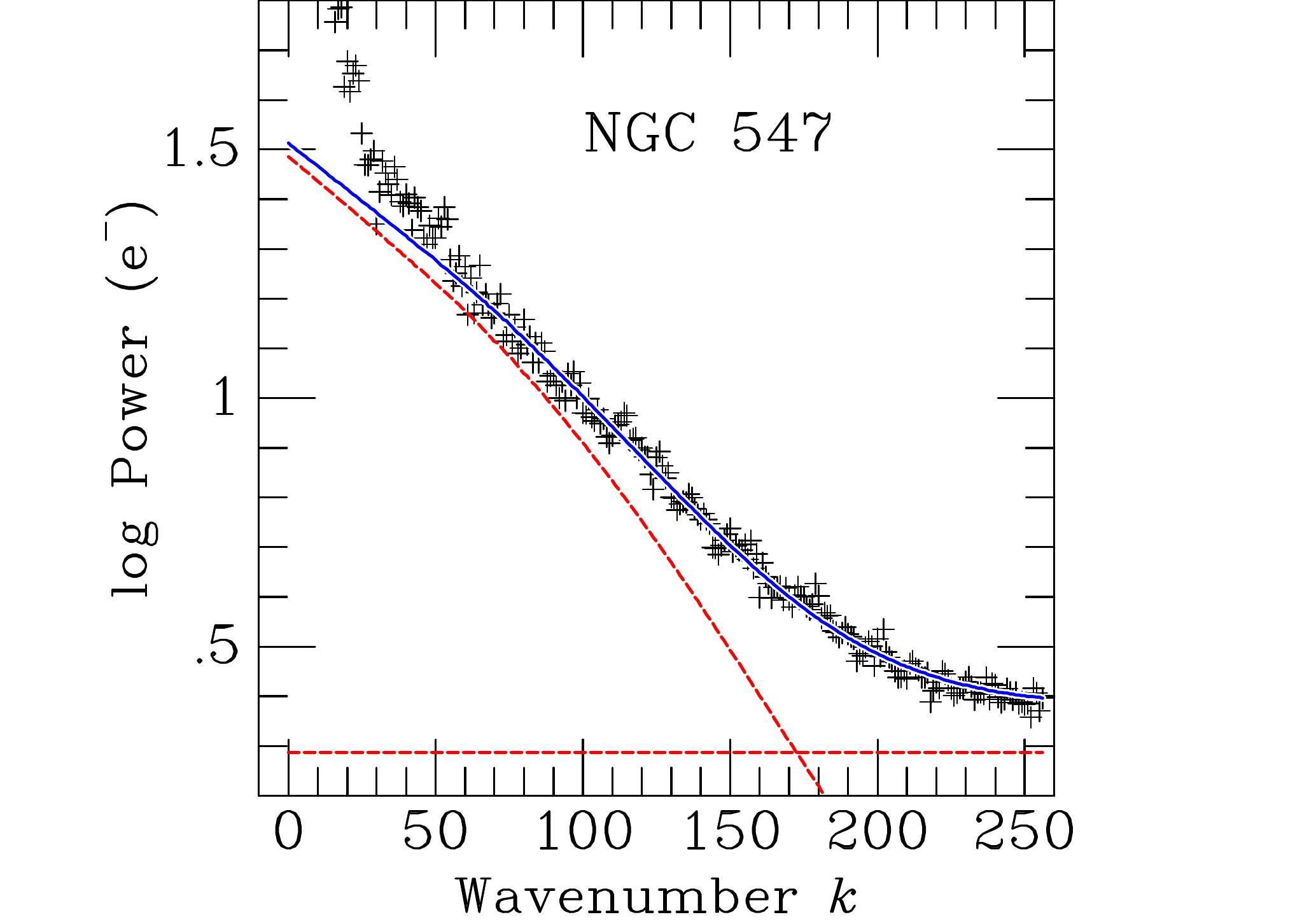}
\hspace{-25pt}
\includegraphics[scale=0.4]{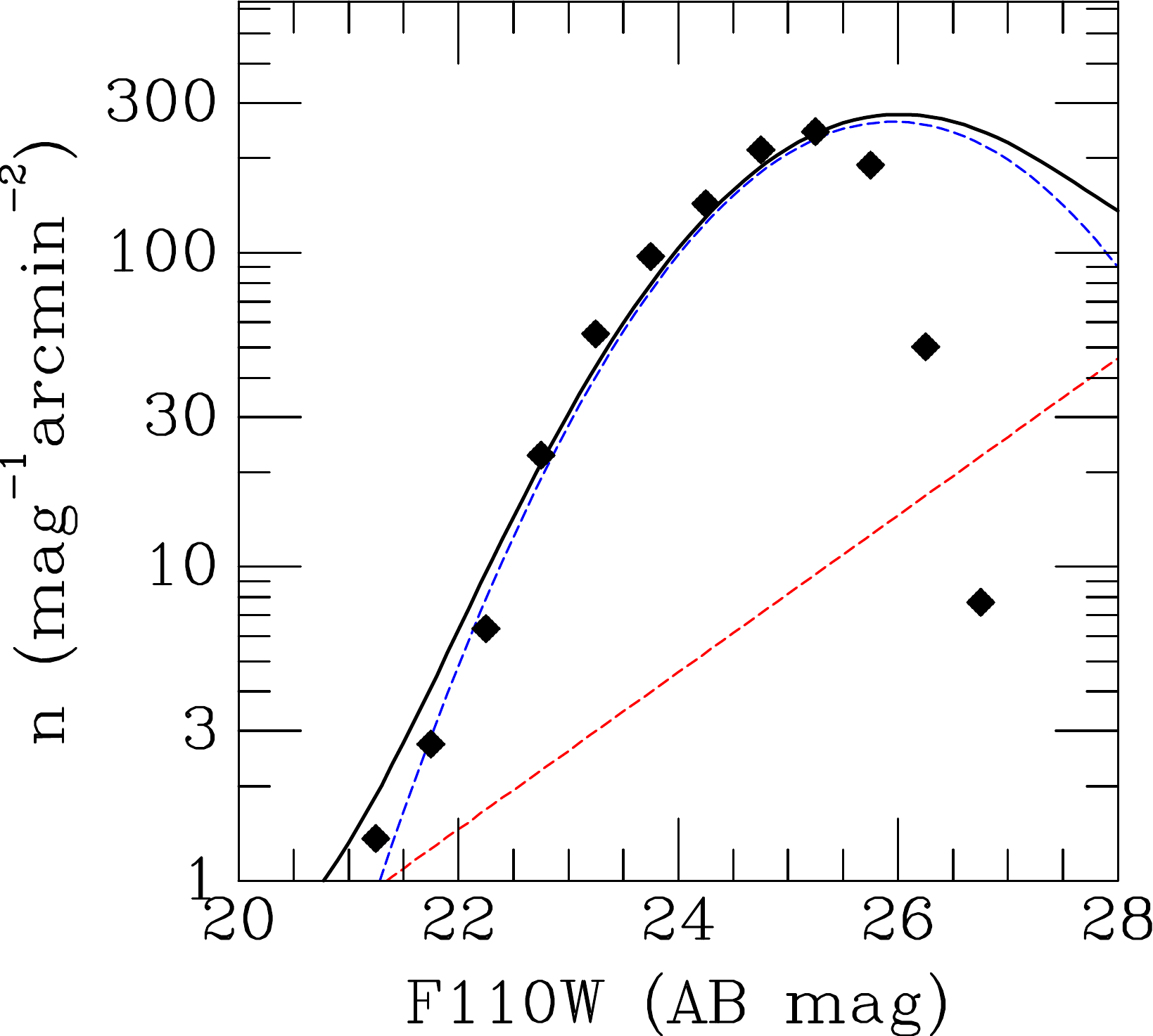}
\caption{Combined figure for NGC~547. (bottom; NGC~545 is also visible.)}
\end{center}
\end{figure*}
\clearpage

\begin{figure*}
\begin{center}
\includegraphics[scale=0.2]{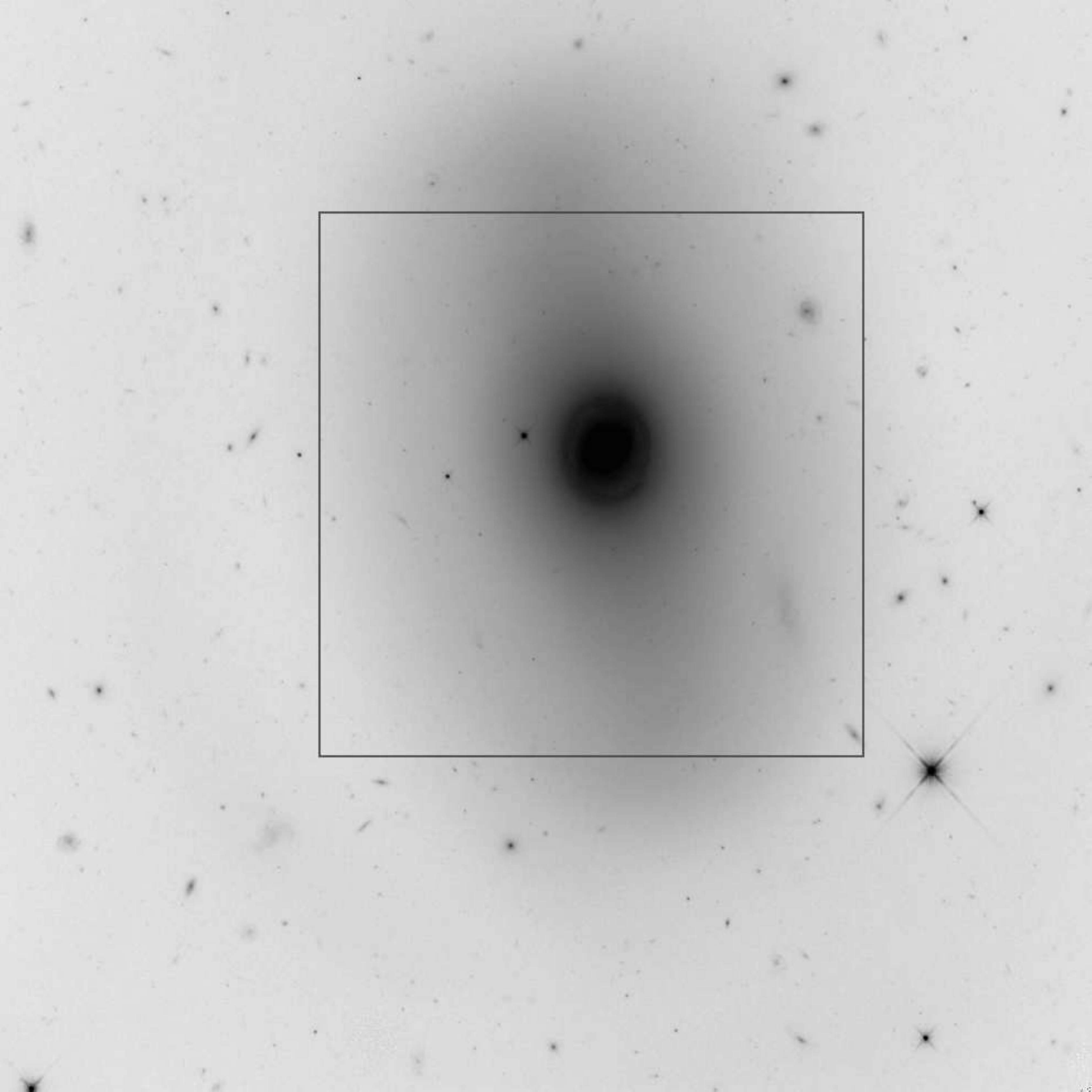}
\includegraphics[scale=0.4]{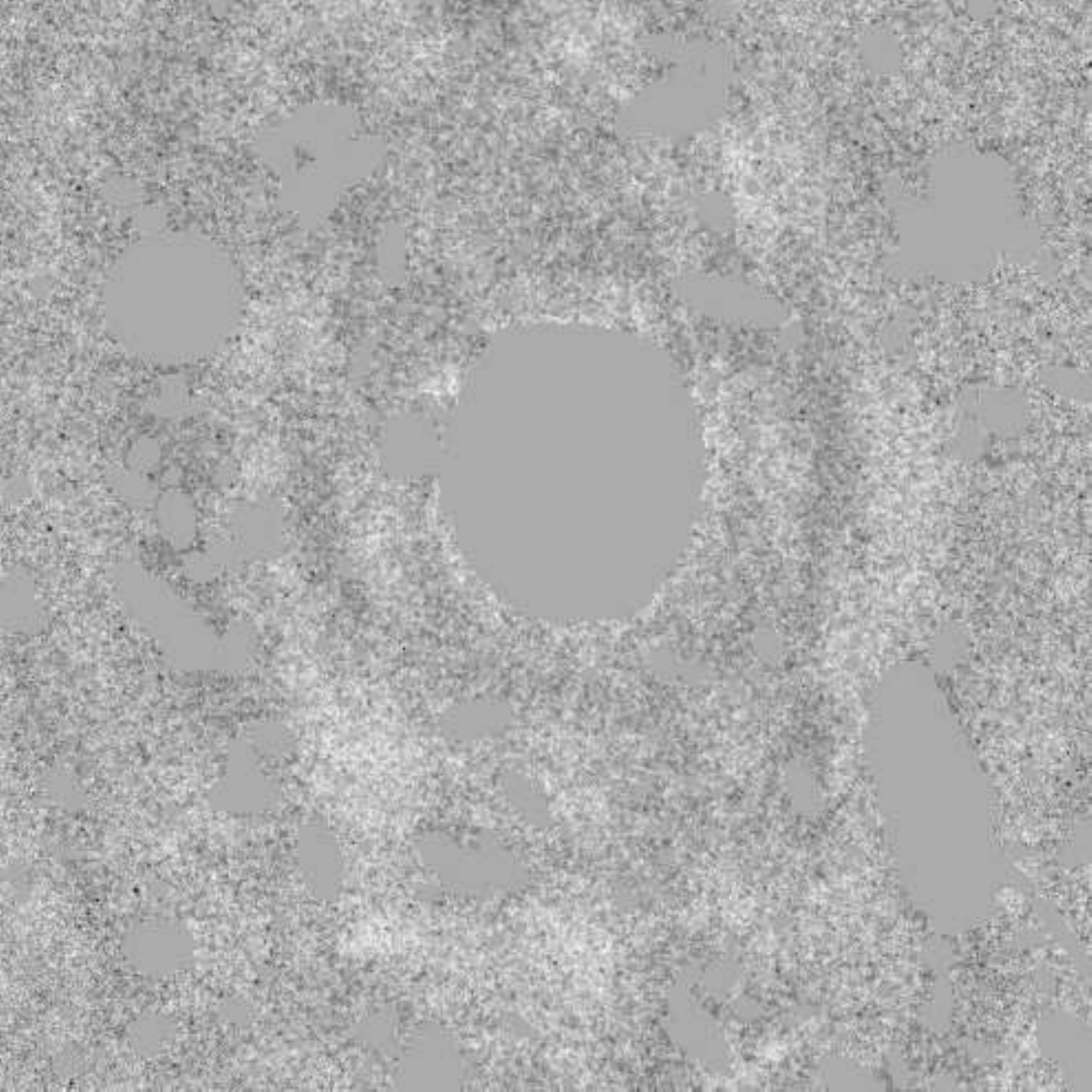} \\
\vspace{10pt}
\includegraphics[scale=0.4]{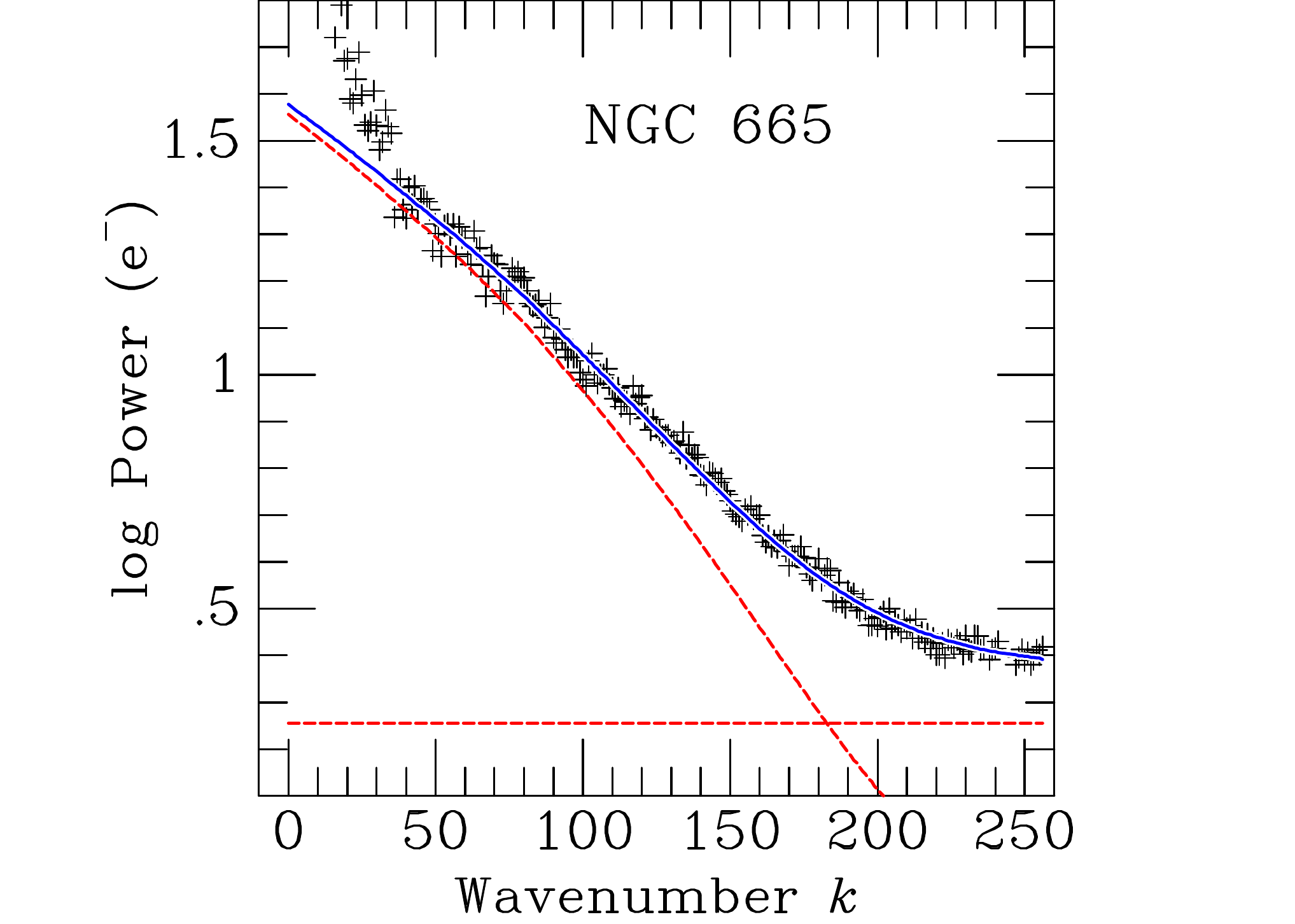}
\hspace{-25pt}
\includegraphics[scale=0.4]{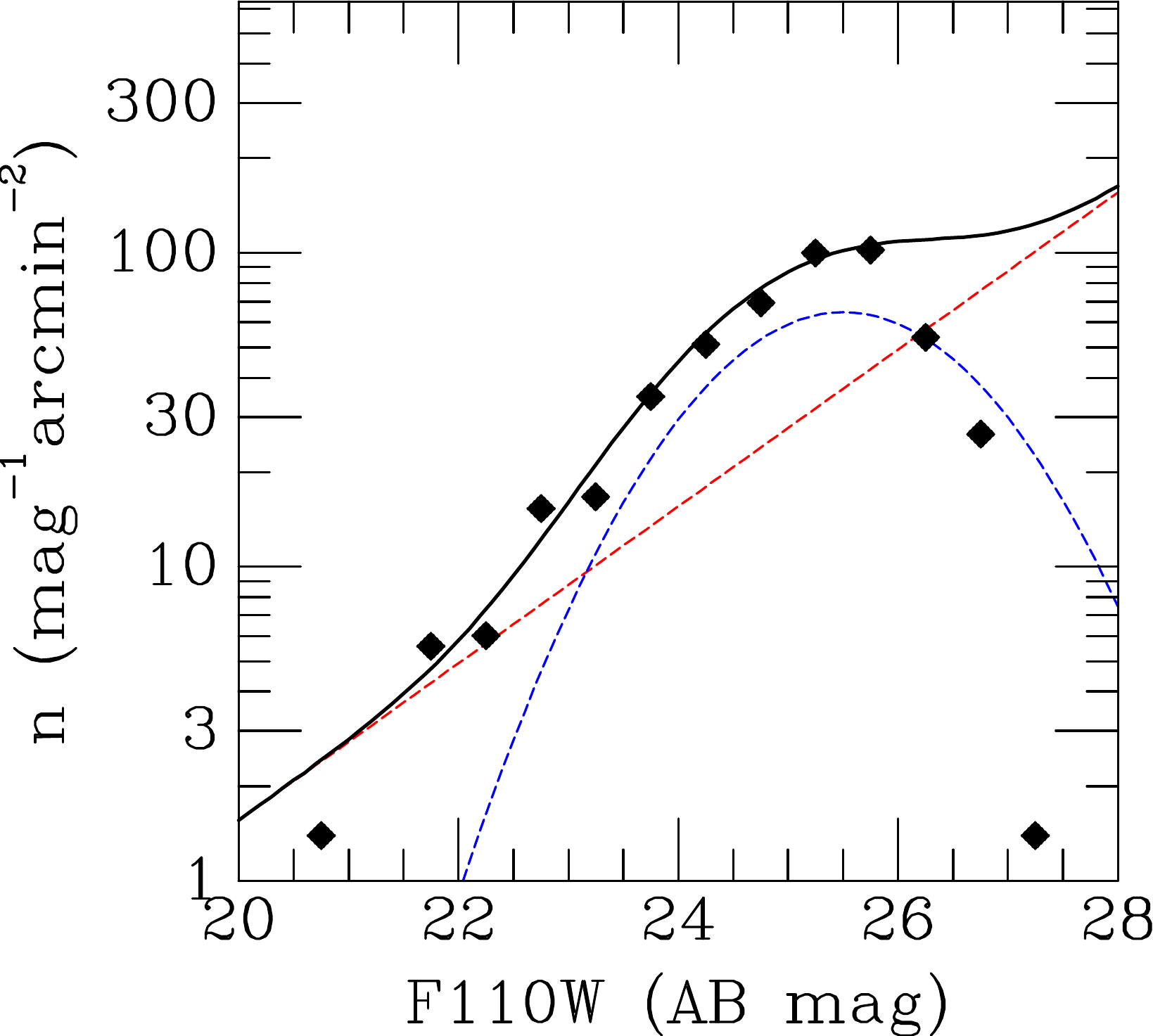}
\caption{Combined figure for NGC~665.}
\end{center}
\end{figure*}
\clearpage

\begin{figure*}
\begin{center}
\includegraphics[scale=0.2]{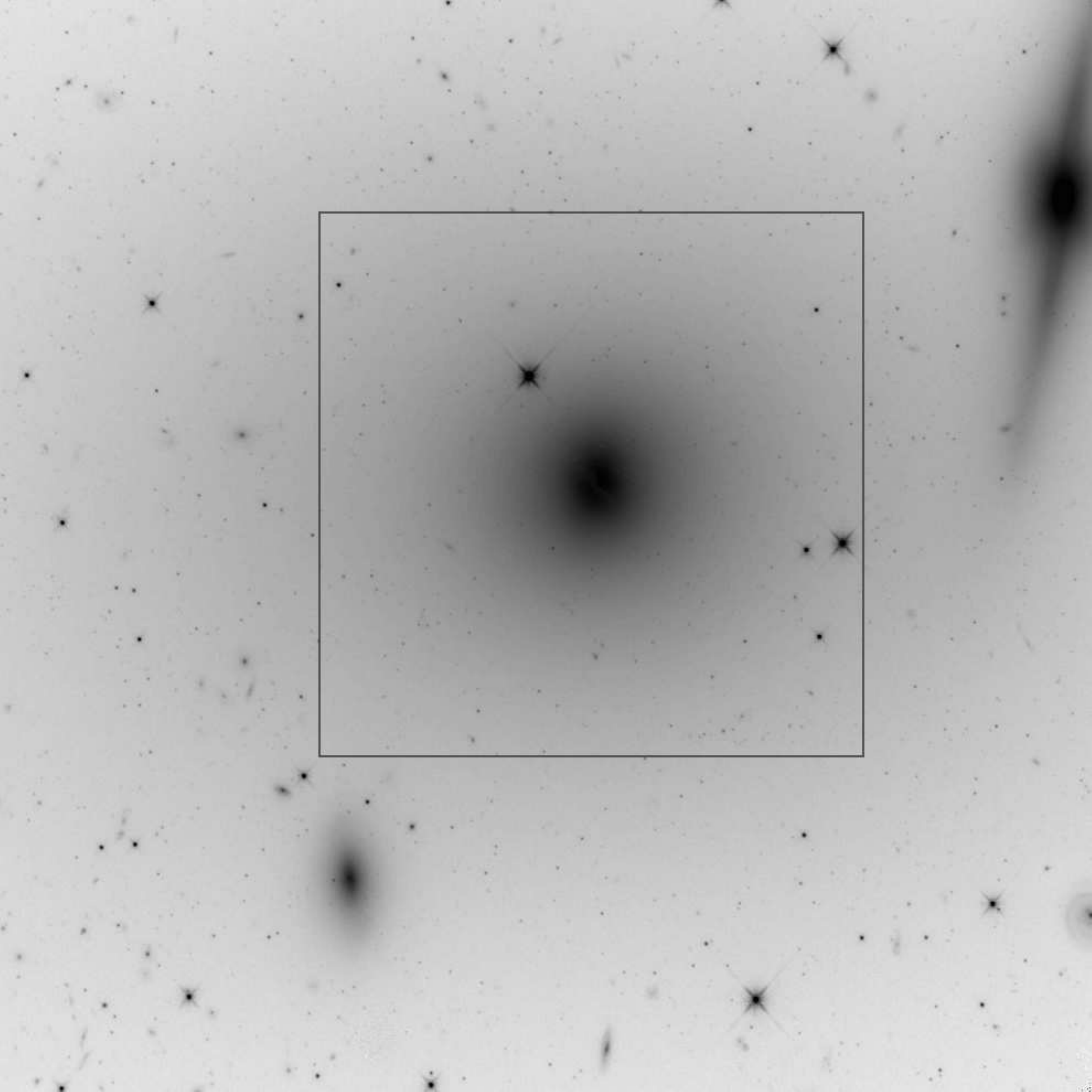}
\includegraphics[scale=0.4]{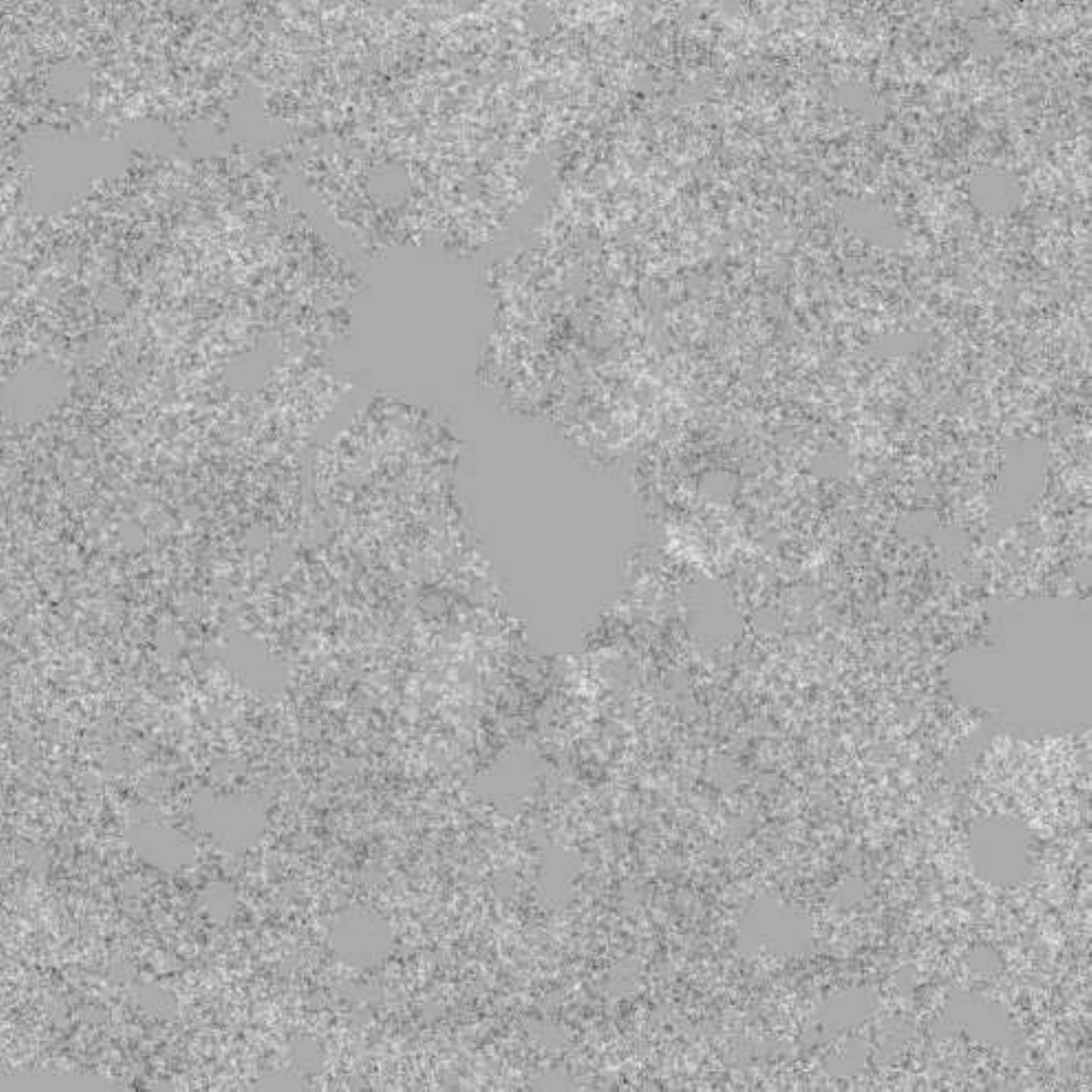} \\
\vspace{10pt}
\includegraphics[scale=0.4]{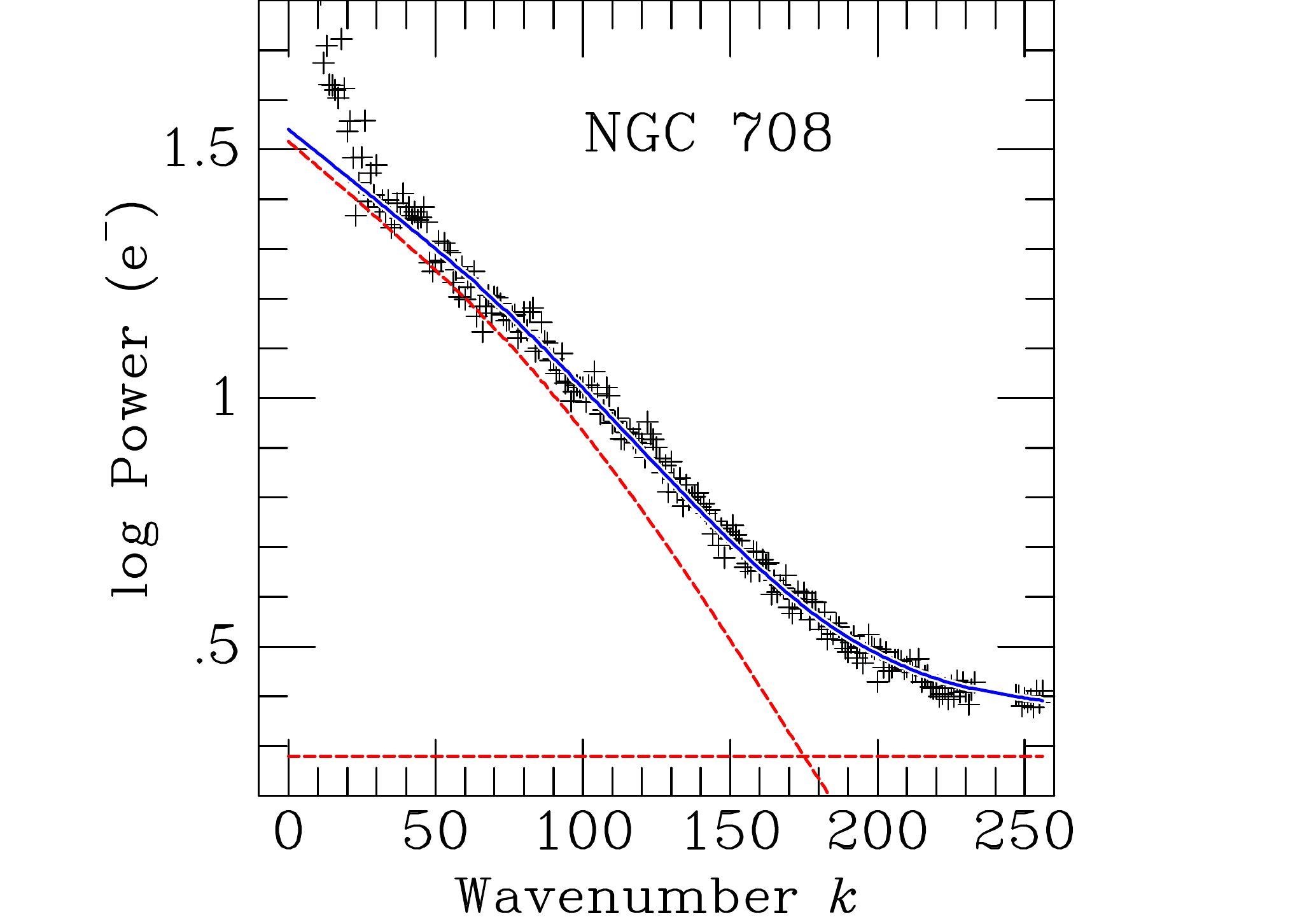}
\hspace{-25pt}
\includegraphics[scale=0.4]{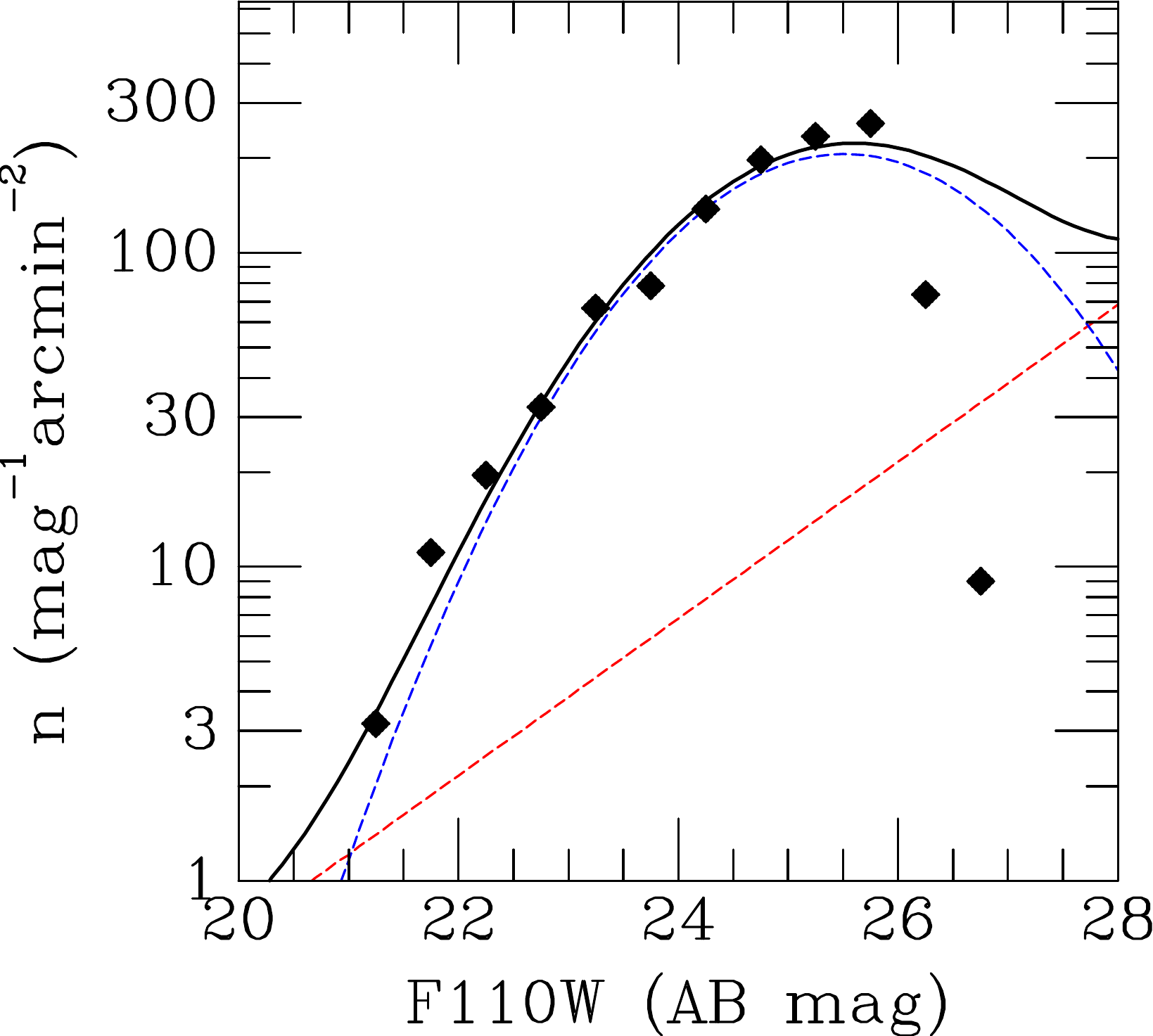}
\caption{Combined figure for NGC~708.}
\end{center}
\end{figure*}
\clearpage

\begin{figure*}
\begin{center}
\includegraphics[scale=0.2]{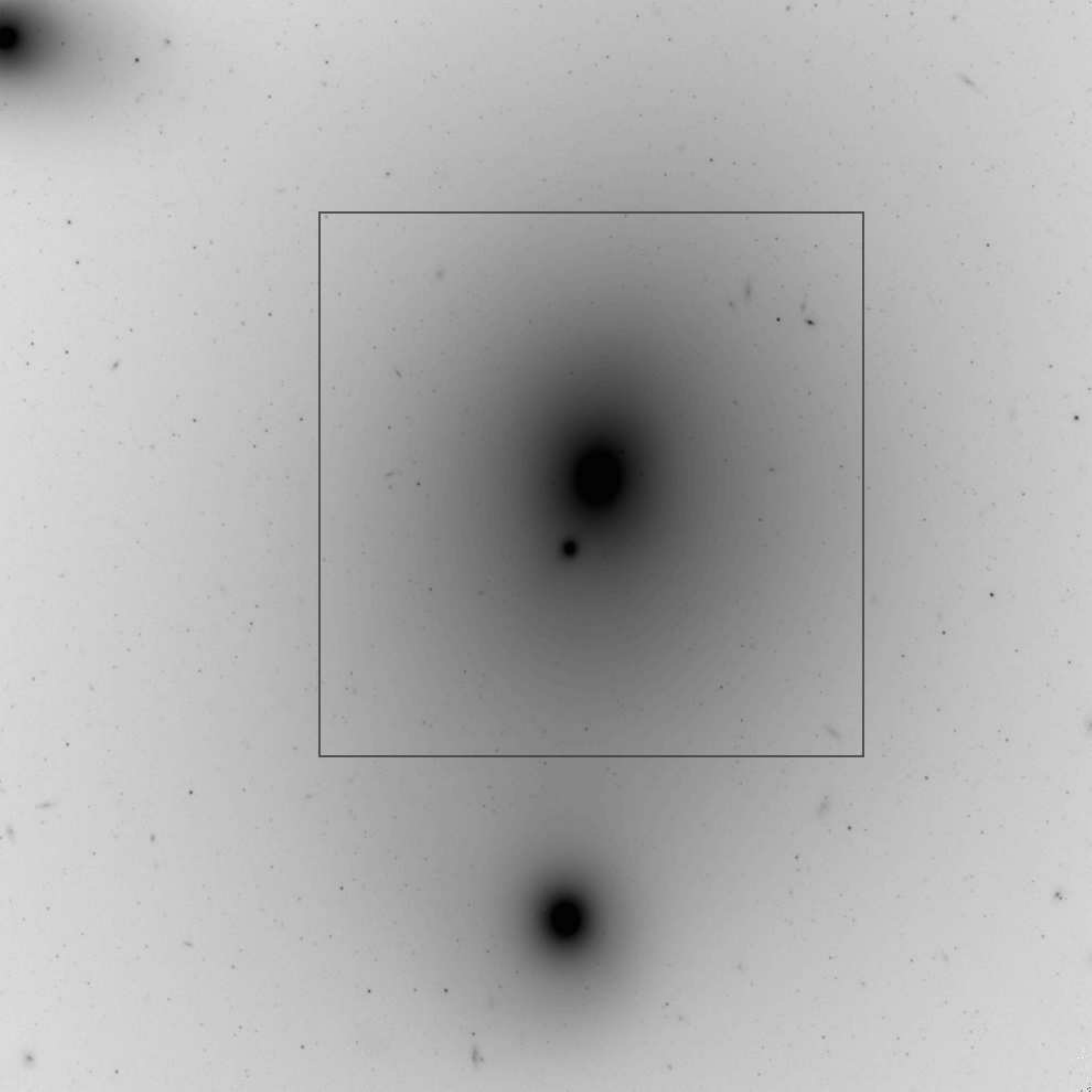}
\includegraphics[scale=0.4]{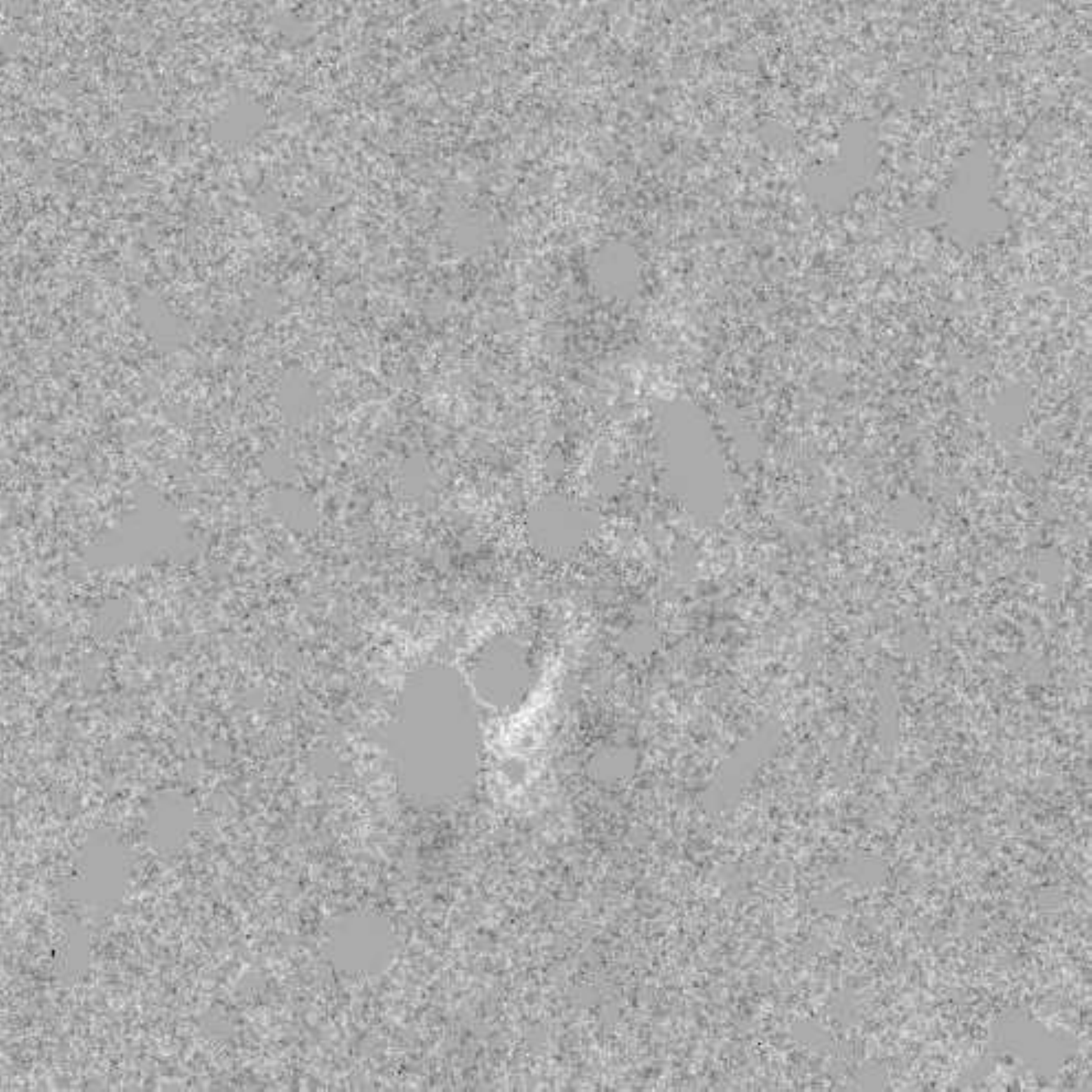} \\
\vspace{10pt}
\includegraphics[scale=0.4]{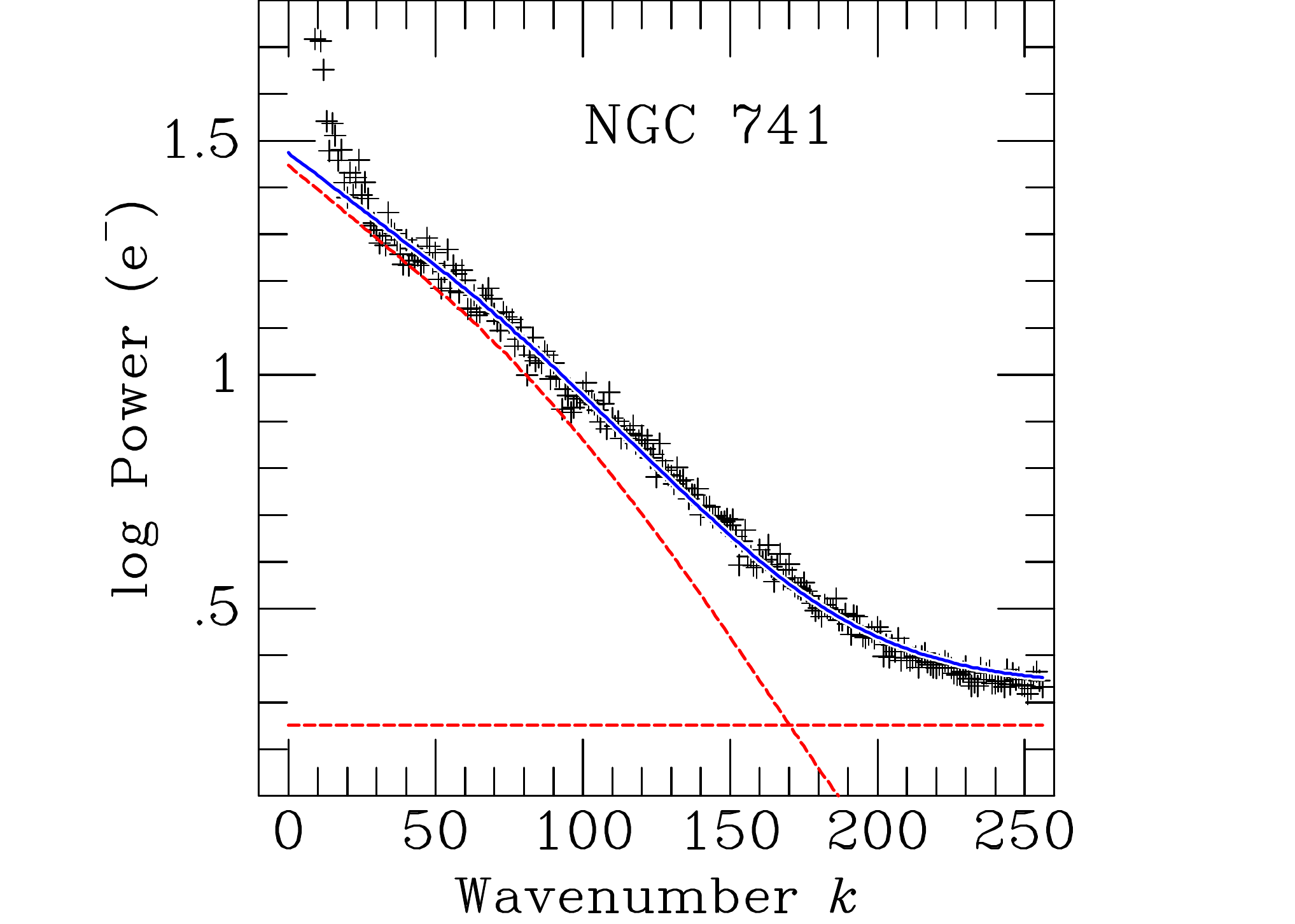}
\hspace{-25pt}
\includegraphics[scale=0.4]{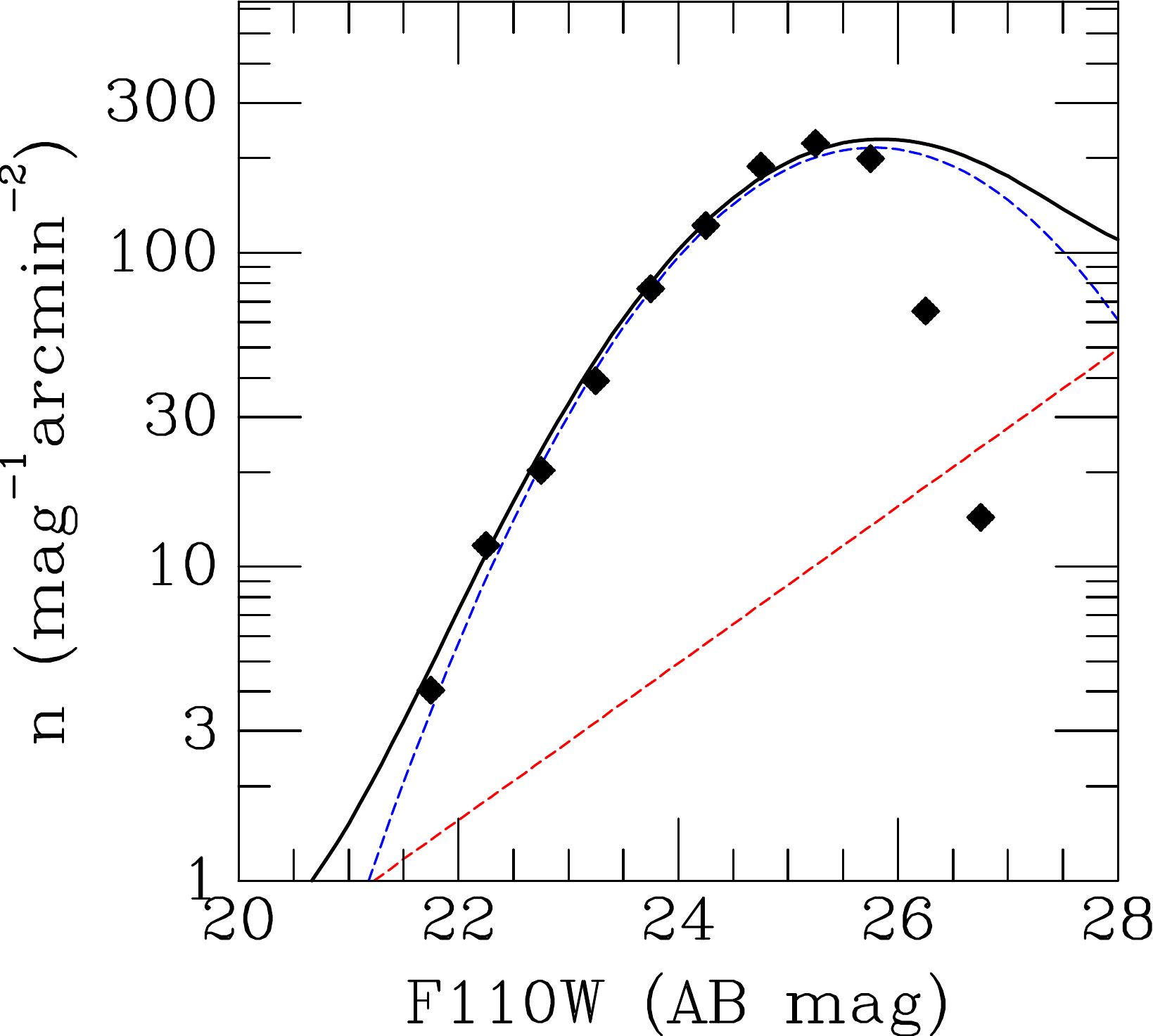}
\caption{Combined figure for NGC~741.}
\end{center}
\end{figure*}
\clearpage

\begin{figure*}
\begin{center}
\includegraphics[scale=0.2]{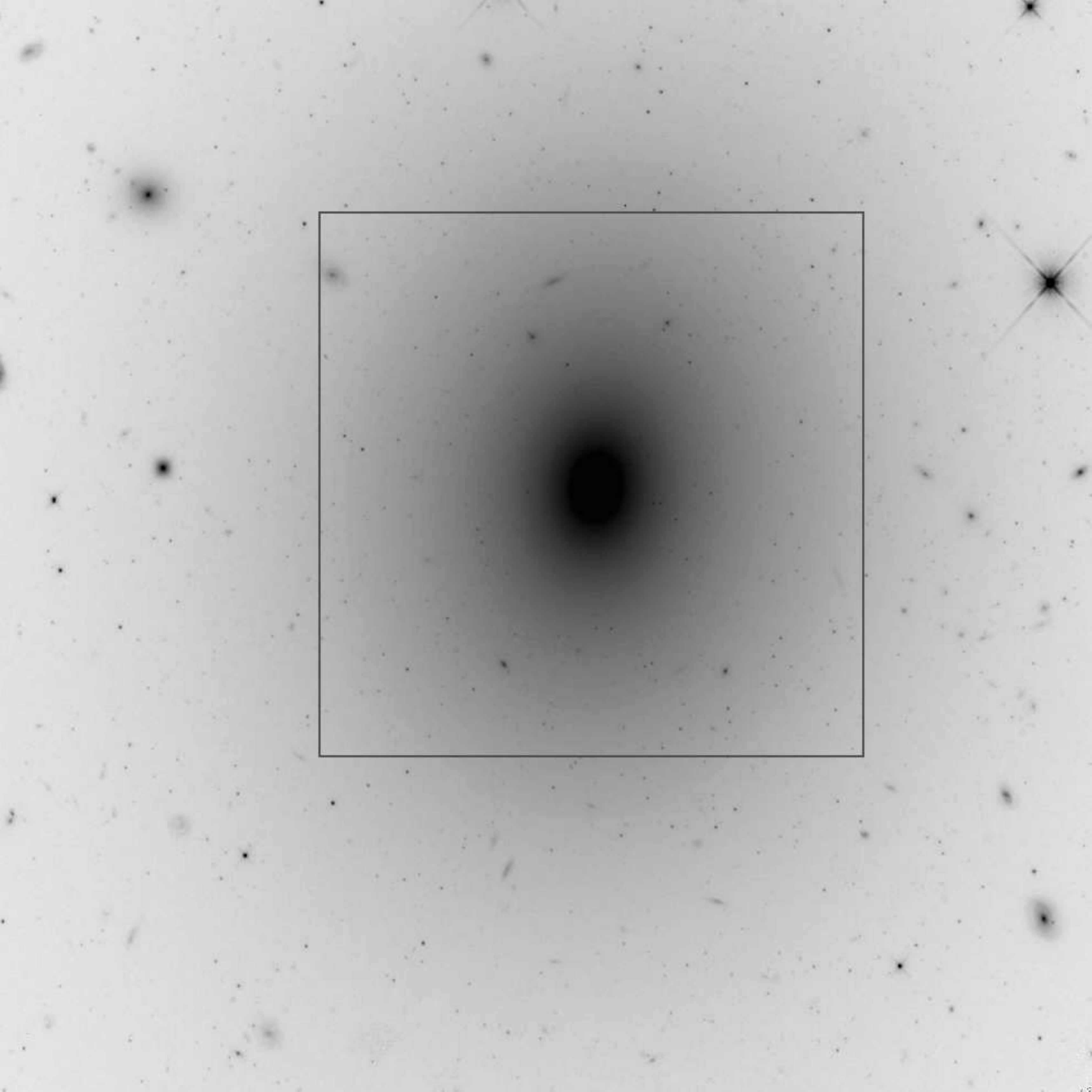}
\includegraphics[scale=0.4]{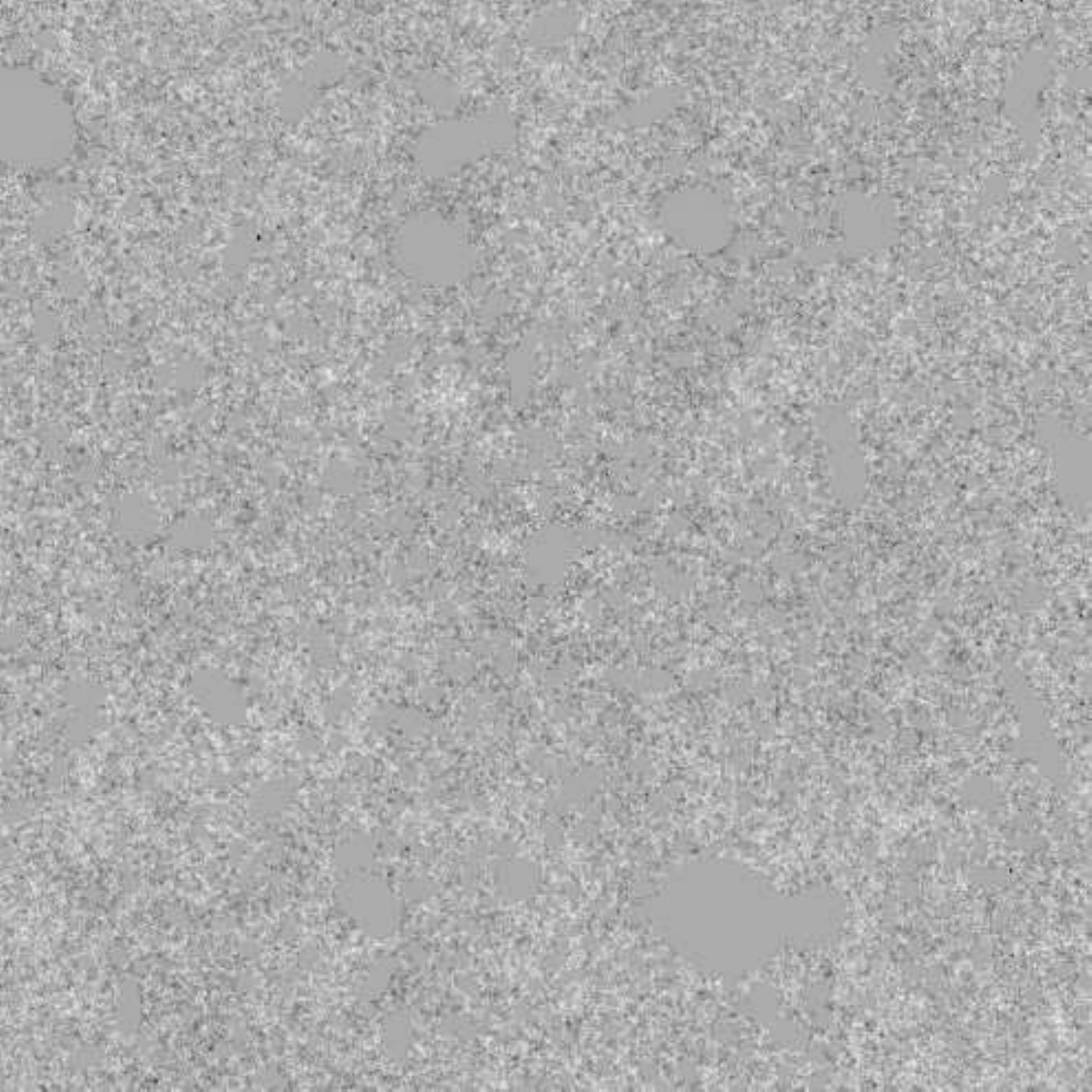} \\
\vspace{10pt}
\includegraphics[scale=0.4]{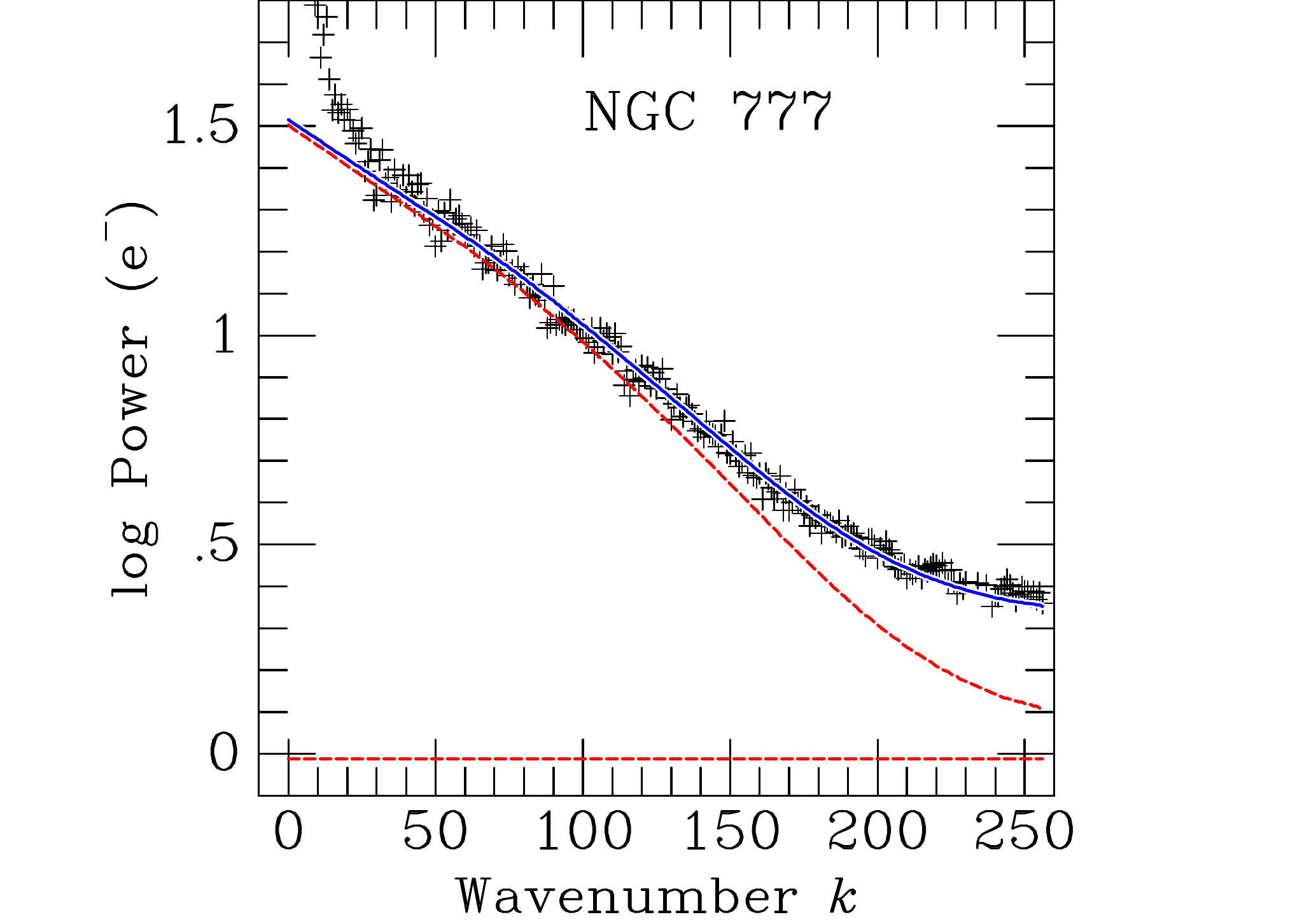}
\hspace{-25pt}
\includegraphics[scale=0.4]{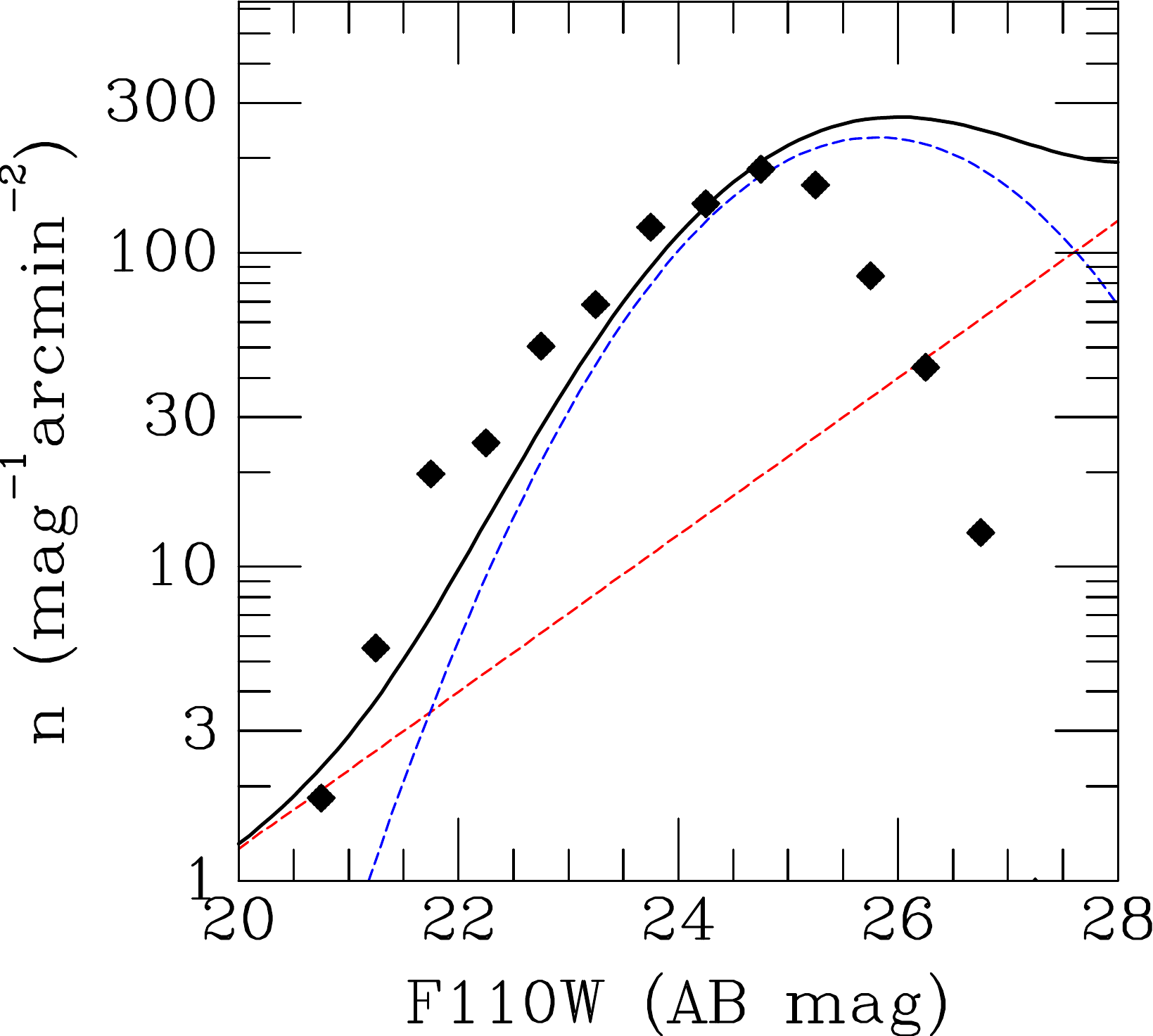}
\caption{Combined figure for NGC~777.}
\end{center}
\end{figure*}
\clearpage

\begin{figure*}
\begin{center}
\includegraphics[scale=0.2]{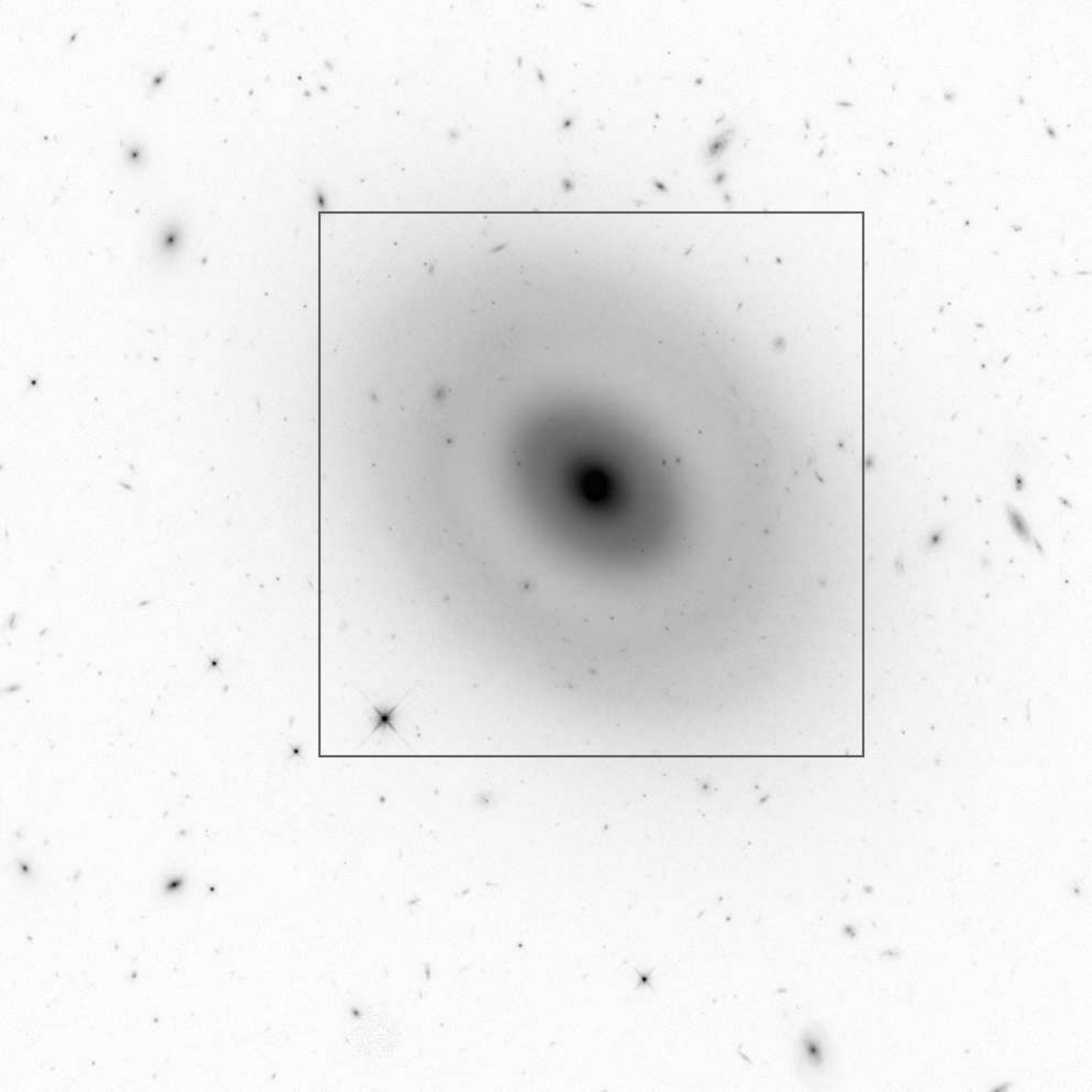}
\includegraphics[scale=0.4]{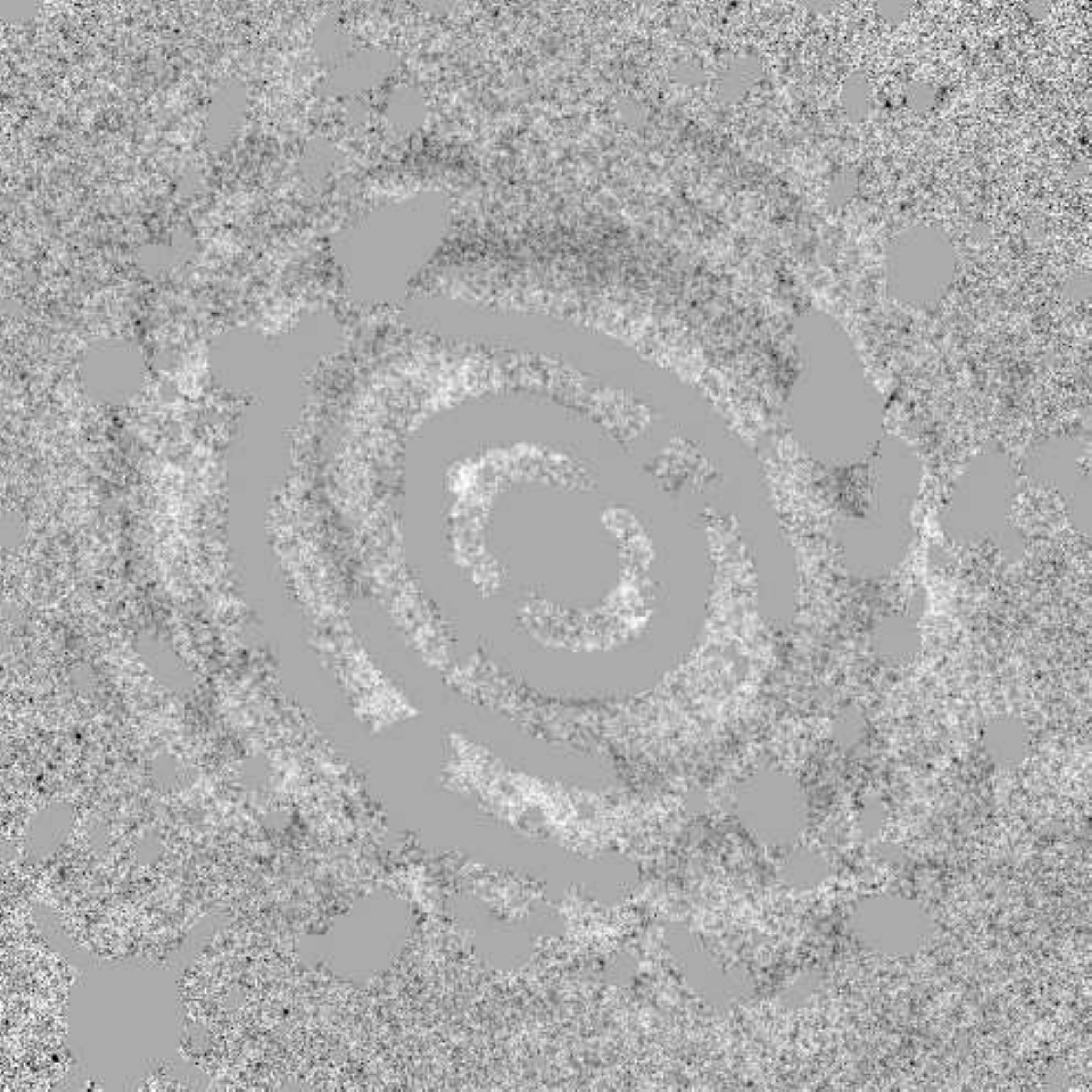} \\
\vspace{10pt}
\includegraphics[scale=0.4]{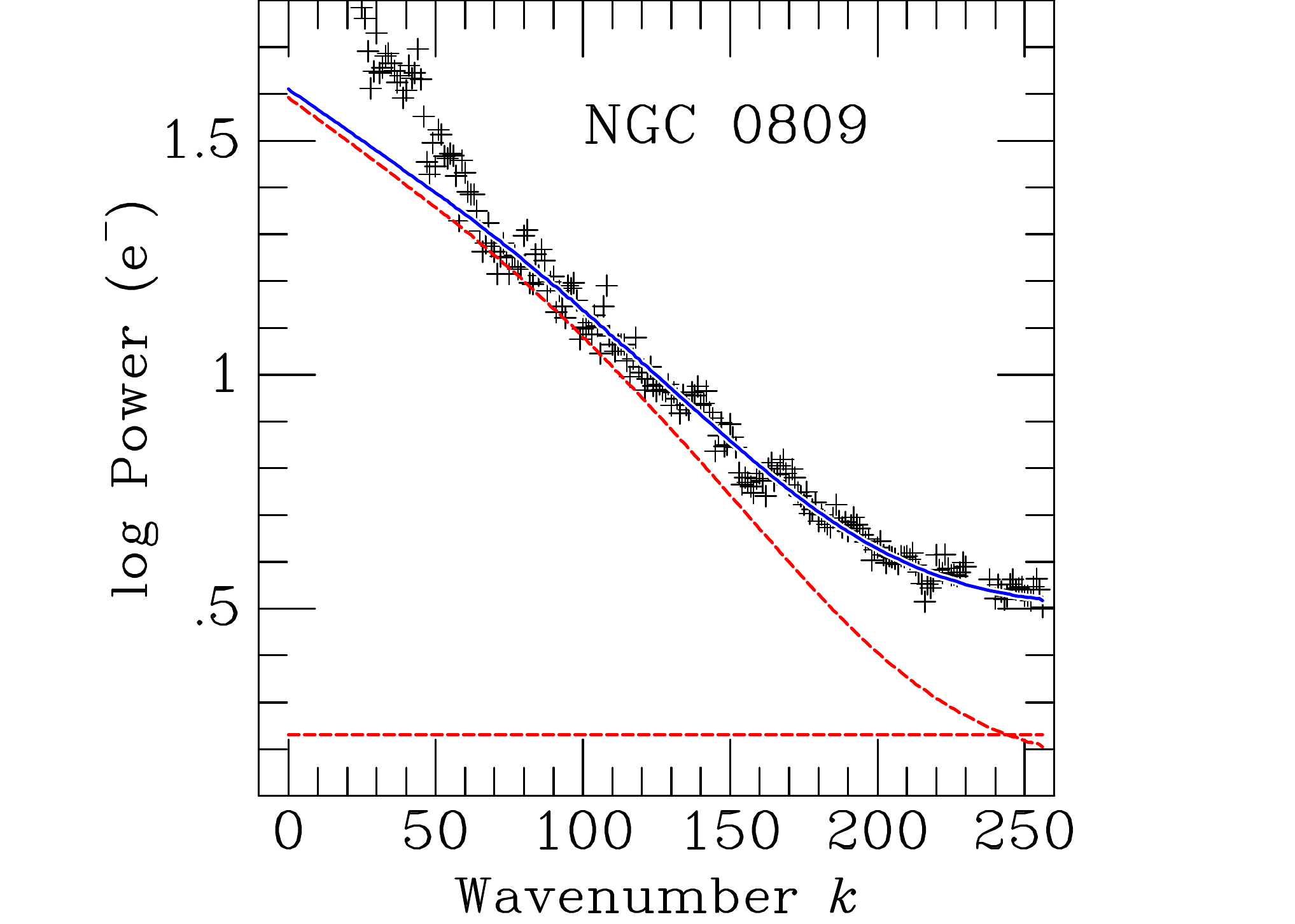}
\hspace{-25pt}
\includegraphics[scale=0.4]{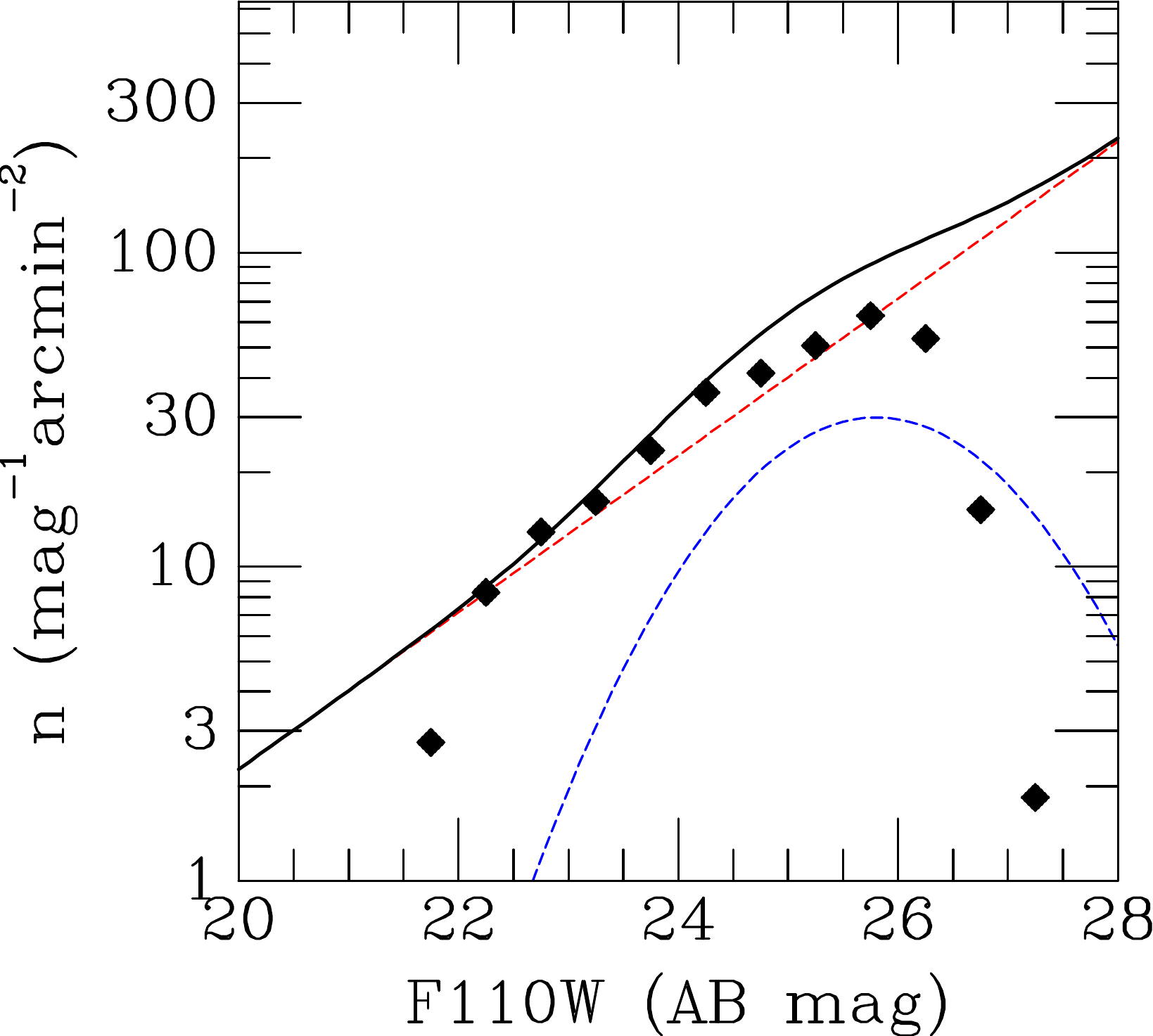}
\caption{Combined figure for NGC~809.}
\end{center}
\end{figure*}
\clearpage

\begin{figure*}
\begin{center}
\includegraphics[scale=0.2]{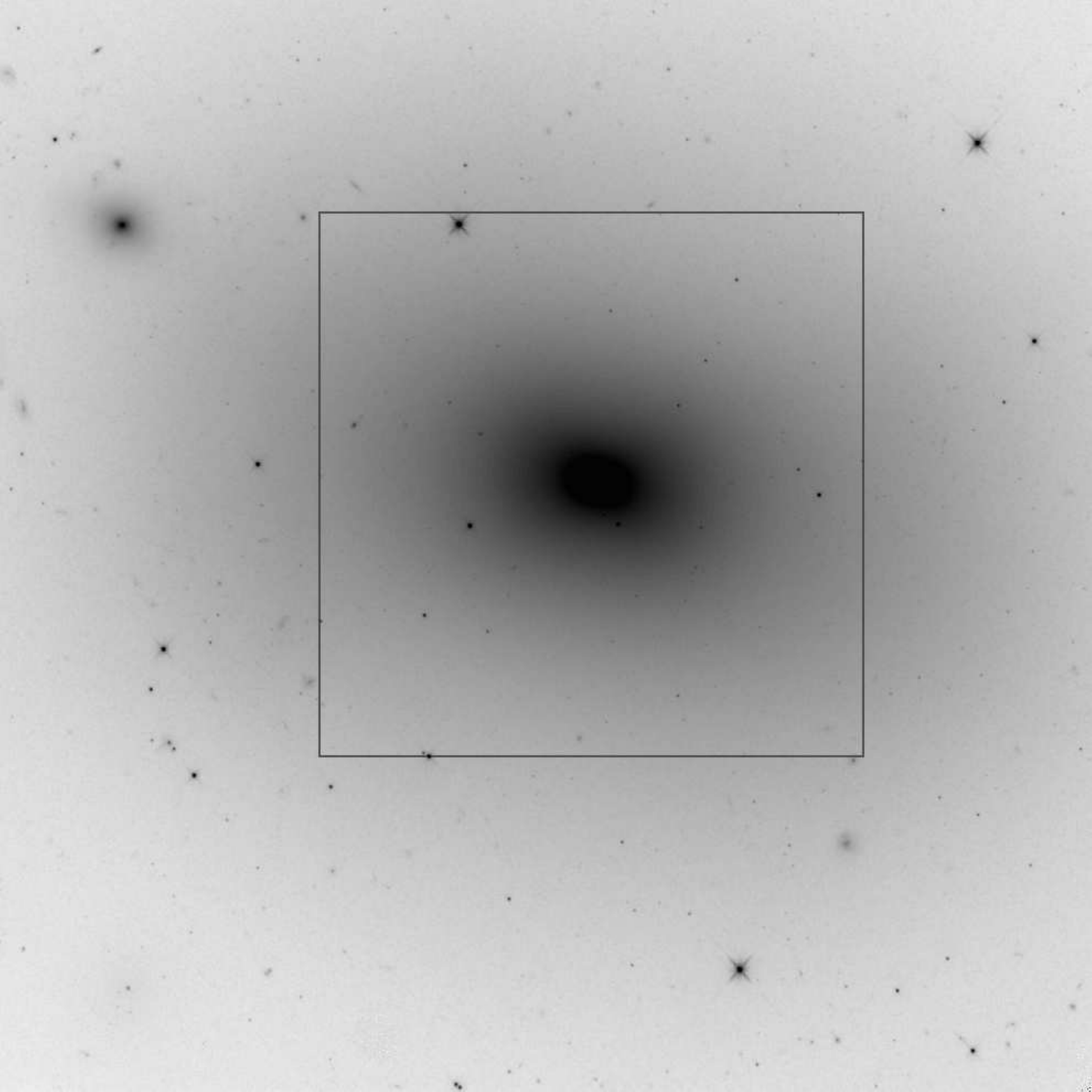}
\includegraphics[scale=0.4]{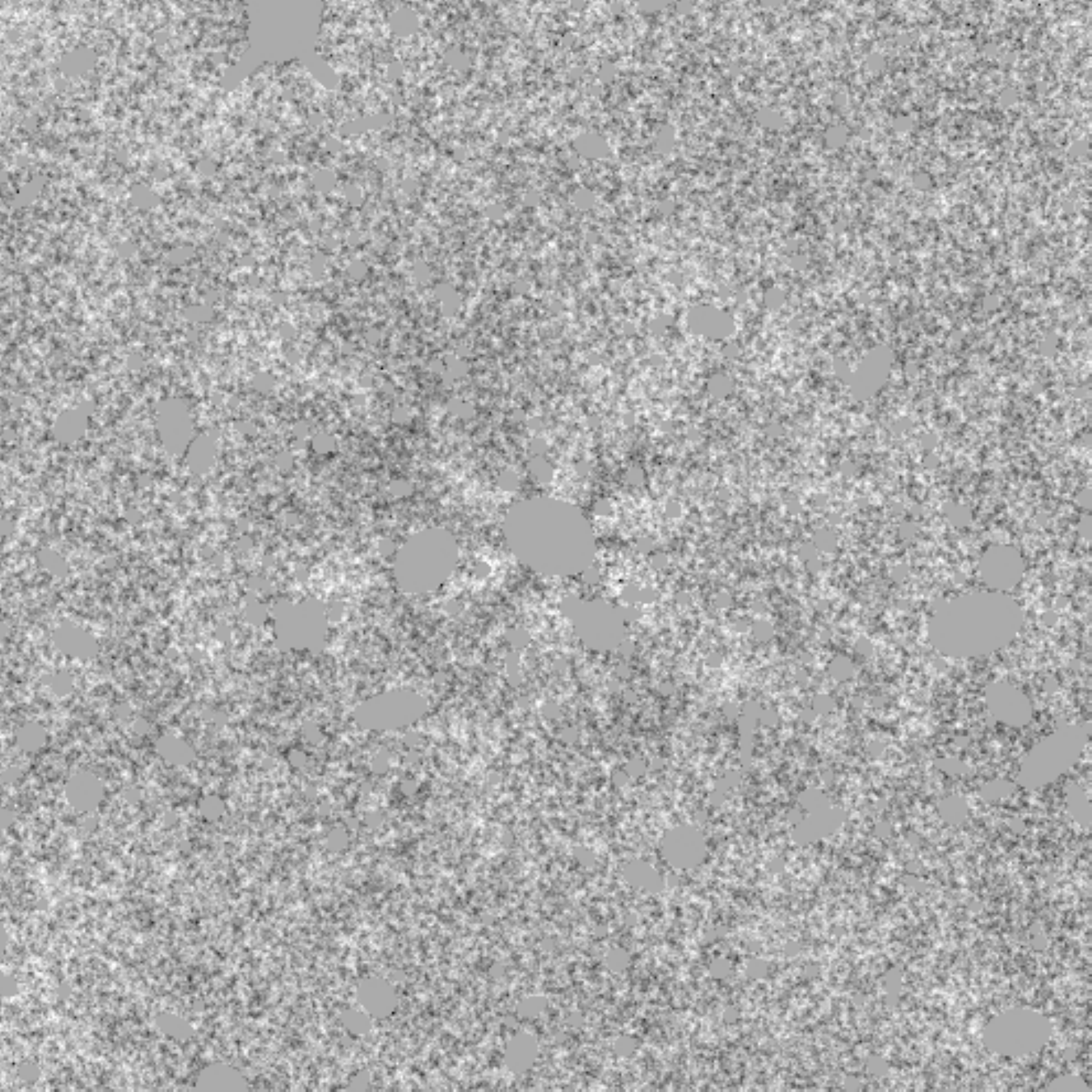} \\
\vspace{10pt}
\includegraphics[scale=0.4]{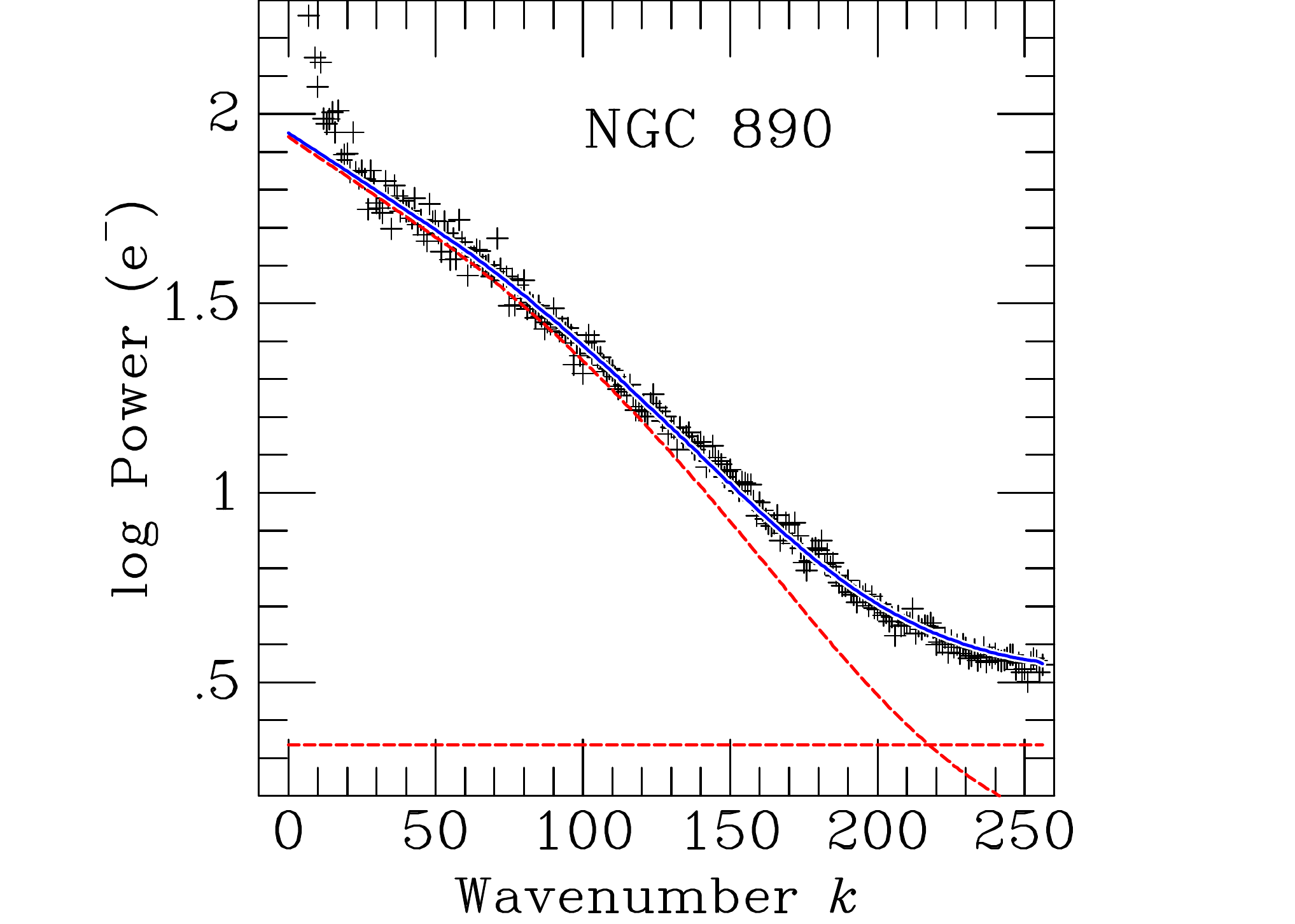}
\hspace{-25pt}
\includegraphics[scale=0.4]{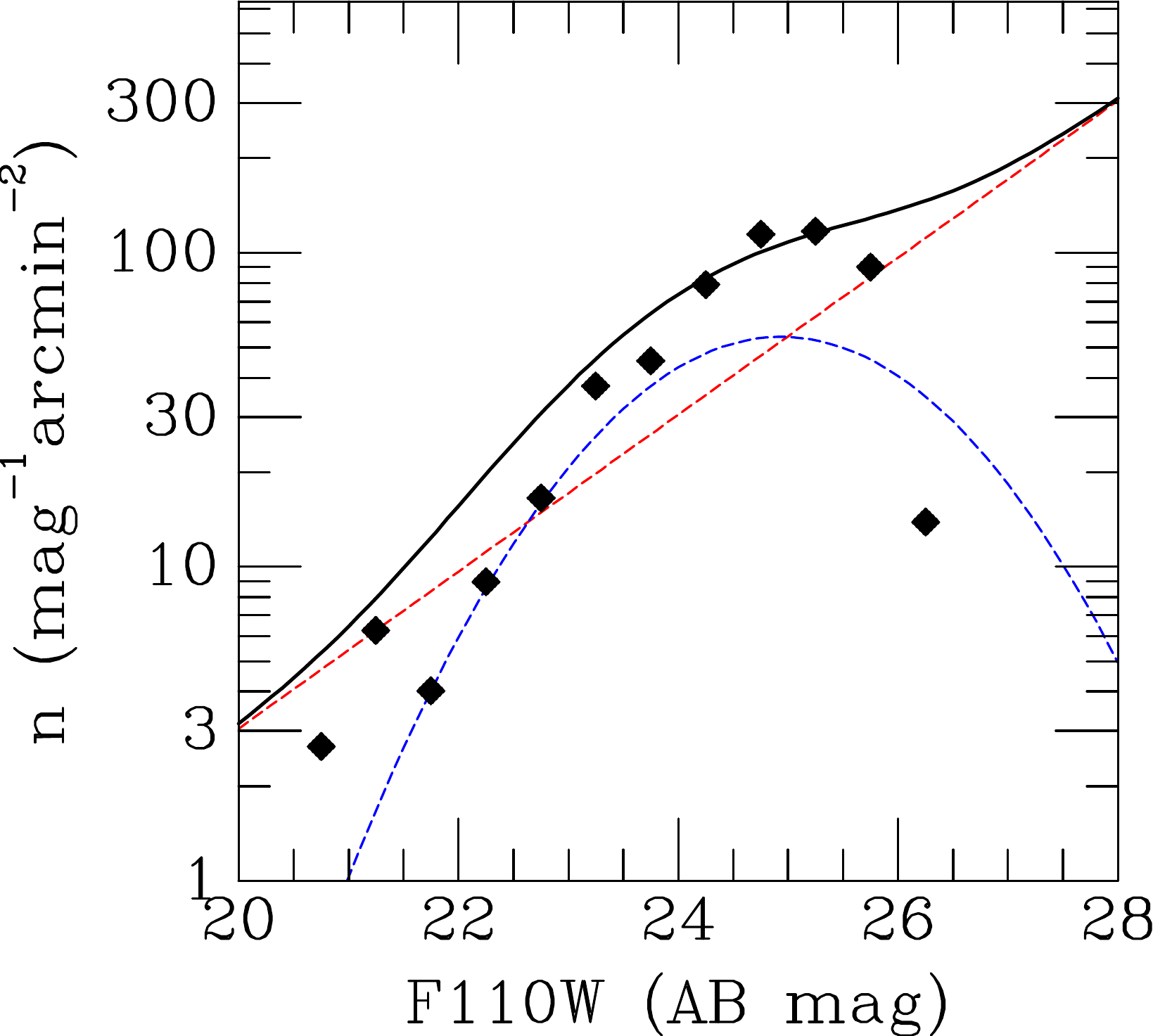}
\caption{Combined figure for NGC~890.}
\end{center}
\end{figure*}
\clearpage

\begin{figure*}
\begin{center}
\includegraphics[scale=0.2]{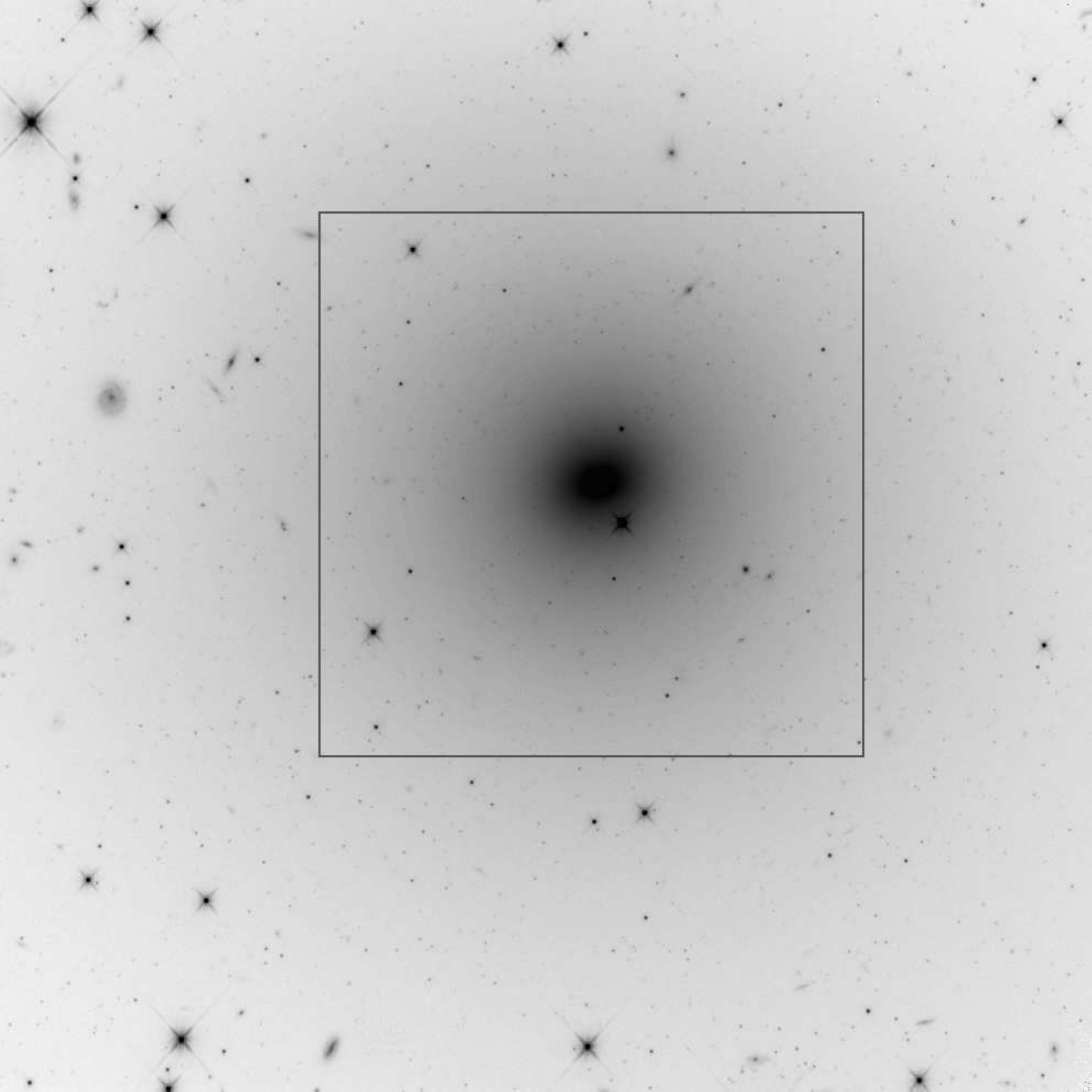}
\includegraphics[scale=0.4]{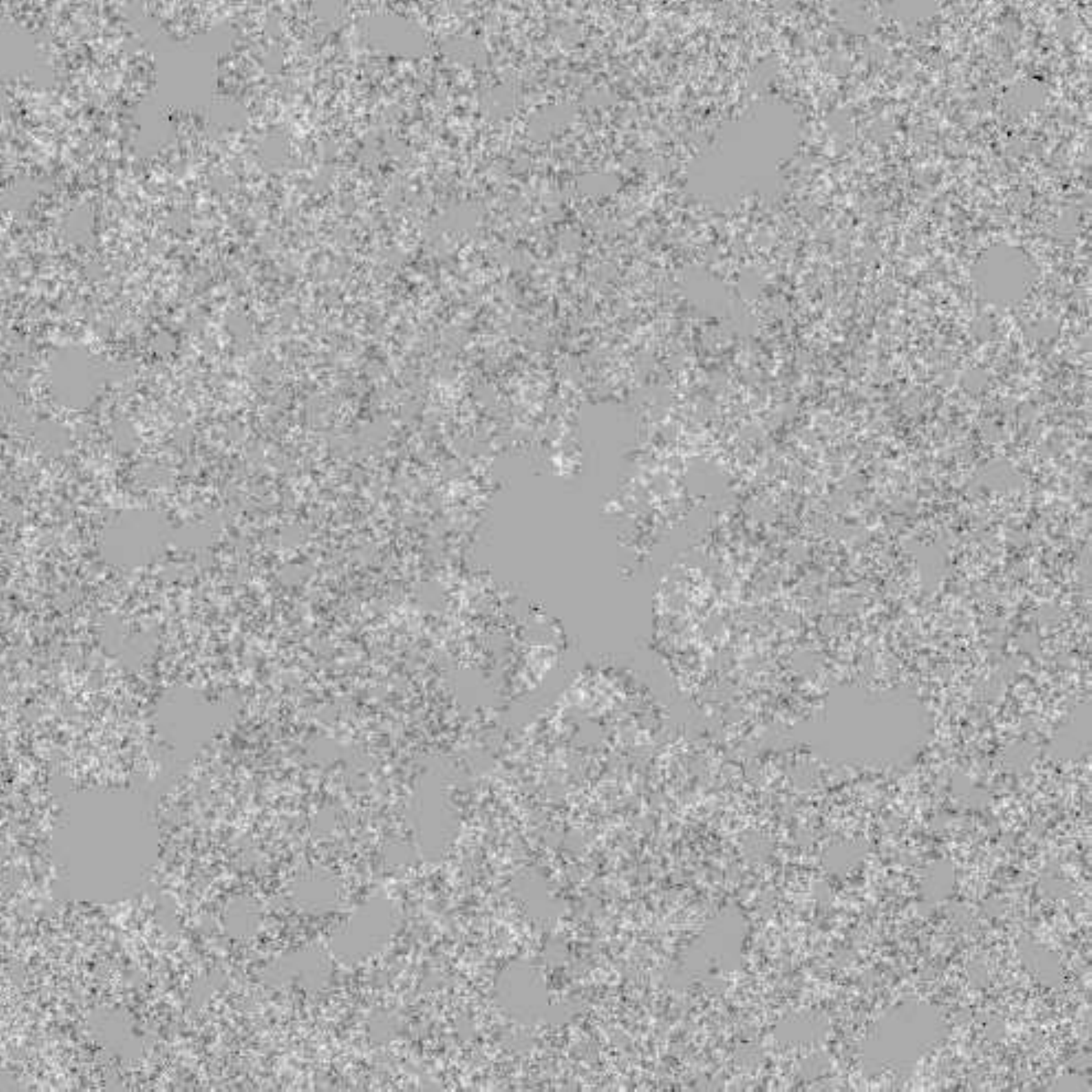} \\
\vspace{10pt}
\includegraphics[scale=0.4]{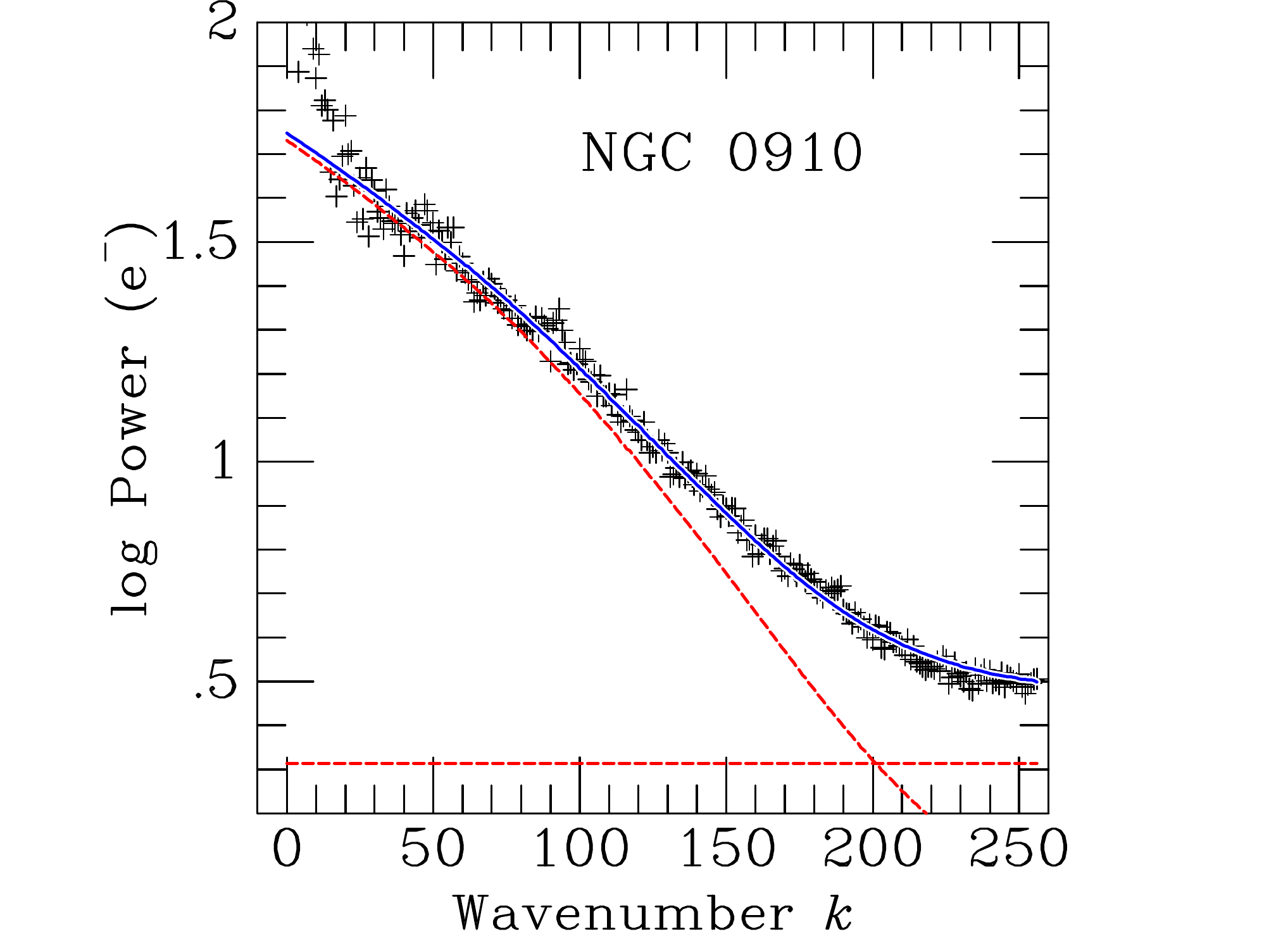}
\hspace{-25pt}
\includegraphics[scale=0.4]{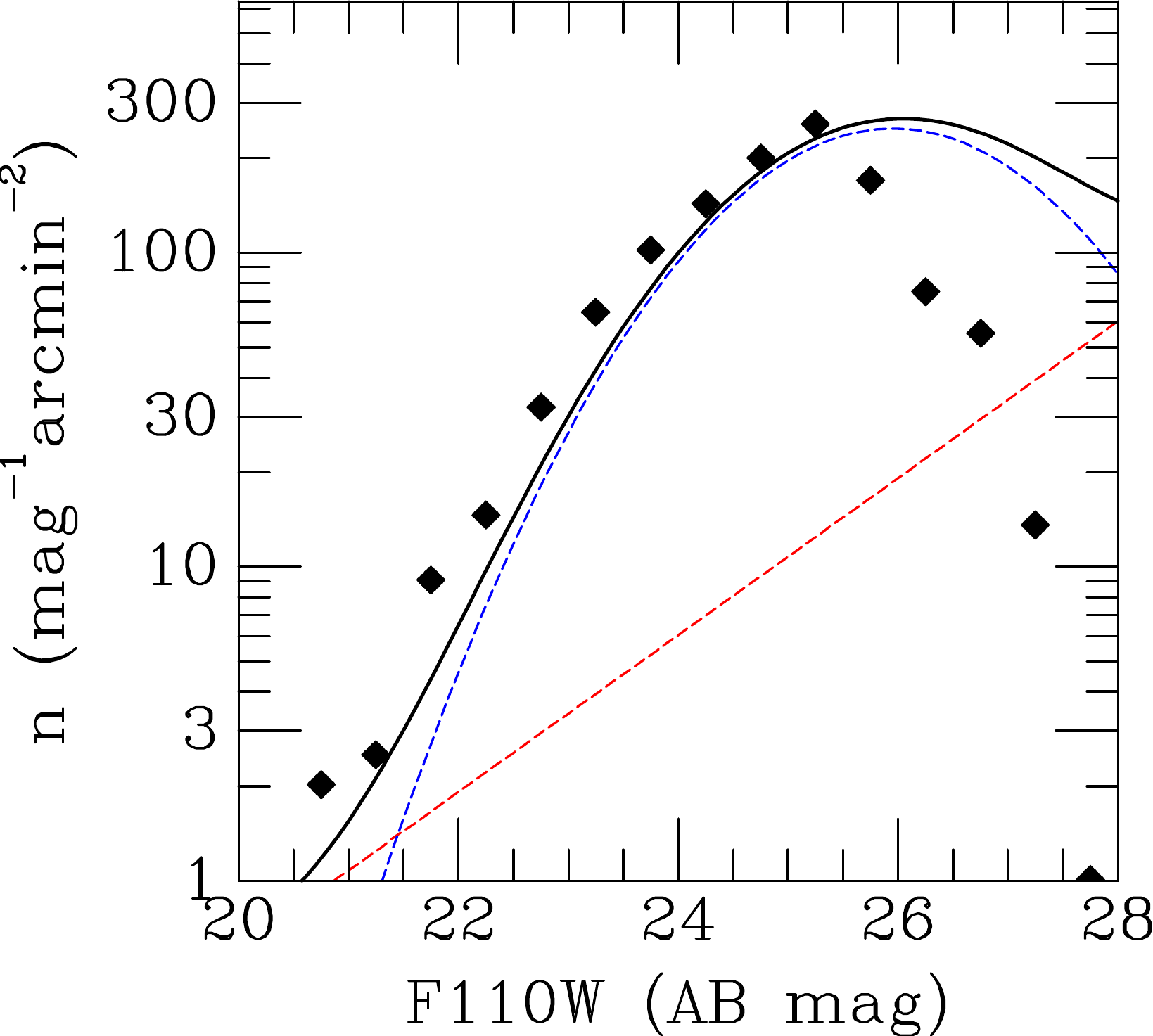}
\caption{Combined figure for NGC~910.}
\end{center}
\end{figure*}
\clearpage

\begin{figure*}
\begin{center}
\includegraphics[scale=0.2]{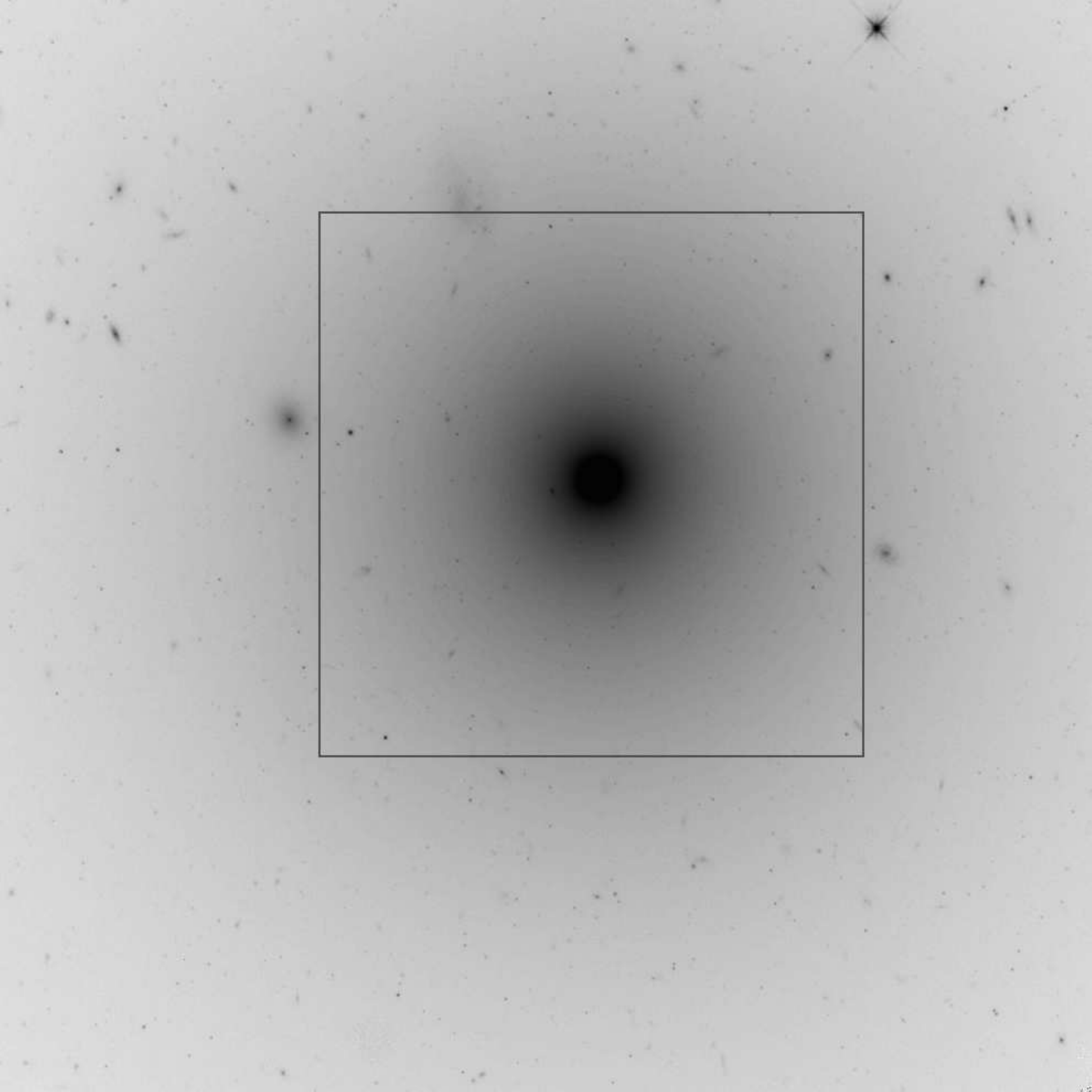}
\includegraphics[scale=0.4]{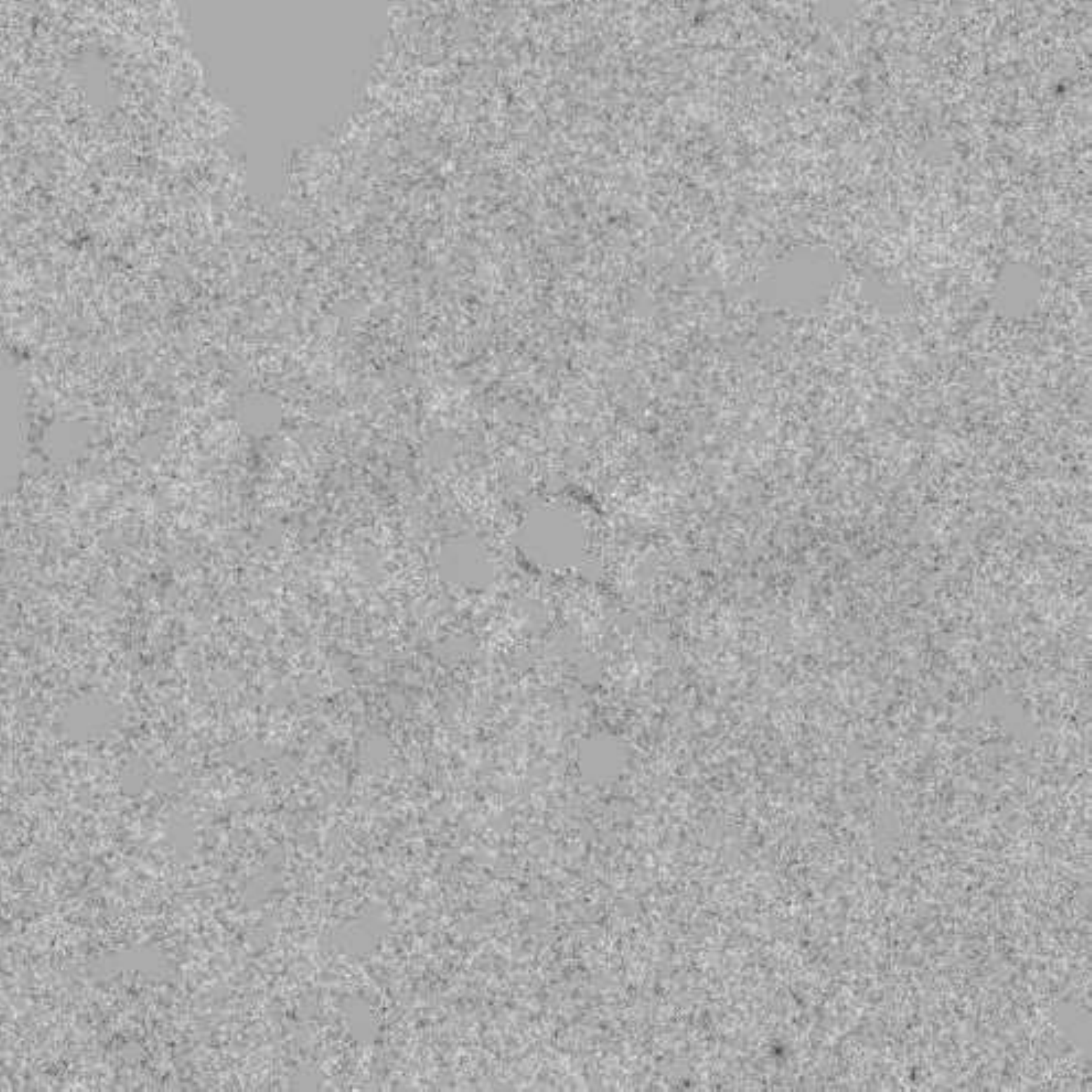} \\
\vspace{10pt}
\includegraphics[scale=0.4]{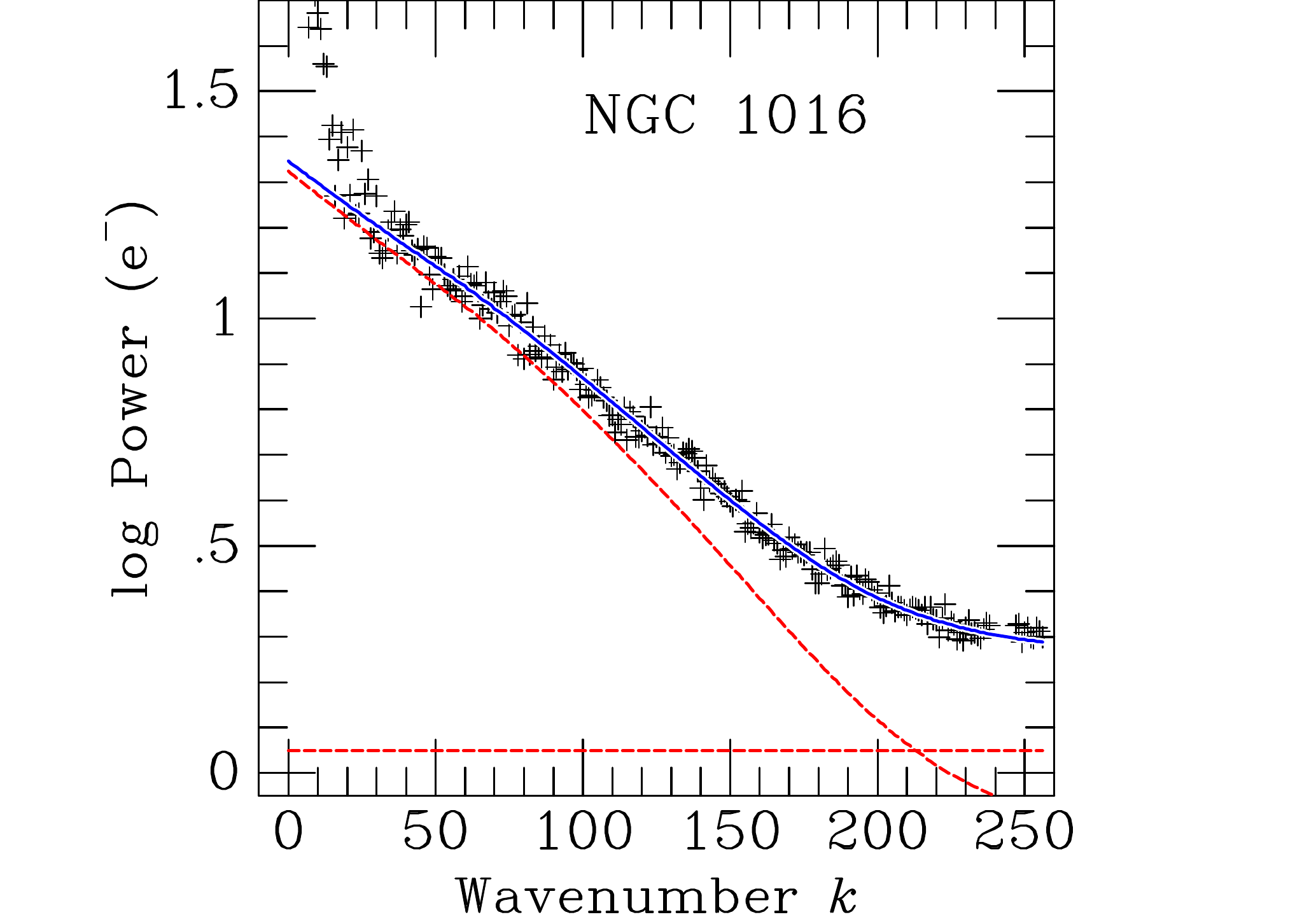}
\hspace{-25pt}
\includegraphics[scale=0.4]{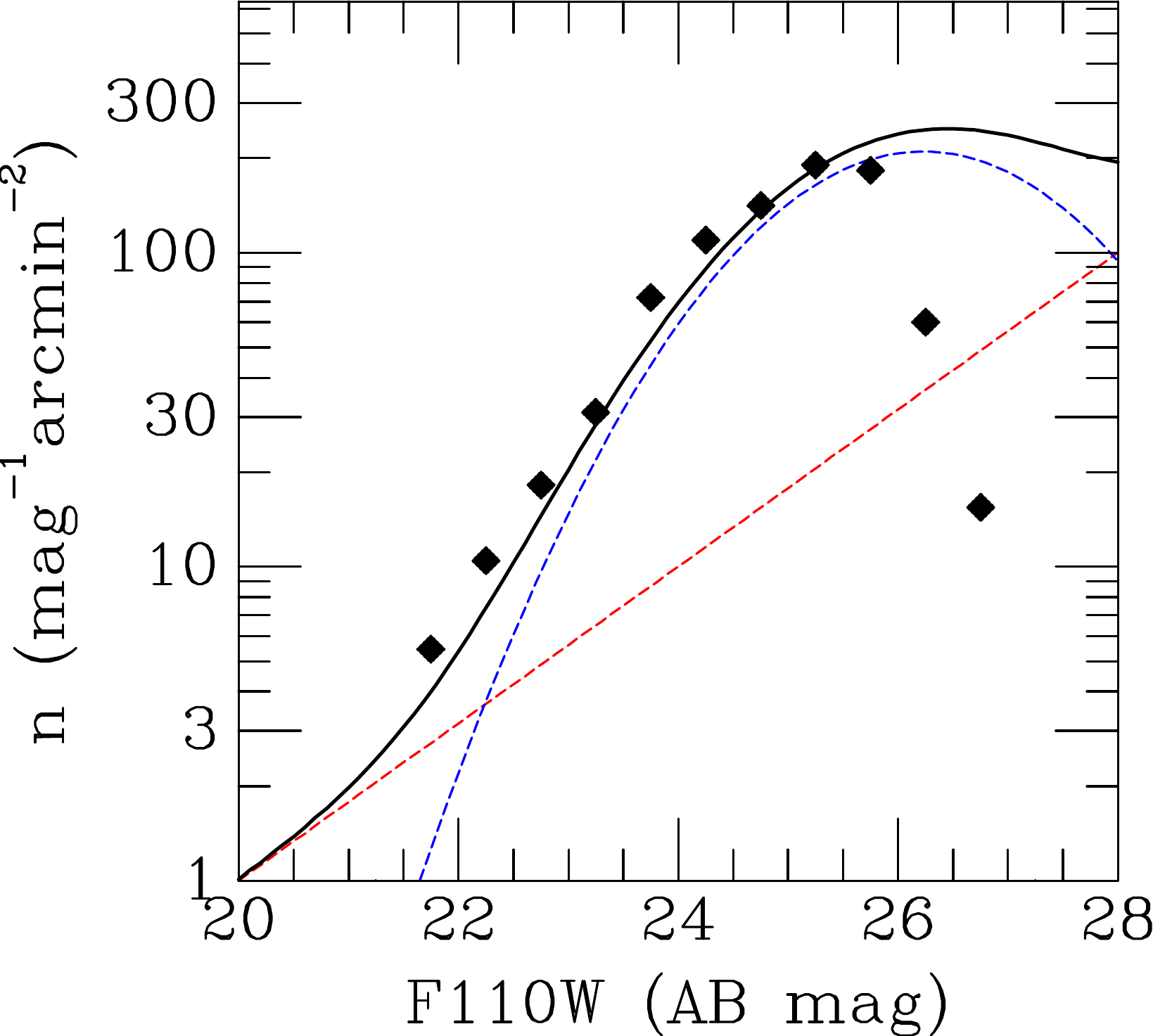}
\caption{Combined figure for NGC~1016.}
\end{center}
\end{figure*}
\clearpage

\begin{figure*}
\begin{center}
\includegraphics[scale=0.2]{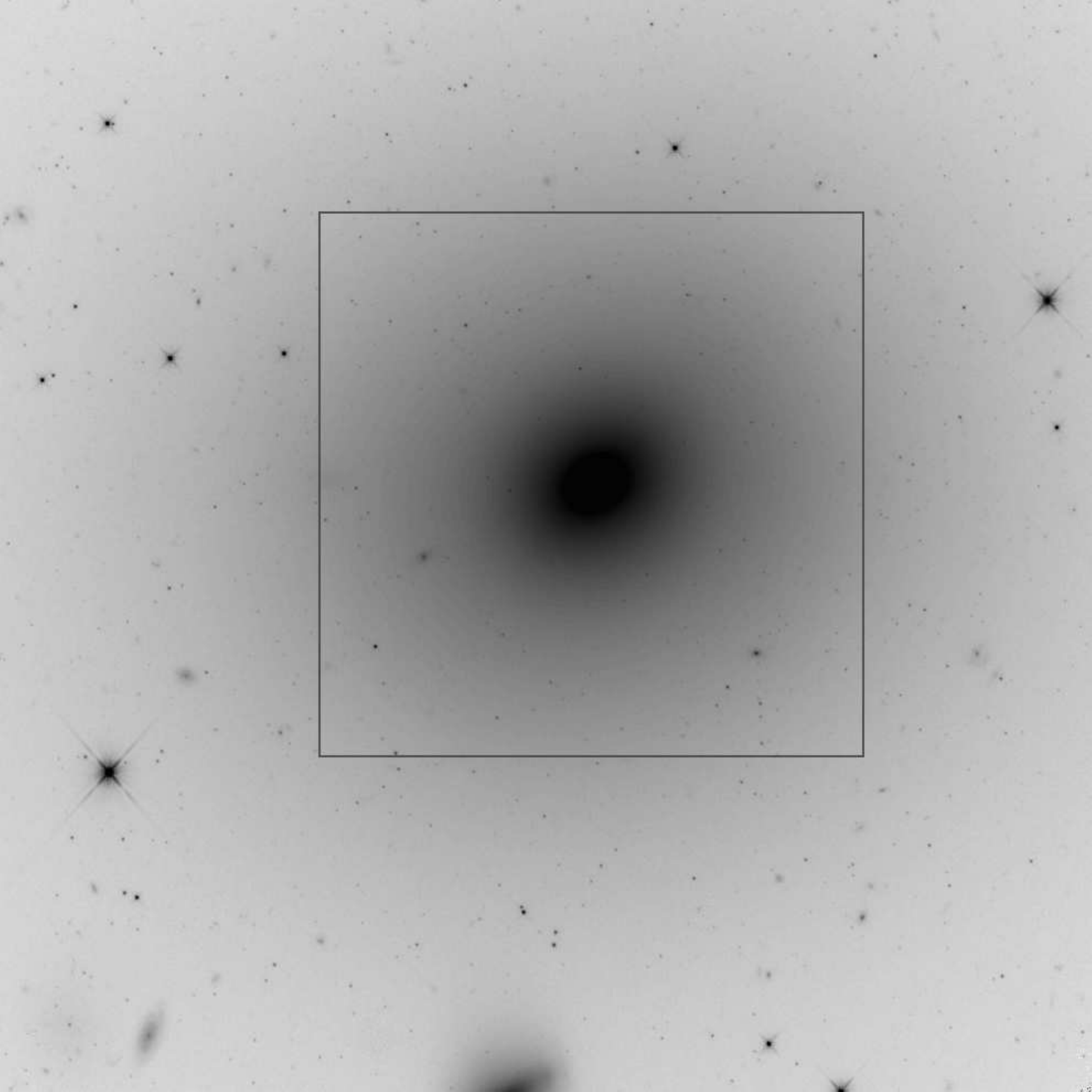}
\includegraphics[scale=0.4]{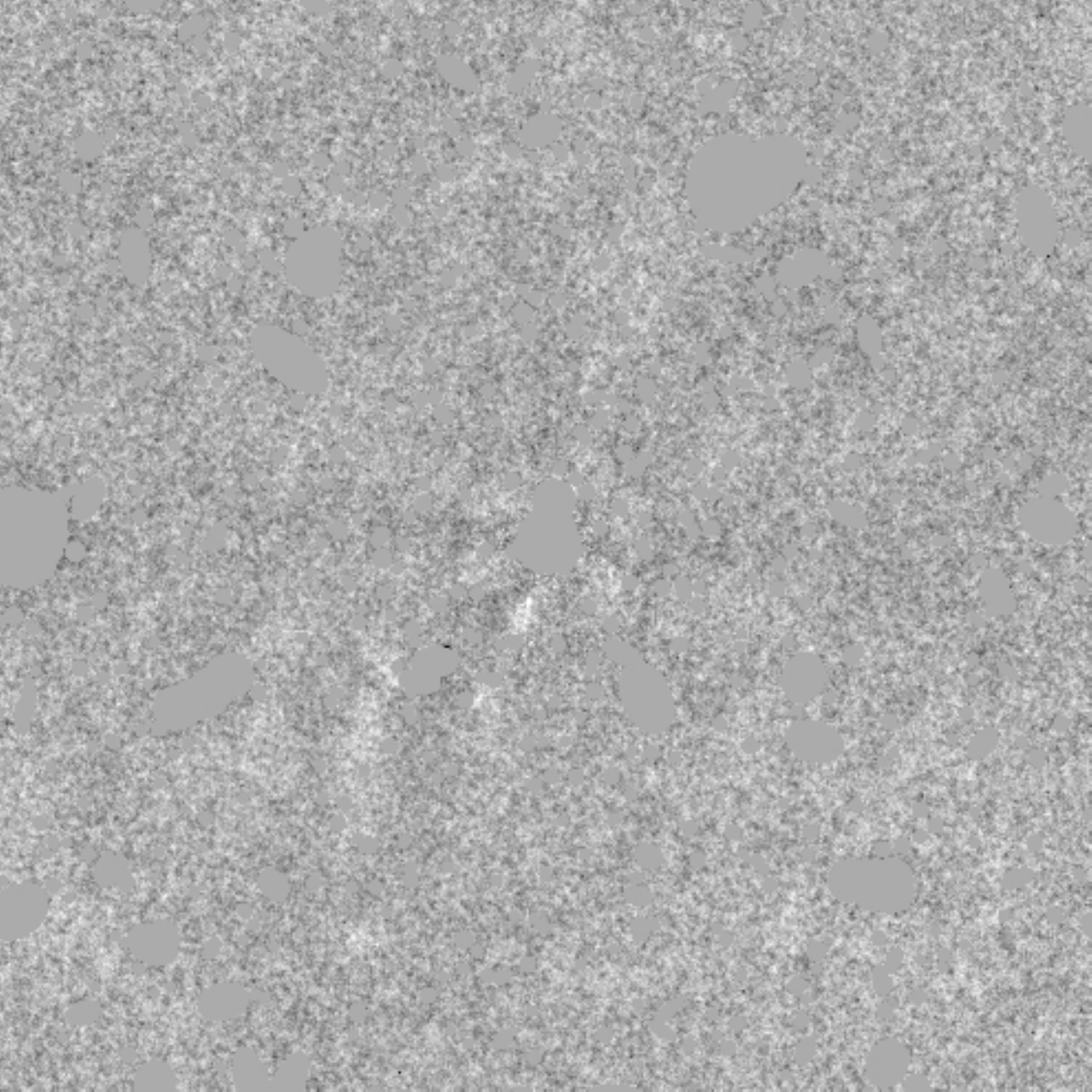} \\
\vspace{10pt}
\includegraphics[scale=0.4]{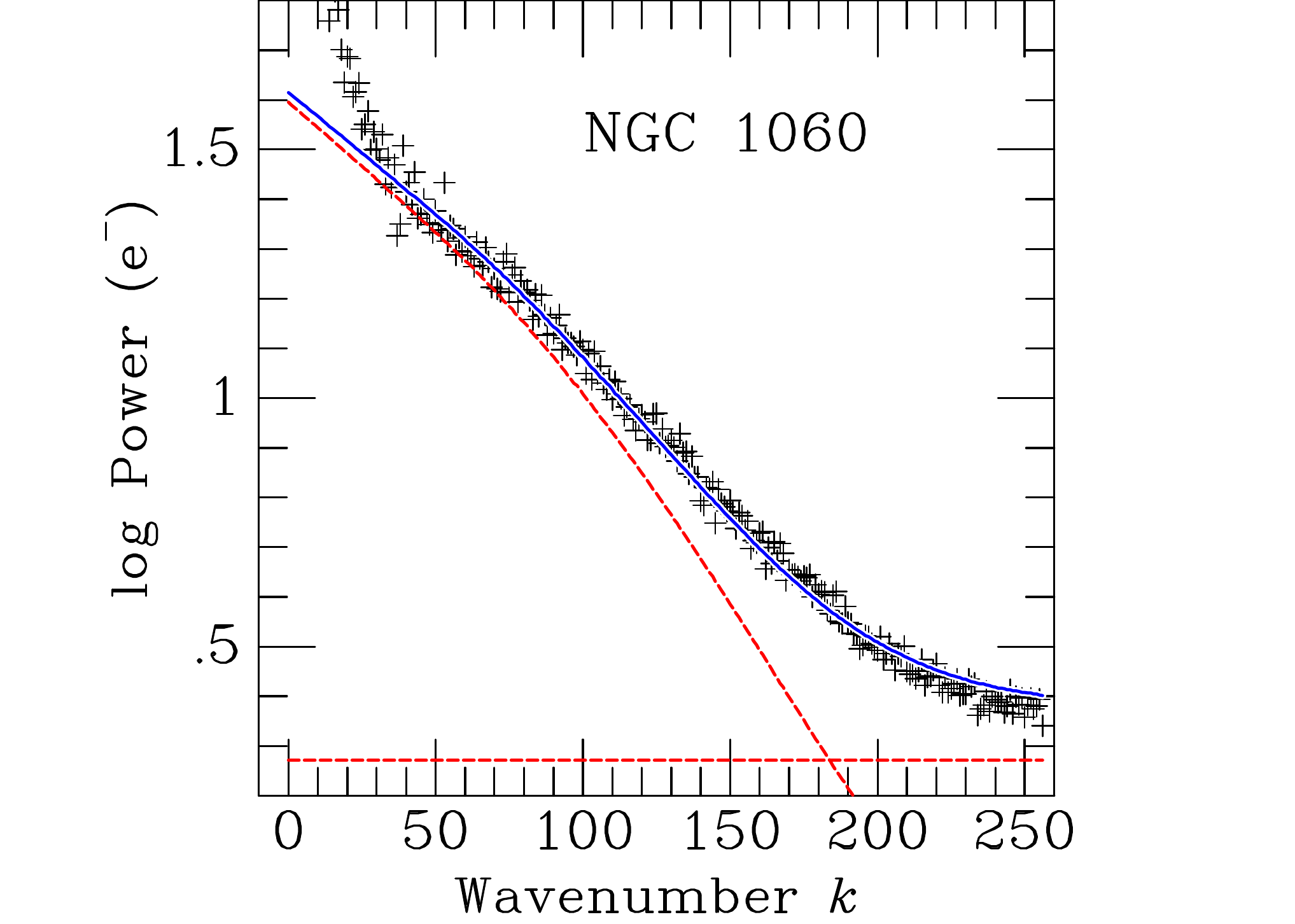}
\hspace{-25pt}
\includegraphics[scale=0.4]{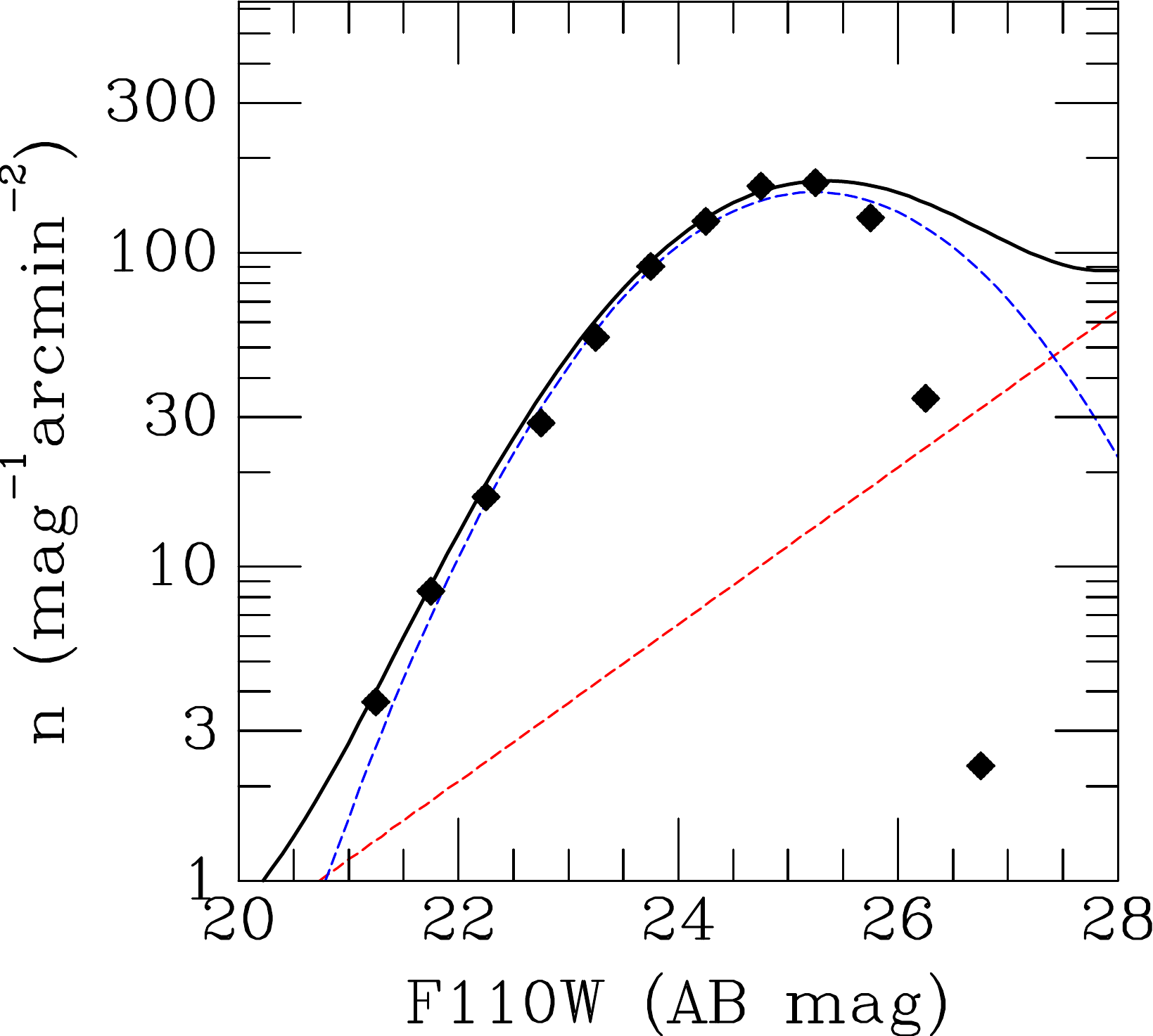}
\caption{Combined figure for NGC~1060.}
\end{center}
\end{figure*}
\clearpage

\begin{figure*}
\begin{center}
\includegraphics[scale=0.2]{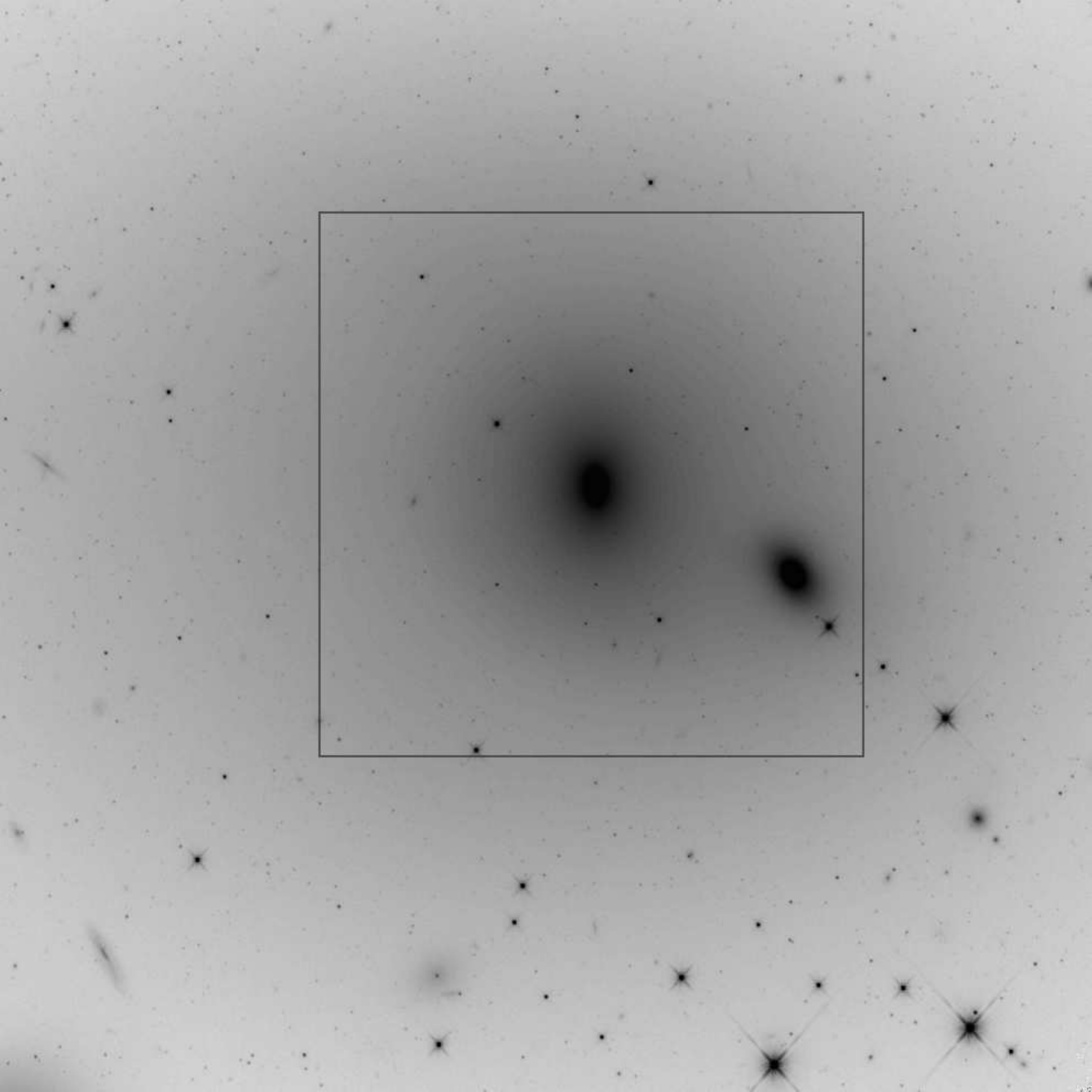}
\includegraphics[scale=0.4]{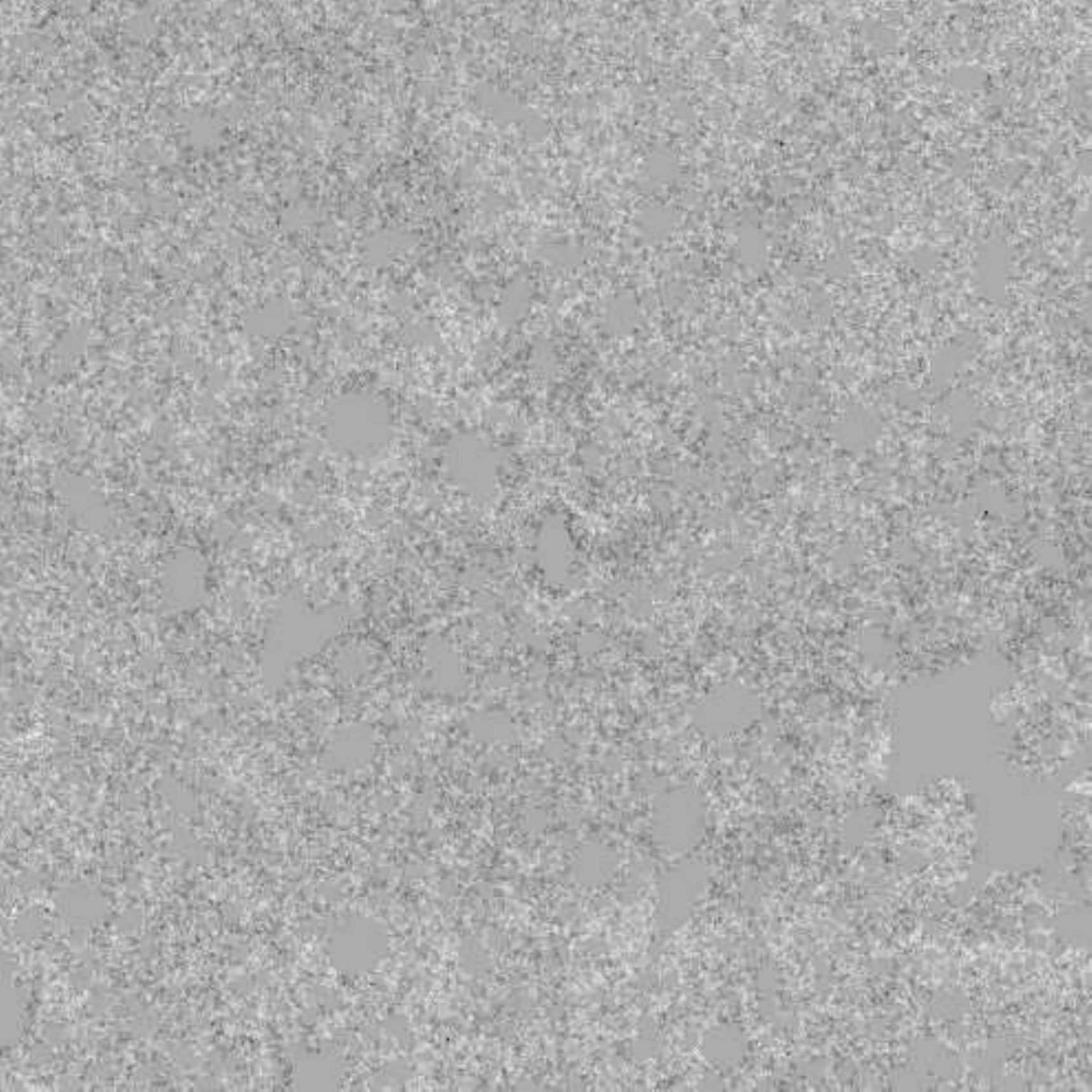} \\
\vspace{10pt}
\includegraphics[scale=0.4]{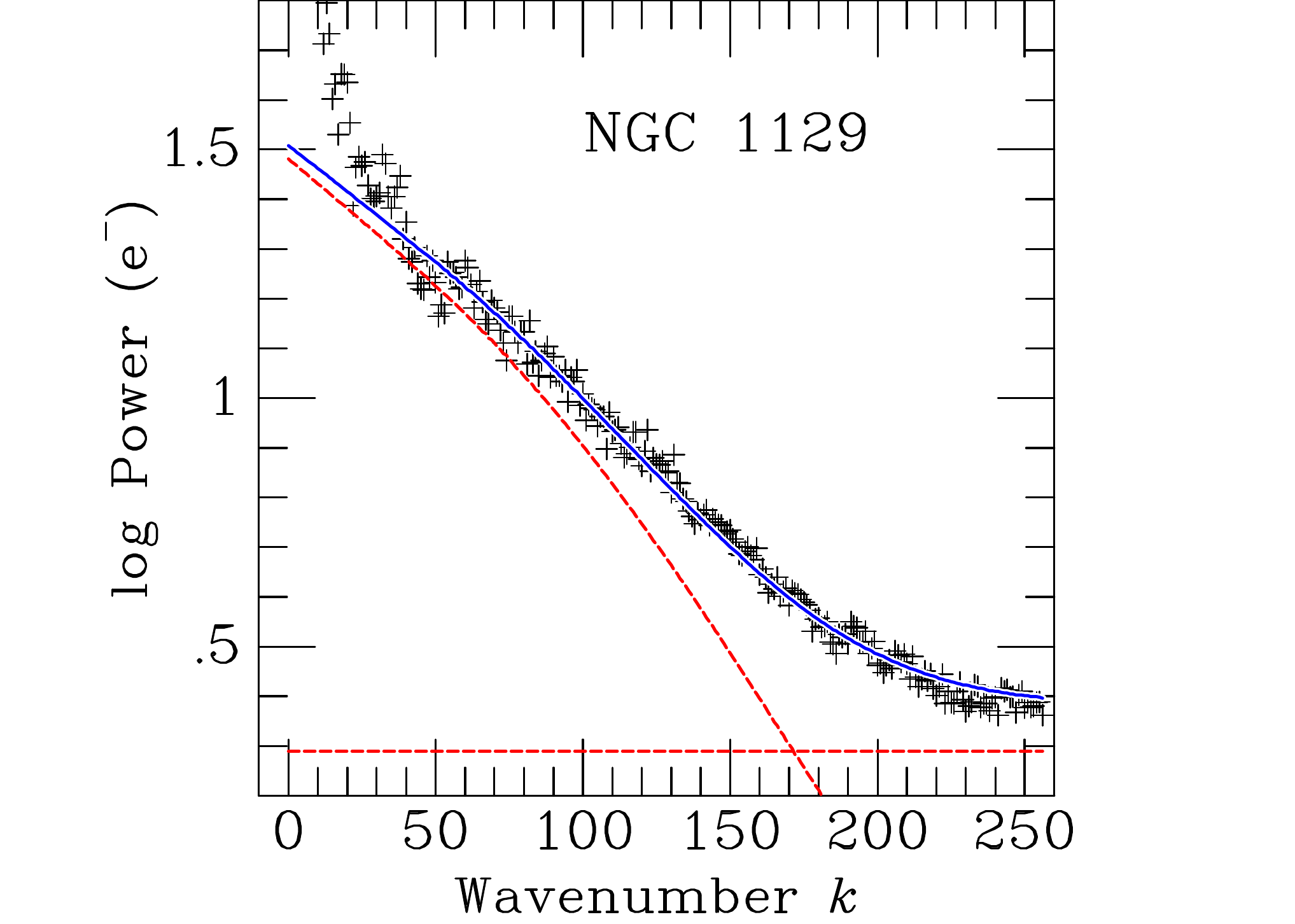}
\hspace{-25pt}
\includegraphics[scale=0.4]{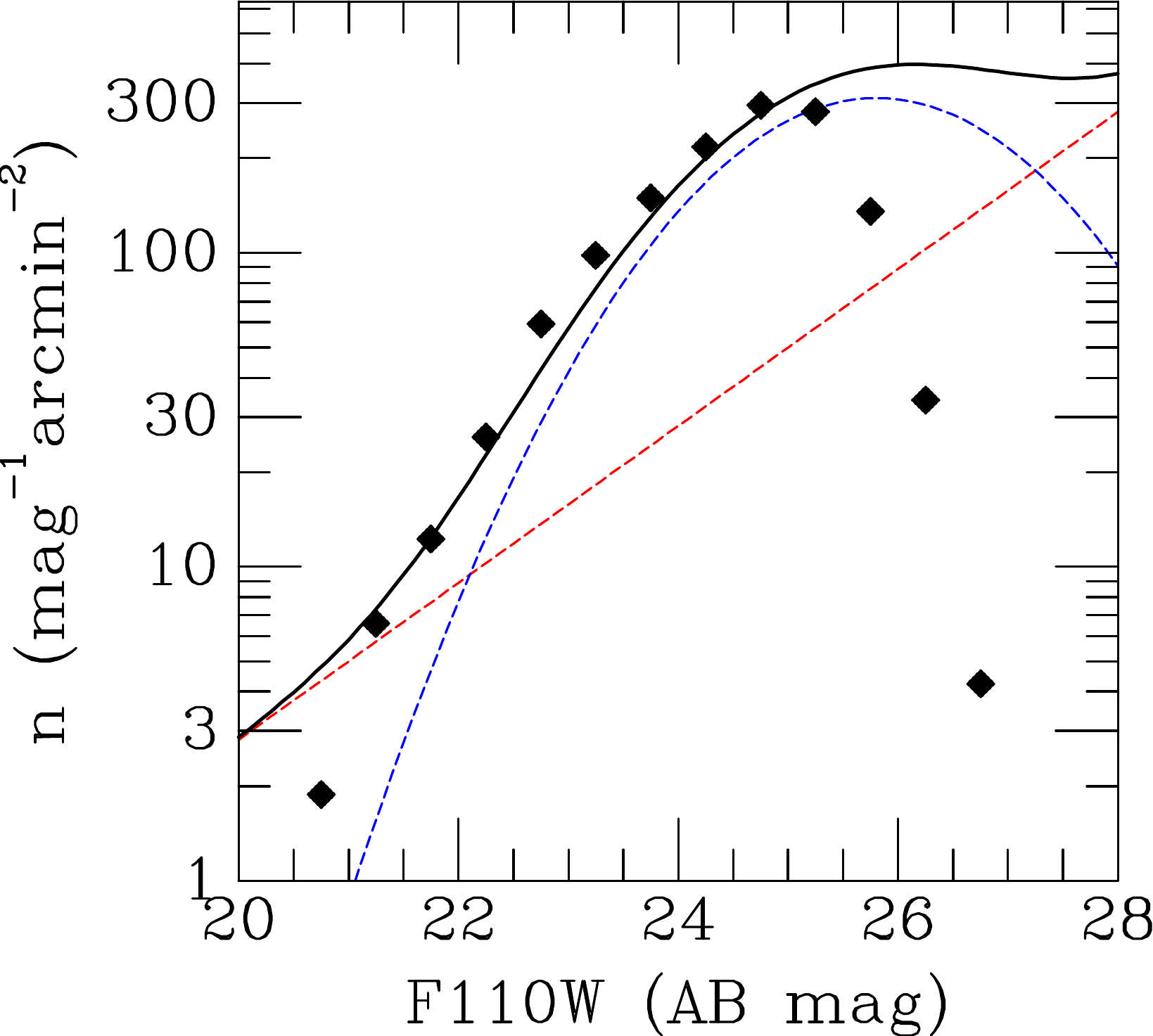}
\caption{Combined figure for NGC~1129.}
\end{center}
\end{figure*}
\clearpage

\begin{figure*}
\begin{center}
\includegraphics[scale=0.2]{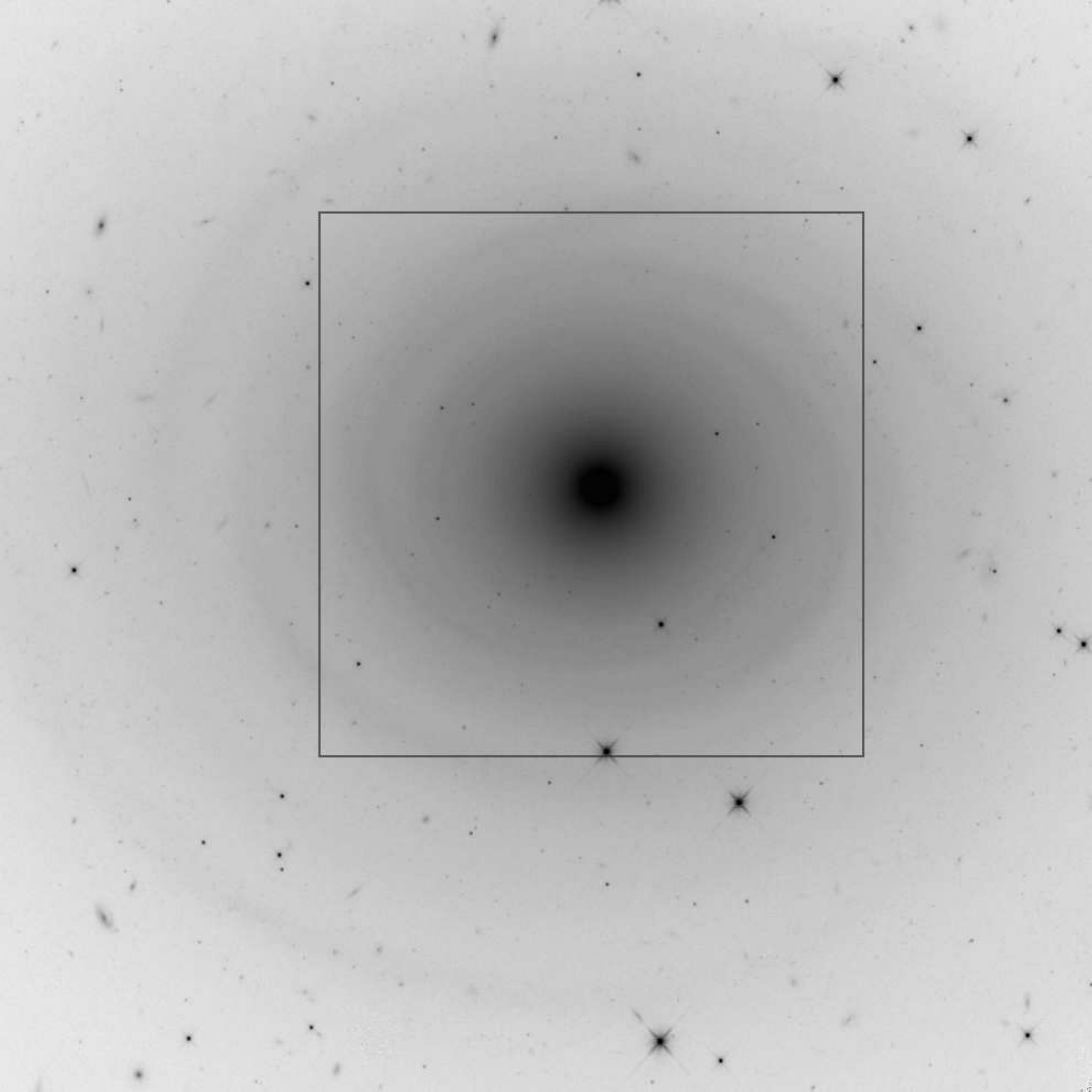}
\includegraphics[scale=0.4]{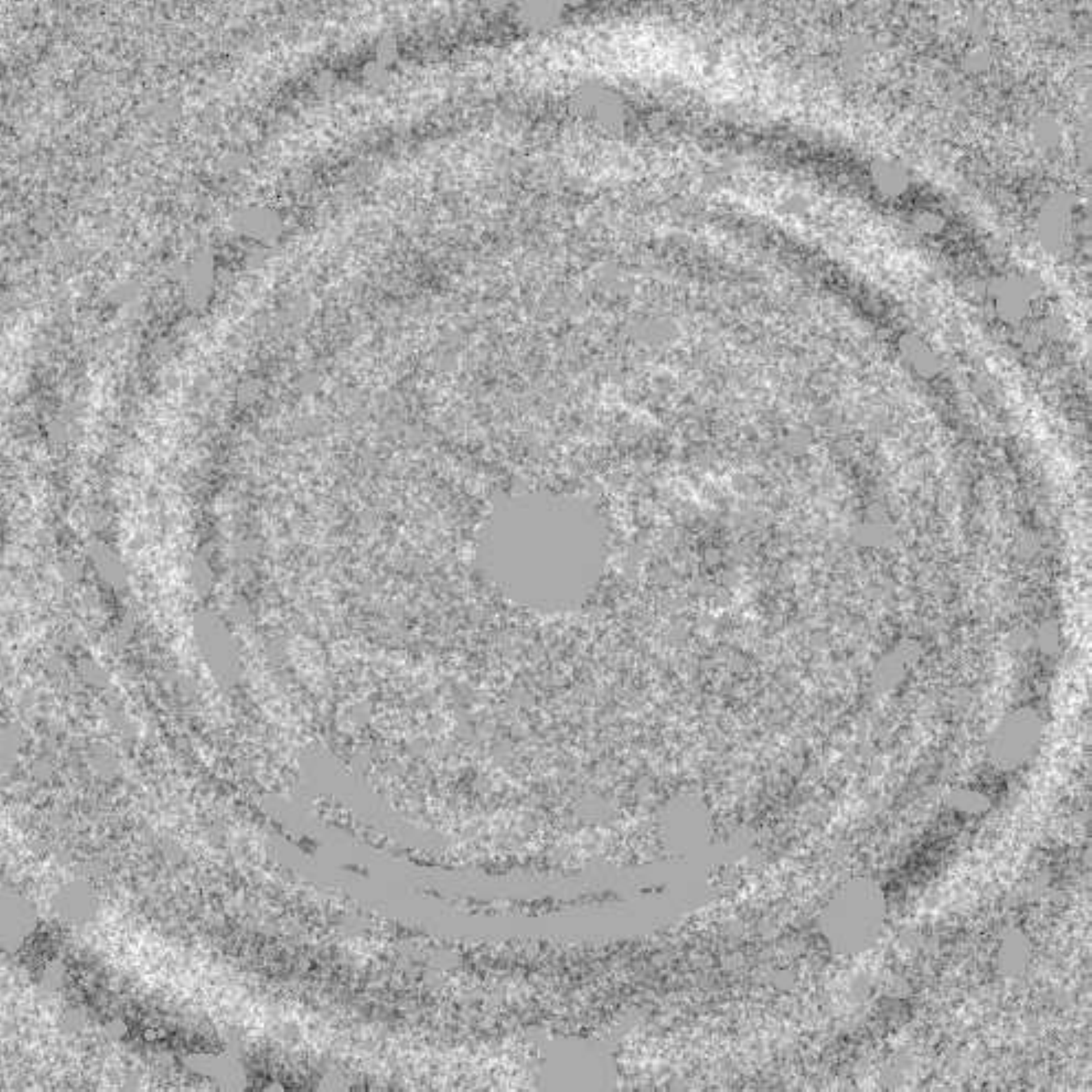} \\
\vspace{10pt}
\includegraphics[scale=0.4]{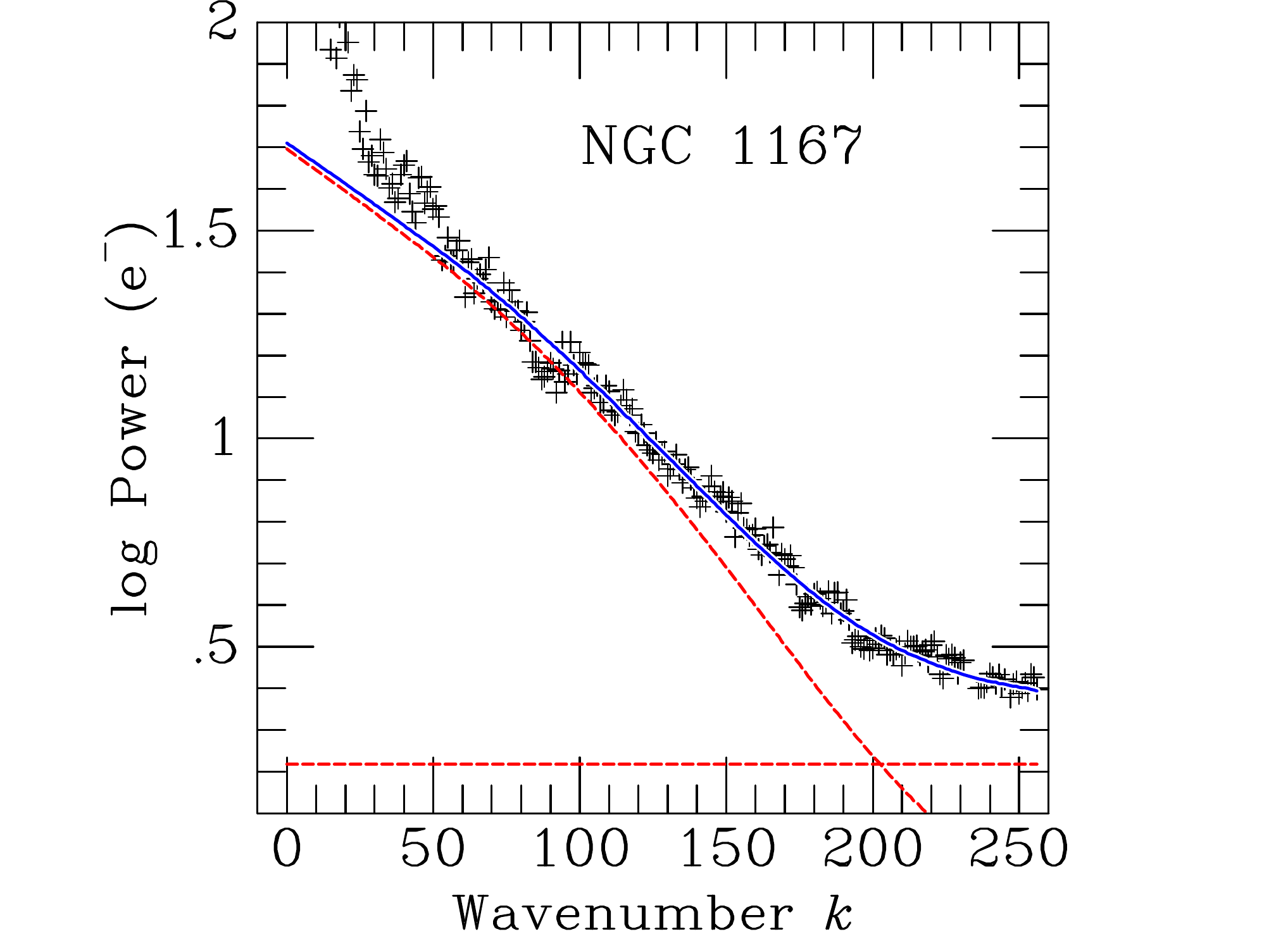}
\hspace{-25pt}
\includegraphics[scale=0.4]{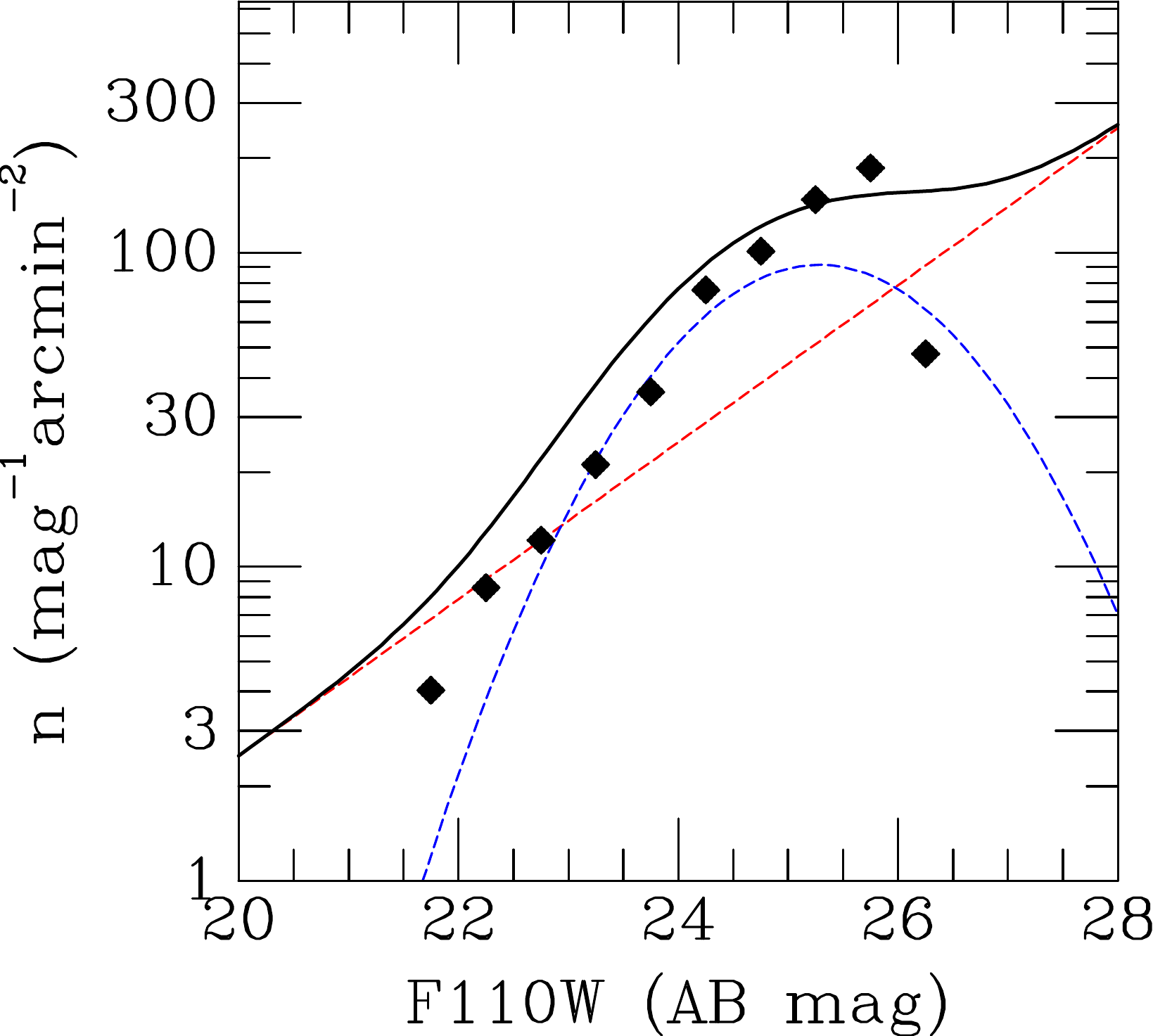}
\caption{Combined figure for NGC~1167.}
\end{center}
\end{figure*}
\clearpage

\begin{figure*}
\begin{center}
\includegraphics[scale=0.2]{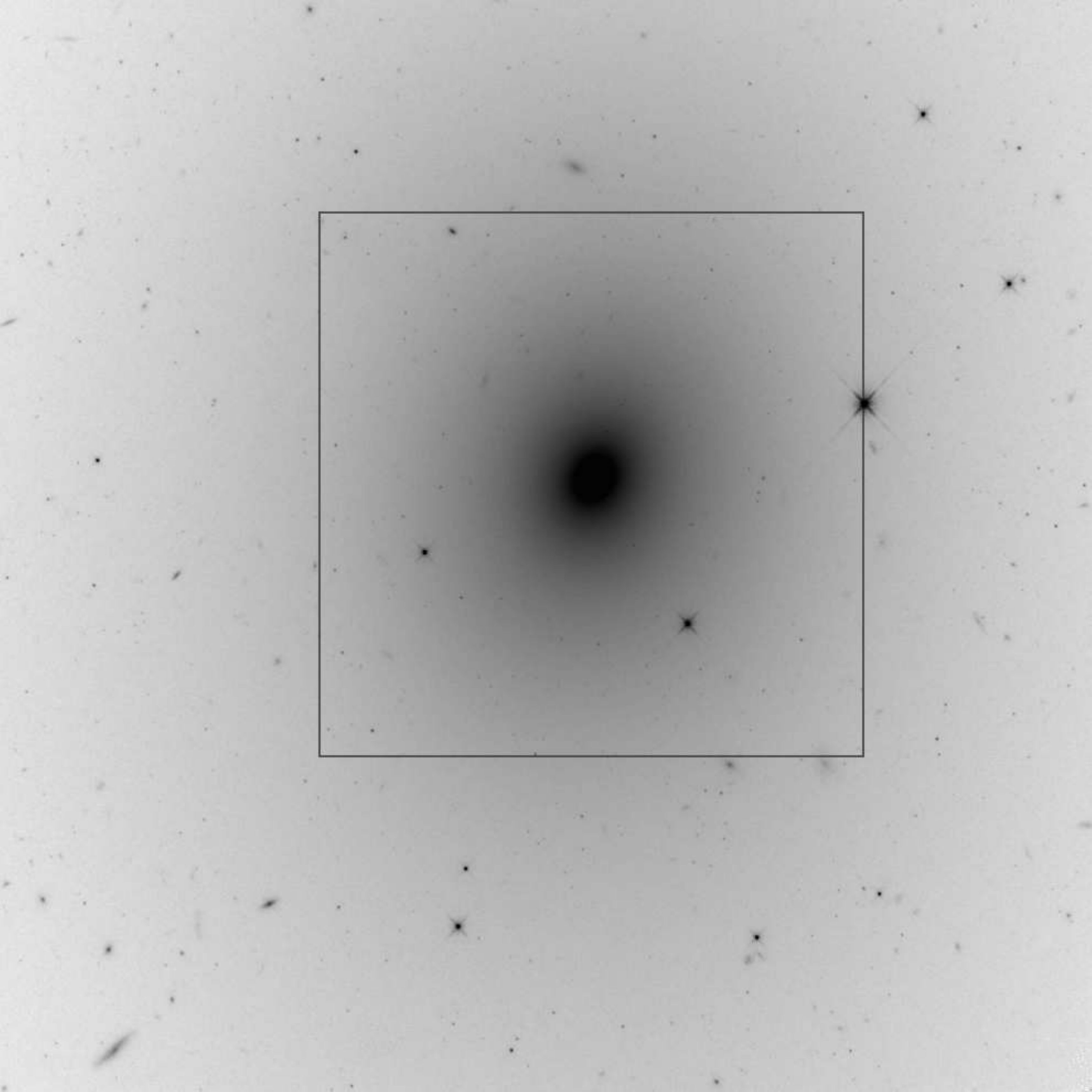}
\includegraphics[scale=0.4]{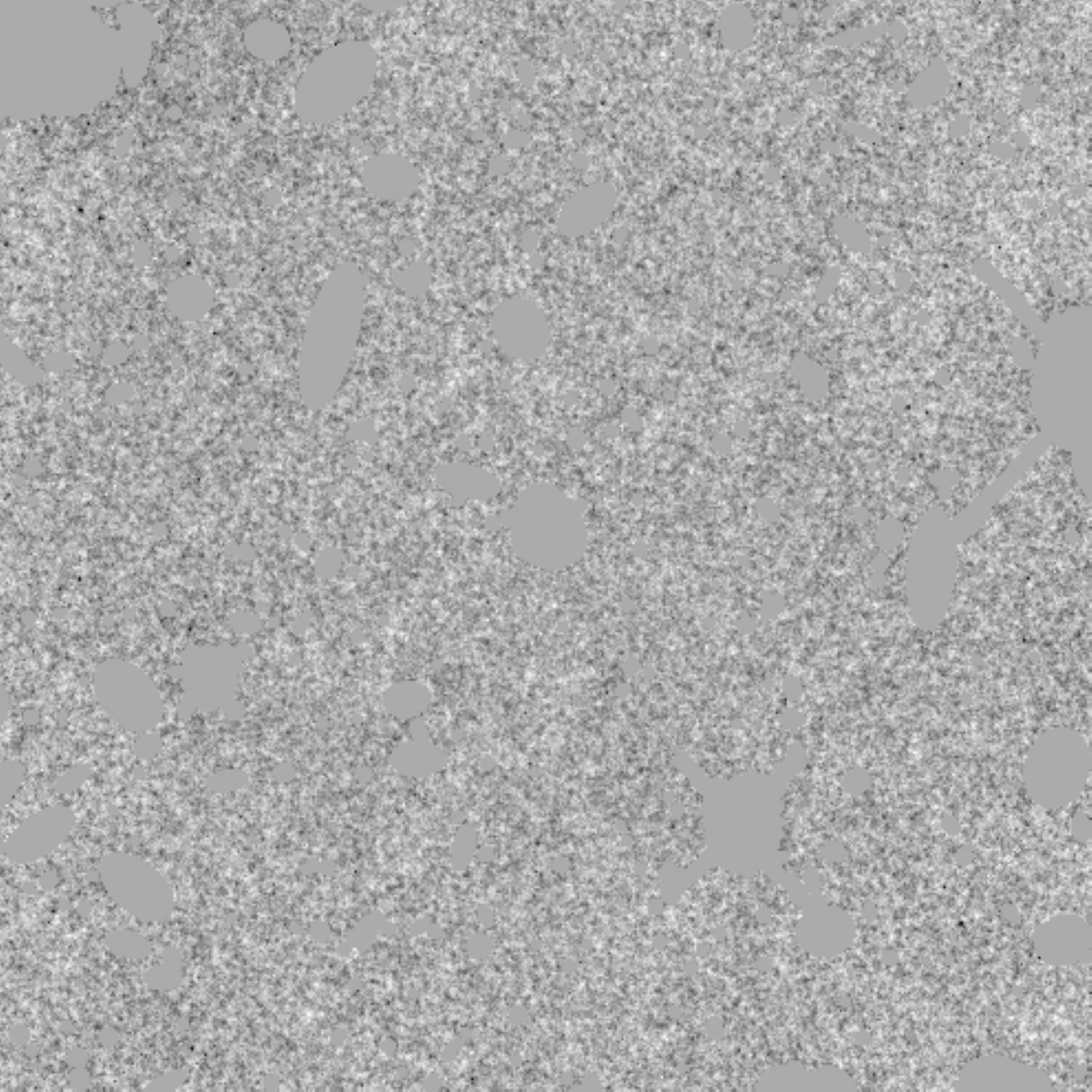} \\
\vspace{10pt}
\includegraphics[scale=0.4]{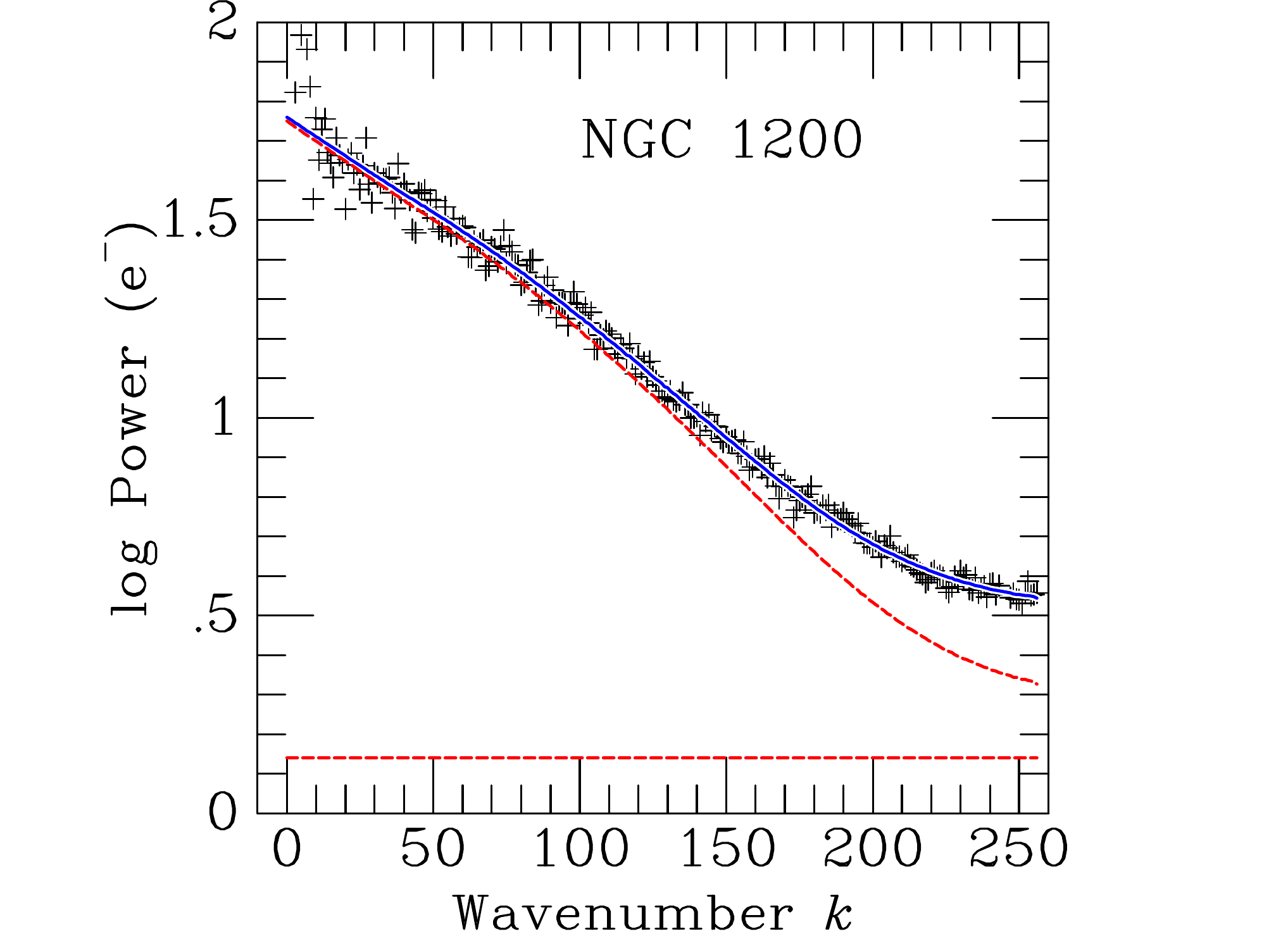}
\hspace{-25pt}
\includegraphics[scale=0.4]{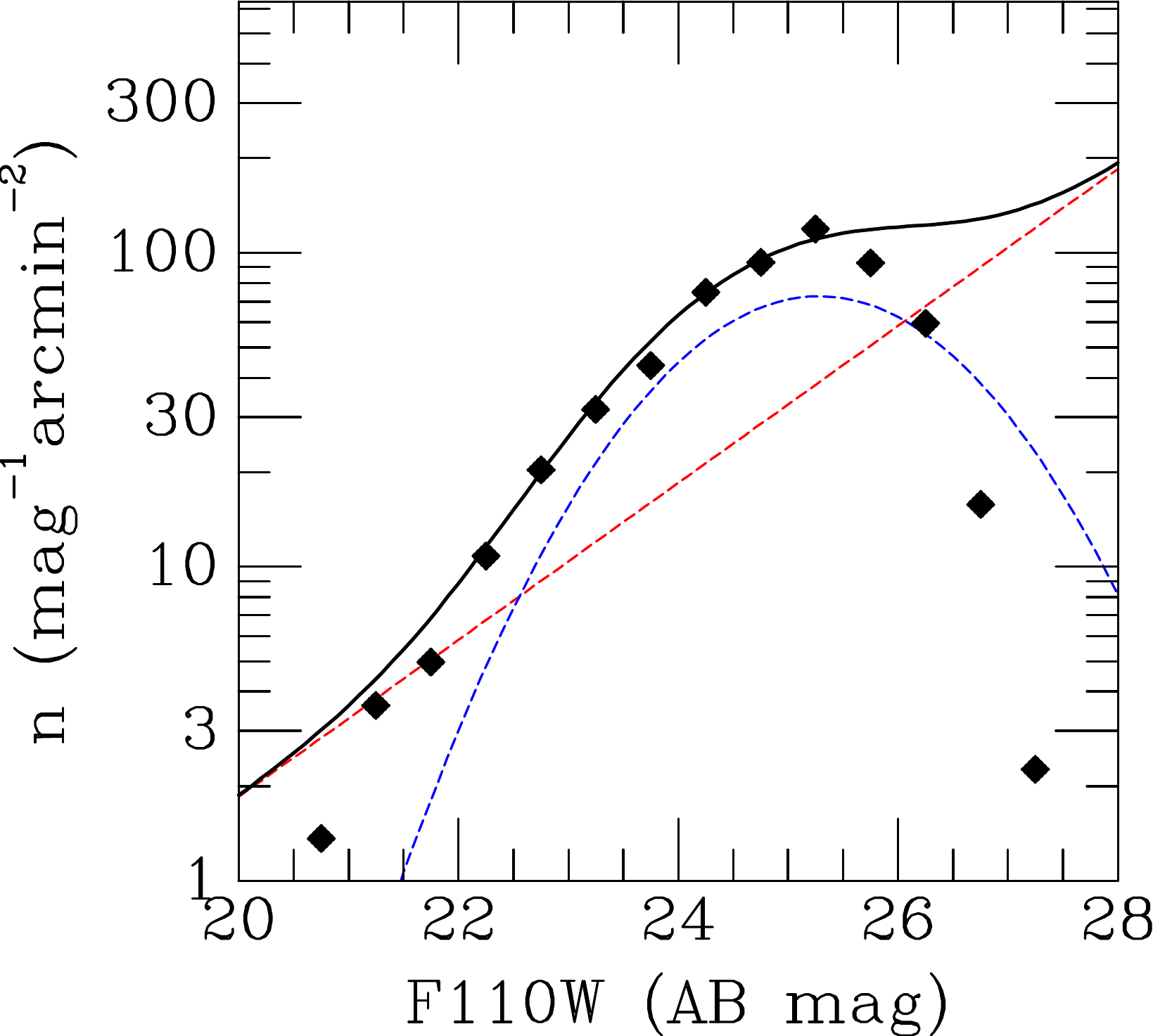}
\caption{Combined figure for NGC~1200.}
\end{center}
\end{figure*}
\clearpage

\begin{figure*}
\begin{center}
\includegraphics[scale=0.2]{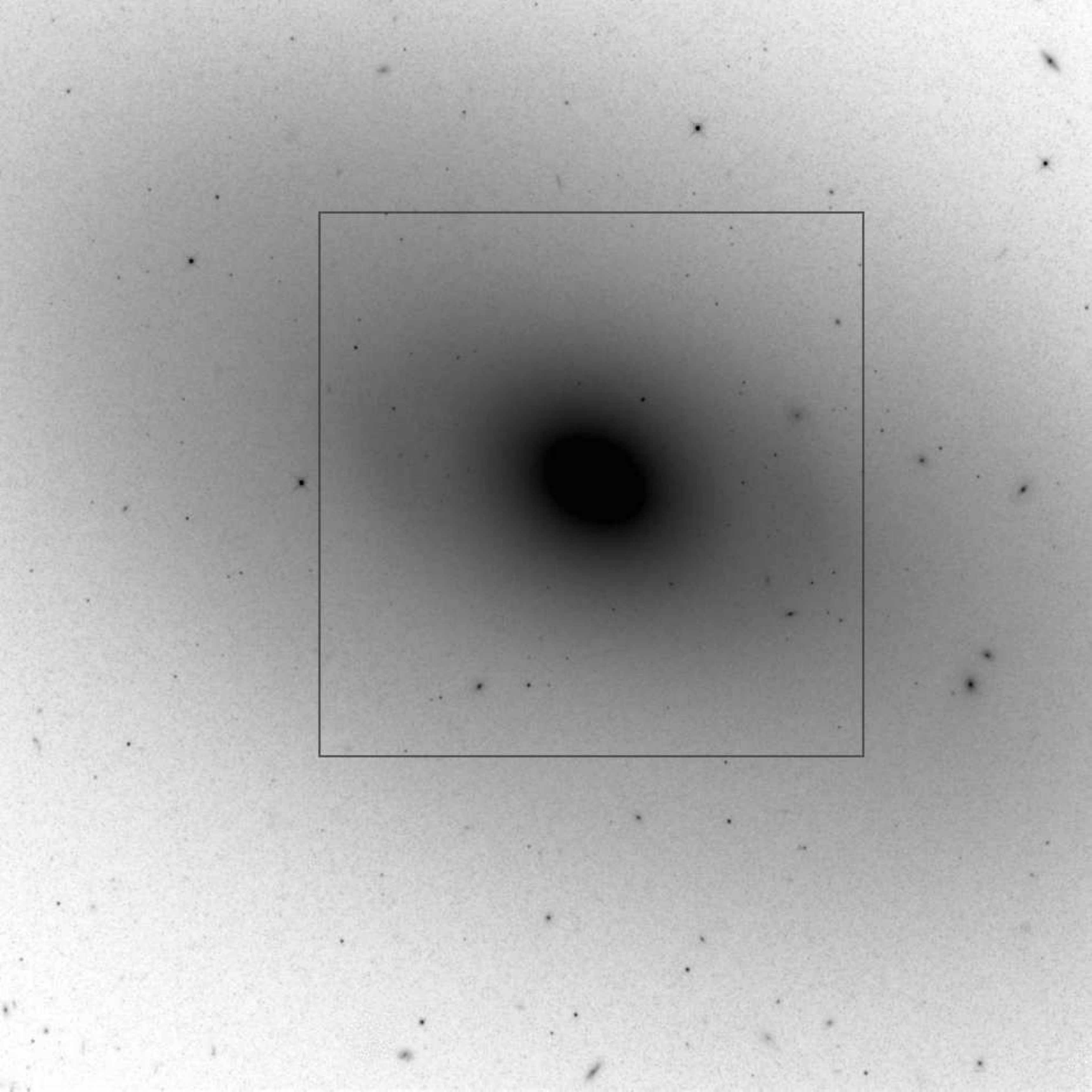}
\includegraphics[scale=0.4]{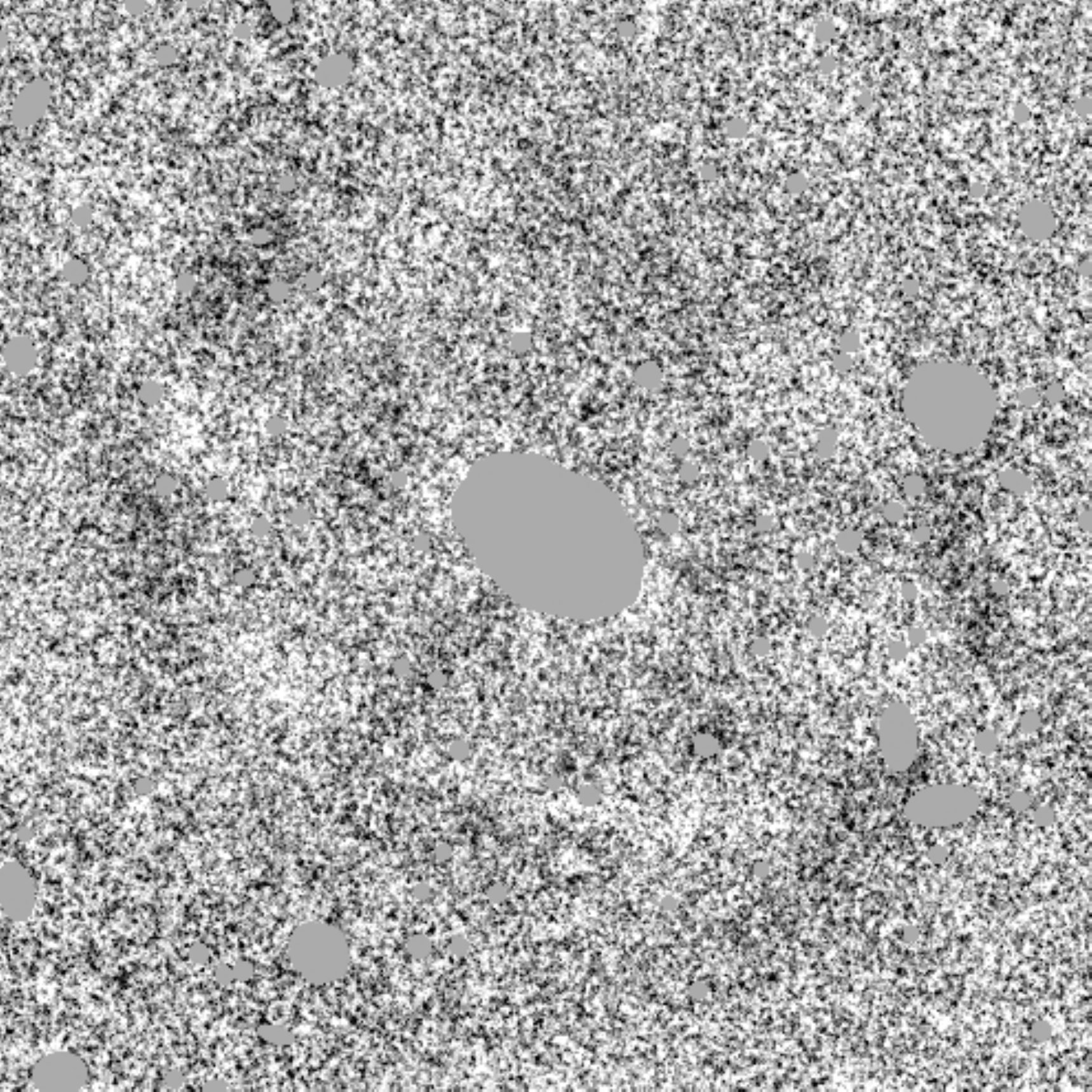} \\
\vspace{10pt}
\includegraphics[scale=0.4]{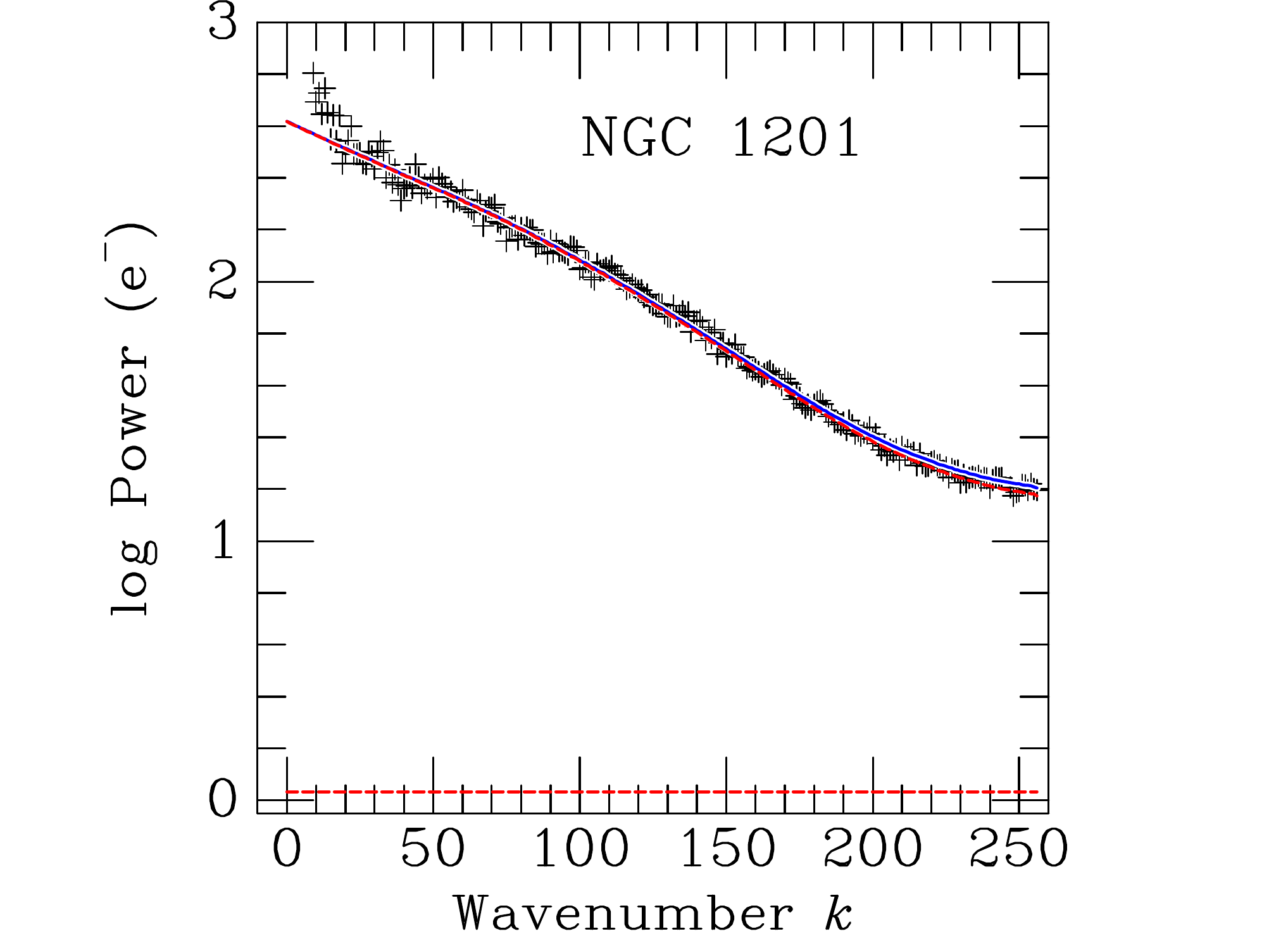}
\hspace{-25pt}
\includegraphics[scale=0.4]{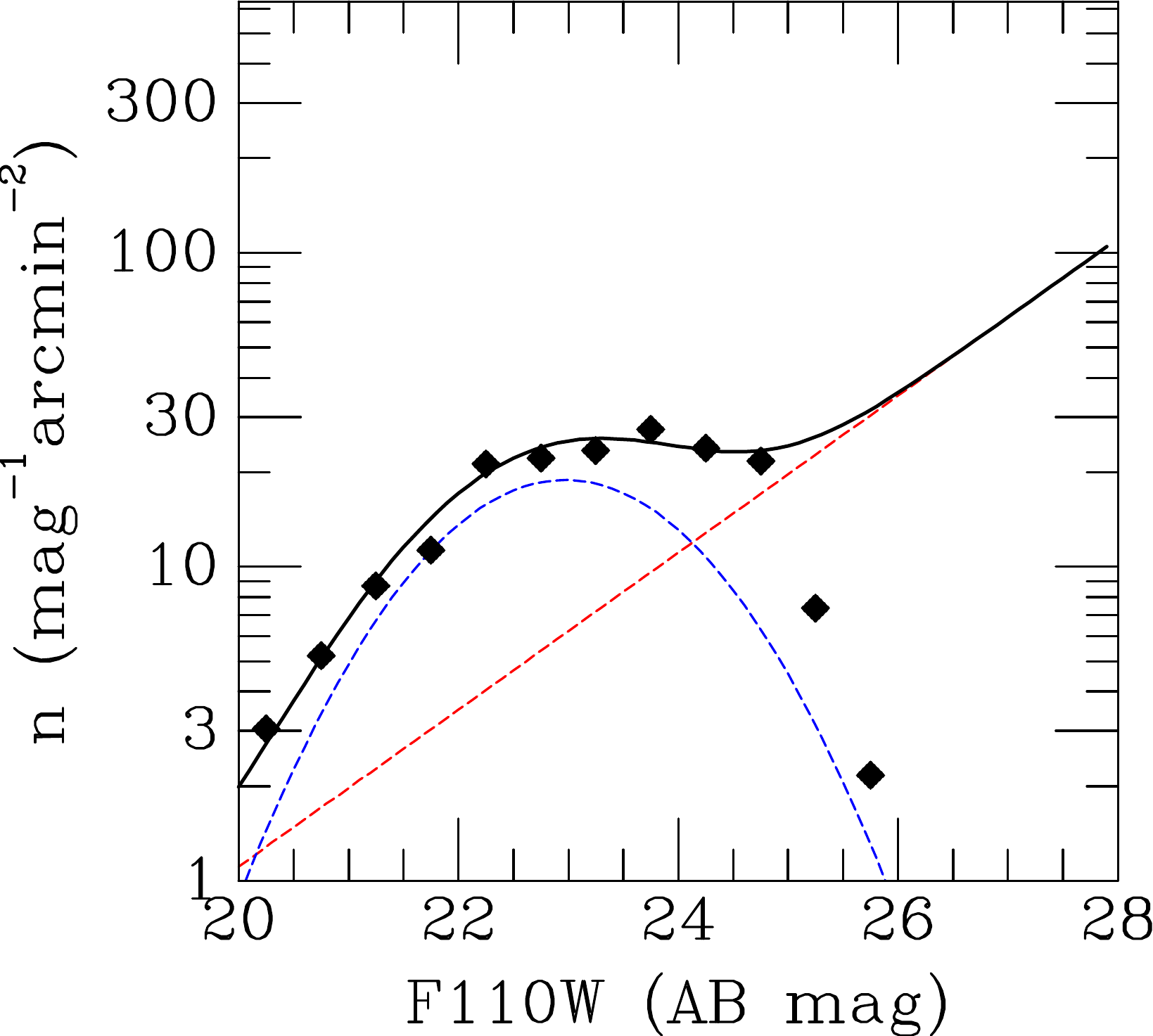}
\caption{Combined figure for NGC~1201.}
\end{center}
\end{figure*}
\clearpage

\begin{figure*}
\begin{center}
\includegraphics[scale=0.2]{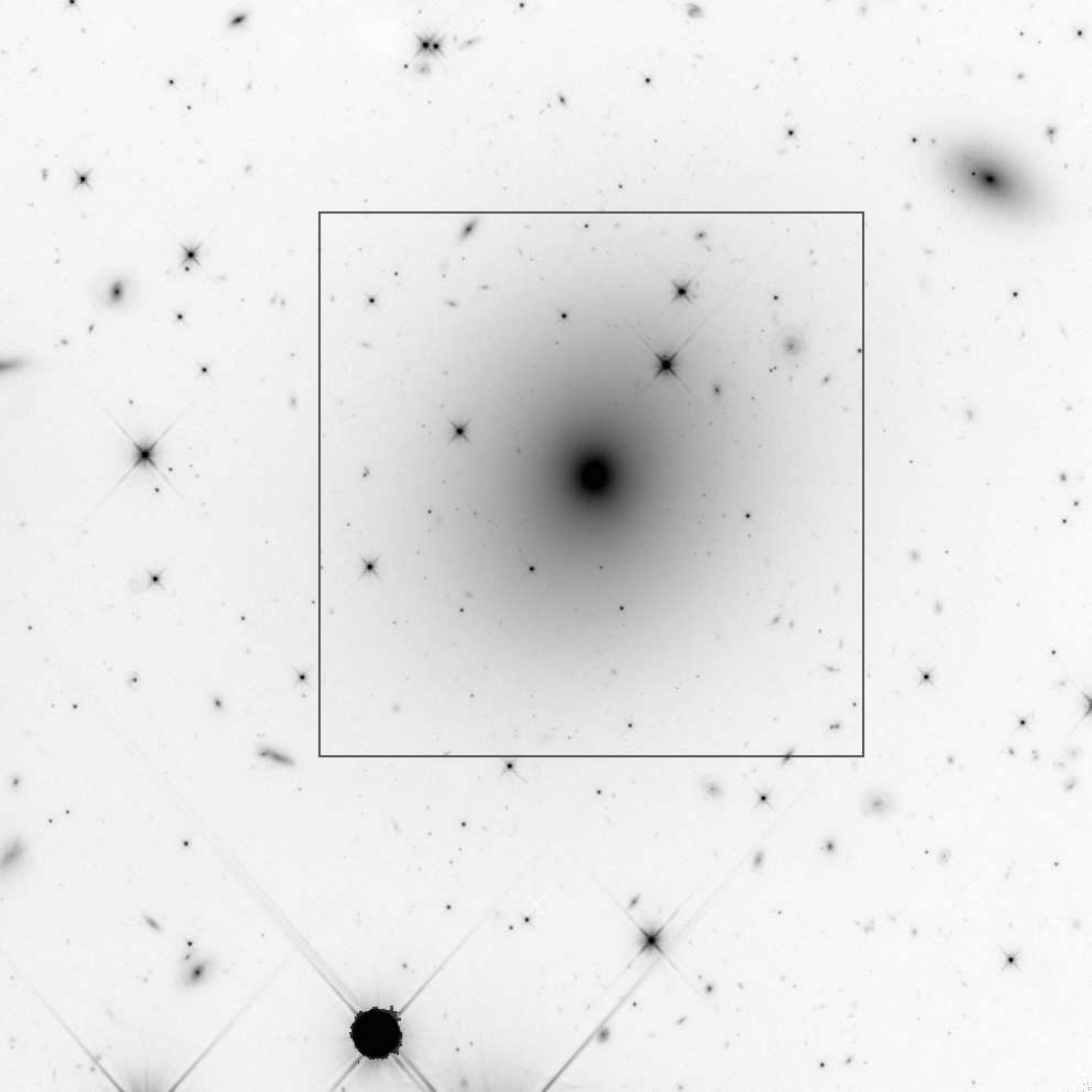}
\includegraphics[scale=0.4]{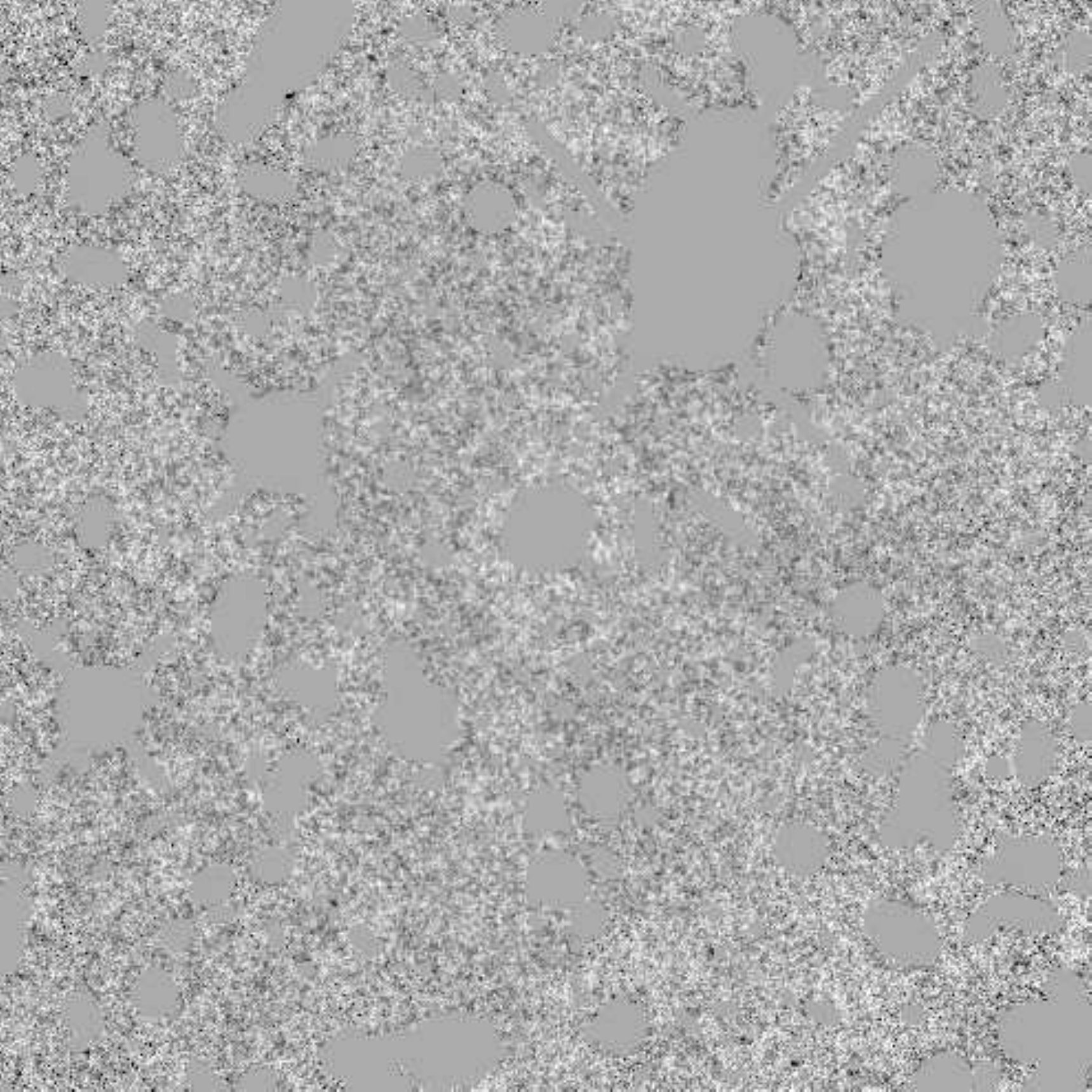} \\
\vspace{10pt}
\includegraphics[scale=0.4]{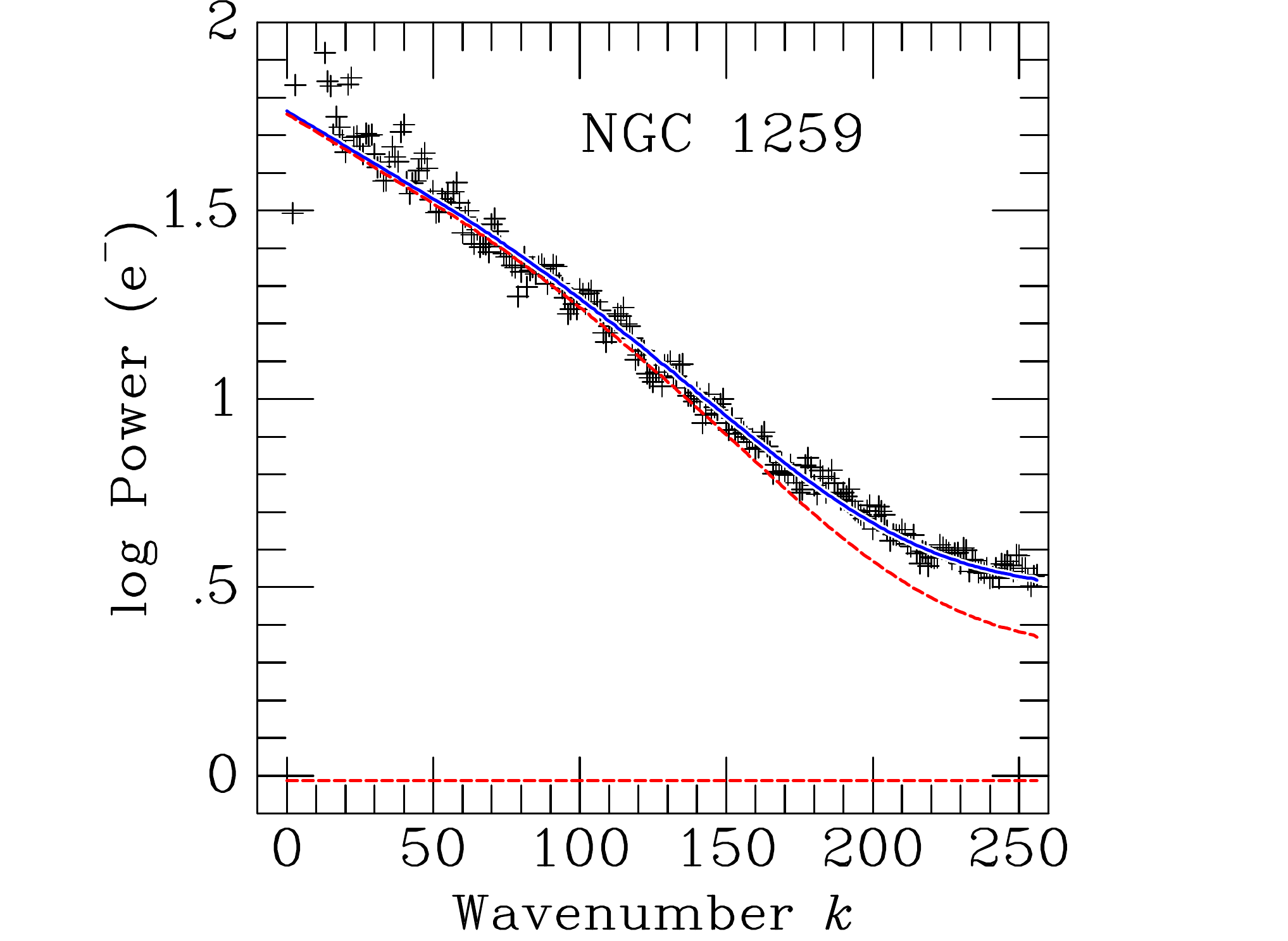}
\hspace{-25pt}
\includegraphics[scale=0.4]{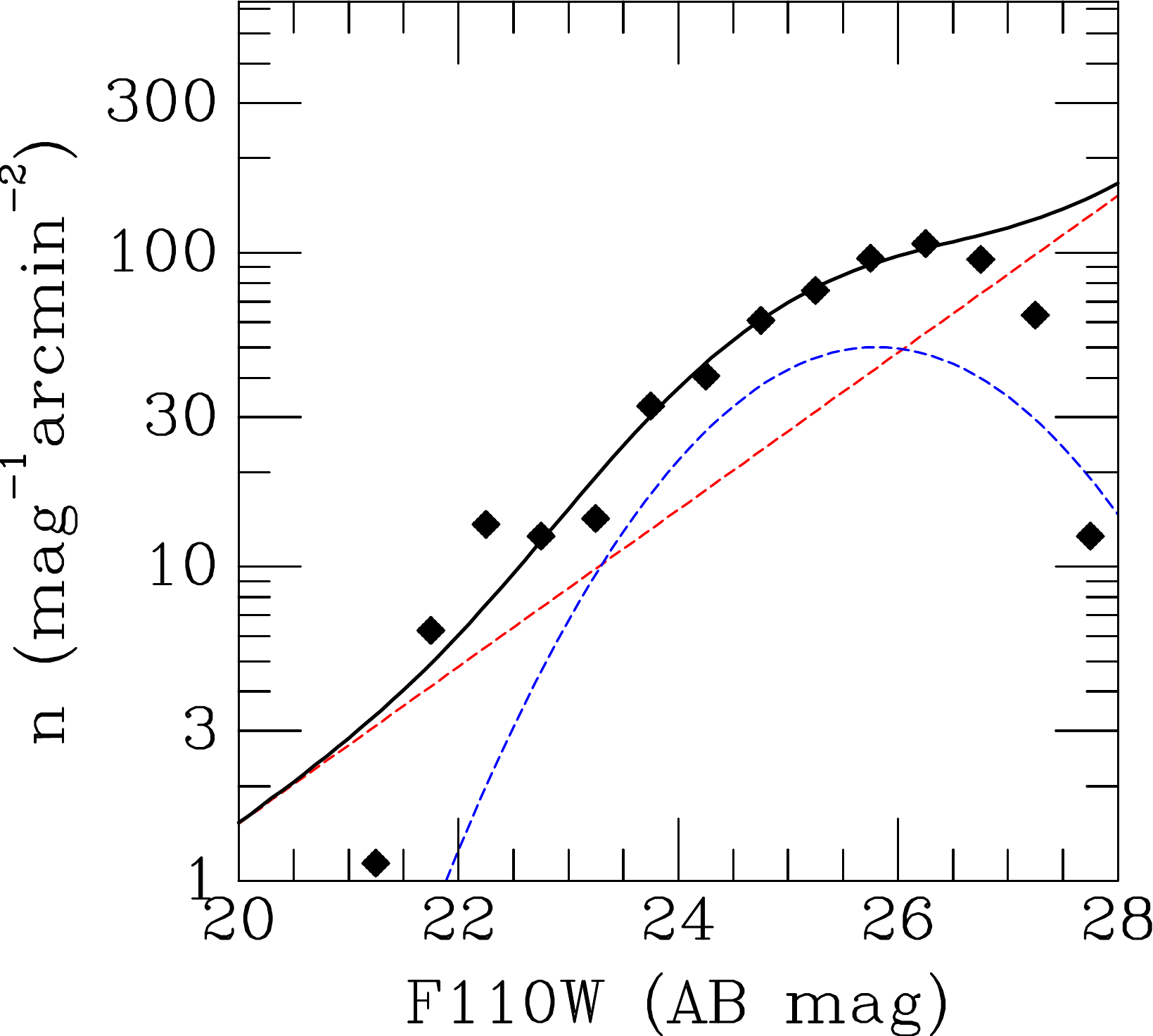}
\caption{Combined figure for NGC~1259.}
\end{center}
\end{figure*}
\clearpage

\begin{figure*}
\begin{center}
\includegraphics[scale=0.2]{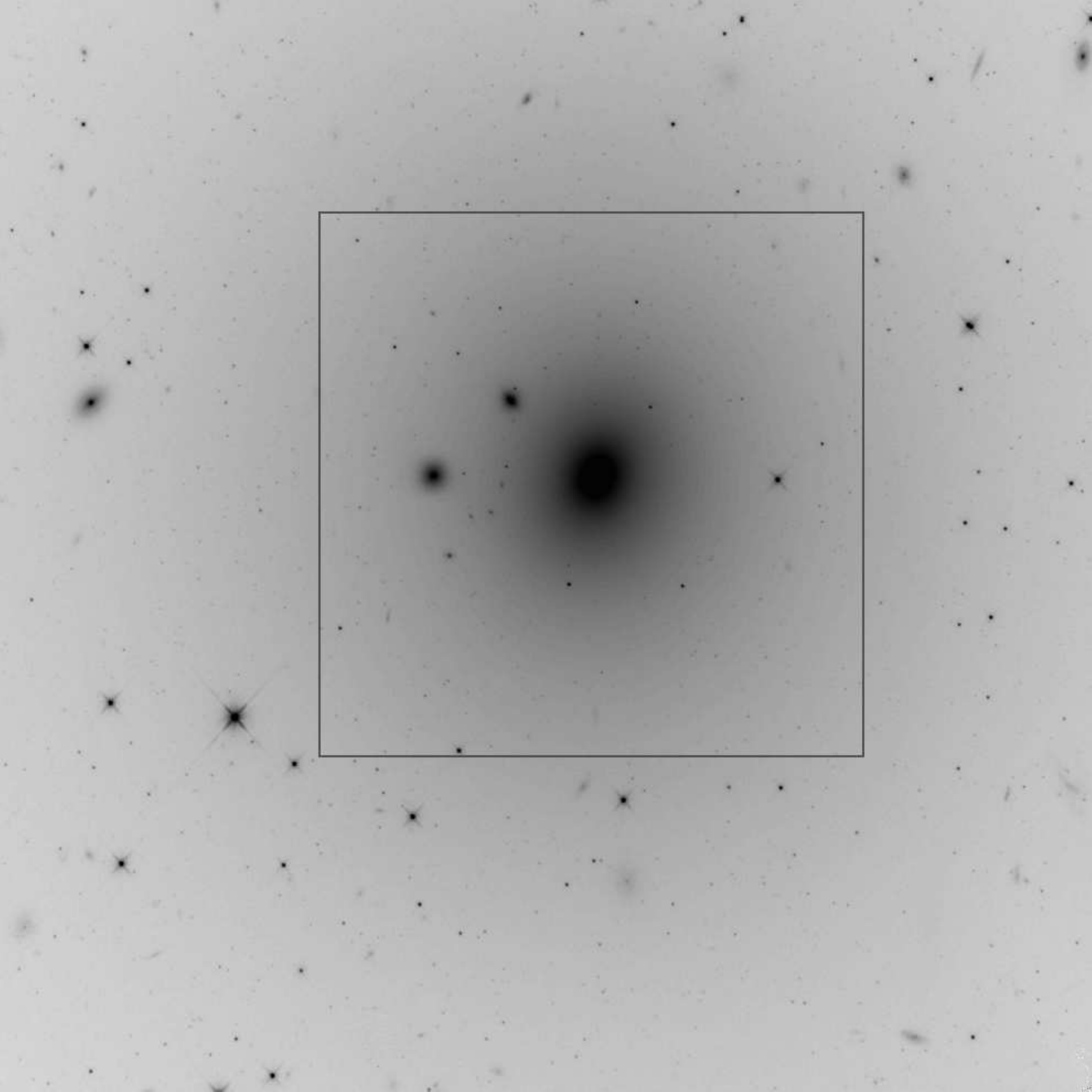}
\includegraphics[scale=0.4]{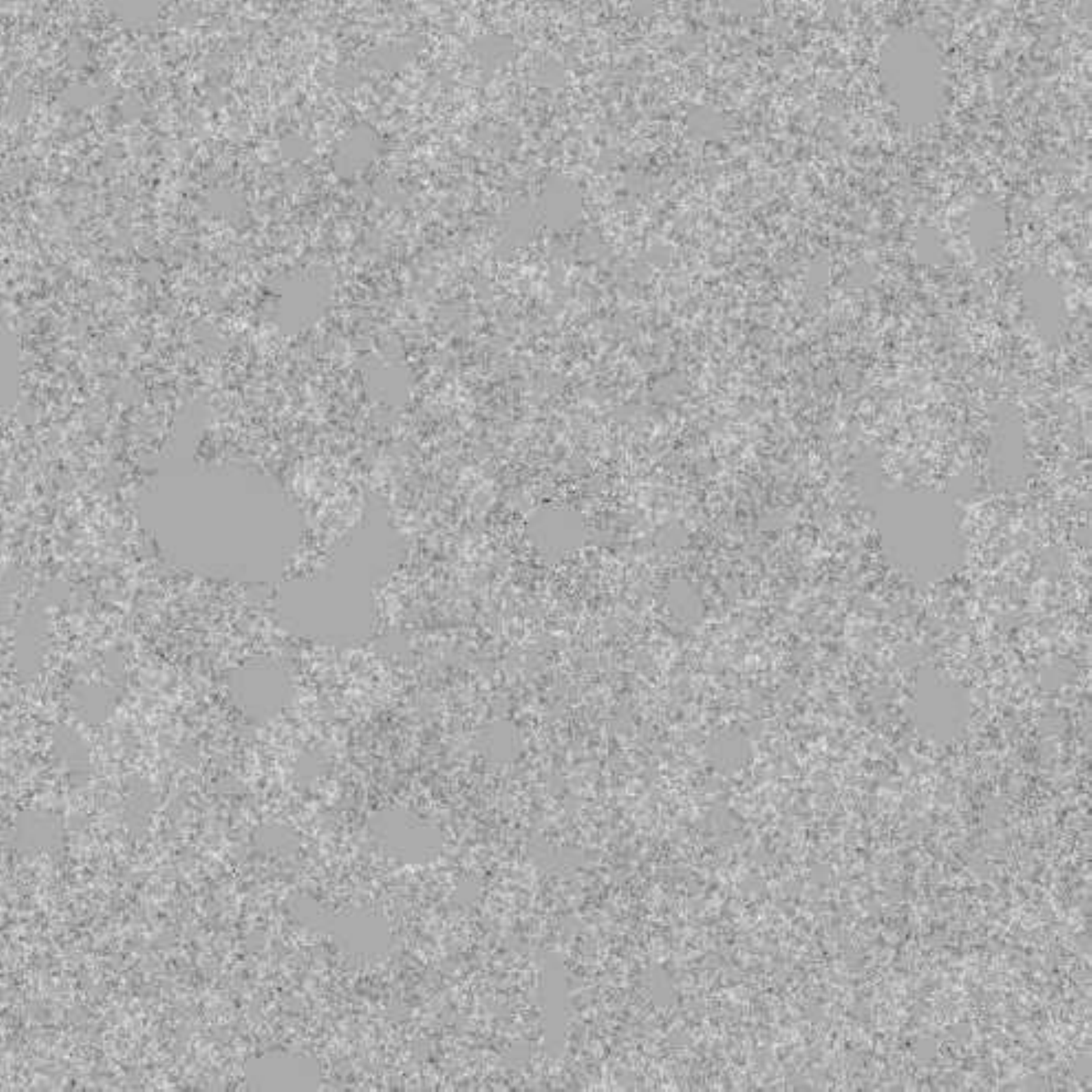} \\
\vspace{10pt}
\includegraphics[scale=0.4]{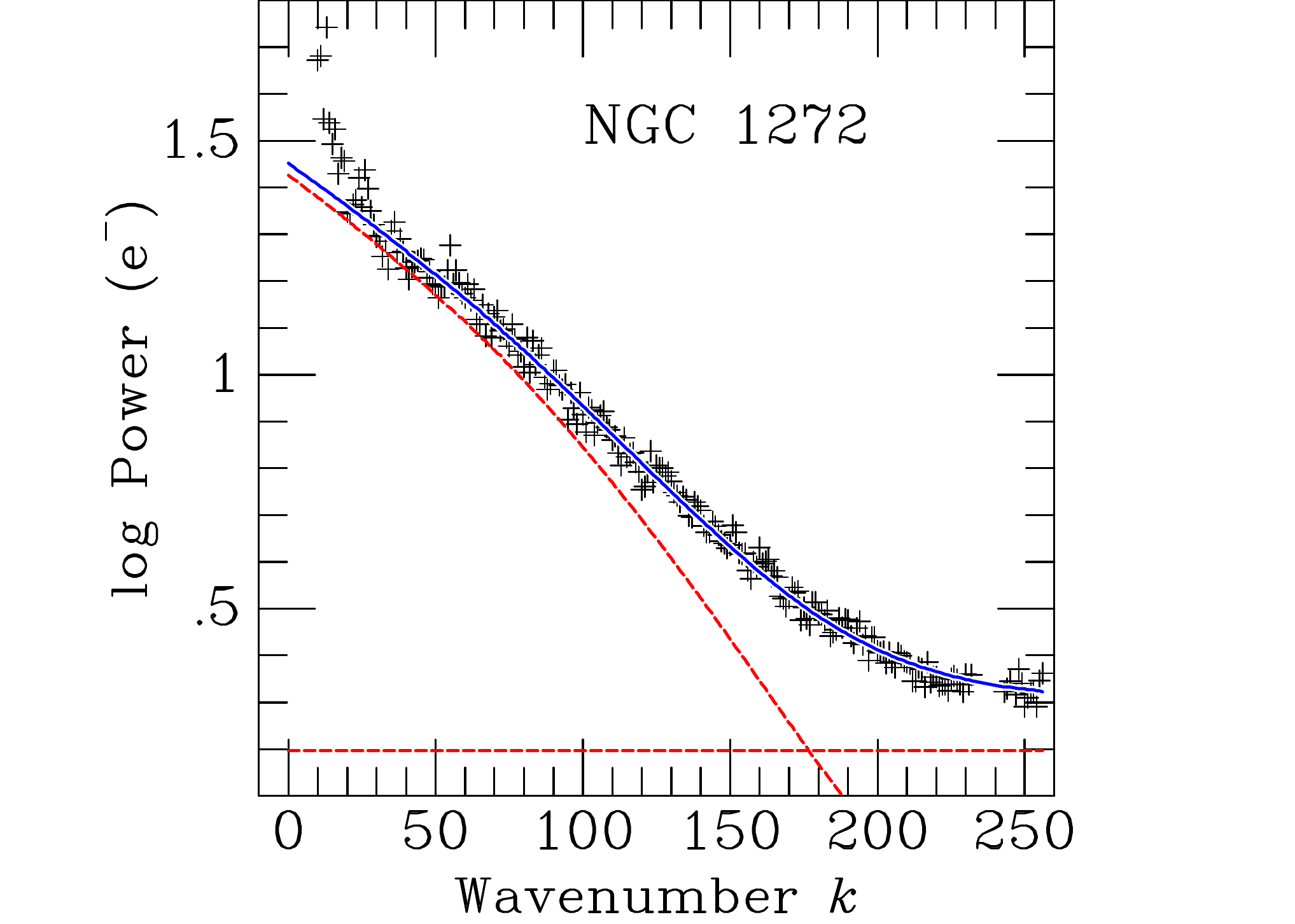}
\hspace{-25pt}
\includegraphics[scale=0.4]{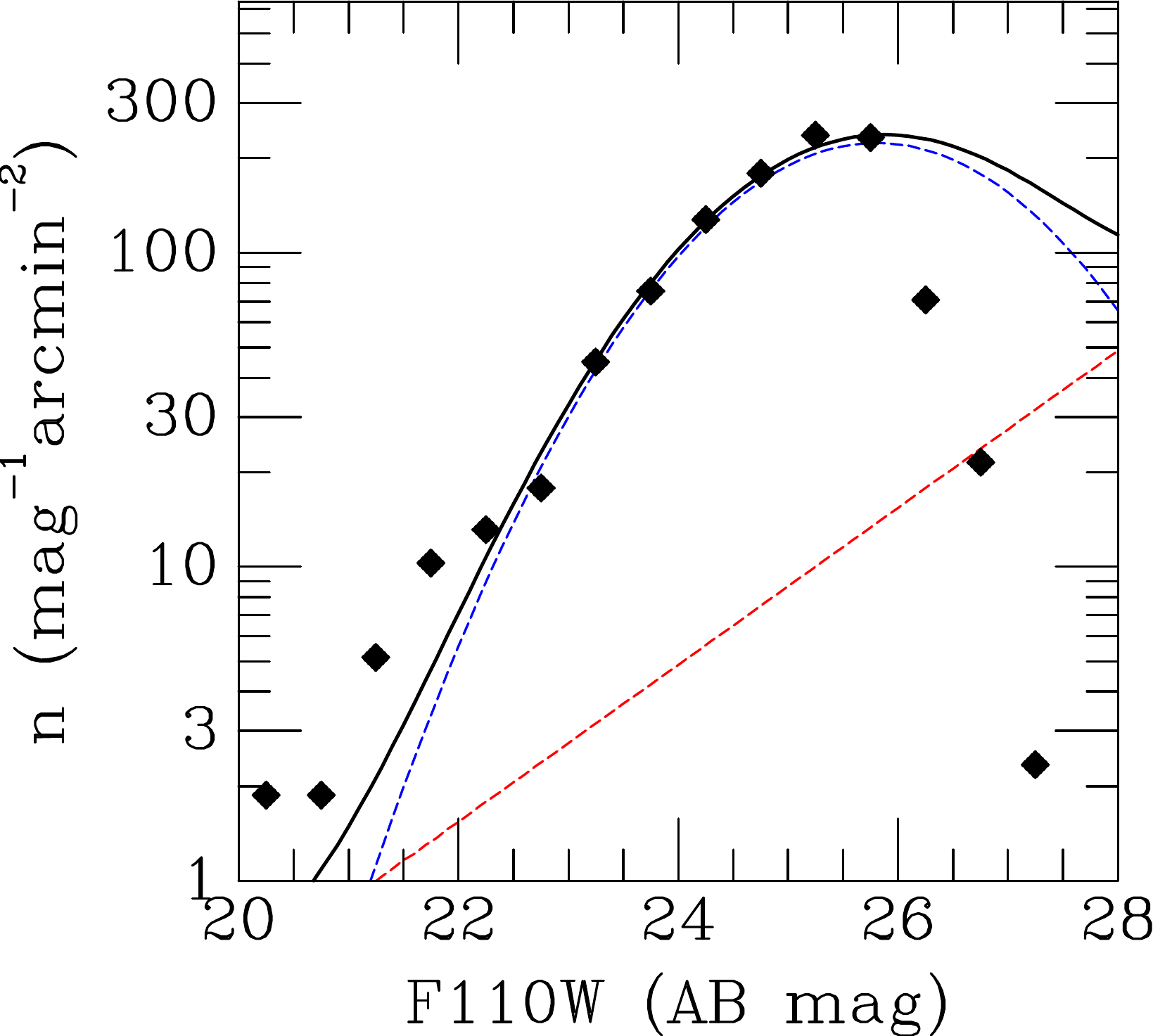}
\caption{Combined figure for NGC~1272.}
\end{center}
\end{figure*}
\clearpage

\begin{figure*}
\begin{center}
\includegraphics[scale=0.2]{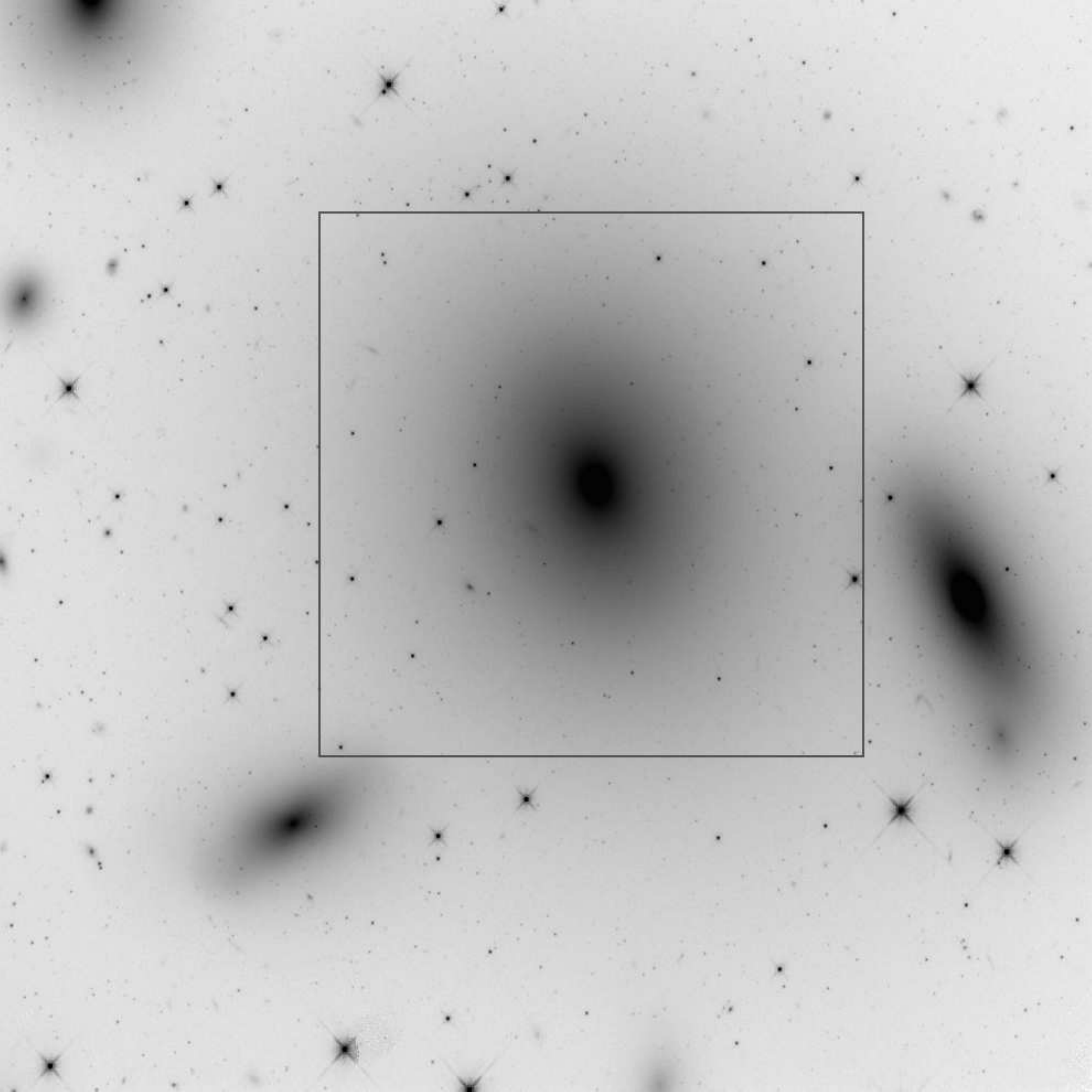}
\includegraphics[scale=0.4]{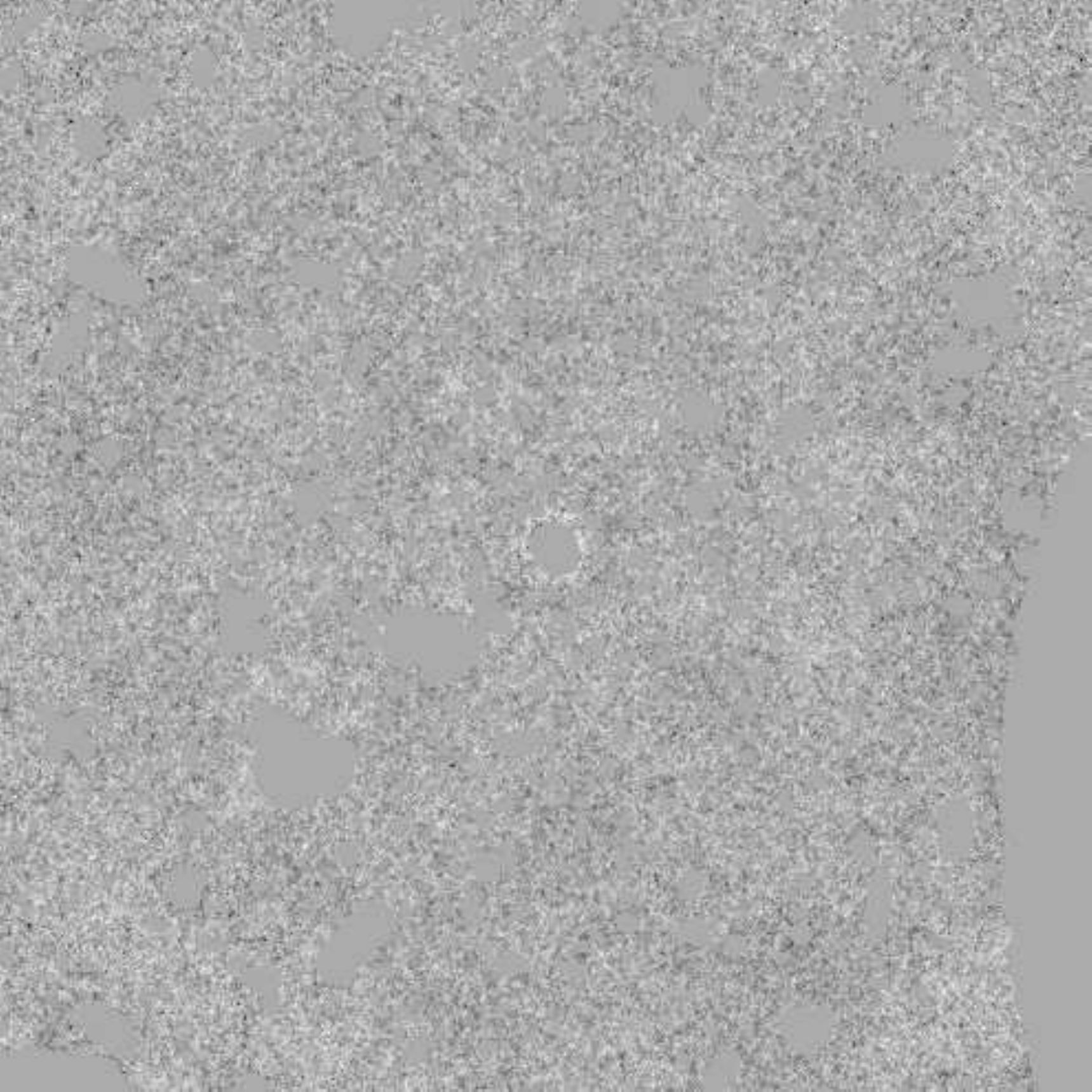} \\
\vspace{10pt}
\includegraphics[scale=0.4]{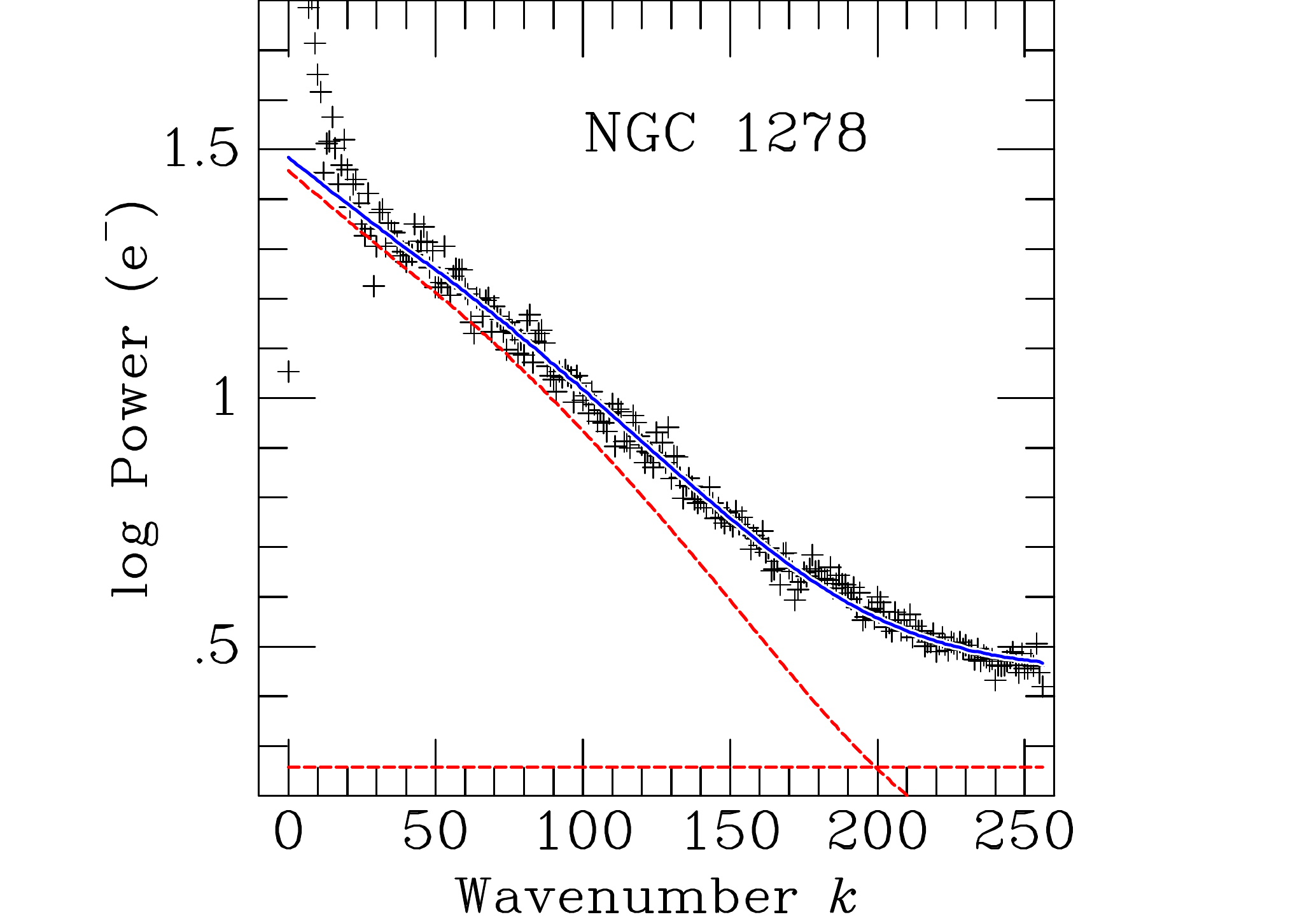}
\hspace{-25pt}
\includegraphics[scale=0.4]{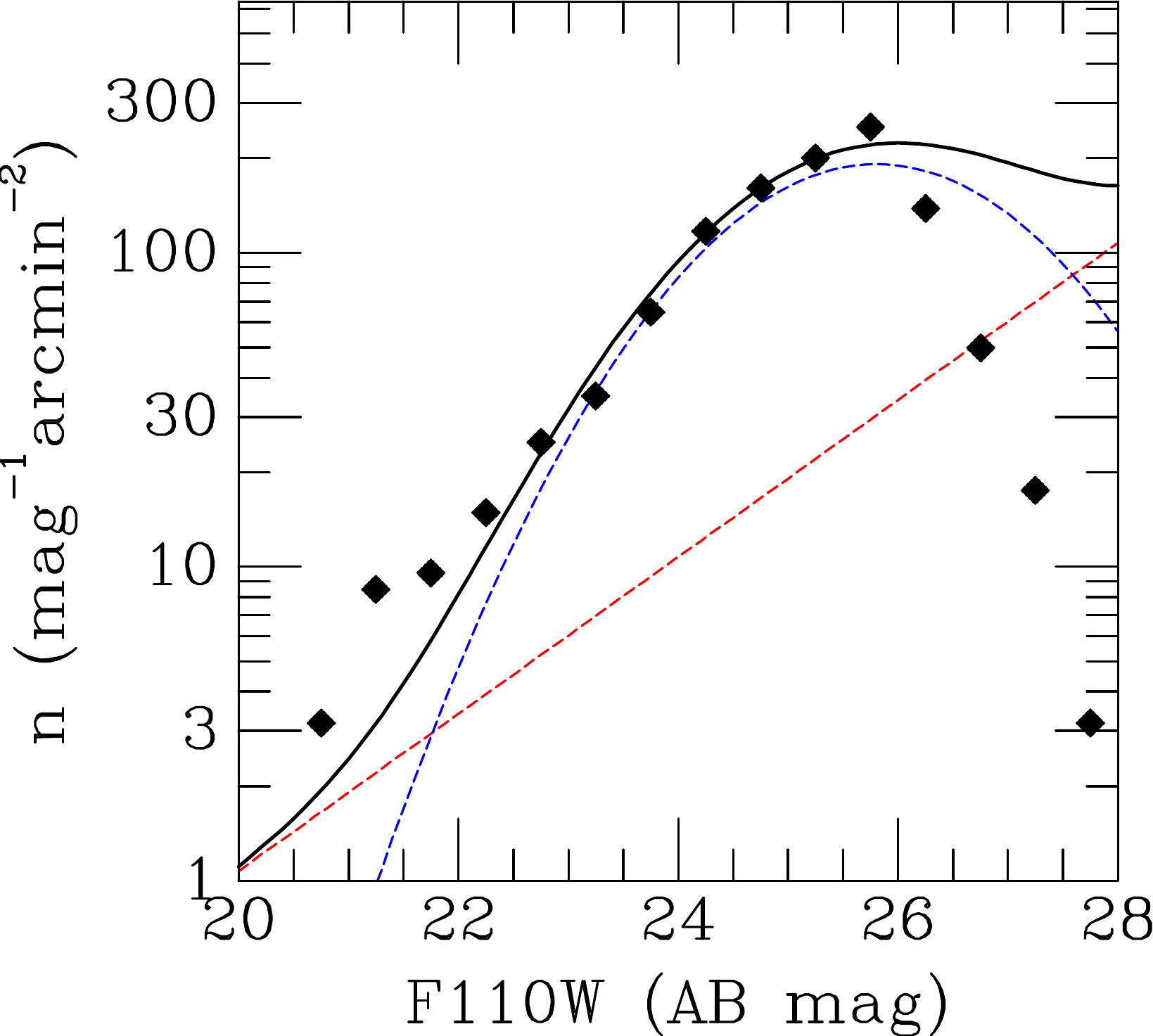}
\caption{Combined figure for NGC~1278.}
\end{center}
\end{figure*}
\clearpage

\begin{figure*}
\begin{center}
\includegraphics[scale=0.2]{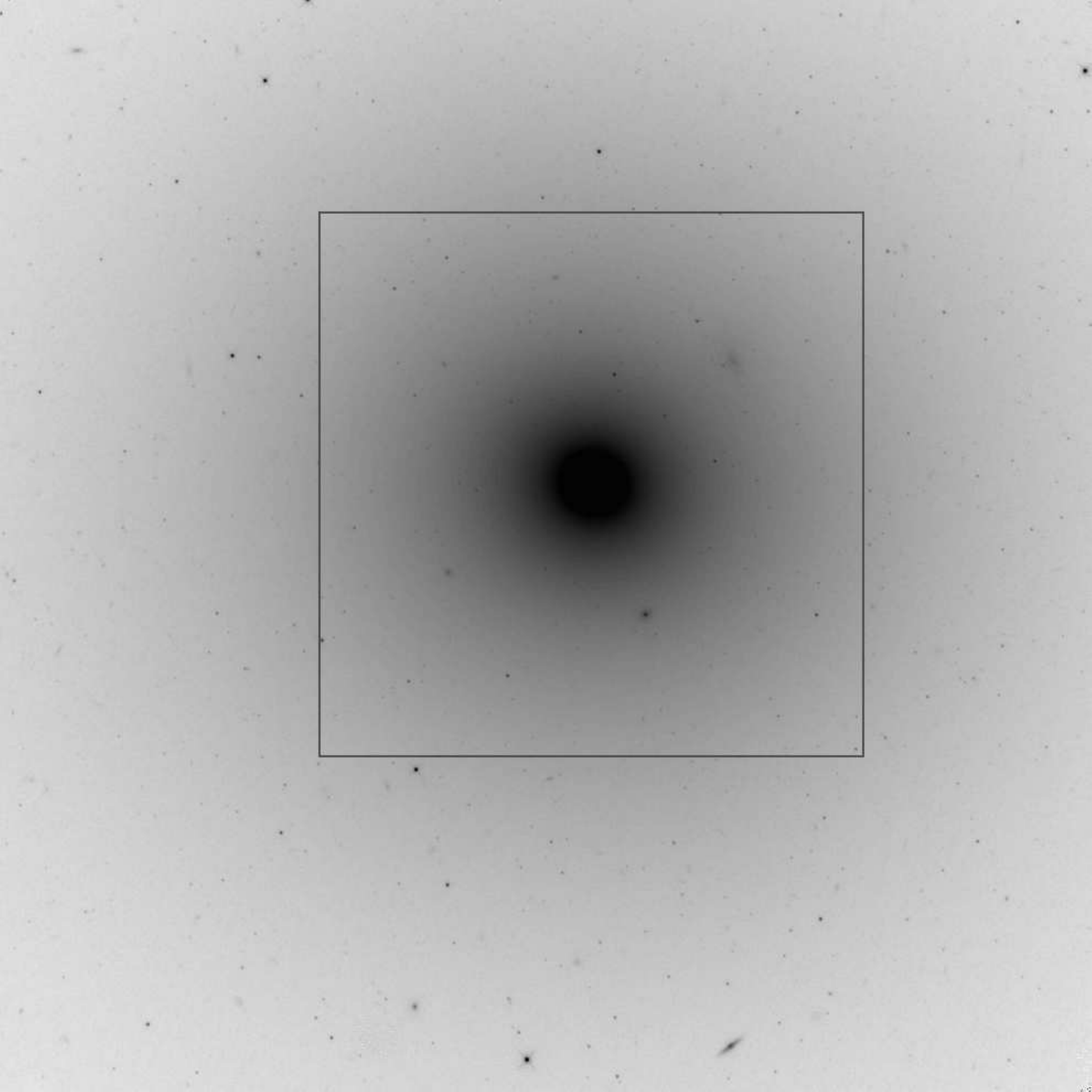}
\includegraphics[scale=0.4]{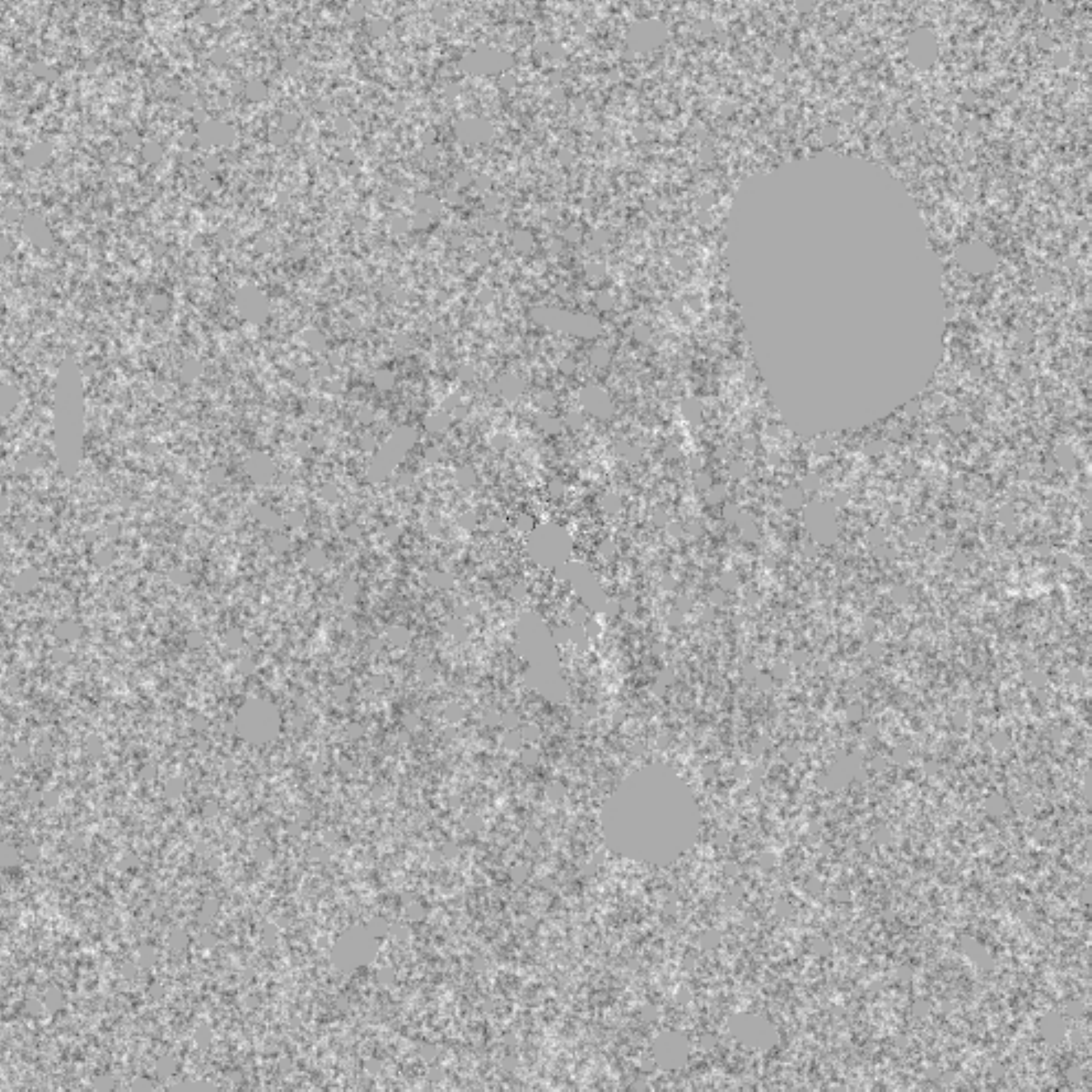} \\
\vspace{10pt}
\includegraphics[scale=0.4]{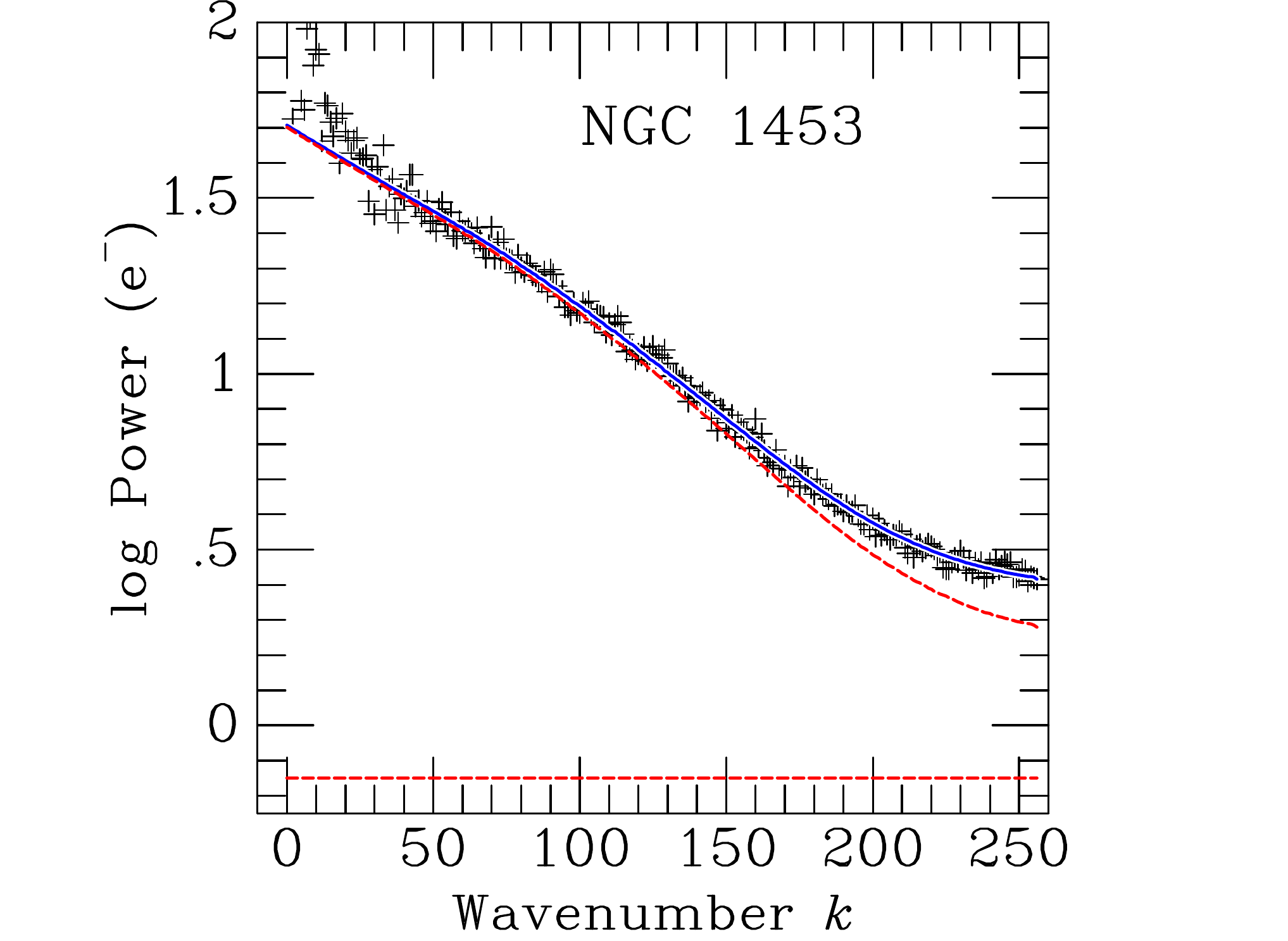}
\hspace{-25pt}
\includegraphics[scale=0.4]{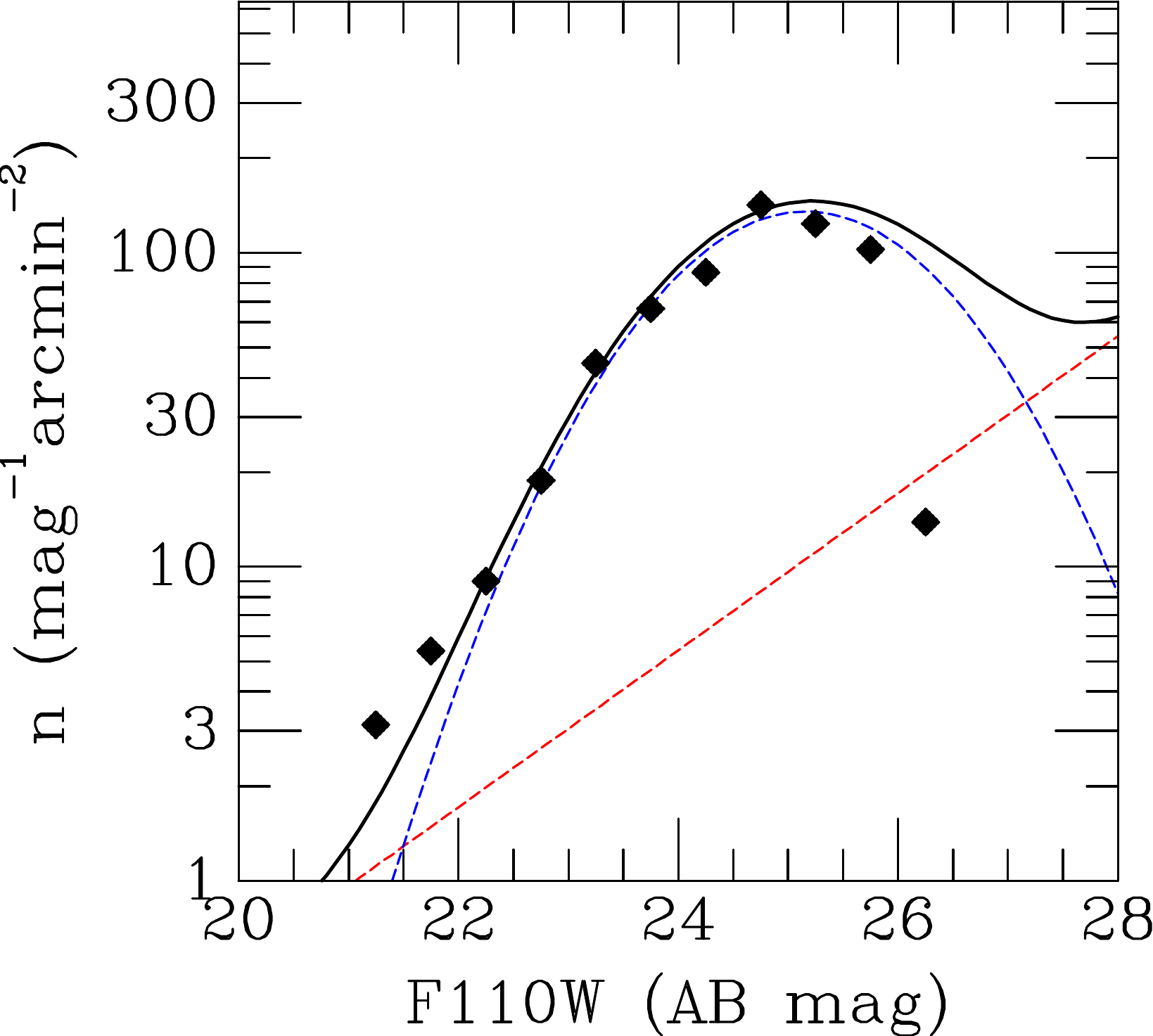}
\caption{Combined figure for NGC~1453.}
\end{center}
\end{figure*}
\clearpage

\begin{figure*}
\begin{center}
\includegraphics[scale=0.2]{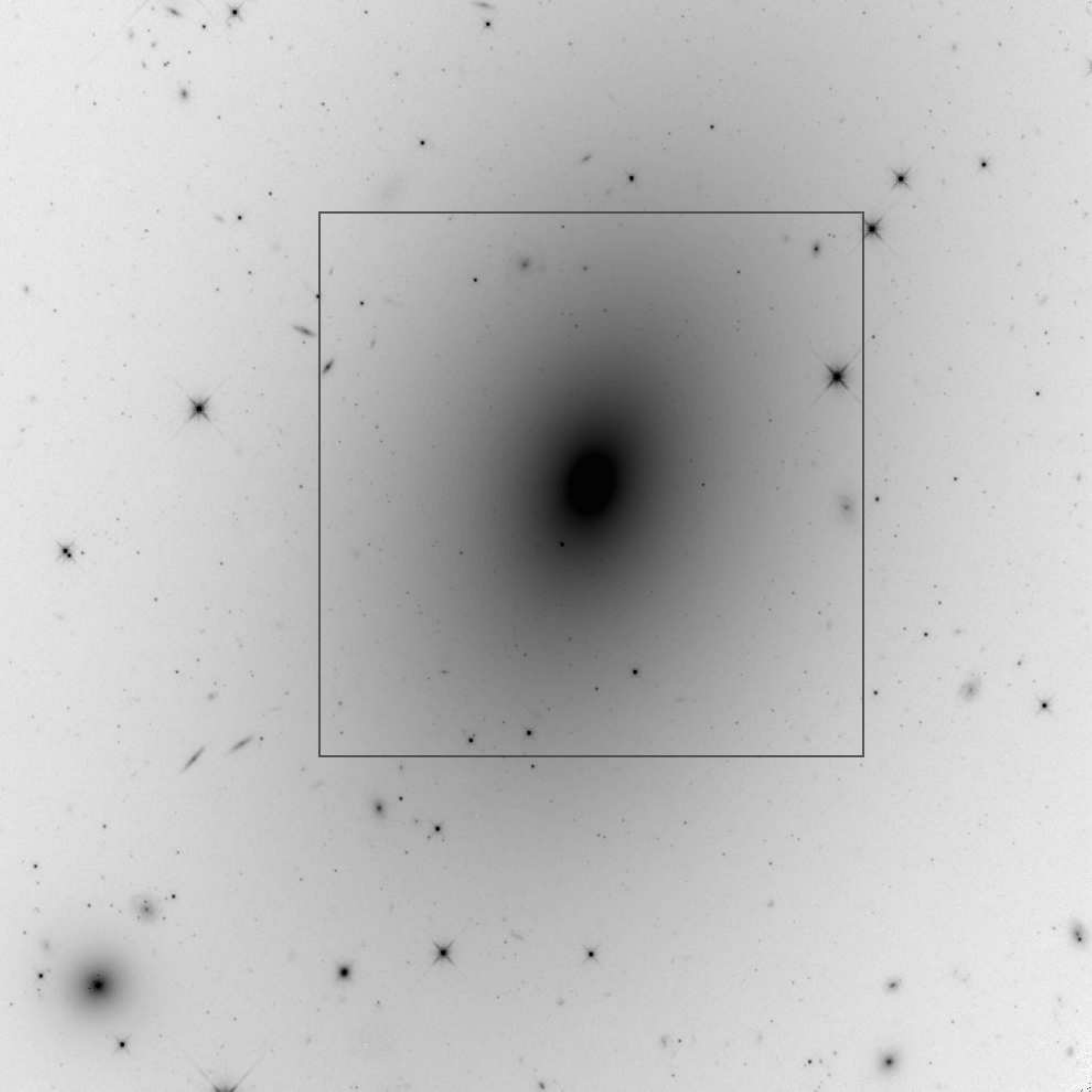}
\includegraphics[scale=0.4]{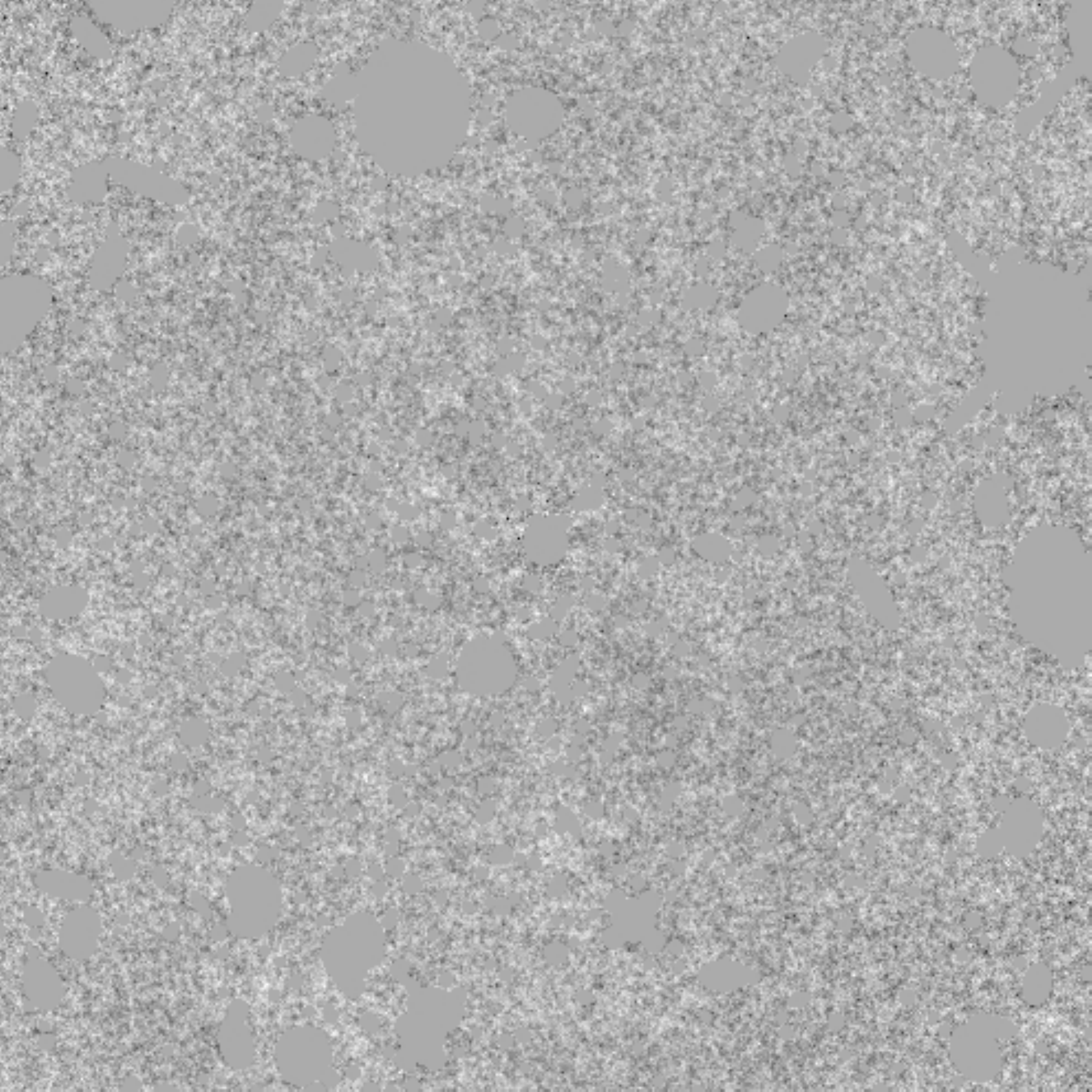} \\
\vspace{10pt}
\includegraphics[scale=0.4]{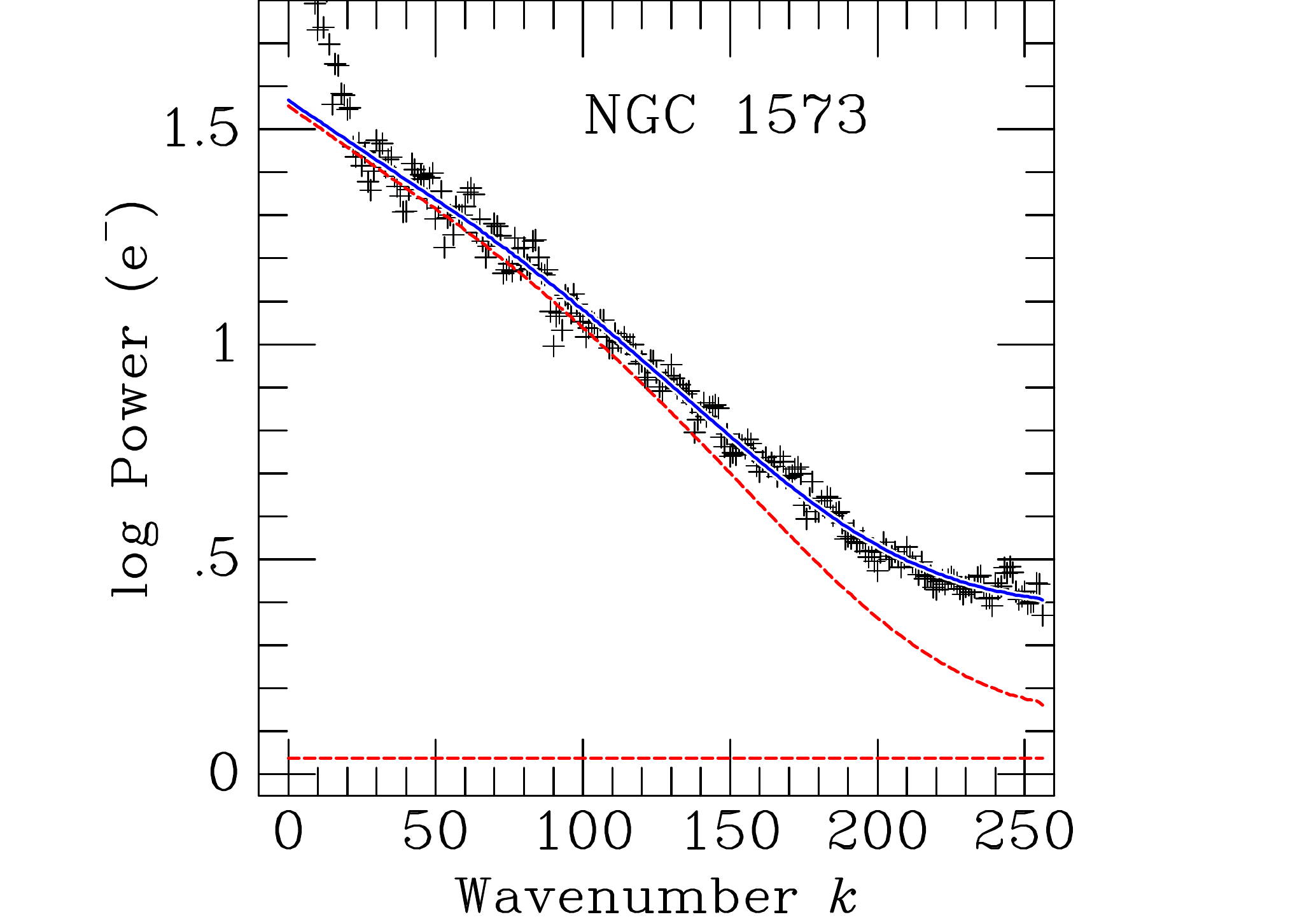}
\hspace{-25pt}
\includegraphics[scale=0.4]{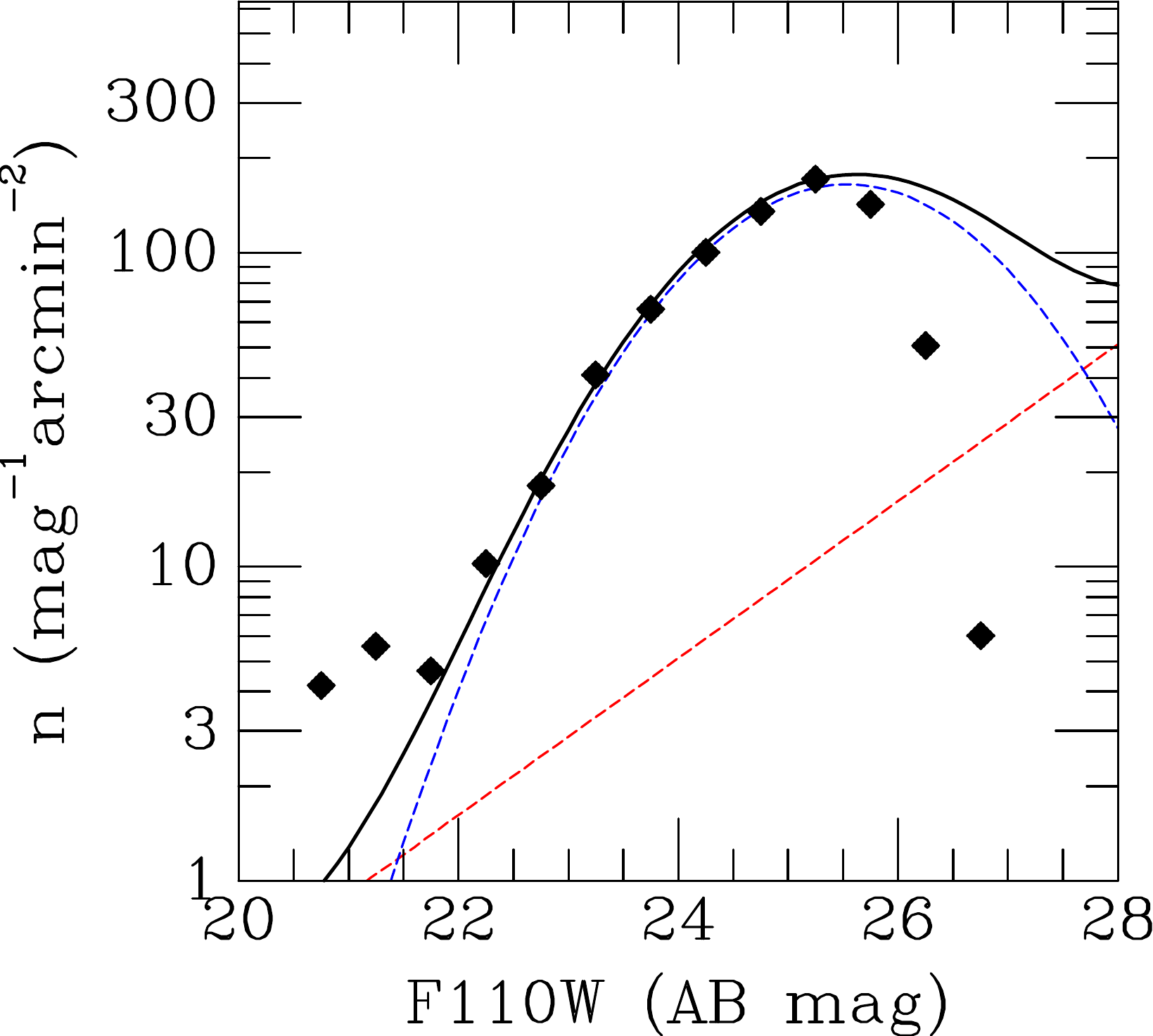}
\caption{Combined figure for NGC~1573.}
\end{center}
\end{figure*}
\clearpage

\begin{figure*}
\begin{center}
\includegraphics[scale=0.2]{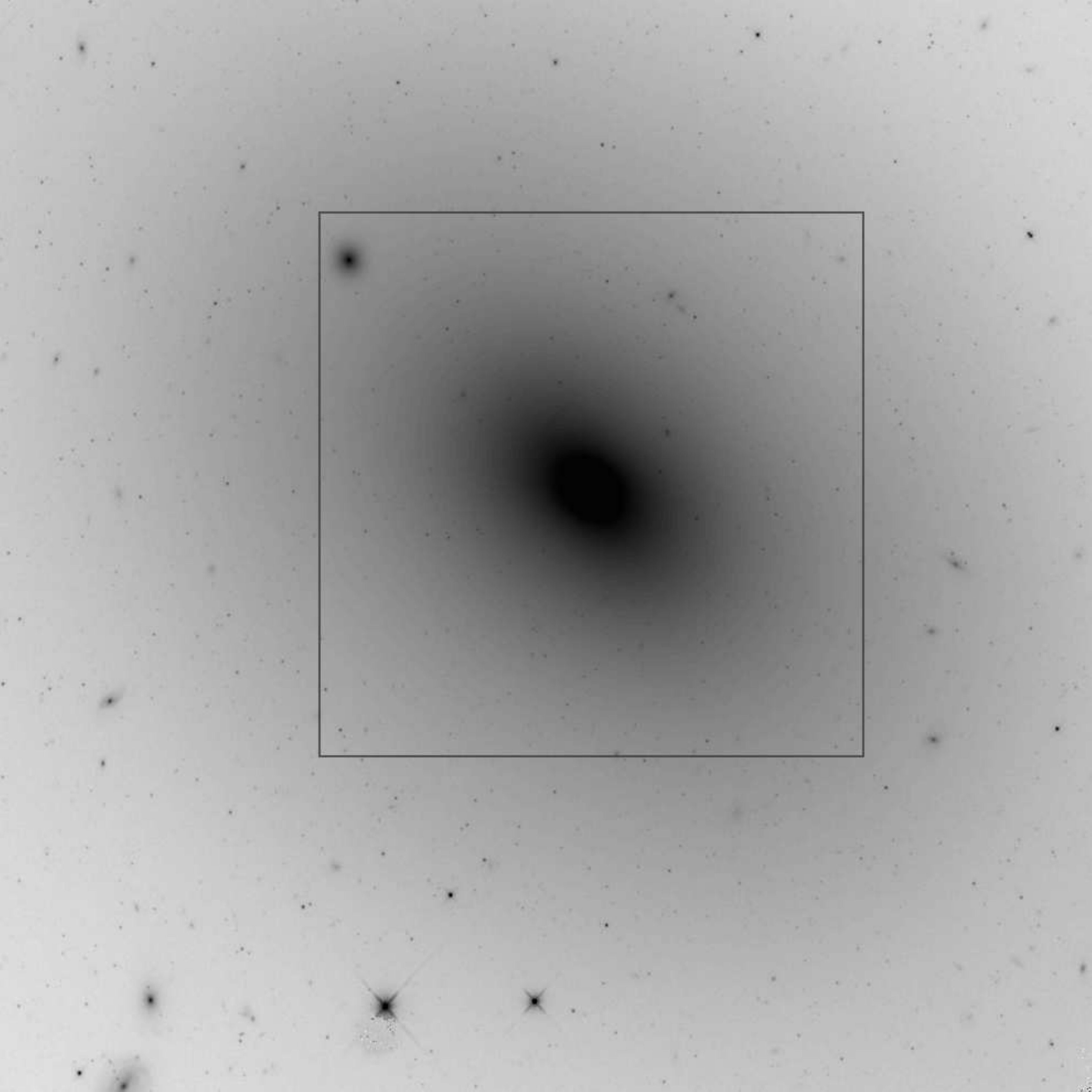}
\includegraphics[scale=0.4]{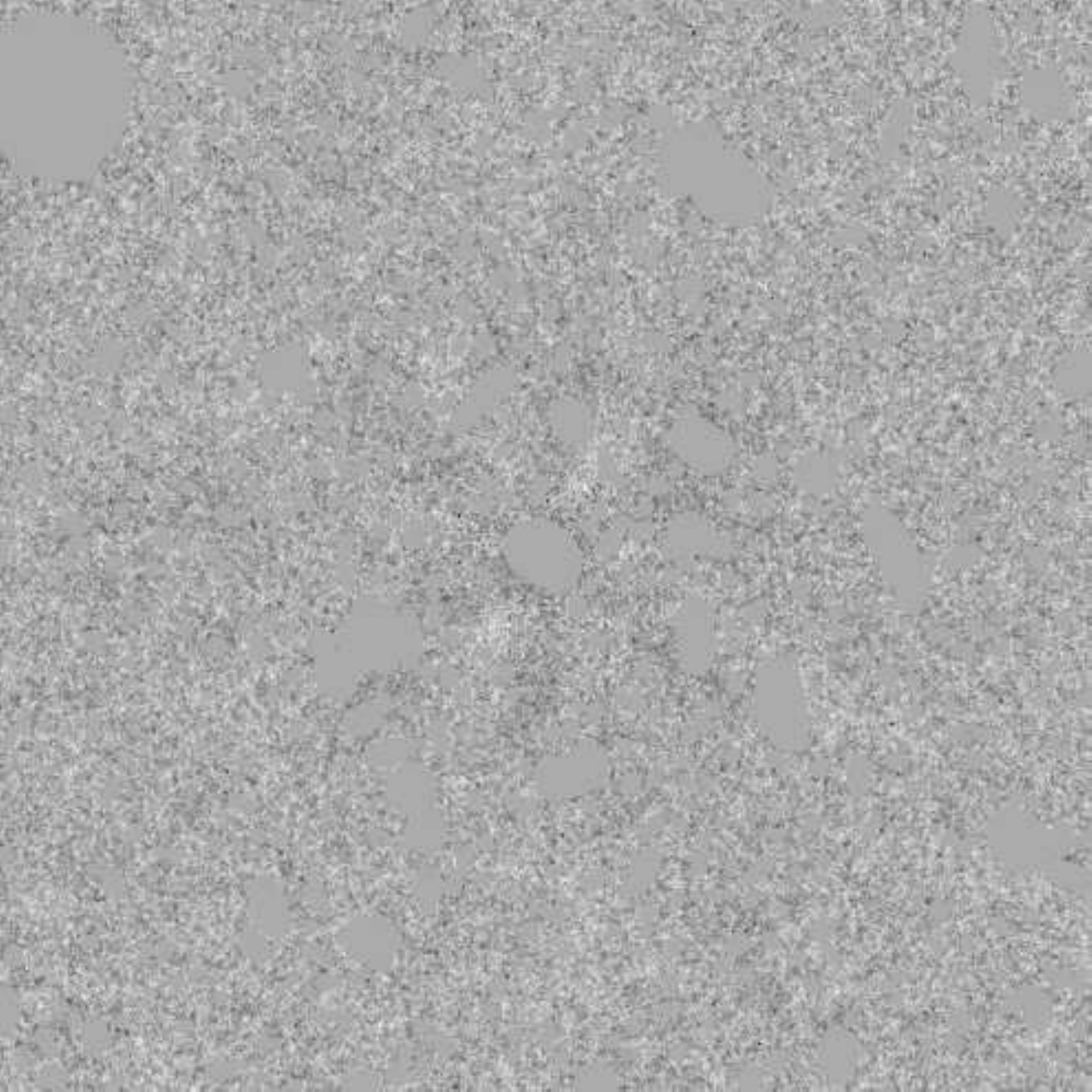} \\
\vspace{10pt}
\includegraphics[scale=0.4]{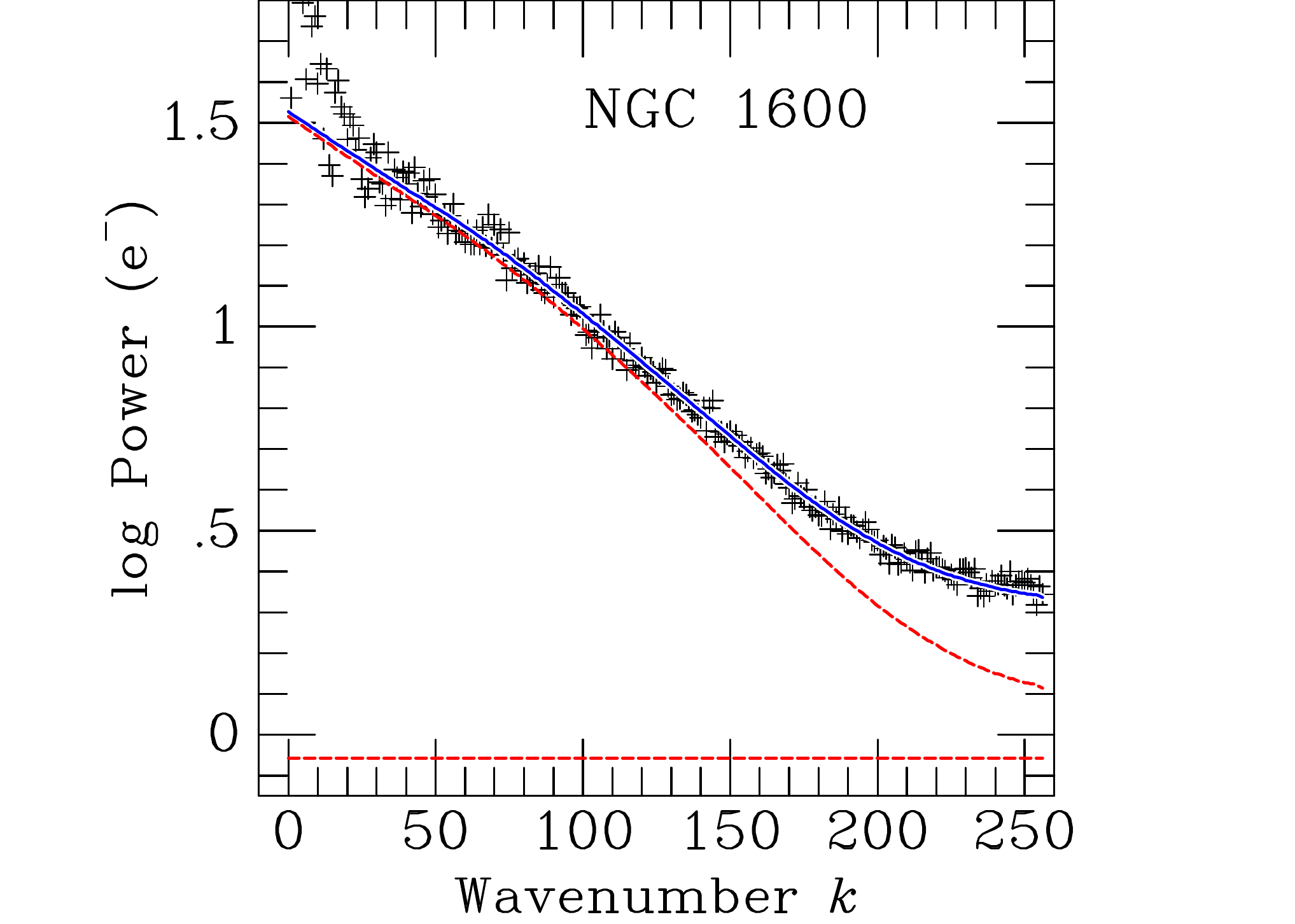}
\hspace{-25pt}
\includegraphics[scale=0.4]{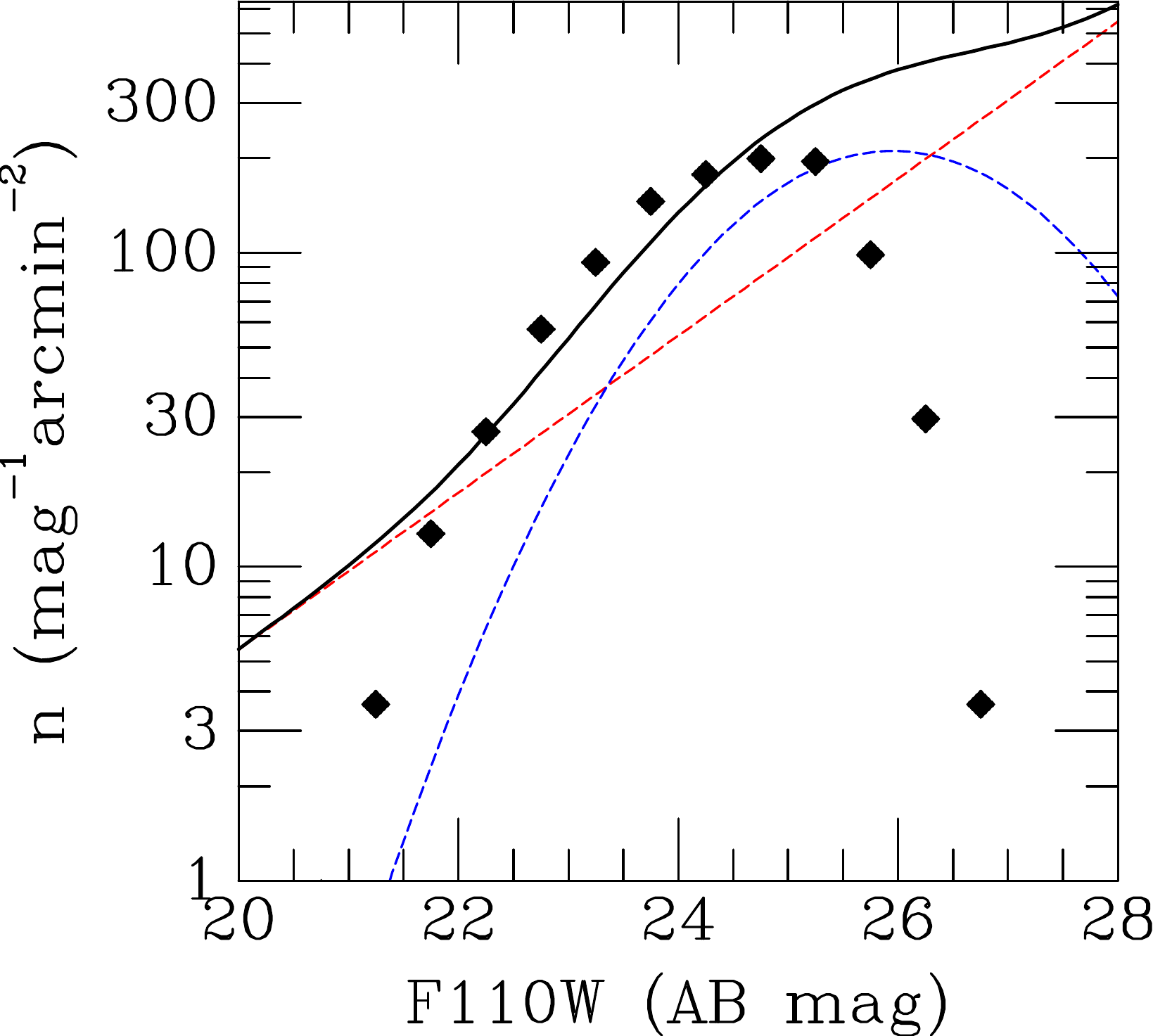}
\caption{Combined figure for NGC~1600.}
\end{center}
\end{figure*}
\clearpage

\begin{figure*}
\begin{center}
\includegraphics[scale=0.2]{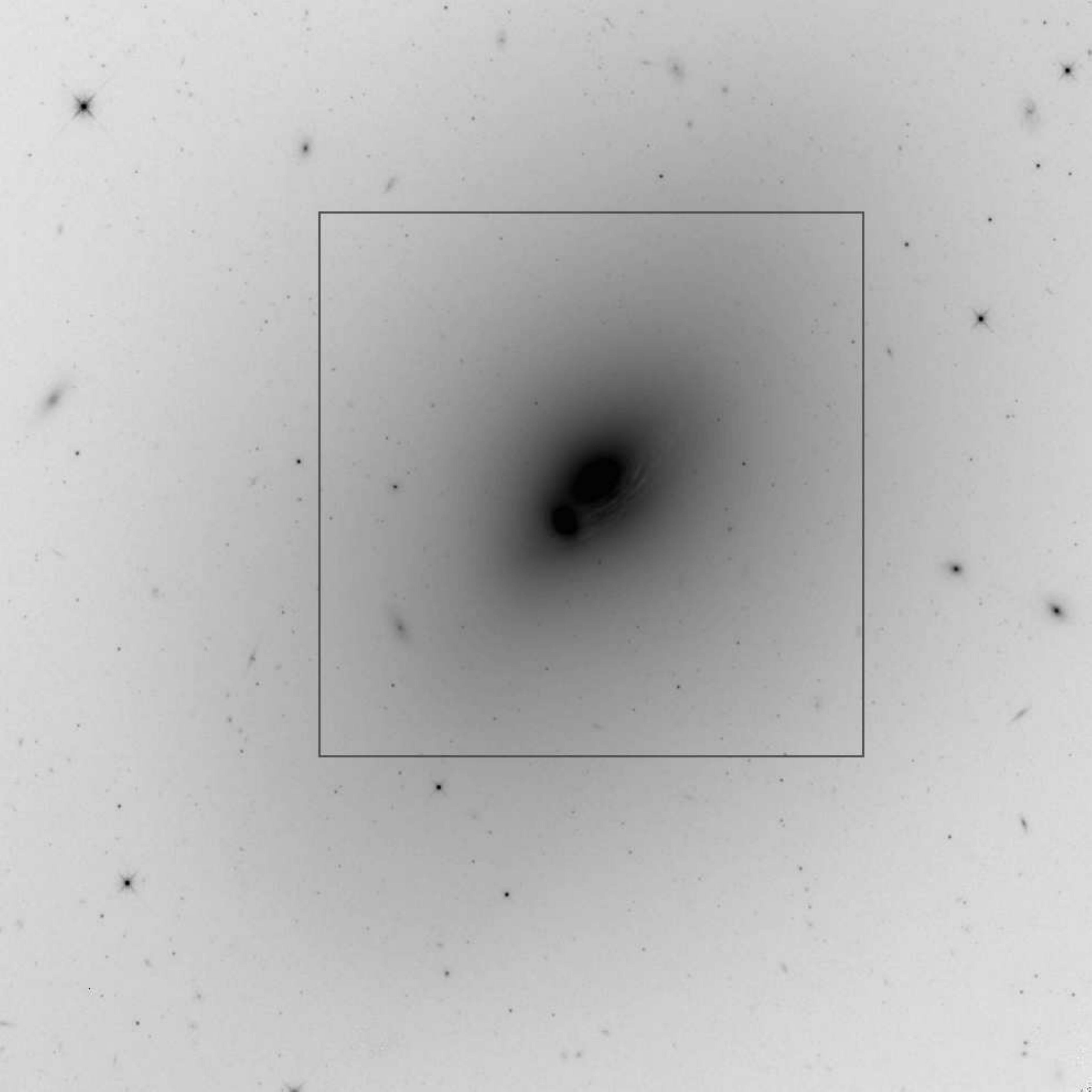}
\includegraphics[scale=0.4]{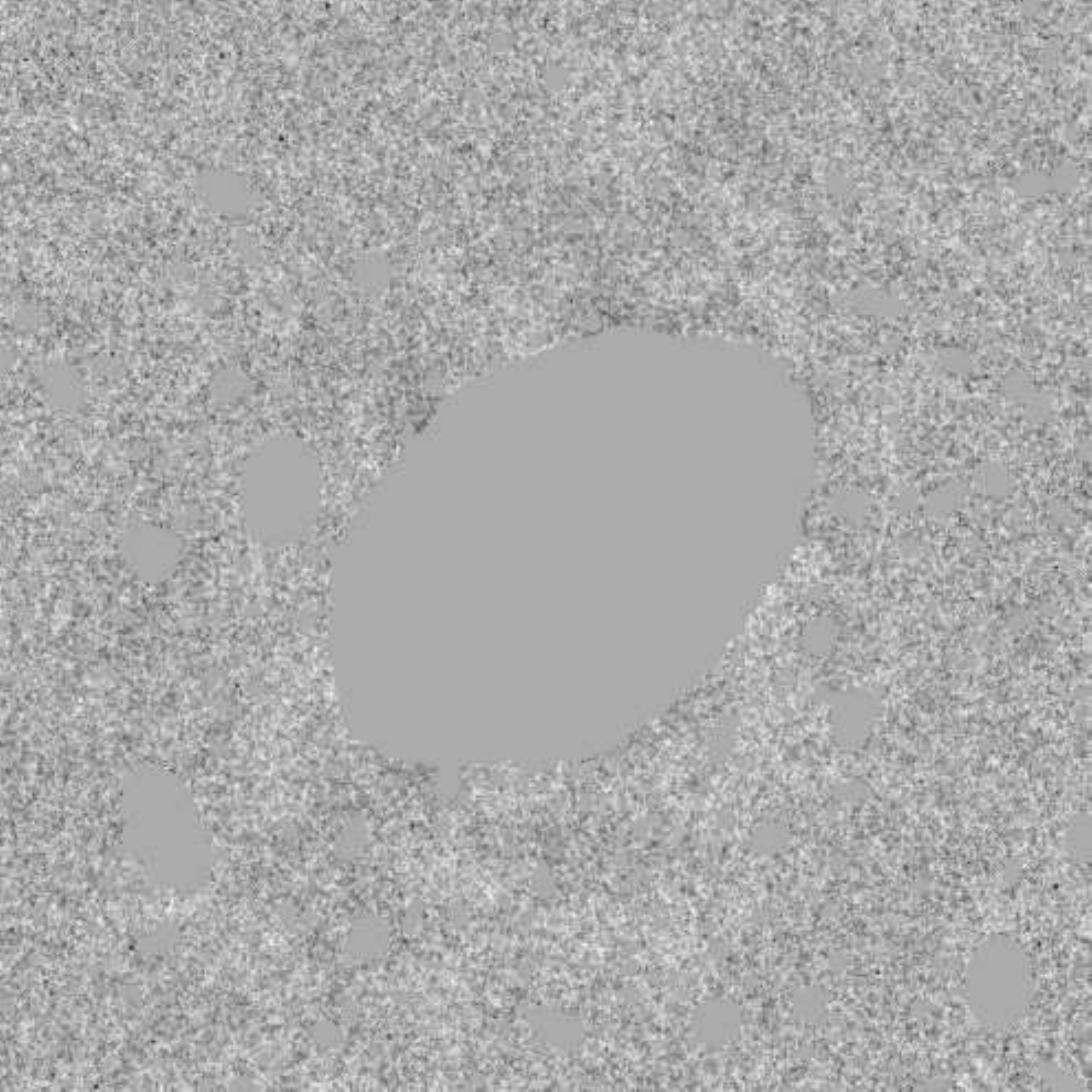} \\
\vspace{10pt}
\includegraphics[scale=0.4]{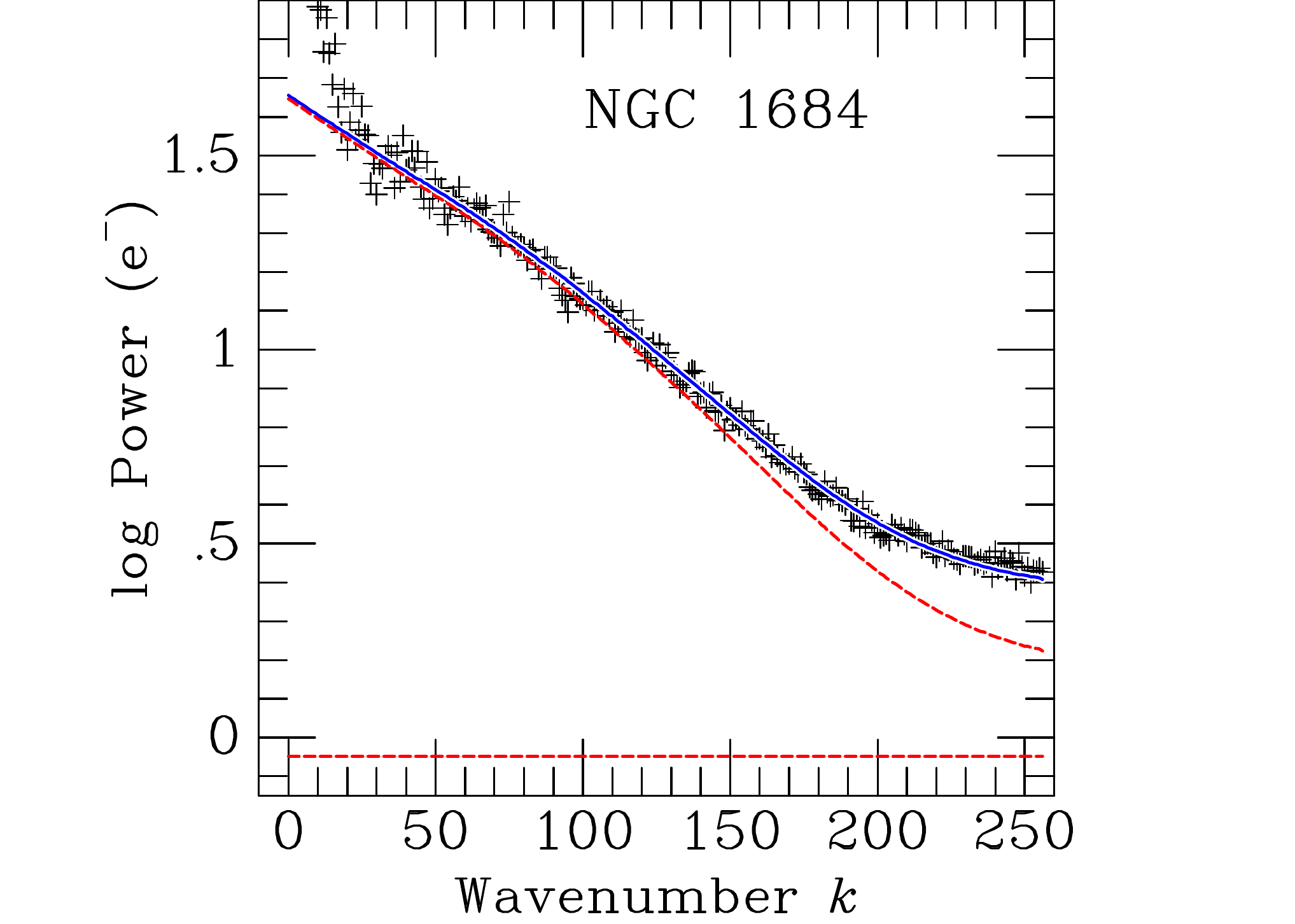}
\hspace{-25pt}
\includegraphics[scale=0.4]{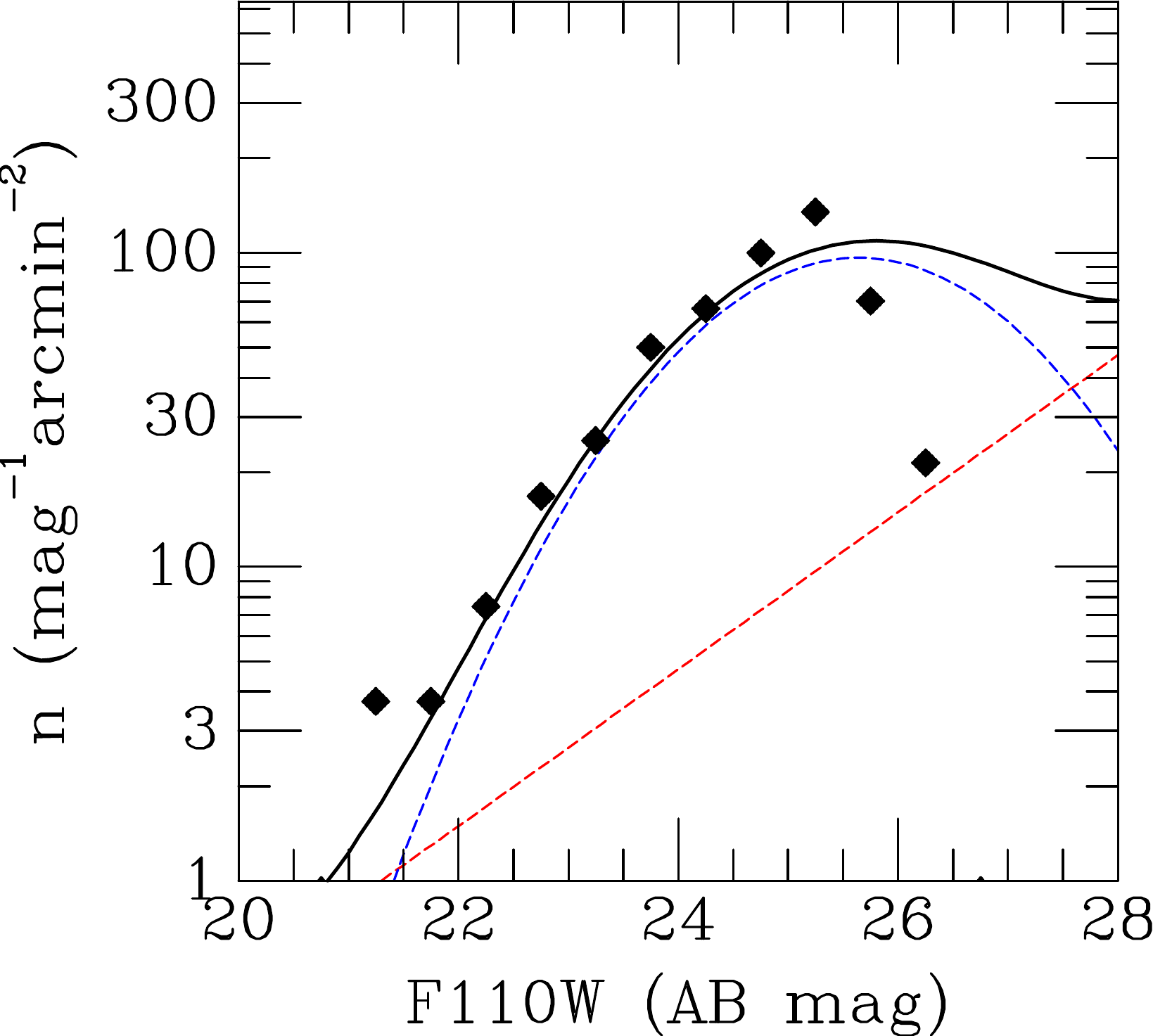}
\caption{Combined figure for NGC~1684.}
\end{center}
\end{figure*}
\clearpage

\begin{figure*}
\begin{center}
\includegraphics[scale=0.2]{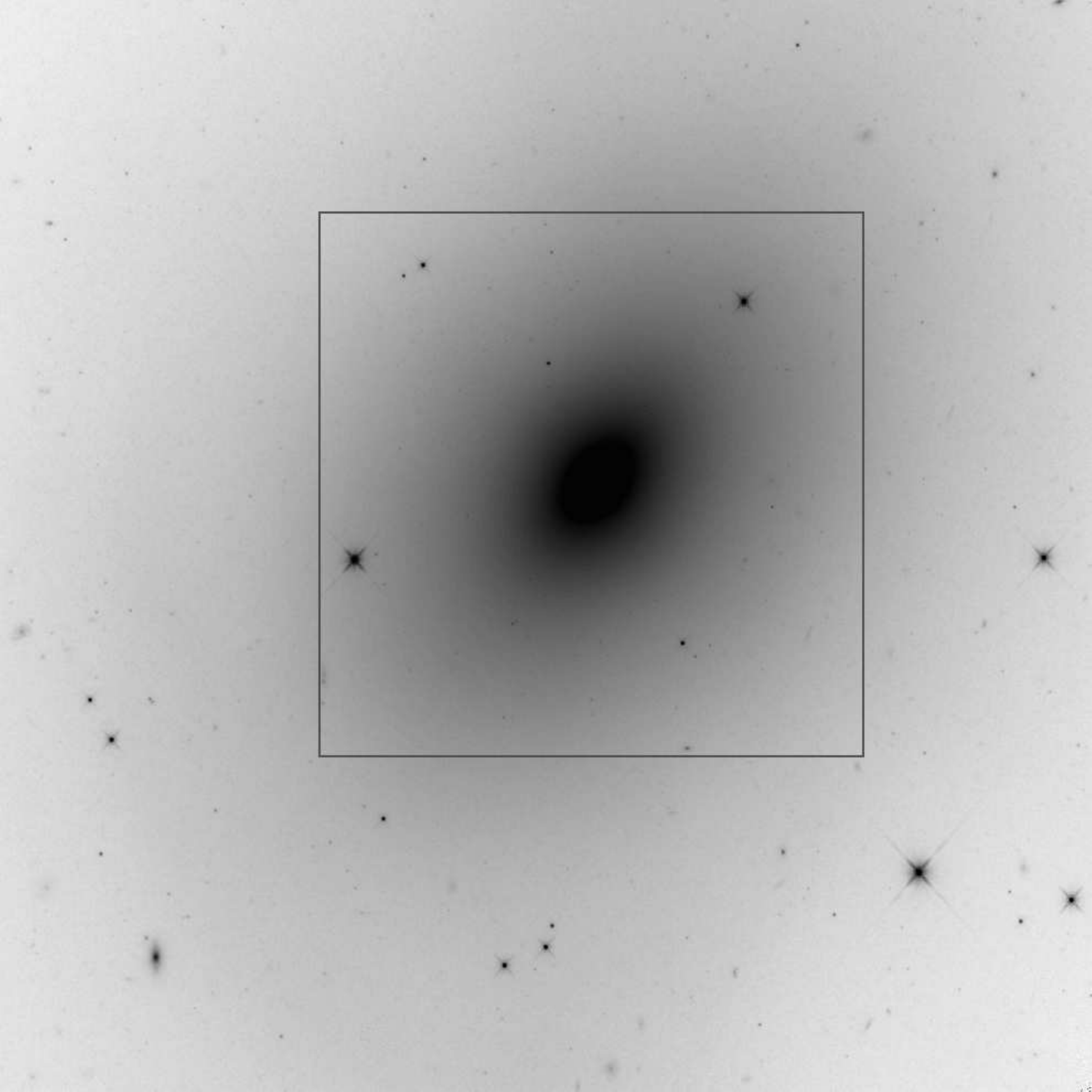}
\includegraphics[scale=0.4]{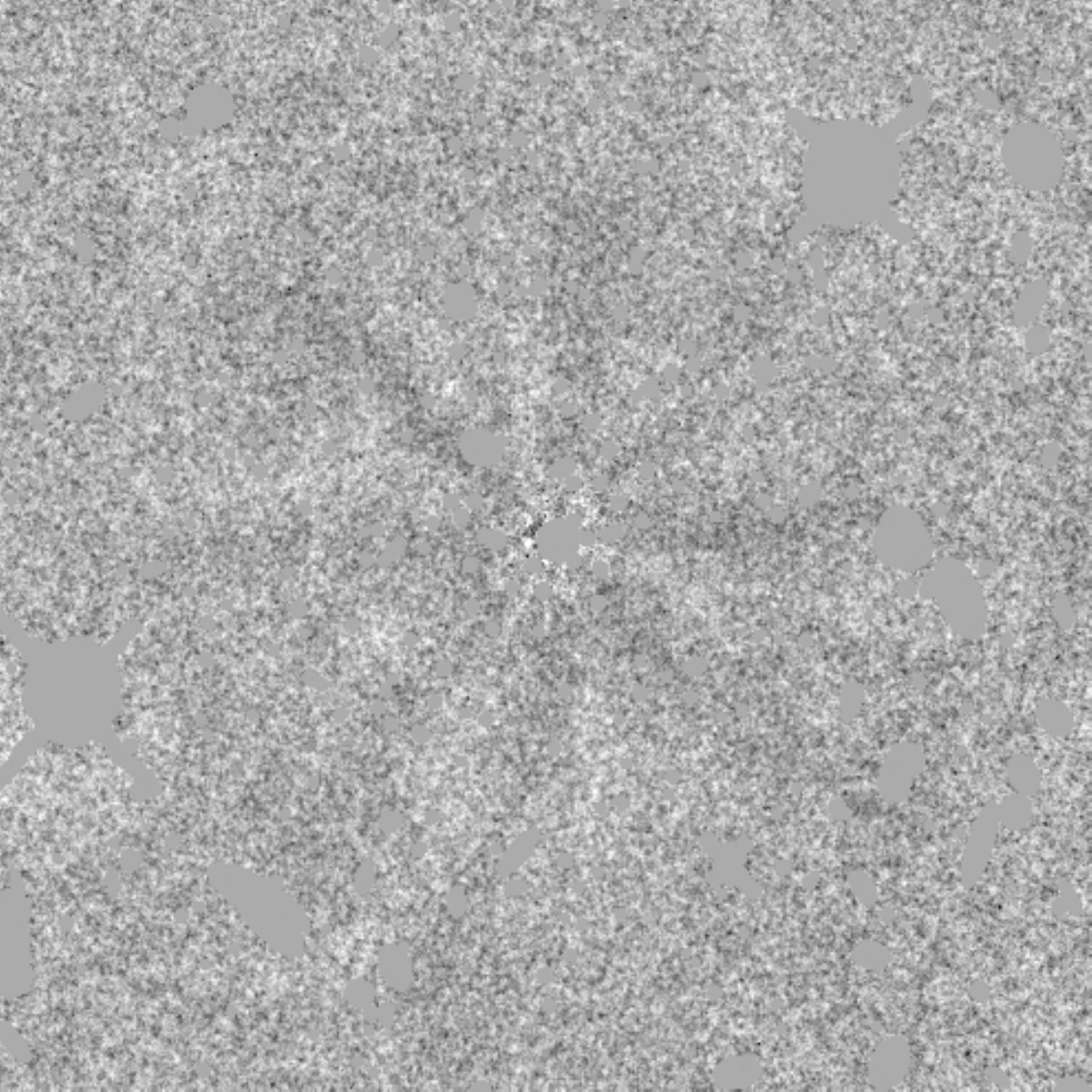} \\
\vspace{10pt}
\includegraphics[scale=0.4]{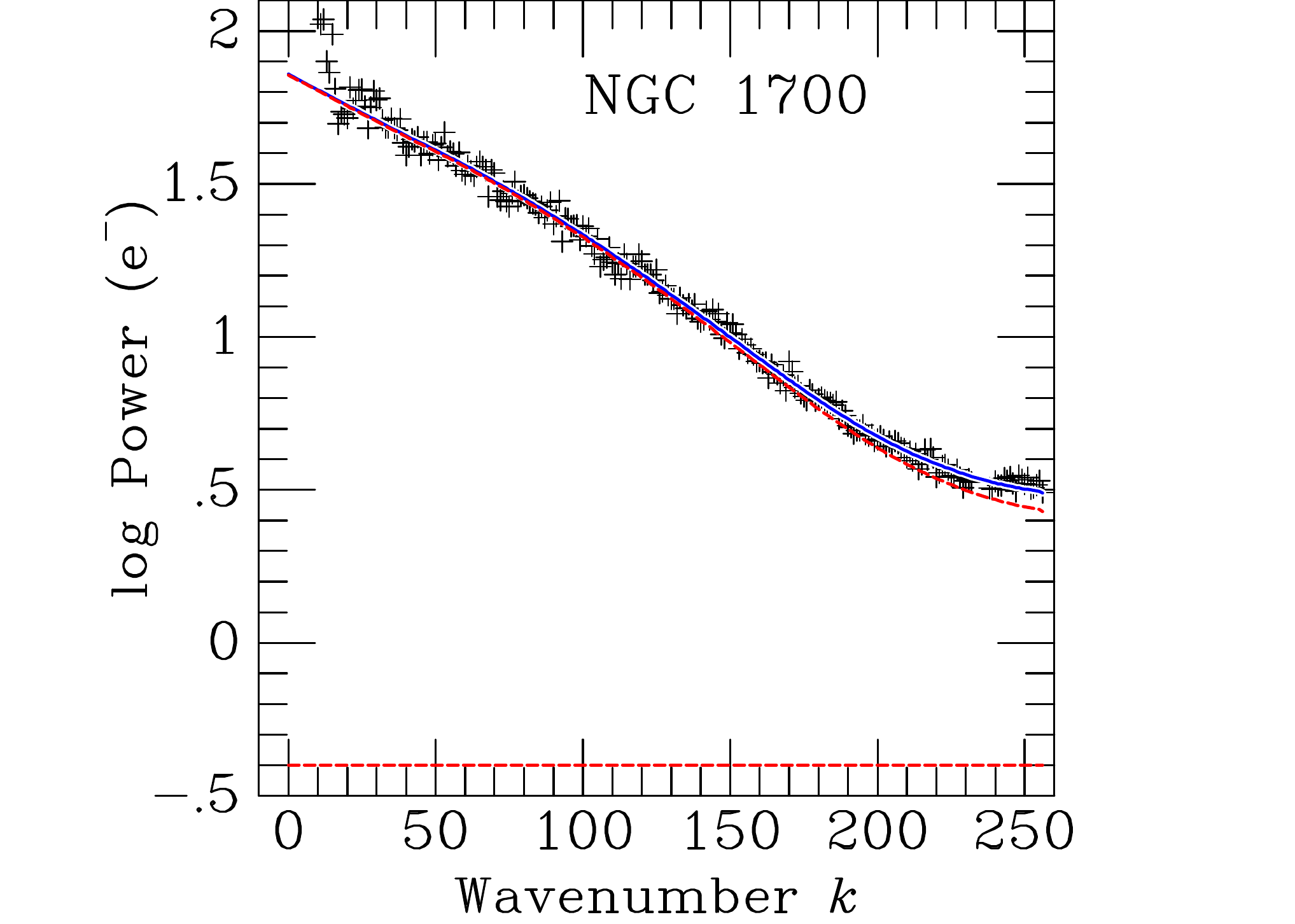}
\hspace{-25pt}
\includegraphics[scale=0.4]{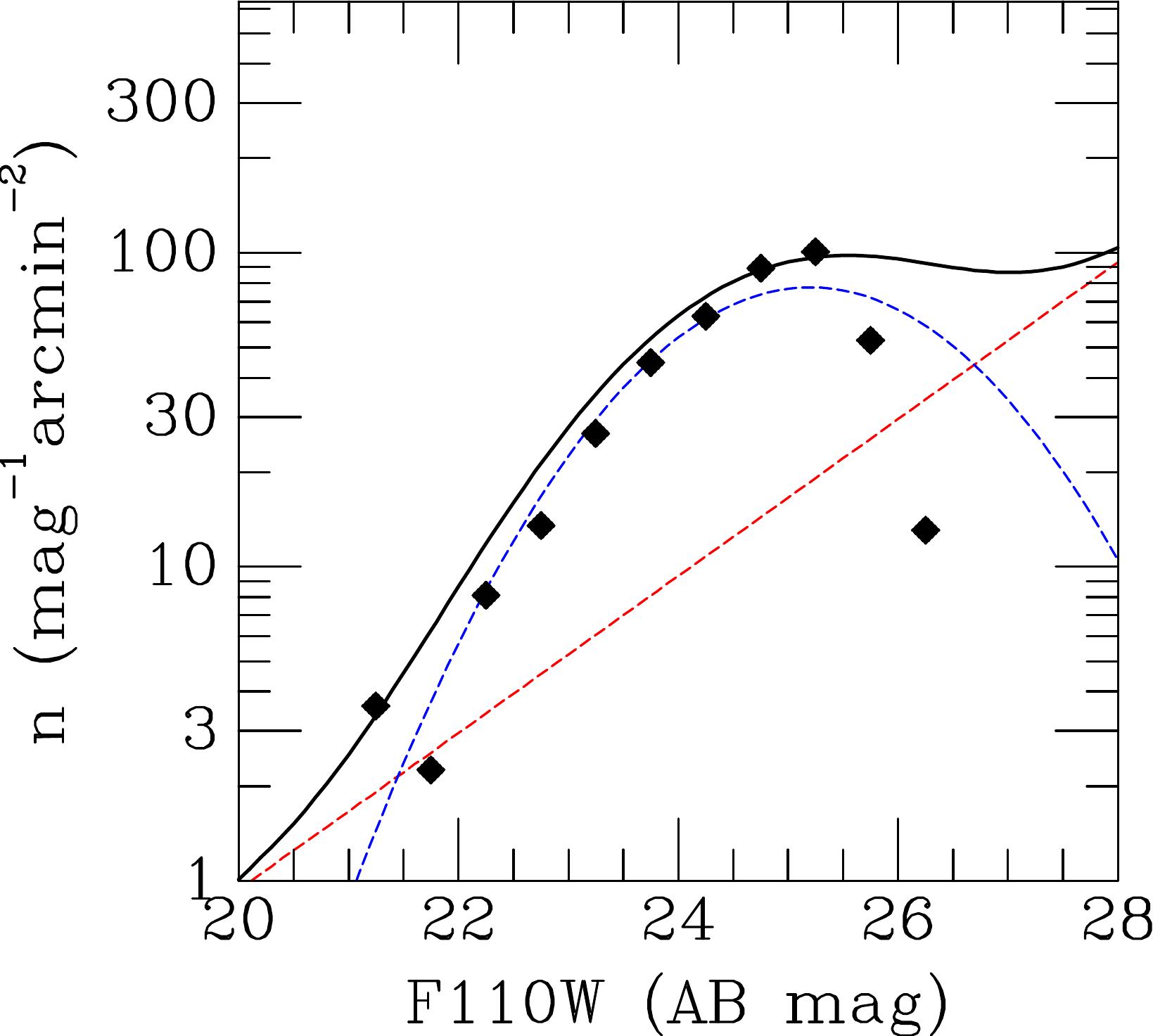}
\caption{Combined figure for NGC~1700.}
\end{center}
\end{figure*}
\clearpage

\begin{figure*}
\begin{center}
\includegraphics[scale=0.2]{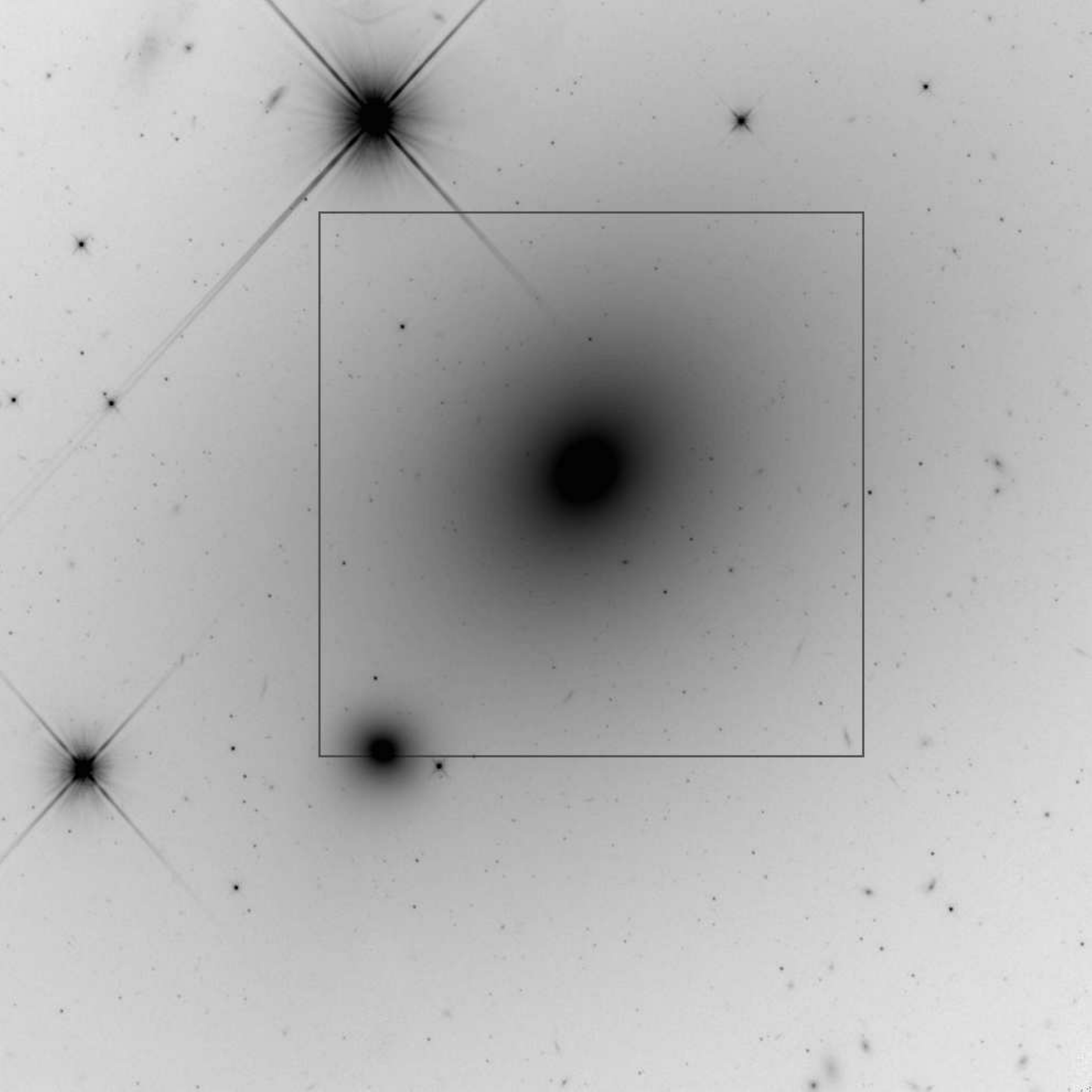}
\includegraphics[scale=0.4]{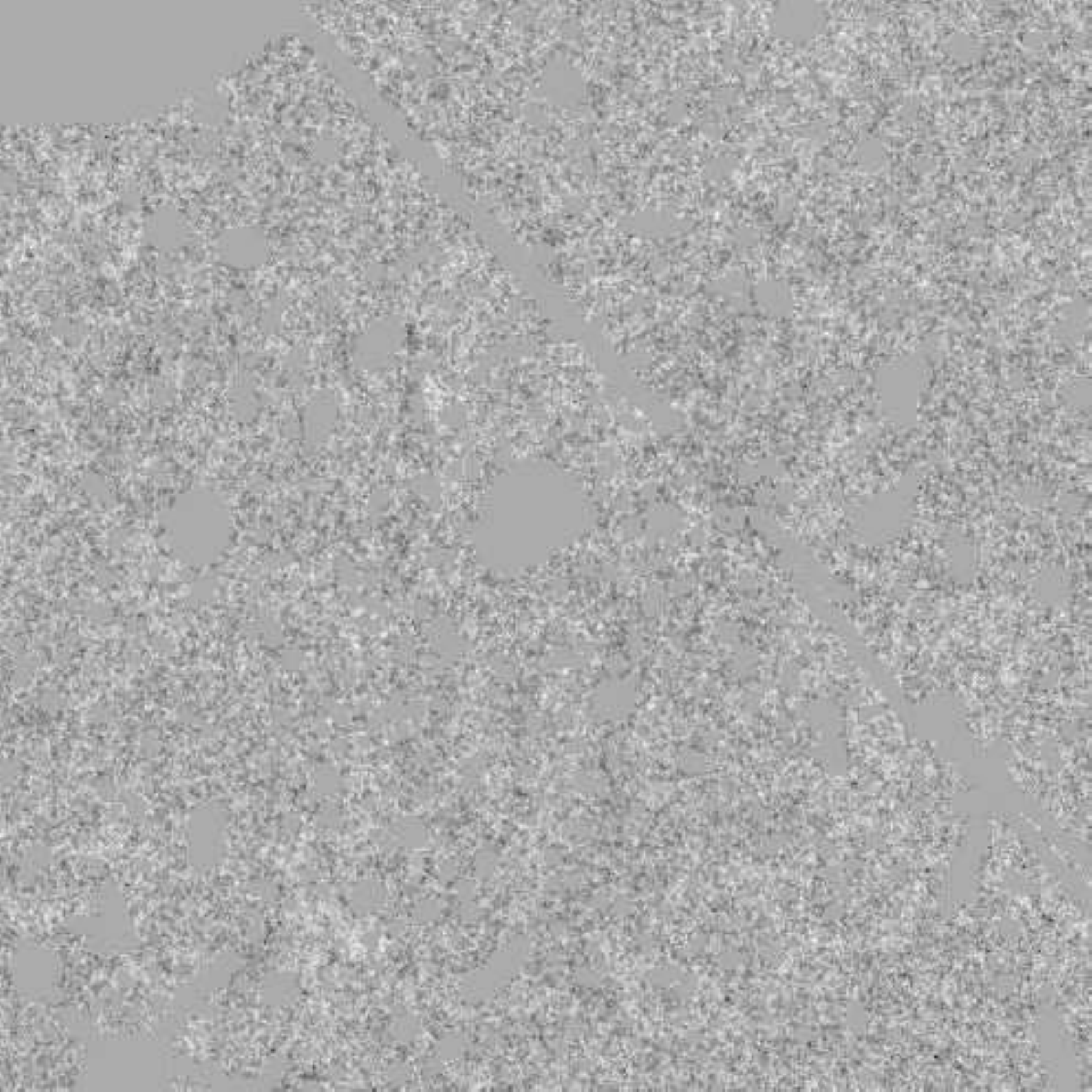} \\
\vspace{10pt}
\includegraphics[scale=0.4]{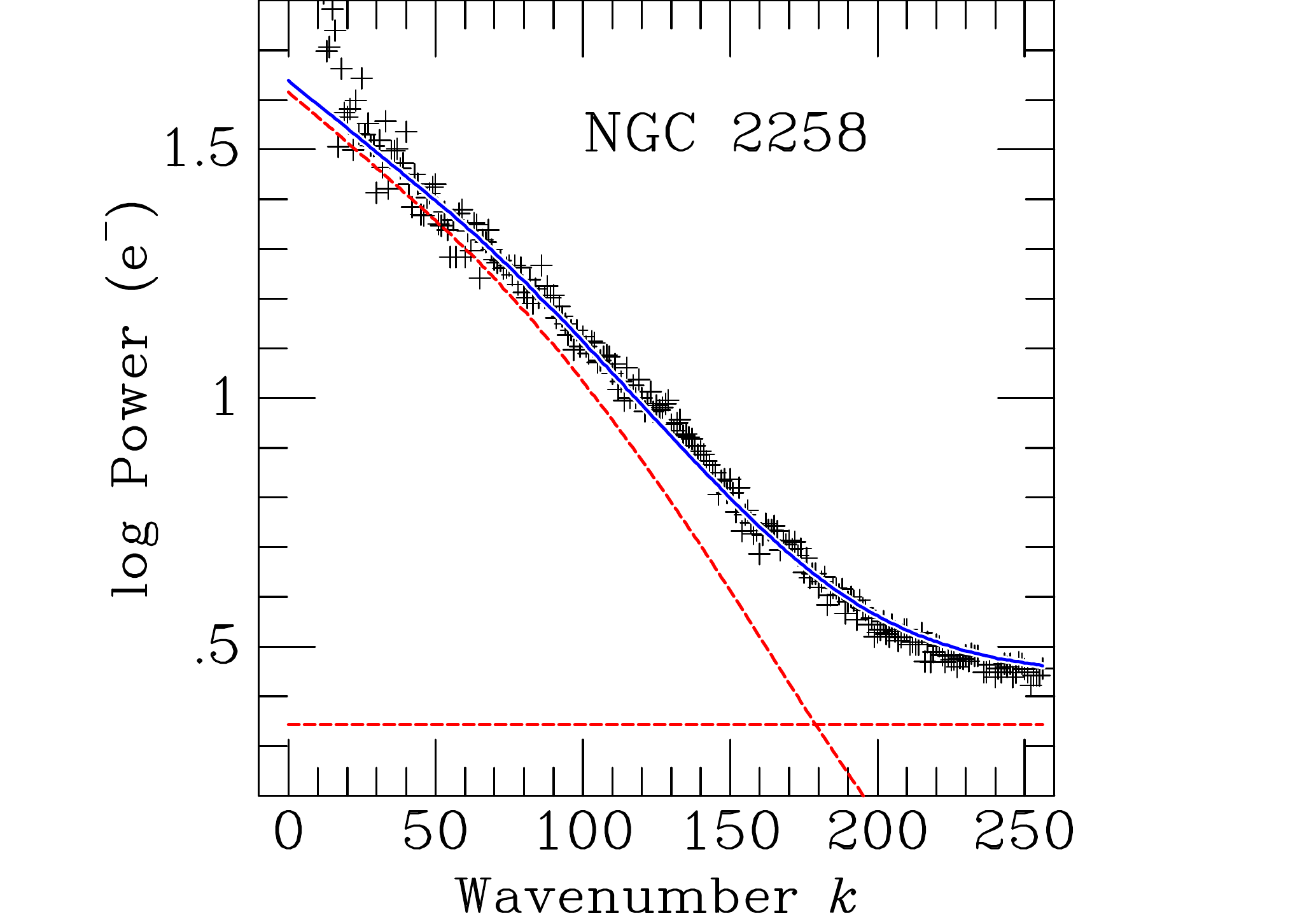}
\hspace{-25pt}
\includegraphics[scale=0.4]{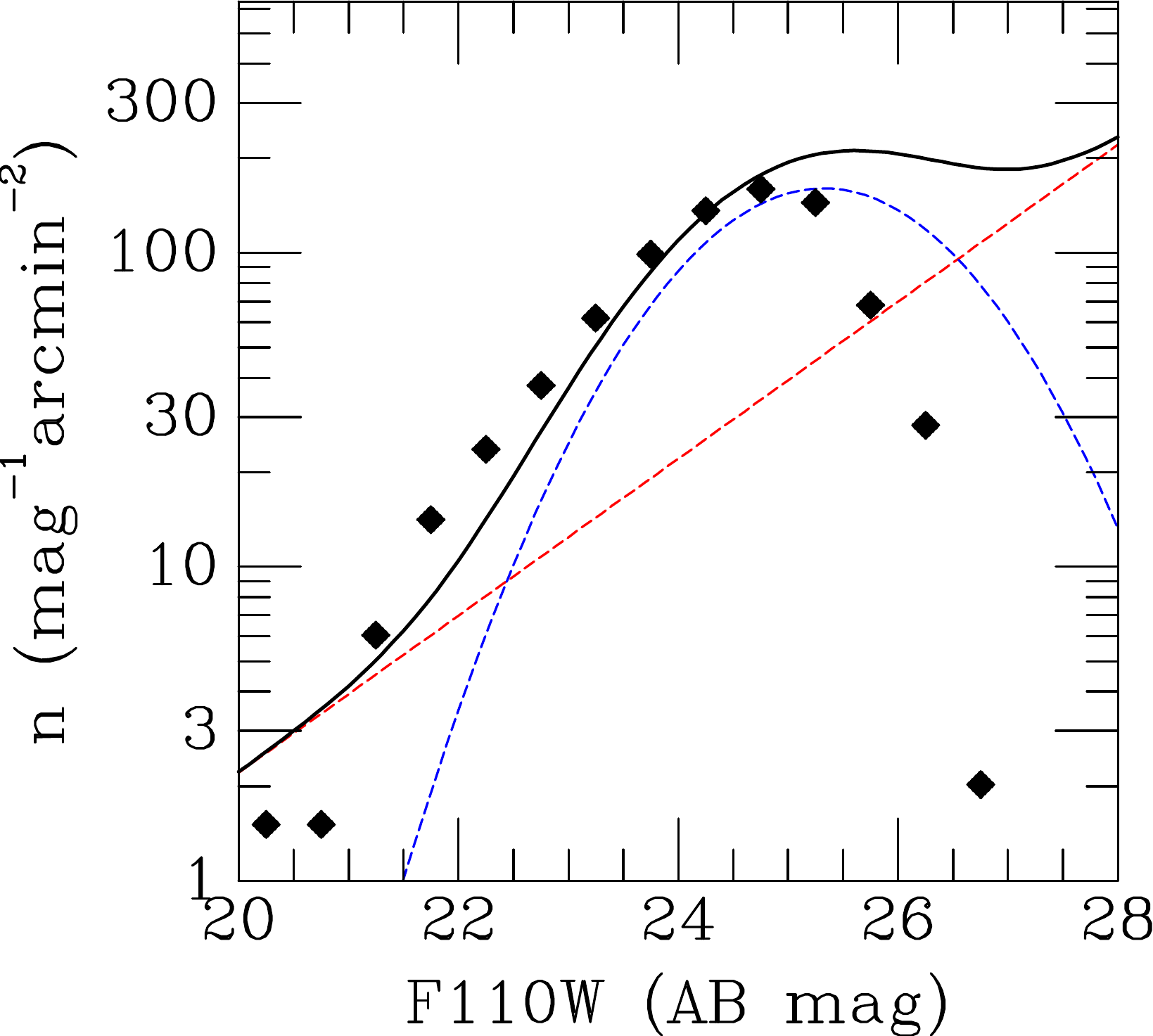}
\caption{Combined figure for NGC~2258.}
\end{center}
\end{figure*}
\clearpage

\begin{figure*}
\begin{center}
\includegraphics[scale=0.2]{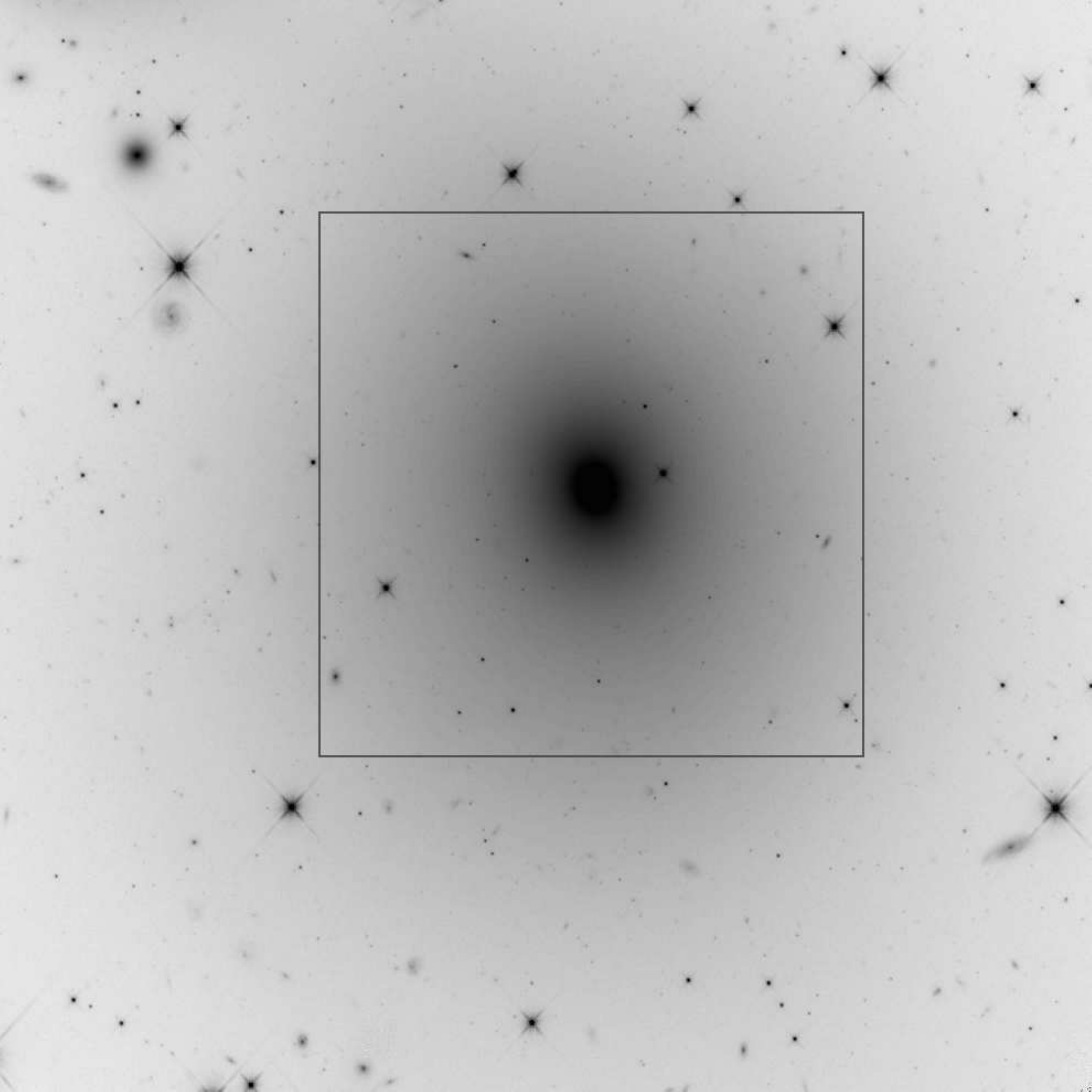}
\includegraphics[scale=0.4]{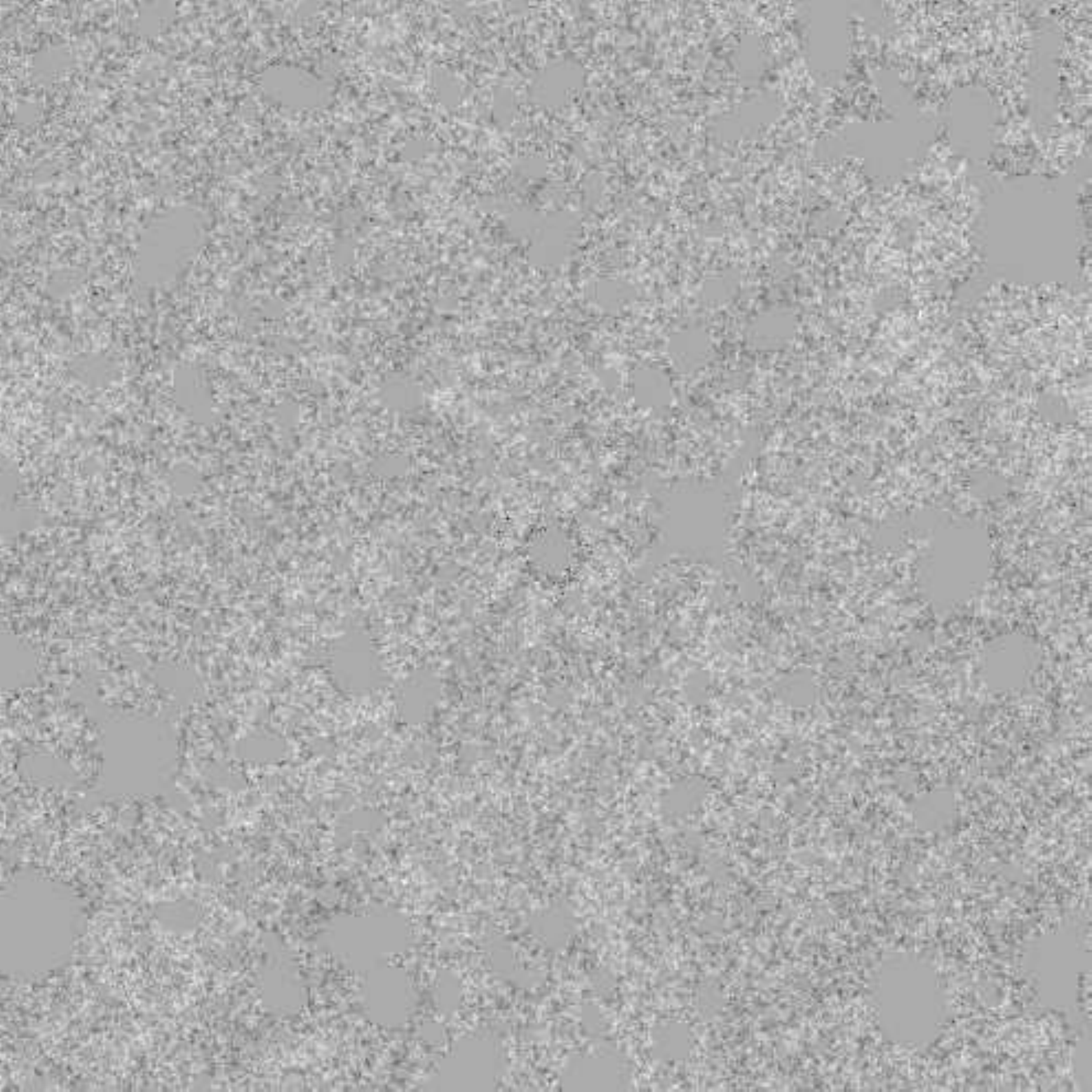} \\
\vspace{10pt}
\includegraphics[scale=0.4]{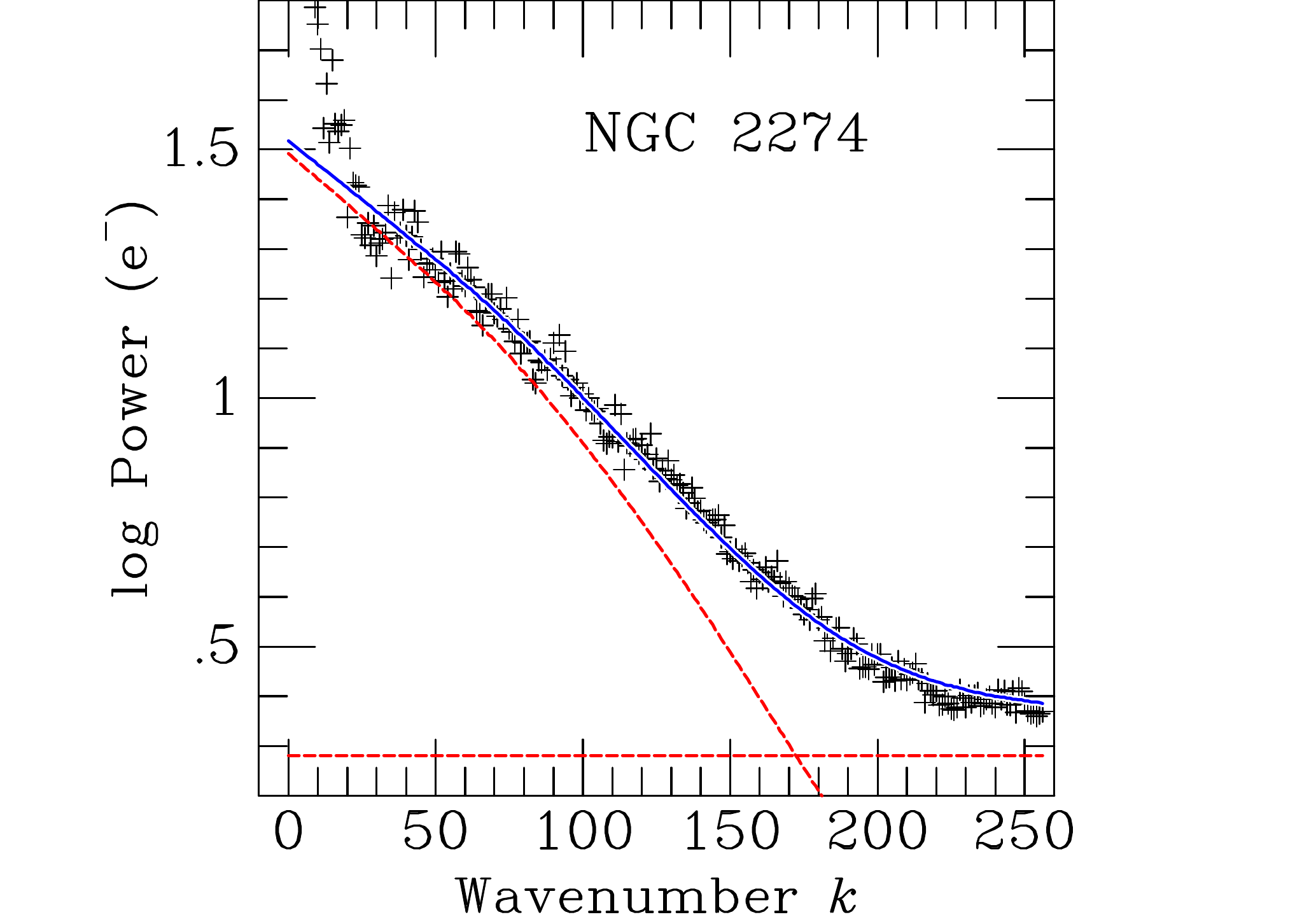}
\hspace{-25pt}
\includegraphics[scale=0.4]{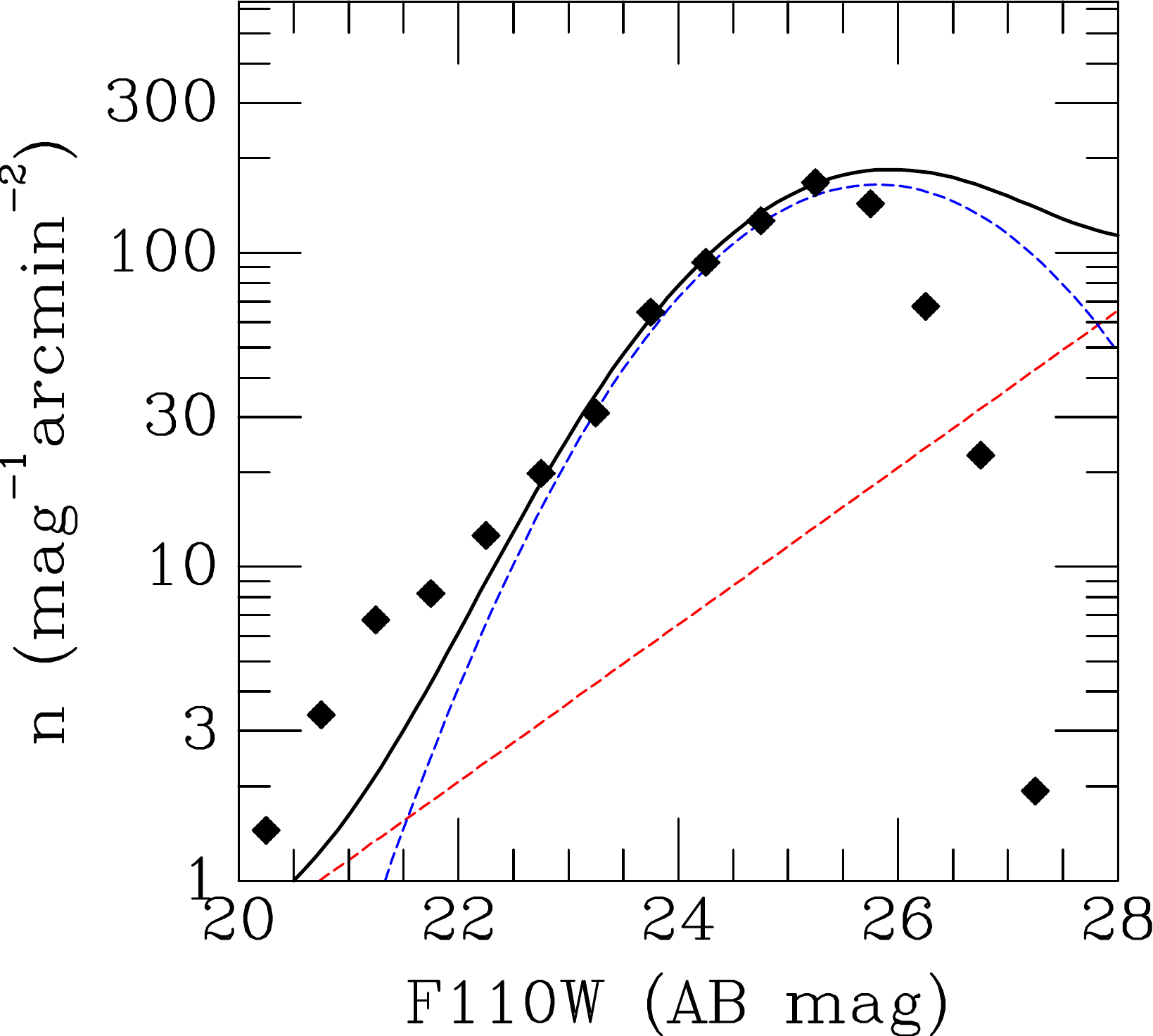}
\caption{Combined figure for NGC~2274.}
\end{center}
\end{figure*}
\clearpage

\begin{figure*}
\begin{center}
\includegraphics[scale=0.2]{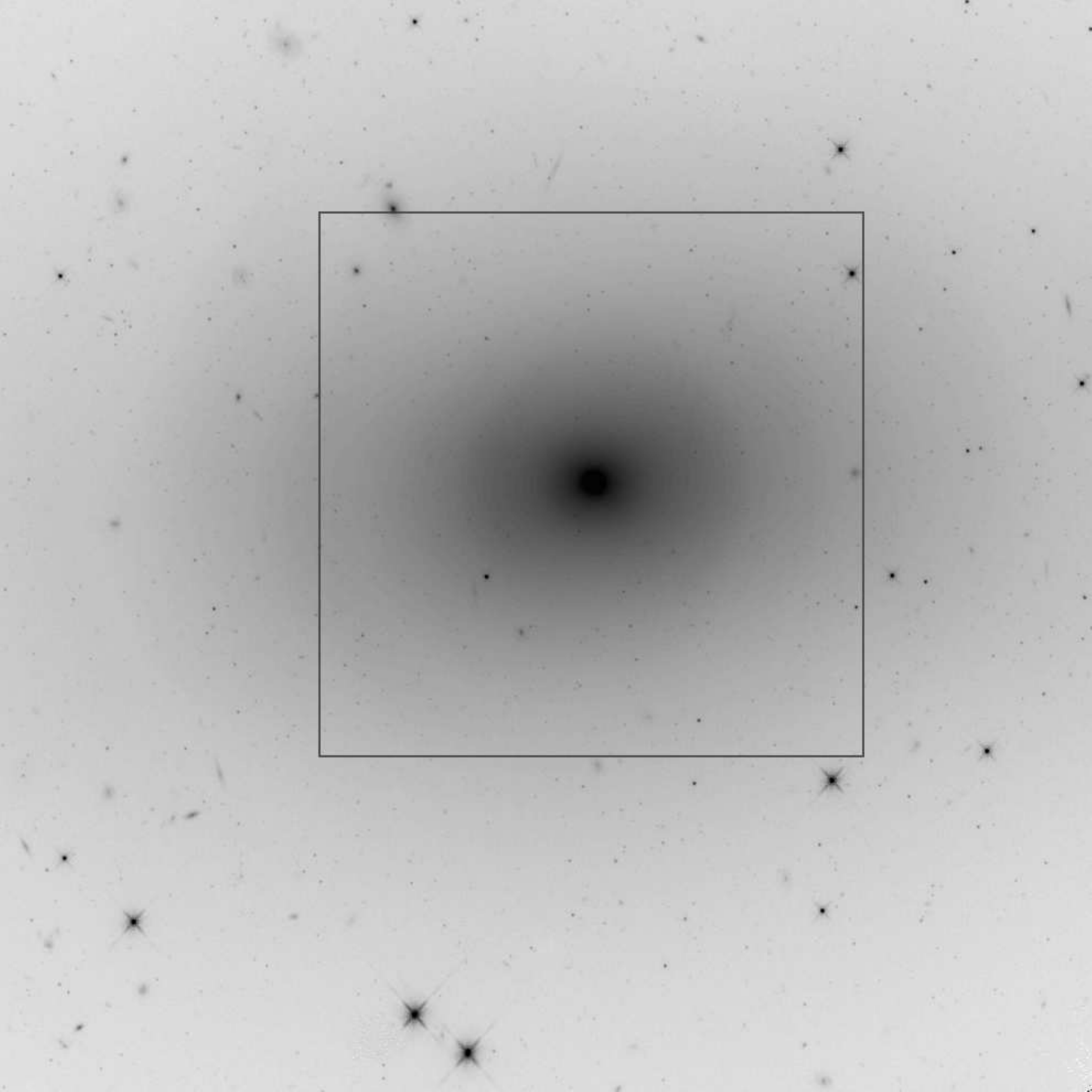}
\includegraphics[scale=0.4]{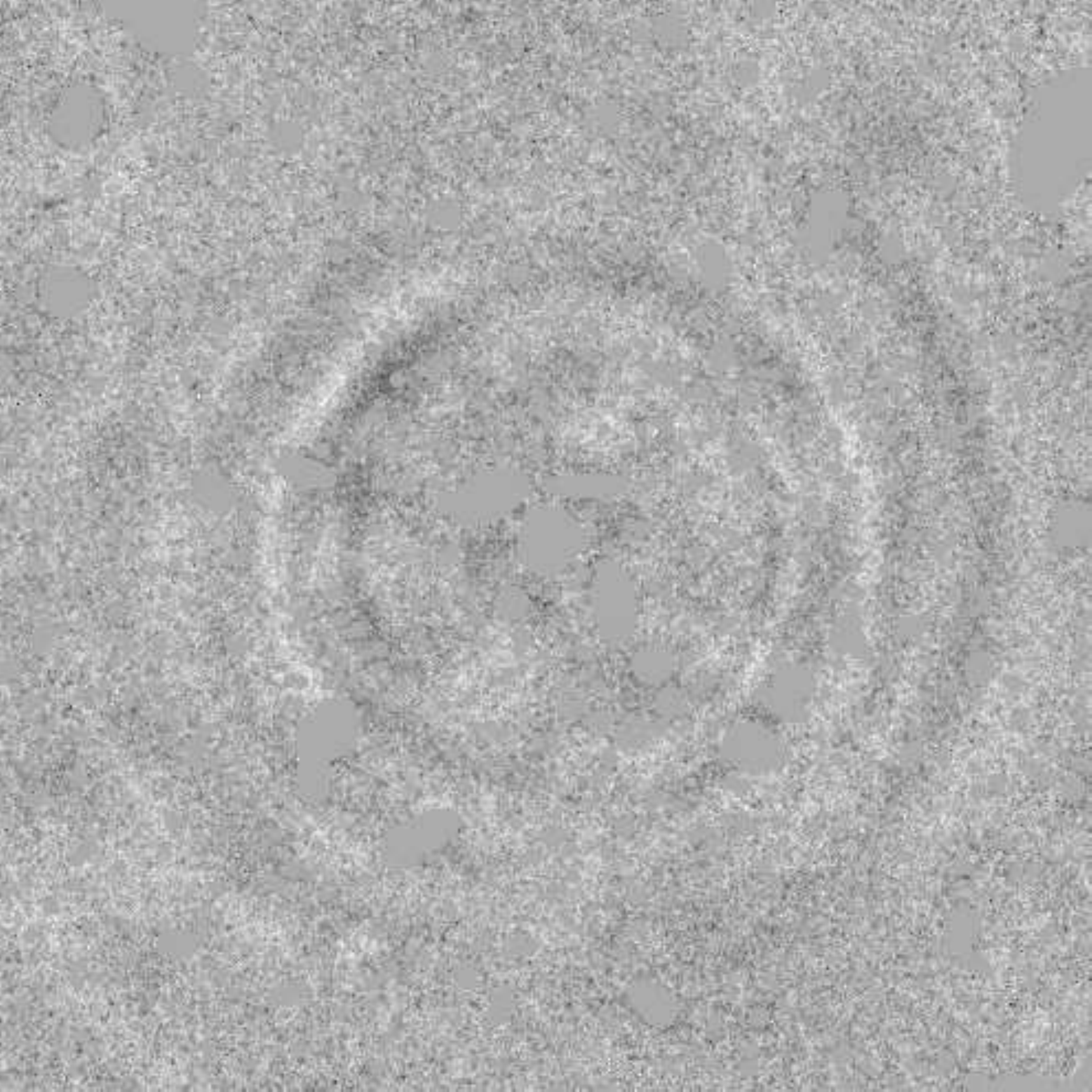} \\
\vspace{10pt}
\includegraphics[scale=0.4]{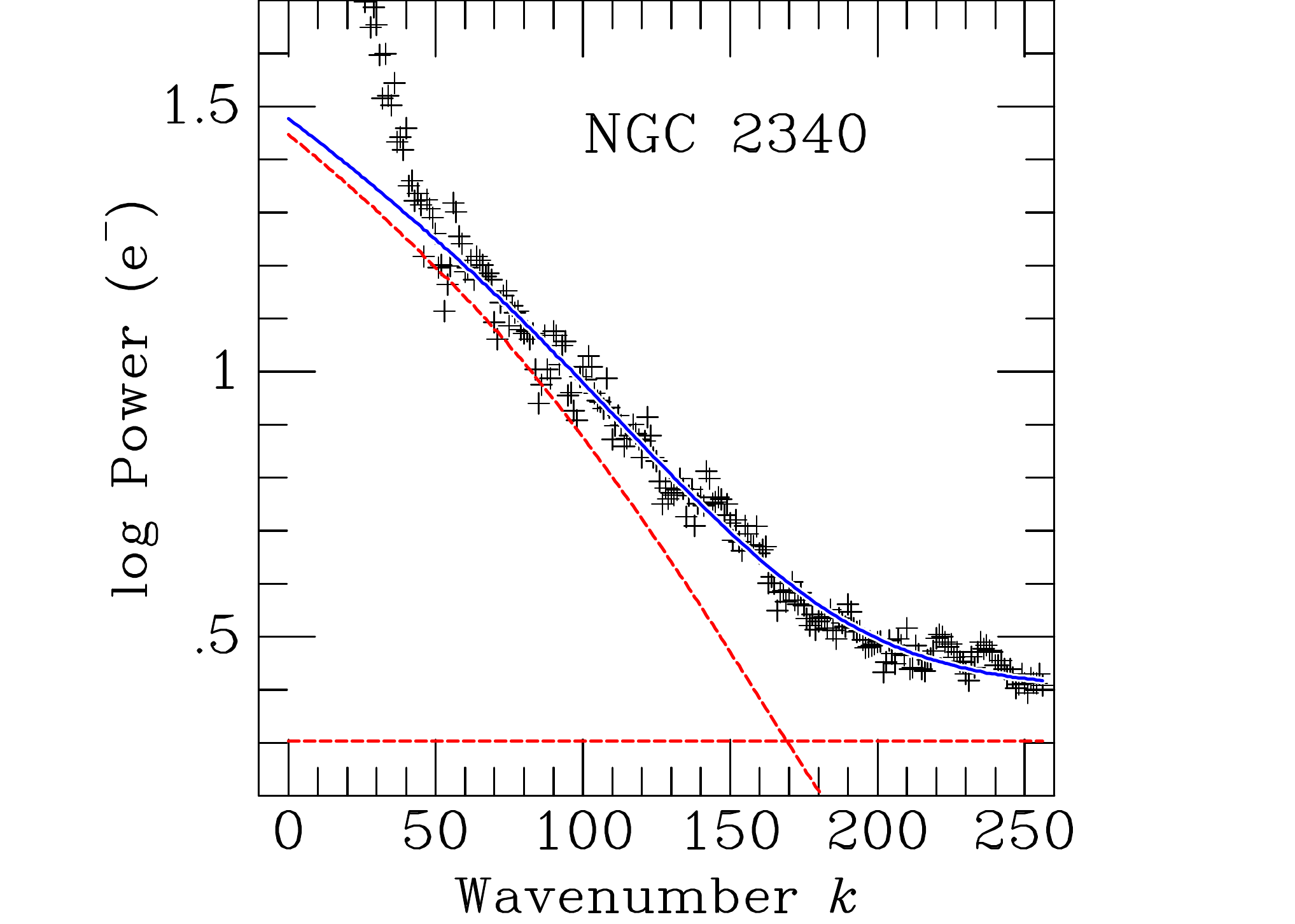}
\hspace{-25pt}
\includegraphics[scale=0.4]{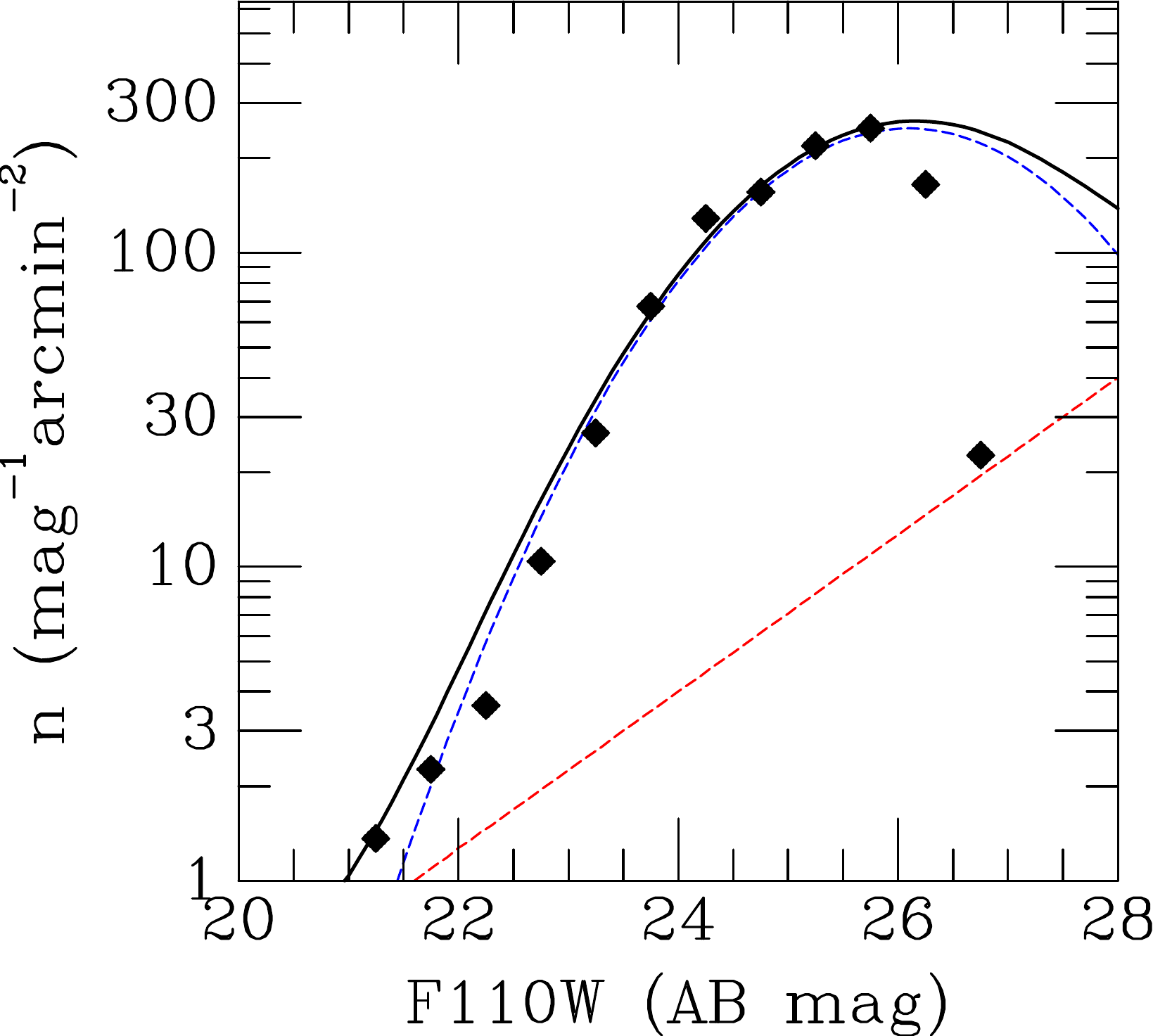}
\caption{Combined figure for NGC~2340.}
\end{center}
\end{figure*}
\clearpage

\begin{figure*}
\begin{center}
\includegraphics[scale=0.2]{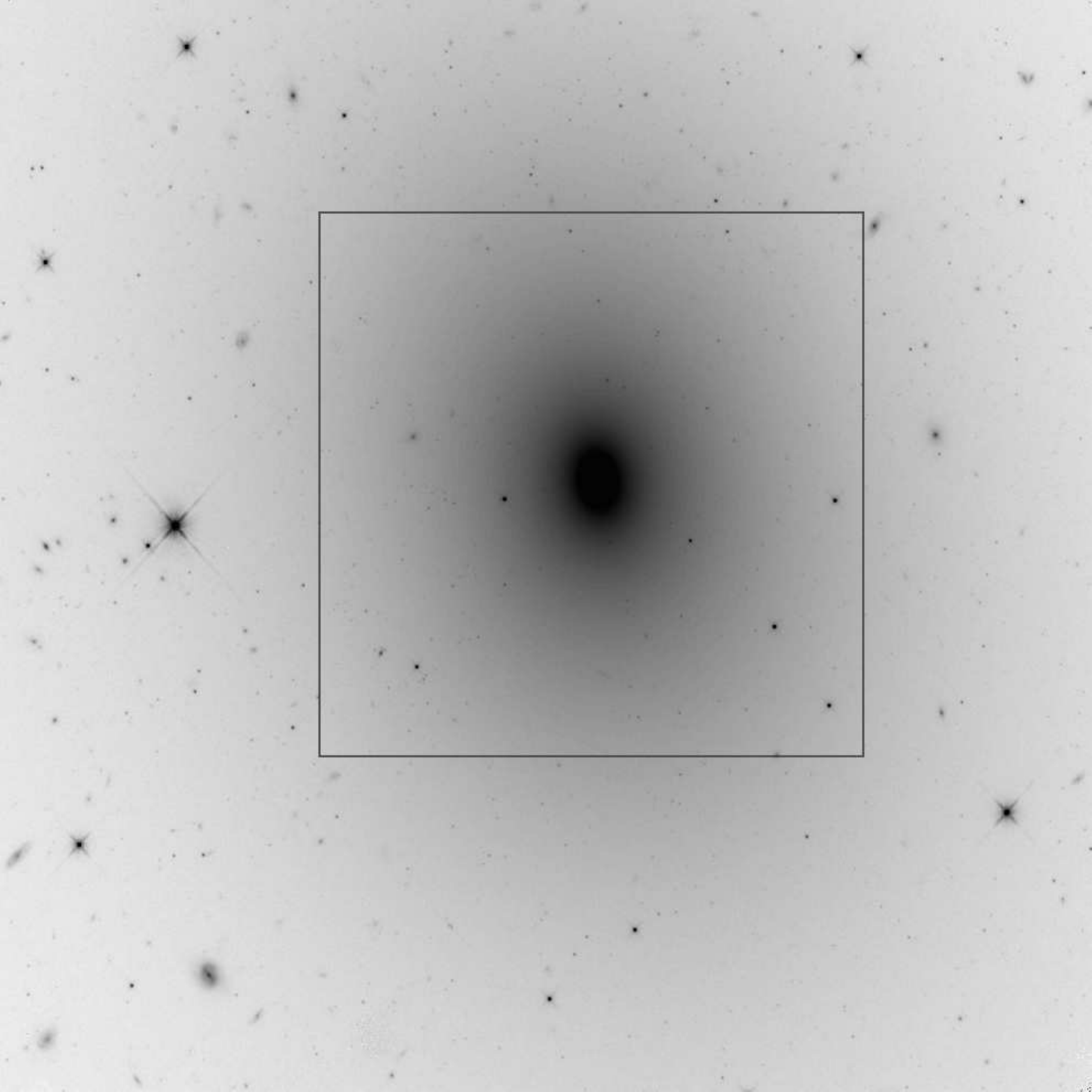}
\includegraphics[scale=0.4]{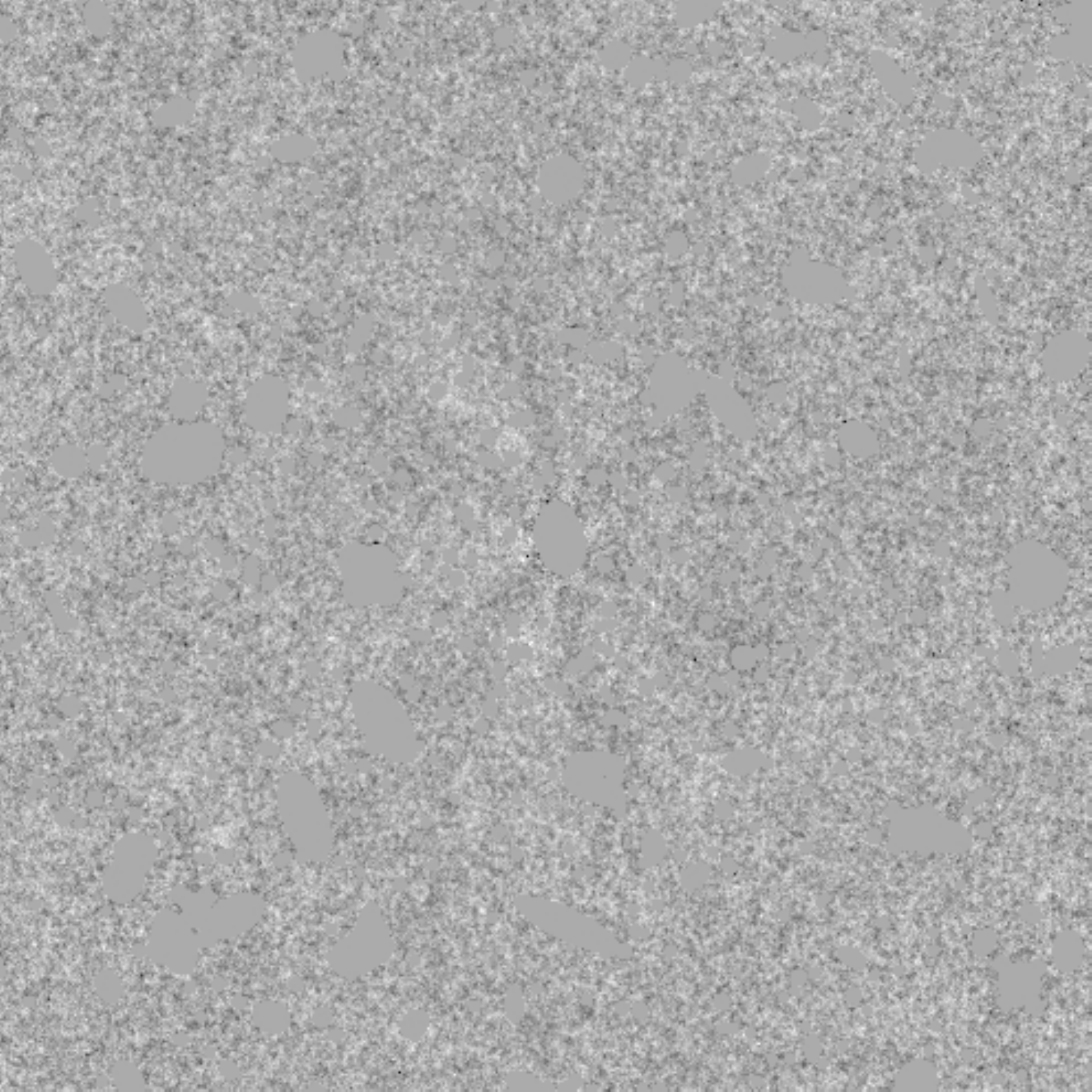} \\
\vspace{10pt}
\includegraphics[scale=0.4]{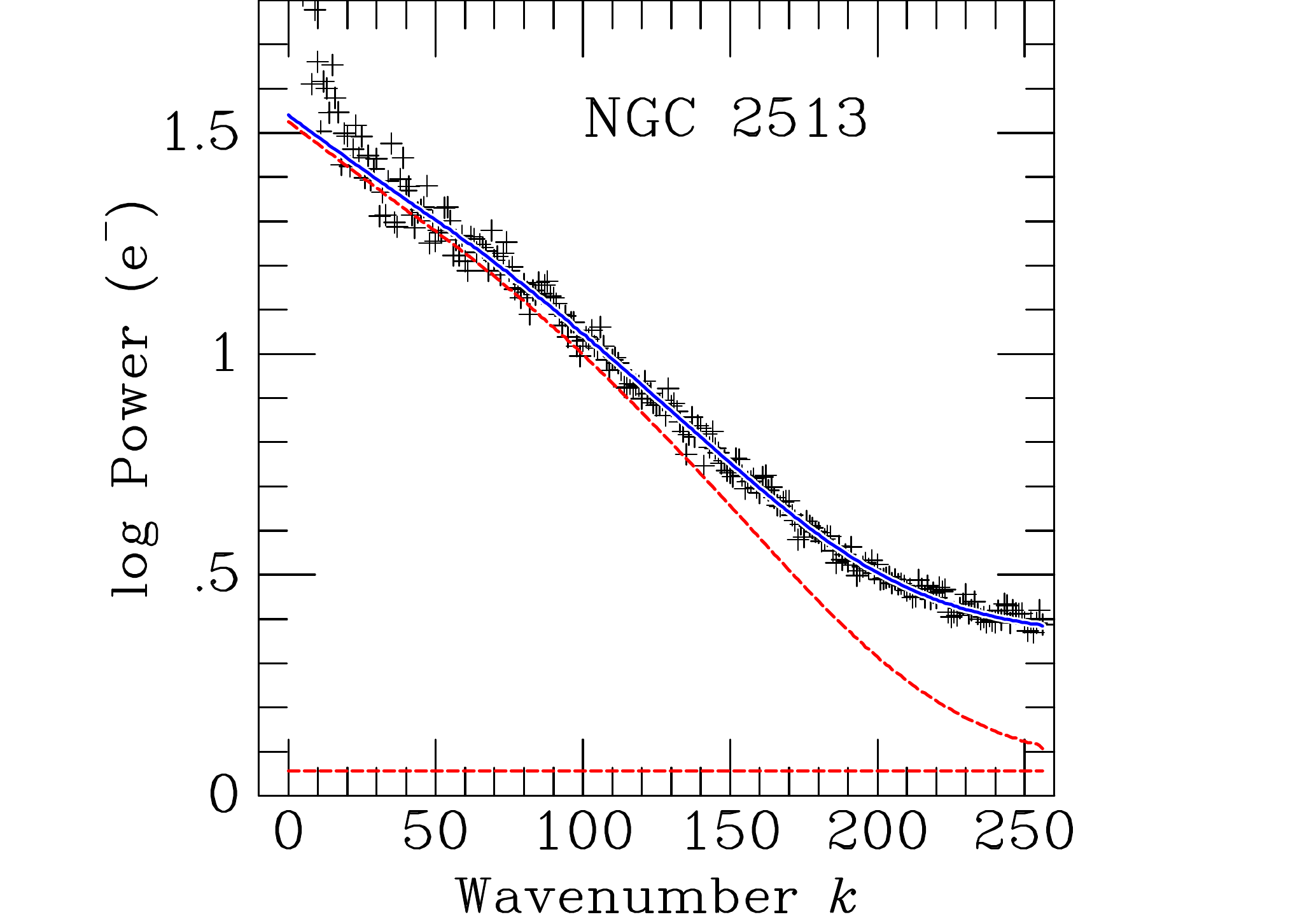}
\hspace{-25pt}
\includegraphics[scale=0.4]{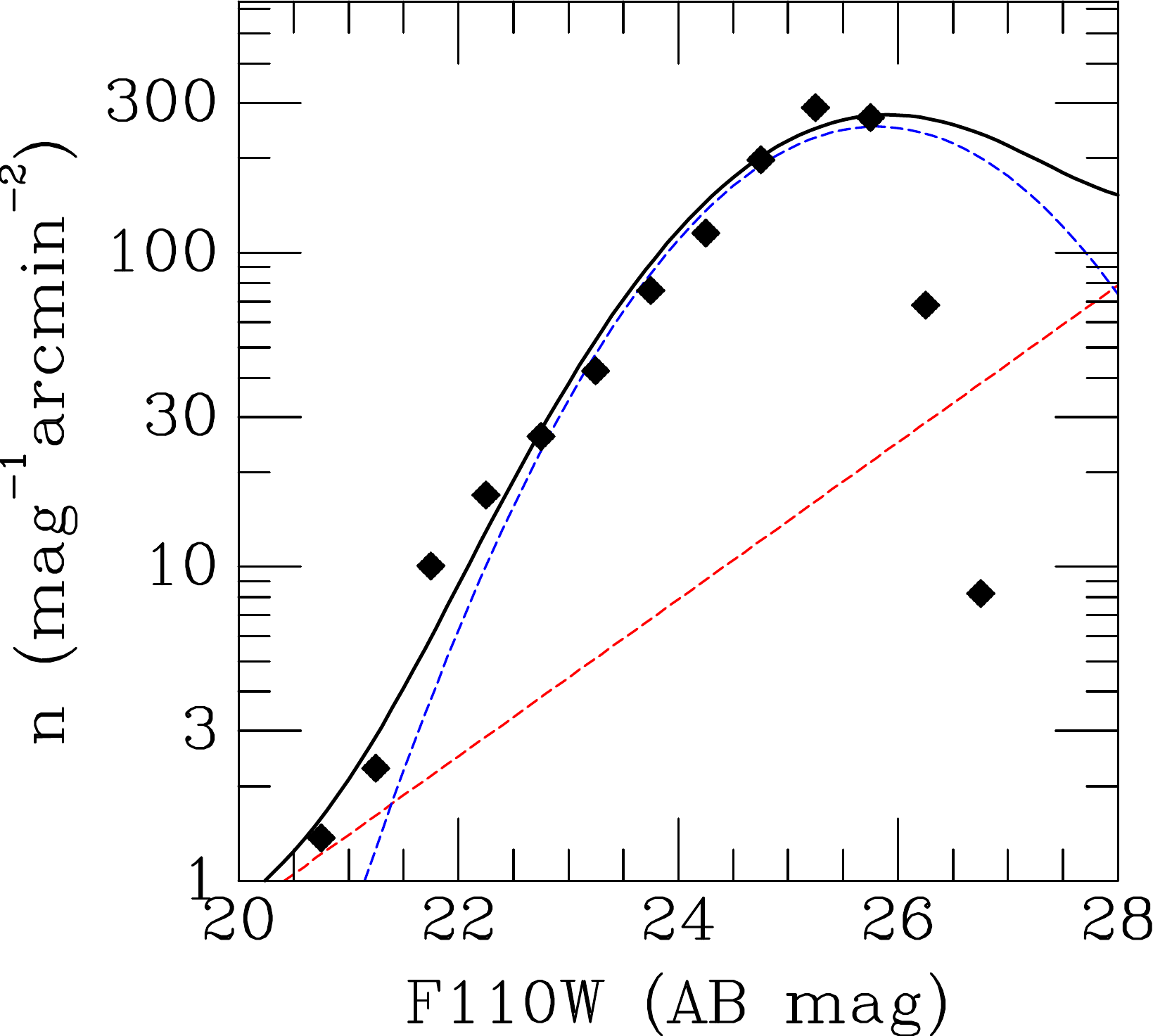}
\caption{Combined figure for NGC~2513.}
\end{center}
\end{figure*}
\clearpage

\begin{figure*}
\begin{center}
\includegraphics[scale=0.2]{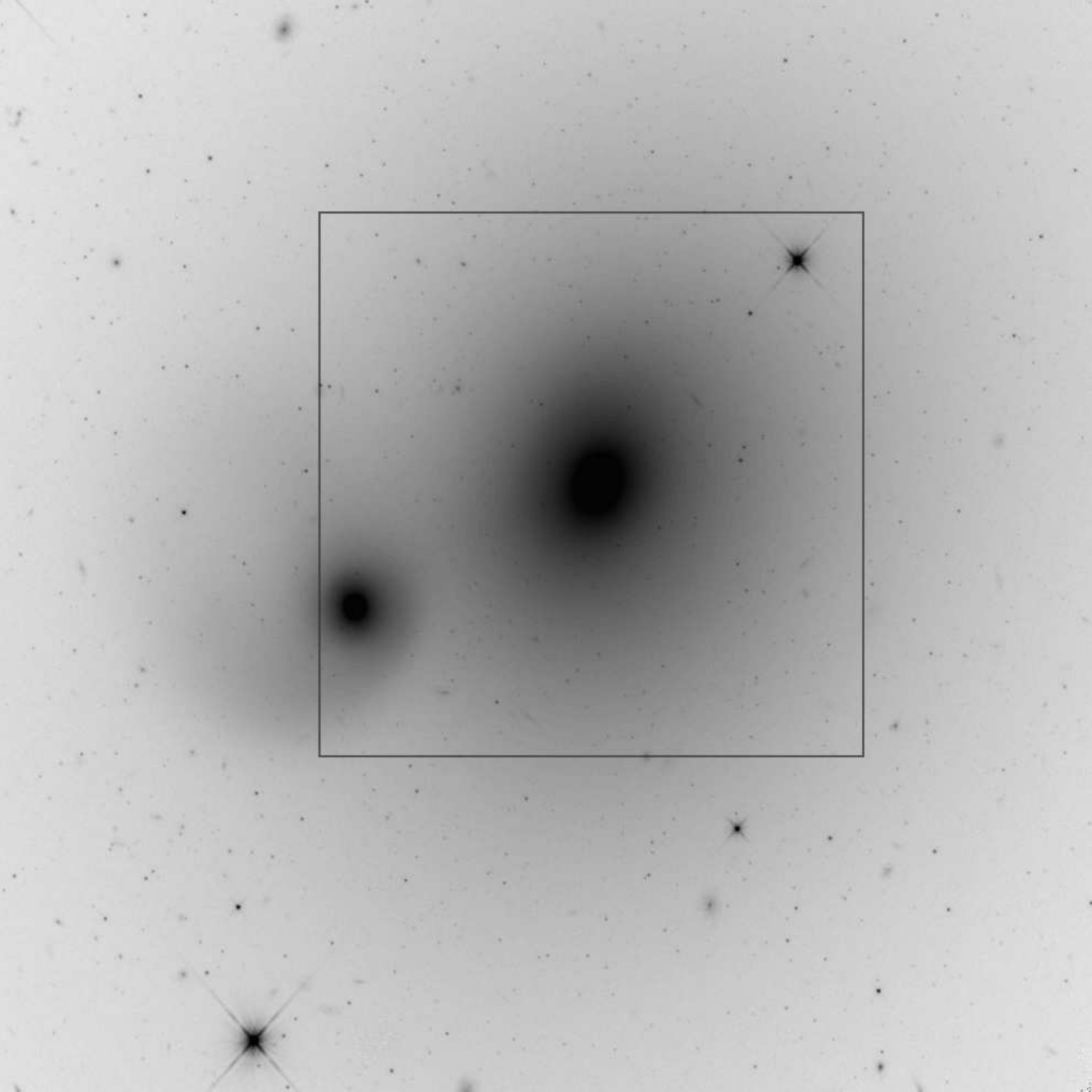}
\includegraphics[scale=0.4]{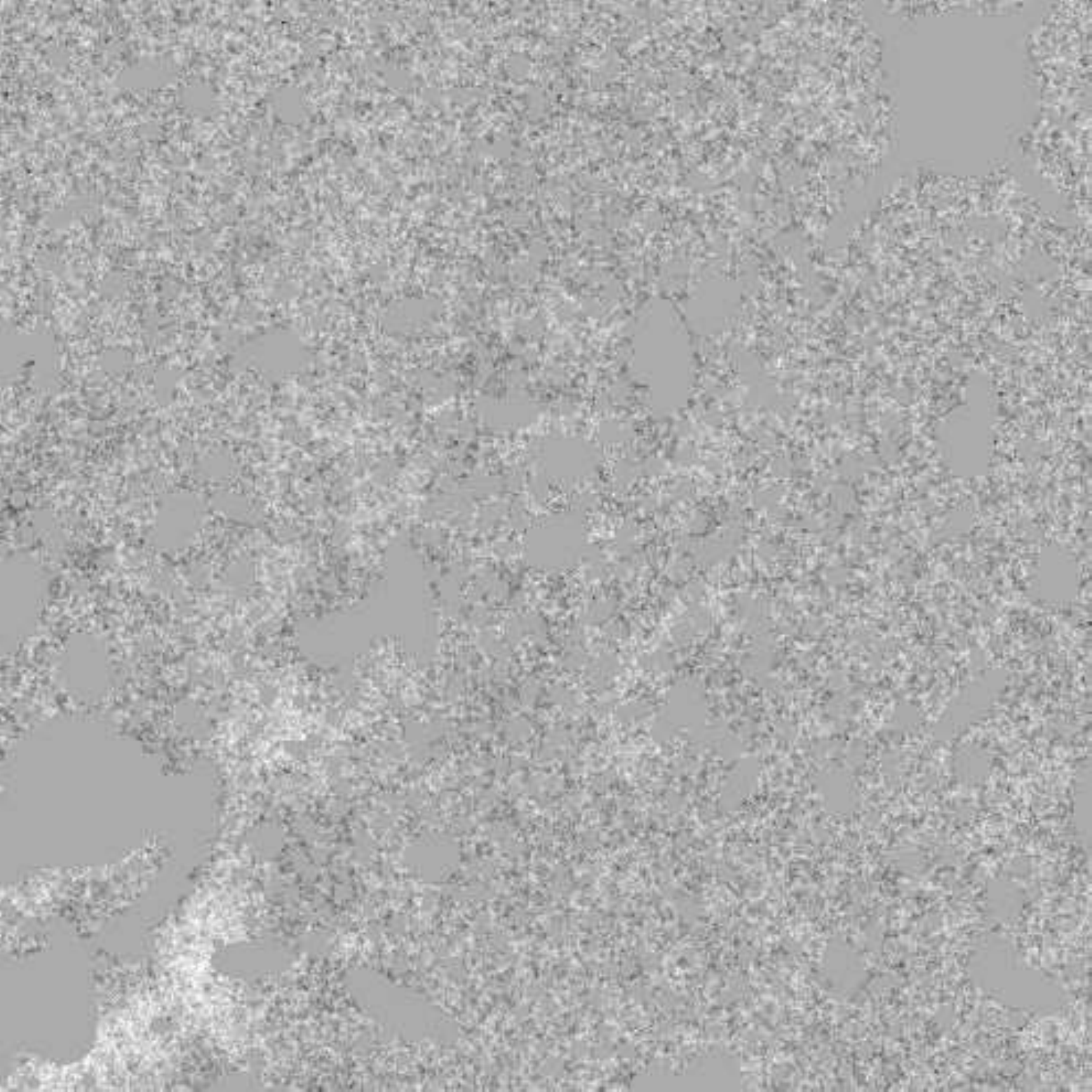} \\
\vspace{10pt}
\includegraphics[scale=0.4]{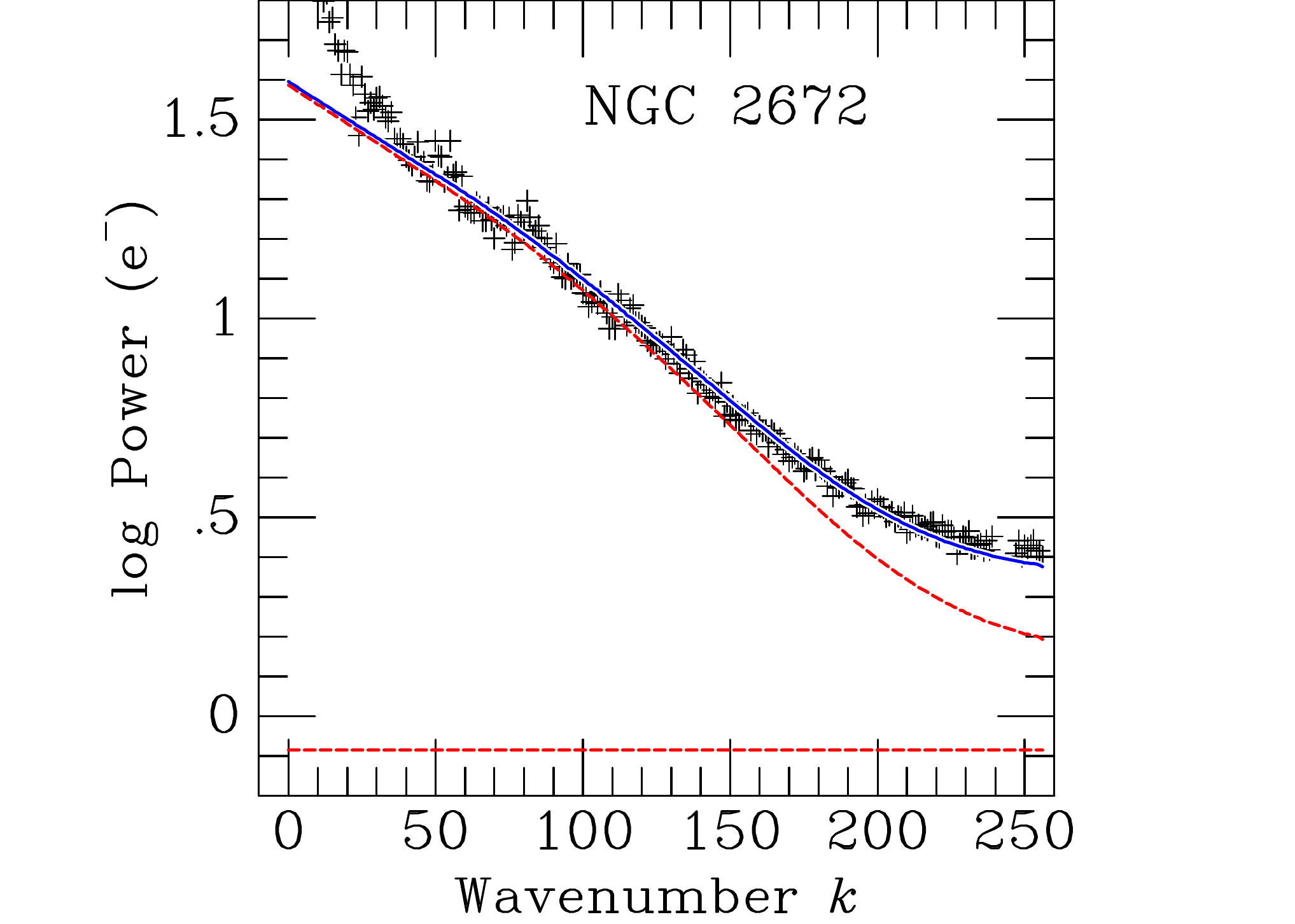}
\hspace{-25pt}
\includegraphics[scale=0.4]{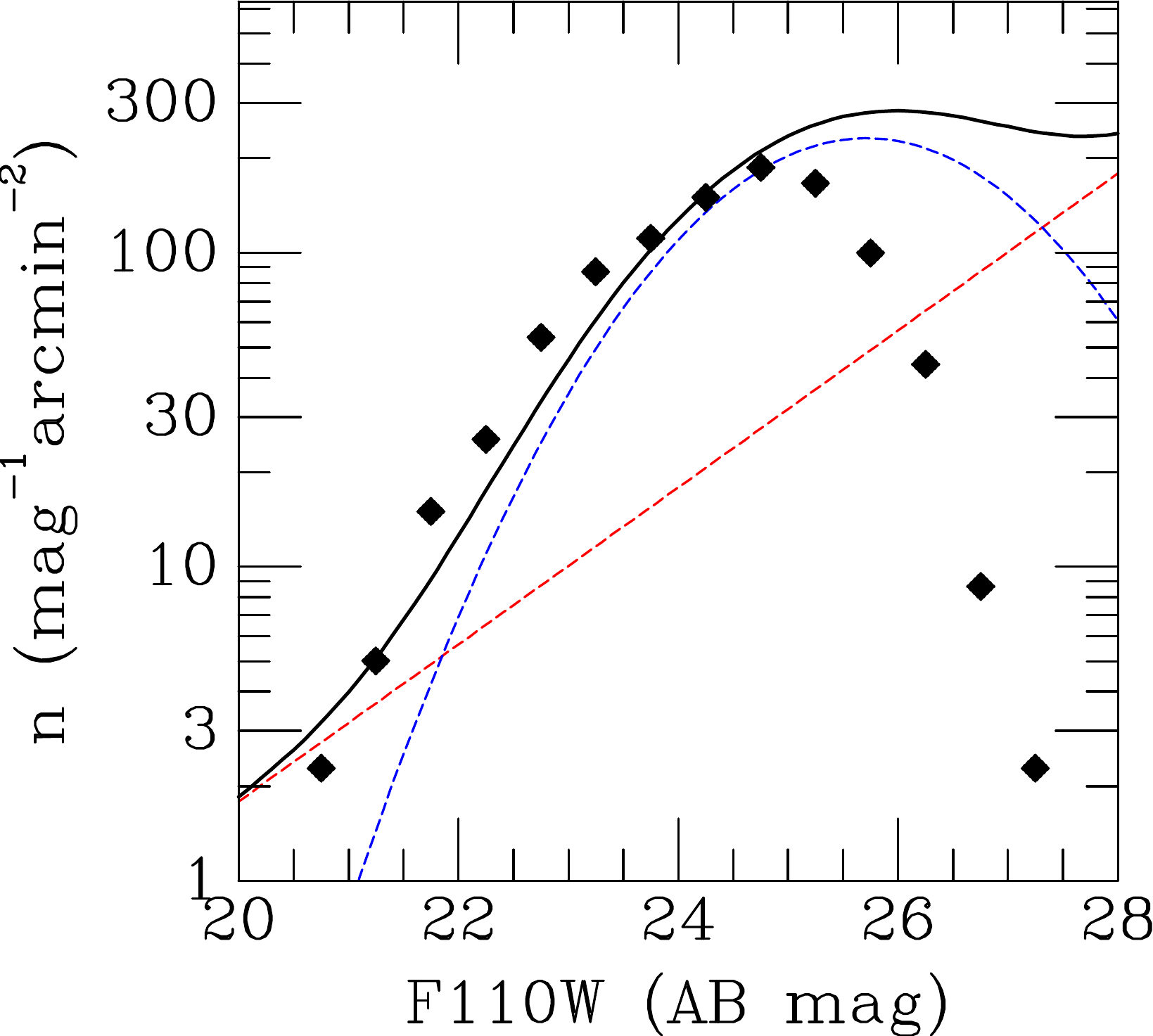}
\caption{Combined figure for NGC~2672.}
\end{center}
\end{figure*}
\clearpage

\begin{figure*}
\begin{center}
\includegraphics[scale=0.2]{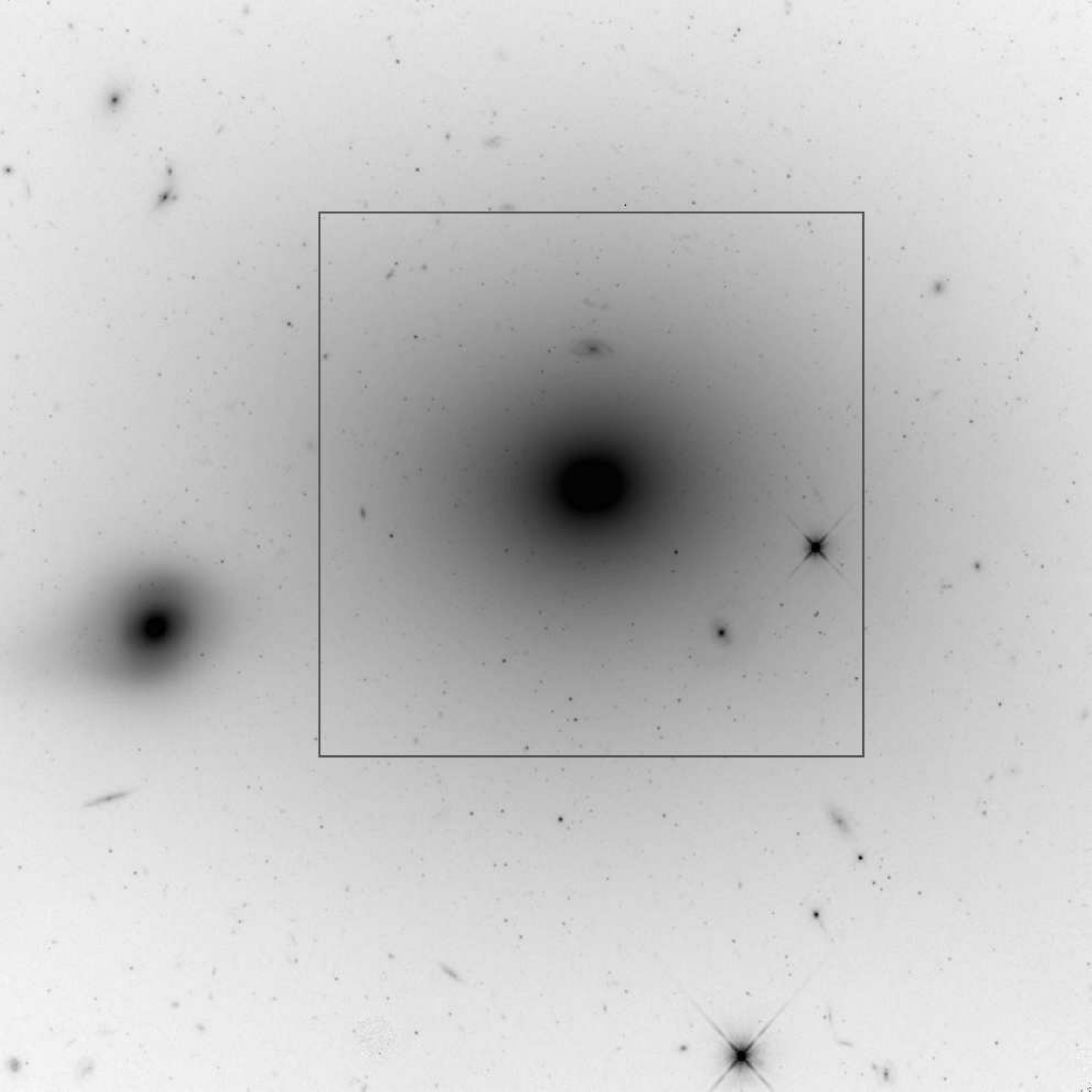}
\includegraphics[scale=0.4]{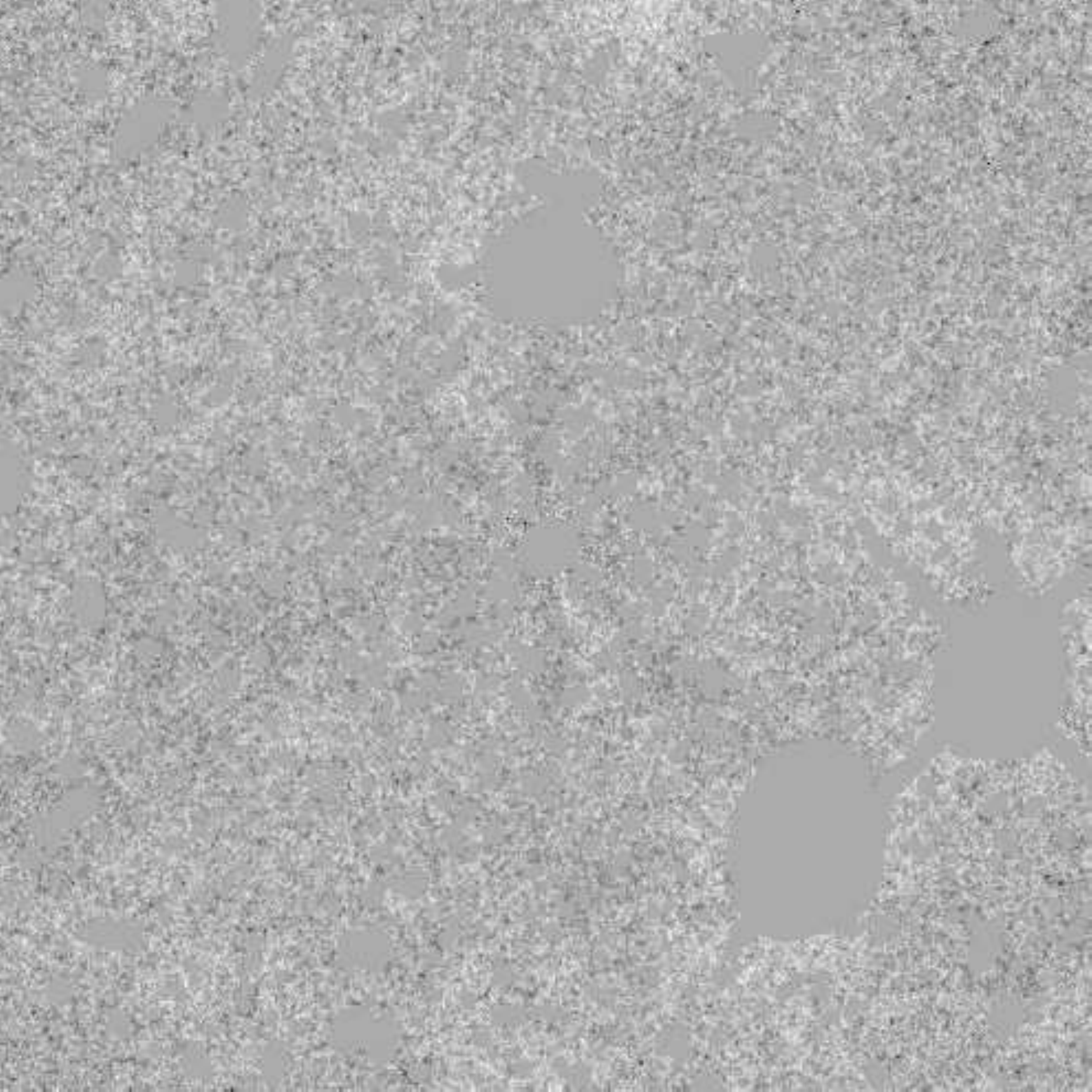} \\
\vspace{10pt}
\includegraphics[scale=0.4]{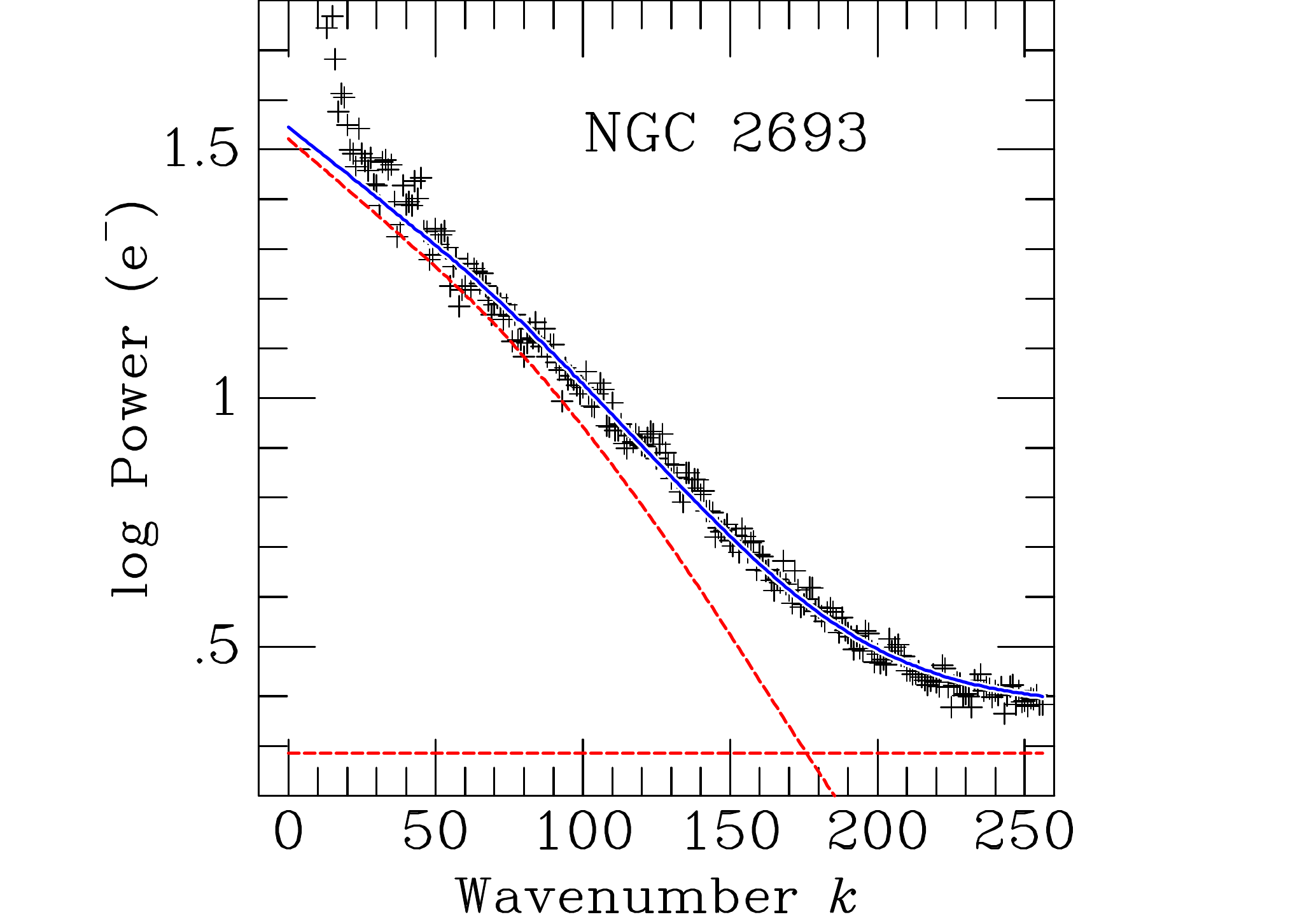}
\hspace{-25pt}
\includegraphics[scale=0.4]{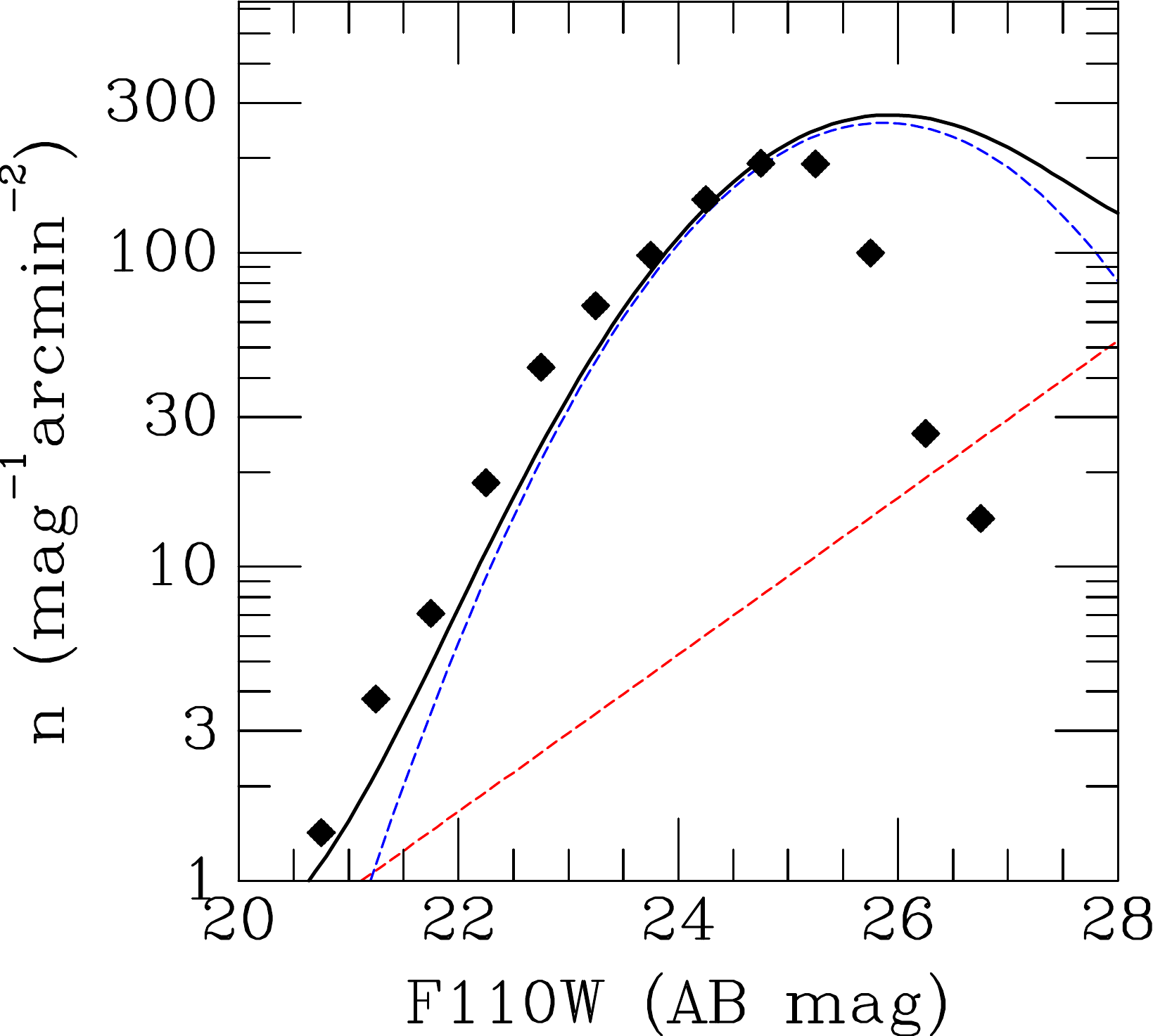}
\caption{Combined figure for NGC~2693.}
\end{center}
\end{figure*}
\clearpage

\begin{figure*}
\begin{center}
\includegraphics[scale=0.2]{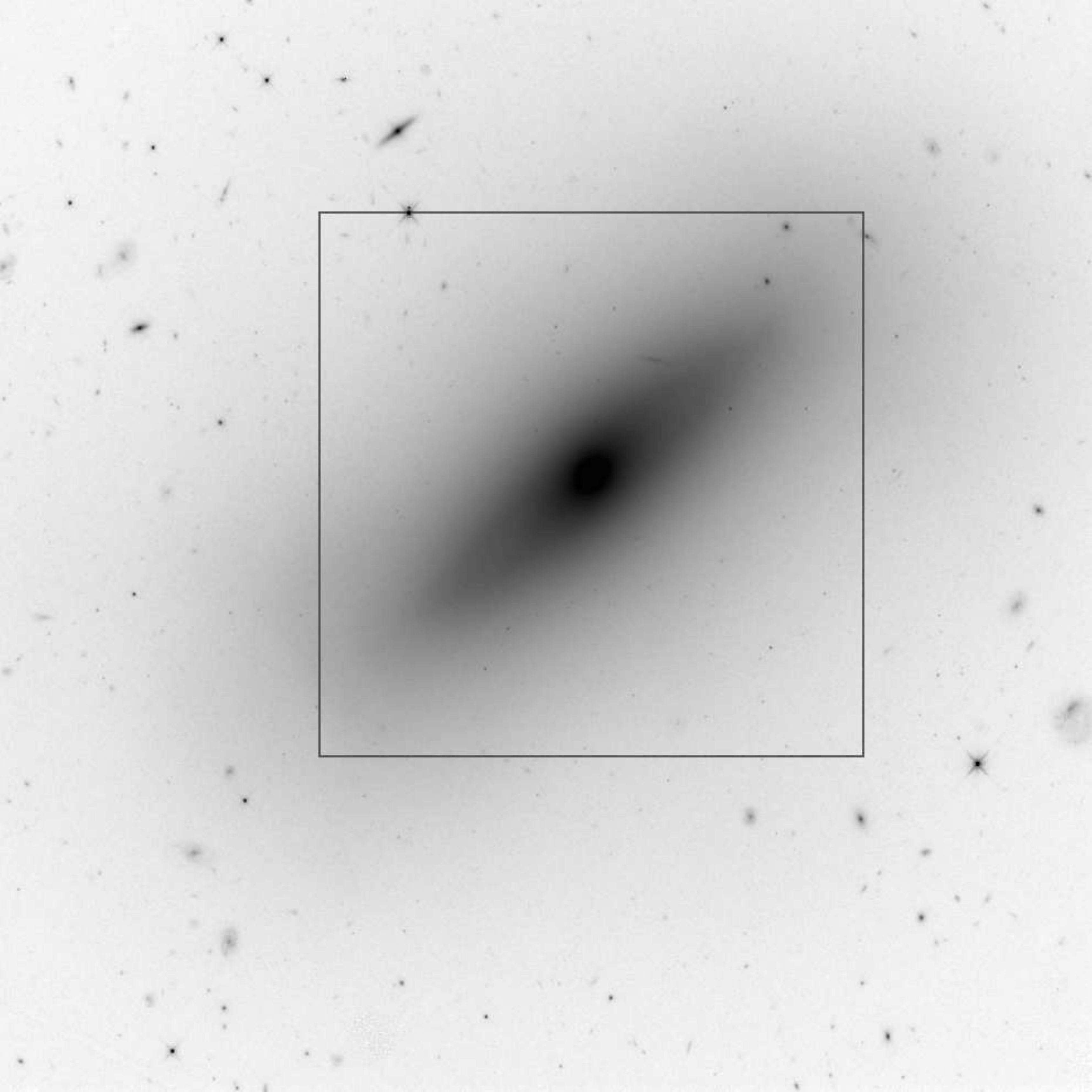}
\includegraphics[scale=0.4]{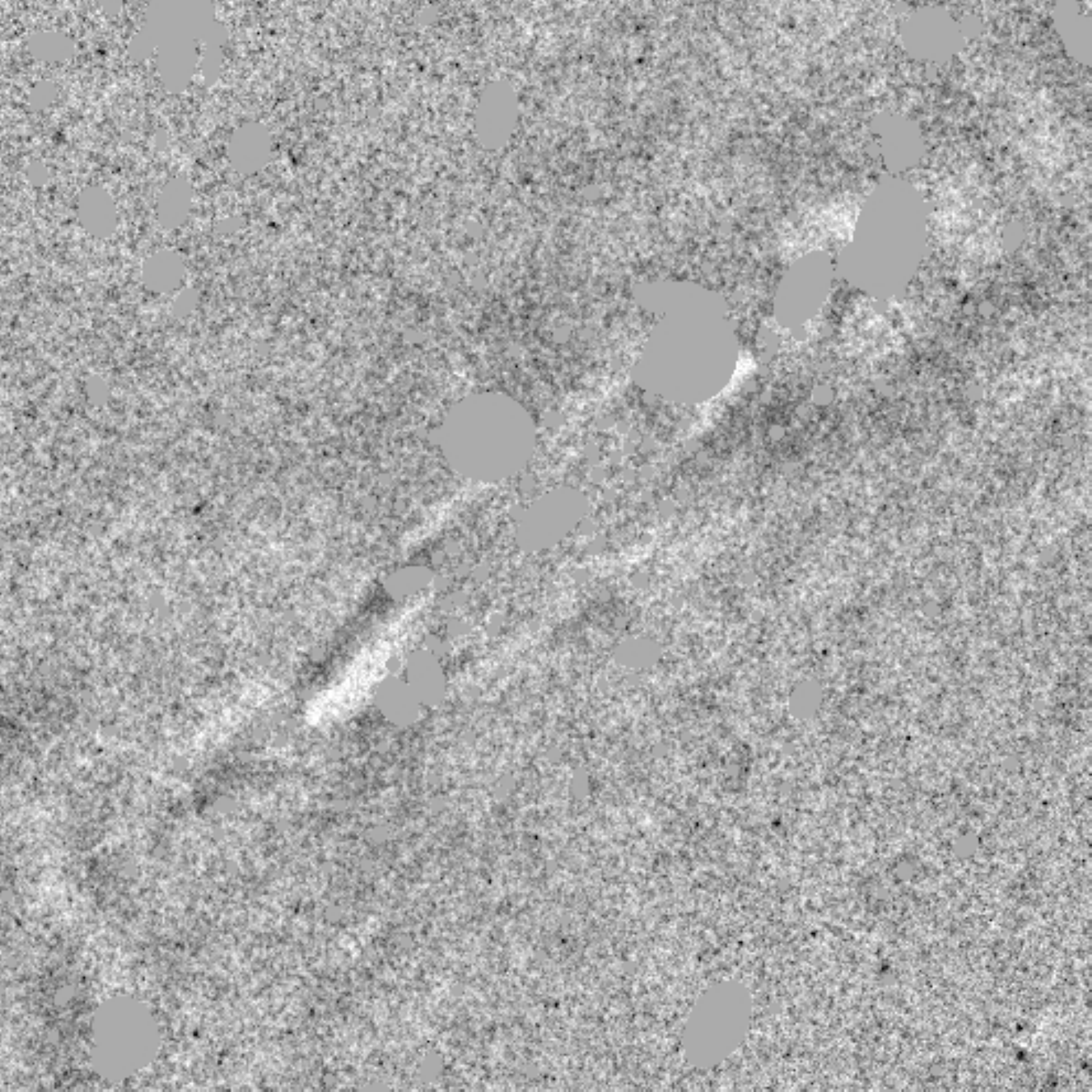} \\
\vspace{10pt}
\includegraphics[scale=0.4]{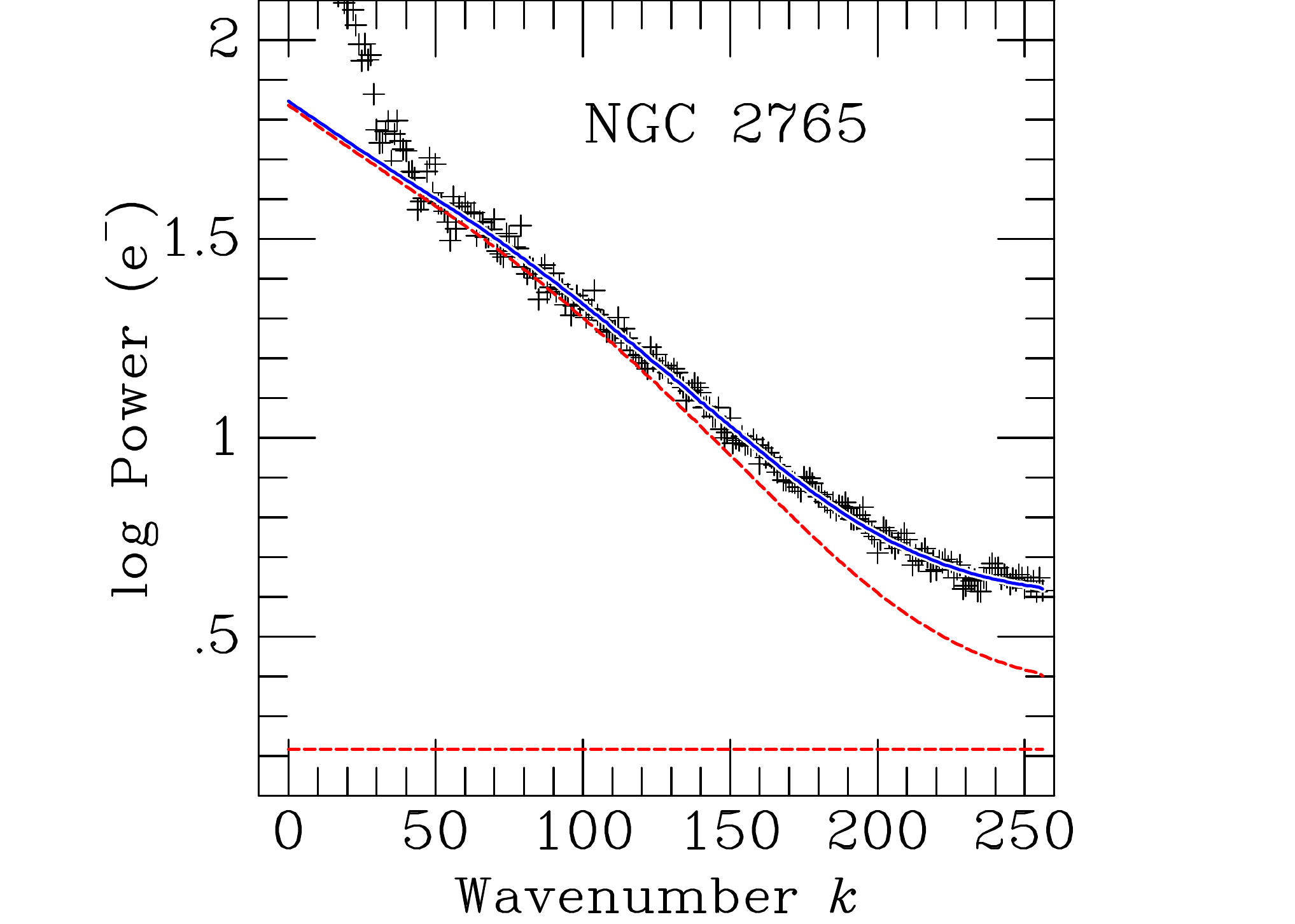}
\hspace{-25pt}
\includegraphics[scale=0.4]{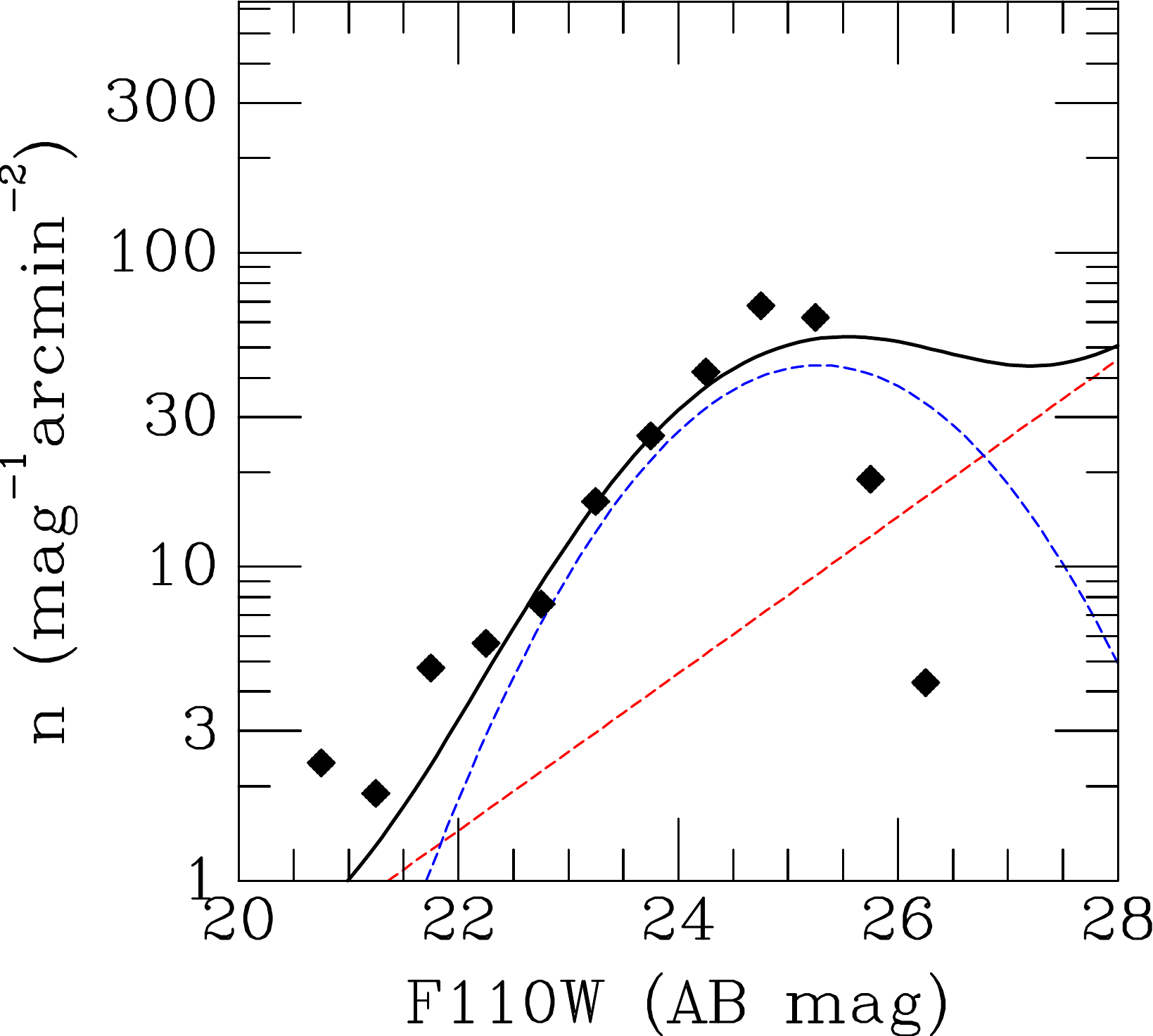}
\caption{Combined figure for NGC~2765.}
\end{center}
\end{figure*}
\clearpage

\begin{figure*}
\begin{center}
\includegraphics[scale=0.2]{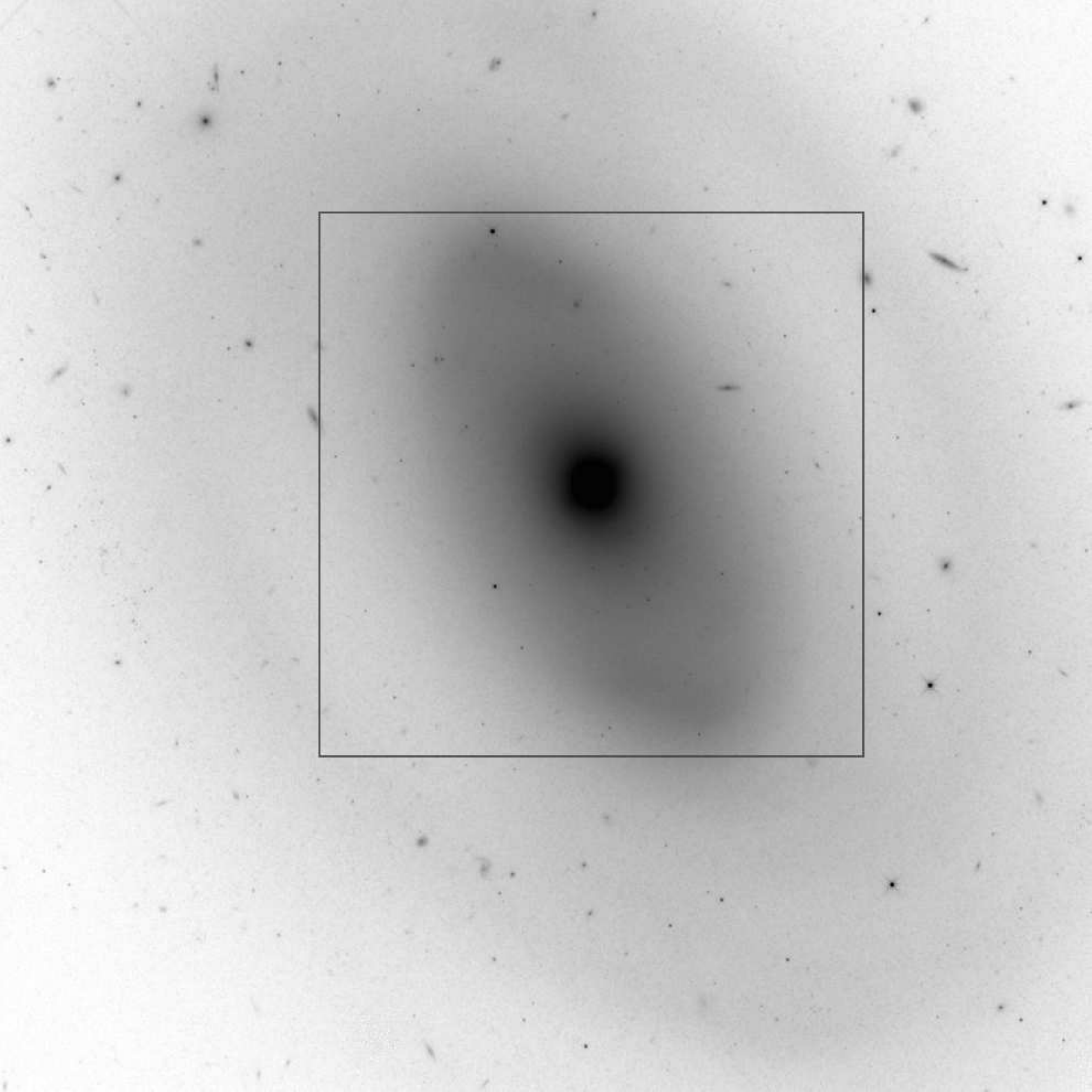}
\includegraphics[scale=0.4]{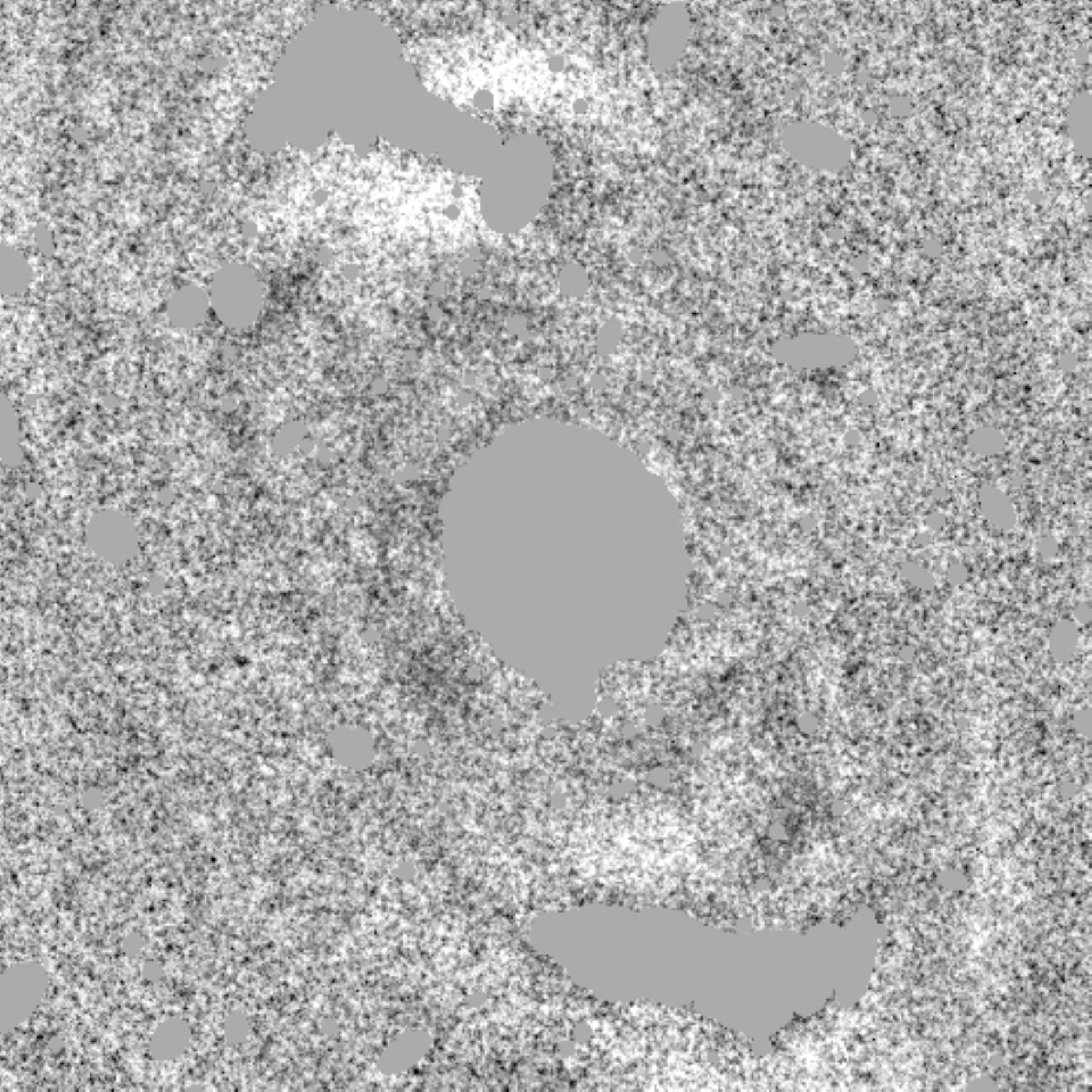} \\
\vspace{10pt}
\includegraphics[scale=0.4]{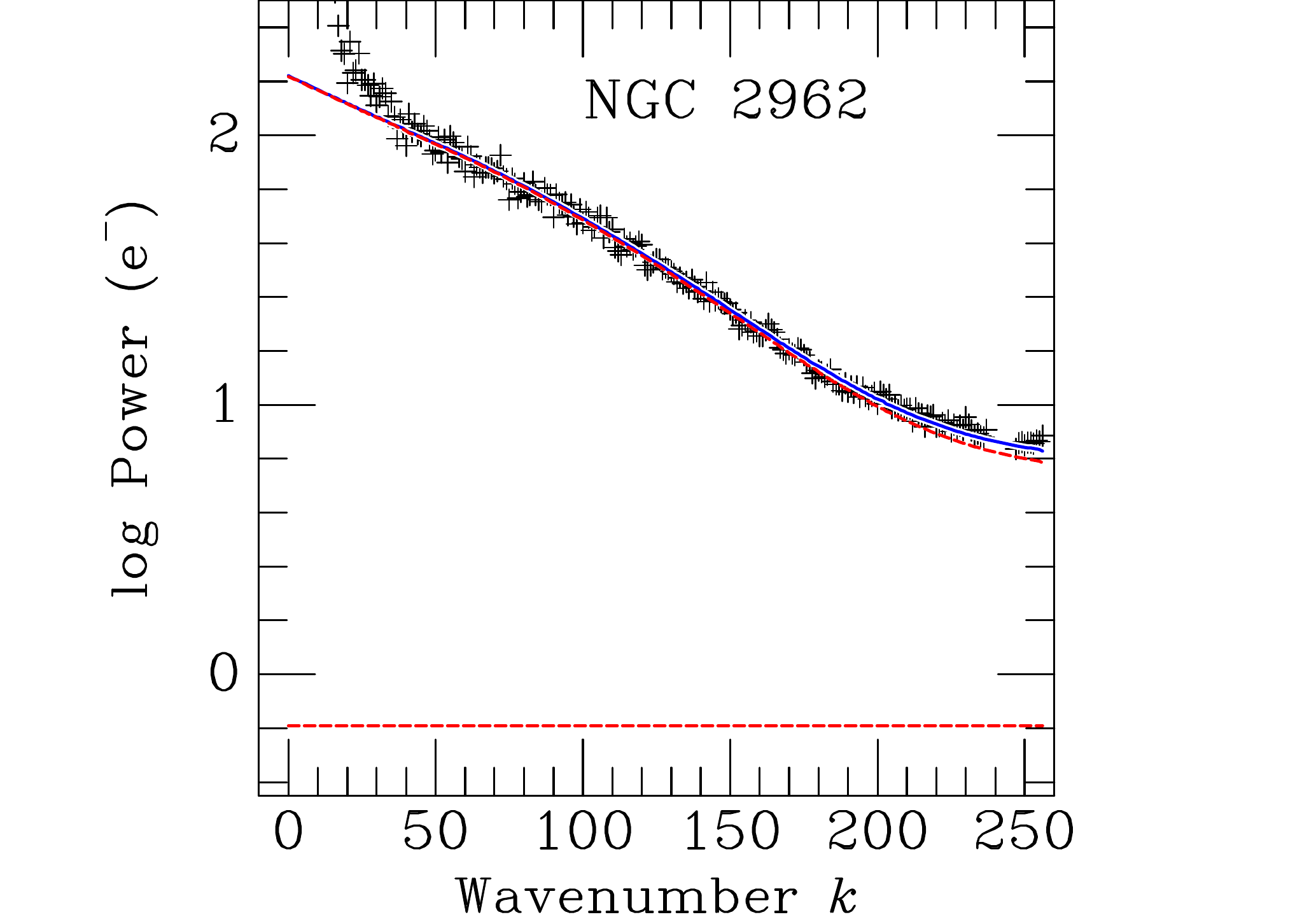}
\hspace{-25pt}
\includegraphics[scale=0.4]{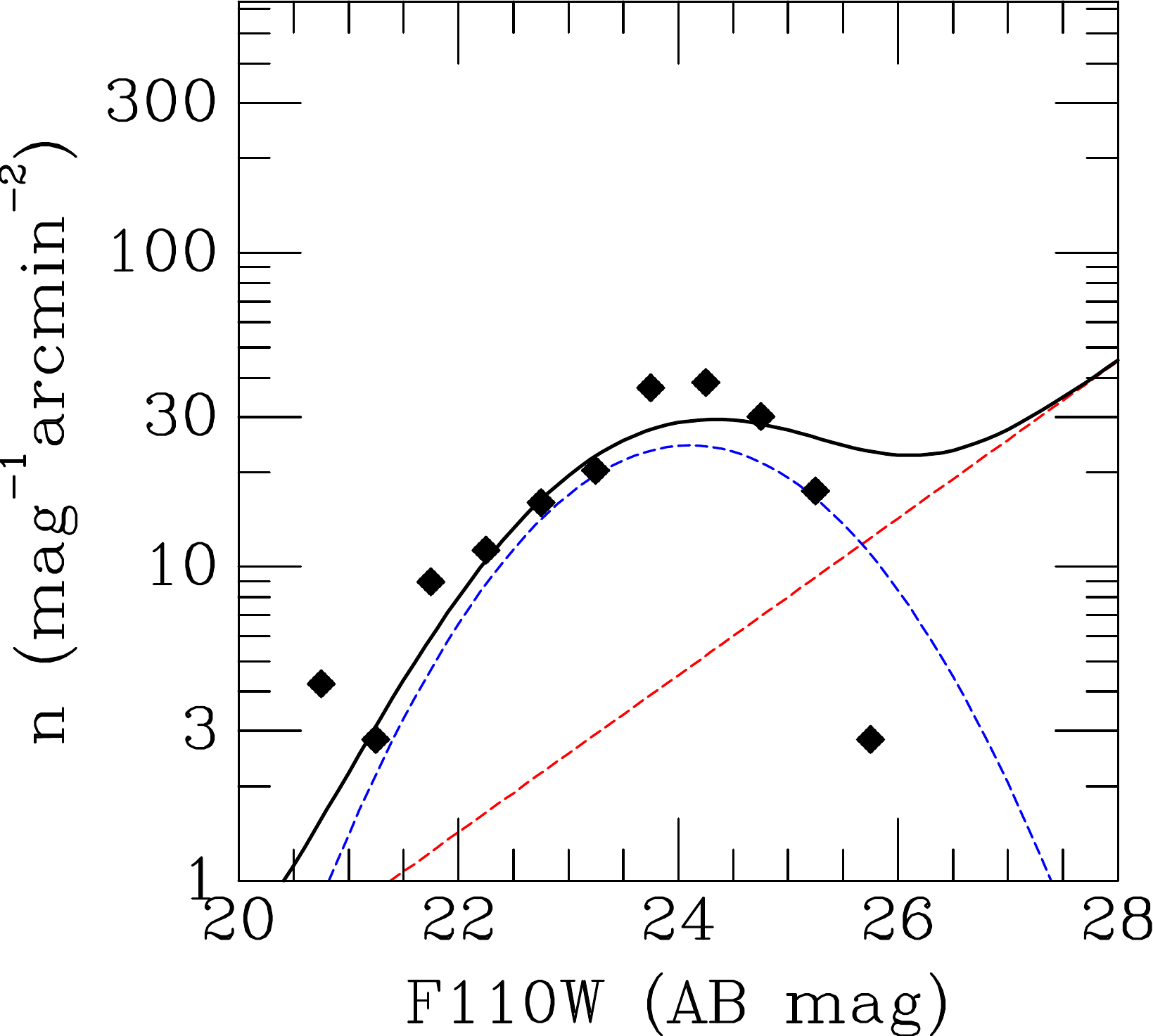}
\caption{Combined figure for NGC~2962.}
\end{center}
\end{figure*}
\clearpage

\begin{figure*}
\begin{center}
\includegraphics[scale=0.2]{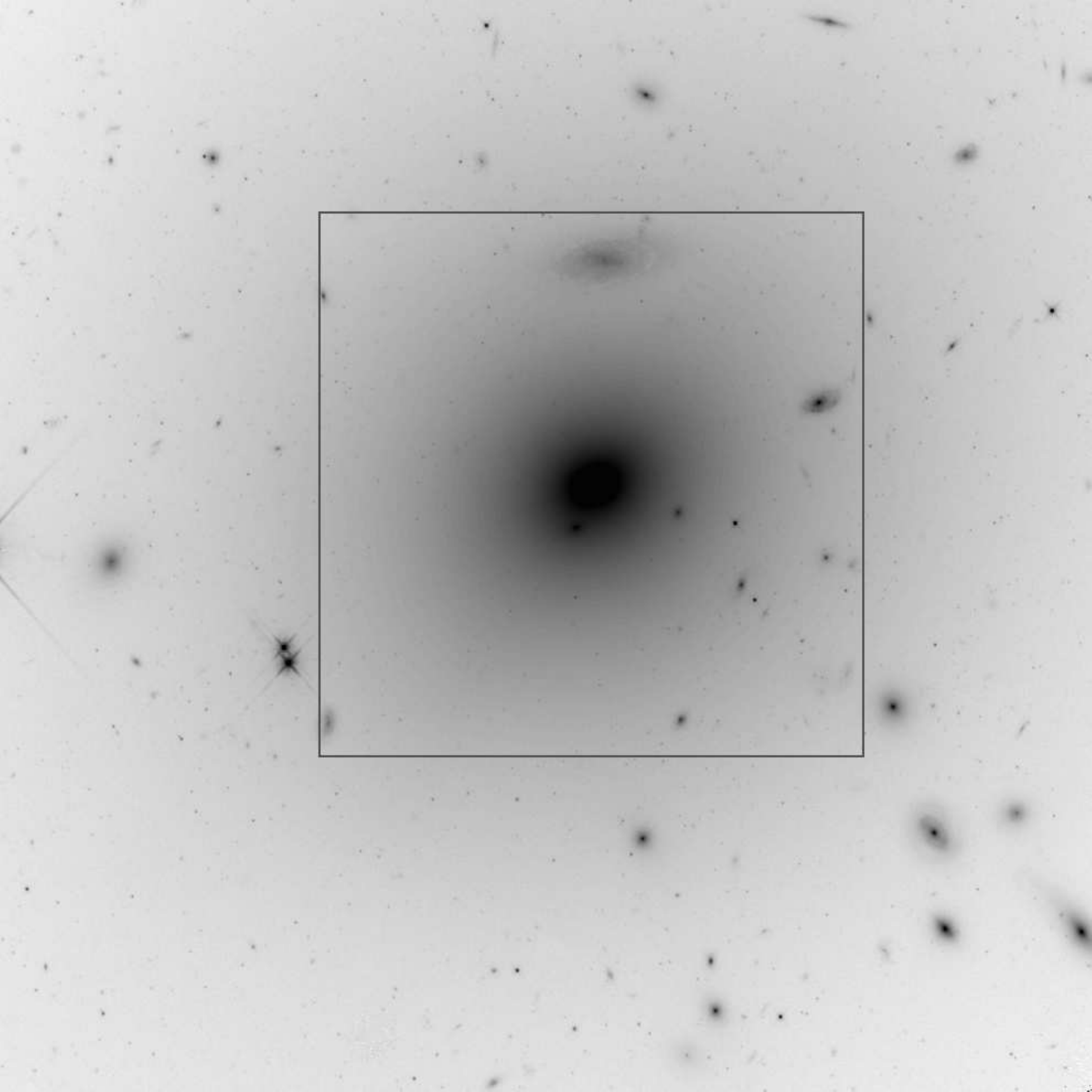}
\includegraphics[scale=0.4]{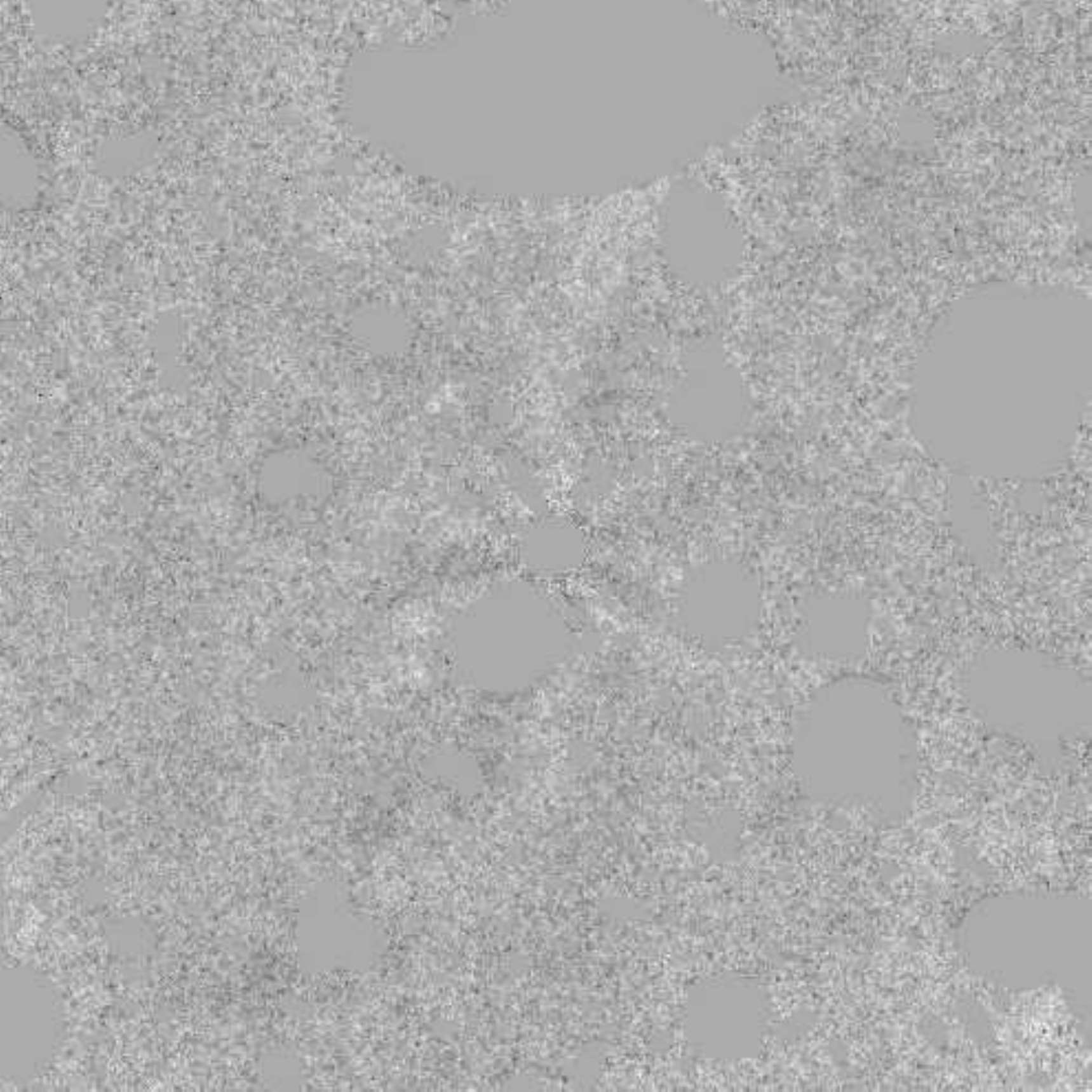} \\
\vspace{10pt}
\includegraphics[scale=0.4]{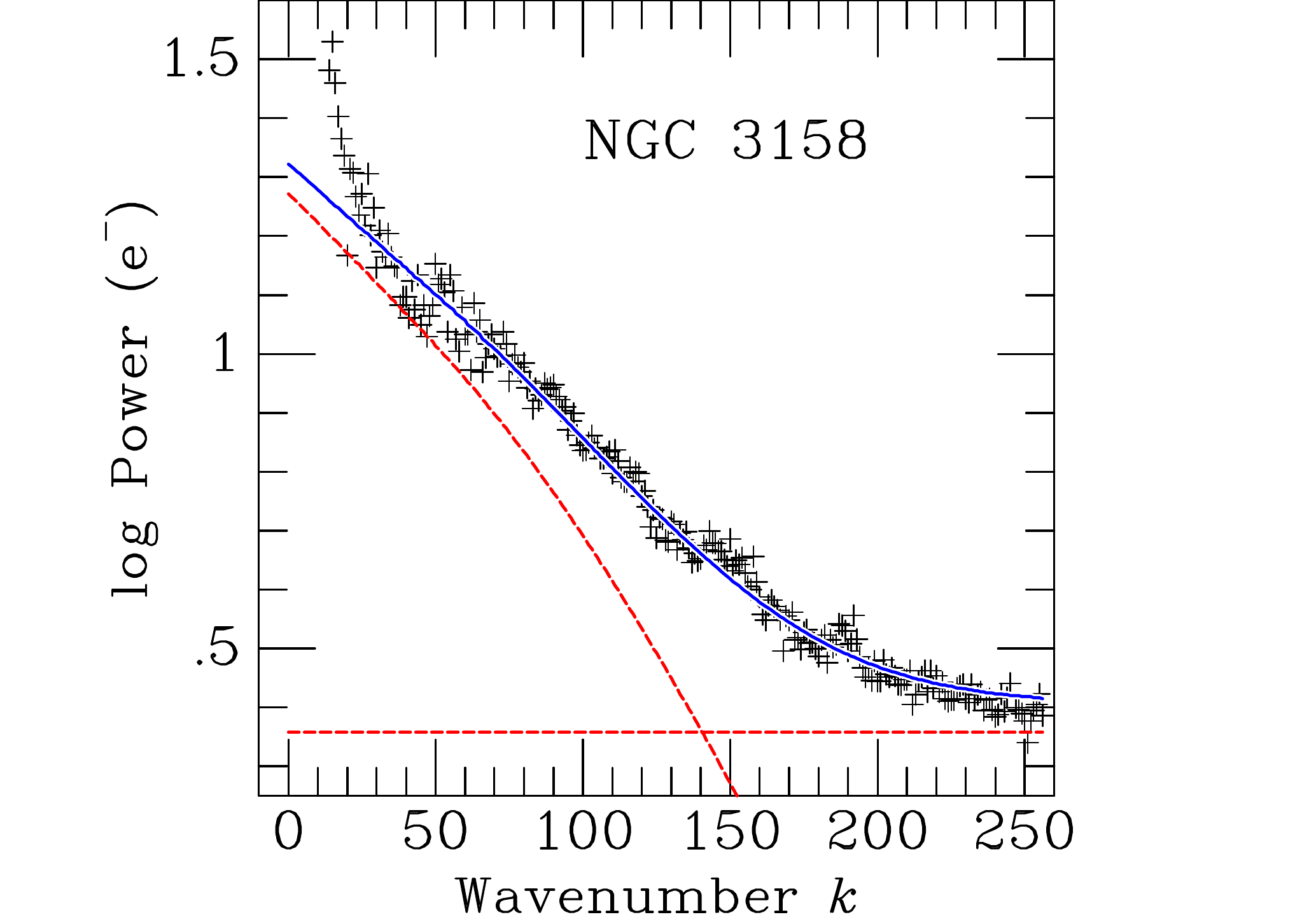}
\hspace{-25pt}
\includegraphics[scale=0.4]{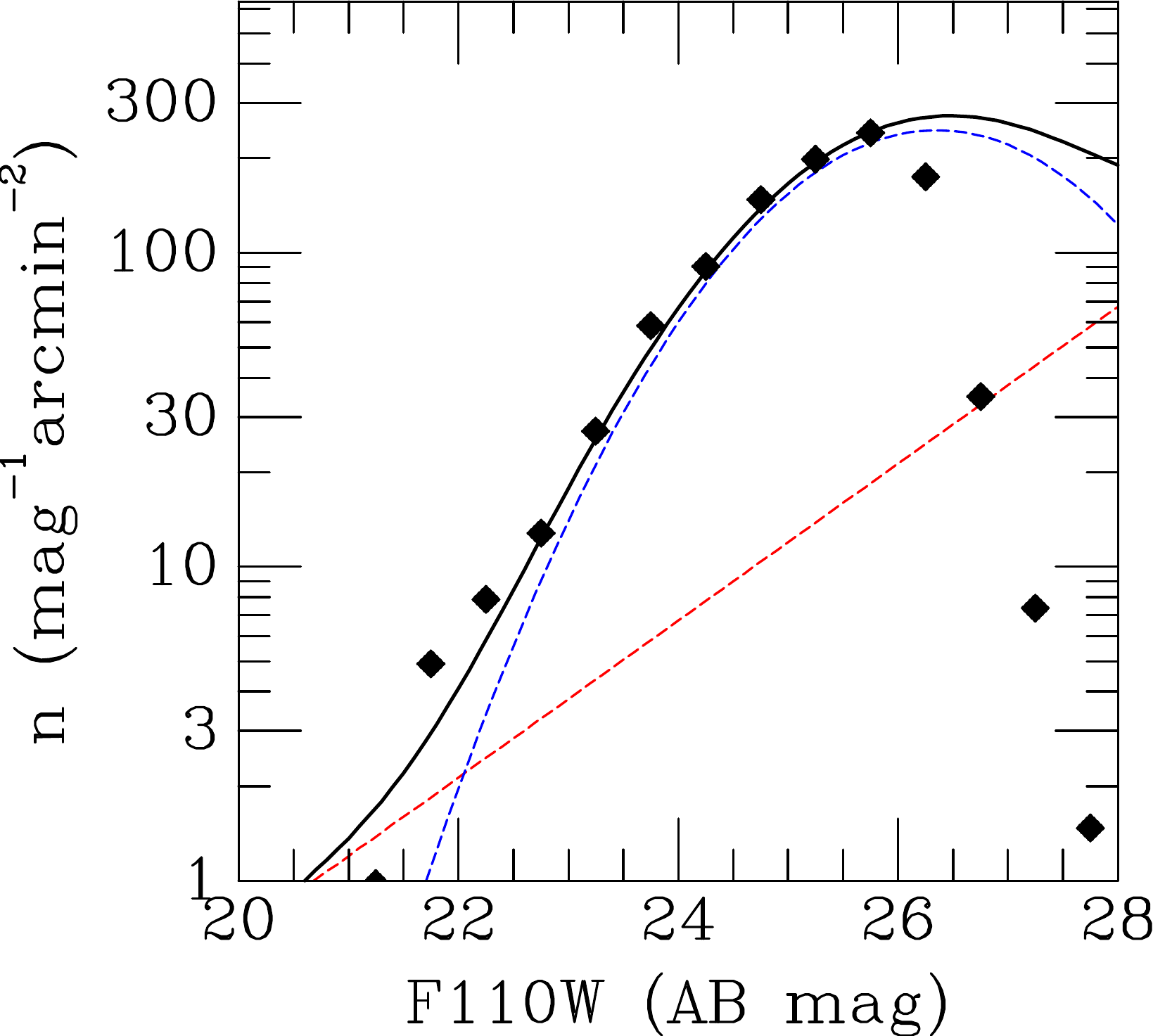}
\caption{Combined figure for NGC~3158.}
\end{center}
\end{figure*}
\clearpage

\begin{figure*}
\begin{center}
\includegraphics[scale=0.2]{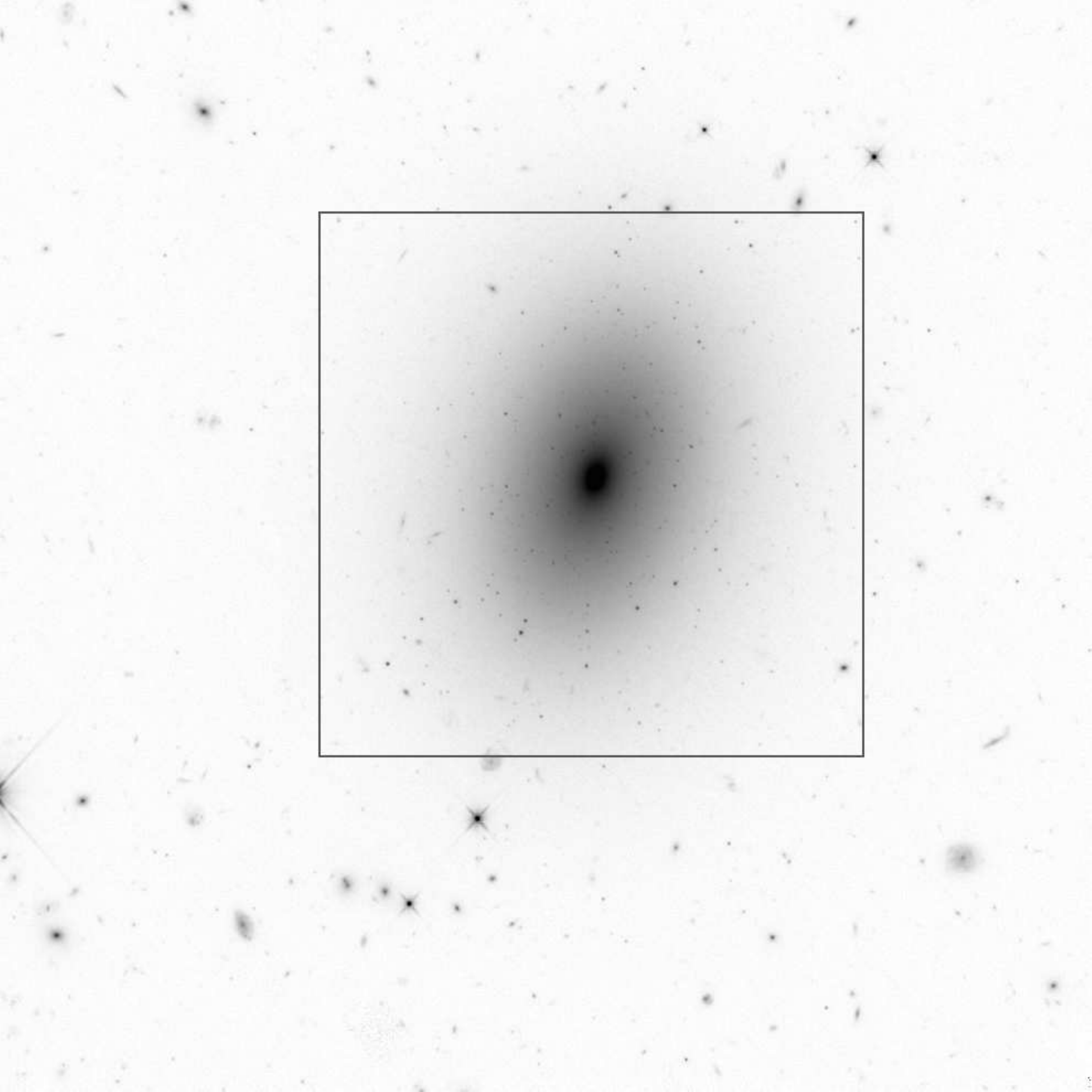}
\includegraphics[scale=0.4]{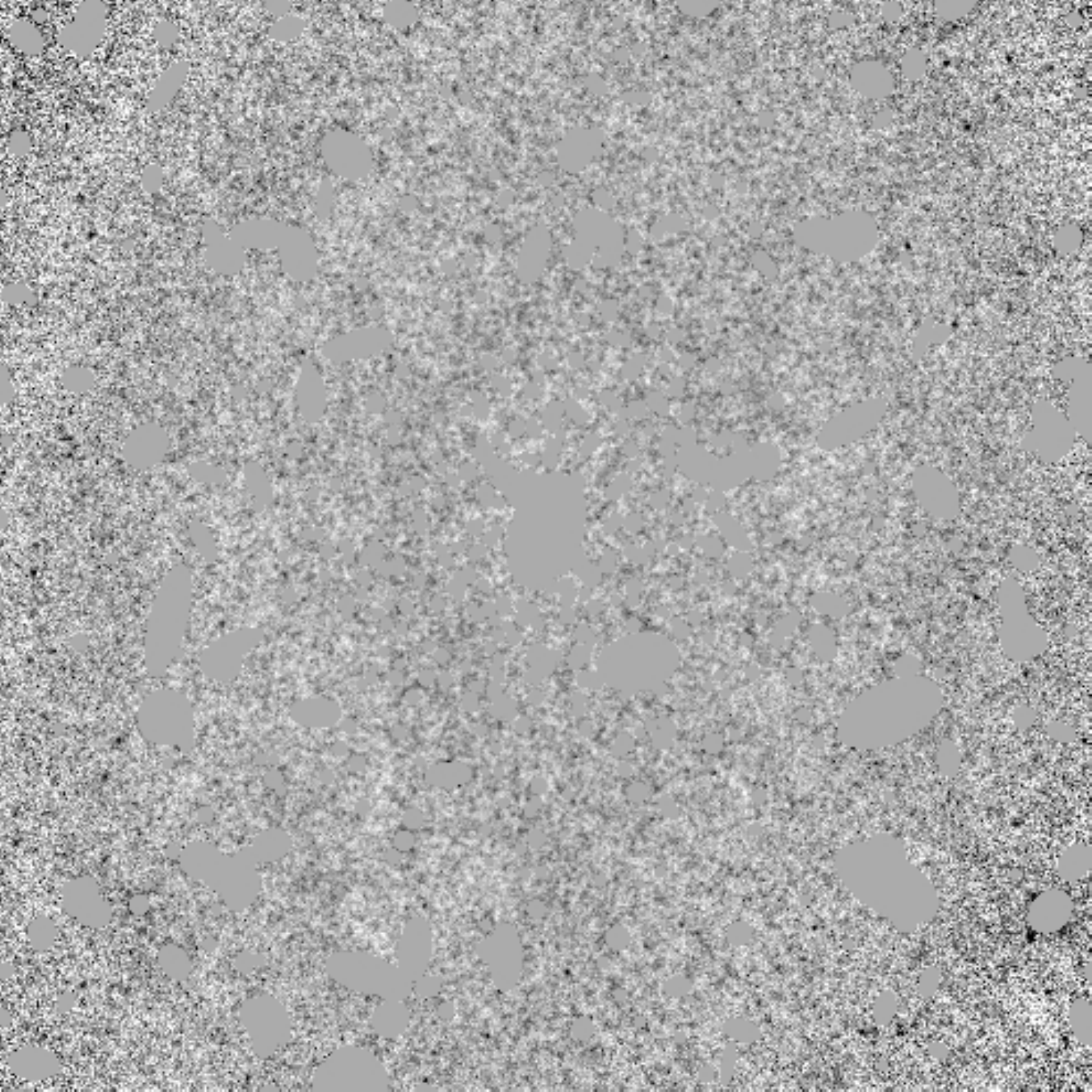} \\
\vspace{10pt}
\includegraphics[scale=0.4]{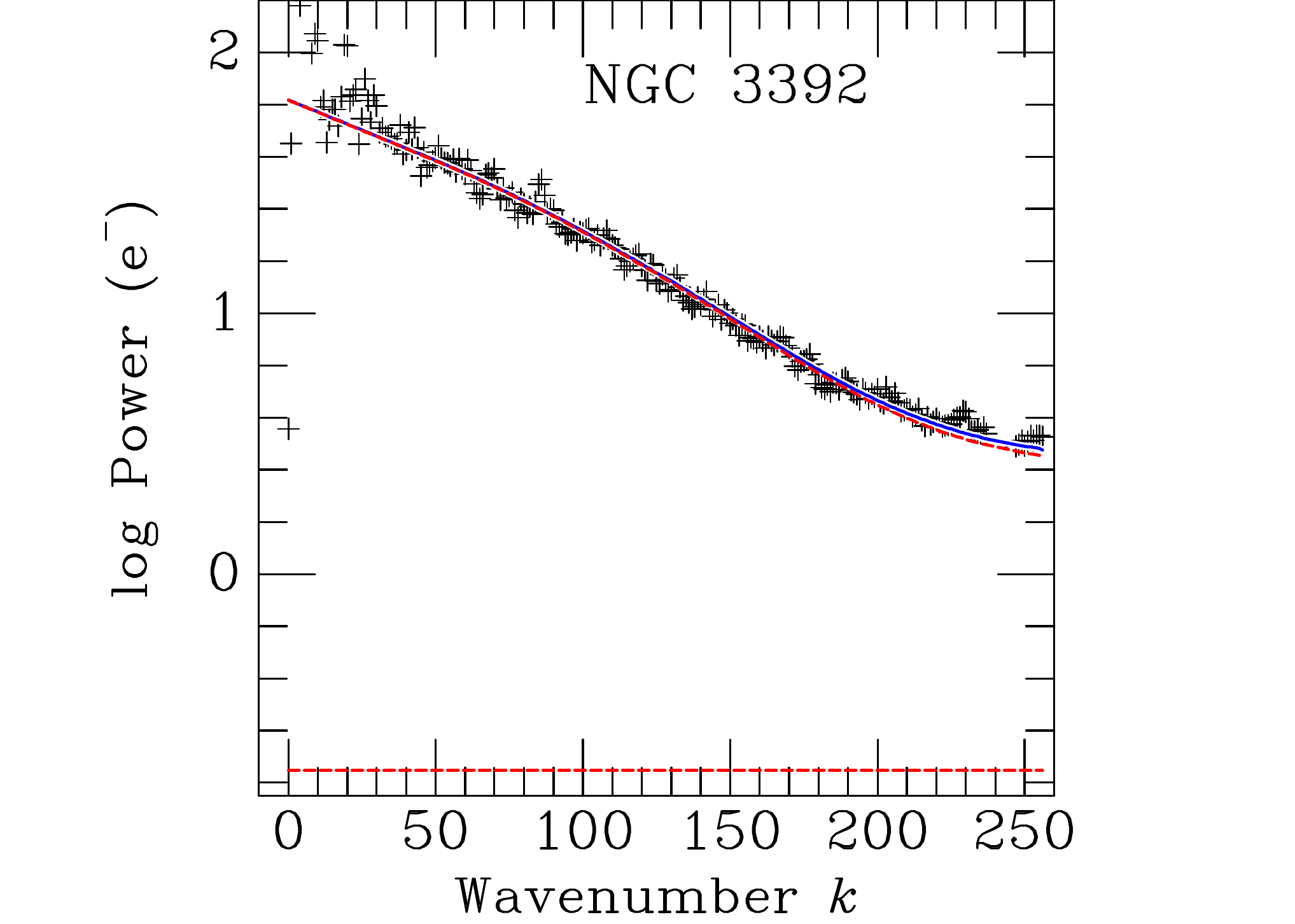}
\hspace{-25pt}
\includegraphics[scale=0.4]{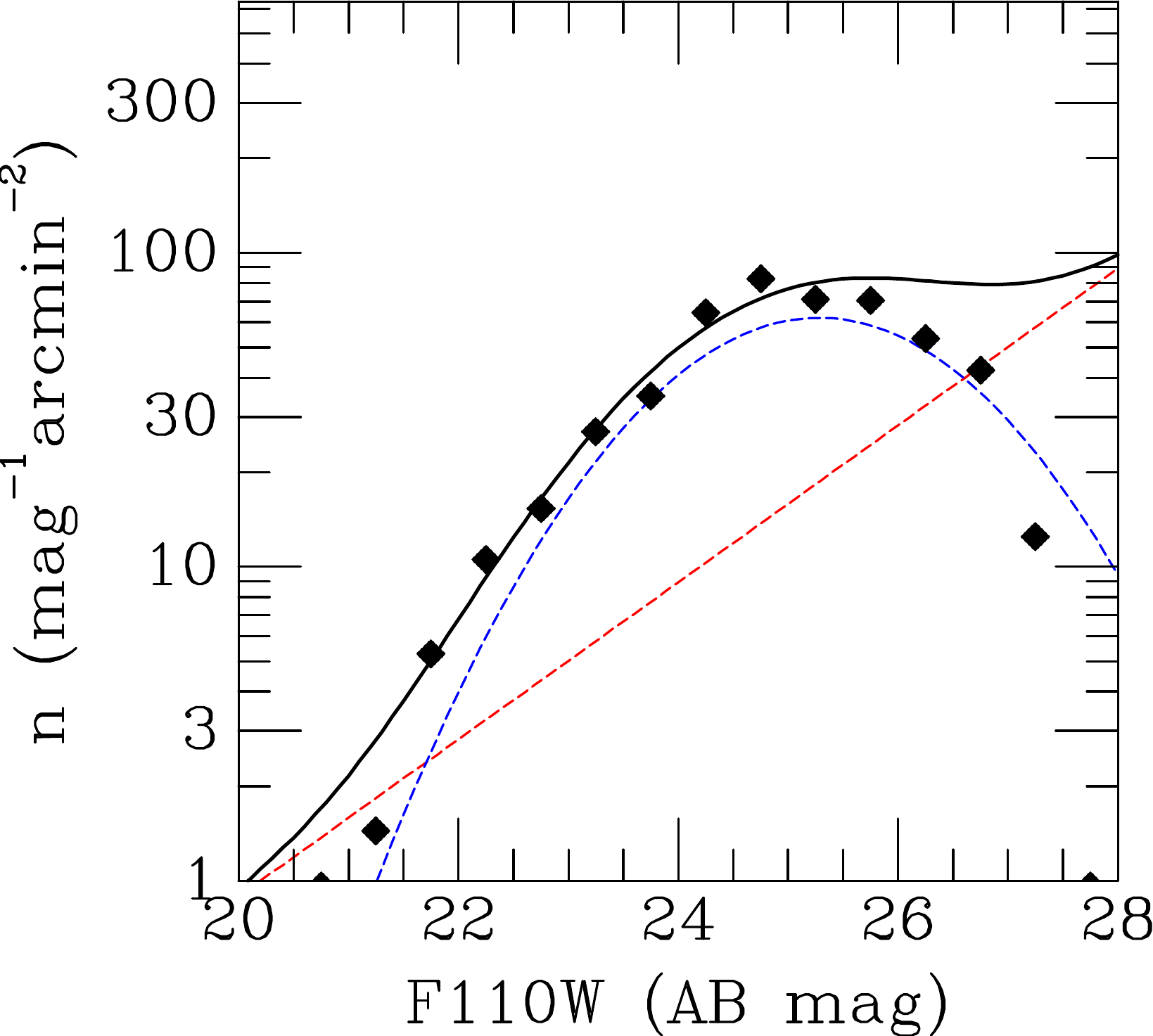}
\caption{Combined figure for NGC~3392.}
\end{center}
\end{figure*}
\clearpage


\begin{figure*}
\begin{center}
\includegraphics[scale=0.2]{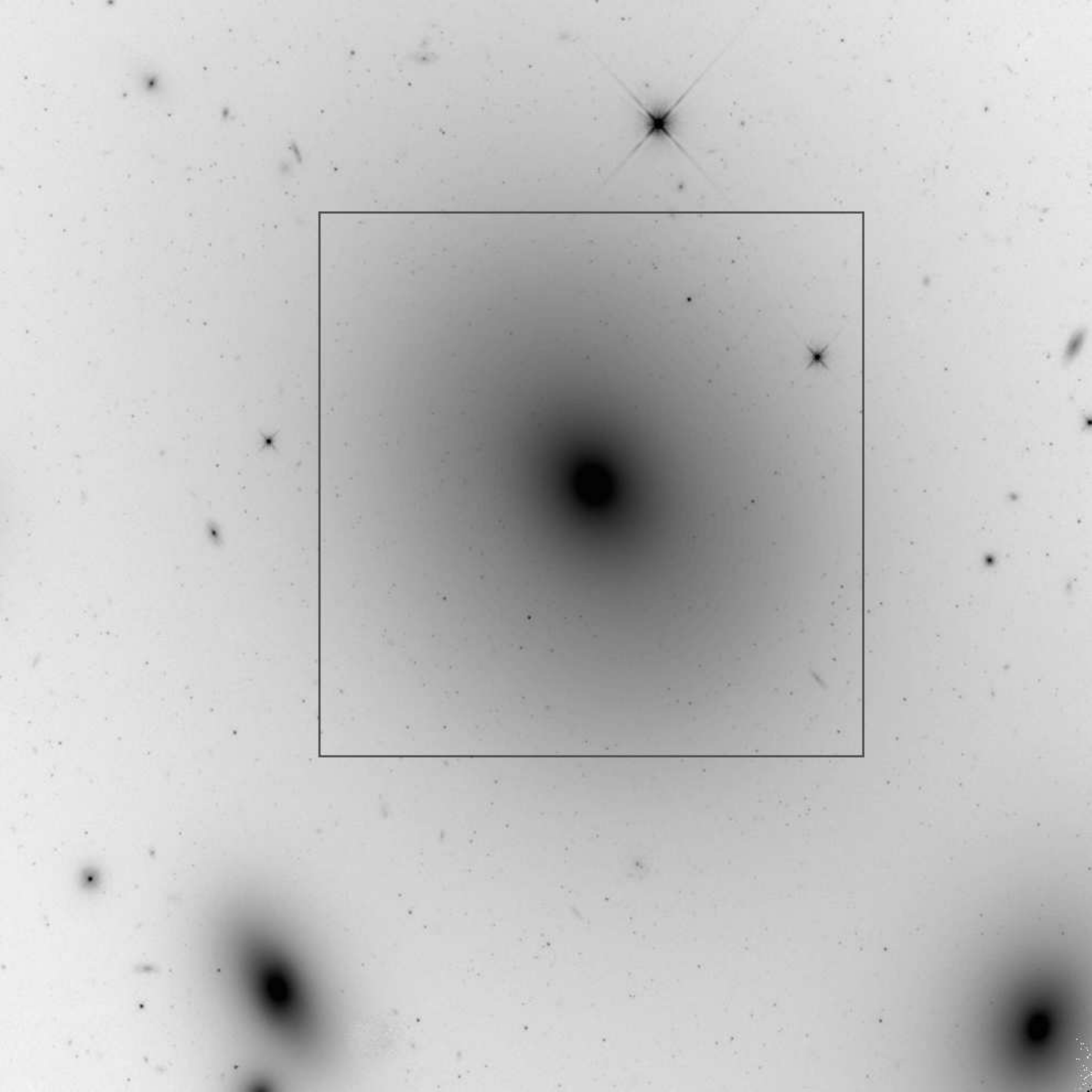}
\includegraphics[scale=0.4]{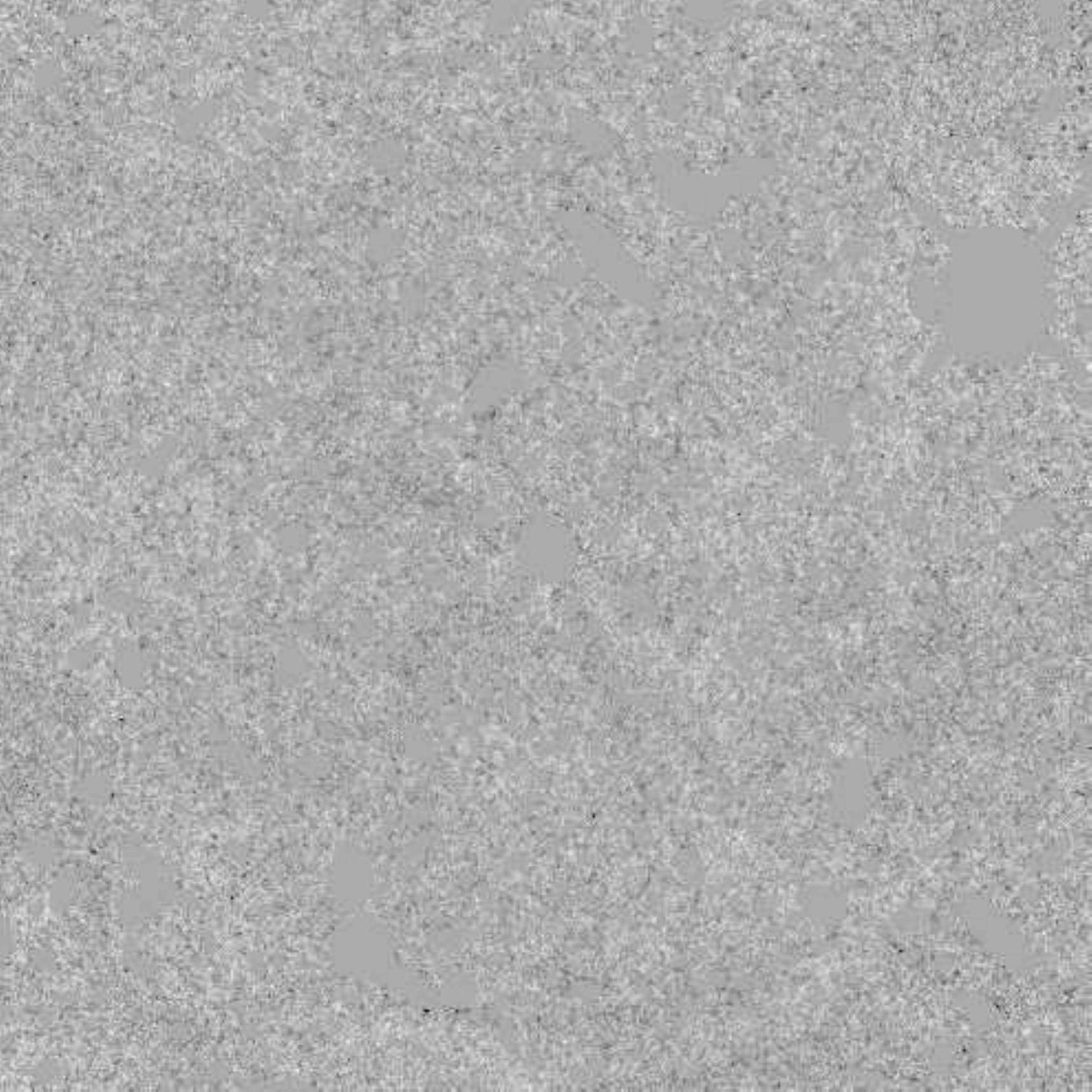} \\
\vspace{10pt}
\includegraphics[scale=0.4]{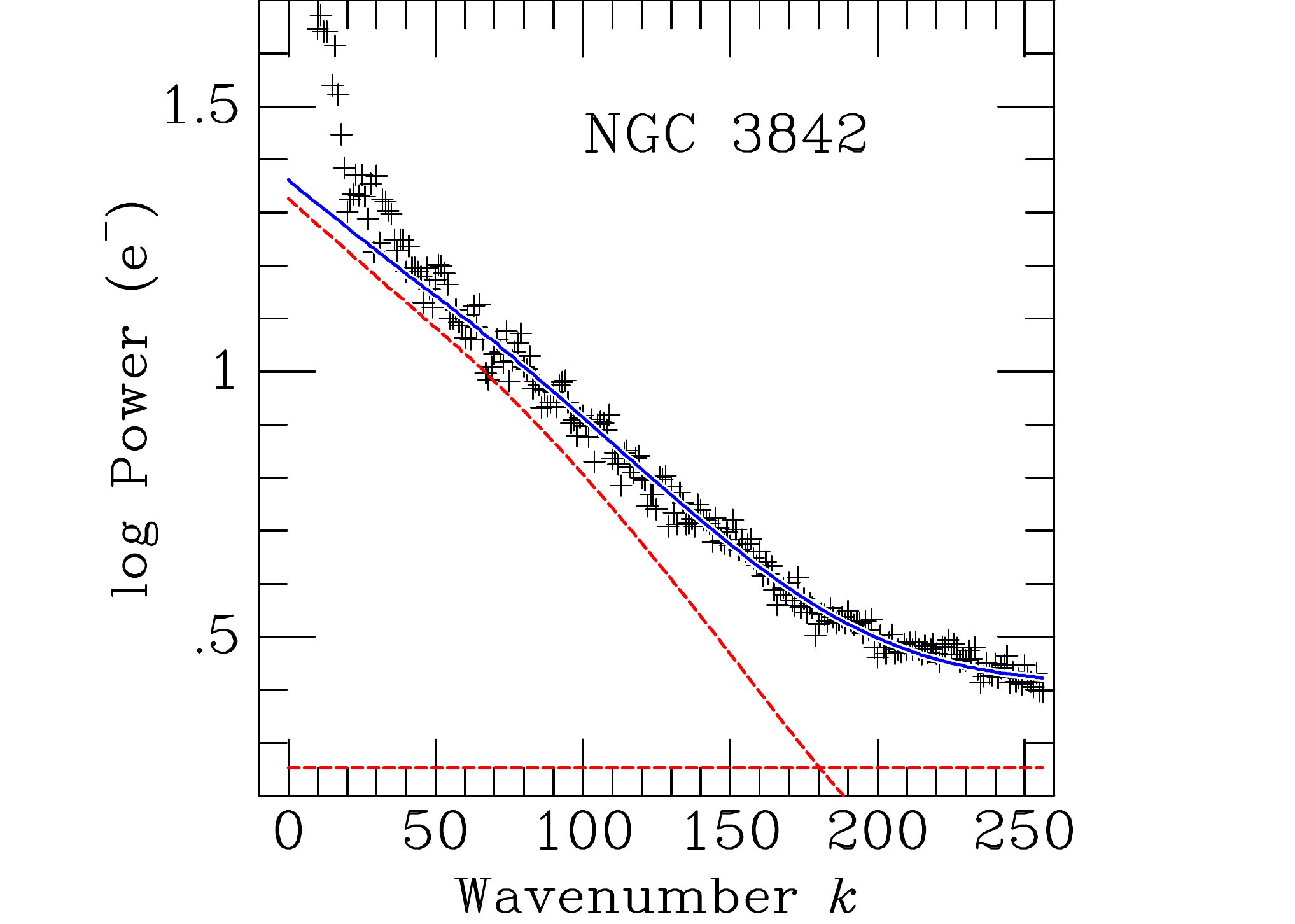}
\hspace{-25pt}
\includegraphics[scale=0.4]{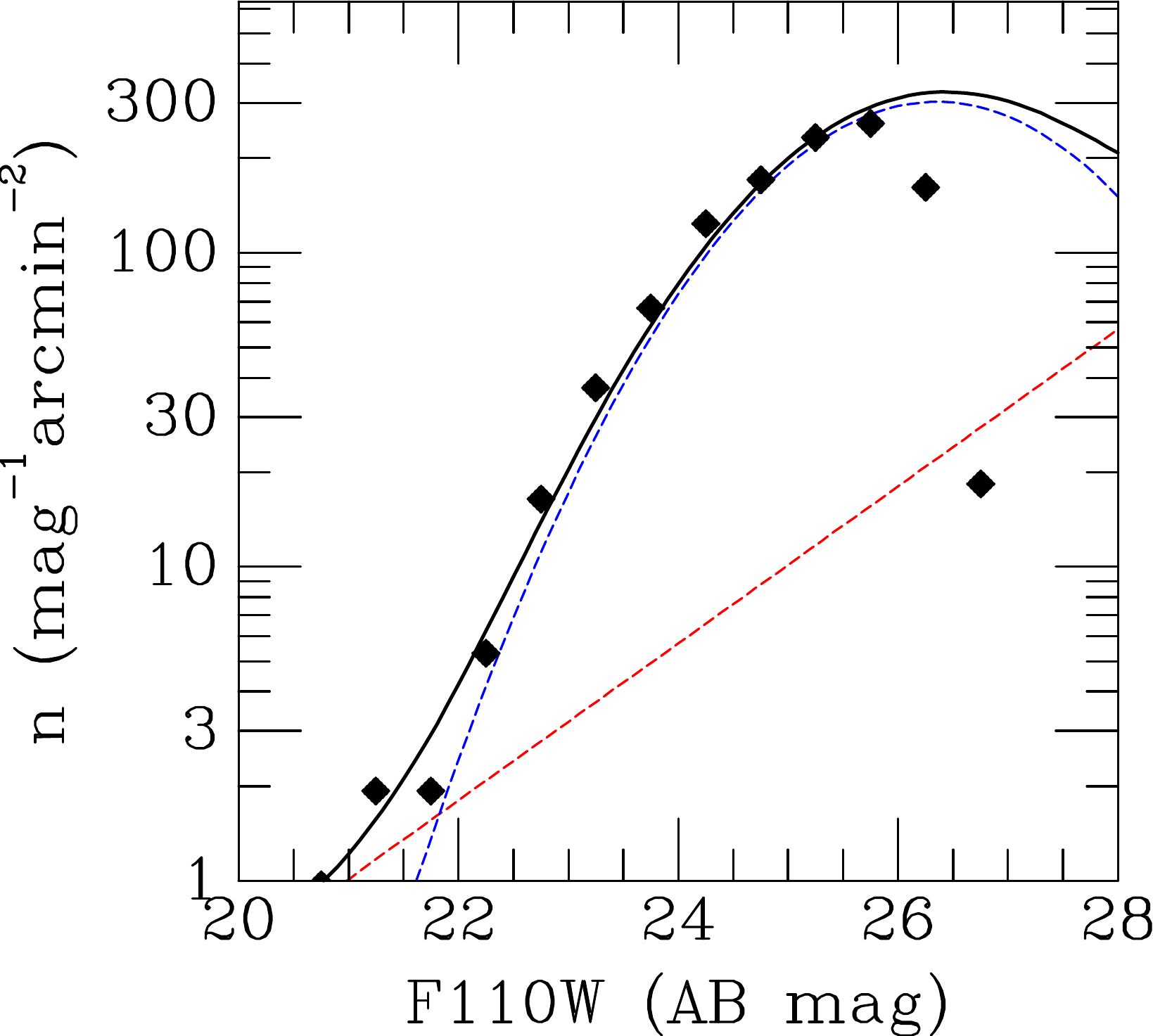}
\caption{Combined figure for NGC~3842.}
\end{center}
\end{figure*}
\clearpage

\begin{figure*}
\begin{center}
\includegraphics[scale=0.2]{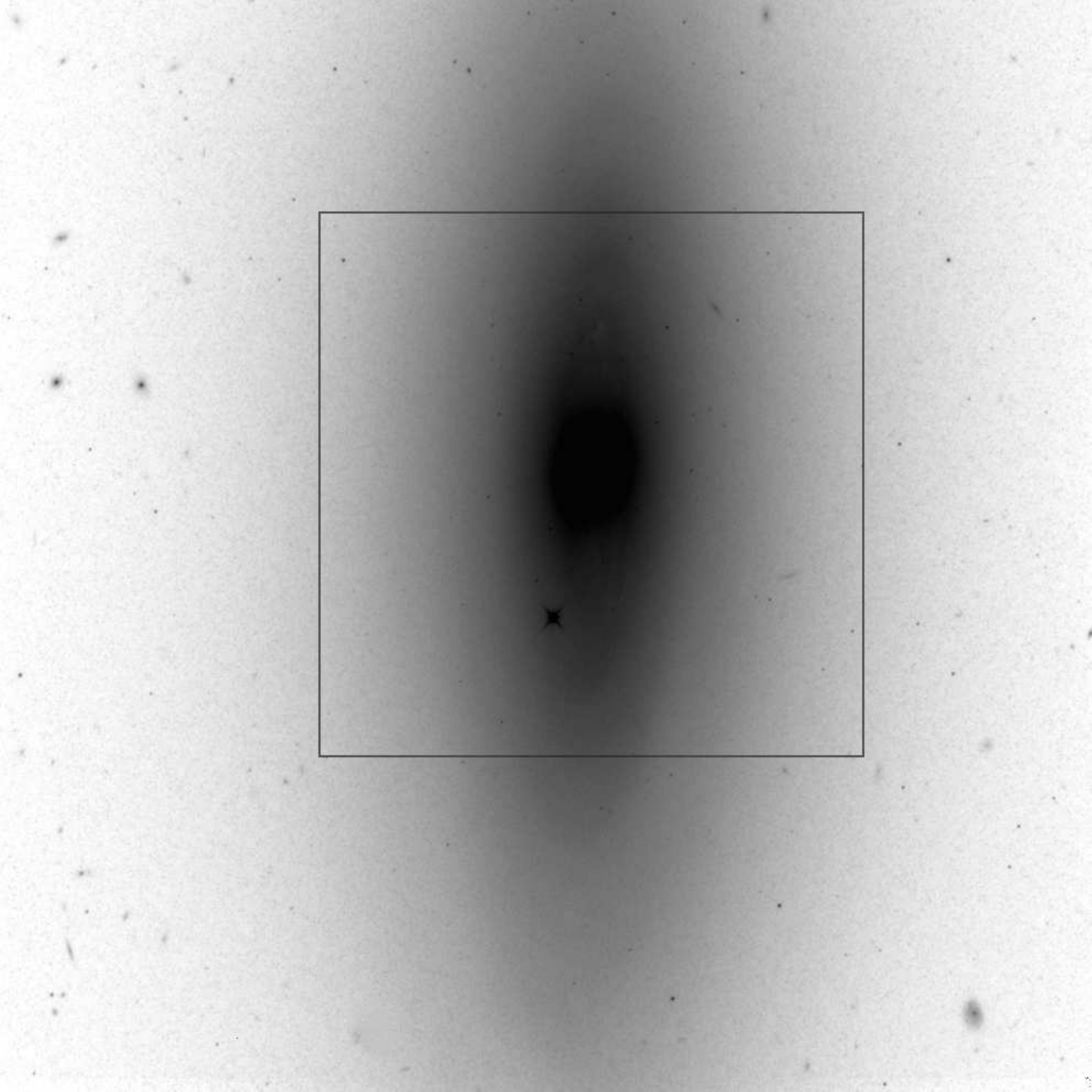}
\includegraphics[scale=0.4]{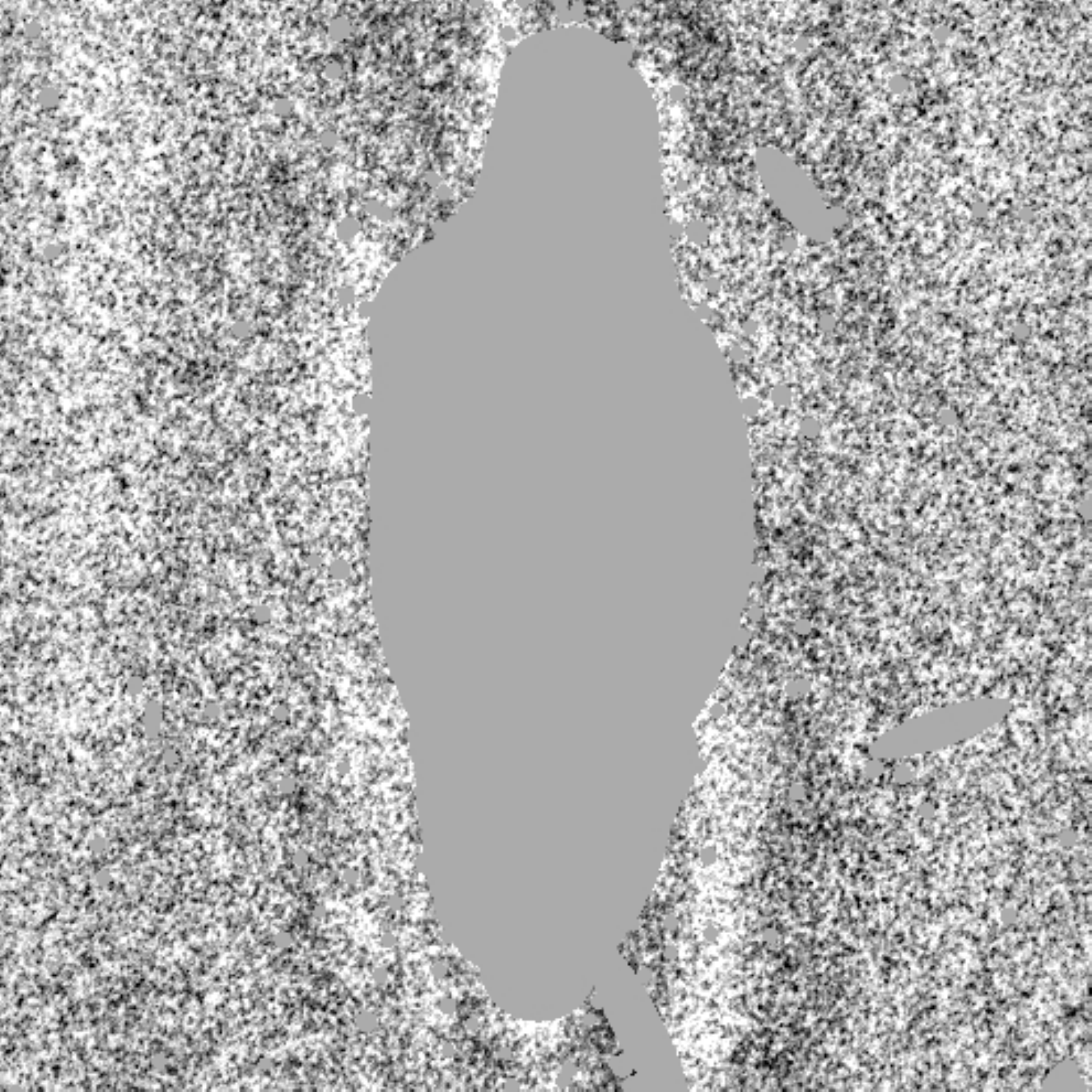} \\
\vspace{10pt}
\includegraphics[scale=0.4]{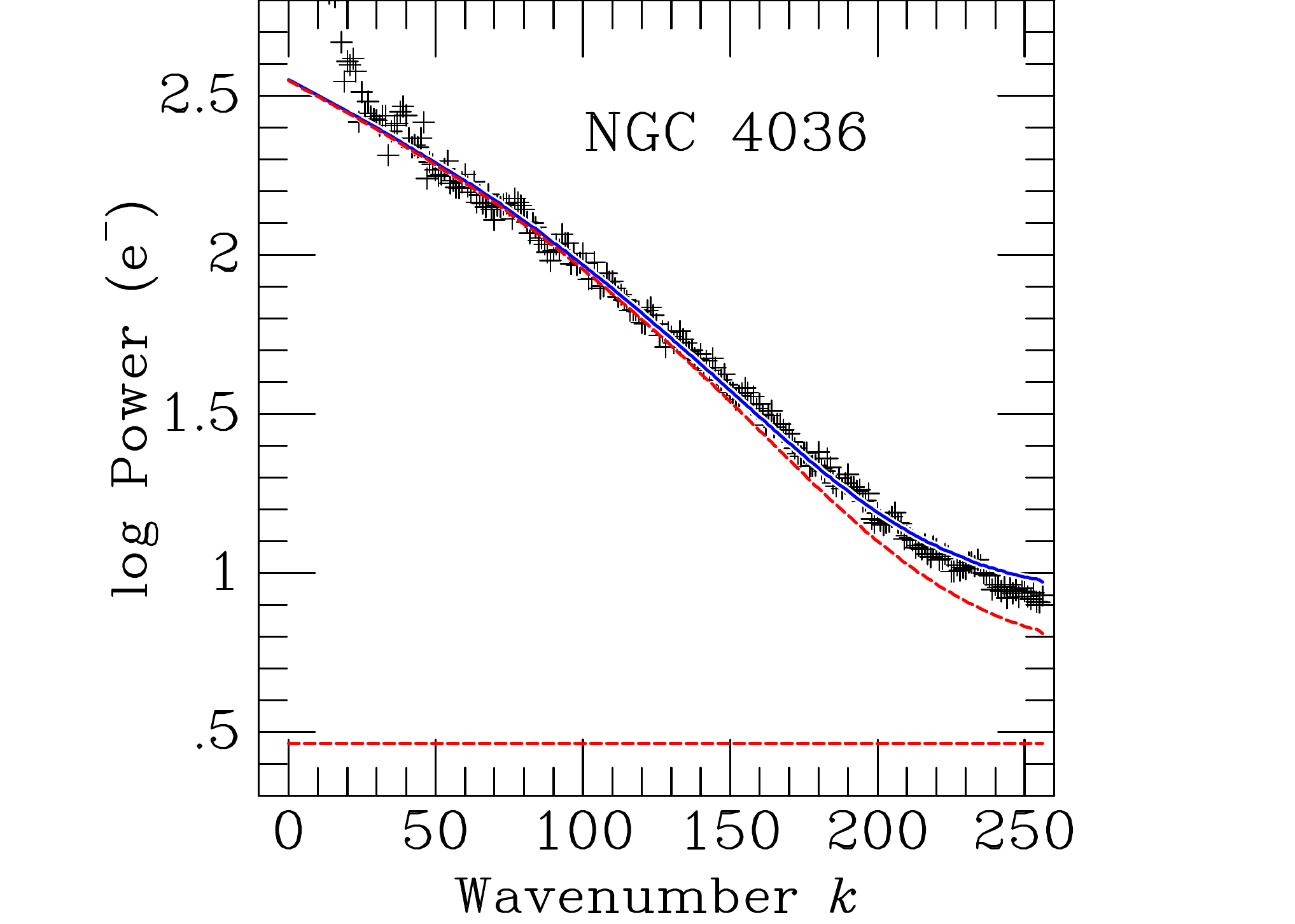}
\hspace{-25pt}
\includegraphics[scale=0.4]{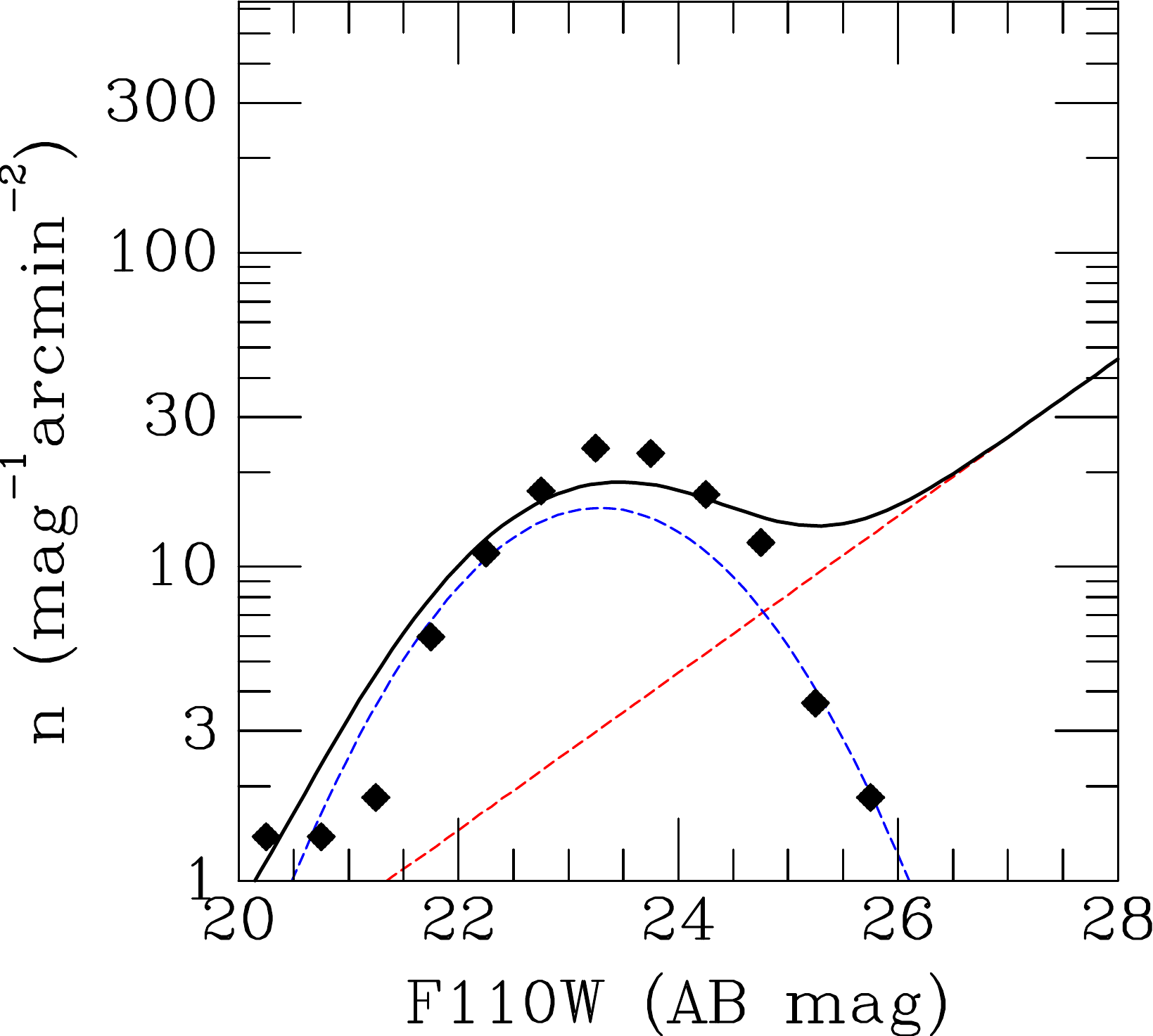}
\caption{Combined figure for NGC~4036.}
\end{center}
\end{figure*}
\clearpage

\begin{figure*}
\begin{center}
\includegraphics[scale=0.2]{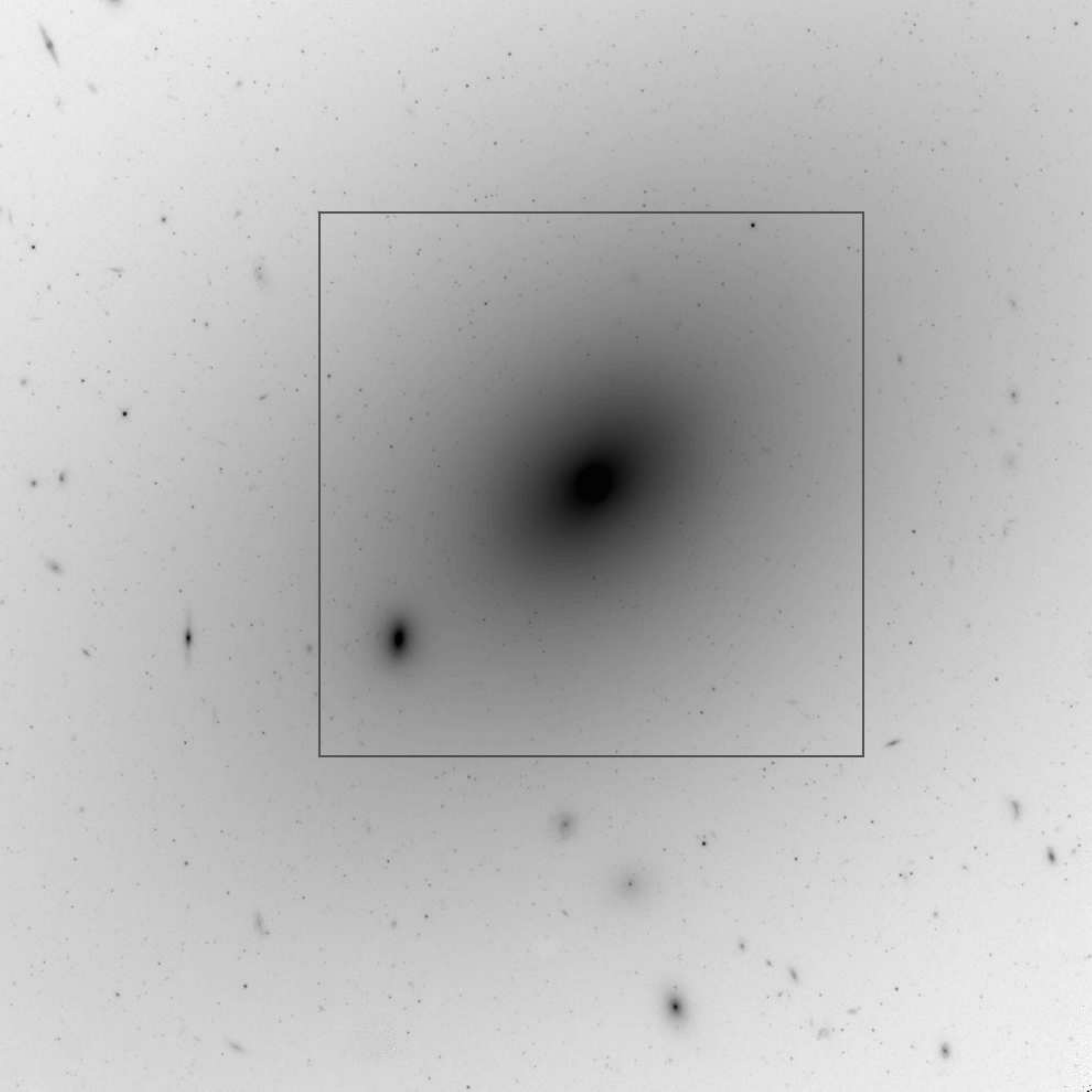}
\includegraphics[scale=0.4]{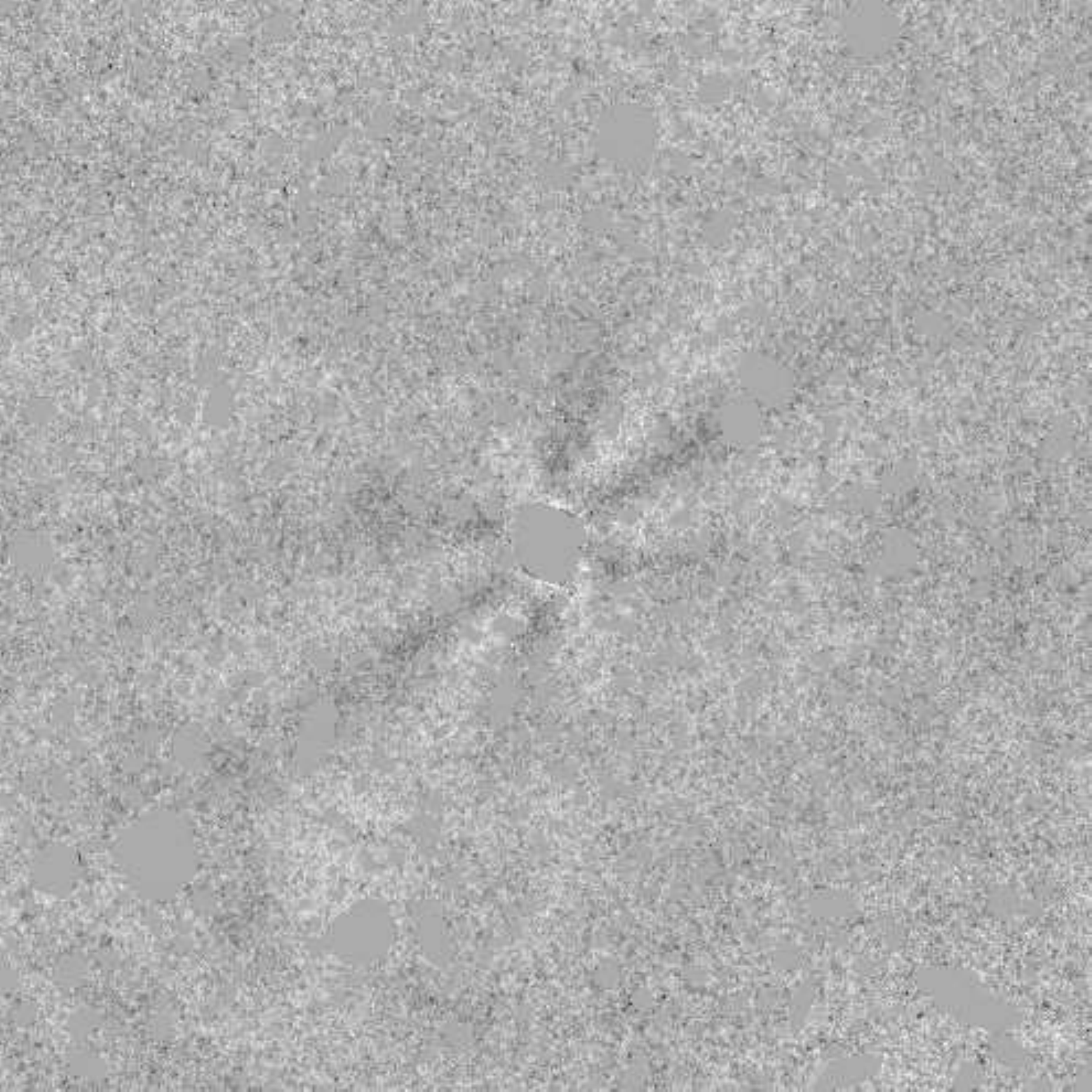} \\
\vspace{10pt}
\includegraphics[scale=0.4]{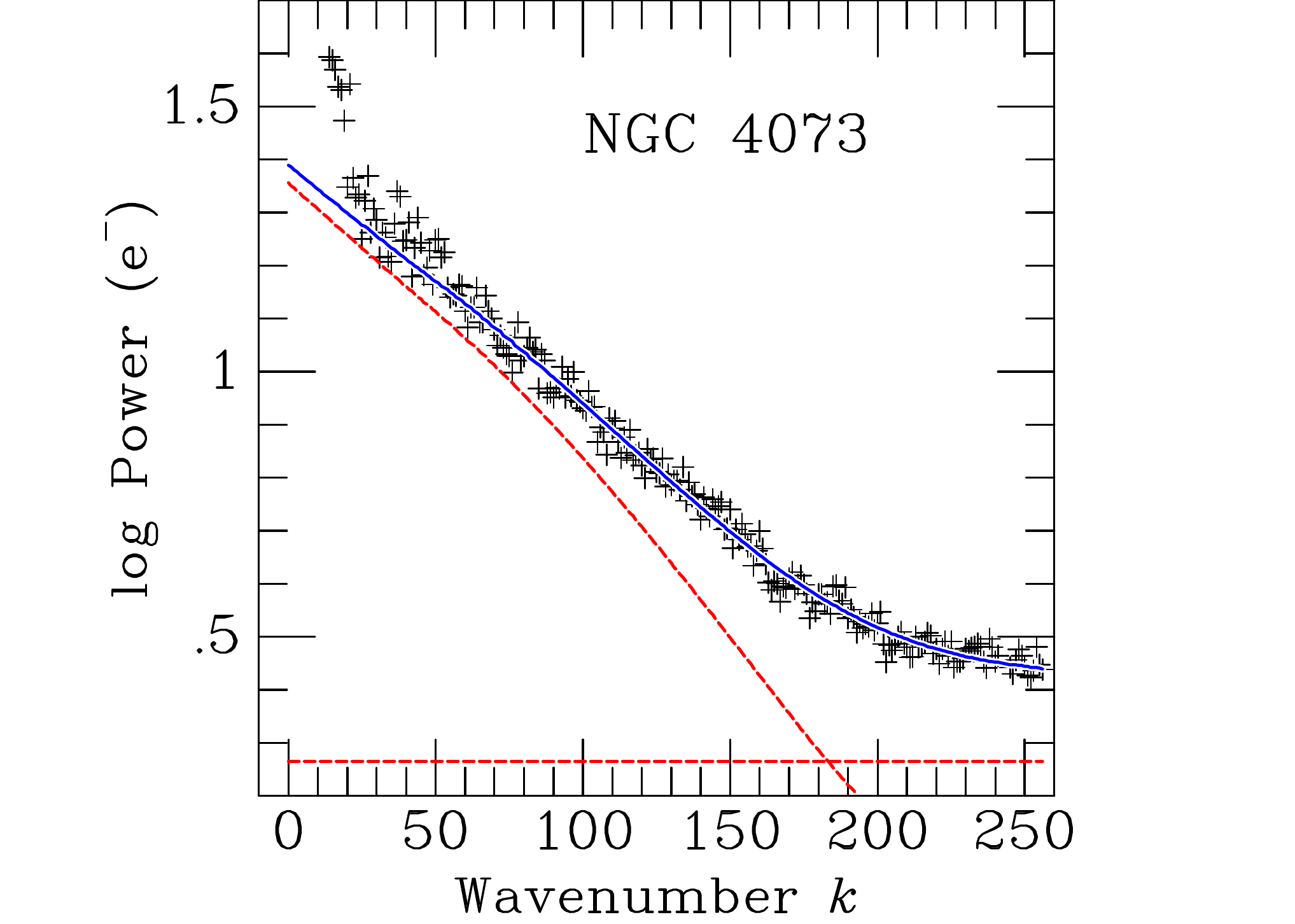}
\hspace{-25pt}
\includegraphics[scale=0.4]{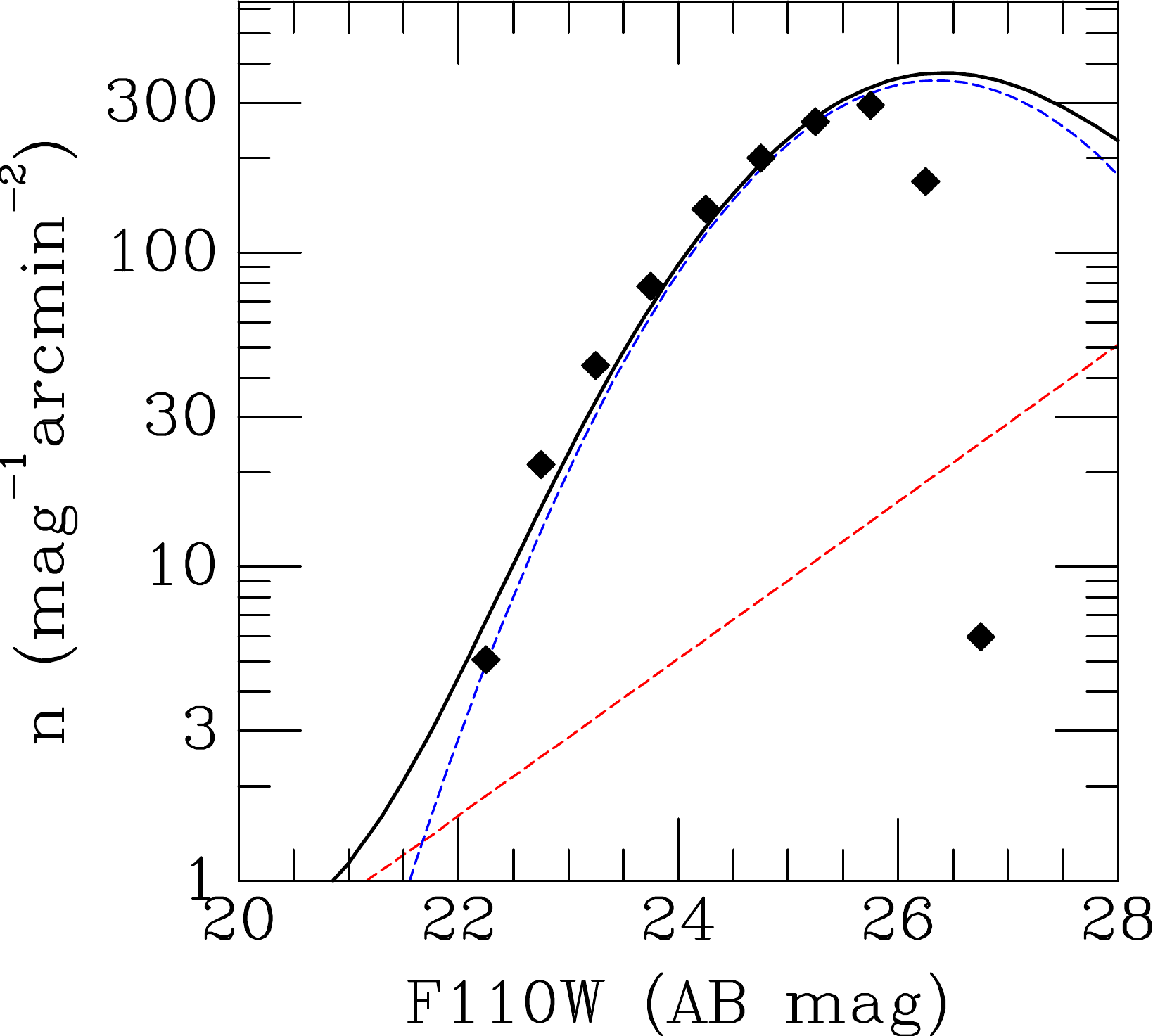}
\caption{Combined figure for NGC~4073.}
\end{center}
\end{figure*}
\clearpage

\begin{figure*}
\begin{center}
\includegraphics[scale=0.2]{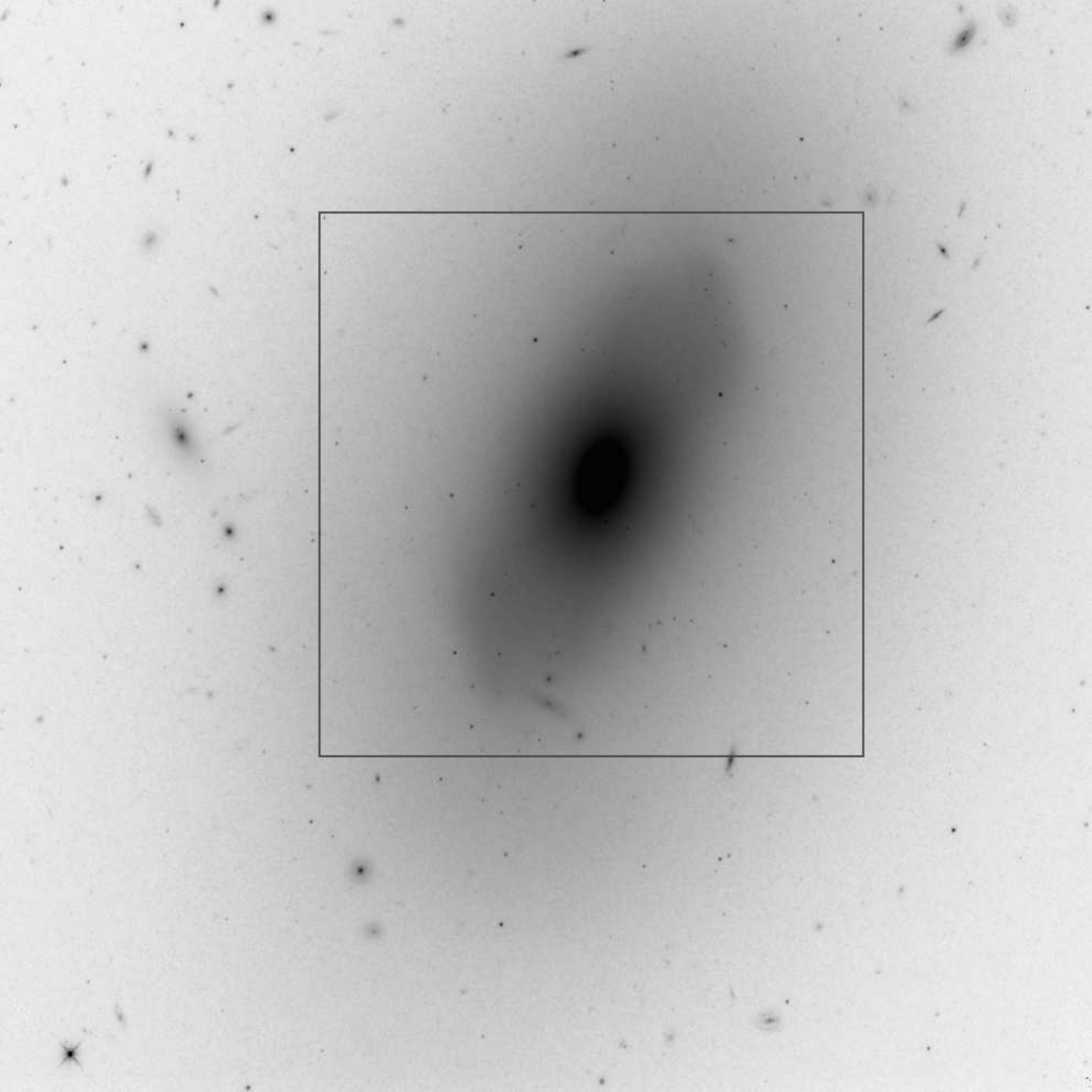}
\includegraphics[scale=0.4]{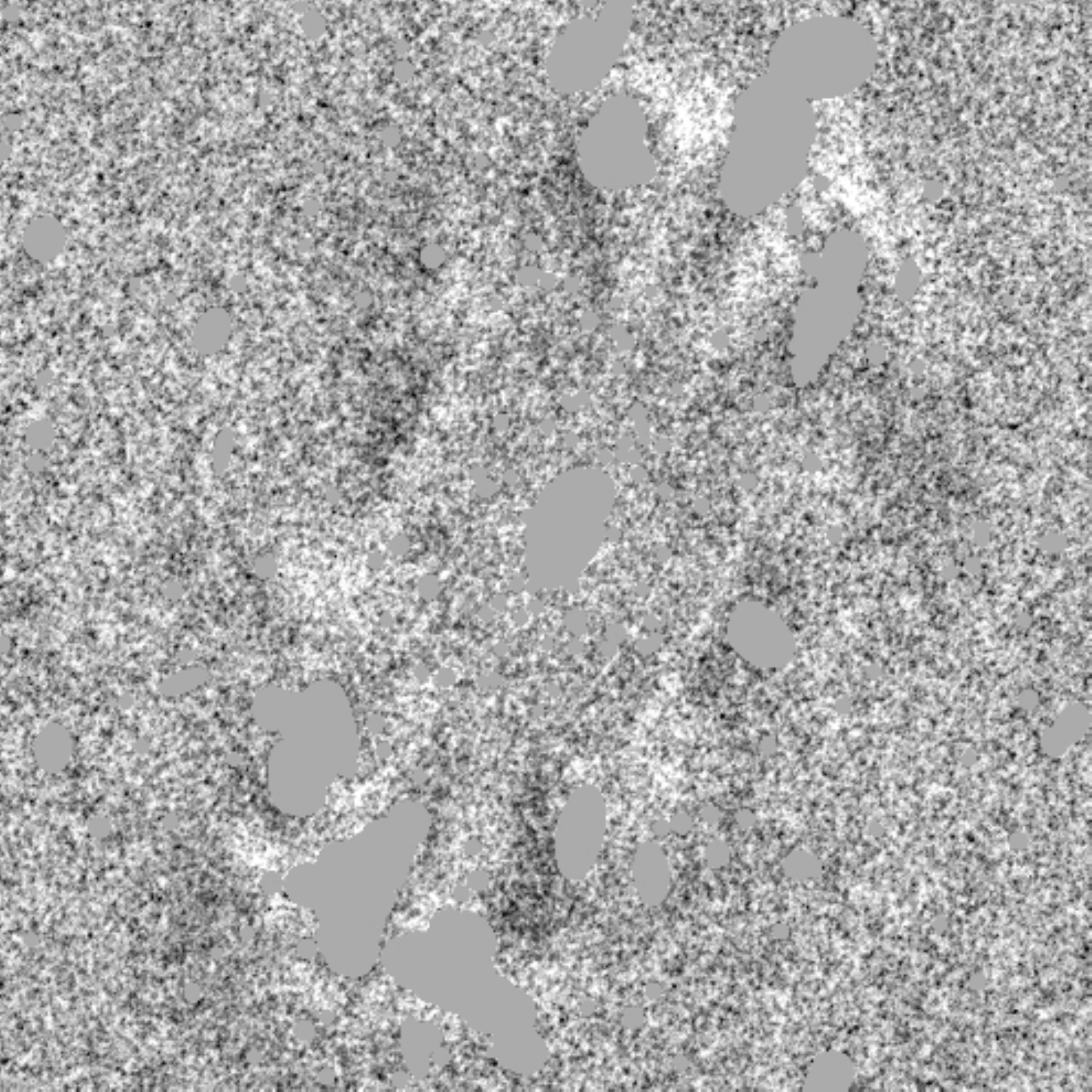} \\
\vspace{10pt}
\includegraphics[scale=0.4]{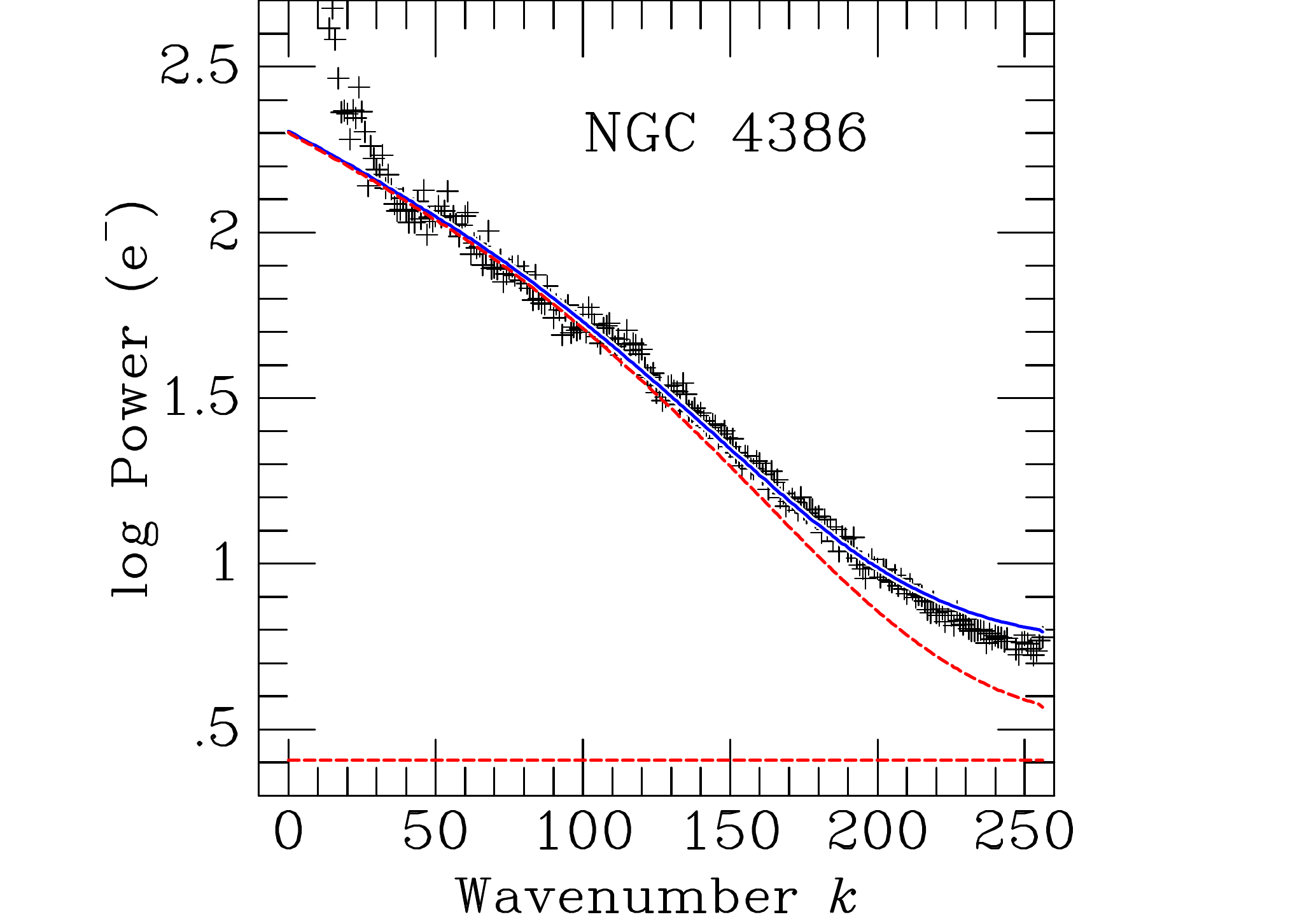}
\hspace{-25pt}
\includegraphics[scale=0.4]{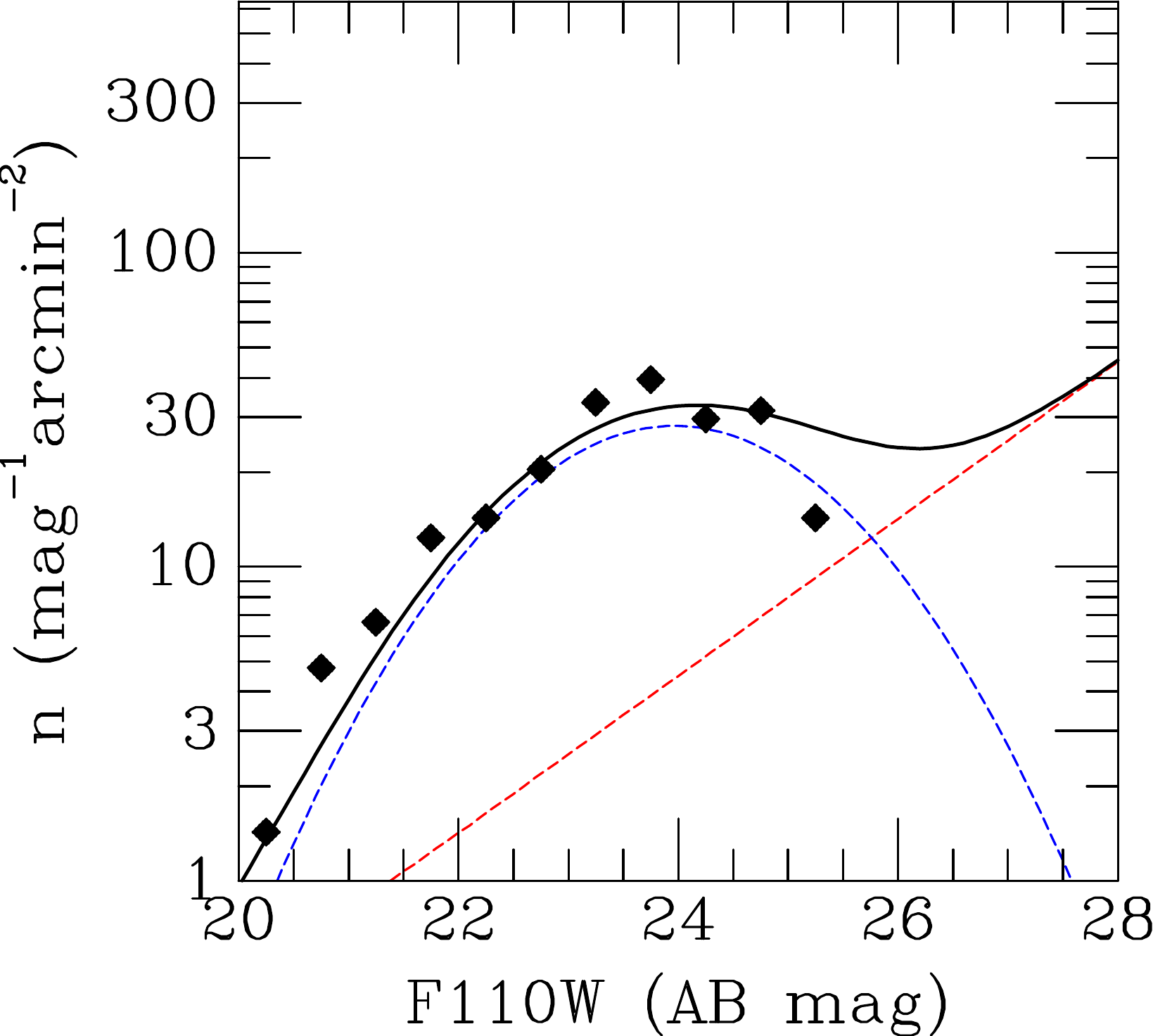}
\caption{Combined figure for NGC~4386.}
\end{center}
\end{figure*}
\clearpage

\begin{figure*}
\begin{center}
\includegraphics[scale=0.2]{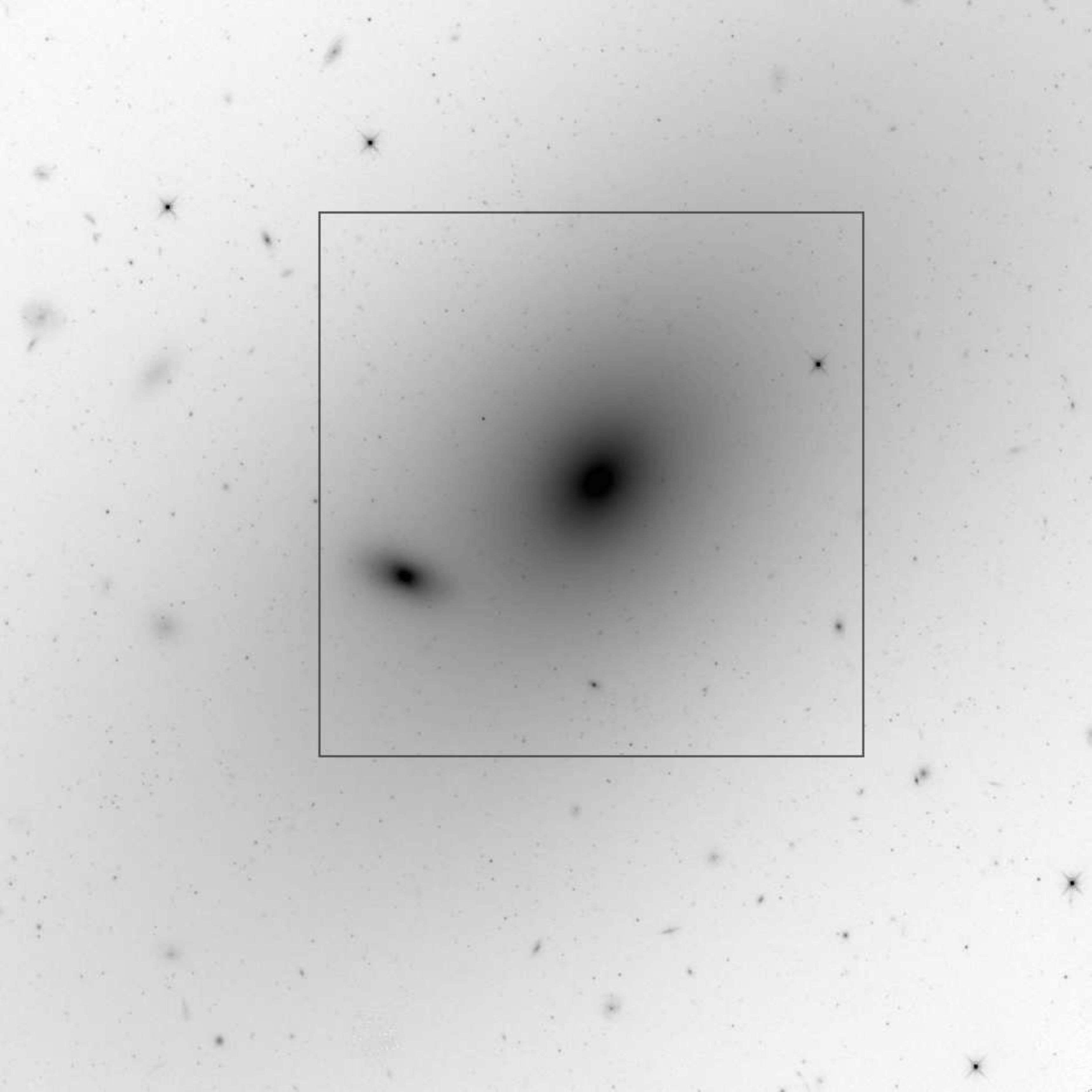}
\includegraphics[scale=0.4]{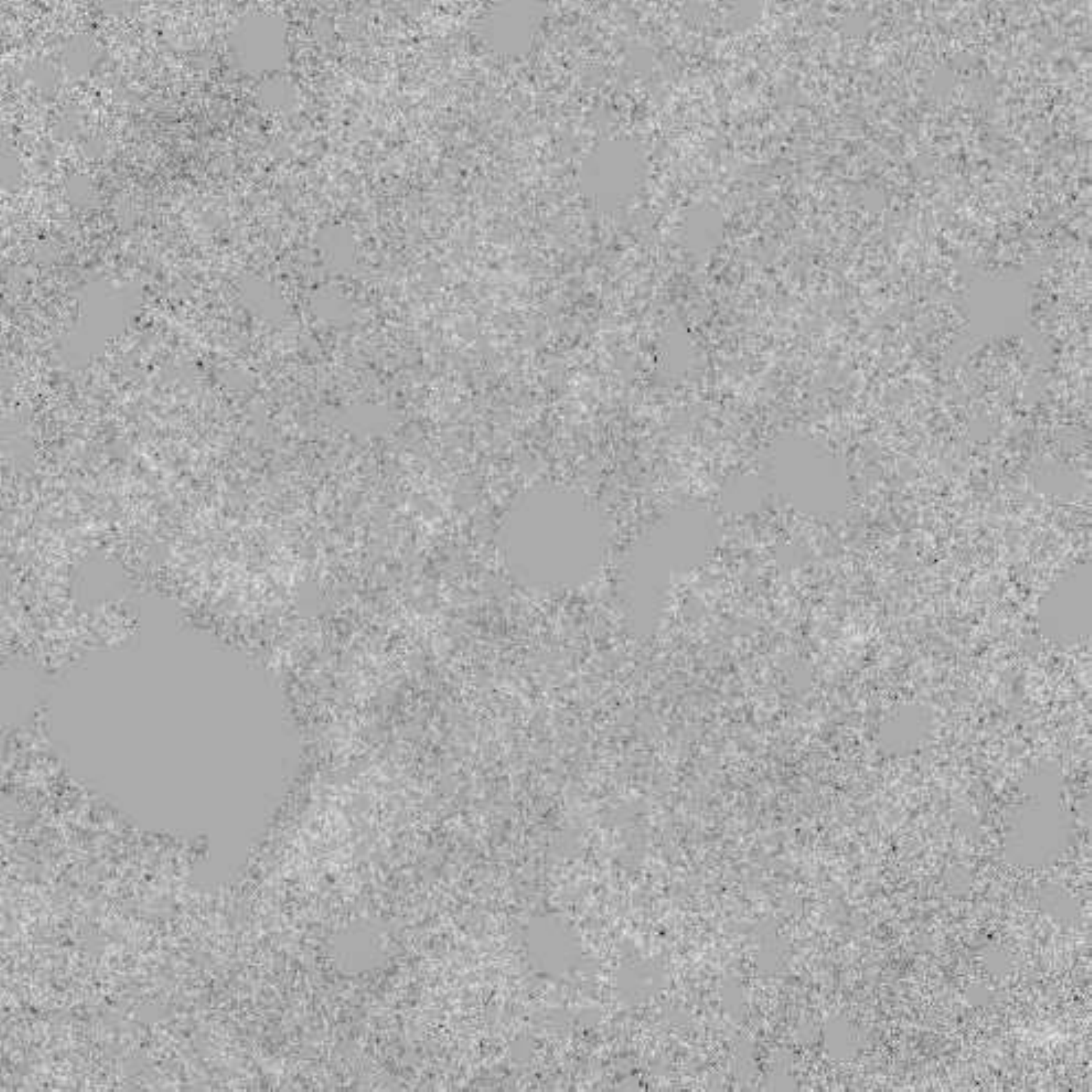} \\
\vspace{10pt}
\includegraphics[scale=0.4]{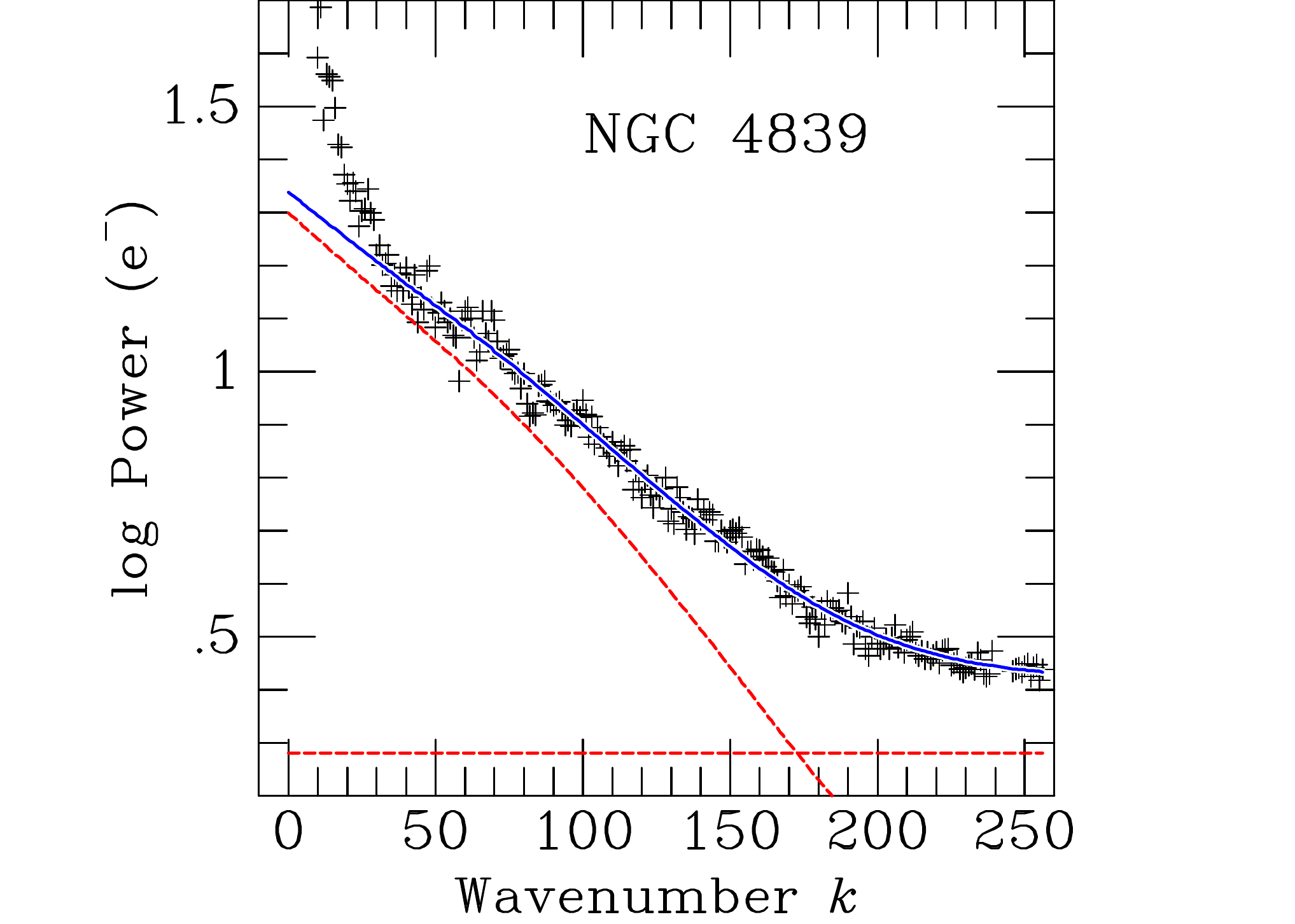}
\hspace{-25pt}
\includegraphics[scale=0.4]{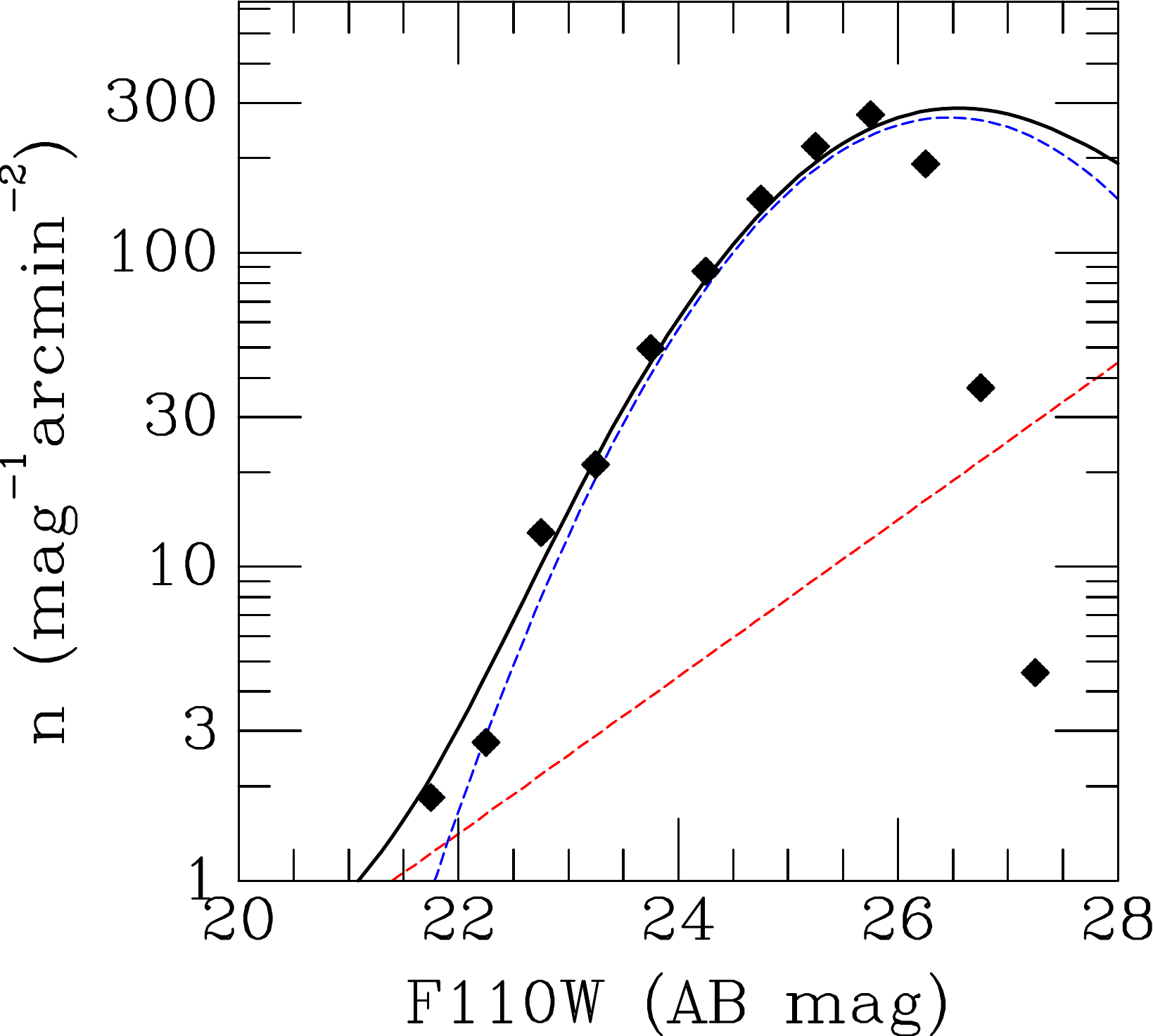}
\caption{Combined figure for NGC~4839.}
\end{center}
\end{figure*}
\clearpage

\begin{figure*}
\begin{center}
\includegraphics[scale=0.2]{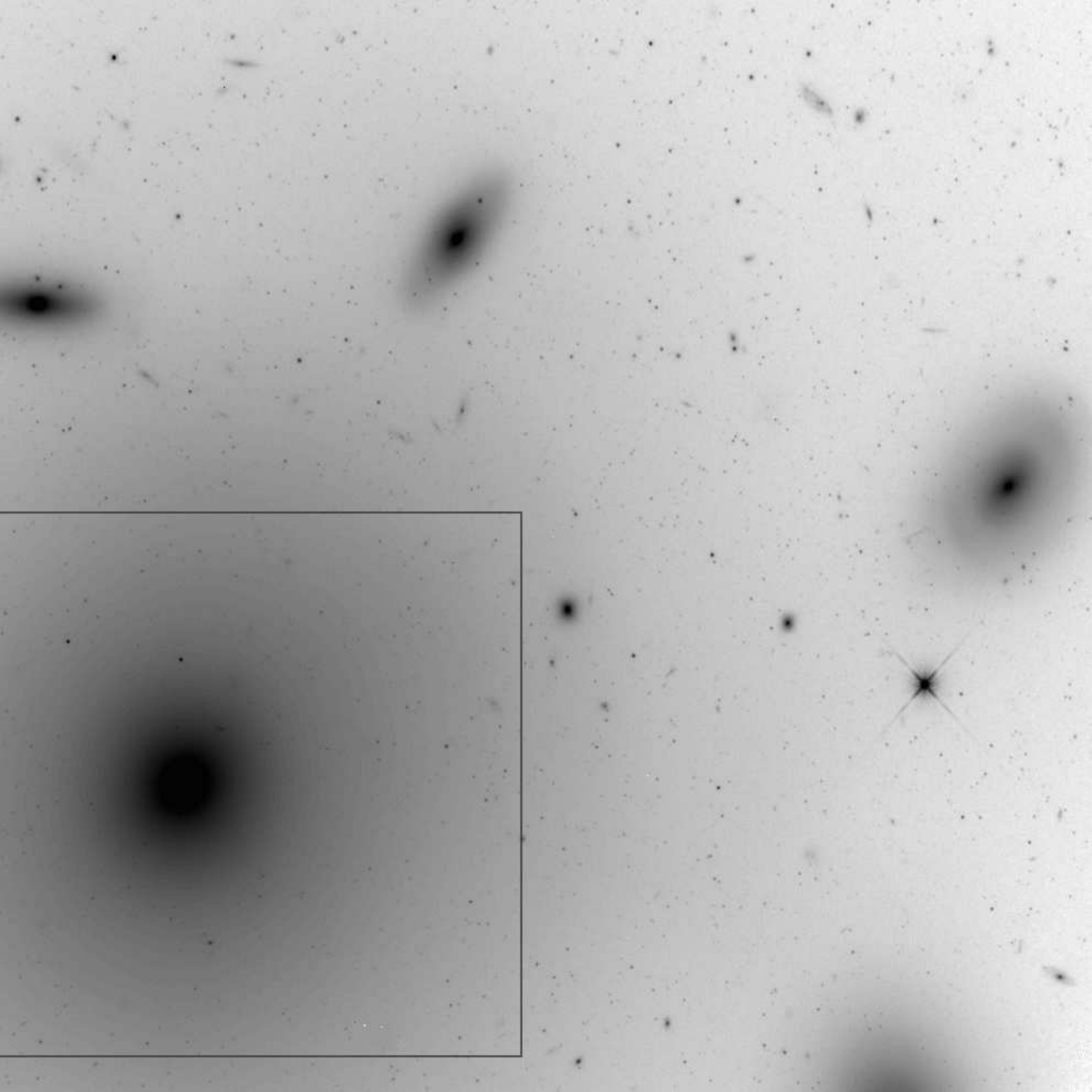}
\includegraphics[scale=0.4]{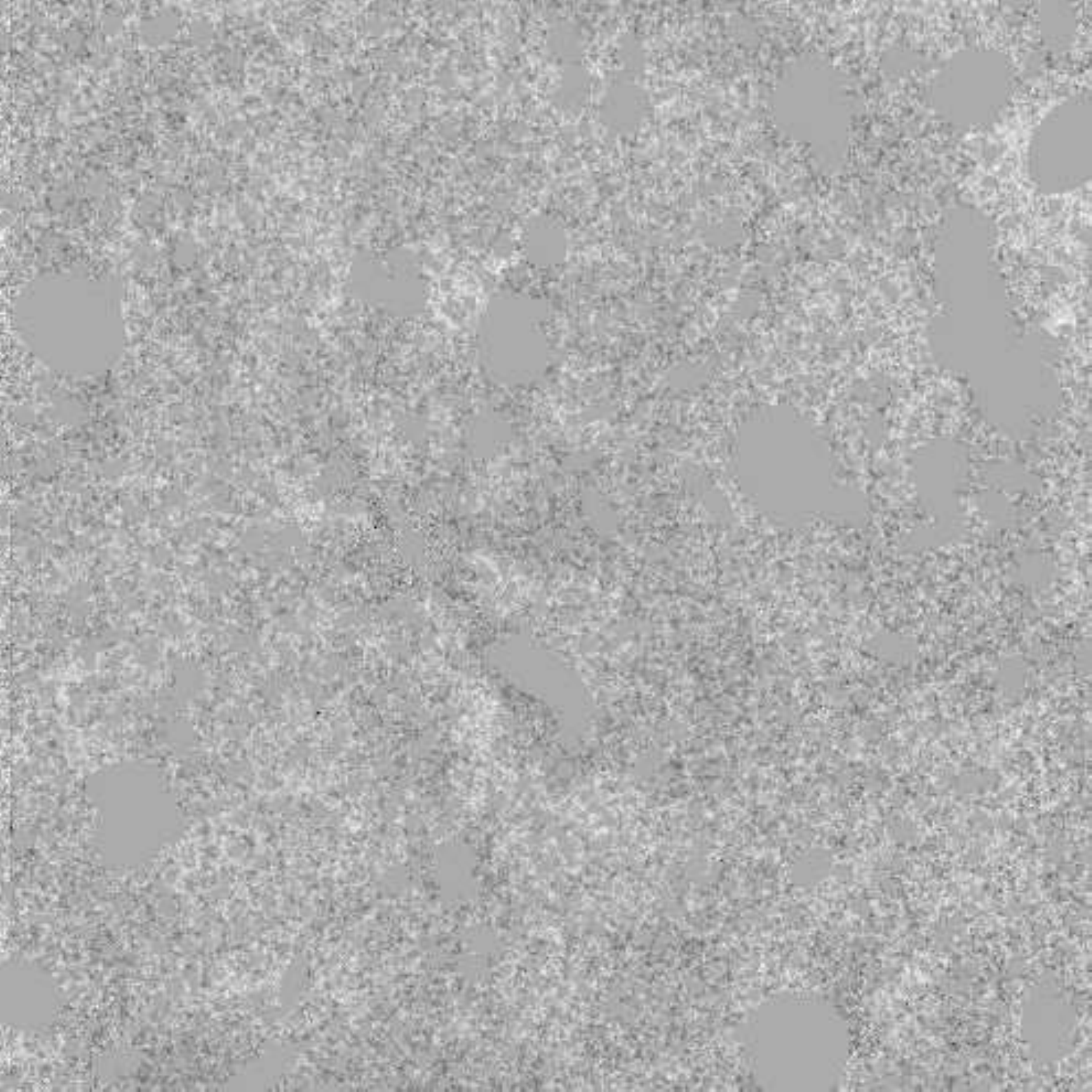} \\
\vspace{10pt}
\includegraphics[scale=0.4]{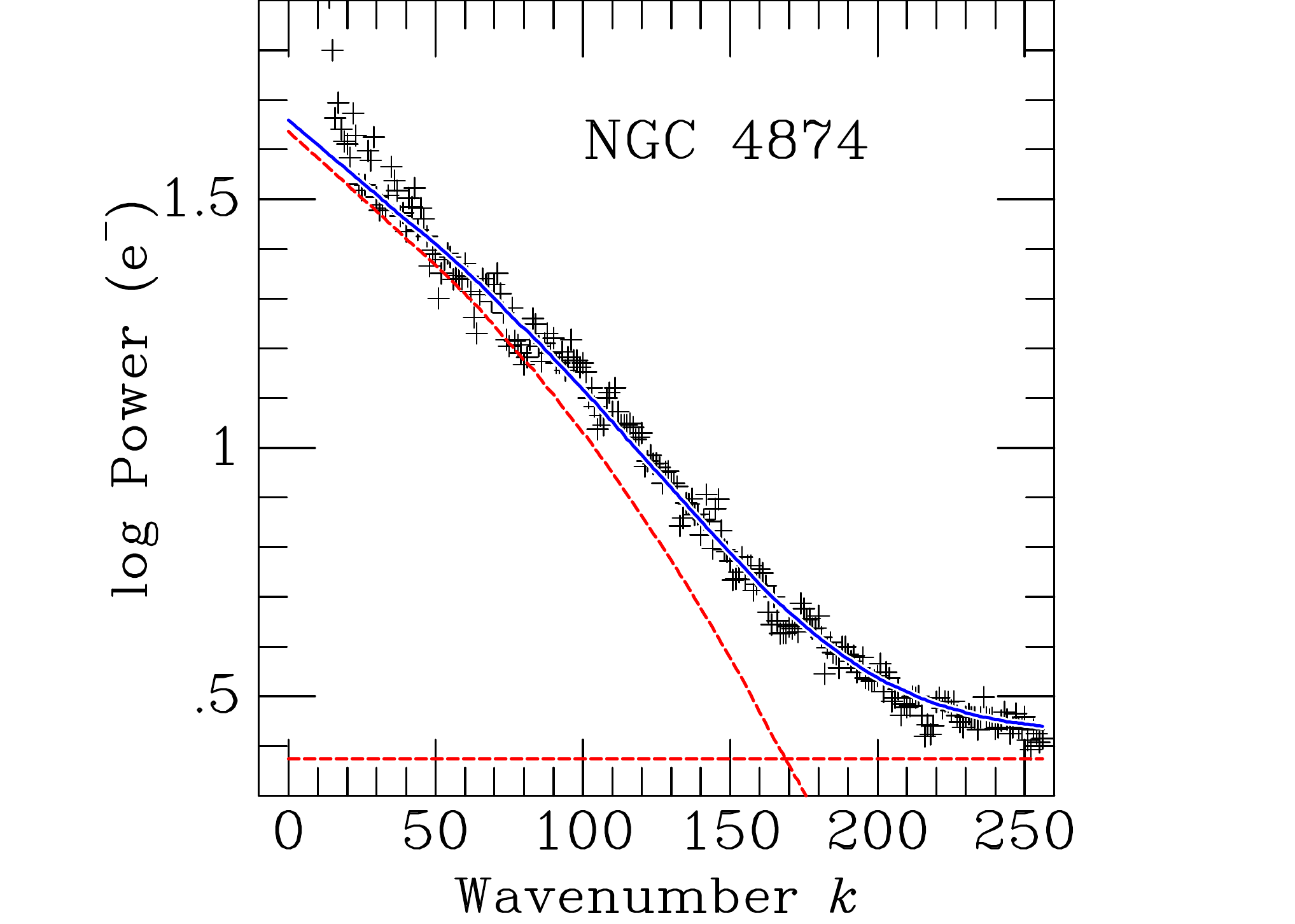}
\hspace{-25pt}
\includegraphics[scale=0.4]{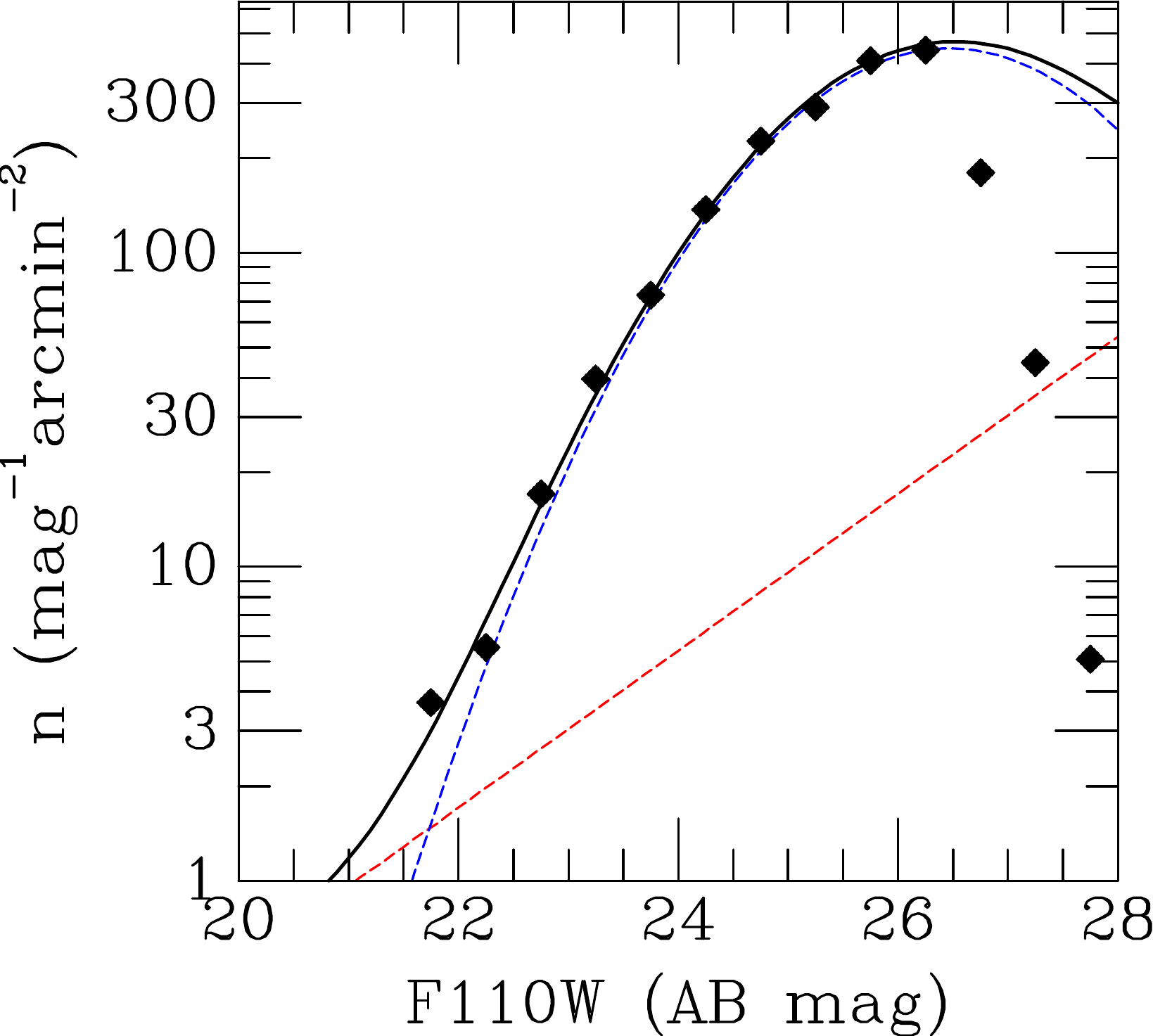}
\caption{Combined figure for NGC~4874.}
\end{center}
\end{figure*}
\clearpage

\begin{figure*}
\begin{center}
\includegraphics[scale=0.2]{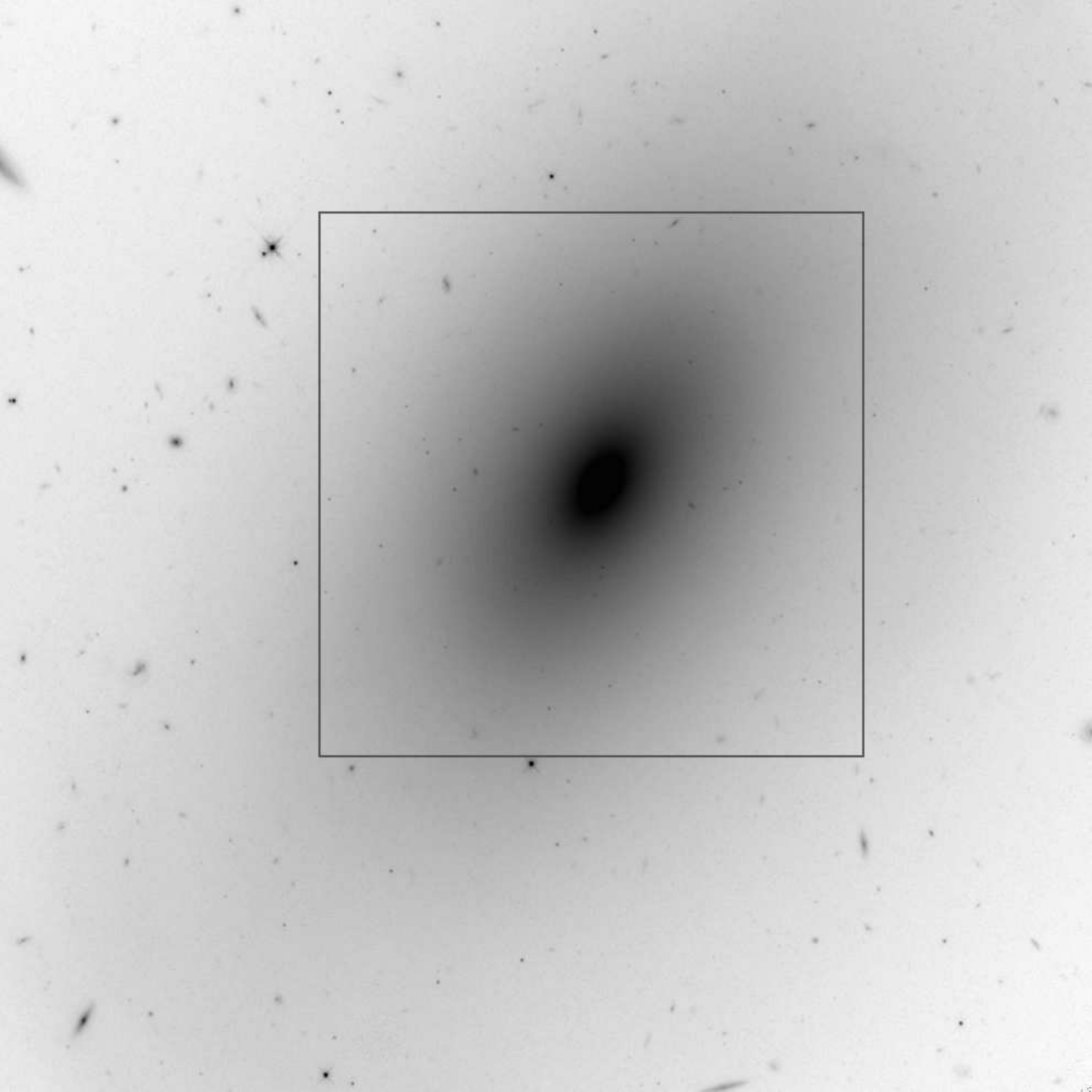}
\includegraphics[scale=0.4]{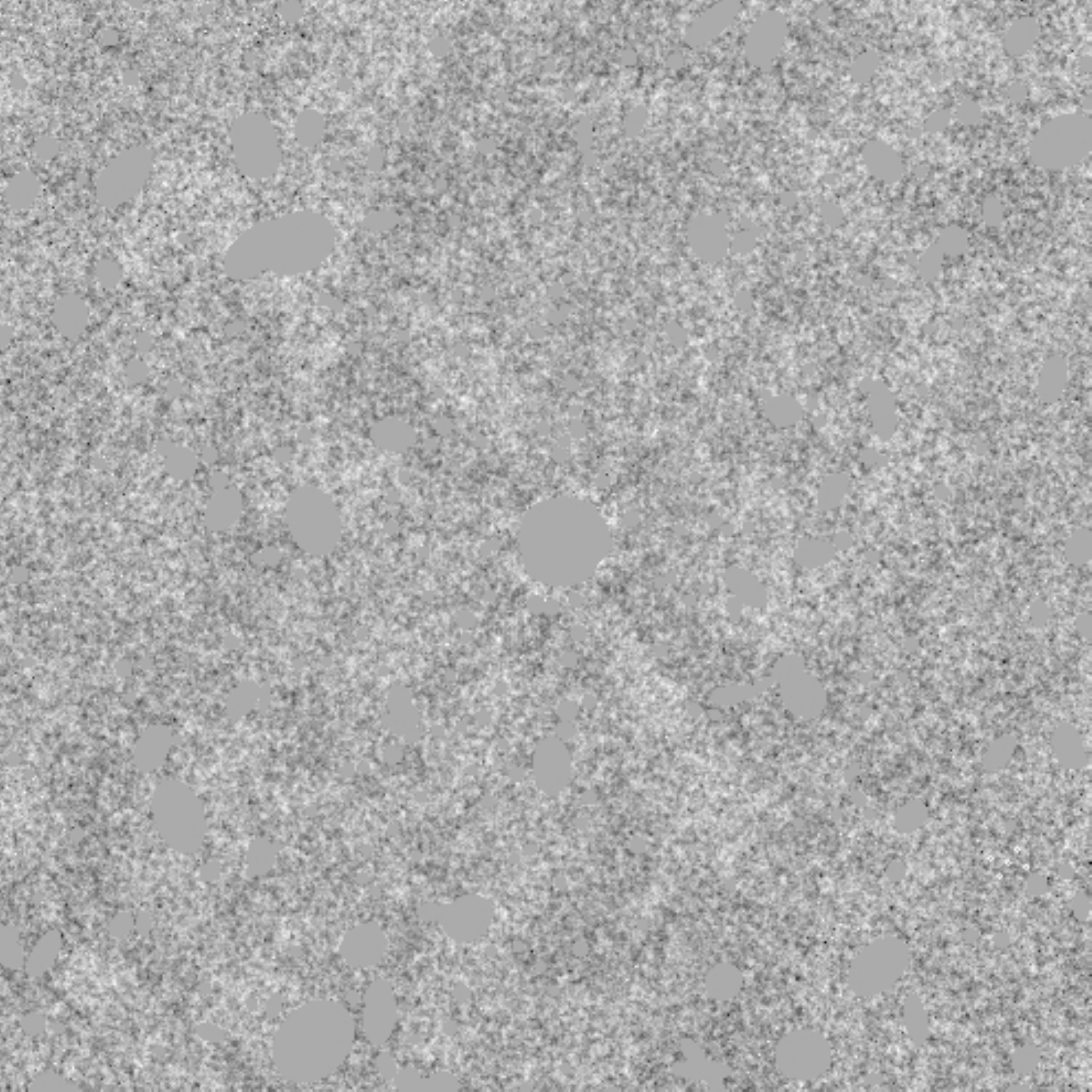} \\
\vspace{10pt}
\includegraphics[scale=0.4]{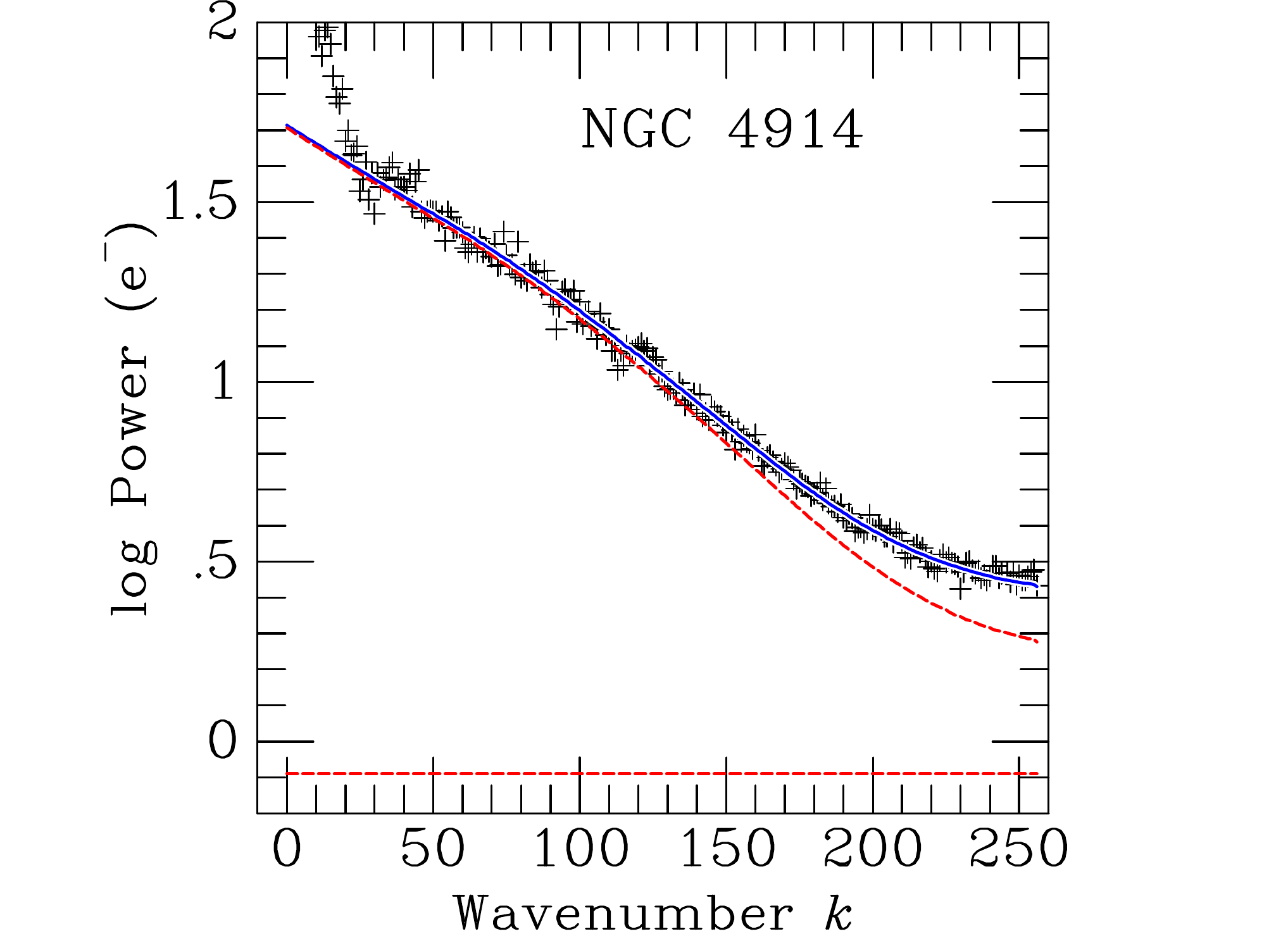}
\hspace{-25pt}
\includegraphics[scale=0.4]{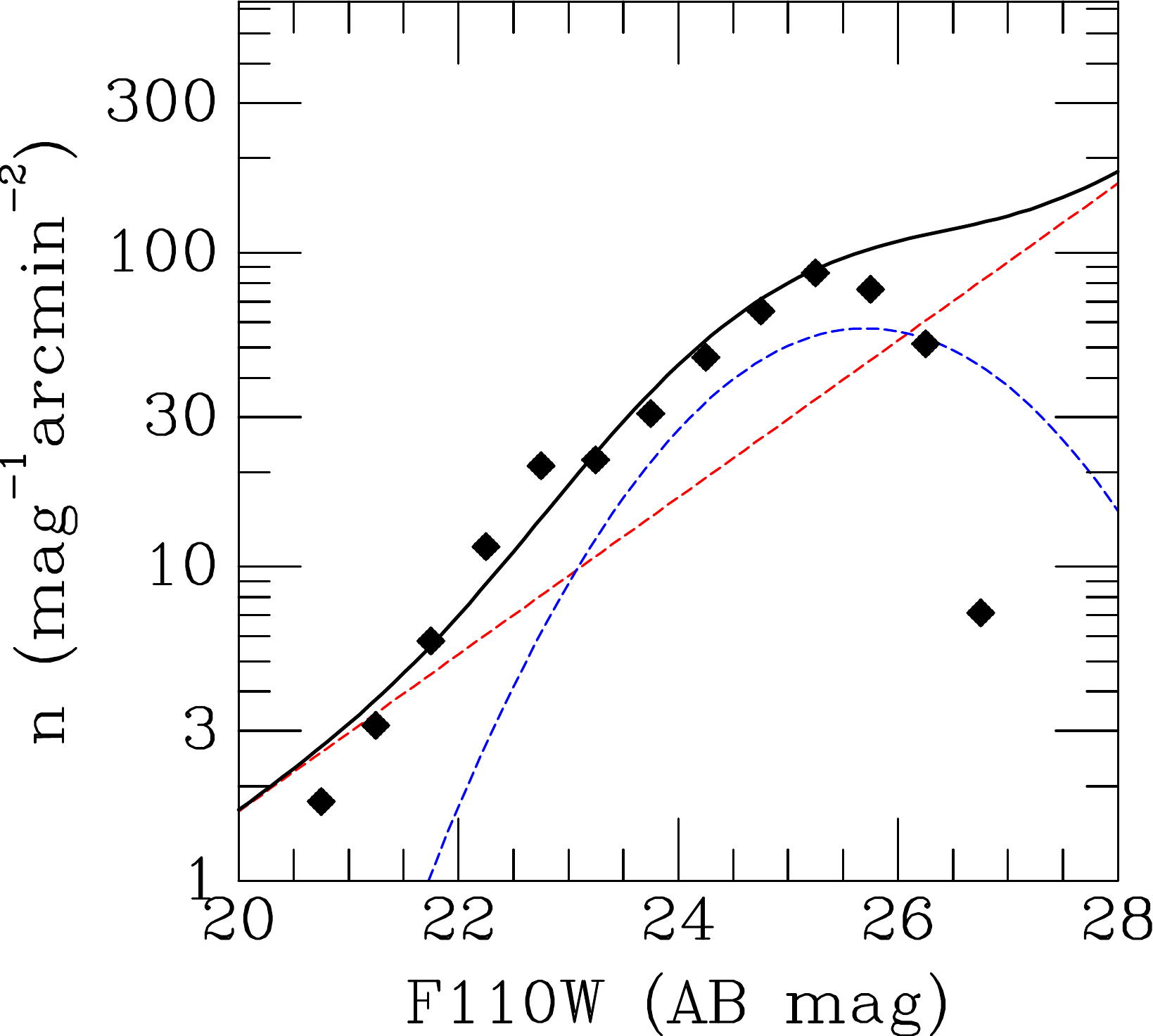}
\caption{Combined figure for NGC~4914.}
\end{center}
\end{figure*}
\clearpage


\begin{figure*}
\begin{center}
\includegraphics[scale=0.2]{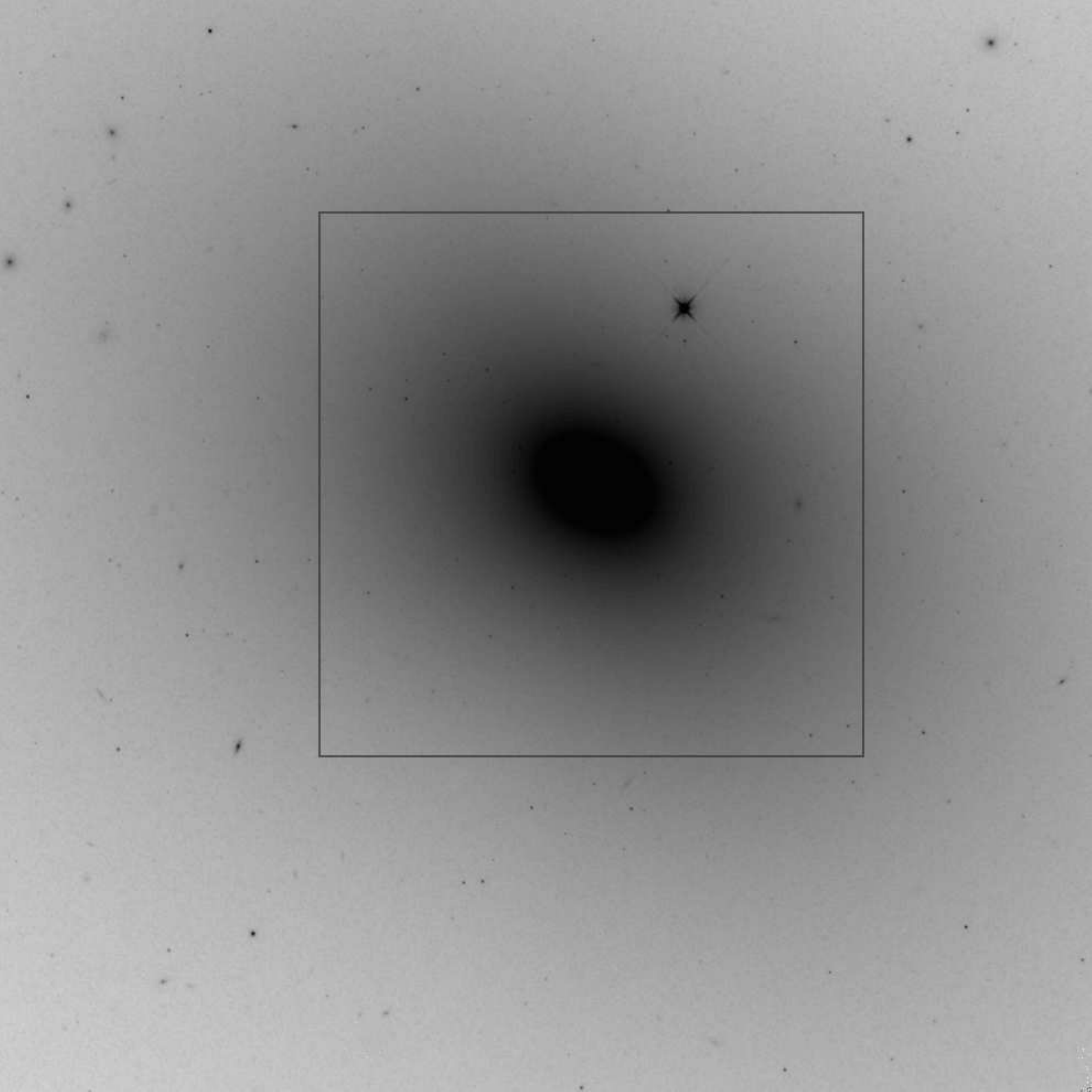}
\includegraphics[scale=0.4]{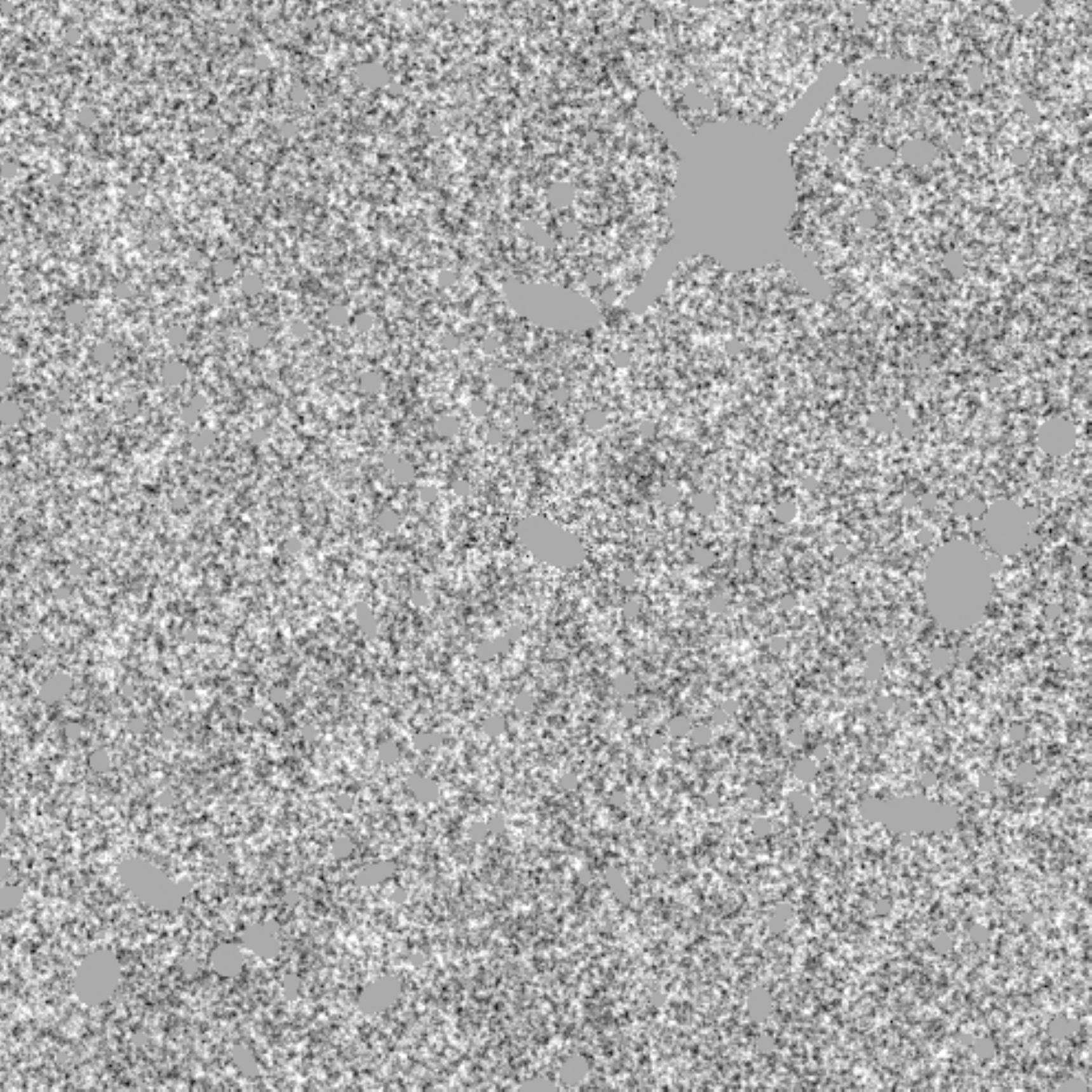} \\
\vspace{10pt}
\includegraphics[scale=0.4]{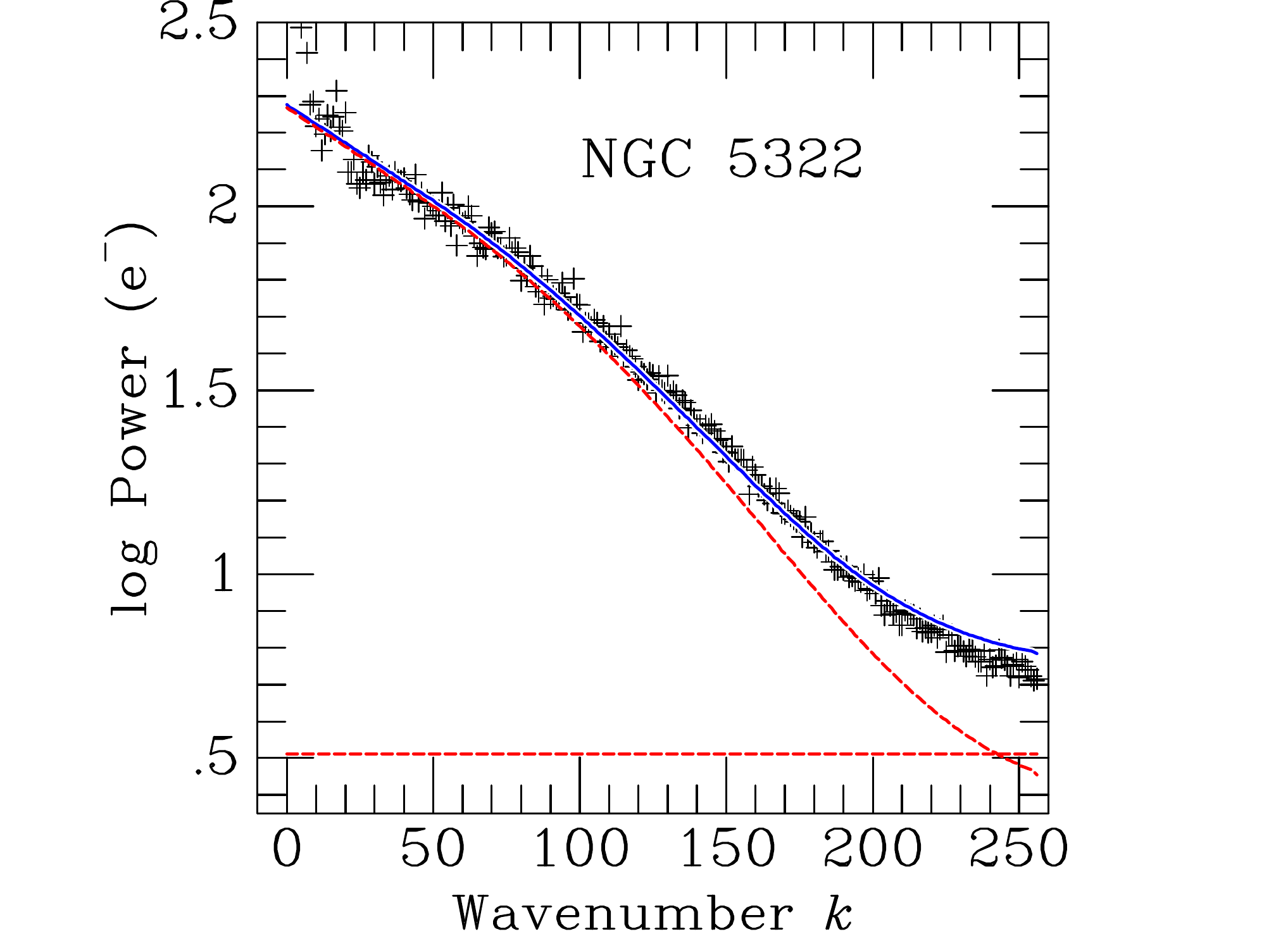}
\hspace{-25pt}
\includegraphics[scale=0.4]{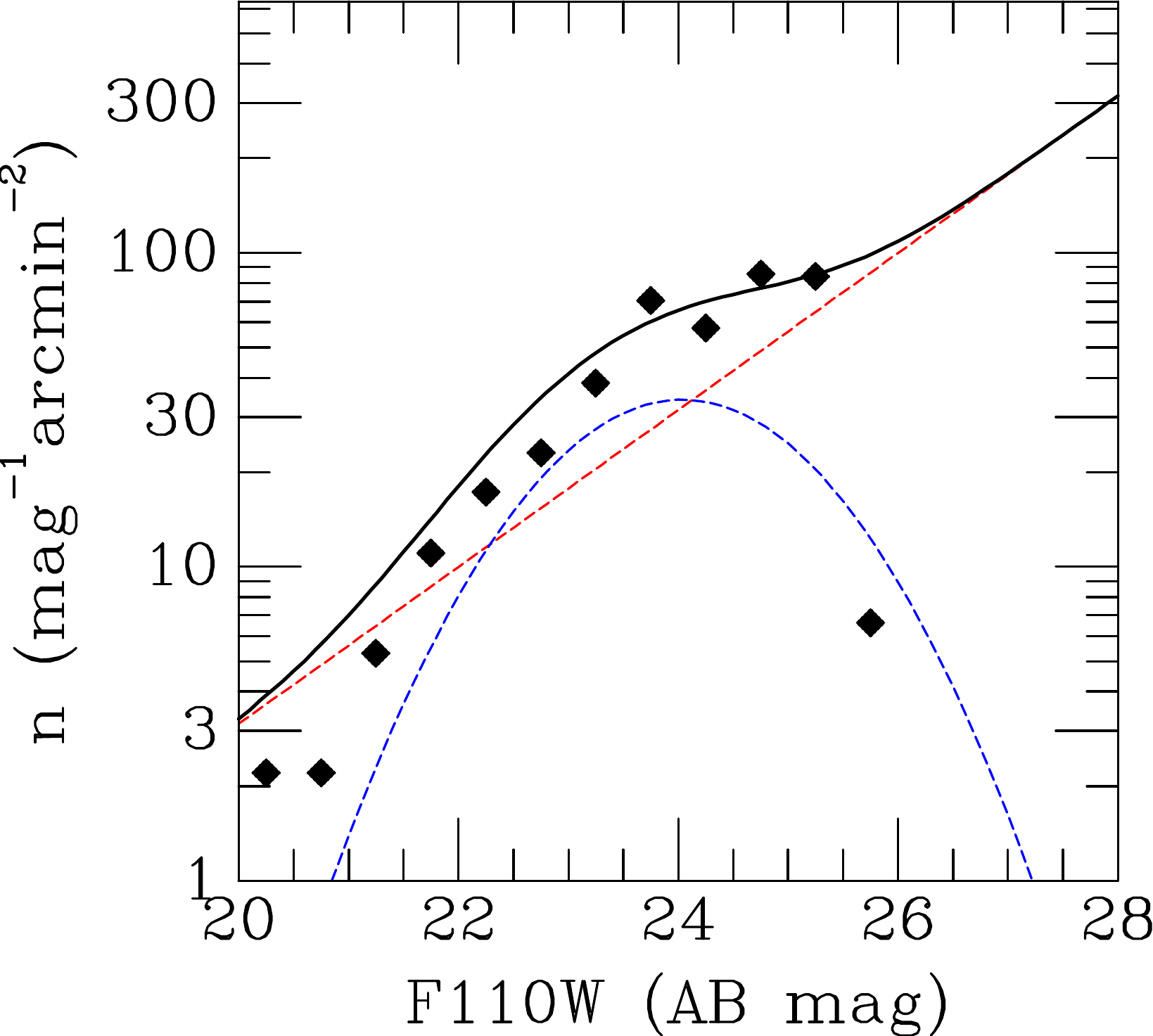}
\caption{Combined figure for NGC~5322.}
\end{center}
\end{figure*}
\clearpage

\begin{figure*}
\begin{center}
\includegraphics[scale=0.2]{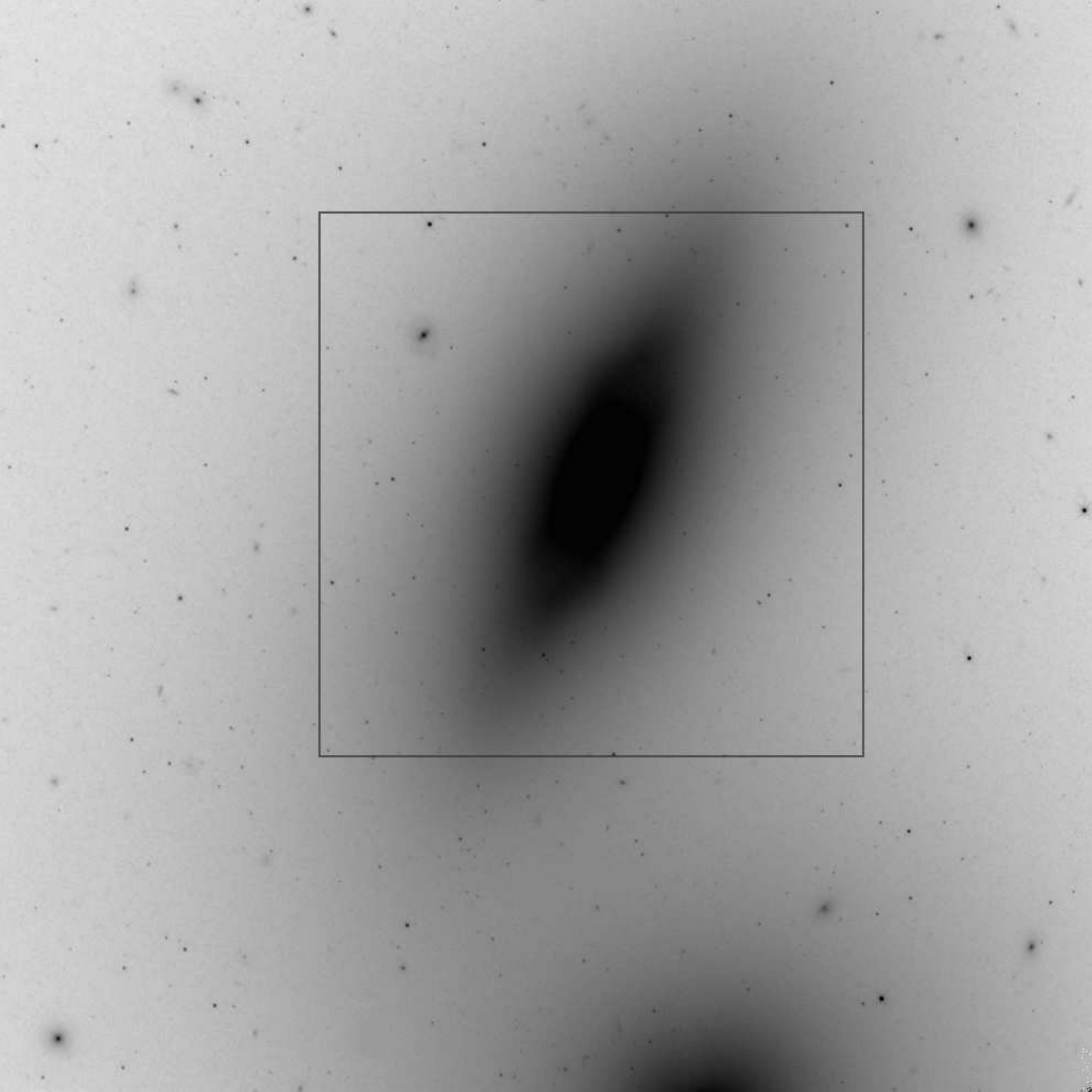}
\includegraphics[scale=0.4]{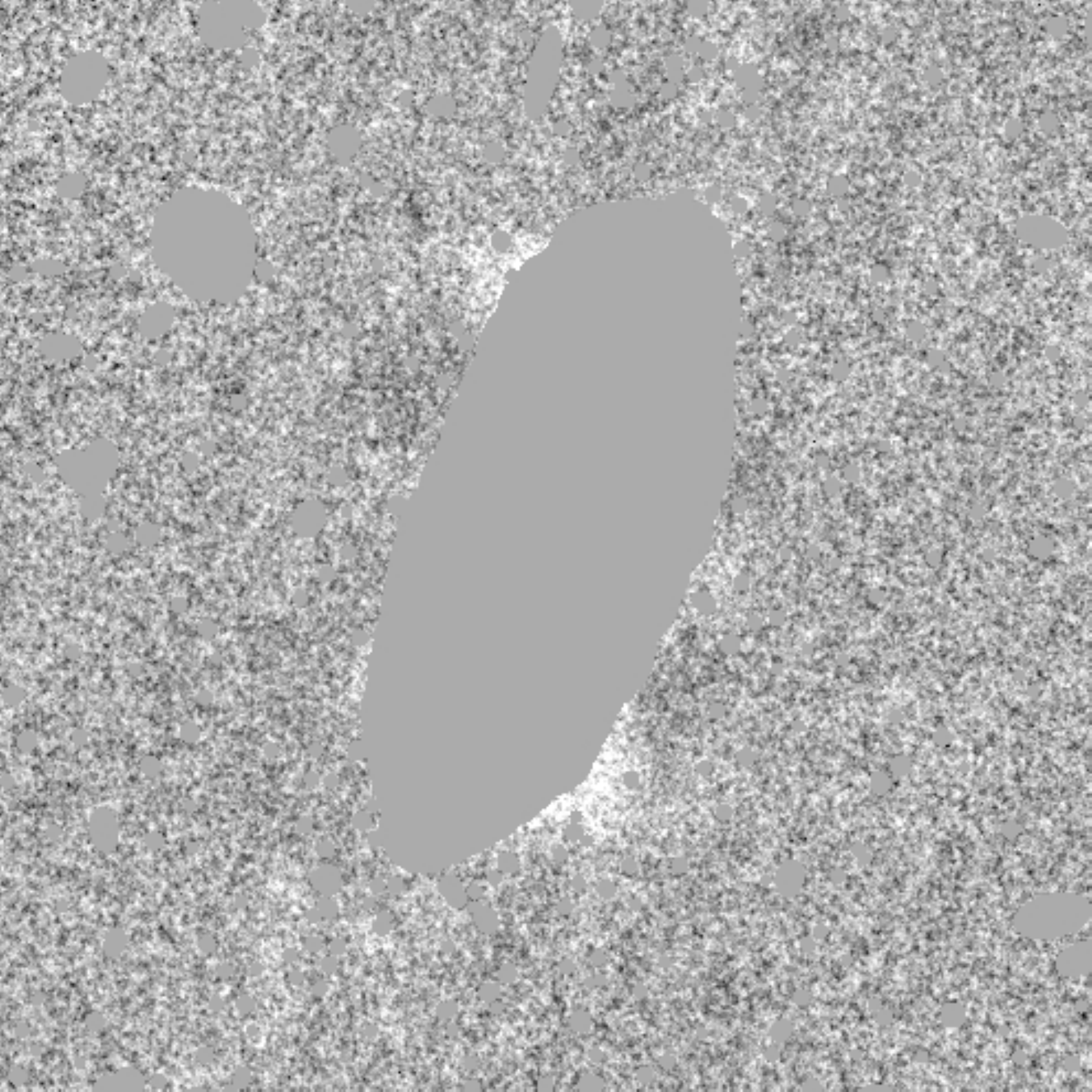} \\
\vspace{10pt}
\includegraphics[scale=0.4]{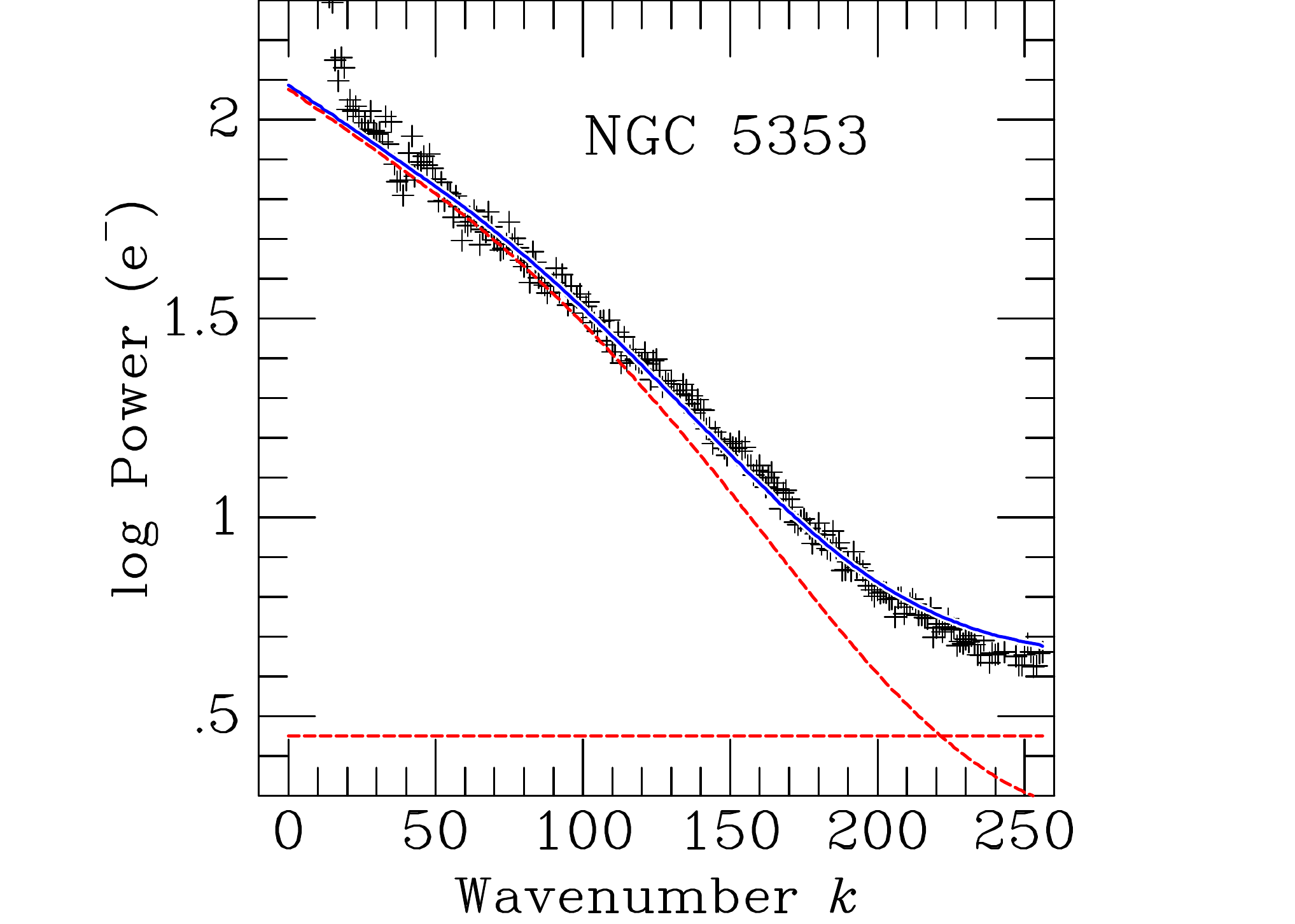}
\hspace{-25pt}
\includegraphics[scale=0.4]{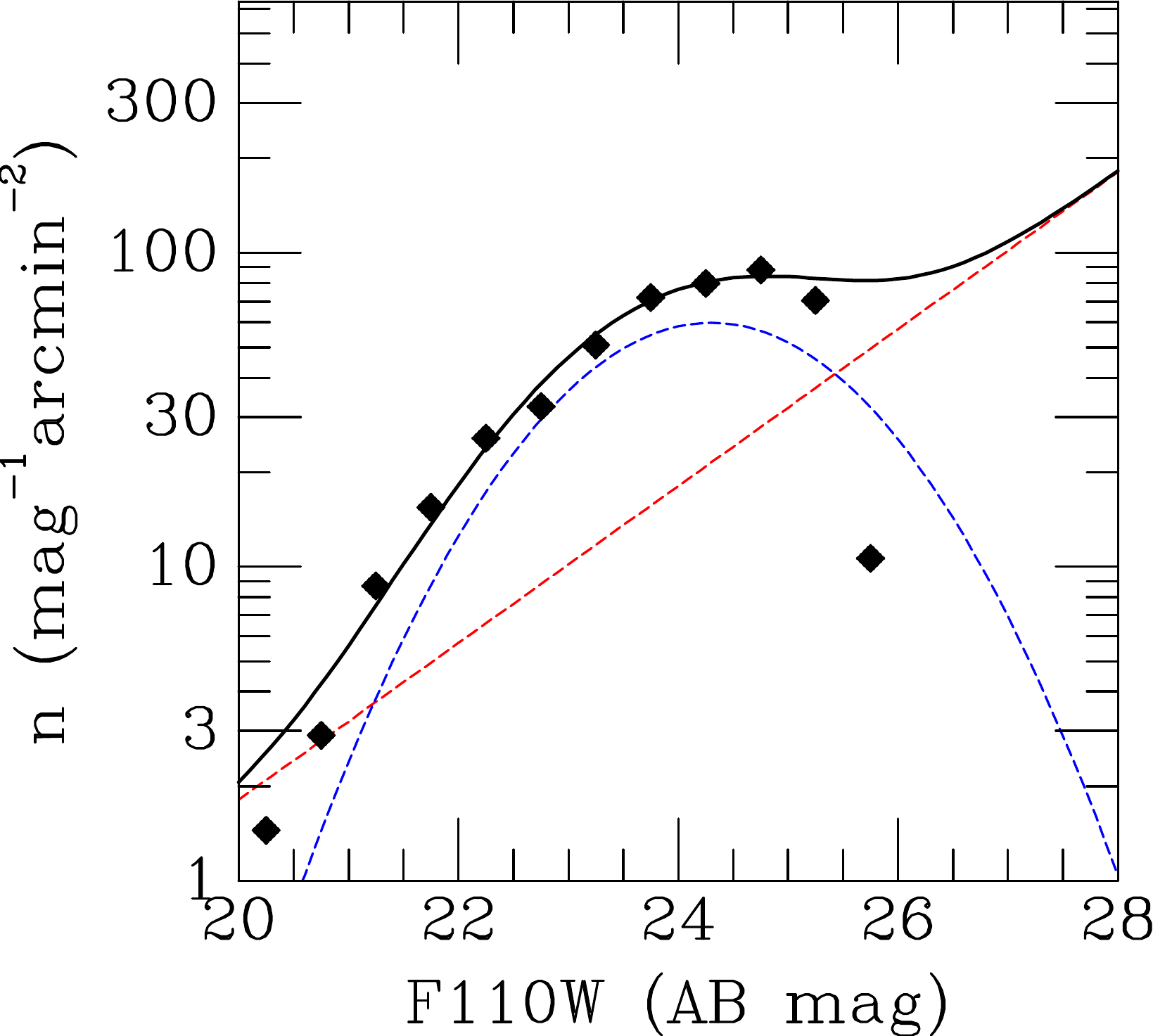}
\caption{Combined figure for NGC~5353.}
\end{center}
\end{figure*}
\clearpage

\begin{figure*}
\begin{center}
\includegraphics[scale=0.2]{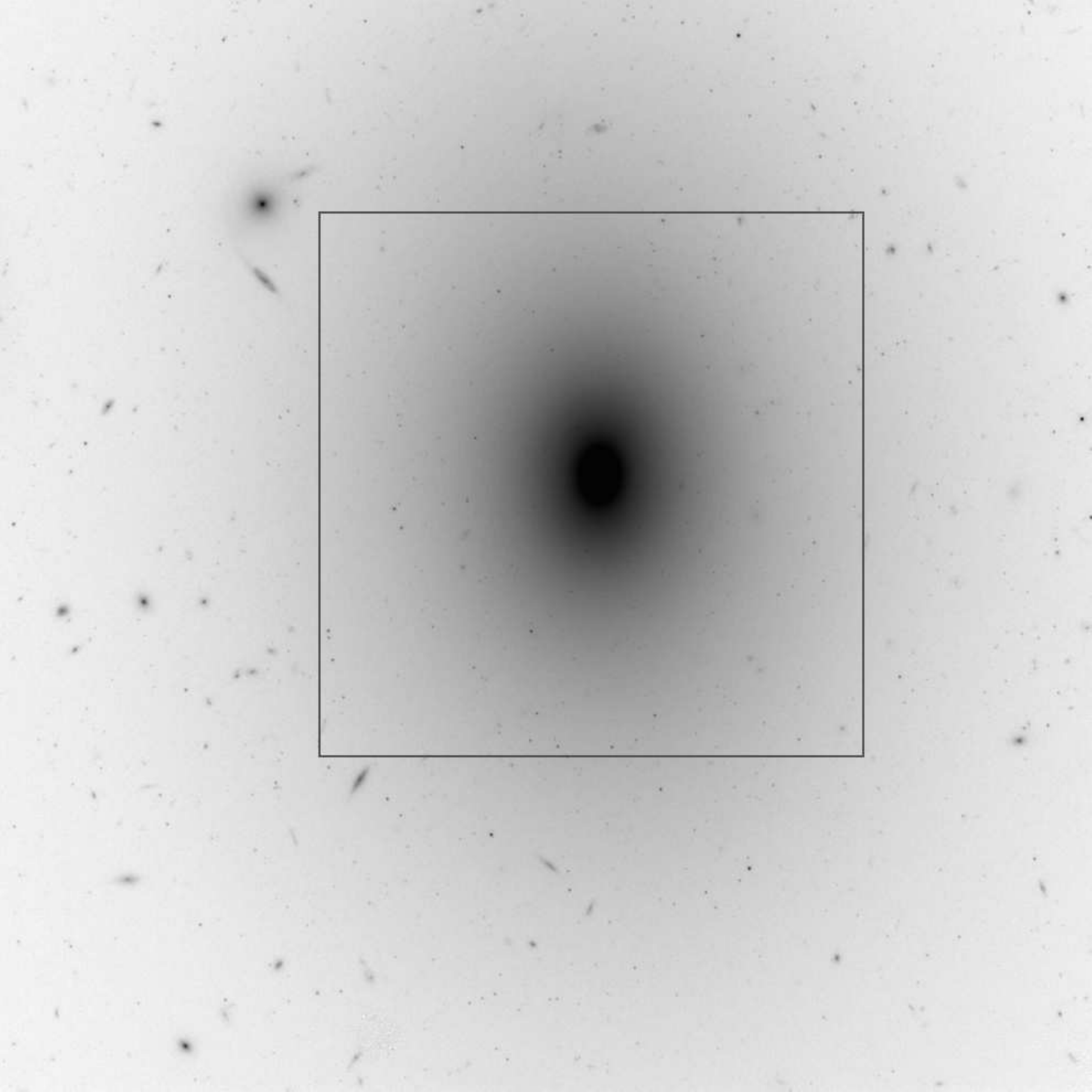}
\includegraphics[scale=0.4]{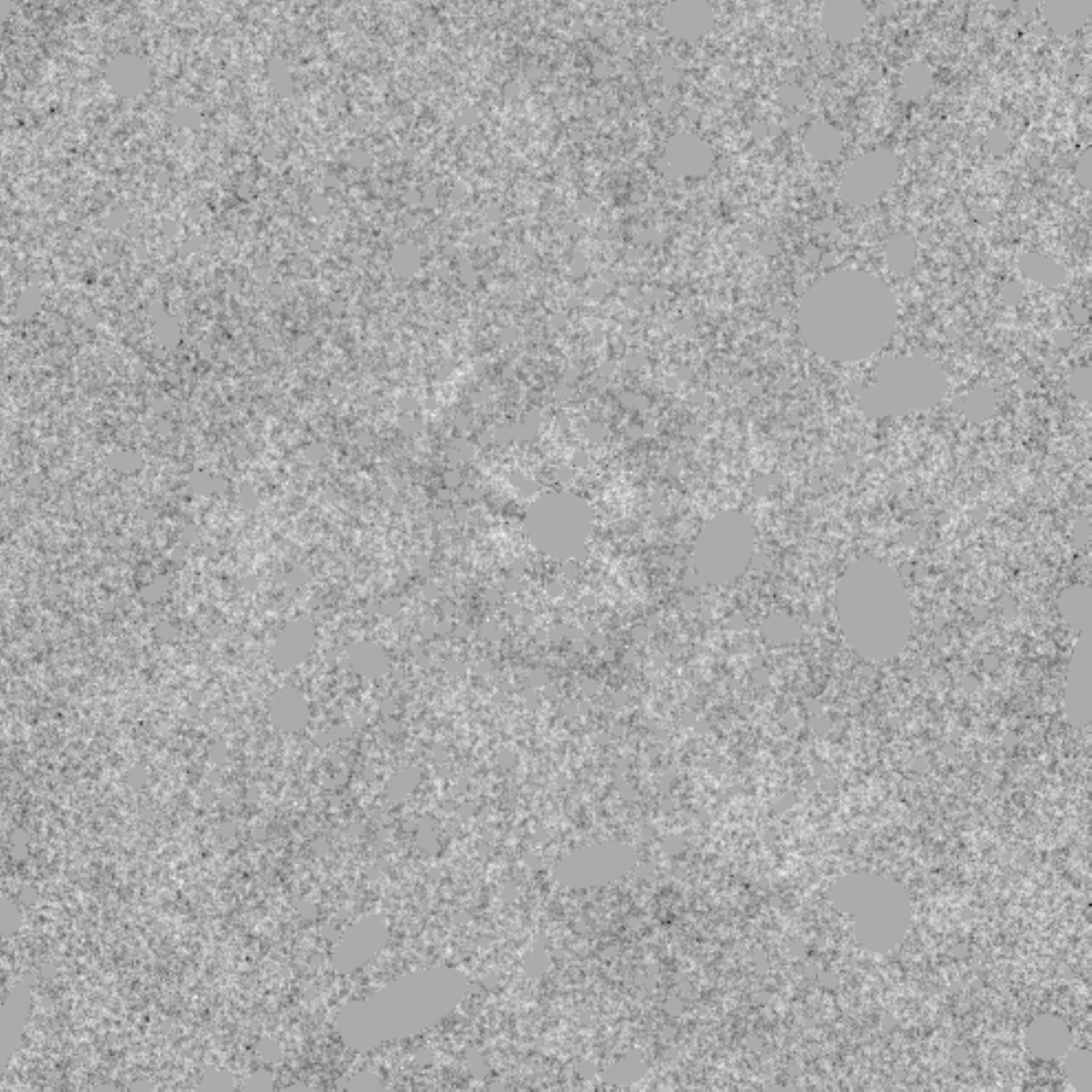} \\
\vspace{10pt}
\includegraphics[scale=0.4]{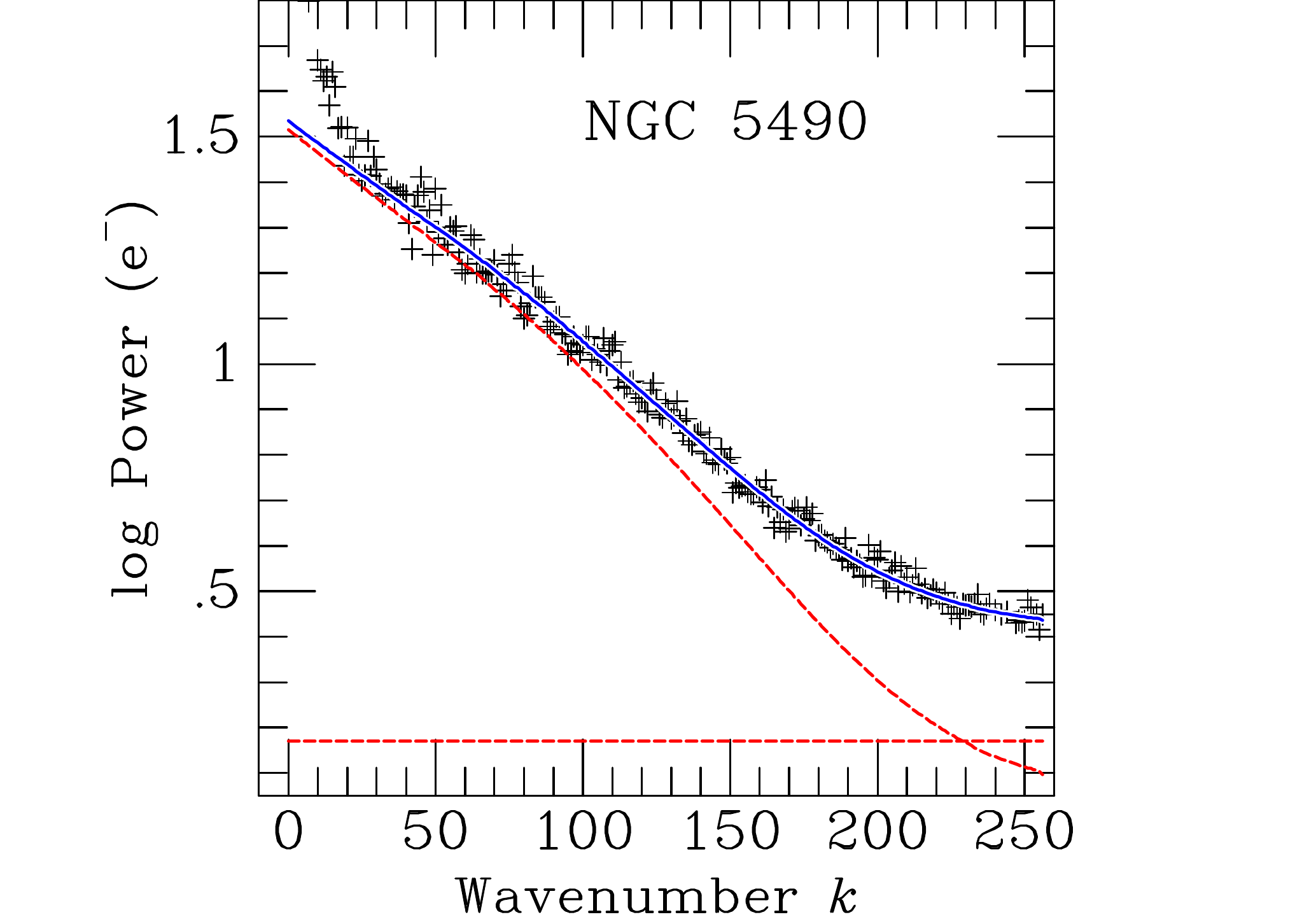}
\hspace{-25pt}
\includegraphics[scale=0.4]{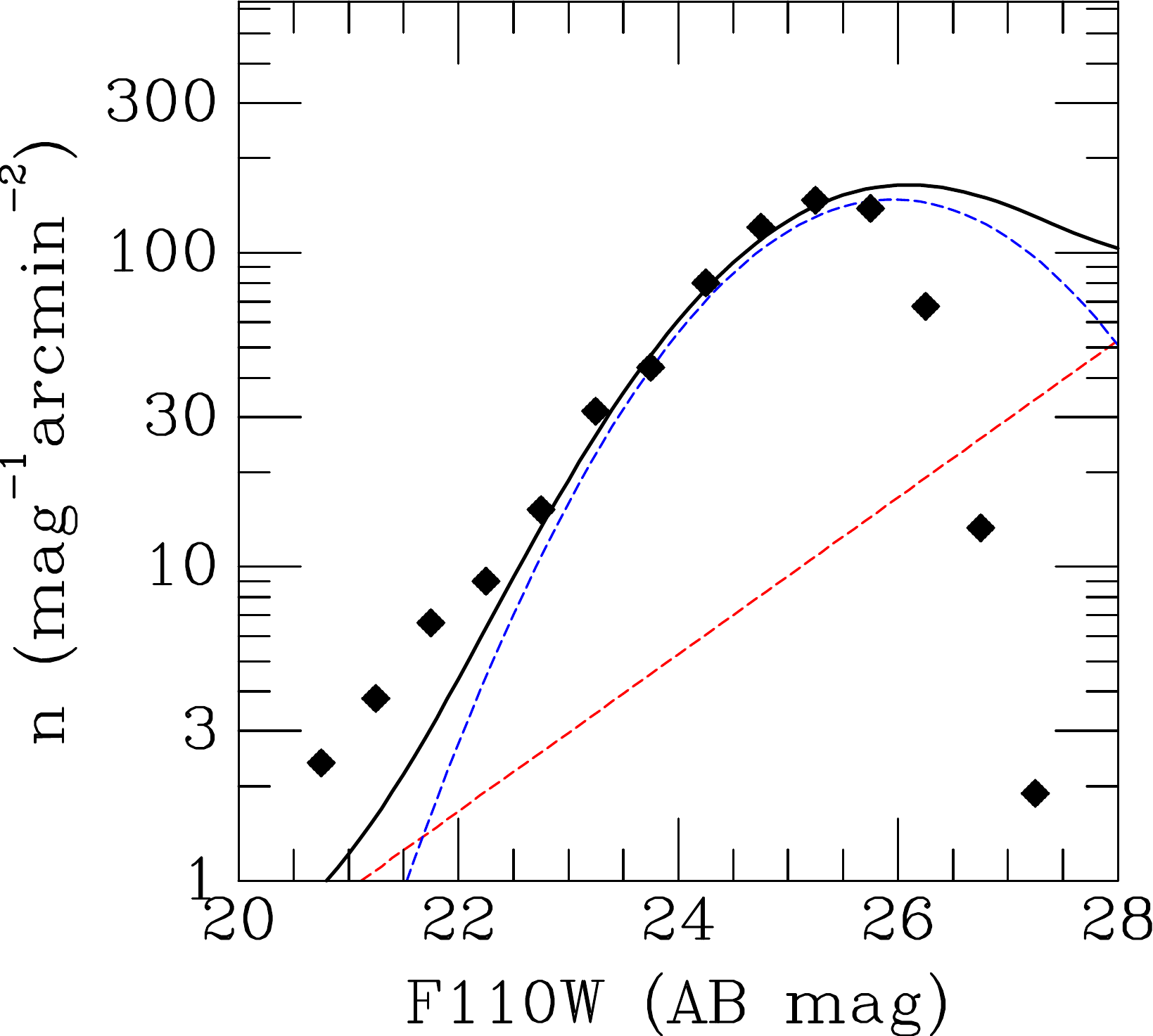}
\caption{Combined figure for NGC~5490.}
\end{center}
\end{figure*}
\clearpage

\begin{figure*}
\begin{center}
\includegraphics[scale=0.2]{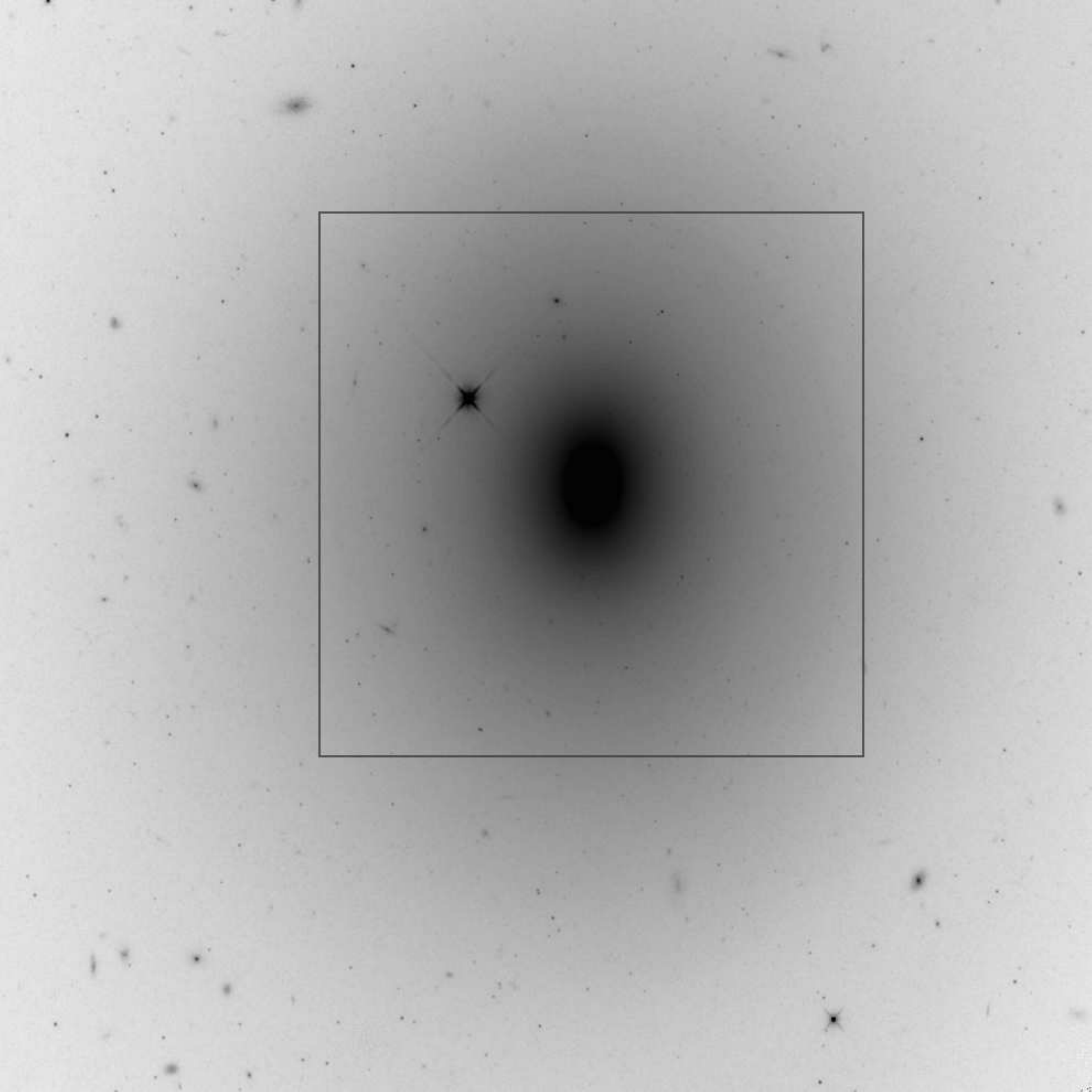}
\includegraphics[scale=0.4]{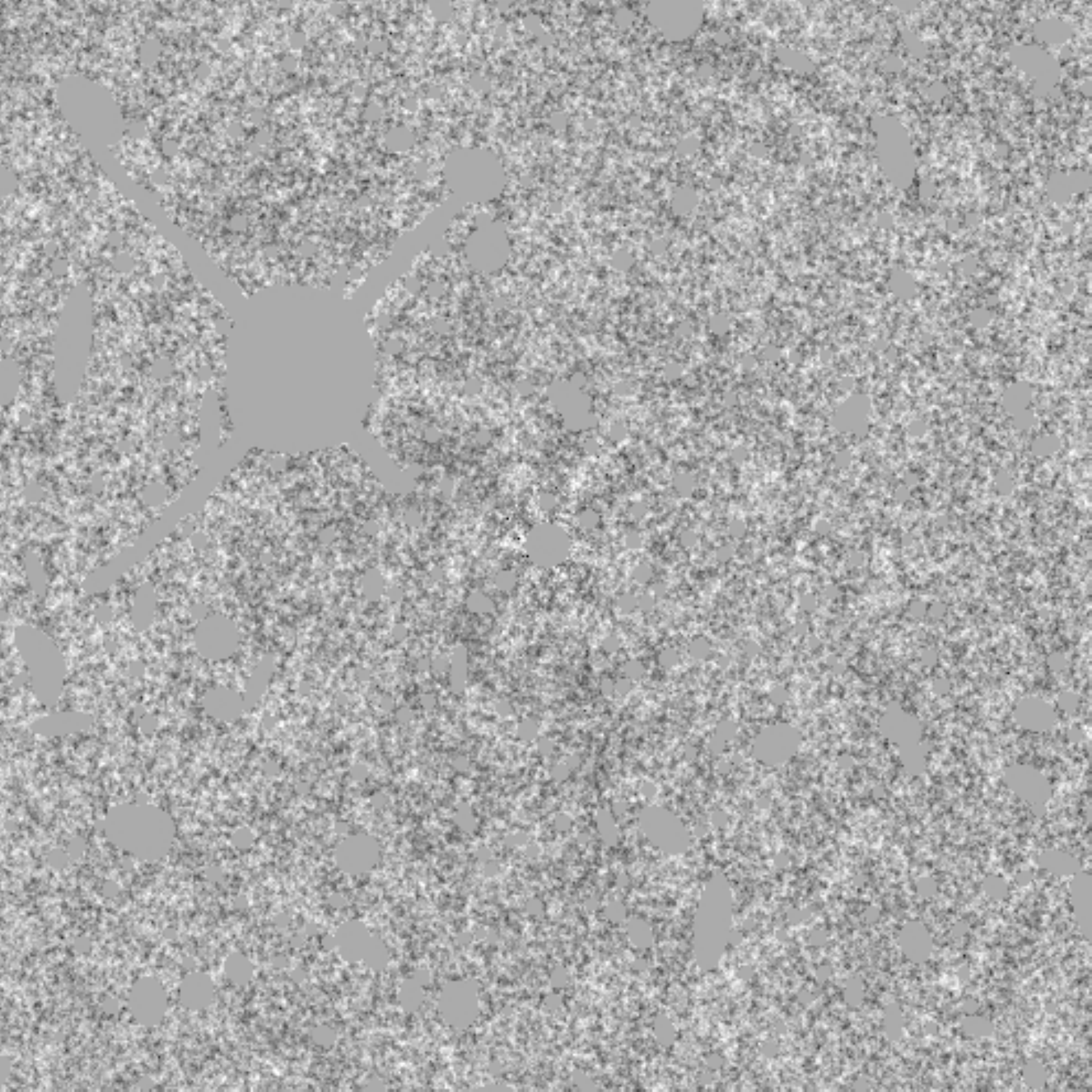} \\
\vspace{10pt}
\includegraphics[scale=0.4]{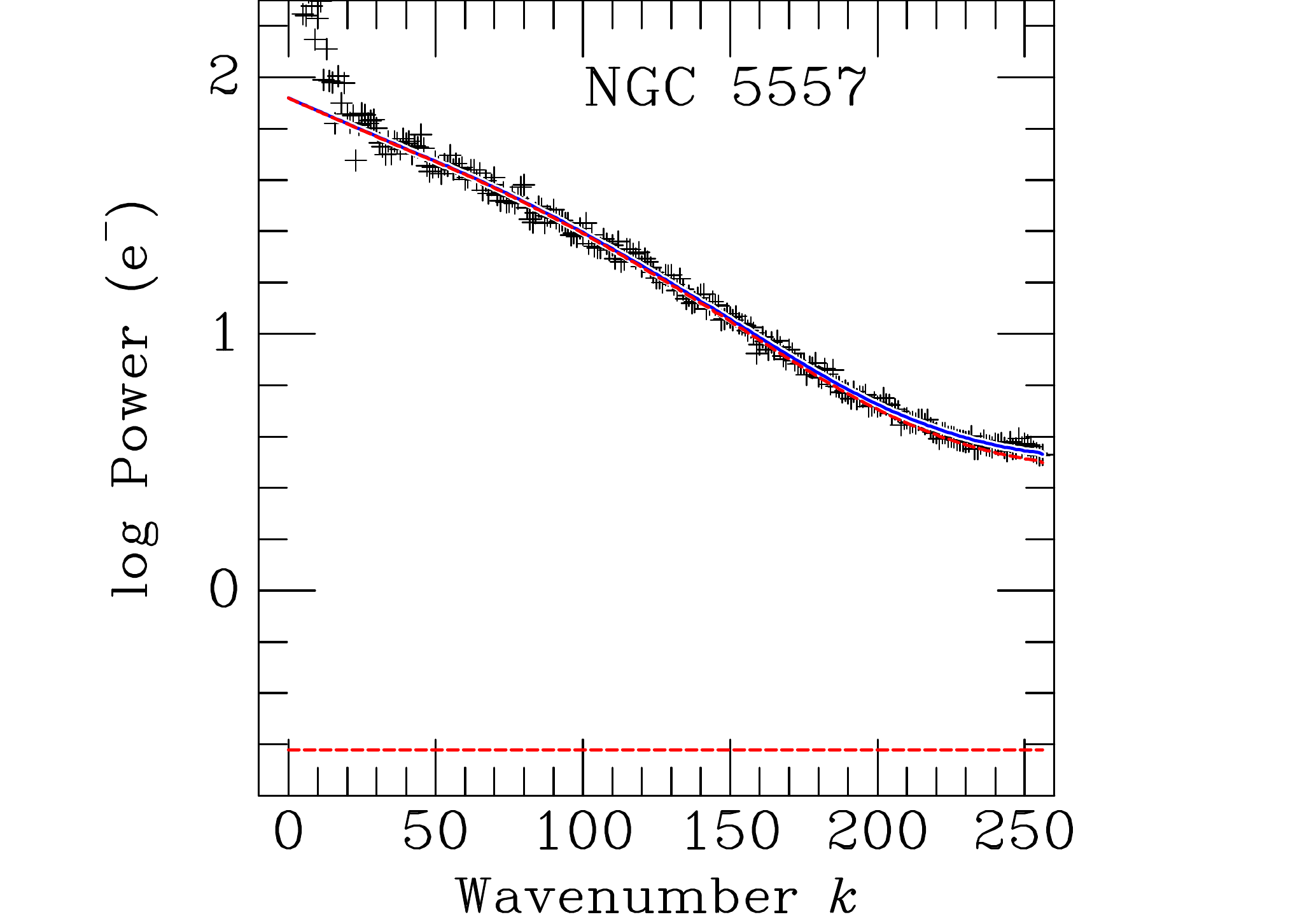}
\hspace{-25pt}
\includegraphics[scale=0.4]{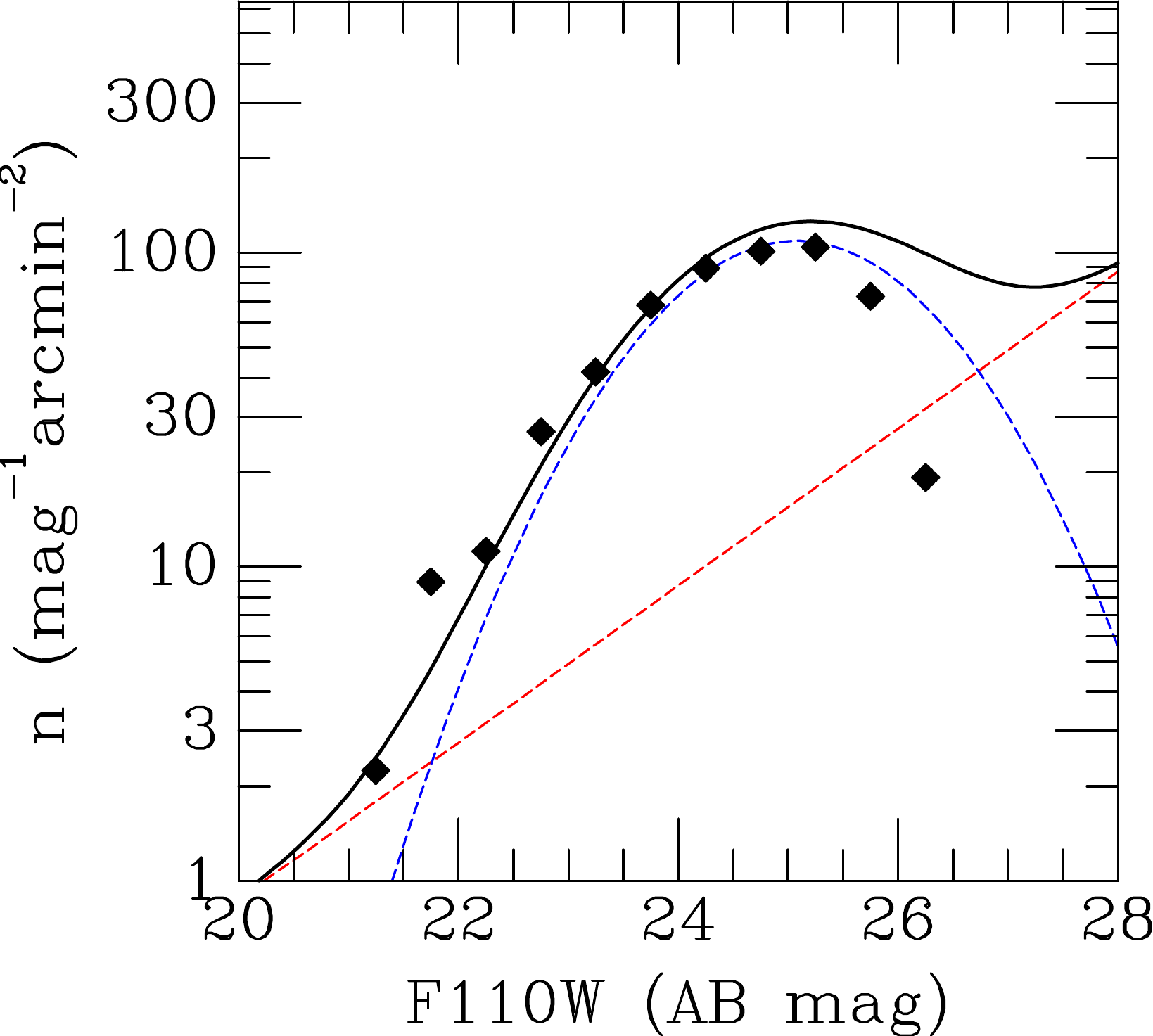}
\caption{Combined figure for NGC~5557.}
\end{center}
\end{figure*}
\clearpage

\begin{figure*}
\begin{center}
\includegraphics[scale=0.2]{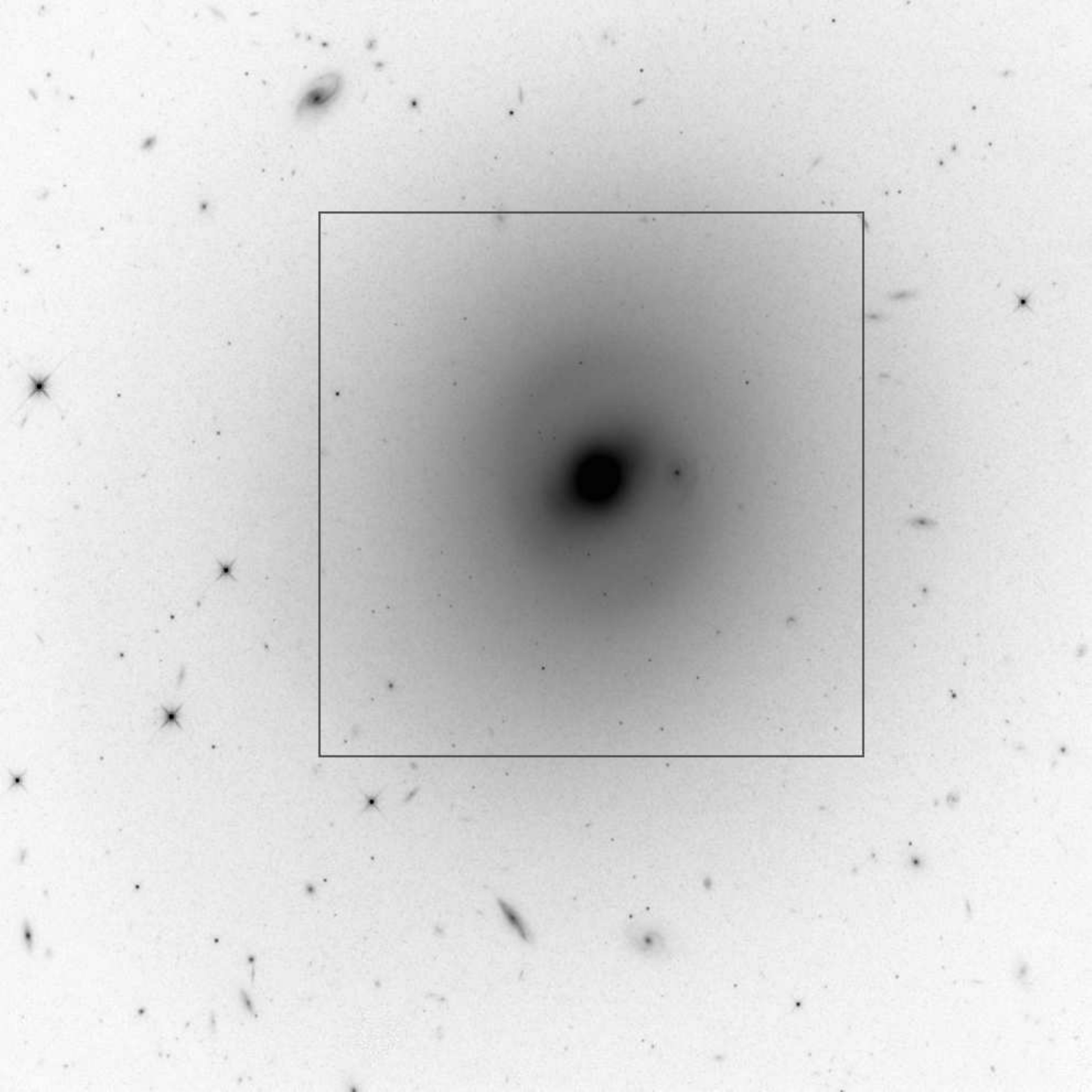}
\includegraphics[scale=0.4]{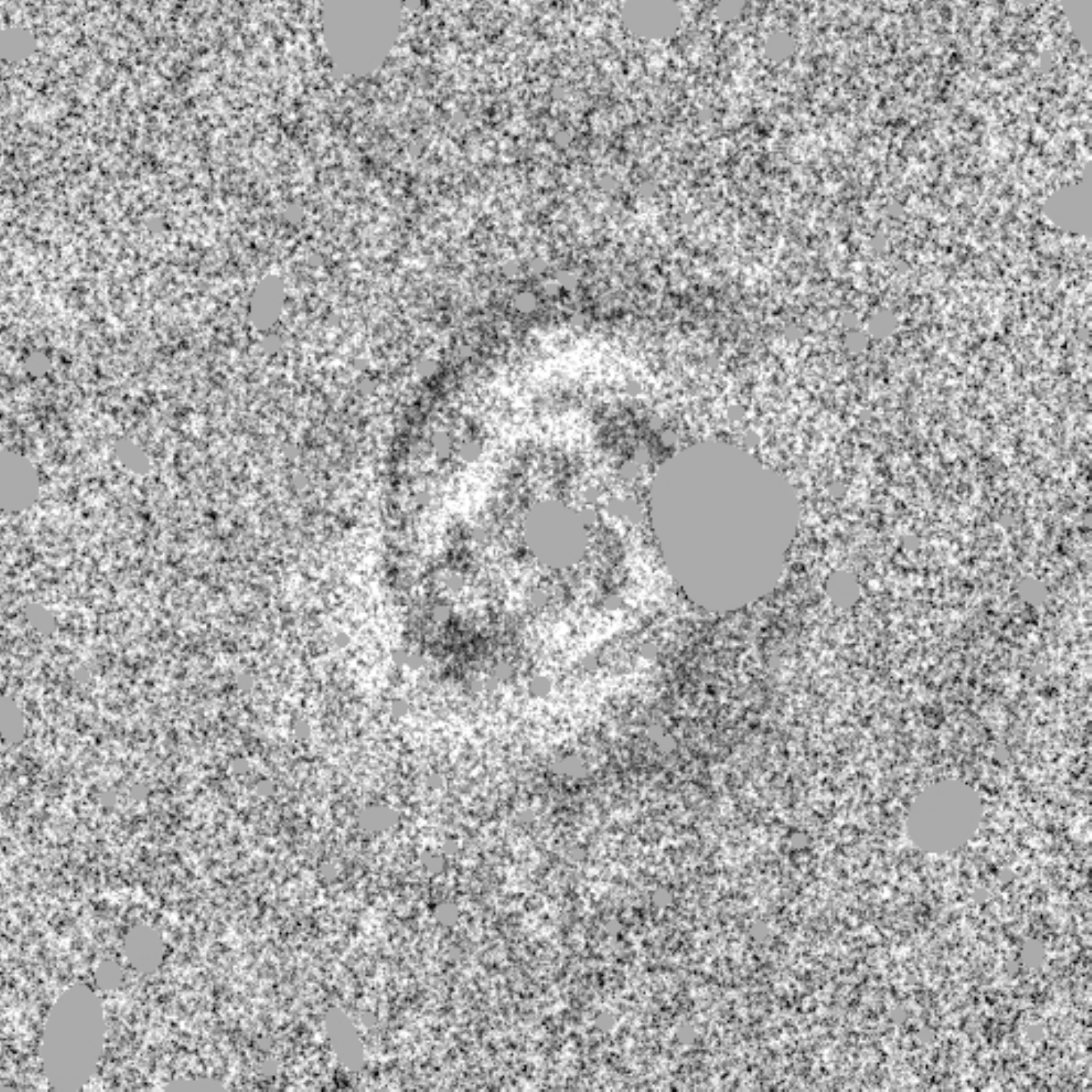} \\
\vspace{10pt}
\includegraphics[scale=0.4]{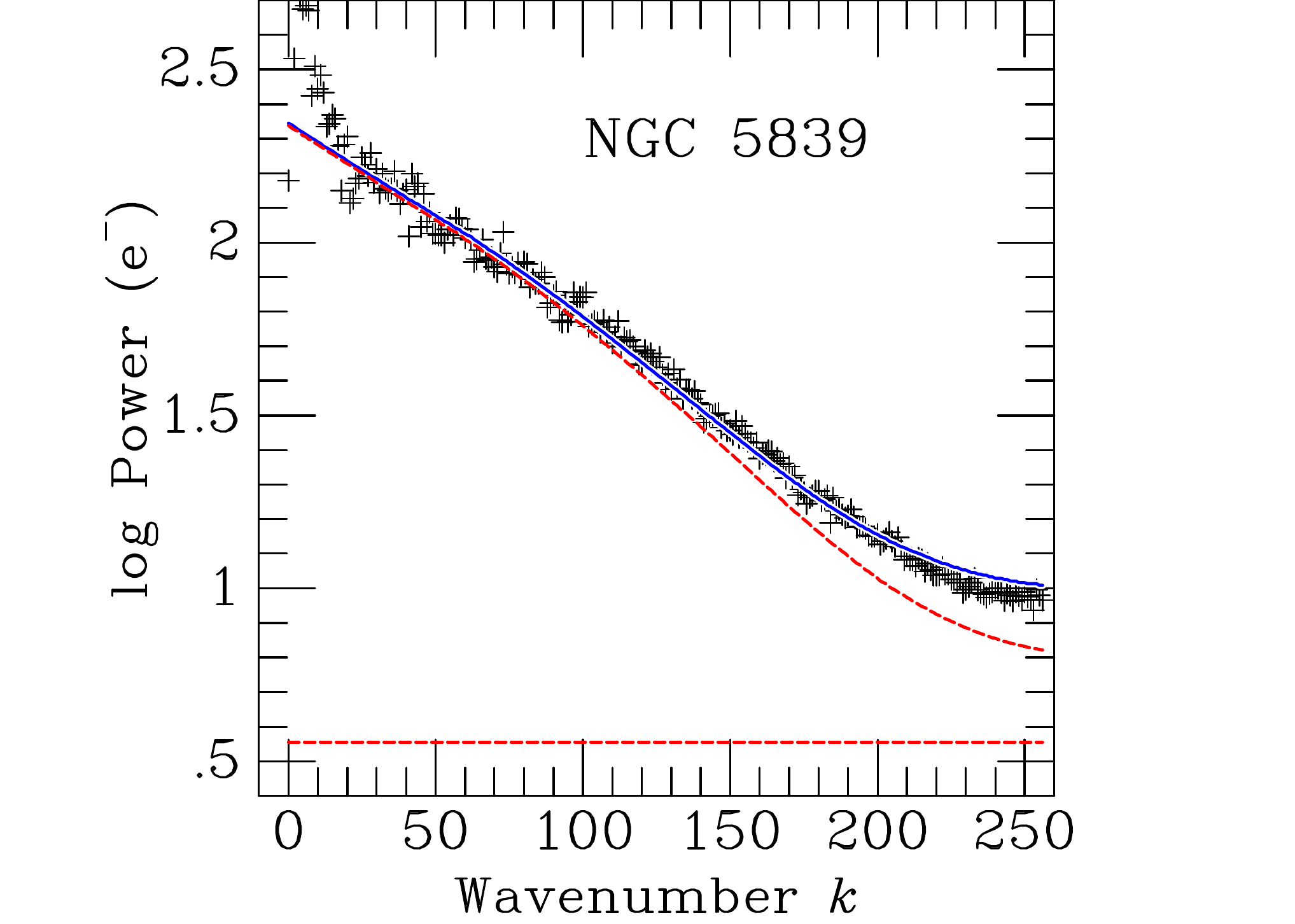}
\hspace{-25pt}
\includegraphics[scale=0.4]{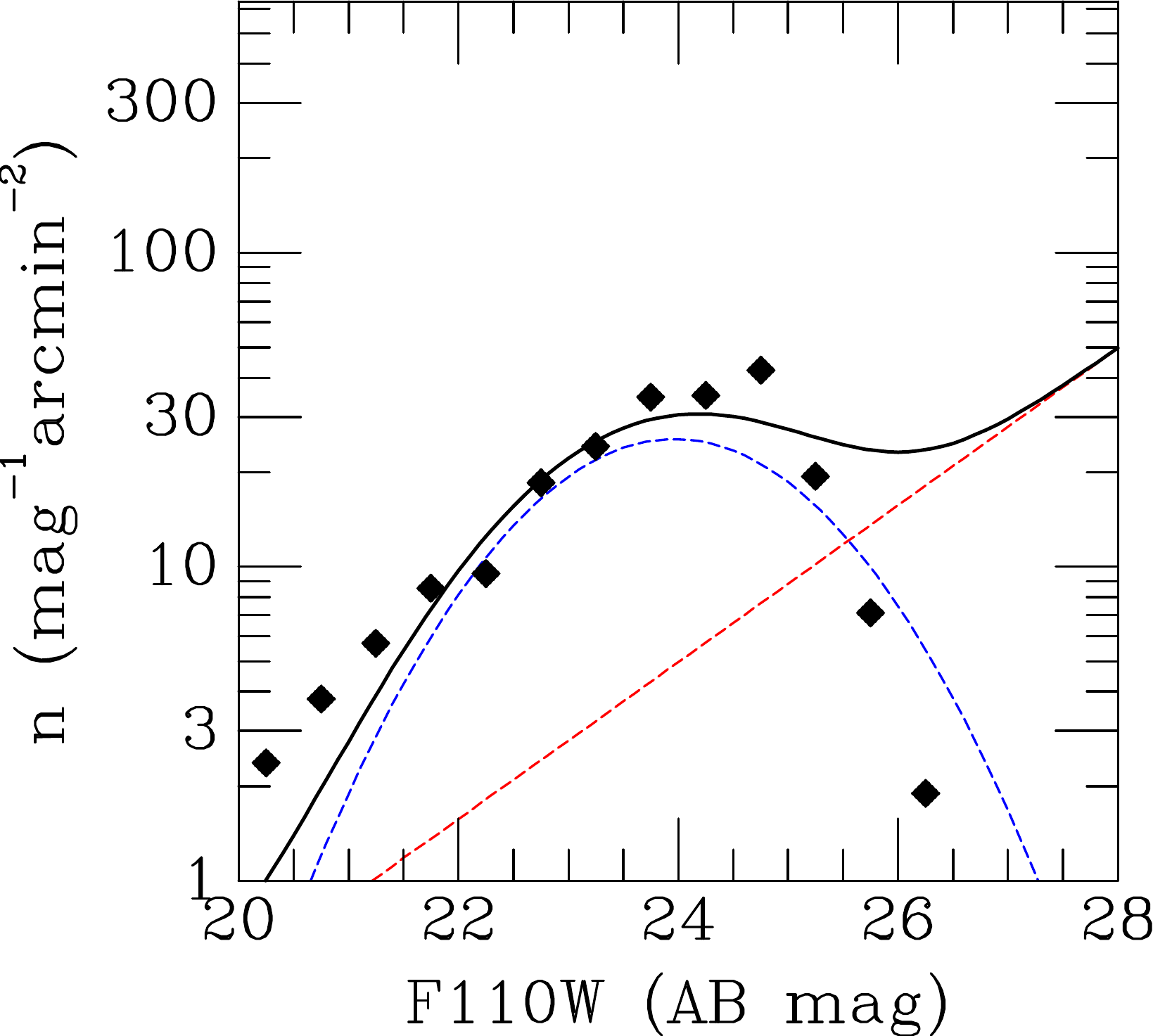}
\caption{Combined figure for NGC~5839.}
\end{center}
\end{figure*}
\clearpage

\begin{figure*}
\begin{center}
\includegraphics[scale=0.2]{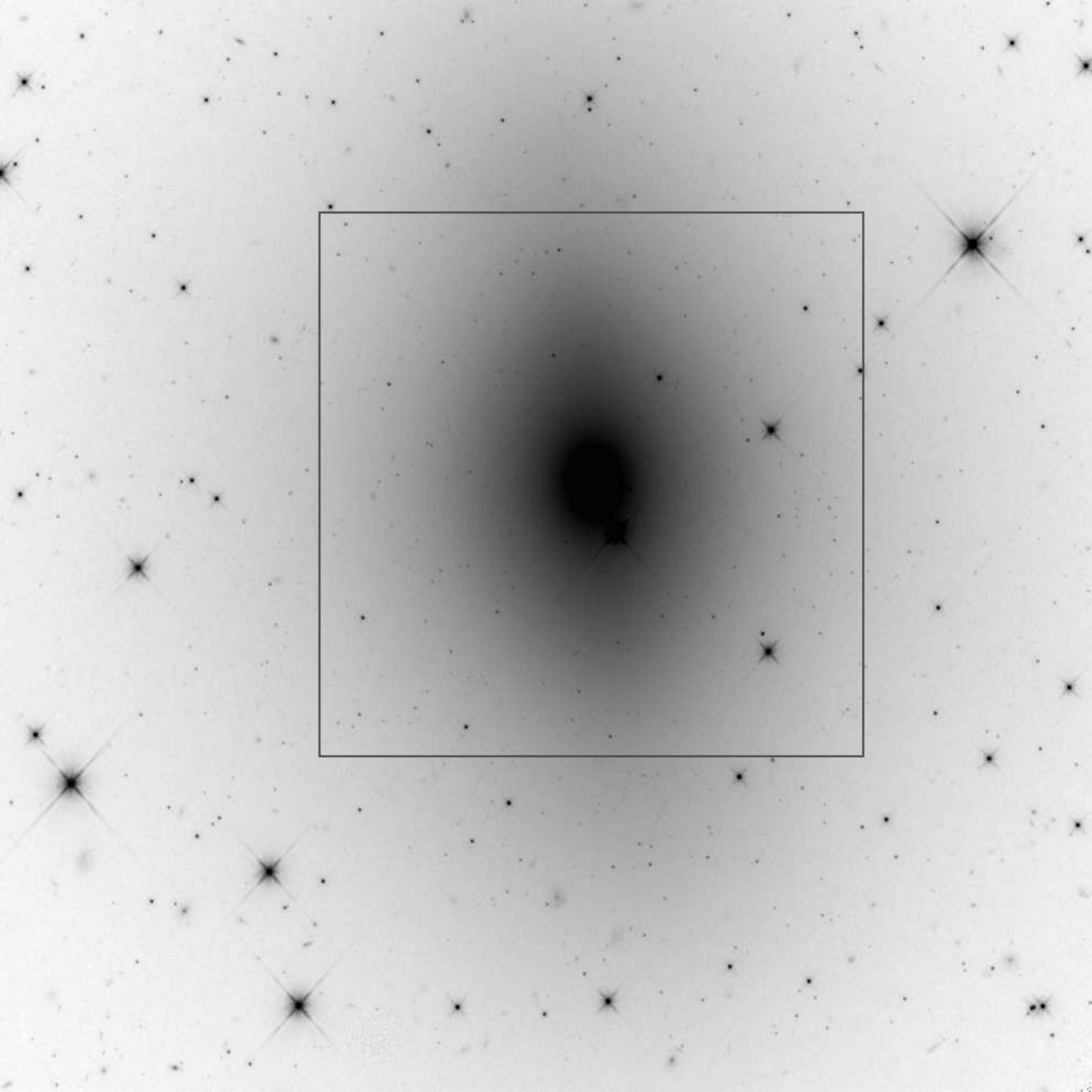}
\includegraphics[scale=0.4]{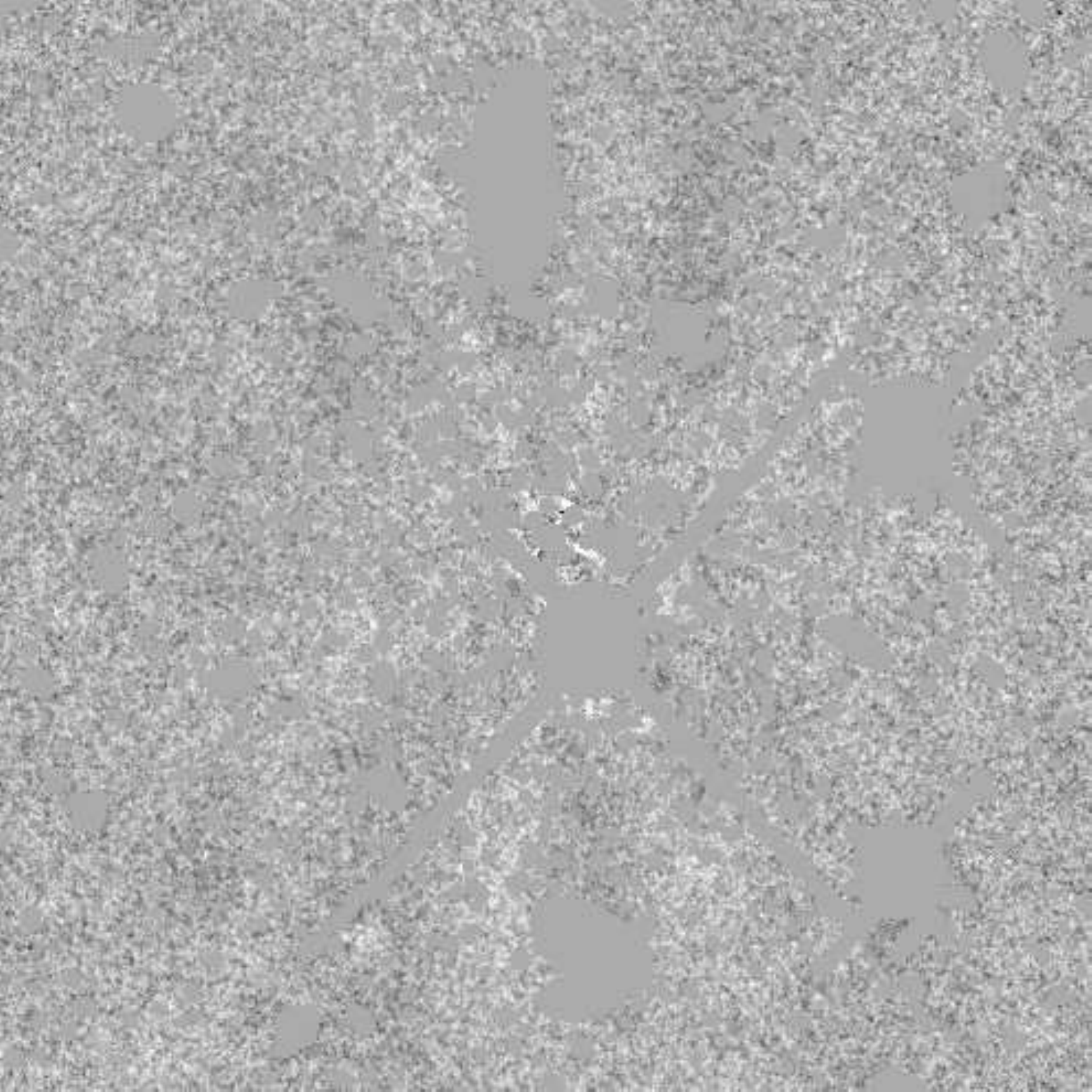} \\
\vspace{10pt}
\includegraphics[scale=0.4]{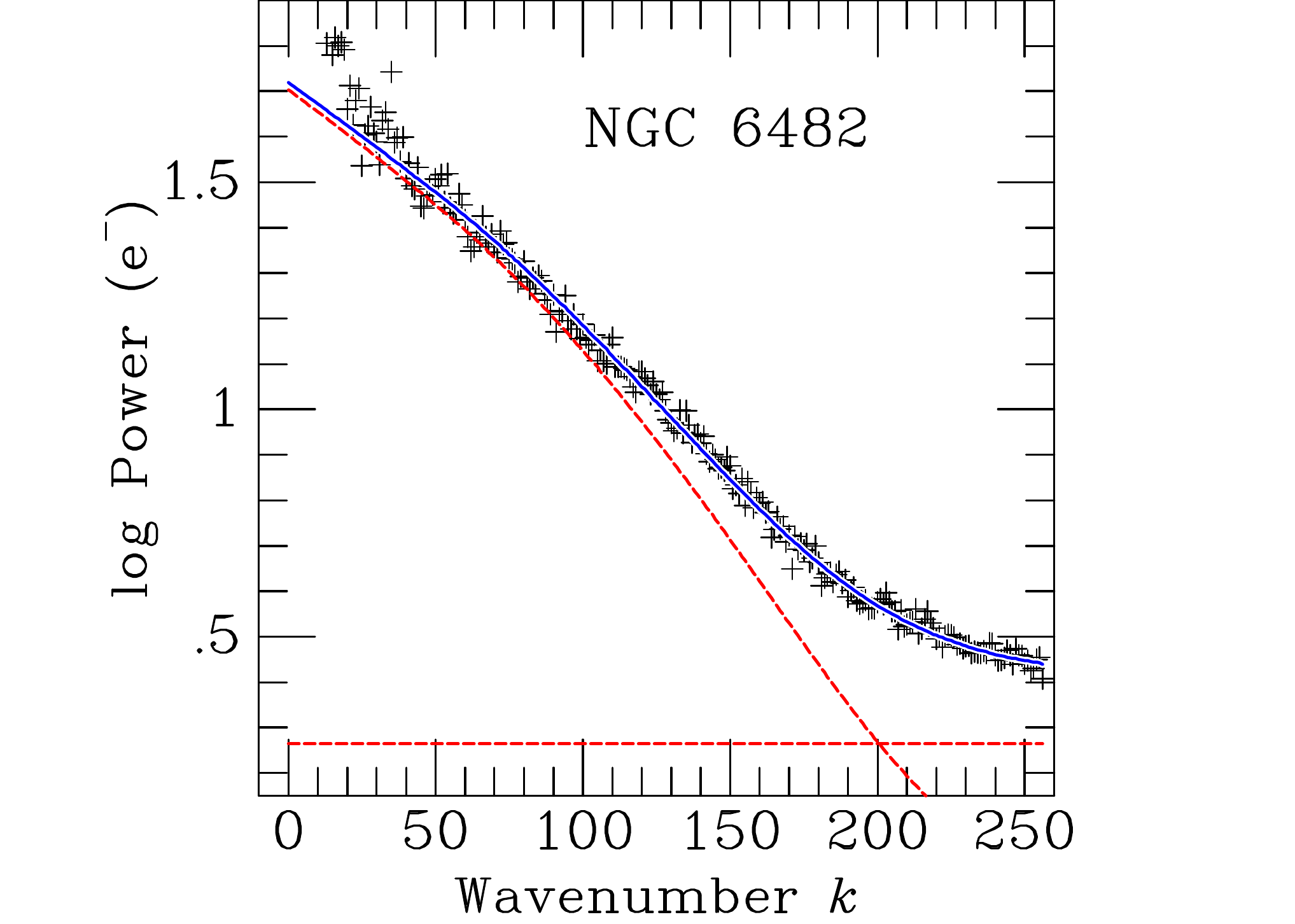}
\hspace{-25pt}
\includegraphics[scale=0.4]{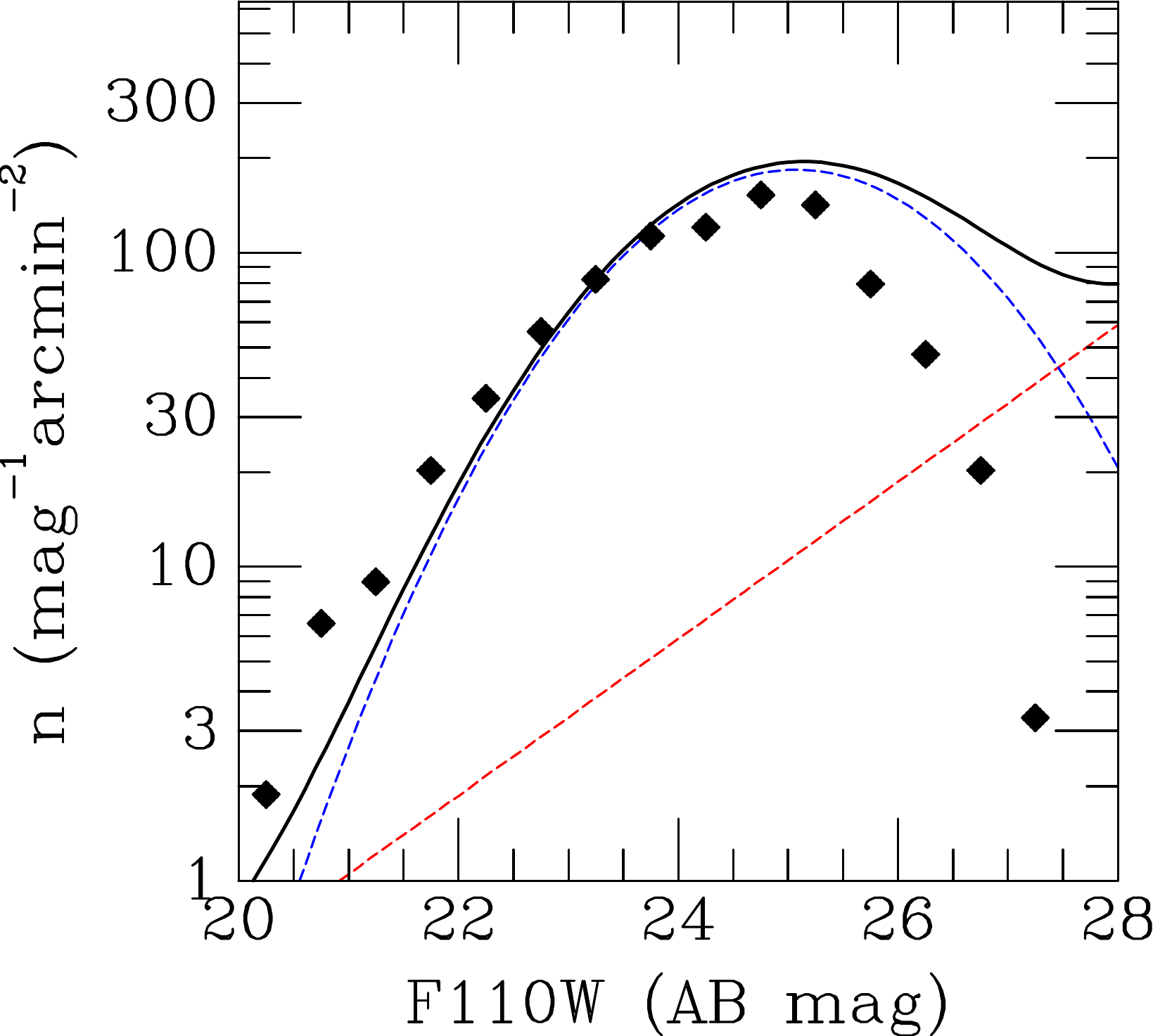}
\caption{Combined figure for NGC~6482.}
\end{center}
\end{figure*}
\clearpage

\begin{figure*}
\begin{center}
\includegraphics[scale=0.2]{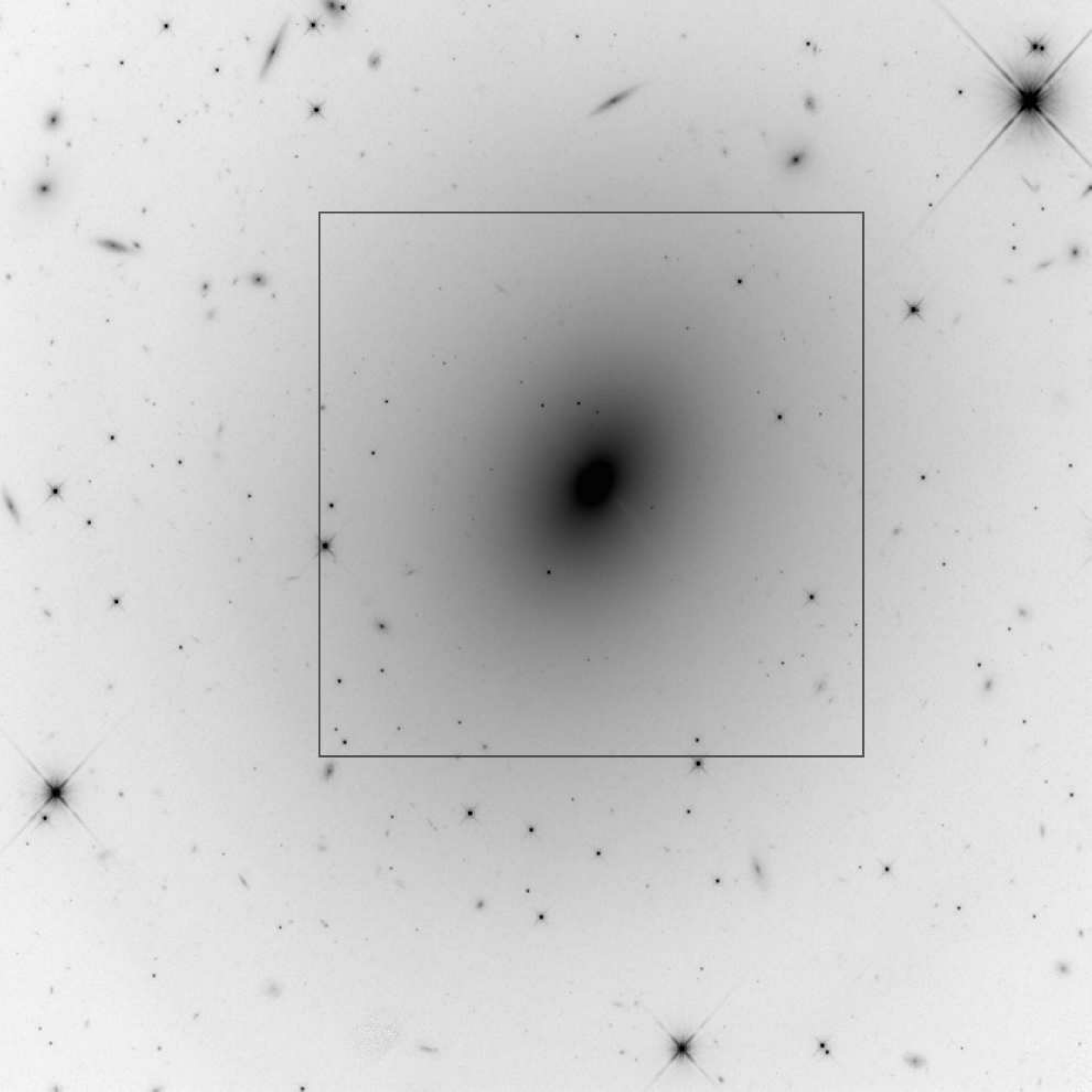}
\includegraphics[scale=0.4]{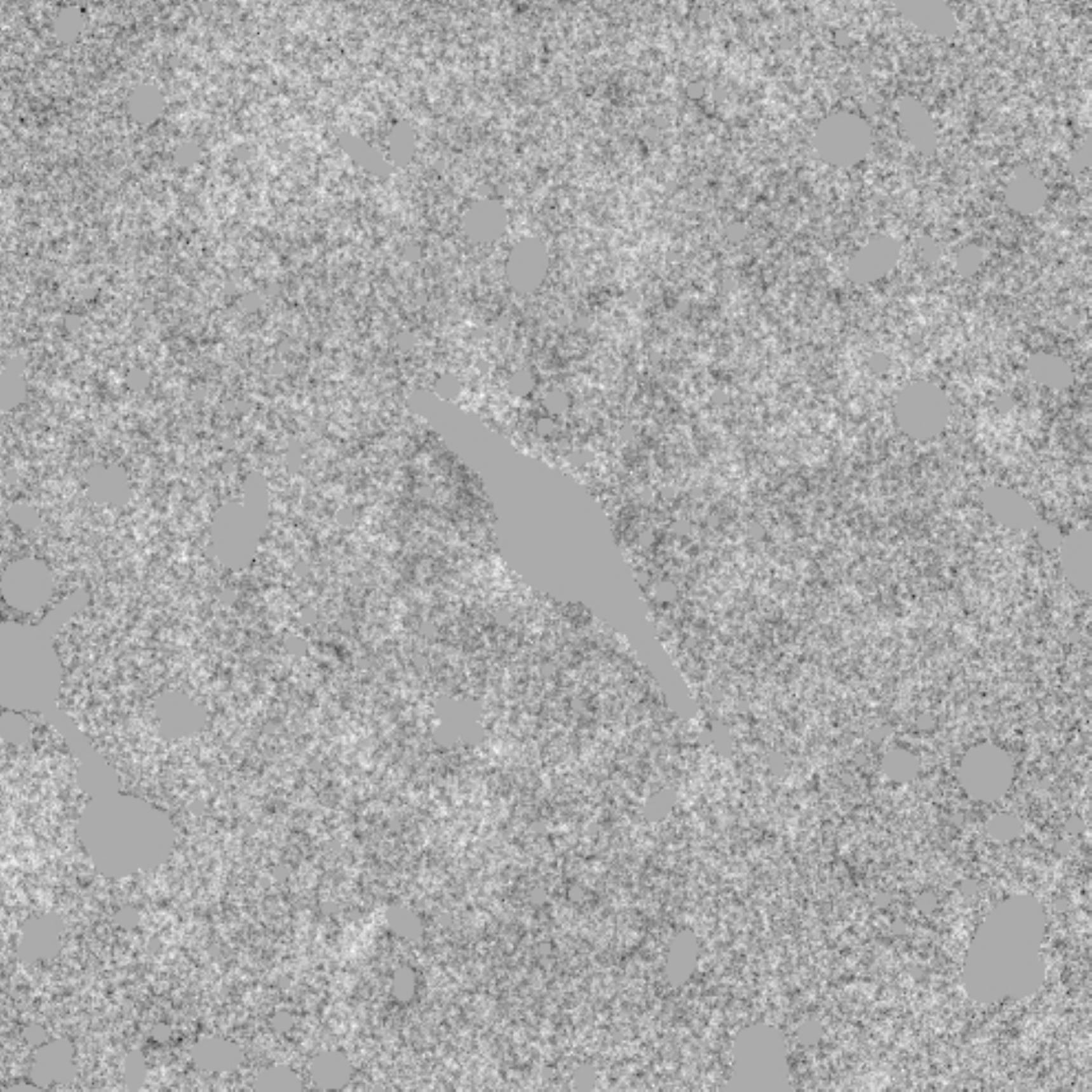} \\
\vspace{10pt}
\includegraphics[scale=0.4]{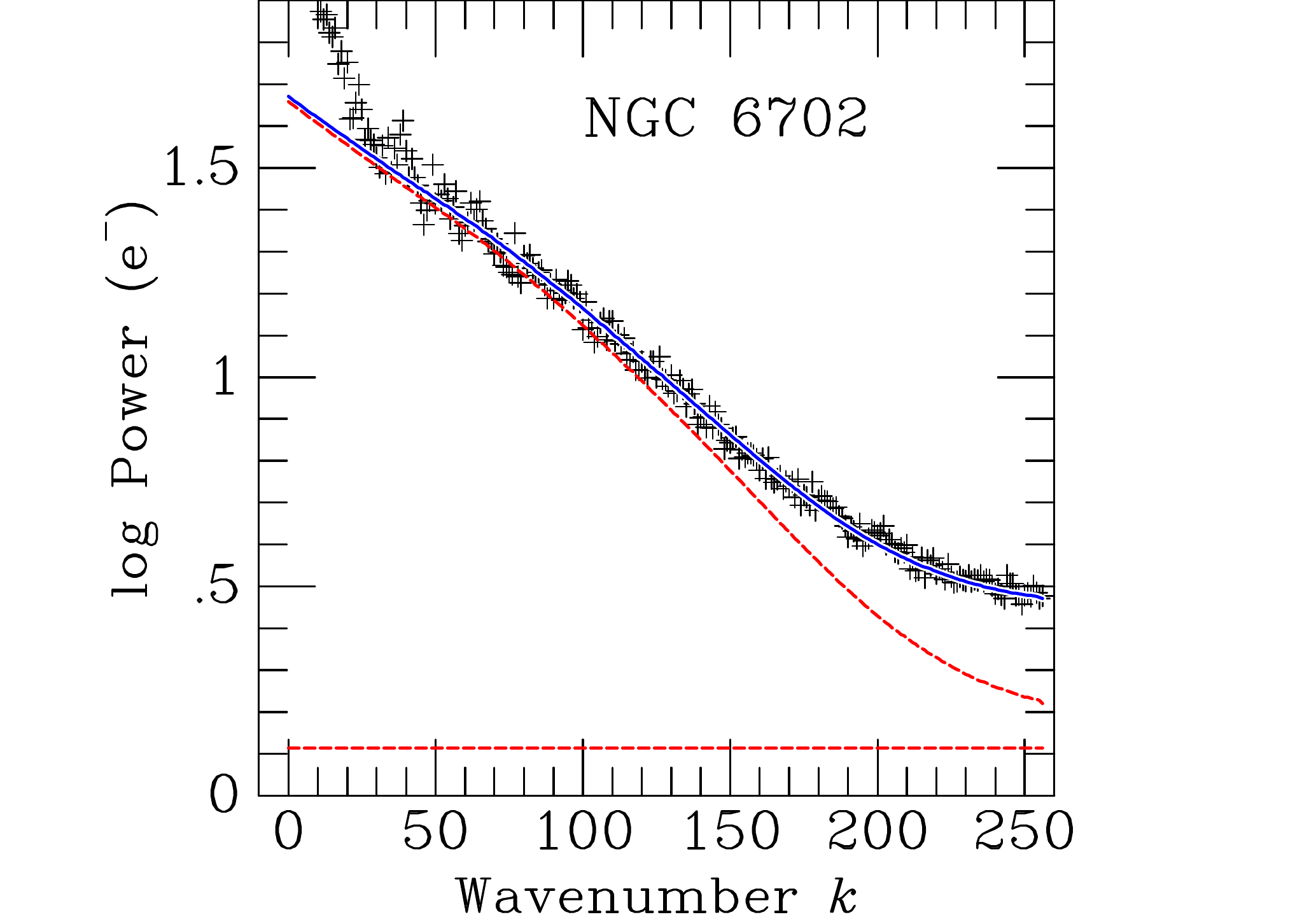}
\hspace{-25pt}
\includegraphics[scale=0.4]{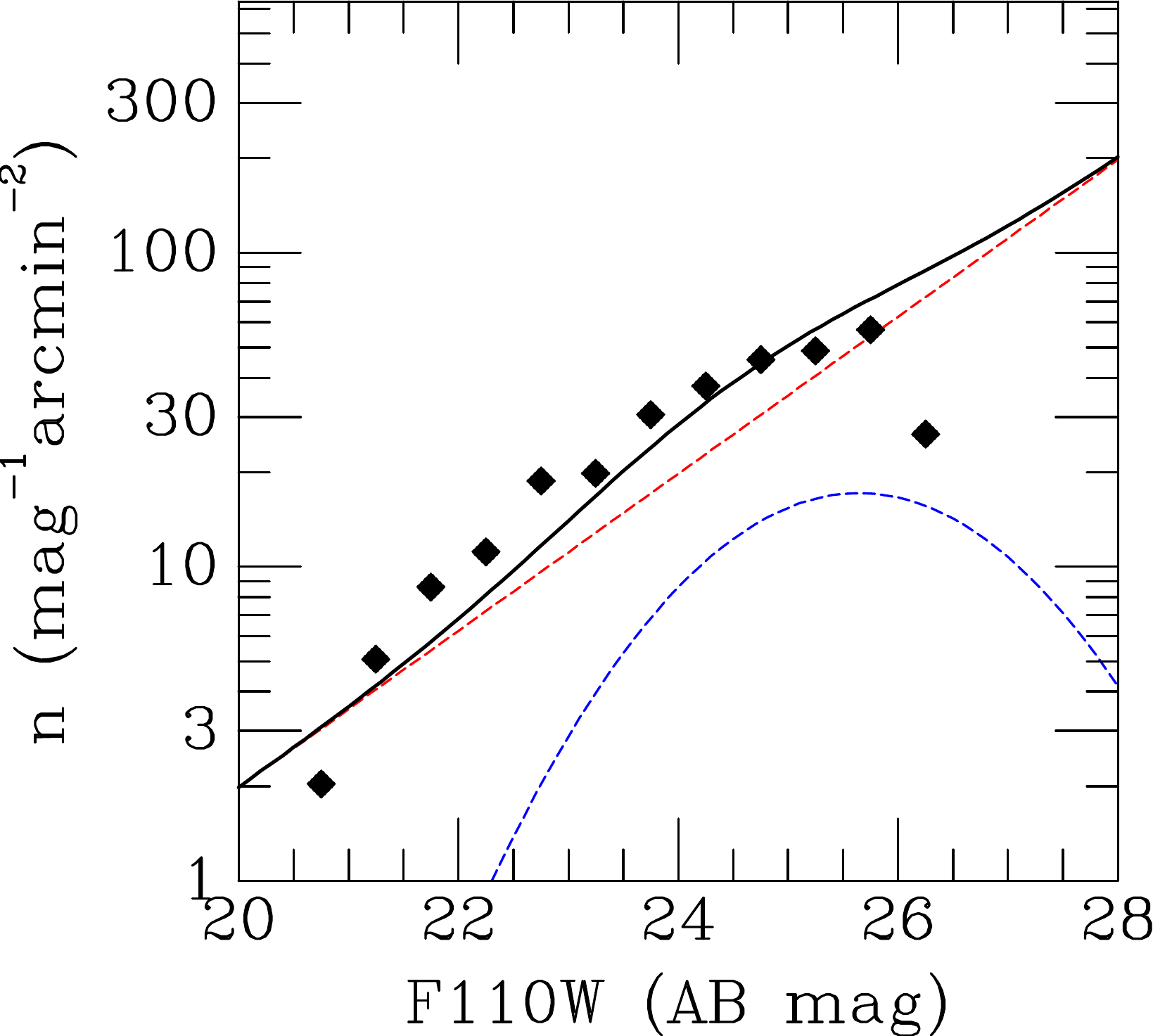}
\caption{Combined figure for NGC~6702.}
\end{center}
\end{figure*}
\clearpage

\begin{figure*}
\begin{center}
\includegraphics[scale=0.2]{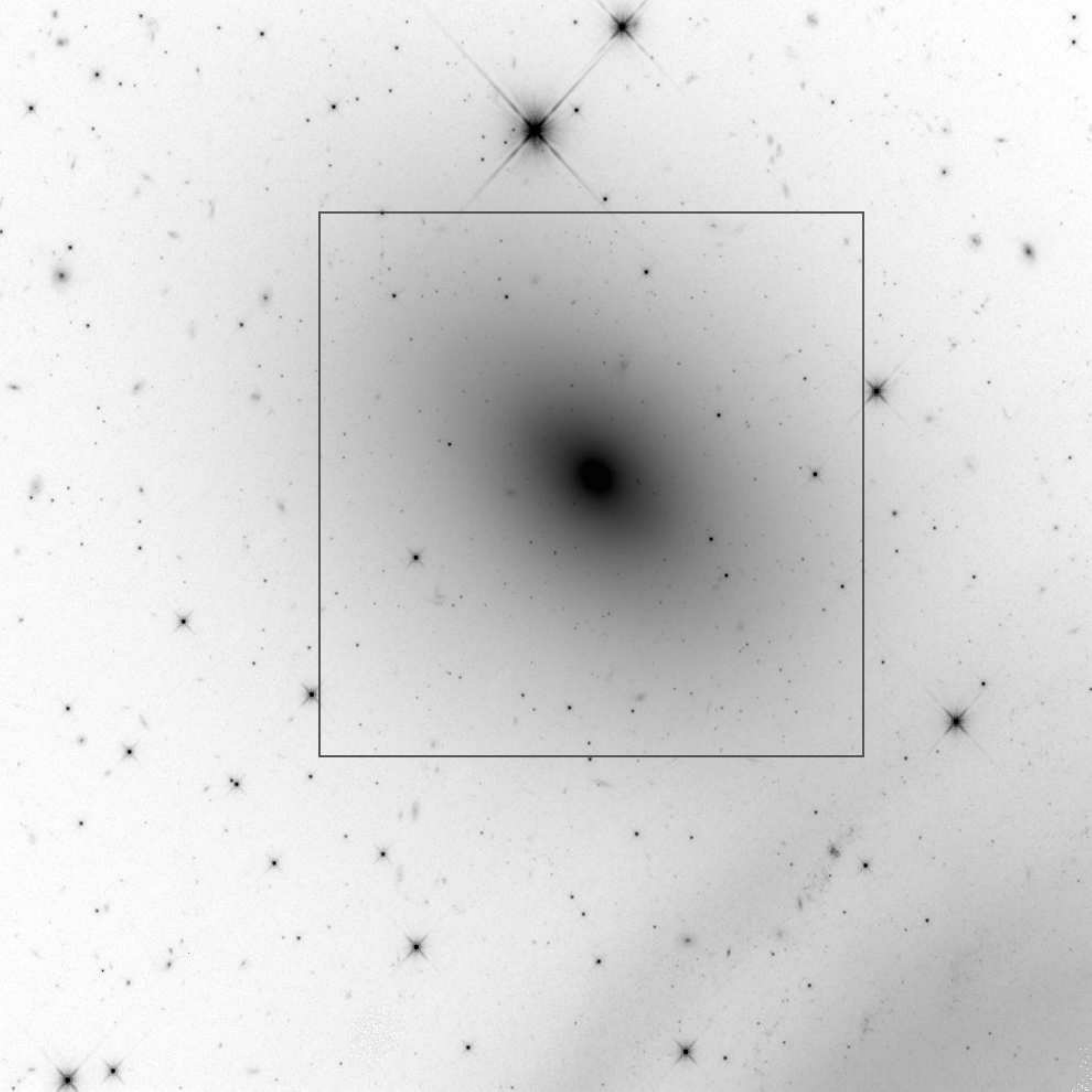}
\includegraphics[scale=0.4]{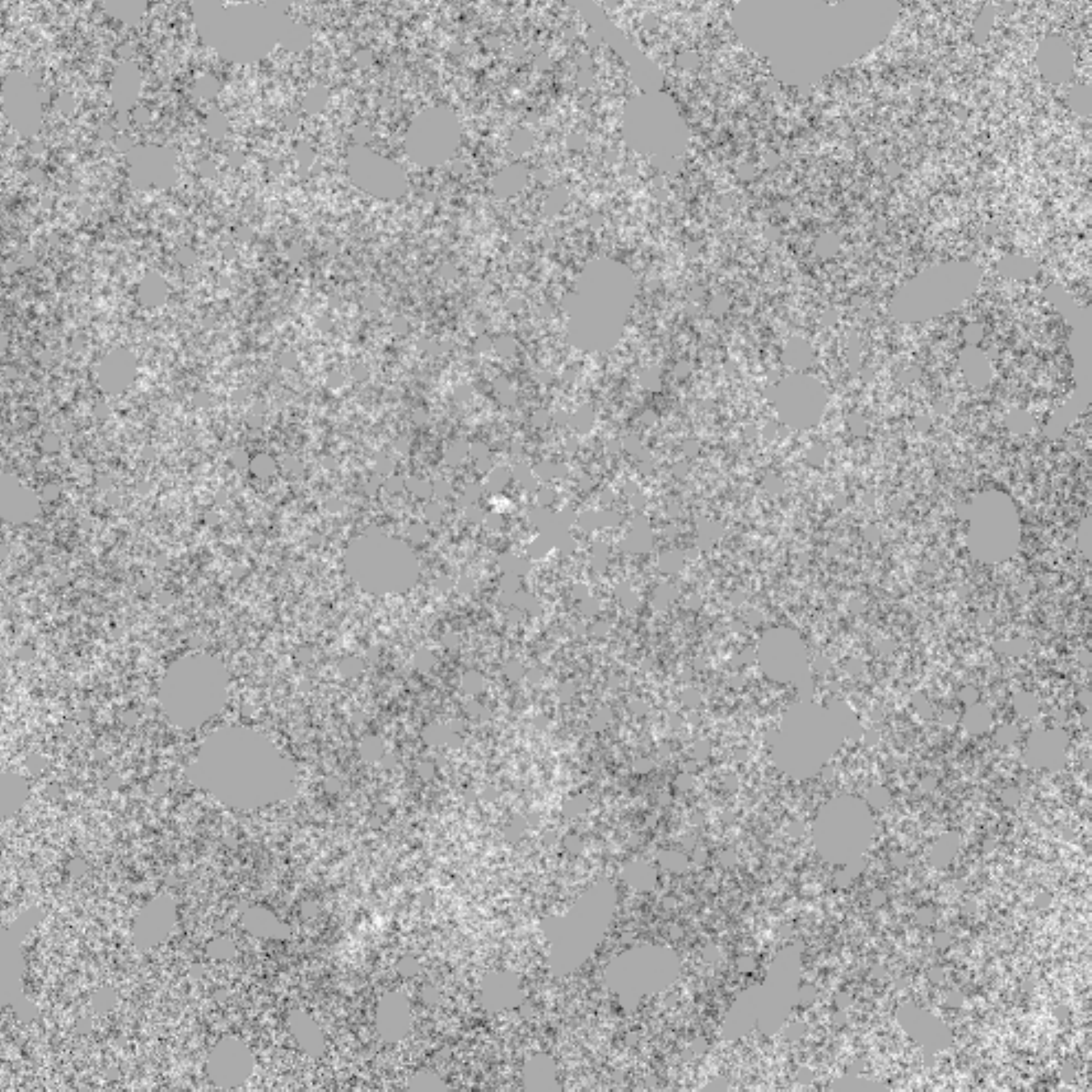} \\
\vspace{10pt}
\includegraphics[scale=0.4]{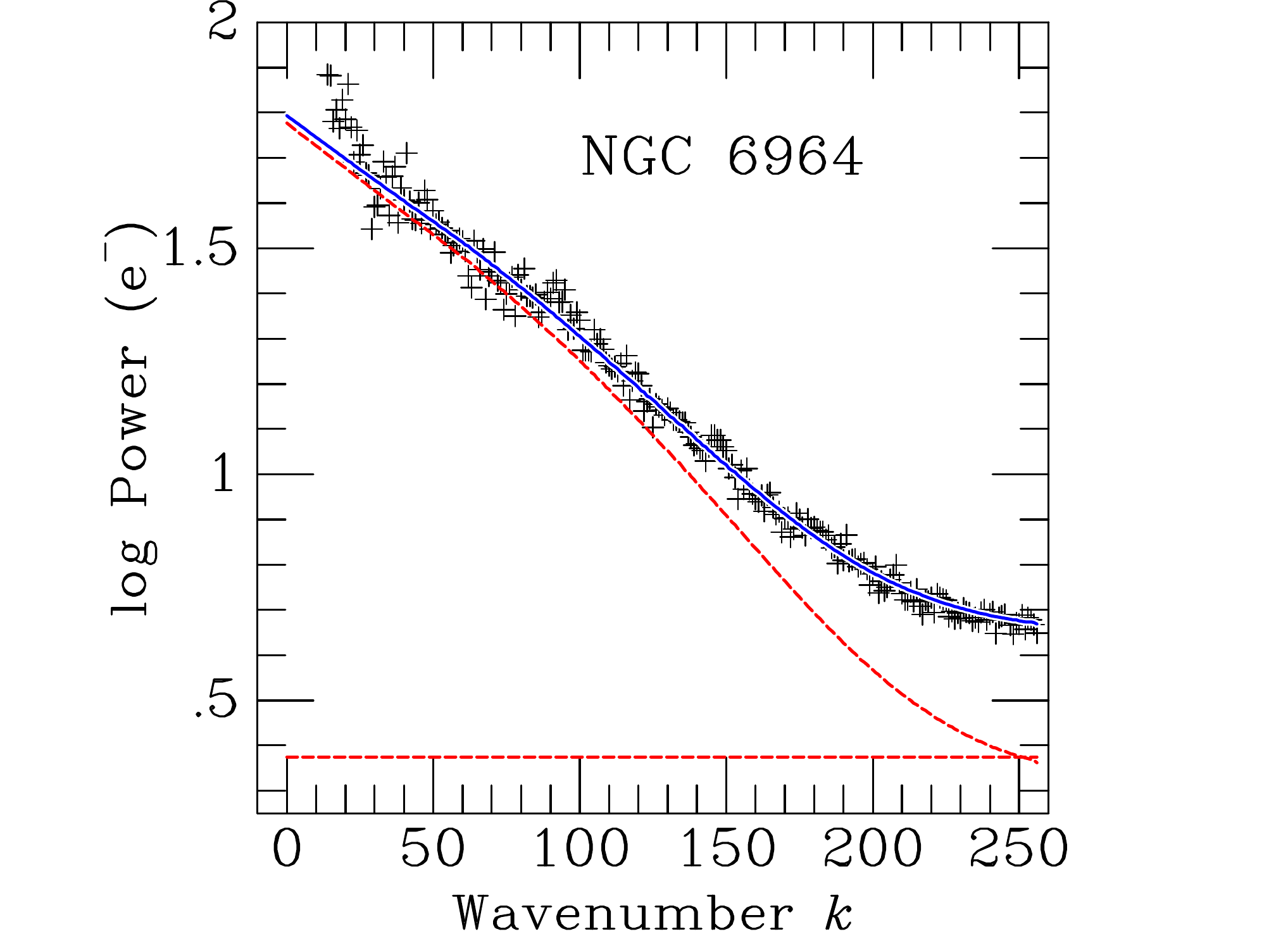}
\hspace{-25pt}
\includegraphics[scale=0.4]{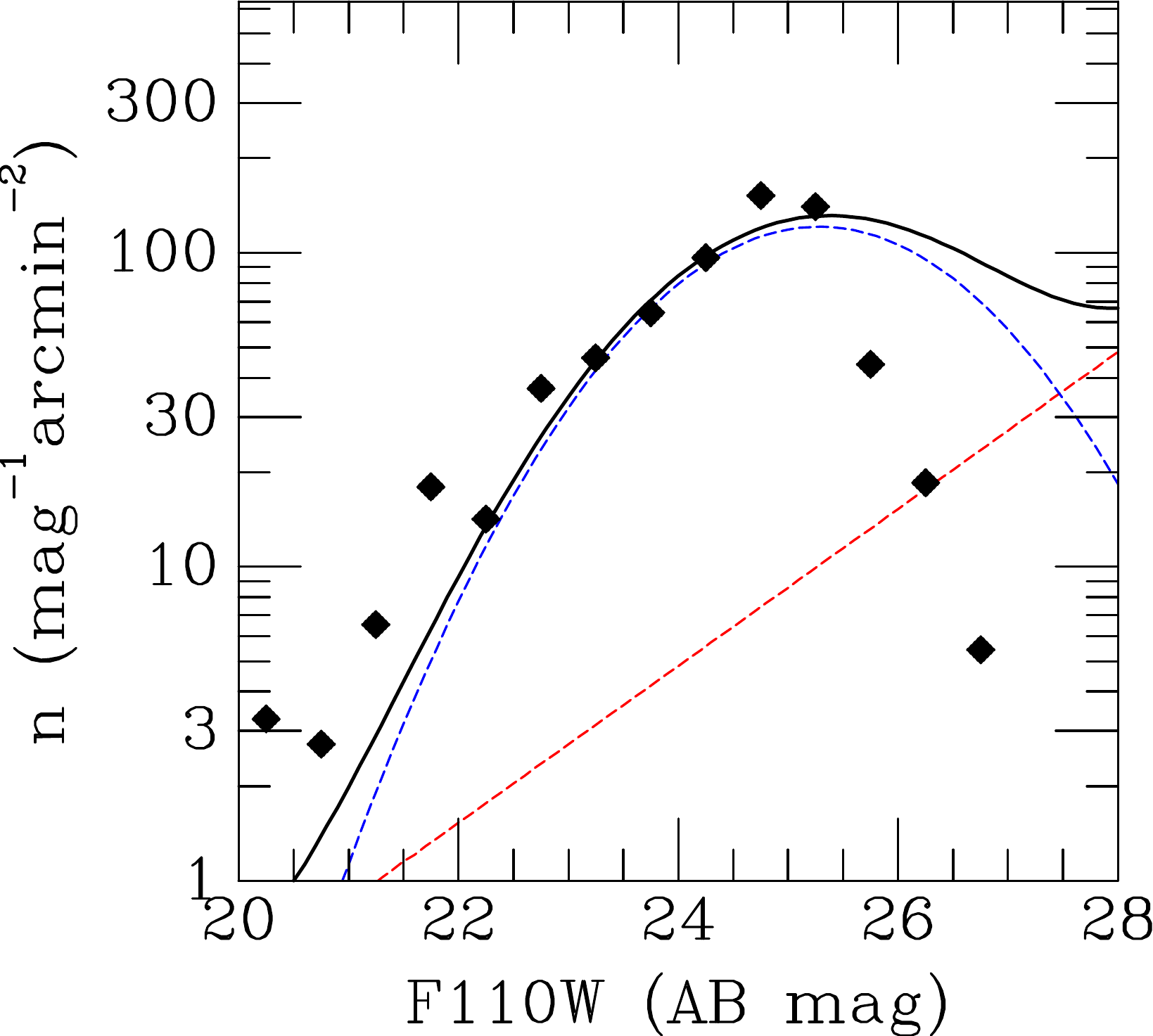}
\caption{Combined figure for NGC~6964.}
\end{center}
\end{figure*}
\clearpage

\begin{figure*}
\begin{center}
\includegraphics[scale=0.2]{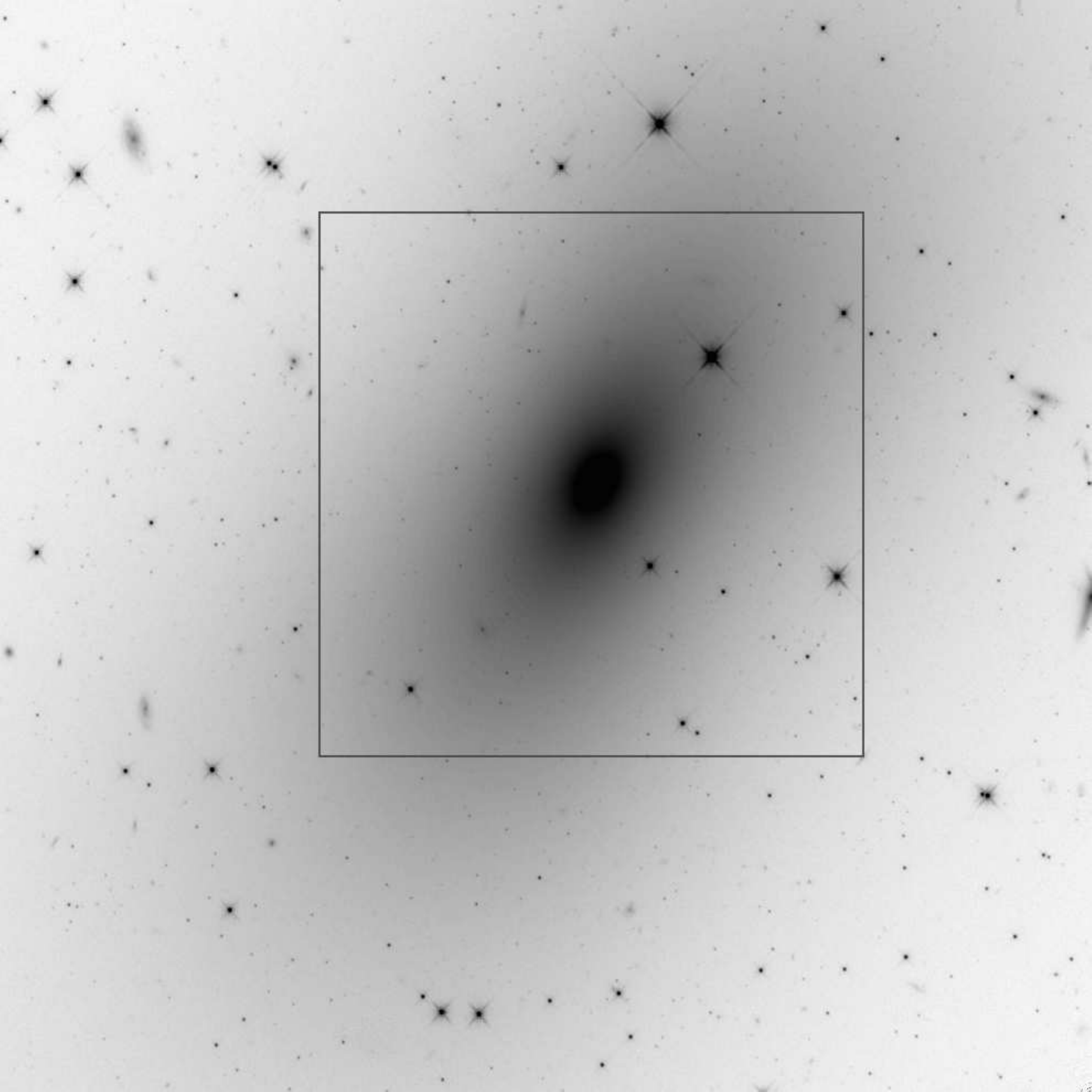}
\includegraphics[scale=0.4]{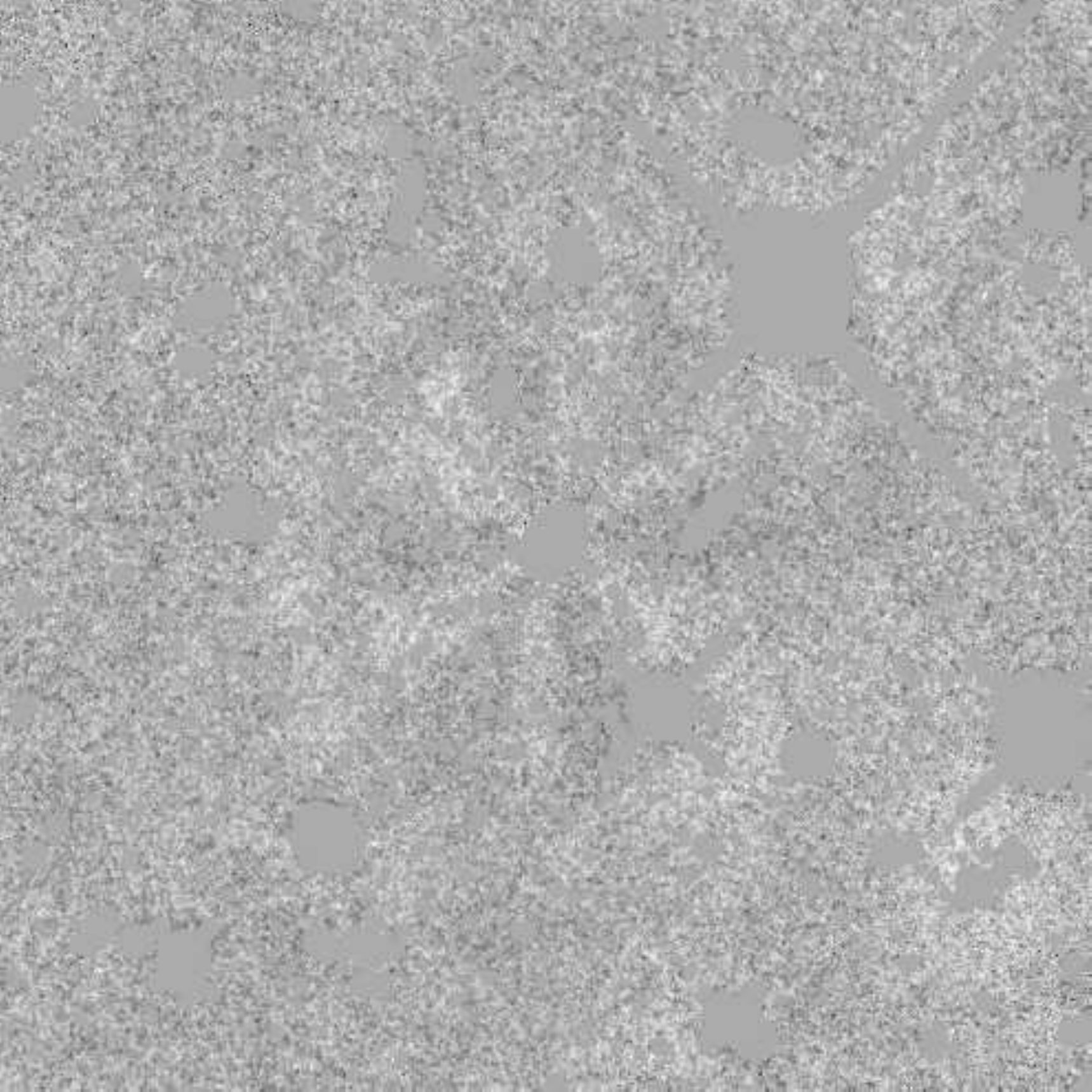} \\
\vspace{10pt}
\includegraphics[scale=0.4]{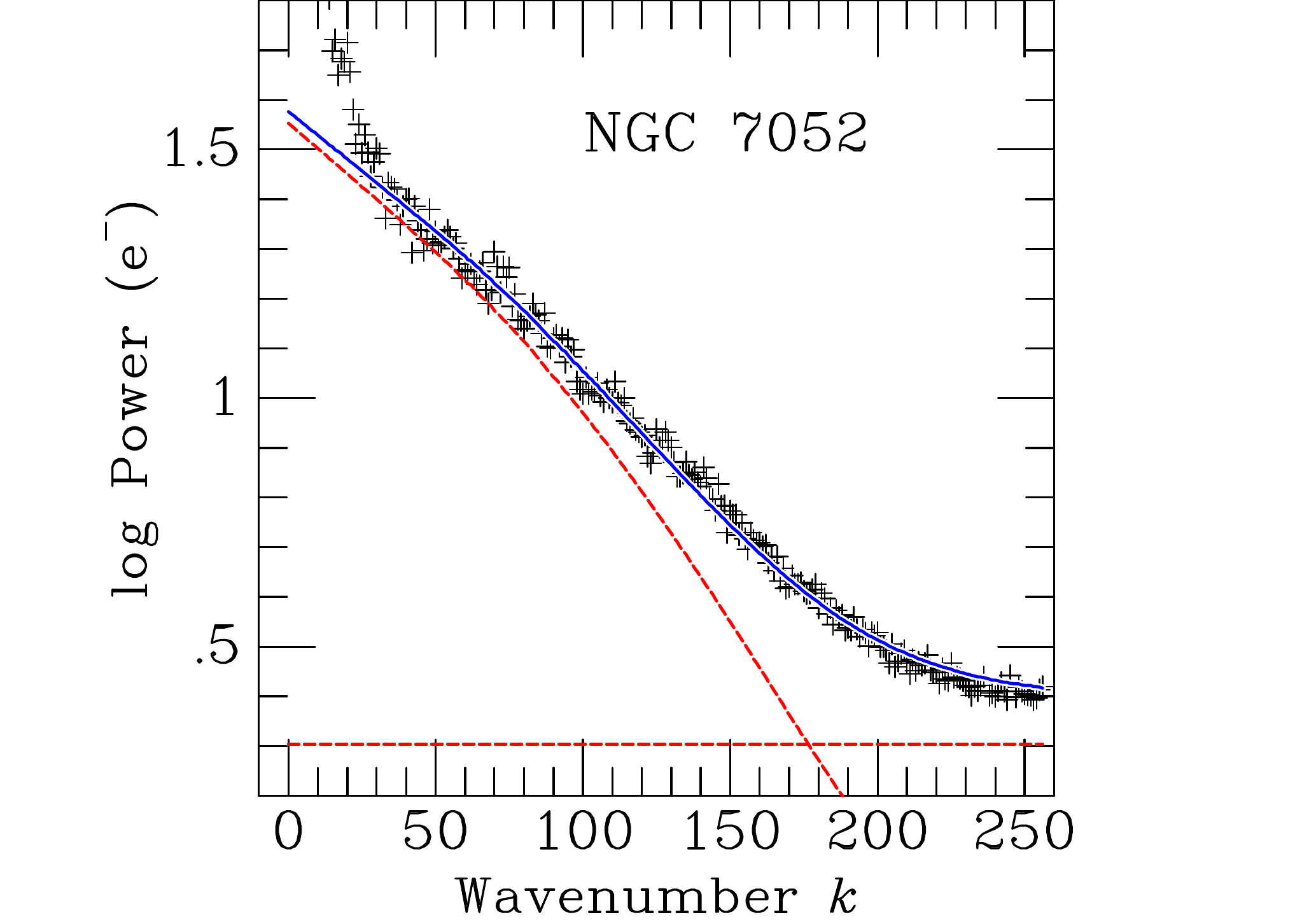}
\hspace{-25pt}
\includegraphics[scale=0.4]{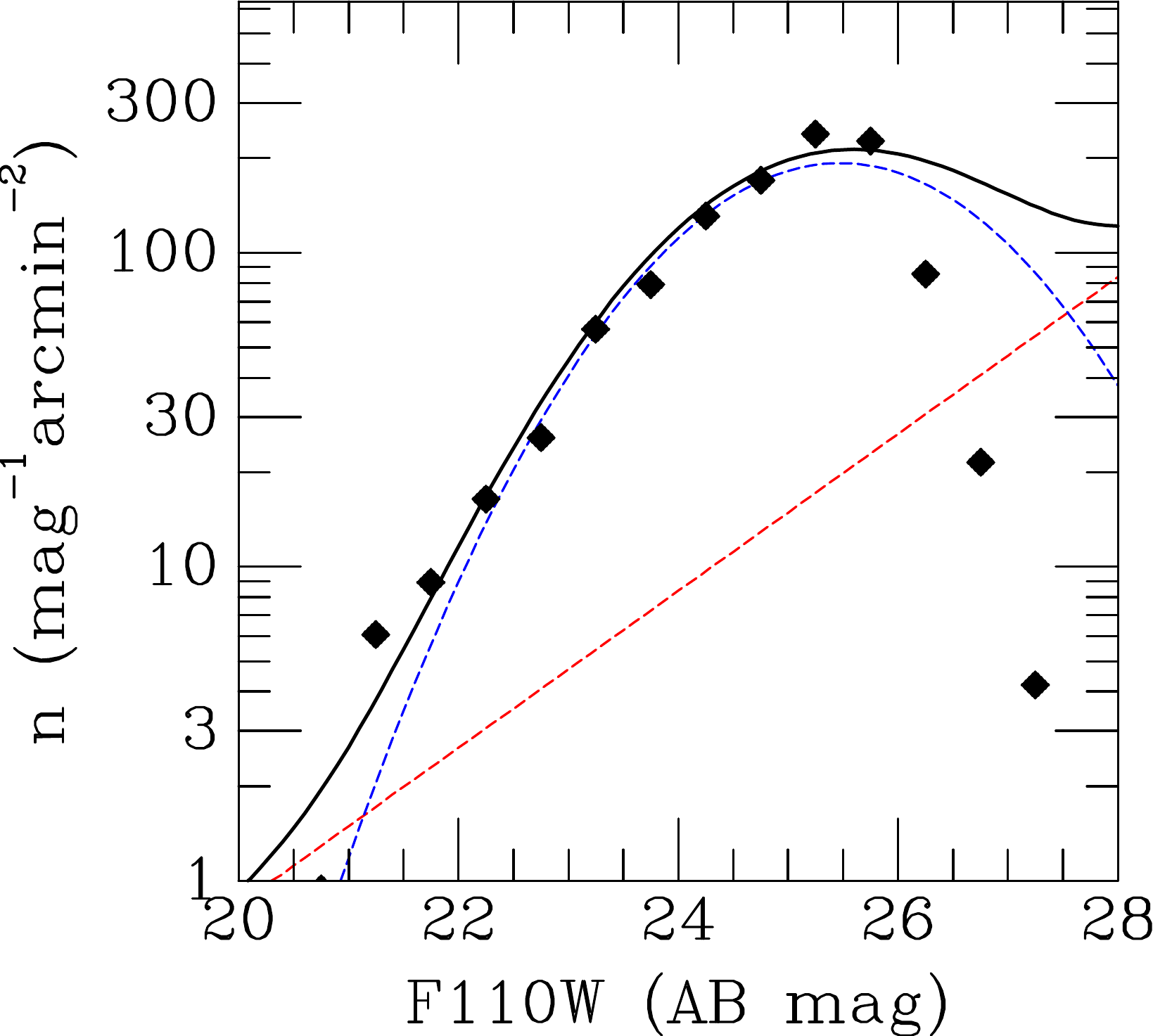}
\caption{Combined figure for NGC~7052.}
\end{center}
\end{figure*}
\clearpage

\begin{figure*}
\begin{center}
\includegraphics[scale=0.2]{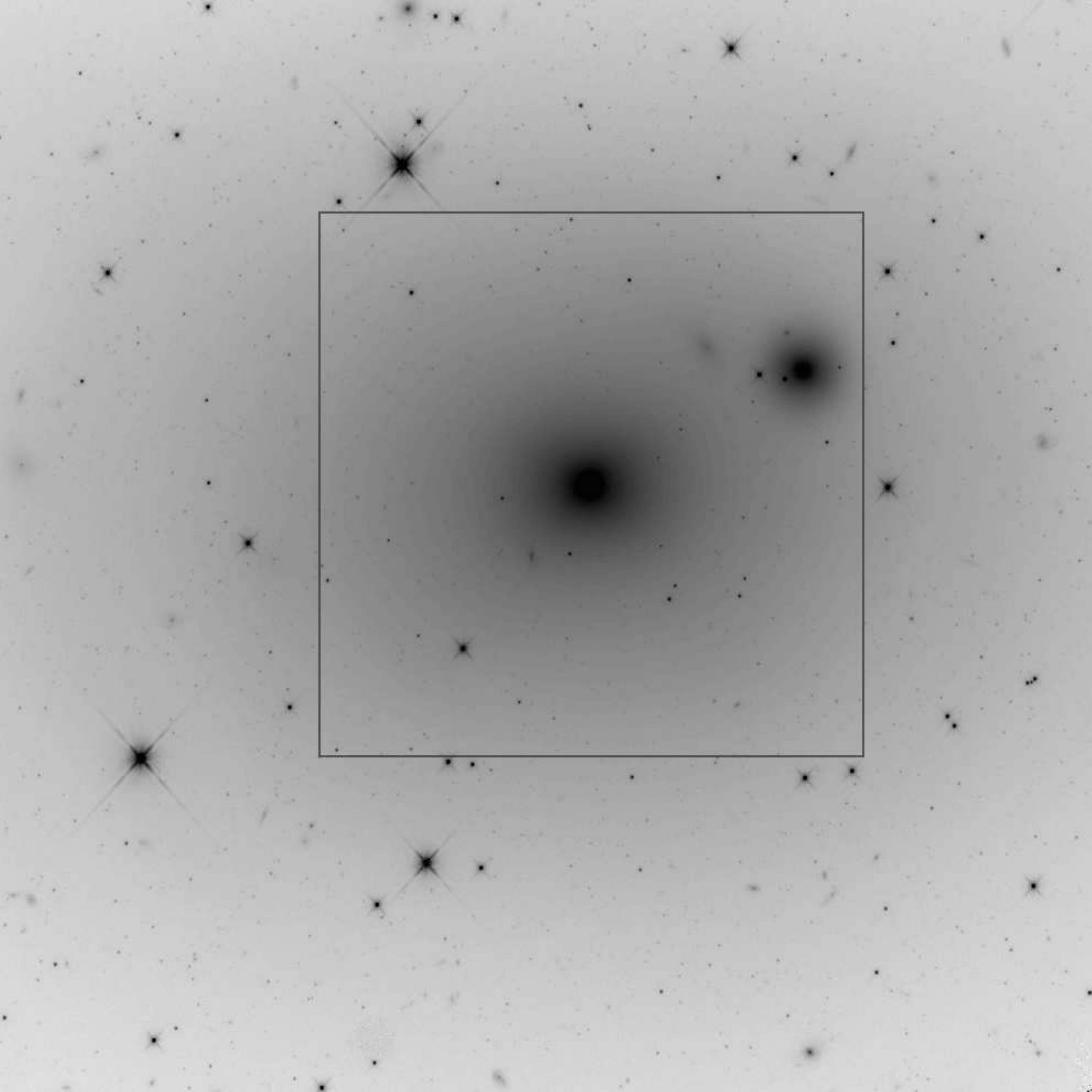}
\includegraphics[scale=0.4]{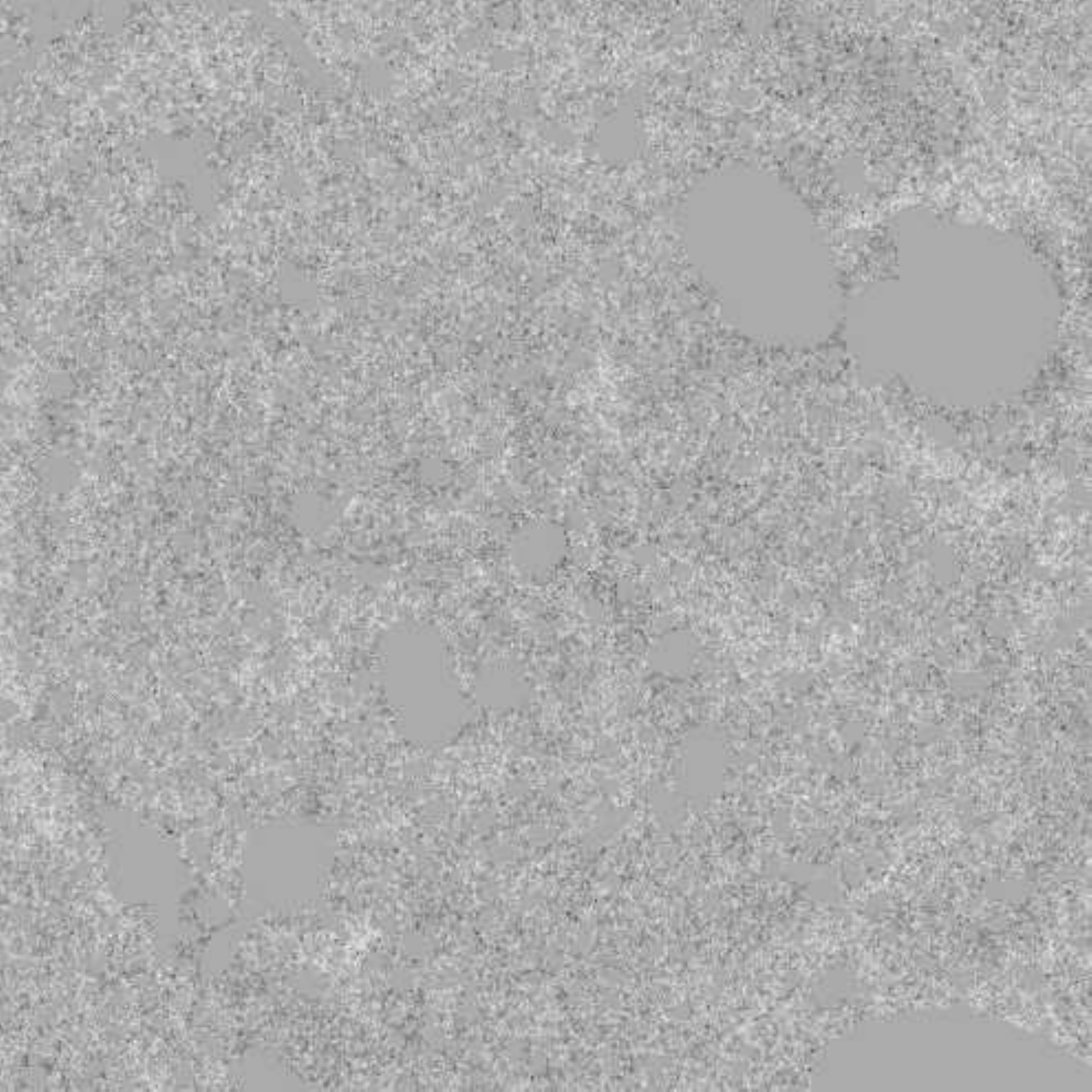} \\
\vspace{10pt}
\includegraphics[scale=0.4]{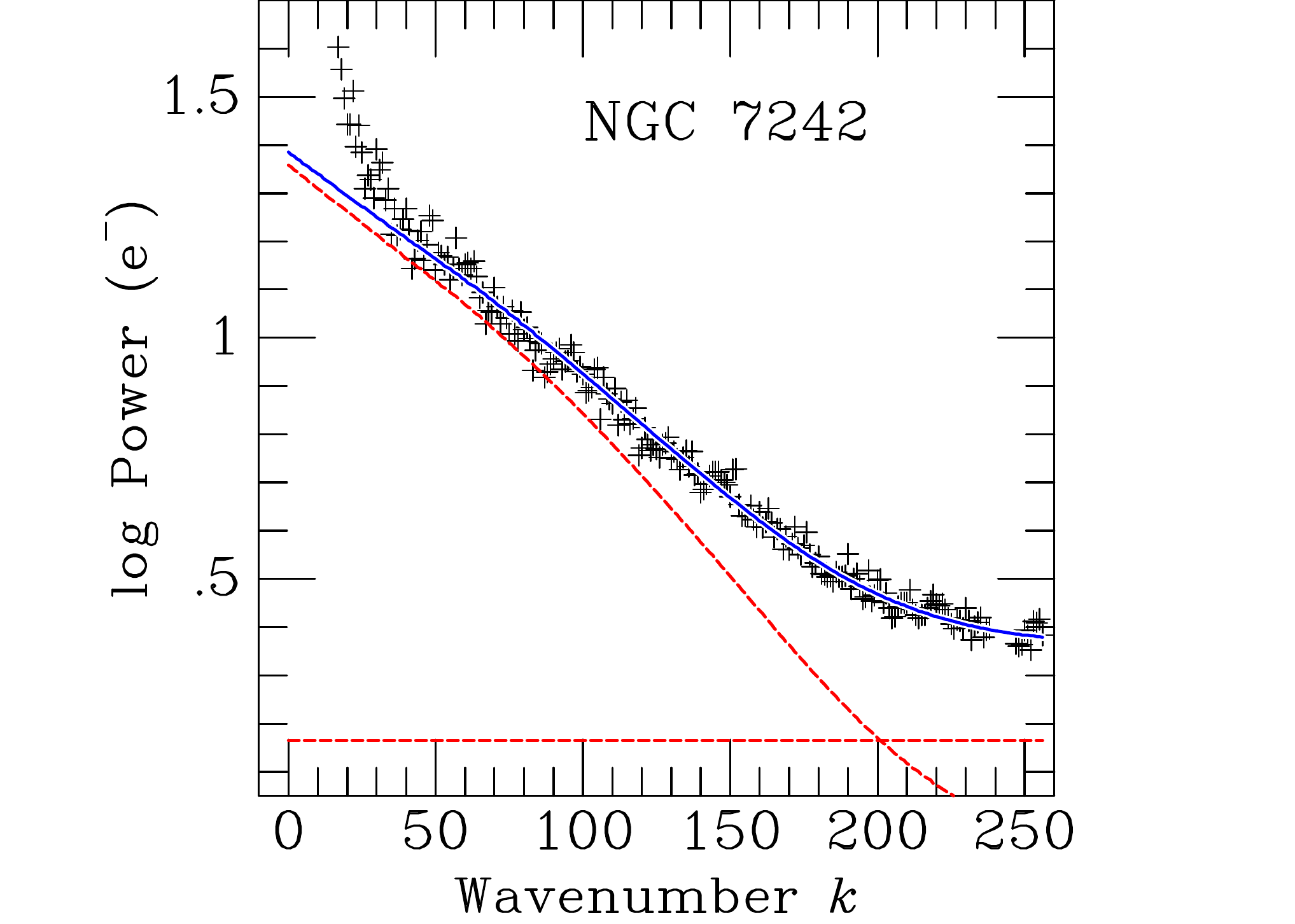}
\hspace{-25pt}
\includegraphics[scale=0.4]{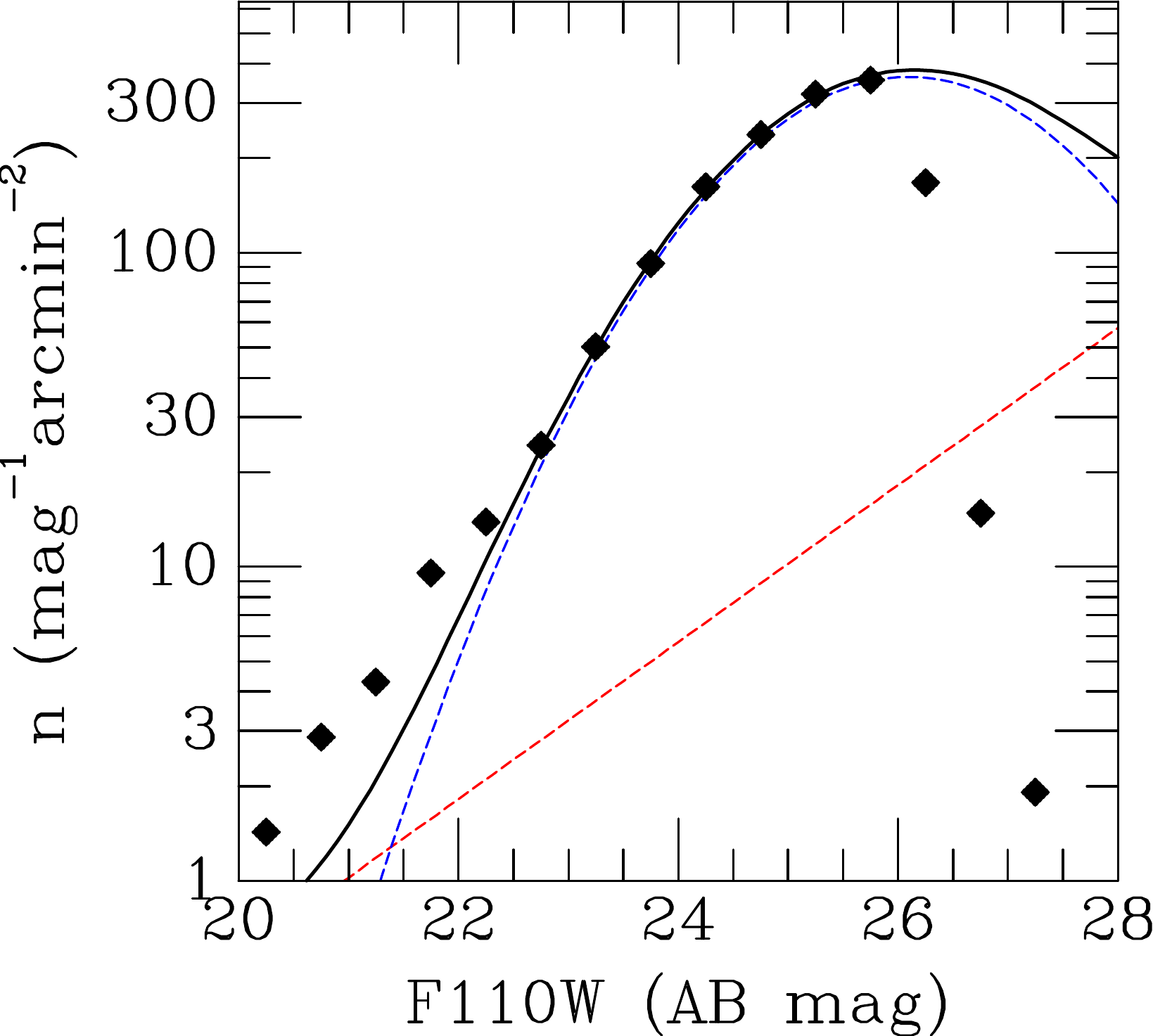}
\caption{Combined figure for NGC~7242.}
\end{center}
\end{figure*}
\clearpage

\begin{figure*}
\begin{center}
\includegraphics[scale=0.2]{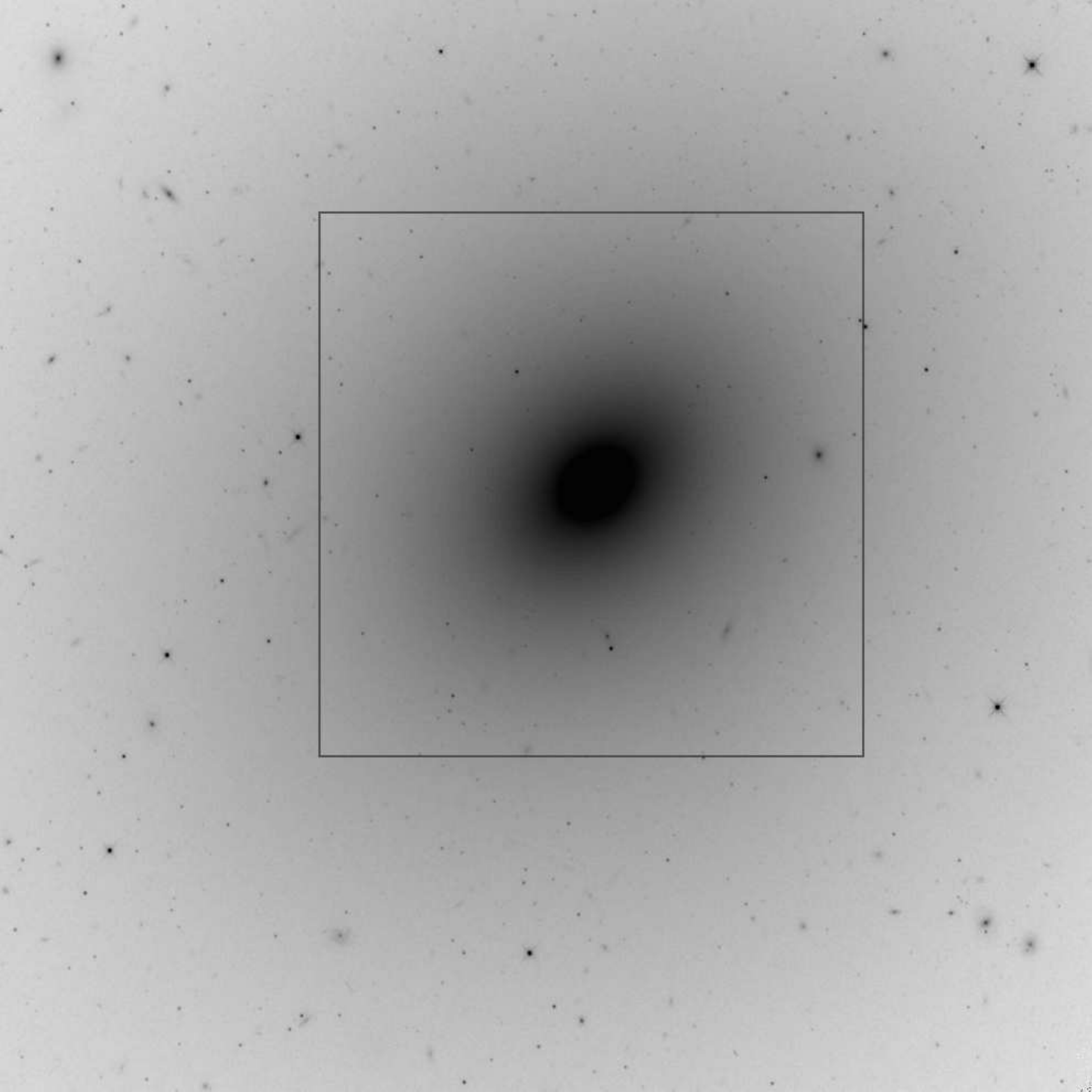}
\includegraphics[scale=0.4]{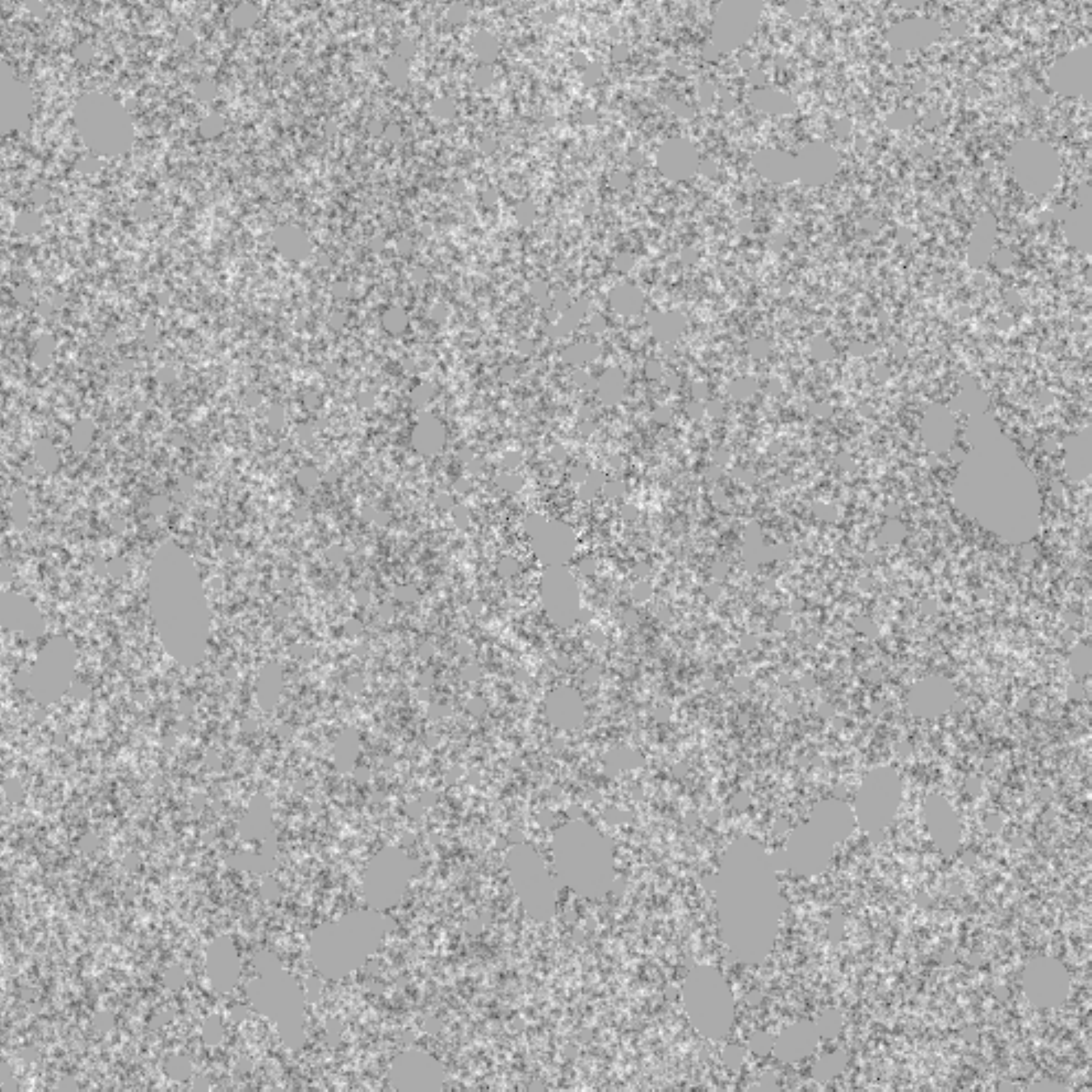} \\
\vspace{10pt}
\includegraphics[scale=0.4]{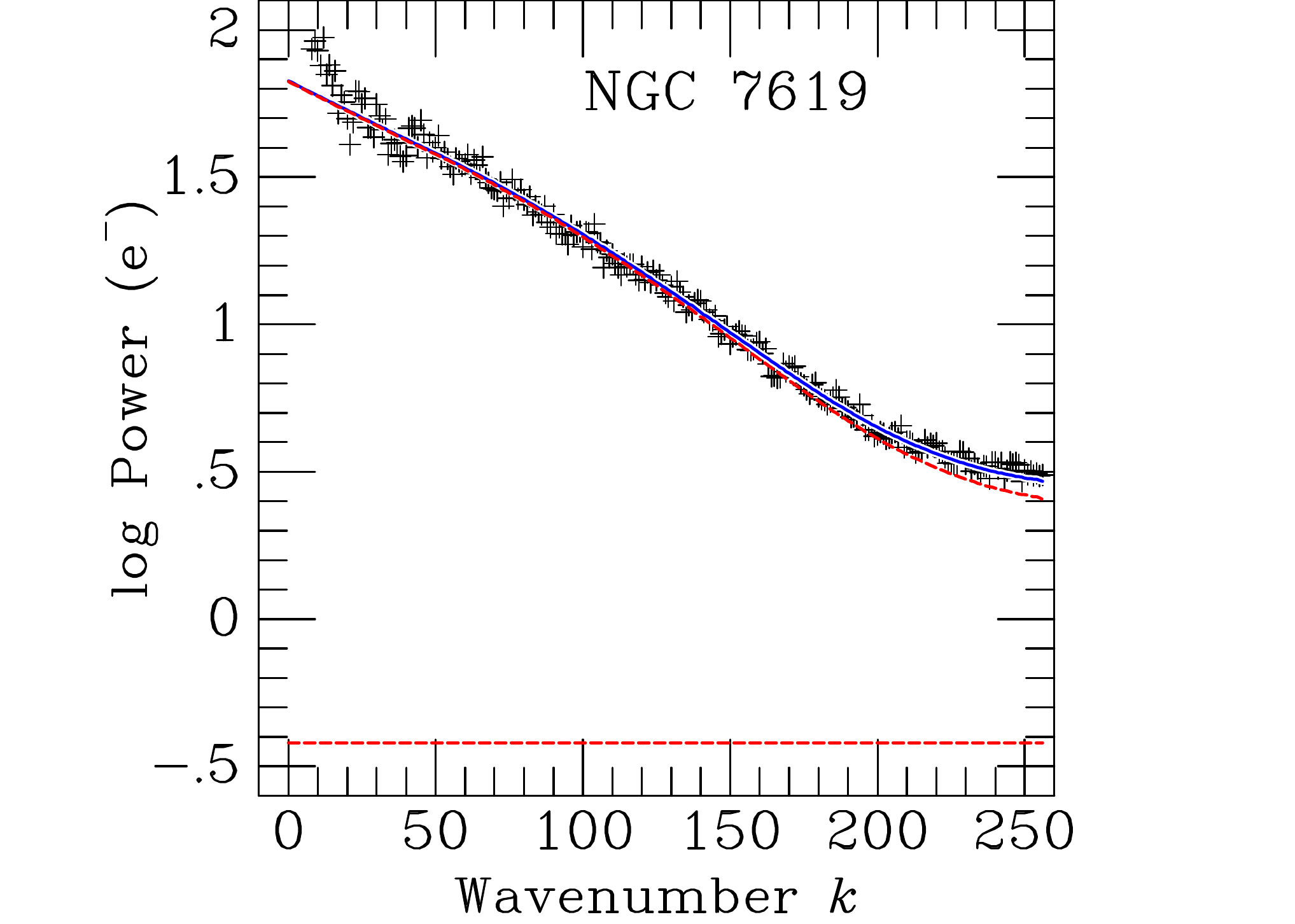}
\hspace{-25pt}
\includegraphics[scale=0.4]{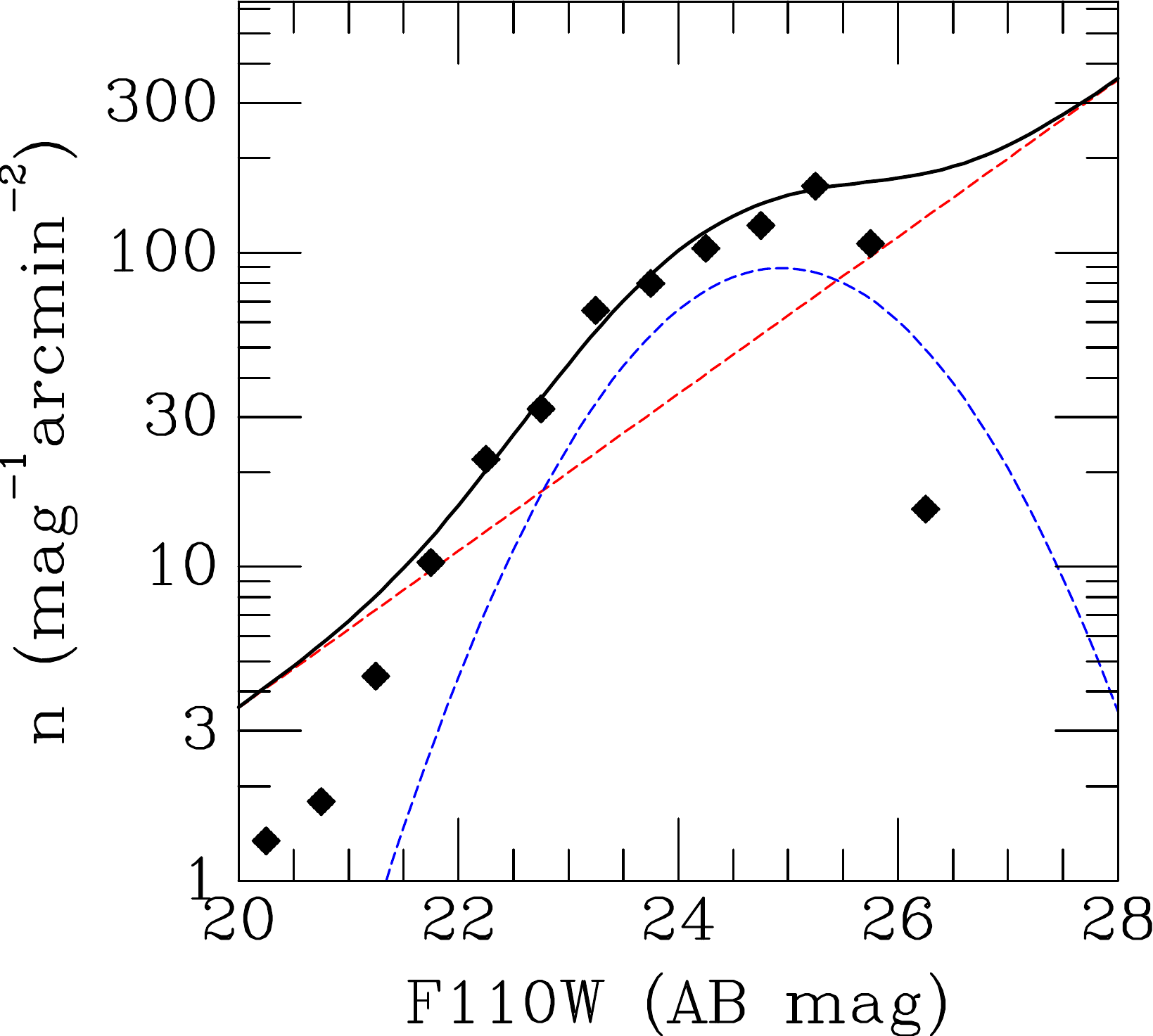}
\caption{Combined figure for NGC~7619.}
\end{center}
\end{figure*}
\clearpage

\begin{figure*}
\begin{center}
\includegraphics[scale=0.2]{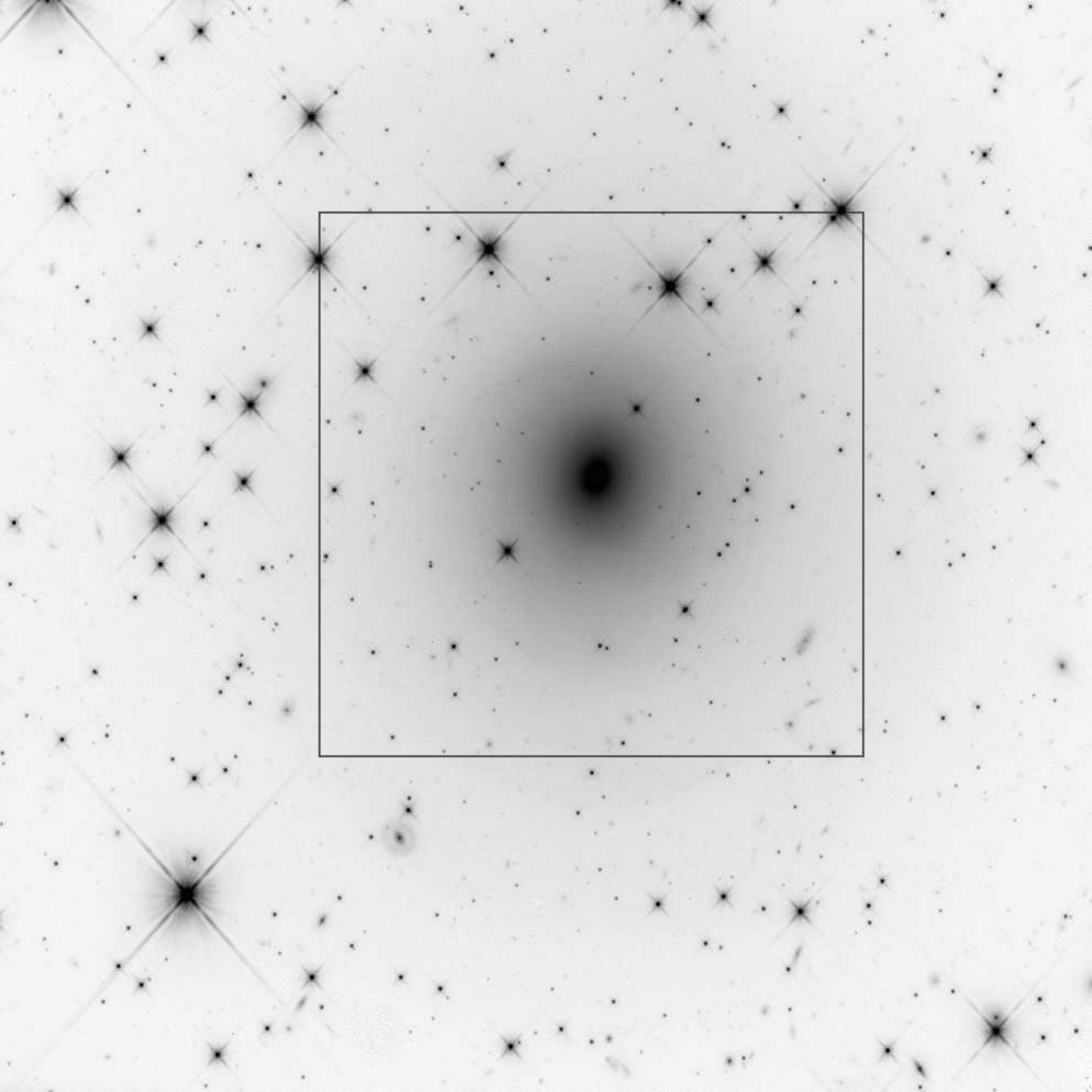}
\includegraphics[scale=0.4]{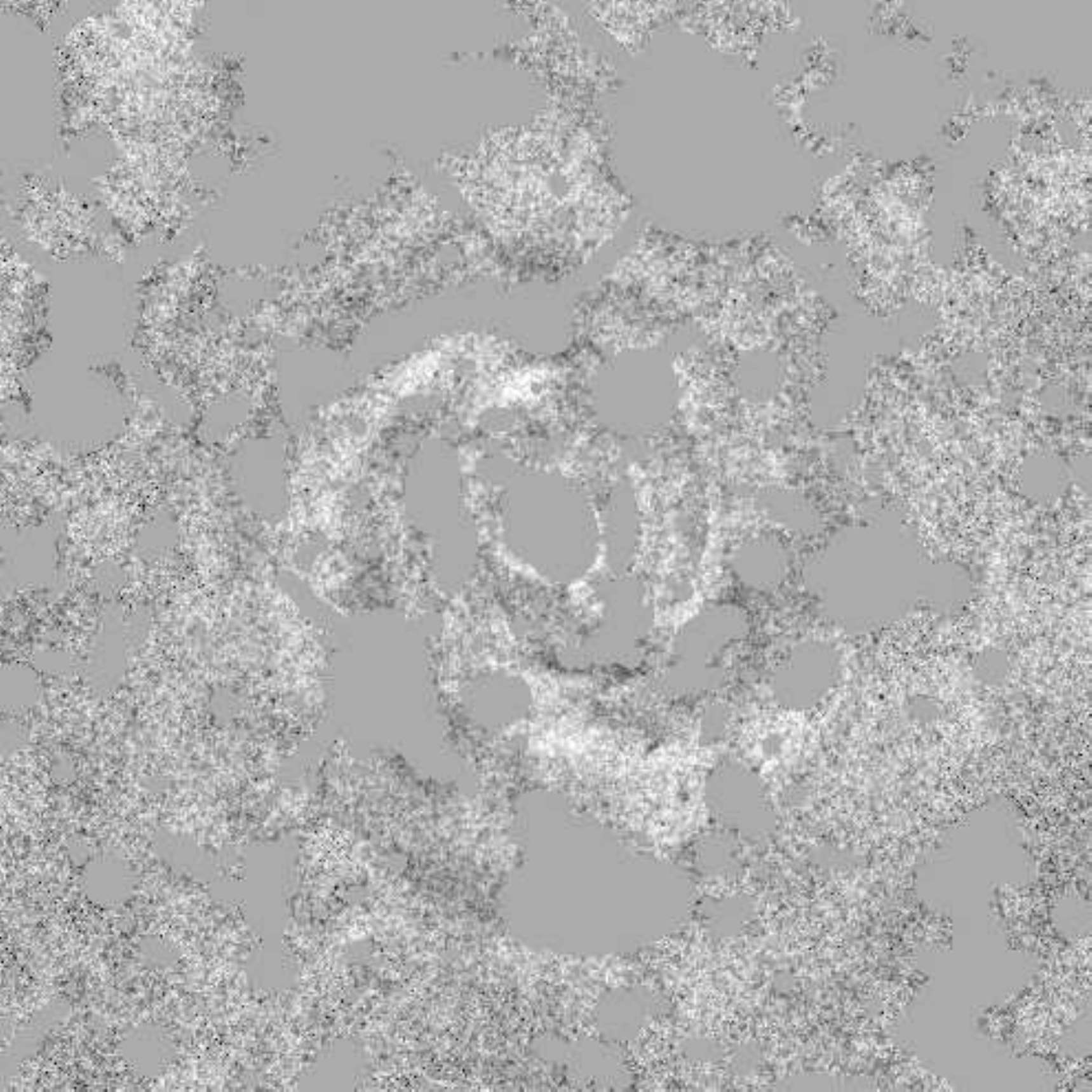} \\
\vspace{10pt}
\includegraphics[scale=0.4]{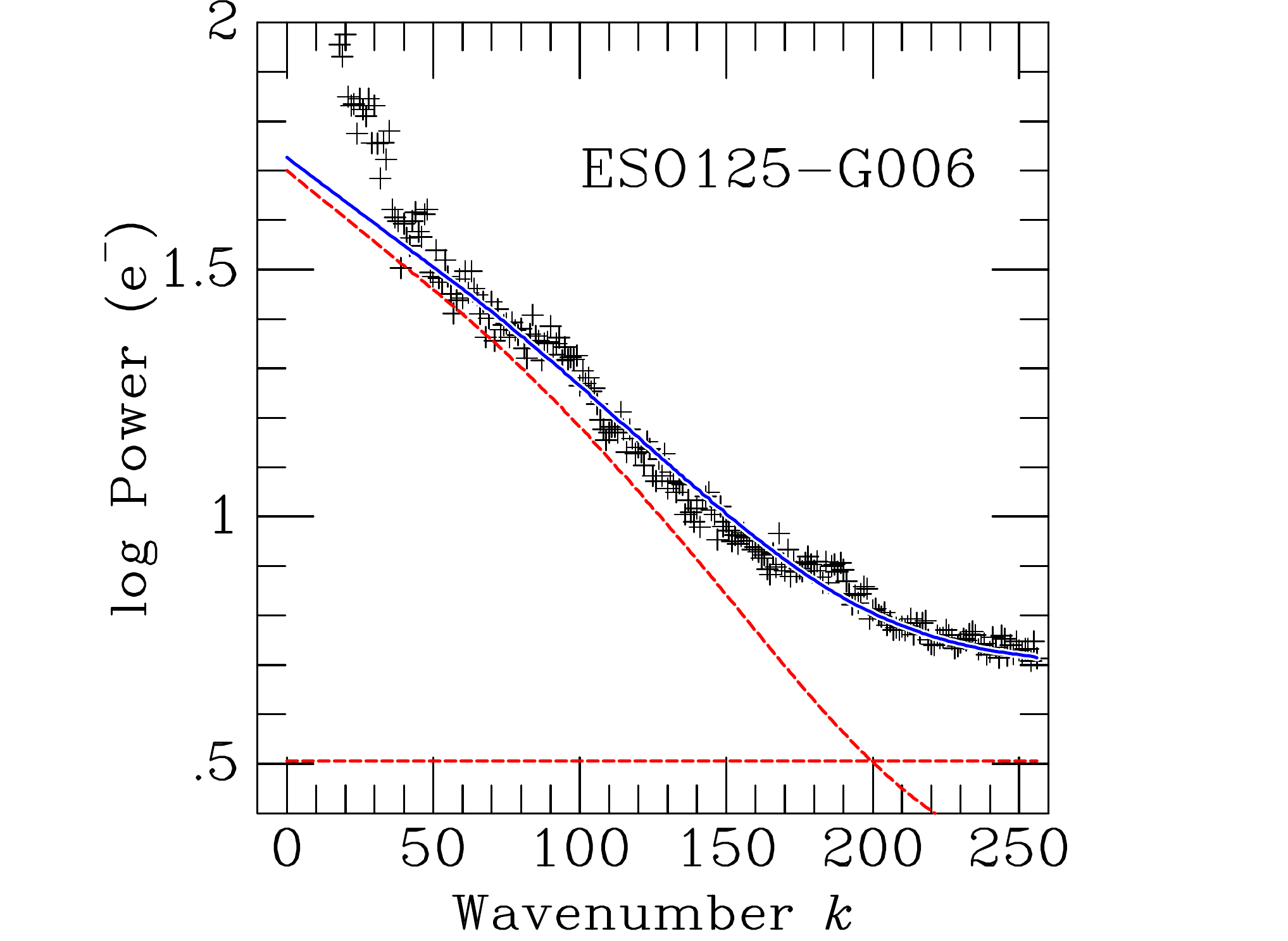}
\hspace{-25pt}
\includegraphics[scale=0.4]{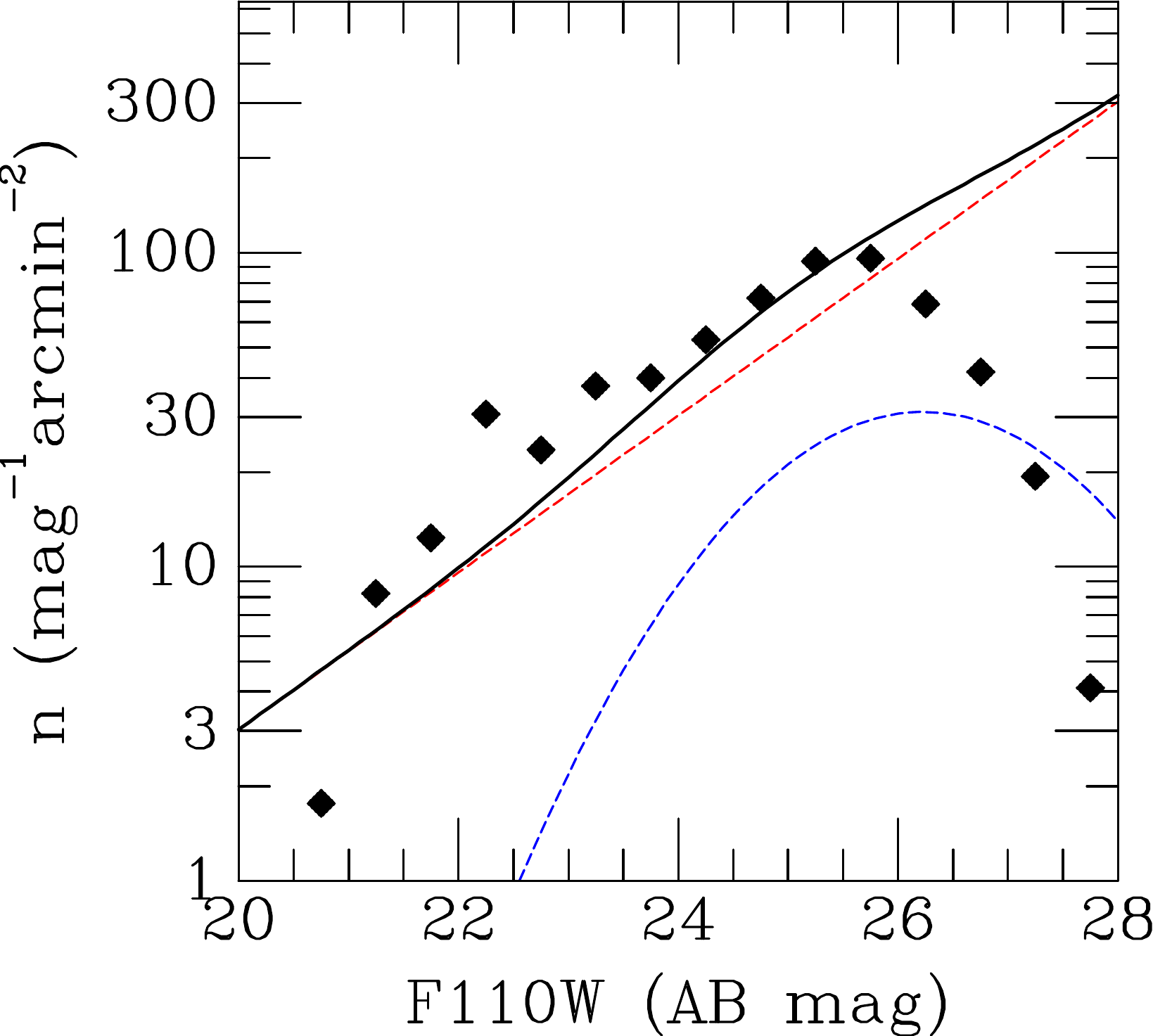}
\caption{Combined figure for ESO125-G006.}
\end{center}
\end{figure*}

\end{document}